%% file: sugra.tex
&latex209
\documentstyle[epsfig,axodraw]{aipproc2}

\catcode`@=11          
\ifx\every@math@size\undefined\else              
  \let\old@expast\@expast
  \def\@expast#1{\old@expast{#1}\let\@tempa\reserved@a}
\fi
\catcode`@=12
\catcode`@=11\def\tableofcontents{\section*{\contentsname
\@mkboth{\uppercase{\contentsname}}{\uppercase{\contentsname}}}%
\@starttoc{toc}}
\catcode`@=12

\begin{document}

\newcommand{\lsim}{\mathrel{\raisebox{-.6ex}{$\stackrel{\textstyle<}{\sim}$}}}
\newcommand{\gsim}{\mathrel{\raisebox{-.6ex}{$\stackrel{\textstyle>}{\sim}$}}}
\let\alt=\lsim
\let\agt=\gsim
\def\lc{\lowercase}

\input{sugra.titlepage}

\clearpage
\pagestyle{plain}
\pagenumbering{roman}
\tableofcontents

\clearpage
\pagenumbering{arabic}

\input{Nath/nath}

\clearpage
\input{Wagner/proceed}

\clearpage
\input{Kao/conventions}

\clearpage
\input{Kao/rges}

%%%%%%%%%%%%%%%%%%%%%%%%%%%%%%%%%%%

%% ii)

\clearpage
\input{Zhang/wgnew}

\clearpage

\input{Feng/feng-new}  %

\clearpage
\input{Pierce/susy-corrs}

\clearpage
\input{Spira/paper}

\clearpage
\input{E.Berger/fnalrep} %% new version 9/21/99 

\clearpage
\input{BaerHan/runii}

%%%%%%%%%%%%%%%%%%%%%%%%%%%%%%%%%%%%%%

%% iii)

\clearpage

\input{Falk/lepbnds}

\clearpage
\input{Kamon-cdf-do/1e2}

\clearpage

\input{Szalap/final}

\clearpage
\input{Falk-cosmo/cosmonew}

\clearpage
\input{Arnowitt-cdm/cdm}

\clearpage

\input{Abel/run-II}  % charge and color breaking constraints

\clearpage
\input{Bhrlik/bsg}

\clearpage
\input{Blazek/g_2wrkshp}

\clearpage
\input{Bhrlik/cp-new}

\clearpage
\input{Cheng/r2report}

\clearpage

\input{Keung/cp-top}

%%%%%%%%%%%%%%%%%%%%%%%%%%%%%%%%%%%%%%%%%%%%%%%%%%%%%%%%%%

%% iv)

\clearpage
\input{Kamon-runii-param/runii-new}

\clearpage
\input{HanTata/file}

\clearpage

\input{Baer-4D/sims}

\clearpage
\input{Eboli-4F/draft04}

\clearpage
\input{Kao/trilepton}  %% with added paragraph from Matchev

\clearpage
\input{Matchev-2l/dilep}  %% new Oct. 99

\clearpage
\input{Nachtman/likesign}

%% new version received 10/4/99
\clearpage
\input{Matchev-tau/tau}

\clearpage
\input{Baer-non-u/non-u}

\clearpage
\input{Arnowitt-trilep/trilep-new}  % new version Oct. 99 

\clearpage
\input{Tata-4E/tata-4E}

%%%%%%%%%%%%%%%%%%%%%%%%%%%%%%%%%%%%%%%%%%%%%%%%%%%%%%%%%%%%%%
%% v)

\clearpage
\input{Tata-summary/summary}

\end{document}

%%%%%%%%%%%%%%%%%%%%%%%%%%%%%%%%%%%%%%%%%%%%%%%%%%%%%%%%%%%%%%

%% file: sugra.titlepage
\vspace*{1in}

\begin{center}
{\LARGE\bf Report of SUGRA Working Group\\[1ex]
 for Run II of the Tevatron}\\[.5in]

\begin{tabular}{l@{\hspace{.5em}}l}
Theory conveners:& V.~Barger$^1$\\
                 & C.E.M.~Wagner$^2$\\[2mm]
CDF contact:& T.~Kamon$^3$\\
D0 contact:& E.~Flattum$^4$\\[2mm]
Physics editors:& T. Falk$^1$\\
                & X. Tata$^5$
\end{tabular}

\bigskip

S.~Abel$^2$,
E.~Accomando$^3$,
G.~Anderson$^6$,
R.~Arnowitt$^3$,
P.~Azzi$^7$,
H.~Baer$^8$,
J.~Bagger$^9$,
W.~Beenakker$^{10}$,
A.~Belyaev$^{11}$,
E.~Berger$^{12}$,
M.~Berger$^{13}$,
M.~Brhlik$^{14}$,
T.~Blazek$^{5}$,
S.~Blessing$^8$,
W.~Bokhari$^{15}$,
N.~Bruner$^{16}$,
M.~Carena$^4$,
D.~Chakraborty$^{17}$,
D.~Chang$^{45}$,
P.~Chankowski$^{18}$,
C.-H.~Chen$^{19}$,
H.-C.~Cheng$^4$,
M.~Chertok$^3$,
G.C.~Cho$^{20}$,
D.~Claes$^{21}$,
R.~Demina$^4$,
J.~Done$^3$,
L.~Duflot$^{44}$,
B.~Dutta$^3$,
O.J.P.~\'Eboli$^{1,11}$,
S.~Eno$^{22}$,
J.~Feng$^{23}$,
G.~Ganis$^{24}$,
M.~Gold$^{16}$,
E.M.~Gregores$^1$,
K.~Hagiwara$^{20}$,
T.~Han$^1$,
B.~Harris$^8$,
K.~Hikasa$^{25}$,
C.~Holck$^{15}$,
C.~Kao$^1$,
Y.~Kato$^{26}$,
M.~Klasen$^{12}$,
W.-Y.~Keung$^{27}$,
M.~Kr\"amer$^2$,
S.~Lammel$^4$,
T.-J.~Li$^1$,
J.D.~Lykken$^4$,
M.~Magro$^1$,
S.~Mani$^{19}$,
K.T.~Matchev$^4$,
M.~Mangano$^2$,     
P.~Mercadante$^{5}$,
S.~Mrenna$^{12}$,
J.~Nachtman$^{28}$,
P.~Nath$^{29}$,
M.M.~Nojiri$^{30}$,  
A.~Nomerotski$^{31}$,
D.~Norman$^3$,
R.~Oishi$^{32}$,
K.~Ono$^{33}$,
F.~Paige$^{34}$,
M.~Paterno$^4$,
S.~Parke$^4$,
D.~Pierce$^9$,
A.~Pilaftsis$^2$,
T.~Plehn$^1$,
A.~Pompos$^{36}$,
N.~Polonksy$^{37}$,
S.~Pokorski$^{18}$,
P.~Quintana$^8$,
M.~Roco$^4$,
D.~Saltzberg$^{28}$,
A.~Savoy-Navarro$^{38}$,
Y.~Seiya$^{32}$,
C.~Smith$^9$,
M.~Spira$^{39}$,
M.~Spiropulu$^{40}$,
Z.~Sullivan$^{12}$,
R.~Szalapski$^{41}$,
B.~Tannenbaum$^{28}$,
T.~Tait$^{12,42}$,
D.~Wackeroth$^{43}$,
Y.~Wang$^{5}$,
J.~White$^3$,
H.H.~Williams$^{15}$,
M.~Worcester$^{28}$,
S.~Worm$^{16}$,             % New Mexico
R.-J.~Zhang$^1$,
M.~Zielinski$^{41}$

\input{sugra.abstract}

\newpage

\begin{tabular}{l}
$^1$Physics Department, University of Wisconsin, Madison, WI 53706\\
$^2$CERN, Theory Division, CH-1211 Geneva 23, Switzerland\\
$^3$Center for Theoretical Physics, Texas A\&M University, College Station, TX 77843\\
$^4$Theory Group, Fermilab, P.O.Box 500, Batavia, IL, 60510\\
$^{5}$Physics Department, University of Hawaii, Honolulu, HA 96822\\
$^6$Department of Physics \& Astronomy, Northwestern University, Evanston, IL 60208, USA\\
$^7$Universita di Padova, Istituto Nazionale di Fisica 
          Nucleare, Sezione di Padova, I-35131 Padova, Italy\\
$^8$Physics Department, Florida State University, Tallahassee, Florida, 32306\\
$^9$Johns Hopkins University, Baltimore, Maryland 21218\\
$^{10}$Department of Physics, University of Durham, Durham DH1 3LE, U.K.\\
$^{11}$Instituto de F\'{\i}sica Te\'orica,
              Universidade  Estadual Paulista,
              Rua Pamplona 145, 01405-900 S\~ao Paulo, Brazil.\\
$^{12}$High Energy Physics Division, Argonne National Laboratory, Argonne, IL 60439\\
$^{13}$Physics Department, Indiana University, Bloomington, IN 47405, USA\\
$^{14}$Physics Department, University of Michigan, Ann Arbor, MI 48109, USA\\
$^{15}$Physics Department, University of Pennsylvania, Philadelphia, PA 19104, USA\\
$^{16}$Department of Physics and Astronomy, University of New Mexico, Albuquerque, NM 87131\\
$^{17}$State University of New York, Stony Brook, NY 11794\\
$^{18}$Institute of Theoretical Physics, Warsaw University, Poland\\
$^{19}$Physics Department, University of California, Davis, CA 95616\\
$^{20}$Theory Group, KEK, Tsukuba, Ibaraki 305-0801, Japan\\
$^{21}$Physics Department, University of Nebraska, Omaha, NE 68182\\
$^{22}$Physics Department, University of Maryland, College Park, MD 20742\\
$^{23}$Lawrence Berkeley National Laboratory, Berkeley, California 94720\\
$^{24}$Max Planck Institute  f\"ur Physik, F\"ohringer Ring~6, D-80505 Munich, Germany\\
$^{25}$Department of Physics, Tohoku University, Sendai 980-8578, Japan\\ 
$^{26}$Physics Department, Osaka City University, Osaka 588, Japan\\
$^{27}$Physics Department, University of Illinois at Chicago, Chicago IL 60607\\
$^{28}$Physics Deparment, University of California, Los Angeles, CA 90024\\
$^{29}$Physics Department, Northeastern University, Boston, MA 02115\\
$^{30}$Yukawa Institute for Theoretical Physics, Kyoto University, Kyoto 606-8502, Japan\\
$^{31}$Physics Department, University of Florida, Gainesville, FL 32611\\
$^{32}$Physics Department, University of Tsukuba, Tsukuba, Ibaraki-ken 305, Japan\\
$^{33}$Physics Department, University of Tokyo, Tokyo 113, Japan\\
$^{34}$Brookhaven National Laboratory, Upton, NY 11973\\
$^{35}$Stanford Linear Accelerator Center, Stanford, CA 94309\\
$^{36}$Physics Department, Purdue University, West Lafayette, Indiana 47907\\
$^{37}$Department of Physics and Astronomy, Rutgers University, Piscataway, NJ 08855\\
$^{38}$University of Paris 6\&7, LPTHE, 2 Place Jussieu, F-75251 Paris, France\\
$^{39}$Institut f\"ur Theoretische Physik, Universit\"at Hamburg,
Luruper Chaussee 149, D-22761 Hamburg, Germany\\
$^{40}$Physics Department, Harvard University, Cambridge MA 02138\\
$^{41}$Department of Physics and Astronomy, University of Rochester,
River Campus, Rochester, NY 14627\\
$^{42}$Michigan State University, East Lansing, Michigan 48824\\
$^{43}$Paul Scherrer Institut, W\"urenlingen und Villigen,
CH-5232 Villigen PSI, Switzerland\\
$^{44}$L.A.L., Universite de Paris-Sud, IN2P3-CNRS, F-91898, Orsay Cedex, France\\
$^{45}$Physics Department, National Tsing-Hua University, Hsinchu 30043, Taiwan, R.O.C.
\end{tabular}

\end{center}

%% file: sugra.abstract
\begin{abstract}
We present an analysis of the discovery reach for supersymmetric
particles at the upgraded Tevatron collider, assuming that SUSY breaking
results in universal soft breaking parameters at the grand unification
scale, and that the lightest supersymmetric particle is stable and
neutral. We first present a review of the literature, including
the issues of unification, renormalization group evolution of the 
supersymmetry breaking parameters and the effect of radiative corrections
on the effective low energy couplings and masses of the theory. We 
consider the  experimental bounds coming from direct searches and
those arising indirectly from precision data, 
cosmology and the requirement of vacuum stability. The issues of flavor
and CP-violation are also addressed. The main subject of this study is to update sparticle production cross
sections, make improved estimates of backgrounds, delineate the
discovery reach in the supergravity framework, and examine how this might vary 
when assumptions about universality of soft breaking parameters are
relaxed. With 30~fb$^{-1}$ luminosity and one detector, charginos and
neutralinos,  as well as  third generation squarks, can be seen if their
masses are  not larger than  200--250~GeV, while 
first and second generation squarks and gluinos can be discovered if 
their masses do not  significantly exceed 400~GeV. We conclude that there are
important and exciting physics opportunities at the Tevatron collider, which 
will be significantly enhanced  by continued Tevatron
operation beyond the first phase of Run II.
This report is organized as follows:  In Sections 1--4 we introduce
the SUGRA model.  In Sections 5--10 we discuss radiative corrections
to masses and couplings and particle production and decay.  In
Sections 11--21 we discuss current constraints on models.
Lastly, Sections 22--32 contain the
analyses for Run II supersymmetry searches at the Tevatron.

\end{abstract}

%% file: Nath/nath.tex
\section{The mSUGRA Paradigm}

\subsection{Introduction}

The purpose of this report is to carry out a study of the
	prospects for testing mSUGRA
	at the upgraded Tevatron. It updates 
	the previous study\cite{amidei}, taking account of new 
	developments since that report. This study is self contained,
	including  a description of the
	model, analyses of mSUGRA predictions and a discussion
	of the prospects for the observation
	of the signals  predicted by the model at the upgraded Tevatron.

The Standard Model of the electro-weak and the strong interactions 
is experimentally very successful. However,  
the model is theoretically unsatisfactory. The unsatisfactory nature 
of the model arises in part due to the existence of 19 arbitrary 
parameters, the fact that the electro-weak symmetry breaking is 
accomplished in an ad hoc fashion by the introduction of
a tachyonic Higgs mass term in the theory, i.e., $V_H=-m^2H^{\dagger}H$,
and the fact that it suffers from a serious fine-tuning problem. The origin of the gauge heirarchy 
problem resides in  
 the loop correction to the Higgs boson mass 
 which is quadratically divergent requiring a cutoff $\Lambda$, i.e.,
$ m_H^2$=$2m^2$+ $c\Lambda^2$,
 where $c$ is a constant. The cutoff represents the
 scale where new physics occurs. 
 If the Standard Model were valid all the way up to the GUT scale 
 without any intervening new physics, then $\Lambda=M_{G}$. 
 In this case the electro-weak scale will be driven to the GUT scale,
 which is obviously wrong. An alternative procedure would be to arrange
 the Higgs mass in the electro-weak region by a cancellations between 
 the $m^2$ and the $\Lambda^2$ terms. However, such a cancellation 
 requires a fine tuning to 22 decimal places, which is
 highly unnatural. This is 
 a very strong theoretical hint for the existence of new physics 
 beyond the Standard Model.  Indeed, requiring no fine-tuning
 already argues for the existence of new physics in the TeV region.
 
Supersymmetry offers a very attractive cure for the fine-tuning problem 
by generating another loop contribution, so that the sum of the loop 
contributions is free of quadratic divergence. Supersymmetry is a 
symmetry which connects bosons and fermions, and its multiplets
contain bose and fermi helicity states in equal numbers. 
[In supersymmetry, the Higgs fields have additional interactions 
involving for example squark loops which also produce a quadratic divergence,
cancelling the quadratic divergence from the quark loops and leaving 
a cutoff dependence of the form ($m_{\tilde q}^2-m_q^2$)
ln($\Lambda^2/m_{\tilde q}^2$). One finds then that the fine tuning 
problem can be avoided if the squark masses are in the TeV region.]

 	The field content of the minimal supersymmetric extension
 	of the Standard Model   
with the $SU(3)_C\times SU(2)_L\times U(1)_Y$ gauge invariance consists of 
three generations of quarks, two Higgs doublets $H_1,H_2$ 
(to give tree-level masses to the
up quarks and to the down quarks and leptons and cancel gauge anomalies) and 
the $SU(3)_C\times SU(2)_L\times U(1)_Y$ gauge bosons, and all their 
superpartners\cite{haber}.
Supersymmetry, if it exists, of course, would not be an exact symmetry 
of nature, as we do not
see squarks which are degenerate with the quarks. One possibility 
 is to break supersymmetry by adding soft breaking terms by hand.
 The number of soft terms one can add is enormous: 105 such terms 
 can be added, making the theory very unpredictive and phenomenologically intractable.

\subsection{Model Description}

Supergravity unification provides a framework for the spontaneous breaking
of supersymmetry, allowing at the same time a cancellation (not fine-tuning) of the cosmological constant, because the potential of the theory is not
positive definite. 
We consider now a class of supergravity grand unified models\cite{can} where
the above mechanism of supersymmetry breaking is used to
break the degeneracy of the quark and squark masses, etc., in the 
physical sector of the theory. The basic elements of this procedure  
consist of breaking supersymmetry in the hidden  sector of the theory and
 communicating this breaking via gravitational interactions to the
 physical sector of the theory. Thus, one writes the total superpotential
 of the theory so that $\tilde W(z_i)=W_h(z)+W(z_a)$, where $W_h$ is the
 superpotential that depends on the hidden sector fields $z$, 
 and $W$ depends on the fields $z_a$ in the visible sector.  
 The simplest possibility for the breaking of supersymmetry in the
 hidden sector is via the
superHiggs mechanism where one assumes, for example, 
that $W_h$ has  
the form $W_h=m^2(z+B)$, where $z$ is a gauge singlet field and $m^2$ and
$B$ are constants. Minimization of the supergravity potential leads to
spontaneous breaking of supersymmetry,  with the gravitino 
developing a mass of  ${\cal O}(\kappa m^2$) 
(where $\kappa^{-1}$$\equiv M_{Pl}$ =$(8\pi G_N)^{-1/2}$=
$2.4 \times 10^{18}$~GeV), while the graviton remains massless.
As will be seen later, soft SUSY breaking masses characterized by the 
scale $M_s\equiv\kappa m^2$ 
lead to spontaneous breaking 
of the electro-weak symmetry, producing the connection\cite{can,applied}
\begin{equation}
M_s={\cal O}(1)\rm~TeV
\end{equation}
  An alternative 
mechanism for the breaking of supersymmetry is by gaugino condensation
arising from $m_C^3\equiv \left<\lambda \gamma^0\lambda\right>\neq 0$, which gives the soft  SUSY breaking scale $M_s\sim \kappa^2 \left<\lambda \gamma^0\lambda\right>\neq 0$.
Here $M_s\sim{\cal O}$(1TeV) requires that $m_C\sim 10^{12-13}$ GeV.   
This mechanism is more difficult to implement explicitly, because  
gaugino condensation is a non-perturbative phenomenon. 
The fact that there are no interactions except gravitational between
the hidden sector fields and the fields of the visible sector protect 
the visible sector from mass growth of size $\left<z\right>={\cal O}(M_{Pl}$), which 
can ruin the mass hierarchy of the theory. 
Also included in the mSUGRA model is a new multiplicative-conserved symmetry called $R$-parity, which can be written $R_P(-1)^{3B + L + 2S}$, and which serves to prevent the rapid decay of the proton via SUSY-mediated interactions.

We consider now models which satisfy the following conditions:
(i) SUSY breaks in the hidden sector via a super Higgs or gaugino
condensation, (ii) The symmetry of the GUT sector breaks so that the
GUT gauge group G$\rightarrow SU(3)_C\times SU(2)_L\times U(1)_Y$ at 
the scale $M_G$, (iii) The K\"ahler potential has no generational
dependent couplings with the super Higgs field. Under these assumptions,
integration of the super Higgs fields and of the superheavy fields gives
an effective potential in the low energy regime, so that
$V_{SB}$=$m_0^2$ $z_az_a^{\dagger}$+
($A_0W^{(3)}$+$B_0W^{(2)}$+h.c.),
where $W=W^{(2)}+W^{(3)}$, with $W^{(2)}$ and $W^{(3)}$ being the
quadratic and cubic part of the observable sector superpotential.  Additionally, one has
a universal gaugino mass term of the form
$L_{mass}^{\lambda}=-m_{1/2}\bar\lambda \lambda$. The effective theory
below the GUT scale $M_G$ contains four soft breaking parameters:
these are the universal scalar mass $m_0$, the universal gaugino mass
$m_{1/2}$, and the universal scaling factors $A_0$ and $B_0$
 of the cubic and quadratic 
couplings. In addition, there is one more parameter in the theory,
the Higgs mixing parameter $\mu_0$, which appears in $W^{(2)}$
=$\mu_0 H_1H_2$. Although $\mu_0$ is not a soft SUSY breaking 
parameter its origin may be linked to soft SUSY breaking. One way to
see the origin of this term is to note that the  $H_1H_2$ term
can naturally appear in the K\"ahler potential as it is a dimension two 
operator. One can use the K\"ahler transformation to move this term from
the K\"ahler potential to the superpotential. A value $\mu_0 \sim M_s$
 then naturally arises after spontaneous breaking of supersymmetry.
The mSUGRA model at the GUT scale is then characterized by the five parameters,
$m_0, m_{1/2}, A_0, B_0, \mu_0$. An essential feature of mSUGRA is\cite{can}
that the soft breaking sector is protected against mass growths 
proportional to $M_G^2/M_{Pl}$, $M_G^3/M_{Pl}^2$,\dots, which all cancel in the 
low energy theory.

	One of the remarkable aspects of  mSUGRA is that it leads to
	the radiative breaking of the $SU(2)_L\times U(1)$ 
	symmetry as a consequence of renormalization group  
	effects\cite{ewb,ross}. 
	As one evolves the soft SUSY breaking parameters 
	from the GUT scale towards the electro-weak scale the
	determinant of the Higgs mass matrix in the Higgs potential
	turns negative generating spontaneous breaking of the electro-weak symmetry. Minimization
	of the potential including loop corrections 
	allows one to compute the two Higgs VEV's
	$v_1=\left<H_1\right>$ and $v_2=\left<H_2\right>$ in terms of the parameters 
	of the theory. Alternately, one can use the minimization 
	equations to eliminate  the parameter $\mu$, where $\mu$
	is the value of $\mu_0$ at the electro-weak scale,
	in terms of the Z boson mass, and eliminate the parameter $B_0$
	in terms of tan$\beta\equiv v_2/v_1$.
	 Including radiative breaking of the 
	electroweak symmetry mSUGRA can be characterized by 
	 four parameters and the sign of $\mu$
\begin{equation}
m_0, m_{1/2}, A_0, \tan\beta,~\rm sign(\mu)\label{nath-e2}
\end{equation}
There are 32 supersymmetric particles in the theory whose masses are
determined in terms of the four parameters of the theory\cite{ross}. 
We list these particles in Table~\ref{nath-t1}. Our notation is defined in the following section.

\begin{table}
\caption{List of physical states. \label{nath-t1}}
\centering\leavevmode
 \begin{tabular}{|l|c|c|}
\hline
particle name & symbol  & spin \\
\hline
gluino & $\tilde g$  & 1/2 \\

 charginos   & $\tilde\chi_1^{\pm}$, $\tilde\chi_2^{\pm}$      & 1/2      \\

 neutralinos    &$ \tilde\chi_1^0$, $\tilde\chi_2^0$, $\tilde\chi_3^0$, $\tilde\chi_4^0$& 1/2 \\
\hline
sleptons    & $\tilde e_L$, $\tilde \nu_{e_L}$, $\tilde e_R$  & 0      \\

     & $\tilde \mu_L$, $\tilde \nu_{\mu_L}$, $\tilde \mu_R$ & 0 \\

      & $\tilde \tau_1$,$\tilde \tau_2$, $\tilde \nu_{\tau_L}$& 0\\
\hline
 squarks & $\tilde u_L$, $\tilde d_L$, $\tilde u_R$,$\tilde d_R$ & 0  \\

         & $\tilde c_L$,$\tilde s_L$, $\tilde c_R$,$\tilde s_R$ & 0  \\

       & $\tilde t_1$, $\tilde t_2$, $\tilde b_1$,$\tilde b_2$ & 0  \\
\hline 
higgs     & h, H, A, $H^{\pm}$      &  0     \\
\hline
\end{tabular}
\end{table}

 Thus many sum
rules exist among the mSUGRA mass spectrum which are experimentally
testable. An interesting property of radiative breaking is that 
over most of the parameter space of the theory one finds that 
$\mu^2/M_Z^2\gg 1$
 and this leads to the approximate relations 
\begin{eqnarray}
2m_{\tilde\chi_{1}}^0&\cong& m_{\tilde\chi_{1}}^{\pm}\cong m_{\tilde\chi_{2}}^0\simeq {1\over 3}
m_{\tilde g}\nonumber\\
m_{\tilde\chi_{3}}^0 &\cong& m_{\tilde\chi_{4}}^0 \cong m_{\tilde\chi_{2}}^{\pm}\simeq |\mu|\gg
m_{\tilde\chi_{1}}^0 \label{nath-e3}
\end{eqnarray}
The above implies that the light neutralino and chargino 
states are mostly gauginos, and 
the heavy 
states mostly higgsinos. It also turns out that under the constraints of
electro-weak symmetry breaking the lightest neutralino is also the
lightest mass supersymmetric particle (LSP) over most of the parameter
space of the theory. 

An interesting aspect of 
mSUGRA model is that it automatically includes a super GIM mechanism for the 
suppression of flavor changing neutral currents for the process
$K_L\rightarrow \mu^+\mu^-$. The mSUGRA boundary conditions give 
the following relation
\begin{equation}
m_{\tilde c}^2-m_{\tilde u}^2=m_c^2-m_u^2
\end{equation}
The super GIM suppression occurs because the squark loop contributions 
in the process $K_L\rightarrow \mu^+\mu^-$  enter in the combination
$m_{\tilde c}^2-m_{\tilde u}^2$ which because of Eq.~(\ref{nath-e5}) is suppressed.
The degeneracy of the squark masses necessary for the super GIM to work
is enforced by the universality condition of Eq.~(\ref{nath-e3}).  
The universality of the gaugino masses at the GUT scale, which is 
enforced in any case when the $SU(3)_C\times SU(2)_L\times U(1)_Y$ gauge 
group is embedded in a simple GUT group, obey the following one loop relation
at scales below the GUT scale 
\begin{equation}
M_i=m_{1/2}\frac{\alpha_i}{\alpha_G}  \label{nath-e5}
\end{equation}
where i=1,2,3 for $U(1)_Y$, $SU(2)_L$, $SU(3)_C$, $\alpha_i$ is the $i^{\rm th}$
fine structure constant, and $\alpha_G=\alpha_i(M_G)$ is the GUT scale
coupling constant. Note that $\alpha_1 = {5\over3}\alpha'$, where $\alpha'$ is the Standard Model hypercharge fine structure constant. There are, however, important 2 loop QCD contributions
for the case i=3\cite{mart}.  The high precision LEP data on the gauge coupling
constants at the Z scale, i.e., $\alpha_i(M_Z)$\cite{alpha} and the experimental 
ratio of $m_b/m_{\tau}$\cite{bbo} appear to be consistent with ideas of 
SUSY and mSUGRA unification. 

In investigating the parameter space of mSUGRA one uses somewhat subjective 
naturalness constraints on the soft SUSY parameters. The simplest approach
 is to set
\begin{equation}
 m_0,~m_{\tilde g}\leq 1 ~TeV \label{nath-e6}
\,;
 \end{equation}
more sophisticated approaches have also been discussed. For 
studies of physics at the Tevatron, the naturalness assumption of 
Eq.~(\ref{nath-e6}) appears sufficient. However, the constraint of Eq.~(\ref{nath-e6}) must be revised upwards for analyses at the LHC which can probe higher regions
of the mSUGRA parameter space.
In investigating the implications of mSUGRA one must also impose 
additional experimental constraints such as those from  
(i) $b\rightarrow s+\gamma$ decay and from (ii) the value of $g_{\mu}-2$.

Constraint (i) arises from the  experimental limit on 
 $b\rightarrow s+\gamma$  from CLEO\cite{cleo}, which gives 
BR($b\rightarrow  s+\gamma$)
	=(3.15$\pm 0.54$)$\times 10^{-4}$, and ALEPH\cite{aleph-nath}, which gives $(3.11\pm0.80_{\rm stat} \pm 0.72_{\rm syst})\times 10^{-4}$. This decay receives 
	contributions in the Standard Model from $W$ boson
	exchange. Here, recent analyses\cite{buras},
	 including the leading and the next to  leading order QCD 
	 corrections and two-loop electroweak corrections, 
	 give the branching ratio
	  BR($b\rightarrow  s+\gamma$) =(3.32$\pm 0.29$)$\times 10^{-4}$.
	 In mSUGRA, there are 
	 additional contributions from the exchange of the charged
	 Higgs, the charginos, the gluinos, and the neutralinos\cite{bert}. While the 
exchange of the charged Higgs gives a constructive intereference with SM amplitudes,
	 the exchange of the charginos and the neutralinos can give
	 contributions with either sign\cite{hewett}. The experimental 
	  $b\rightarrow  s+\gamma$ branching ratio puts a stringent 
	  constraint on the parameter space of the theory. As will
	  be discussed later the $b\rightarrow  s+\gamma$ constraint
	  affects in a very significant way dark matter analyses for
	  one sign of $\mu$\cite{na}. 
	  A further reduction of the experimental
	  error in this decay mode will certainly constrain the
	  parameter space further and may even reveal the existence of
	  new physics if a significant
	  deviation from the SM results are confirmed.
	  
	Constraint (ii) is relevant because supersymmetric contributions 
	to ($g_{\mu}-2$) can be very significant\cite{kosower}. The current 
	experimental value of $a_{\mu} \equiv (g_{\mu}/2		   	
 	-1)$ is $a_{\mu}^{exp}= 1.1659230(84) \times 10^{-10}$
 	while the Standard Model result for $a_{\mu}$ is given by
 	$a_{\mu}^{theory}(SM)= 11659162.8(7.7) \times 10^{-10}$\cite{davier}. 
 	In mSUGRA,  additional contributions to $(g-2)$  
 	arise from the exchange of the charginos and the neutralinos.
 	One finds that the supersymmetric electro-weak contributions
 	can be as large or even larger than the Standard Model 
 	electro-weak 
 	contributions. In fact, supersymmetric contributions can be large
	enough that even the current experiment puts a constraint on
	the mSUGRA parameter space. In the near future, 
	the Brookhaven
	experiment E821 will begin collecting data and is expected
	to increase the sensitivity of the ($g_{\mu}-2$) 
	measurement  by a factor
	of 20, to $a_{\mu}\sim 4\times 10^{-10}$. The improved 
	measurement may reveal the existence of new physics beyond the
	Standard Model, or
	if no effect is seen, would constrain the 
	mSUGRA parameter space even further. In either
	 case, the (g$_{\mu}-2$) experiment is an
	important test of mSUGRA.
	
	We can supplement the mSUGRA analysis with further  
	contraints which involve additional assumptions.
	Thus, for example, we can consider the constraints of 
\begin{itemize}
\item[](iii) relic density
\item[](iv) $b-\tau$ unification 
\item[](v) proton lifetime limits
\end{itemize}

	Constraint (iii) applies when $R$-parity is conserved. This possibility is very attractive, in that 
in this case the lightest neutralino 
 becomes a candidate for cold dark matter (CDM) over much of the
 parameter space of the model. Currently there exists a whole array  of
 cosmological models such as HCDM, $\Lambda$CDM, $\Lambda$HCDM,
 $\tau$CDM,\dots etc.,  which all require some component of CDM.
 At the very minimum one has the constraint
 that the supersymmetric dark matter not overclose the universe,
 i.e., $\Omega_{\tilde\chi_1}^0 <1$, where $\Omega=\rho_{\tilde\chi_1^0}/\rho_c$,
 where  $\rho_{\tilde\chi_1^0}$ is the neutralino matter density and 
  $\rho_c$ is the critical matter density needed to close the 
  universe.  Of course, more stringent constraints on 
  $\Omega_{\tilde\chi_1}^0 h^2$, 
  which is the quantity computed theoretically (where $h$ is the Hubble 
  parameter in units of 100km/sec\,Mpc.) 
  would ensue if one assumed a  specific cosmological model\cite{relic}.
   The density contraints can be very severe in limiting the parameter 
  space of mSUGRA. These results have also important implications 
  for the search for dark matter\cite{detection}.
   Constaints  (iv) and 
	(v) are more model-dependent as compared to the constraints  
	(i)--(iii).  Thus, for example, 
	the predictions of $m_b/m_{\tau}$ mass ratio depends on 
	the GUT group and on the textures\cite{bbo}. 
	Similarly, the nature of the
	GUT group and textures also enter in the analysis 
	of proton lifetime\cite{pdecay}. It should be noted that a tiny amount of $R$-parity violation at a level irrelevant for collider searches could negate any constraints from the cosmological relic density.

\subsection{Extensions of mSUGRA}

	We discuss now some possible generalizations of mSUGRA.
\begin{itemize}
\item[](a) CP violation
\item[](b) Non-universalities of soft terms
\item[](c) R parity violation
\item[](d) Corrections from Planck scale physics 
\item[](e) Connection of mSUGRA to M theory
\end{itemize}
	We discuss briefly each of these items and more discussion 
	will follow in the subparts later.

	 (a) The mSUGRA formalism allows for complex phases for the
	soft parameters. However, not all the phases are independent.
	One can remove all but two phases, which can be chosen
	to be $\theta_{\mu_0}$ (the phase of $\mu_0$)
	 and $\alpha_{A_0}$ (the phase of $A_0$), so that the 
	mSUGRA parameter space with CP violation expands to six
	parameters, i.e., Eq.~(\ref{nath-e2}) is 
	replaced by 
\begin{equation}
m_0, m_{1/2}, A_0, tan\beta,\theta_{\mu_0}, \alpha_{A_0}\,. \label{nath-e7}
\end{equation}	
One of the important constraints on SUSY models with CP violation arises from the experimental limits on the neutron 
edm and on the electron edm. The current experimental limits on these
are $d_n<1.1\times 10^{-25}$e$\cdot$cm for the neutron, and $d_e<4.3\times 
10^{-27}$e$\cdot$cm for the electron. These limits produce  a strong constraint
on the  parameter space of Eq.~(\ref{nath-e7})\cite{cp}.

(b) As mentioned earlier, 
the universality of the scalar soft breaking terms arises from the
assumption that the K\"ahler potential does not have generational 
dependent couplings with the hidden sector fields. A relaxation of this 
constraint  leads to non-universalities of
the soft breaking terms\cite{soni}, which must then be restricted by 
group symmetries and the phenomenological constraints of flavor
changing neutral currents (FCNC). One of the sectors which is not very 
strongly constrained by FCNC is the Higgs sector, and one can
introduce non-universalities of the type 
\begin{equation}
m_{H_1}^2(M_G)=m_0^2(1+\delta_1), ~m_{H_2}^2(M_G)=m_0^2(1+\delta_2)
\end{equation}	
where the typical range considered for the $\delta_i$ is $|\delta_i|\leq 1$. 
Similarly, FCNC constraints are also insensitive to 
the non-universalities in the third generation sector and one 
may consider non-univesralities in this sector along with the
non-universalities in the Higgs sector. The non-universalities produce
identifiable signals at low energy\cite{soni}.

(c) Analysis of signatures of supersymmetry in mSUGRA depend importantly 
on whether or not one assumes  $R$-parity invariance. If one assumes 
that $R$-parity is conserved, then the LSP is stable and 
one will have supersymmetric particle decays which result in lots of 
missing energy. If $R$-parity is 
 violated, then the LSP is not stable and will decay with possible
 signatures of 2 charged leptons ($l_il_j\nu_k$),
 lepton and 2 jets ($l_iu_j\bar d_k$, $\nu_i d_j \bar d_k$),
 and three jets ($d_id_ju_k$). Thus the signatures of SUSY events 
 at colliders would be very different if $R$-parity is 
 violated\cite{rparity}.

(d) There can be important corrections to mSUGRA predictions from
Planck scale terms\cite{planck}. This possibility arises because $M_G$ is
only two 1--2 orders of magnitude away from the string/Planck scale and 
thus corrections of O($M_G/M_{Pl}$) could be relevant. 
 One example of such corrections is the 
 Planck contribution to the gauge kinetic energy function 
 $f_{\alpha\beta}$ which produces splittings of the 
 $SU(3)_C$, $SU(2)_L$, and $U(1)_Y$  gauge couplings
 at the GUT scale. The same Planck  correction 
 can generate non-universal contributions to the gaugino masses at
 the GUT scale.
 Planck  corrections may also be responsible for the generation 
 of textures which control the hierarchy of quark-lepton masses at low energy.

 (e) Although currently one does not have a phenomenologically viable
 string model, it is our hope that such a model exists and that 
 perhaps mSUGRA is
 its low energy limit below the string scale
$M_{str}\approx 5\times 10^{17}$ GeV. 
  The underlying structure of mSUGRA,
  i.e., N=1 supergravity 
 coupled to matter and gauge fields,
 is what one expects in the low energy limit of a compactified string model. 
 It is also possible to envision how the soft breaking
 sector of mSUGRA can arise in strings where the fields governing 
 SUSY breaking are the dilaton (S) and the 
  moduli  ($T_i, \bar T_i$).
  Of course, the problem of  SUSY breaking in string theory is 
  as yet unsolved, and consequently one cannot  make serious predictions 
  either in the weakly coupled heterotic string or in its strongly coupled
  M-theory limit. However, when one has a viable 
  string model with the right SUSY breaking, it would be possible to
  make connection with mSUGRA at the string scale by matching the
  boundary conditions at $M_{str}$.

Other extensions beyond (a)--(e) are discussed elsewhere in this volume.

\subsection{Signals of supersymmetry in mSUGRA}

  Aside from indirect signals that might appear in the 
  precision experimental determination of $b\rightarrow s+\gamma$
  and in the $g_{\mu}-2$ measurements, one can have signals via
  decay of the proton and via the direct detection of a neutralino
  in dark matter detectors.  However, the most convincing evidence of
  supersymmetry will be the direct observation of supersymmetric 
  particles at colliders. The purpose  of this report is to study
  the reach of the upgradraded Tevatron for SUSY particles in 
  various channels. One of the signals is the production
  of the Higgs in direct collisions, i.e., $q\bar q, gg\rightarrow$
  $h, H, A, H^{\pm}$. The tree level mass of the lightest  Higgs is 
  governed by gauge interactions, with important modifications arising 
  from one- and two-loop corrections. Generally, one expects $m_h$ 
  to obey\cite{higgs} 
 \begin{equation} 
 m_h \leq 120\mbox{--}150 \rm\ GeV \label{nath-e9}
 \end{equation}
 The upper limit on the Higgs is somewhat model dependent because
 the soft parameters enter in the loop corrections to
 the Higgs mass. However, Eq.(\ref{nath-e9}) represents a fair upper limit on
 $m_h$ for any reasonable naturalness assumption on $m_0$ and $m_{1/2}$.
 The upper limit on the Higgs mass is lowered if one includes 
 additional constraints discussed earlier. 
 There are several processes which give pair production of 
 sparticles at hadron colliders. Thus, 
 squarks and gluinos can be pair produced via processes such as
   $q\bar q$,
    $q g$, 
    $ g g$
     $\rightarrow \tilde q  \bar{\tilde q}$, 
     ${\tilde q}{\tilde g}$, 
 ${\tilde g}{\tilde g}$. 
 Similarly, one can have pair  production of chargino and neutralino
 final states, i.e., $\tilde\chi_i^{\pm}\tilde\chi_j^{\mp}$, $\tilde\chi_i^{\pm}\tilde\chi_j^0$,
 $\tilde\chi_i^0\tilde\chi_j^0$ as well as final states such as 
 $\tilde q\tilde\chi_i^{\pm}$, $\tilde g\tilde\chi_i^{\pm}$, $\tilde q \tilde\chi_i^0$,
 $\tilde g\tilde\chi_i^0$. Other SUSY states that can be pair produced are
 $\tilde l\tilde l$, $\tilde l\tilde \nu$, $\tilde \nu \tilde \nu$.

   With $R$-parity invariance, sparticles must decay to other sparticles until this decay chain terminates in the neutral stable LSP which escapes detection.
Typical SUSY 
  signals all involve large missing energy events with 
  the neutralinos and neutrinos carrying the missing energy. For instance, 
  the chargino decay involves 
   \begin{equation}
 \tilde\chi_1^{-}\rightarrow e^-\bar\nu \tilde\chi_1^0
 \end{equation}
 which exhibits the signature  $e^-+ E_T$(missing) in the final 
 state. Similarly, the decay of the squark involves $\tilde q\rightarrow$
 $q+\tilde\chi_1^0$ as one of its modes which in the final state will 
 give  jet+$E_T$(missing). A signal of particular interest is the
 trileptonic signal, which arises from the decay of the final states 
 $\tilde\chi_1^{\pm}\tilde\chi_2^0$ via the channels $\tilde\chi_1^{\pm}\rightarrow$
 $l\nu\tilde\chi_1^0$ and $\tilde\chi_2^0\rightarrow l\bar l\tilde\chi_1^0$. In this 
 case one finds $\bar l_1l_1l_2+E_T$(missing). This channel is fairly
 clean, with no hadronic activity expected from QCD radiations, and is thus a promising channel 
 for the detection of supersymmetry.

%\vspace{2cm}
%{\bf References}

%\end{enumerate} 

%\end{document}

%% file: Wagner/proceed.tex
\newcommand{\st}{\scriptstyle}
\newcommand{\sst}{\scriptscriptstyle}
\newcommand{\mco}{\multicolumn}
\newcommand{\epp}{\epsilon^{\prime}}
\newcommand{\vep}{\varepsilon}
\newcommand{\ra}{\rightarrow}
\newcommand{\ppg}{\pi^+\pi^-\gamma}
\newcommand{\vp}{{\bf p}}
\newcommand{\ko}{K^0}
\newcommand{\kb}{\bar{K^0}}
\newcommand{\al}{\alpha}
\newcommand{\ab}{\bar{\alpha}}
\def\simlt{\stackrel{<}{{}_\sim}}
\def\simgt{\stackrel{>}{{}_\sim}}
\def\be{\begin{equation}}
\def\ee{\end{equation}}
\def\bea{\begin{eqnarray}}
\def\eea{\end{eqnarray}}
\def\simlt{\stackrel{<}{{}_\sim}}
\def\simgt{\stackrel{>}{{}_\sim}}
\def\CPbar{\hbox{{\rm CP}\hskip-1.80em{/}}}%temp replacement due to no font
\def\ap#1#2#3   {{\em Ann. Phys. (NY)} {\bf#1} (#2) #3.}
\def\apj#1#2#3  {{\em  Astrophys. J.} {\bf#1} (#2) #3.}
\def\apjl#1#2#3 {{\em Astrophys. J. Lett.} {\bf#1} (#2) #3.}
\def\app#1#2#3  {{\em Acta. Phys. Pol.} {\bf#1} (#2) #3.}
\def\ar#1#2#3   {{\em Ann. Rev. Nucl. Part. Sci.} {\bf#1} (#2) #3.}
\def\cpc#1#2#3  {{\em Computer Phys. Comm.} {\bf#1} (#2) #3.}
\def\err#1#2#3  {{\em Erratum} {\bf#1} (#2) #3.}
\def\ib#1#2#3   {{\em ibid.} {\bf#1} (#2) #3.}
\def\jmp#1#2#3  {{\em J. Math. Phys.} {\bf#1} (#2) #3.}
\def\ijmp#1#2#3 {{\em Int. J. Mod. Phys.} {\bf#1} (#2) #3.}
\def\jetp#1#2#3 {{\em JETP Lett.} {\bf#1} (#2) #3.}
\def\jpg#1#2#3  {{\em J. Phys. G.} {\bf#1} (#2) #3.}
\def\mpl#1#2#3  {{\em Mod. Phys. Lett.} {\bf#1} (#2) #3.}
\def\nat#1#2#3  {{\em Nature (London)} {\bf#1} (#2) #3.}
\def\nc#1#2#3   {{\em Nuovo Cim.} {\bf#1} (#2) #3.}
\def\nim#1#2#3  {{\em Nucl. Instr. Meth.} {\bf#1} (#2) #3.}
\def\np#1#2#3   {{\em Nucl. Phys.} {\bf#1} (#2) #3.}
\def\pcps#1#2#3 {{\em Proc. Cam. Phil. Soc.} {\bf#1} (#2) #3.}
\def\pl#1#2#3   {{\em Phys. Lett.} {\bf#1} (#2) #3.}
\def\prep#1#2#3 {{\em Phys. Rep.} {\bf#1} (#2) #3.}
\def\prev#1#2#3 {{\em Phys. Rev.} {\bf#1} (#2) #3.}
\def\prl#1#2#3  {{\em Phys. Rev. Lett.} {\bf#1} (#2) #3.}
\def\prs#1#2#3  {{\em Proc. Roy. Soc.} {\bf#1} (#2) #3.}
\def\ptp#1#2#3  {{\em Prog. Th. Phys.} {\bf#1} (#2) #3.}
\def\ps#1#2#3   {{\em Physica Scripta} {\bf#1} (#2) #3.}
\def\rmp#1#2#3  {{\em Rev. Mod. Phys.} {\bf#1} (#2) #3.}
\def\rpp#1#2#3  {{\em Rep. Prog. Phys.} {\bf#1} (#2) #3.}
\def\sjnp#1#2#3 {{\em Sov. J. Nucl. Phys.} {\bf#1} (#2) #3.}
\def\spj#1#2#3  {{\em Sov. Phys. JEPT} {\bf#1} (#2) #3.}
\def\spu#1#2#3  {{\em Sov. Phys.-Usp.} {\bf#1} (#2) #3.}
\def\zp#1#2#3   {{\em Zeit. Phys.} {\bf#1} (#2) #3.}

\def\NPB#1#2#3{{\it Nucl.~Phys.} {\bf{B#1}} (19#2) #3}
\def\PLB#1#2#3{{\it Phys.~Lett.} {\bf{B#1}} (19#2) #3}
\def\PRD#1#2#3{{\it Phys.~Rev.} {\bf{D#1}} (19#2) #3}
\def\PRL#1#2#3{{\it Phys.~Rev.~Lett.} {\bf{#1}} (19#2) #3}
\def\ZPC#1#2#3{{\it Z.~Phys.} {\bf C#1} (19#2) #3}
\def\PTP#1#2#3{{\it Prog.~Theor.~Phys.} {\bf#1}  (19#2) #3}
\def\MPL#1#2#3{{\it Mod.~Phys.~Lett.} {\bf#1} (19#2) #3}
\def\PR#1#2#3{{\it Phys.~Rep.} {\bf#1} (19#2) #3}
\def\RMP#1#2#3{{\it Rev.~Mod.~Phys.} {\bf#1} (19#2) #3}
\def\HPA#1#2#3{{\it Helv.~Phys.~Acta} {\bf#1} (19#2) #3}

%%%%%%%%%%%%%%%%%%%%%%%%%%%%%%%%%%%%%%%%%%%%%%%%%%
%                                                %
%    BEGINNING OF TEXT                           %
%                                                %
%%%%%%%%%%%%%%%%%%%%%%%%%%%%%%%%%%%%%%%%%%%%%%%%%%

\section{Status of Unification of Couplings in  the MSSM}

Although the Minimal Supersymmetric extension
of the Standard Model (MSSM)  enables a natural solution to the
hierarchy problem of the Standard Model (SM), it is thought to provide
only a low energy effective description of a more fundamental, unified,
theory which will become manifest at much higher energy scales.  One
of the possible experimental tests of this framework is the unification
of the renormalization group evolved gauge couplings. The scale
at which the couplings unify gives a hint of the relevant energy at
which the low energy description should be replaced by the more
fundamental one. The condition of unification~\cite{gut,unif} 
is a non-trivial one,
since it depends on the exact relation between the $\beta$ function
coefficients and on the low energy values of the three gauge
couplings measured experimentally. The idea of unification can be tested
quantitatively, but it is always associated with large theoretical
uncertainties, related to the unknown spectrum of supersymmetric
particles at low energies, as well as the exact physical thresholds
at scales close to the grand unification scale. A possible approach
to treat these uncertainties is to develop a bottom-up 
approach: 
\begin{enumerate} \item Obtain the low energy values
of the three gauge couplings  including the
corrections induced by one-loop diagrams of standard model
particles and their superpartners.
\item Compute the high energy threshold necessary to achieve
the unification of gauge couplings by extrapolating
their low energy values to high energies via two-loop renormalization
group evolution.
\end{enumerate}
In this approach, one assumes the absence of new physics affecting
the evolution of the gauge couplings, up to scales of order of the
grand unification scale. 

The value of the gauge couplings in the modified $\overline{MS}$ scheme
in the Standard Model are well known. The largest uncertainty is
associated with the strong gauge coupling, whose
value is known only at the level of 3\%, $\alpha_3(M_Z) \simeq
0.119 \pm .003$~\cite{PDB}. The values of the weak gauge couplings, instead,
are known with high precision. In particular, the value of
the weak mixing angle can be given as a function of the electroweak
parameters $G_F$, $M_Z$, $1/\alpha_{em}(M_Z) (\simeq 127.9$~\cite{LP}),
the pole top quark mass $M_t$ and the Higgs mass $m_h$.
For a Higgs mass of order of
100 GeV, as is appropriate in low energy supersymmetric models,
and a pole top quark mass of order of 170 GeV, $\sin^2\theta_W(M_Z)$ in
the modified $\overline{MS}$ scheme
is given by~\cite{LP,CPP},
%
%\begin{equation}
\bea
 \sin^2\theta_W(M_Z) & \simeq  &  0.2315 + 5.4 \times 10^{-6} (m_h - 100)   
  -   2.4 \times 10^{-8} (m_h -100)^2 
\nonumber \\
         &  & -  3. \times 10^{-5} (M_t - 170)
   -  8.4 \times 10^{-8}  (M_t - 170)^2 \pm 0.0003.
\label{eq:1}
\eea
All the masses are in GeV units.  The above expressions
take into account all one-loop corrections within the
Standard Model. 

Supersymmetric one-loop diagrams lead to logarithmic corrections,
as well as corrections  proportional to the
inverse of the supersymmetric particle masses that become 
negligible when the masses are pushed towards large values with respect
to $M_Z$. The decoupling of the non-logarithmic corrections
is very fast and, within the present experimental limits, 
these corrections become relevant only if there are light,
left-handed doublets which appear in the spectrum. The logarithmic
corrections, instead, are very important for obtaining the
exact supersymmetric predictions. Their effect can be studied
by renormalization group methods, using a step function decoupling
of the supersymmetric particle contributions, at energies below the
relevant supersymmetric particle mass. This program leads to
the following expressions:
\begin{eqnarray}
\frac{1}{\alpha_i(M_G)} & = & \frac{1}{\alpha_i(M_Z)}
- \frac{b_i}{2\pi} \ln \left(\frac{M_G}{M_Z} \right) 
+ \gamma_i + \frac{1}{\alpha_i^{\rm thr.}}
\nonumber\\
\frac{1}{\alpha_i^{\rm thr.}} & = & \sum_{\eta}
\frac{b_i^{\eta}}{2\pi}\ln \left(\frac{M_{\eta}}{M_Z} \right)  
+ {\rm h.e.t.}
\end{eqnarray}
where $\gamma_i$ represents the two-loop corrections, as well 
as the corrections factors to convert from the modified $\overline{MS}$
scheme to the $\overline{DR}$ scheme, $M_G$ is the scale at which
the weak gauge couplings unify,
$b_i^{\eta}$ is the contribution to the $\beta_i$ function
coefficient of the superparticle $\eta$ with mass $M_{\eta}$
and h.e.t.\ denotes the corrections coming from the unknown high
energy theory at scales of order of $M_G$.
Using these equations, a simple formula for the low energy
value of the strong gauge coupling can be obtained as
a function of the weak gauge couplings under the assumption
of exact unification of gauge couplings at the scale $M_G$ (i.e., neglecting GUT-scale threshold effects),
\begin{eqnarray}
\frac{1}{\alpha_3(M_Z)}  =  
(1 + B) \left[ \frac{1}{\alpha_2(M_Z)} + \gamma_2 \right]
- B \left[ \frac{1}{\alpha_1(M_Z)} + \gamma_1 \right] - \gamma_3
 +  \frac{1+B}{\alpha_2^{{\rm thr.}}} - \frac{B}{\alpha_1^{{\rm thr.}}}
- \frac{1}{\alpha_3^{{\rm thr.}}},
\end{eqnarray}
where
\begin{equation}
B = \frac{b_2 - b_3}{b_1 - b_2}.
\end{equation}
As mentioned above, the predictions coming from gauge
coupling unification depend strongly on the low energy values for
the gauge couplings, as well as on the relation between the different
$\beta_i$ coefficients.  As a first test of the unification relations,
one can ignore the threshold corrections, as well as the small $\gamma_i$
correction factors. One can then deduce the value of $B$ needed
in order to obtain a good unification prediction,
\begin{equation}
\left.
B \right|_{\rm unif.} = \frac{ (\alpha_2(M_Z))^{-1} - (\alpha_3(M_Z))^{-1}}
         { (\alpha_1(M_Z))^{-1} - (\alpha_2(M_Z))^{-1}} \simeq 0.7.
\end{equation}
In the MSSM, $B = 5/7$, leading to an excellent agreement between theory
and experiment at the one-loop level. 

The above procedure can be extended to include the effect of the
threshold of low energy supersymmetric particles. These 
can be described in terms of one single scale $T_{SUSY}$~\cite {LP}. 
Considering different characteristic mass scales for left-handed
squarks ($m_{\tilde{Q}}$), right-handed squarks ($m_{\tilde{U}}$),
gluinos ($m_{\tilde{g}}$), left-handed sleptons ($m_{\tilde{L}}$),
right-handed sleptons ($m_{\tilde{E}}$),
 electroweak gauginos ($m_{\tilde{W}}$), 
Higgsinos ($m_{\tilde{H}}$) and the heavy
Higgs doublet ($m_H$), $T_{SUSY}$ is given as~\cite{CPW}
\be
T_{SUSY} = m_{\tilde{H}} \left( \frac{m_{\tilde{W}}}{m_{\tilde{g}}}
    \right)^{\frac{28}{19}} 
\left[ \left( \frac{m_{H}}{m_{\tilde{H}}} \right)^{\frac{3}{19}}
\left( \frac{m_{\tilde{W}}}{m_{\tilde{H}}} \right)^{\frac{4}{19}}
\left( \frac{m^3_{\tilde{L}} m^7_{\tilde{Q}}}
{m^2_{\tilde{E}} m^5_{\tilde{U}} m^3_{\tilde{D}}} 
\right)^{\frac{3}{19}}
\right]
\label{eq:TSUSY}
\ee
Assuming exact unification at the scale $M_G$ and ignoring threshold
corrections induced by physics at the grand unification scale,  the value of the strong gauge coupling at low energies is 
determined as a function of $\sin^2\theta_W$, $\alpha_{em}$ and 
$T_{SUSY}$. The result is 
\be 
\left.
\frac{1}{\alpha_3(M_Z)} = \frac{1}{\alpha_3(M_Z)} \right|_{\rm SUSY}
+ \frac{19}{28 \pi} 
                             \ln \left(\frac{T_{SUSY}}{M_Z} \right),
\label{Ts}
\ee
where $ \alpha_3(M_Z)|_{SUSY}$ would be the  
value of the strong gauge coupling 
at $M_Z$ if the theory were exactly supersymmetric down to the scale $M_Z$.

The effective threshold scale $T_{SUSY}$, Eq.~(\ref{eq:TSUSY}), 
has only a mild dependence on
the squark and slepton masses. This can be traced
to the fact that squarks and
sleptons come in complete representations of $SU(5)$ which modify all
$\beta_i$ coefficients in the same way, keeping the value of $B$ constant.
The scale $T_{SUSY}$ depends strongly on the overall Higgsino mass,
as well as on the ratio of masses of the weakly and strongly 
interacting gauginos. 
 In models with universal gaugino masses at the grand unification scale, 
\begin{equation}
T_{SUSY}
 \simeq  m_{\tilde{H}} \left( \frac{\alpha_2(M_Z)}{\alpha_3(M_Z)}
\right)^{3/2} \simeq \frac{|\mu|}{7}  \;,
\label{tsusyapp}
\end{equation}
where $|\mu|$ characterizes the Higgsino mass in the case of negligible
mixing in the neutralino/chargino sector. Hence, even
if all supersymmetric masses are of the order of 1 TeV, the effective
supersymmetric scale $T_{SUSY}$ can still be of the order of the weak
scale, or smaller.

It is also interesting to investigate the variation of the scale
at which the gauge couplings $\alpha_1$ and $\alpha_2$ unify as
a function of the low energy supersymmetric spectrum. From the
above equations, we obtain~\cite{CPP}
\begin{equation}
\left.
M_G = M_G \right|_{SUSY} \times \left(\frac{M_Z}{G_{SUSY}} \right)^{2/7}
\label{mgut}
\end{equation}
where
\begin{equation}
G_{SUSY} = \left(M_H m_{\tilde{H}}^4 m_{\tilde{W}}^{20} \right)^{1/25}
\times \frac{ \left( M_{\tilde{Q}} M_{\tilde{L}} \right)^{1/8} }
            { \left( M_{\tilde{U}}^4 M_{\tilde{D}} 
             M_{\tilde{E}}^3 \right)^{1/8} }
\end{equation}
where $ \left. M_G \right|_{SUSY}$ is the unification scale value that would
be obtained if the theory is supersymmetric up to the scale
$M_G$. Observe that, due to the small value of the exponent in 
Eq.~(\ref{mgut}), the grand unification scale is quite stable under 
variations of the low energy superymmetry breaking mass parameters. In
particular, it cannot be reconciled with the string scale in weakly
coupled string theory, $M_S \simeq 5 \times 10^{17}$ GeV by 
means of $G_{SUSY}$.

 For a top quark mass of the order of 170 GeV, assuming that all sparticles
are heavy, so that non-logarithmic corrections can be ignored,
and ignoring corrections induced by particles with masses
of order of $M_{G}$,
 the unification condition implies  the following numerical
 correlation~\cite{LP,Nir2,BCPW},
\be
\alpha_3(M_Z) \simeq 0.128 + 1.2 \times 10^{-4} 
\left(M_t[GeV] - 170 \right)
- 0.0035 \ln\left(\frac{T_{SUSY}}{M_Z}\right)
\ee

The values quoted above take already into account the negative
corrections obtained from a large top Yukawa coupling, like
the ones obtained for low values of 
$\tan\beta \simeq 2$ ($h_t(m_t) \simeq 1.1$ at the weak scale). 
For moderate values of $\tan\beta \simeq 5$, 
the value of  $\alpha_3(M_Z)$ increases
by one percent, due to the slightly lower values
of the top quark Yukawa coupling, while for large values
of $\tan\beta$ it can decrease by one percent compared to the
values given in Table~\ref{wagner-t1}, if both the
top and the bottom Yukawa coupling become large. 

\begin{table}[h]
\caption[]{Gauge coupling unification 
predictions for $\alpha_s(M_Z)$, for 
given values of $\sin^2\theta_W$ (correlated with $M_t$),  $m_h = M_Z$
 and $T_{SUSY}= 1$~TeV, 400~GeV, and $(M_Z)$. \label{wagner-t1}}
\medskip
\centering\leavevmode
\begin{tabular}{|c|c|c|c|c|}
\hline
$M_t$[GeV] & $\sin^2\theta_W(M_Z)$ & $\alpha_3(M_Z)$ & $\alpha_3(M_Z)$
& $\alpha_3(M_Z)$  \\
        &               &           $T_{SUSY}$ = 1 TeV 
& $T_{SUSY} = 400$ GeV & $T_{SUSY} = M_Z$ \\ 
\hline
% & &  \\
% 160 & 0.2318   & 0.11  (0.125) \\
& &  & & \\
170 & 0.2315   & 0.119 & 0.123 &  0.128 
\\
\hline
% & &  \\
% 180 & 0.2306 &  0.122   (0.131)\\ \hline\hline
\end{tabular}
\end{table}

The predicted value of $\alpha_s(M_Z)$ from unification  may be further
modified if some sparticle masses are ${\cal{O}}(M_Z)$.
 Indeed, not only the leading-log
 contributions but the full one-loop threshold contributions from
SUSY loops should be included when extracting the couplings from
the data~\cite{CPP,all,BMP}. 
The main additional
effects come from light sfermions and  are given by \cite{Nir2}--\cite{all}
\begin{equation}
\frac{\delta \sin^2 \theta_W}{\sin^2 \theta_W}
\simeq \frac{\cos^2 \theta_W}{\sin^2 \theta_W - \cos^2 \theta_W}
\left( \frac{\delta \alpha_{em}}{\alpha_{em}} +
\frac{\Pi_{WW}(0)}{M_W^2} - \frac{\Pi_{ZZ}(M_Z^2)}{M_Z^2} 
\right),
\label{eq:delatsin}
\end{equation}
where 
%${\rm s}^2 = \sin^2 \theta_W$, ${\rm c}^2 = \cos^2 \theta_W$ and
$\Pi_{ij}$ are the vacuum polarization contributions to the gauge 
bosons.  The above corrections to $\sin^2\theta_W$ 
are dominated by the logarithmic corrections discussed above, which
are induced by a correction to the electromagnetic coupling and to
the external momentum dependent part of $\Pi_{ZZ}(q^2)$. For particle
masses far from the $Z$-production threshold, the dominant non-logarithmic
corrections are approximately described by the parameter 
$\Delta \rho(0) = \Pi_{WW}(0)/M_W^2 - \Pi_{ZZ}(0)/M_Z^2$. In the MSSM
it follows that $\Delta \rho(0)^{SUSY} \geq 0$ and,  after considering the 
additional terms in Eq.~(\ref{eq:delatsin}), one still obtains
that the correction to
$\sin^2\theta_W$ induced by the non-logarithmic
corrections $\delta_{\rm non-LL} \sin^2\theta_W \simlt 0$  in all the 
parameter space consistent with the present
experimental constraints. This translates into an increase, 
with respect to the  results from Table~\ref{wagner-t1}, in the
values of $\alpha_3(M_Z)$  predicted from supersymmetric grand unification.
One should also note that large
corrections to $\Delta \rho (0)$ are disfavoured by  present experimental
data, particularly for large values of the top quark mass
 $M_t \simgt 170$ GeV, ruling out large non-logarithmic corrections to
the unification predictions.

\begin{figure}
%\psdraft
\centerline{
\psfig{figure=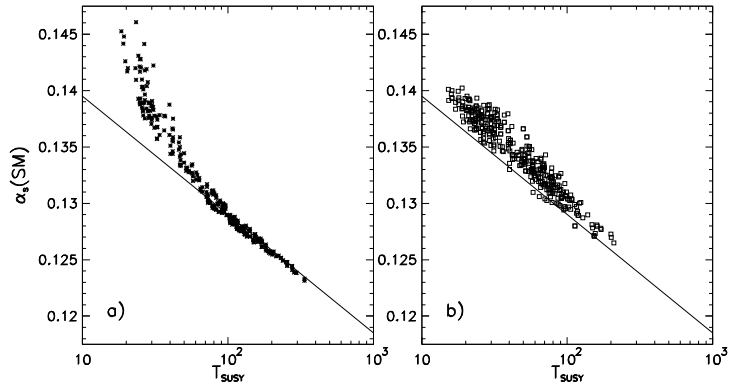,width=10 cm}}
\caption[]{Predictions of $\alpha_3(M_Z)$ as a function of
the effective scale $T_{SUSY}$ for $M_t = 175$ GeV and a generic supergravity
spectrum (Fig.~\ref{wagner-f1}.a),
as well as for a superparticle spectrum,
which gives a good fit to the precision electroweak
measurement data (Fig.~\ref{wagner-f1}b),
as described in the text.
The solid lines represent the prediction for $\alpha_3(M_Z)$ in
the leading logarithmic approximation, Eq.~(\ref{Ts}). \label{wagner-f1}}
\end{figure}

%% added from new version
Figure \ref{wagner-f1} shows the predicted value of $\alpha_3(M_Z)$ as a function
of $T_{SUSY}$ for the case of a generic minimal supergravity spectrum
(Fig.~\ref{wagner-f1}a) and for a low energy supersymmetry spectrum  that
gives a good fit to the precision electroweak measurement data (Fig.~\ref{wagner-f1}b). In the
minimal supergravity case, there is a strong correlation between the
parameter $|\mu|$ (or, equivalently, $T_{SUSY}$, Eq.~(\ref{tsusyapp}))
and the squark spectrum. This implies that for low values of $T_{SUSY}$
the non-logarithmic correction to $\sin^2\theta_W$ tend to become large,
leading to a departure of the predicted value of $\alpha_3(M_Z)$ with
respect to the one obtained in the leading-logarithmic approximation.
The lowest values of $T_{SUSY}$ shown in Fig.~\ref{wagner-f1}a  correspond to values of the sparticle masses close to the present experimental bounds,
which tend to
worsen the standard model fit to the precision measurement data.
In  Fig.~\ref{wagner-f1}b, instead, a good fit to the precision measurement data is obtained by relaxing the universality assumption and taking
large values of the left-handed stop mass parameters,
while keeping small values of those for the right-handed stop. The largeness
of the left-handed stop mass ensures the smallness of the non-logarithmic
stop-induced corrections to $\sin^2\theta_W$, implying a good agreement
between the predicted value of $\alpha_3(M_Z)$ and the one
obtained in the leading-log approximation.

It is interesting to observe that the prediction for $\alpha_3(M_Z)$
is in excellent agreement with the experimental values if:
\begin{enumerate}
\item The supersymmetric threshold scale is in the range:
400 GeV $\simgt T_{SUSY} \simgt 37$ eV. 
\item Finite, non-logarithmic corrections induced by supersymmetric
particle loops are suppressed. 
\end{enumerate}
The predicted value of $\alpha_3(M_Z)$ coming from the
condition of exact unification of couplings coincides with the
experimental central value of $\alpha_3(M_Z)$  for a scale 
$T_{SUSY} \simeq$~1~TeV.

In supergravity models, the low energy values of the supersymmetry
breaking parameters may be computed by the knowledge of their
boundary conditions at the grand unification scale and their
renormalization group equations. Flavor changing neutral 
currents are efficiently suppressed if sparticles with the
same quantum numbers under the standard model gauge group are
assumed to obtain equal supersymmetry breaking masses at the 
GUT scale. 
% Moreover, particles belonging to the same SU(5) multiplet
% may be assumed to acquire equal masses. The same can be done with the
% superpartners of the standard model gauge bosons. Quite generally,
% however,  there are threshold corrections to these mass 
% relations that one should take into account together with other
% threshold corrections at the GUT scale. 
For small or moderate
values of $\tan\beta$, assuming that no other particles
affect the evolution of the gauge couplings, and taking $\alpha_3(M_Z) = 0.12$,
%and ignoring the
% effects of the stop-Higgs trilinear coupling at the
% unification scale, $A_0 = 0$, 
the masses at the weak scale
are approximately given by~\cite{five}~\footnote{The coefficients quoted below
have been obtained by running the renormalization group equations
of all dimensionless couplings, as well as of all supersymmetry breaking
parameters, up
to the scale $M_Z$. This amounts to ignore
one-loop threshold corrections to these parameters, which would
lead to small modifications to the supersymmetric particle masses}:
\begin{eqnarray}
M_i & \simeq & \frac{M_i(0)}{\alpha_i(0)} \alpha_i
\nonumber\\
m_{\tilde{f}_i}^2 
& \simeq & m_{\tilde{f}_i}^2(0)\left[1-\frac{y_t}{6} (\delta_{Q_3,i} 
          + 2 \delta_{U_3,i}) \left(m_{Q}^2(0) + m_U^2(0) + m_{H_2}^2(0)
           \right)\right] 
\nonumber\\
&  & +
             0.78 \; n^3_{f_i} M_3^2 + 0.88 \; n_2^{f_i} M_2^2 +
            \frac{y_{f_i}^2}{4} 1.55 M_1^2  - \Delta m^2(M_i,y_t)
             \left(\frac{\delta_{Q_3,i}}{6}
            + \frac{\delta_{U_3,i}}{3}\right)
\nonumber\\
m_{H_i}^2 & = & m_{H_i}^2(0) \left[1 - \frac{y_t}{2} \delta_{i,2}
              \left(m_{Q}^2(0) + m_U^2(0) + m_{H_2}^2(0)
           \right)\right]  
% \nonumber\\
 + 0.8 M_2^2
            + 0.23 M_1^2 -  \delta_{i,2} \frac{\Delta m^2(M_i,y_t)}{2},
\end{eqnarray}
where $M_i$, with $i=1$--3, are the three gaugino masses at the weak
scale, $n^N_{f_i}$ is equal to 1 if the particle $f_i$ belongs
to the fundamental representation of $SU(N)$ and is zero otherwise,
$y_{f_i}^2/4 = 3/5 (q_i - T_3)^2$, $q_i$ is the electric charge of the
particle $f_i$,  $Q_3$ and $U_3$ denote the third generation squark
$SU(2)$ doublet and up squark $SU(2)$ singlet, respectively, and 
$y_t$ denote the square of the ratio of the top quark yukawa coupling to its
fixed point value.  In the above, 
a vanishing argument always implies a function
evaluated at the grand unification scale.  The function $\Delta m^2(M_i,y_t)$ 
is  given by
\begin{eqnarray}
\Delta m^2(M_i,y_t) & \simeq & y_t \left(1.25 M_3^2  + 1.46 M_2^2 + 0.6 M_1^2 
+ 0.8 M_3 M_2 + 0.23 M_3 M_1 + 0.18 M_2 M_1 \right) 
\nonumber\\
& & - y_t^2 \left(0.65 M_3 + 0.5 M_2 + 0.16 M_1 \right)^2
%\nonumber\\ & & 
- y_t \left(1 - y_t \right) A_0 \left( 0.32 M_1 + M_2 + 1.29 M_3 - A_0
\right)
\end{eqnarray}
where $A_0$ is the stop-Higgs trilinear coupling at the
unification scale,
and, for convenience, we have written all gaugino factors as a
function of their  values at the weak scale. The low energy value of
the stop-Higgs trilinear coupling $A_t$ is given by
\begin{eqnarray}
A_t = A_0 ( 1 -y_t) -1.15 M_3 - 0.8 M_2 - 0.16 M_1 
+ y_t \left( 0.65 M_3 - 0.5 M_2 - 0.16 M_1 \right)
\end{eqnarray}
Observe that
no unification assumption or relation between the high energy values
of the soft supersymmetry breaking parameters  has been made in the 
above formulae. The formulae, however, lose 
their validity for large values of 
$\tan\beta$, at which the bottom- and $\tau$-Yukawa coupling effects
can no longer be ignored.

Finally, $\mu^2$ may be approximately obtained from
the tree-level condition of radiative electroweak symmetry breaking
\begin{equation}
\mu^2 \simeq \frac{m_{H_1}^2 - m_{H_2}^2 \tan^2\beta}{\tan^2\beta -1}
- \frac{M_Z^2}{2}.
\label{esb}
\end{equation}
As discussed above,
the effective threshold scale $T_{SUSY}$ is strongly dependent on the
value of the mass parameter $\mu$ as well as on the ratio of the gluino
and wino masses. 
% If the unification of gaugino masses is assumed at
% high energies, the effective threshold scale $T_{SUSY}$ tends to be
% suppressed (see Eq. (\ref{tsusyapp})) by a factor of order 7 compared
% to the $\mu$ parameter. This suppression is partially compensated
% by an enhancement of the $\mu$ parameter due to the condition
% of radiative electroweak symmetry breaking, Eq. (\ref{esb}), for
% $\tan\beta \simeq 2$, 
% $\mu^2 \simeq {\cal O}(3. M_{1/2}^2 +  0.8 m_0^2)$ 
% (somewat larger (lower) values may be
% obtained for smaller (larger) values  of $\tan\beta$). 
For values of the gaugino masses at the scale $M_G$ 
of order of 250 GeV 
(500 GeV), a common scalar mass of order of the gaugino masses,
and $\tan\beta \simeq {\cal O}(4)$,
the effective scale $T_{SUSY}$ that is obtained from the above equations
is of the order of 75 GeV (150 GeV), implying that the value of $\alpha_3(M_Z)$
will be more than 2 standard deviations above  the experimentally
measured value. For $\tan\beta \simeq 4$, $T_{SUSY}$ becomes slightly
lower, of order 60 GeV (120 GeV) and, as mentioned before 
$\left. \alpha_3(M_Z)\right|_{\rm SUSY}$ increases by one percent.
The situation improves, however, if large values
of $m_0$ are considered, since $\mu$ can be enhanced in this case.
For instance, for values of $m_0$ of order 1 TeV (2 TeV), 
$\tan\beta \simeq 2$ and $M_{1/2}$
of order 250 GeV, as considered below, the value of $T_{SUSY}$ can
be raised to be close to 130 GeV (220 GeV). This implies that lower
high energy threshold corrections are needed for larger values of $m_0$.
%(although the fine-tuning problem is obviously worse in this case).  

The strong restrictions above can be partially
ameliorated in models in which gaugino unification at the
grand unification scale is relaxed~\cite{RS}. A simple way of doing
this is to assume that the wino mass is 
of the order of or larger than the gluino mass (see Eq. (\ref{eq:TSUSY})).
For instance, taking again a value
of $M_2$ of the order of 200 GeV (400 GeV), and a value of 
$m_0 \simeq 250$ GeV,
but now assuming that all gaugino
masses at low energies are of the same order, the above equations show
that the effective scale $T_{SUSY}$ can be raised to values of order of
350 GeV (700 GeV). Hence, in such models
the low energy values are in better agreeement
with the exact unification predictions. It is important to note that
this pattern of gaugino masses at low energies is not  
obtained  within grand unified models, unless supersymmetry 
breaking originates via  F-terms which break the underlying
grand unification symmetry, or it
is transmitted to the observable sector at scales lower than 
or of the order of $M_G$.

That this somewhat drastic remedy may not be essential becomes clear if we recognize that any unified theory at the scale
$M_{G}$ will lead to threshold corrections. A valid question
is what is the size of the threshold corrections at the scale
$M_{G}$ necessary to bring the low energy prediction for 
$\alpha_3(M_Z)$ in agreement with the experimental value.
\begin{eqnarray}
\left. \left.
\frac{1}{\alpha_3(M_Z)}\right|_{\rm exp} -
\frac{1}{\alpha_3(M_Z)}\right|_{\rm SUSY} & = & \frac{19}{28 \pi} 
\ln \left(\frac{T_{SUSY}}{M_Z}\right)  
%\nonumber\\
 +  \frac{1 + B}{\alpha_2^{\rm h.e.t.}}
+ \frac{B}{\alpha_1^{\rm h.e.t.}}
- \frac{1}{\alpha_3^{\rm h.e.t.}}
\label{het}
\end{eqnarray}
where $1/\alpha_i^{\rm h.e.t.}$ are the threshold effects
induced by the particles with masses of order $M_{G}$. These
threshold corrections are highly model dependent. 
The sum of the terms in the second line of Eq.~(\ref{het})
can be parametrized as the variation of the prediction of
$\alpha_3(M_Z)$ due to high energy physics,
$\Delta(1/\alpha_3^{\rm h.e.t.})$.

\begin{figure}[h]
%\psdraft
%\vspace*{.3in}
 
\centering\leavevmode
\psfig{figure=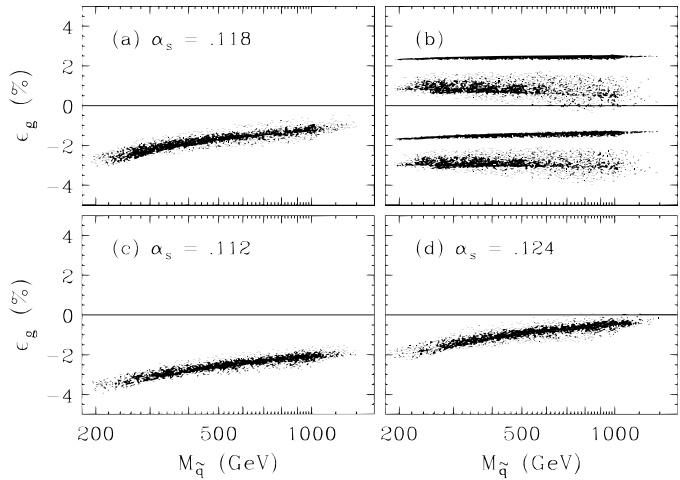,width=9cm}
\caption[]{(a), (c), (d) Threshold corrections at the grand unification scale,
$\epsilon_g = (1 - g_1/g_3)(M_{GUT})$,  necessary to 
achieve a value of the low energy strong gauge coupling, as
a function of $M_{\tilde{q}}$ for a generic supergravity
spectrum. (b) The maximum and minimum $\epsilon_g$ allowed
in minimal SU(5) (two top regions) and missing doublet SU(5)
(bottom two regions). \label{wagner-f2}}
\end{figure}

Due to the evolution of the strong gauge coupling,
the size of the high energy threshold corrections 
at the scale $M_{GUT}$ necessary to
achieve correct unification predictions should be of order 
1--3 $\%$ for $400 \simgt T_{SUSY} \simgt 50$ GeV, which
according to our discussion above, corresponds to characteristic
squark masses between a few hundred GeV and a few TeV (see
Figure \ref{wagner-f2}). This is a
natural size for these corrections, which depend strongly
on the model. For instance,
in SU(5) models
these threshold corrections are given by~\cite{BMP}
\begin{eqnarray}
\Delta \alpha_3^{\rm h.e.t.} & \simeq & 
\frac{18}{28\pi} 
\ln \left(\frac{m_{H^T}}{M_G}\right) \;\;\;\;\;\;\;\;\;\;\;\;\;\;\;
\;\;\;\;\;\;({\rm Minimal\ SU(5)})
\nonumber\\
& \simeq &
\frac{18}{28\pi}  \left[
\ln \left(\frac{m_{H^T}}{M_G}\right)  - 9.8 \right] \;\;\;\;\;\;\;\;\;\;\;\;\;\;\;
({\rm Missing \; Partner \; SU(5)})
\end{eqnarray}
where $m_{H^T}$ is the effective mass of the colour triplet Higgs boson,
leading to proton decay processes via dimension five operators. 
It is clear from the above
equations that in missing partner SU(5) models, for an effective
colour triplet mass $m_{H^T}\simeq 10^{18}$~GeV, which suppresses
the proton decay processes, the value of
$\alpha_3(M_Z)$ can be easily brought in agreement with the
low energy observed values, even for $T_{SUSY} \simeq 20$ GeV.
In general, the required pattern of corrections is strongly restricted
by proton decay constraints. The situation is similar for SU(5) as
for  SO(10) models~\cite{RL}. 

Apart from the possibilities of having a
$T_{SUSY} \simeq 1$ TeV, or a complicated structure of GUT
threshold corrections~\cite{GUTthr}, one may also consider the existence of
non-renormalizable operators of the kind $\Sigma F_{\mu\nu}F^{\mu\nu}/M_{Pl}$,
which in minimal models, with one adjoint $\Sigma$, are unique at leading
order in powers of $1/M_{Pl}$.
Tevatron searches have the potential  of testing squark and gaugino
masses consistent with the low energy values shown in Figure~\ref{wagner-f2}. 

Finally, let us consider an interesting property of the renormalization
group evolution of the gauge couplings pointed out 
by Shifman~\cite{Shifman}. 
In the $\overline{DR}$ scheme, the two-loop
contribution of a supermultiplet to the renormalization
group evolution of the gauge couplings from the scale $M_G$ up
to the scale $Q >M_{\eta}$ can be given by
\begin{equation}
\Delta^{\eta} \alpha_i^{-1}(Q) = \frac{b_i^{\eta}}{2\pi} 
\ln\left(\frac{M_G}{Q Z_{\eta}}\right),
\end{equation}
where  $Z_{\eta}$ is the one-loop wave function renormalization constant
associated with the supermultiplet $\eta$. Observe that this is a
compact expression that parametrizes the two-loop contributions
to the evolution of the gauge couplings,
by knowledge of $Z_{\eta}$ at the one-loop order.
Now, assume  a vector like supermultiplet $\eta$+$\overline{\eta}$,  
which acquires an explicit supersymmetry conserving mass $M_{\eta}^G$
at the GUT scale (via an explicit term in the superpotential). Its
renormalized mass will be given by $M_{\eta}^G/Z_{\eta}$. Hence,
after decoupling of this particle at the scale $M_{\eta} >M_Z$,
\begin{equation}
\Delta^{\eta+\bar{\eta}} \alpha_i^{-1}(M_Z) = \frac{b_i^{\eta}}{\pi} 
ln\left(\frac{M_G}{M_{\eta}^G}\right).
\end{equation}
Observe that
the above formula does not mean that vector like multiplets do
not affect the evolution of the gauge couplings beyond the one-loop
order, 
since their presence affects the value of the wave function 
renormalization constant of all the other particles 
present in the spectrum~\cite{Ross}.
Indeed, the overall effect of vector like particles belonging
to complete representations of $SU(5)$, whose effect in
the prediction of $\alpha_3(M_Z)$ vanishes at one-loop order,
is to enhance the two-loop
effects on the unification relations, leading to a value of 
$\alpha_3(M_Z)$ larger than in the MSSM.

\subsection{Yukawa Coupling Unification}

The unification of the bottom and $\tau$ Yukawa couplings at
the grand unification scale is a property of the simplest
unfication scenarios.
For the present experimental allowed range of values
 for the top quark mass~\cite{PDB}, 
$M_t = 173.8 \pm 5.2$  GeV, and the strong gauge coupling,
the condition of bottom-tau  Yukawa coupling 
unification implies either low values of
$\tan \beta$, $ 1.5\simlt \tan \beta \simlt 3$, or very large values 
of $\tan \beta \simgt 20$~\cite{Ramond}--\cite{largetb}.
Most interesting is the fact that to achieve b-$\tau$ unification,
$h_b(M_{GUT}) =  h_{\tau}(M_{GUT})$, large values
of the top Yukawa coupling, $h_t$, at $M_{GUT}$
are necessary in order to compensate for the effects of the strong 
interaction renormalization in the running of the bottom Yukawa coupling. 
These large 
values of $h_t^2(M_{GUT})/4 \pi \simeq$ 0.1--1 are exactly those that 
ensure the attraction towards the infrared (IR) 
fixed point solution of the top
quark mass~\cite{CPW,Yukawa,LP1}. 
The exact value of the running bottom mass $m_b(M_b) \simeq $~4.15--4.35~GeV,
is very important to determine the top quark mass predictions~\cite{BCPW,KKRW}.
A larger $m_b$, for instance, will
be associated with a larger bottom Yukawa coupling at low energies,
allowing unification
for smaller values of the top Yukawa coupling, an effect that can be
enhanced by relaxing the exact unification conditions~\cite{KKRW,EPS,NIRlast}. 

For $M_{GUT} \simeq 2 \times 10^{16}$ GeV, the infrared fixed point value
of the top Yukawa coupling
$(h_t^{IR})^2/ 4 \pi \simeq
(8/9) \alpha_s(M_Z)$, and the corresponding running top quark mass
$m_t^{IR} \simeq h_t^{IR} v \sin \beta$, with 
$v \simeq 174$ GeV.
After the computation of the proper low energy threshold 
corrections~\cite{NIRlast,NIRref}, 
the infrared fixed point solution associates to 
each value of $M_t$ the lowest possible value of $\tan \beta$ consistent
with the validity of perturbation theory up to scales of order 
$M_{GUT}$. The most interesting consequence of the 
IR fixed point $M_t$--$\tan\beta$ relation is associated with
the lightest CP-even  Higgs mass predictions in the MSSM~\cite{COPW,BABE}. 
Indeed, for $\tan\beta$ larger than 1, the lowest tree level value of the 
lightest Higgs mass, $m_h$,
is obtained  at the lowest value of $\tan\beta$, a property that holds
even after including two-loop corrections. For
squark masses below or of the order of 1 TeV, the present experimental
bounds on the Higgs mass begin to constraint in a relevant way
the low $\tan\beta$ solution to bottom-tau Yukawa coupling
unification~\cite{CEH,CCPW}.

\begin{figure}[h]
%\psdraft
\vspace*{.5in}

\centerline{
\psfig{figure=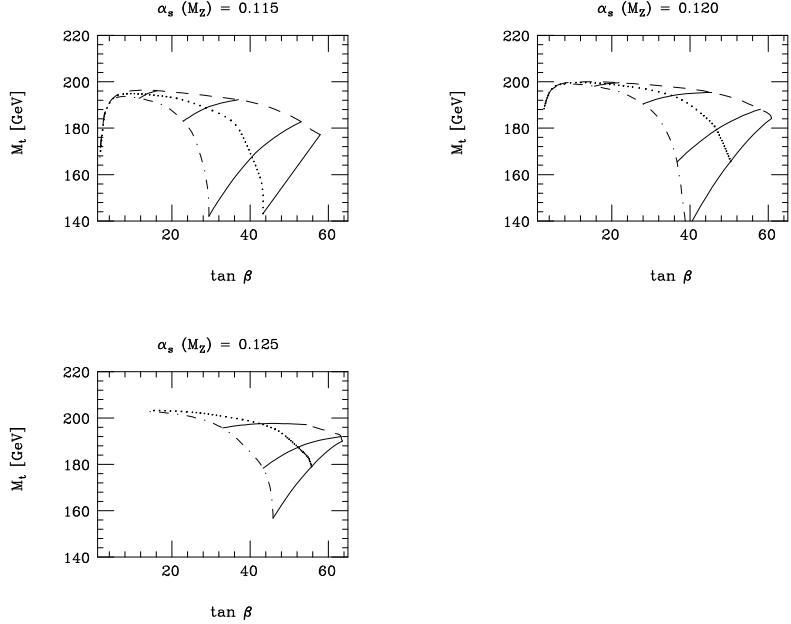,width=12cm,angle=90}}
\caption[]{Top quark mass predictions coming from the
condition of bottom-tau Yukawa unification, for 
different values of the bottom mass corrections, 
parametrized by $\Delta m_b = -K_c \tan\beta$, for
$K_c = 0$ (dashed-line), 0.003 (dotted line) and
0.006 (dot-dashed ine). The solid line represent  the
ratio $\omega$ of the bottom-Yukawa coupling to the top-Yukawa
coupling at the GUT scale, $\omega =$ 1, 0.6 and
0.2, respectively.  The curves are cut when
the value of $h_t^2(M_{GUT})/4\pi \geq 1$.\label{wagner-f3}}
\end{figure}

In the context of $b$--$\tau$ Yukawa coupling unification 
possible large  radiative corrections to the bottom
 mass are crucial in determining
 the top quark mass and  $\tan \beta$ predictions.
In general, one assumes that the top and bottom quarks couple each to only one
of the Higgs doublets and hence $m_t (M_t) =  h_t (M_t) v_2$
and $m_b (M_t) =  h_b (M_t) v_1$,
with $v_i$ the vacuum expectation value of the Higgs $H_i$. 
 However, a coupling of the bottom (top) quark to the 
neutral component of the Higgs $H_{2(1)}$ may
be generated at the one-loop level, and since $v_2 \gg v_1$
for large values of $\tan \beta > 10$,  large    
 corrections to the bottom mass may 
be present~\cite{Hall,wefour,CDWR,Steve}, 
\bea
m_b &=& h_b \; v_1 + \Delta h_b \; v_2 \equiv h_b v_1 (1 + K \tan \beta).
\eea
$\Delta m_b = K \tan \beta$
receives contributions from stop-chargino and sbottom-gluino loops,
the latter being the dominant ones.
The magnitude of  $\Delta m_b$ is strongly dependent on the supersymmetric
spectrum and its sign is generally 
governed by the overall sign of $\mu \times m_{\tilde{g}}$,
\bea
K &  =&  \mu \; m_{\tilde{g}} \;\tan \beta \;\;
 \left[ \frac{2 \alpha_s}{3 \pi}
I_1 ( m^2_{\tilde{b}_1}, m^2_{\tilde{b}_2},  m^2_{\tilde{g}})
 +   \frac{A_t}{ m_{\tilde{g}}} \frac{h_t^2}{(4 \pi)^2}
I_2 ( m^2_{\tilde{t}_1}, m^2_{\tilde{t}_2}, \mu^2) \right],
\label{eq:Deltamb}
\eea
where $ m_{\tilde{b}_i}$ and $m_{\tilde{t}_i}$ 
are the sbottom and stop mass eigenstates, respectively, the integral
factor is $I_i = C_i/a_{max}$,
with $a_{max}$ the maximum of the squared masses, and
$C_i$ = 0.5--0.9 depending on the mass splitting. Using the relation
$ m_{\tilde{g}} \simeq $ 2.6--2.8 $M_{1/2}$ and the fact that from the 
renormalization group  equations, $A_t$ is in general of  opposite  
sign to $M_{1/2}$ and of ${\cal{O}} (M_{1/2})$, it follows that there is a partial 
cancellation between the two terms in Eq.~(\ref{eq:Deltamb}). Although 
important, such partial cancellation 
is in general not sufficient to render the bottom mass corrections small.
Hence, in the large $\tan \beta$ region, the bottom mass corrections
need to be appropriately computed to extract the proper value of the
bottom Yukawa coupling at low energies.
The predictions
from $b$--$\tau$ Yukawa coupling unification will therefore 
depend on the particular
supersymmetric spectrum under consideration. In particular, the
exact value of $\tan\beta$ at which the unification of Yukawa couplings
is achieved depends strongly on the size of the $\Delta m_b$ 
corrections~\cite{CDWR,MP,CEP}. In Figure~\ref{wagner-f3} the effect of the bottom mass
corrections is displayed. A value of the running bottom mass 
$m_b(M_b) = 4.15$ GeV has been  used. For
values of the strong gauge coupling $\alpha_3(M_Z) \simgt 
0.12$ unification in the low $\tan\beta$ regime demands slightly
larger values of the running bottom mass, $m_b(M_b) \simgt 4.25$ GeV. 

The sign of the bottom mass corrections is also related to the
one of the supersymmetric contributions to the amplitude of
the $b \rightarrow s \gamma$ decay process in models with universal
soft supersymmetry breaking parameters at high energies.
In these models, the most important supersymmetric
contribution to the $b \rightarrow s\gamma$  decay rate
comes from the chargino-stop one-loop diagram~\cite{bsga}. 
The chargino contribution to the $b \rightarrow s \gamma$
decay amplitude depends on the soft supersymmetry breaking
mass parameter $A_t$ and on the supersymmetric mass parameter
$\mu$, and for very large values of $\tan\beta$, it is
given by~\cite{Diaz,BOP},
\be
A_{\tilde{\chi}^+} \simeq \frac{m_t^2}{m_{\tilde{t}}^2}
\frac{A_t \mu}{m_{\tilde{t}}^2} \tan\beta \; G\left(
\frac{m_{\tilde{t}}^2}{\mu^2}\right),
\ee
where $G(x)$ is a function that takes values of order 1
when the characteristic stop mass $m_{\tilde{t}}$
is of order $\mu$ and grows for lower values
of $\mu$. One can show that, for positive (negative)
values of $A_t \times \mu$ the chargino contributions are of
the same (opposite) sign as the charged Higgs ones~\cite{wefour}. 
Hence, to partially cancel the light charged Higgs contributions, rendering
the  $b \rightarrow s \gamma$
decay rate acceptable, negative values for $A_t \times \mu$ are required.
As  follows from Eq.~(\ref{eq:Deltamb}) and the discussion below,
 this requirement
has direct implications on the corrections to the bottom mass
and,  after a detailed analysis, one concludes that it puts
strong constraints on models with Yukawa 
coupling unification~\cite{wefour,BOP}. It is important to remark,
however, that these constraints can be evaded if small flavor
violating up or down squark mixing effects are present in the
low energy spectrum~\cite{CEP,Steve,Masiero}.

Finally, it is important to remark that, in the context of
SO(10) unification scenarios which predict the unification of
the top and the neutrino Yukawa couplings at the scale $M_G$,
the unification predictions depend strongly on the mass of
the right handed neutrinos. In particular, a value of the
right handed $\tau$ neutrino mass smaller than $10^{15}$ GeV would
put very strong constraints on bottom-$\tau$ Yukawa coupling
unification in the small $\tan\beta$ regime~\cite{neutrinos}.

%% file: Kao/conventions.tex
%------------------------------------------------------------------------
% Conventions for the mSUGRA working Group
%------------------------------------------------------------------------

\newcommand{\tchi}{\tilde{\chi}}
\newcommand{\what}{\widehat}
\newcommand{\wtilde}{\widetilde}

\section{The Minimal Supersymmetric Model}

%------------------------------------------------
% The Lagrangian Density
%------------------------------------------------
\subsection{The Lagrangian Density}

The minimal supersymmetric standard model (MSSM) 
is the minimal supersymmetric extension of the standard model (SM) with 
\begin{enumerate} 
\item two Higgs doublets $H_1$ and $H_2$ such that 
$H_1$ couples to the fermions with $t_3 = -1/2$ and 
$H_2$ couples to the fermions with $t_3 = +1/2$, 
\item a supersymmetric partner for each SM particle and every Higgs boson,  
\item a conserved $R$-parity $R \equiv (-1)^{3B+L+2S} = (-1)^{3(B-L)+2S}$, 
\item a Higgs mixing parameter $\mu$ assumed to be real, 
\item three real symmetric $3\times 3$ matrices 
      of squark mass-squared parameters, 
      $M_{\wtilde{Q}}^2$, $M_{\wtilde{U}}^2$ and $M_{\wtilde{D}}^2$, 
\item two real symmetric $3\times 3$ matrices 
      of slepton mass-squared parameters, 
      $M_{\wtilde{L}}^2$ and $M_{\wtilde{E}}^2$, and 
\item three real $3\times 3$ matrices of trilinear Yukawa parameters, 
      $A_U$, $A_D$ and $A_E$. 
\end{enumerate}

The fields in the MSSM are summarized in a table as the following:

\begin{center}
\begin{tabular}{lll} 
Superfield      & Bosonic Fields         & Fermionic Fields          \\
\hline
$\what{G}$      & Gluons: $g^a, a = 1,...,8$ & Gluinos: $\tilde{g}^a$    \\ 
$\what{W}$      & $W$ Bosons: $W^\alpha, \alpha = 1,2,3$ 
                & Winos: $\wtilde{W}^\alpha$    \\ 
$\what{B}$      & $B$ Boson: $B$         & Bino: $\wtilde{B}$          \\ 
$\what{Q}$      & Squarks: $\wtilde{Q} = (\tilde{u}_L,\tilde{d}_L)$
                & Quarks:  $Q = (u_L,d_L)$ \\ 
$\what{U}$      & Squarks: $\wtilde{U} = \tilde{u}_R^*$
                & Quarks:  $U = u_L^c$ \\ 
$\what{D}$      & Squarks: $\wtilde{D} = \tilde{d}_R^*$
                & Quarks:  $D = d_L^c$ \\ 
$\what{L}$      & Sleptons: $\wtilde{L} = (\tilde{\nu}_L,\tilde{\ell}_L)$
                & Leptons:  $L = (\nu_L,\ell_L)$ \\ 
$\what{E}$      & Sleptons: $\wtilde{E} = \tilde{\ell}_R^*$
                & Leptons:  $E = \ell_L^c$ \\ 
$\what{H}_1$    & Higgs bosons: $H_1 = (H_1^0,H_1^-)$ 
                & Higgsinos: $\wtilde{H}_1 = (\wtilde{H}_1^0,\wtilde{H}_1^-)$ \\ 
$\what{H}_2$    & Higgs bosons: $H_2 = (H_2^+,H_2^0)$
                & Higgsinos: $\wtilde{H}_2 = (\wtilde{H}_2^+,\wtilde{H}_2^0)$
\end{tabular} 
\end{center}

\noindent 
where $U,u = u,\; c$ and $t$; $D,d = d,\; s$ and $b$; 
and $E,\ell = e, \mu$ and $\tau$. 
In this model, the lightest supersymmetric particle (LSP) is stable. 

%------------------------
% Lagrangian density
%------------------------

The MSSM Langrangian density with soft supersymmetry breaking terms 
has the following form
\begin{eqnarray}
{\cal L}_{\rm soft} 
% Higgs
& = &-m_{H_1}^2|H_1|^2 -m_{H_2}^2|H_2|^2 
     +\mu B \epsilon_{ij}\left( H_1^iH_2^j +H.c. \right) \nonumber \\
% Gauginos
&   &-{1\over 2}M_1\overline{\wtilde{B}}{B}
     -{1\over 2}M_2\overline{\wtilde{W}^\alpha} \wtilde{W}^\alpha
     -{1\over 2}M_3\overline{\tilde{g}^a} \tilde{g}^a \nonumber \\
% Sfermions
&   &-M_{\wtilde{Q}}^2[\tilde{u}_L^*\tilde{u}_L+\tilde{d}_L^*\tilde{d}_L]
     -M_{\wtilde{U}}^2\tilde{u}_R^*\tilde{u}_R
     -M_{\wtilde{D}}^2\tilde{d}_R^*\tilde{d}_R \nonumber \\
&   &-M_{\wtilde{L}}^2[ \tilde{\ell}_L^*\tilde{\ell}_L
                       +\tilde{\nu}_L^*\tilde{\nu}_L]
     -M_{\wtilde{E}}^2\tilde{\ell}_R^*\tilde{\ell}_R \nonumber \\
% Trilinear Terms
&   &-\epsilon_{ij}\left( 
     -\lambda_u A_U H_2^i \wtilde{Q}^j \tilde{u}_R^* 
     +\lambda_d A_D H_1^i \wtilde{Q}^j \tilde{d}_R^* 
     +\lambda_{\ell} A_E H_1^i \wtilde{L}^j \tilde{\ell}_R^* 
    \right)\;,
\end{eqnarray}
where
\begin{eqnarray}
Q   & = & \left( \begin{array}{c} u_L \\ d_L
               \end{array} \right), \;\;
L     = \left( \begin{array}{c} \nu_L \\ \ell_L
               \end{array} \right), \;\;
H_1   = \left( \begin{array}{c} H_1^0 \\ H_1^-
               \end{array} \right), \;\; {\rm and} \;\;
H_2   = \left( \begin{array}{c} H_2^+ \\ H_2^0
               \end{array} \right), \nonumber \\
\wtilde{Q} & = & \left( \begin{array}{c} \tilde{u}_L \\ \tilde{d}_L
               \end{array} \right), \;\;
\wtilde{L}   = \left( \begin{array}{c} \tilde{\nu}_L \\ \tilde{\ell}_L
               \end{array} \right), \;\;
\wtilde{H}_1 = \left( \begin{array}{c} \wtilde{H}_1^0 \\ \wtilde{H}_1^-
               \end{array} \right), \;\; {\rm and} \;\;
\wtilde{H}_2 = \left( \begin{array}{c} \wtilde{H}_2^+ \\ \wtilde{H}_2^0
               \end{array} \right)\;, 
\end{eqnarray}
$U, u = u,\; c$ and $t$; $D, d = d,\; s$ and $b$;
$E, \ell = e,\; \mu$ and $\tau$; 
and $\epsilon_{ij}$ is the antisymmetric tensor 
with $i,j = 1,\; 2$ and $\epsilon_{12} = 1$. 
The tree-level Yukawa couplings are defined by
\begin{eqnarray}
\lambda_u={{\sqrt{2}m_u}\over {v\sin \beta}}\;, \qquad
\lambda_d={{\sqrt{2}m_d}\over {v\cos \beta}}\;, \qquad
\lambda_{\ell}={{\sqrt{2}m_{\ell}}\over {v\cos \beta}}\;,
\end{eqnarray}
where $\tan \beta =v_2/v_1$ is the ratio of the vacuum expectation values
of $H_2^0$ and $H_1^0$.
The $\mu $ term in the superpotential contributes to the Higgs potential which
at tree level is
\begin{eqnarray}
V_{\rm Higgs}
& = &(m_{H_1}^2+\mu ^2)|H_1|^2+(m_{H_2}^2+\mu^2)|H_2|^2
-m_3^2(\epsilon_{ij}{H_1}^i{H_2}^j+{\rm H.c.}) \nonumber \\
&   &+{1\over 8}(g^2+g^{\prime 2})\left [|H_1|^2-|H_2|^2\right ]^2
+{1\over 2}g^2|H_1^{i*}H_2^i|^2\;, \label{tree}
\end{eqnarray}
where $m_{H_1}^{}$, $m_{H_2}^{}$, and
$m_3$ are soft-supersymmetry breaking parameters. We shall
define as usual the soft Higgs mass parameters
\begin{eqnarray}
m_1^2 & = & m_{H_1}^2+\mu ^2\;, \nonumber \\
m_2^2 & = & m_{H_2}^2+\mu ^2\;, \nonumber \\
m_3^2 & = & B\mu\;. 
\end{eqnarray}
Of the eight degrees of freedom in the two Higgs doublets, 
three Goldstone bosons ($G^{\pm}$, $G^0$) 
are absorbed to give the $W^{\pm}$ and the $Z$ masses, 
leaving five physical Higgs bosons: 
the charged Higgs bosons $H^{\pm}$, 
the CP-even Higgs bosons $h^0$ (lighter) and $H^0$ (heavier), and
the CP-odd Higgs boson $A^0$.

%------------------------------------------------
% The Mass Matrices
%------------------------------------------------
\subsection{The Mass Matrices}

In this section, the mass matrices of gauginos, top squarks, bottom squarks 
and the tau sleptons are presented in the gauge eigenstates. 
%------------------------
% Charginos
%------------------------
The chargino mass matrix in the ($\wtilde{W}^+,\wtilde{H}^+$) basis is
\begin{equation}
{\cal M}_C=\left( \begin{array}{c@{\quad}c}
M_2 & \sqrt{2}M_W\sin \beta \\
\sqrt{2}M_W\cos \beta & +\mu
\end{array} \right)\;.
\end{equation}
This mass matrix is not symmetric and must be diagonalized 
by two matrices \cite{Haber}.

%------------------------
% Neutralinos
%------------------------

The neutralino mass matrix in the basis of 
($\wtilde{B},\wtilde{W}^3,\wtilde{H}^0_1,\wtilde{H}^0_2$) is
\begin{equation}
\arraycolsep=0.01in
{\cal M}_N=\left( \begin{array}{cccc}
M_1 & 0 & -M_Z\cos \beta \sin \theta_W^{} & M_Z\sin \beta \sin \theta_W^{}
\\
0 & M_2 & M_Z\cos \beta \cos \theta_W^{} & -M_Z\sin \beta \cos \theta_W^{}
\\
-M_Z\cos \beta \sin \theta_W^{} & M_Z\cos \beta \cos \theta_W^{} & 0 & -\mu
\\
M_Z\sin \beta \sin \theta_W^{} & -M_Z\sin \beta \cos \theta_W^{} & -\mu & 0
\end{array} \right)\;.
\end{equation}
This mass matrix is symmetric and can be diagonalized 
by a single matrix \cite{Haber}.

%------------------------
% Sfermions
%------------------------

The top squark, bottom squark and the tau slepton mass-squared matrices 
in the ($\tilde{f}_L,\tilde{f}_R$) basis are expressed as 
\begin{equation}
\arraycolsep=0.01in
{\cal M}_{\tilde{t}}^2 =  
\left( \begin{array}{c@{\quad}c}
m_{\wtilde{Q}}^2 + m_t^2+{1\over 6}
(4M_W^2-M_Z^2)\cos 2\beta & m_t(A_t-\mu \cot \beta )\\
m_t(A_t-\mu \cot \beta ) & 
m_{\tilde{t}_R}^2 + m_t^2-{2\over 3}(M_W^2-M_Z^2)\cos 2\beta
\end{array} \right)\;,
\end{equation}
\begin{equation}
\arraycolsep=0.01in
{\cal M}_{\tilde{b}}^2 =  
\left( \begin{array}{c@{\quad}c}
m_{\wtilde{Q}}^2 + m_b^2-{1\over 6}
(2M_W^2+M_Z^2)\cos 2\beta & m_b(A_b-\mu \tan \beta )\\
m_b(A_b-\mu \tan \beta ) & 
m_{\tilde{b}_R}^2 + m_b^2+{1\over 3}(M_W^2-M_Z^2)\cos 2\beta
\end{array} \right)\;,
\end{equation}
\begin{equation}
\arraycolsep=0.01in
{\cal M}_{\tilde{\tau}}^2 =  
\left( \begin{array}{c@{\quad}c}
m_{\wtilde{L}}^2 + m_{\tau}^2-{1\over 2}(2M_W^2-M_Z^2)\cos 2\beta &
m_{\tau}(A_{\tau}-\mu \tan \beta )\\
m_{\tau}(A_{\tau}-\mu \tan \beta ) & 
m_{\tilde{\tau}_R}^2+ m_{\tau}^2+(M_W^2-M_Z^2)\cos 2\beta
\end{array} \right)\;,
\end{equation}
which are diagonalized by orthogonal matrices with mixing angles
$\theta_{\tilde{t}}$, $\theta_{\tilde{b}}$, and $\theta_{\tilde{\tau}}$.
The mass eigenstate for the massive third generation sneutrino is
\begin{eqnarray}
M_{\tilde{\nu}}^2 &=& M_{\wtilde{L}}^2 + {1\over 2}{M_Z^2}\cos 2\beta\;.
\end{eqnarray}

In the mass eigenstates, we label the charginos, the neutralinos 
and the sfermions such that
\begin{itemize}
\item $m_{\tchi_1^\pm} < m_{\tchi_2^\pm}$, 
\item $m_{\tchi_1^0} < m_{\tchi_2^0} < M_{\tchi_3^0} < m_{\tchi_4^0}$, and
\item $m_{\tilde{t}_1} < m_{\tilde{t}_2}$, 
      $m_{\tilde{b}_1} < m_{\tilde{b}_2}$, 
  and $m_{\tilde{\tau}_1} < m_{\tilde{\tau}_2}$. 
\end{itemize}

%-----------------------------------------------------------------------
%   Mass Eigenstates
%-----------------------------------------------------------------------
\subsection{Mass Eigenstates of Top Squarks}

The mass eigenstates of the top squarks are defined as 
\begin{eqnarray}
\left( \begin{array}{c} \tilde{t}_1 \\ \tilde{t}_2 \end{array} \right)
& \equiv & 
\left( \begin{array}{rr} 
 \cos\theta_{\tilde{t}} & \sin\theta_{\tilde{t}} \\
-\sin\theta_{\tilde{t}} & \cos\theta_{\tilde{t}} 
\end{array} \right) 
\left( \begin{array}{c} \tilde{t}_L \\ \tilde{t}_R \end{array} \right), \\
\tilde{t}_L & = & \cos\theta_{\tilde{t}} \tilde{t}_1 
                 -\sin\theta_{\tilde{t}} \tilde{t}_2, \nonumber \\
\tilde{t}_R & = & \sin\theta_{\tilde{t}} \tilde{t}_1 
                 +\cos\theta_{\tilde{t}} \tilde{t}_2 
\label{eq:stlr} 
\end{eqnarray}

The mass matrix in the basis of ($\tilde{t}_L,\tilde{t}_R$) 
and its eigenvalues are 
\begin{eqnarray}
{\cal M}_{\tilde{t}}^2 & \equiv & \left( \begin{array}{ll} 
 m_{LL} & m_{LR} \\
 m_{LR} & m_{RR} 
\end{array} \right), \\
m_{\pm}^2 & = & \frac{1}{2} [ (m_{LL}+m_{RR}) \pm \Delta ], \nonumber \\
\Delta    & = & [ ( m_{LL}-m_{RR} )^2 +4m_{LR}^2 ]^{1/2} 
\end{eqnarray}
where
\begin{eqnarray} 
m_{LL} 
& = & m_{\tilde{Q}_L}^2 +m_t^2 +M_Z^2 \cos 2\beta ( +C_L ), \nonumber \\ 
m_{RR} 
& = & m_{\tilde{t}_R}^2 +m_t^2 +M_Z^2 \cos 2\beta ( -C_R ), \nonumber \\ 
m_{LR} & = & m_t( A_t -\mu\cot\beta ),
\label{eq:malr}
\end{eqnarray}
$C_L = 1/2 -\frac{2}{3} \sin^2 \theta_W$ and 
$C_R = -\frac{2}{3} \sin^2 \theta_W$.

Requiring $m_{\tilde{t}_1} < m_{\tilde{t}_2}$, we have 
\begin{eqnarray} 
m_{\tilde{t}_1}^2 
& = & m_-^2  
  =   m_t^2 +\frac{1}{2}( m_{\tilde{Q}_L}^2 +m_{\tilde{t}_R}^2 ) 
     +\frac{1}{4}( M_Z^2\cos 2\beta )
     -\frac{\Delta}{2}, \nonumber \\
m_{\tilde{t}_2}^2
& = & m_+^2  
  =   m_t^2 +\frac{1}{2}( m_{\tilde{Q}_L}^2 +m_{\tilde{t}_R}^2 ) 
     +\frac{1}{4}( M_Z^2\cos 2\beta )
     +\frac{\Delta}{2}, \\
\Delta
& = & \{ [ ( m_{\tilde{Q}_L}^2 -m_{\tilde{t}_R}^2 ) 
          +\frac{\cos 2\beta}{6}( 8M_W^2-5M_Z^2 ) ]^2 
          +4m_t^2( A_t -\mu\cot\beta )^2 \}^{1/2} \nonumber 
%\label{eq:mst2} 
\end{eqnarray}

The mixing angle $\theta_{\tilde{t}}$ 
can be evaluated from $\tan\theta_{\tilde{t}}$,
\begin{equation}
\tan \theta_{\tilde{t}} = \frac{  m_{\tilde{t}_1}^2 -m_{LL} }{ m_{LR} }. 
\label{eq:angle} 
\end{equation}
The conventions for the top squark mass eigenstates and the mixing angle 
can be generalized to other scalar fermions. 

%------------------------------------------------
% Bridging Various Conventions
%------------------------------------------------
\subsection{Bridging Various Conventions}

In this section, we list various conventions previously used 
by members in the mSUGRA working group.
To obtain the same SUSY mass spectrum with the conventions in this report, 
the following modifications can be applied for $\mu$ and $A_t$,  

\medskip

\begin{center}
\begin{tabular}{lll} 
\hline
Collaboration             & $\mu$           & $A_t$ \\
\hline
ISAJET                    &  the same        & the same \\
SPYTHIA                   &  the same        & the same \\
\hline
Arnowitt and Nath         &  the same        & $A_t \to -A_t$ \\
Baer and Tata             &  the same        & the same \\
Barger, Berger and Ohmann &  $\mu \to -\mu$  & the same \\
Carena and Wagner         &  the same        & the same \\
Chankowski and Pokorski   &  the same        & the same \\
Drees and Nojiri          &  the same        & $A_t \to -A_t$ \\
Ellis et al.              &  the same        & $A_t \to -A_t$ \\
Haber and Kane            &  the same        & the same \\
Langacker and Polonsky    &  $\mu \to -\mu$  & the same \\
Nilles                    &  $\mu \to -\mu$  & the same \\
Martin and Ramond         &  the same        & $A_t \to -A_t$ \\
Pierce et al.             &  $\mu \to -\mu$  & the same \\
Kunszt and Zwirner        &  $\mu \to -\mu$  & the same 
\end{tabular}
\end{center}
%------------------------------------------------
%   REFERENCES
%------------------------------------------------

%-----------------------------------------------------------------------
% END DOCUMENT
%-----------------------------------------------------------------------

%% file: Kao/rges.tex
%=======================================================================
% Rges.tex for the mSUGRA group of Physics at RUN II
% This manuscript is in LaTeX
%=======================================================================

\def\m0{$m_0$}                         %m_0
\def\mhalf{$m_{1/2}$}                     %m_{1/2}
\def\a0{$A_0$}                         %a_0
\def\sgnmu{sign($\mu$)}              %sign of mu
\def\yratio  {{Y_t \over Y_t^{ir}}}
\def\yratlo  {{Y_t/Y_t^{ir}}}
\def\ppbar{$p\overline{p} $}            %ppbar
\def\pbarp{$\overline{p}p $}            %pbarp
\def\qqbar{$q\overline{q}$}             %qqbar
            \def\ttbar{$t\overline{t}$}             %ttbar
\def\bbbar{$b\overline{b}$}             %bbbar
\def\epm{$e^+e^-$}                      %e+e-
\def\mpm{$\mu^+\mu^-$}                      %mu+mu-
\def\ino{\widetilde \chi}               %Ino
\def\winog{\widetilde \chi^\pm}               %Wino
\def\zinog{\widetilde \chi^0}               %Zino
\def\winol{\widetilde \chi^\pm_1}             %Wino1
\def\winoop{\widetilde \chi^+_1}              %Wino1+
\def\winoom{\widetilde \chi^-_1}              %Wino1-
\def\zinol{\widetilde \chi^0_1}               %Zino1
\def\winoh{\widetilde \chi^\pm_2}             %Wino2
\def\zinoh{\widetilde \chi^0_2}               %Zino2
\def\zinot{\widetilde \chi^0_3}               %Zino3
\def\zinof{\widetilde \chi^0_4}               %Zino4
\def\w3{\widetilde W_3}               %W3
\def\Bino{\widetilde B}               %Bino
\def\Wino{\widetilde W}               %Wino
\def\Zino{\widetilde Z}               %Zino
\def\photino{\widetilde \gamma}               %Photino
\def\higgsino{\widetilde H}               %Photino
\def\sbottom{\tilde b}             %sbottom
\def\stop{\tilde t}             %stop
\def\stau{\tilde\tau}             %stau
\def\slepton{\tilde\ell}             %slepton
\def\selectron{\tilde e}           %selectron
\def\squark{\widetilde Q}             %squark
\def\sneutrino{\tilde \nu}          % neutrino
\def\gluino{\tilde g}             %gluino
\def\gravitino{\widetilde G}          % gravitino
\def\LSP{LSP}                         %LSP
%\setlength{\parskip}{10pt}
%\setlength{\baselineskip}{11pt}

%\renewcommand{\baselinestretch}{1.1}

%-----------------------------------------------------------------------
% BEGIN DOCUMENT
%-----------------------------------------------------------------------

%=======================================================================
% TITLE PAGE
%=======================================================================

%-----------------------------------
%   Title
%-----------------------------------
\section{Renormalization Group Equation Evolution
and SUSY Particle Mass Spectra}

%=======================================================================
%   BEGIN MAIN TEXT
%=======================================================================

\subsection{Renormalization Group Equations}

The mSUGRA model is characterized by relatively few parameters (five) 
at a high energy near the Planck scale. To translate these parameters into
masses and mixings for the supersymmetric particles that might be observed 
in an experiment, one must apply the Renormalization Group (RG) equations 
to the parameters of the model. 
The five parameters are treated as boundary conditions for the coupled
set of RG equations, and the resulting gauge, Yukawa, and soft supersymmetry
breaking parameters defined with the running scale near the electroweak scale
enter into the MSSM Lagrangian, providing predictions and correlations 
between the various particles in the SUSY spectrum. Some recent analyses of 
the mass patterns in the mSUGRA model can be found in Refs.~\cite{masses,bbo2}.

The renormalization equations for the gauge couplings\footnote{ 
In the RG equations $g_1^2 = (5/3) (g')^2$, where $g'$ 
is the $U(1)$ gauge coupling in the standard model.}
and the Yukawa couplings to one-loop order are (with $t=\ln (Q/Q_0)$)
\begin{equation}
{{dg_i}\over {dt}}={b_ig_i^3\over{16\pi^2 }}\;,
\end{equation}
\begin{eqnarray}
{{d{\bf U}}\over {dt}}&=&{1\over {16\pi ^2}}
\Big [-\sum c_ig_i^2+3{\bf UU^{\dagger }}+{\bf DD^{\dagger }}
+{\bf Tr}[3{\bf UU^{\dagger }}]\Big ]{\bf U}\;,
\label{dUdt}
\end{eqnarray}
\begin{eqnarray}
{{d{\bf D}}\over {dt}}&=&{1\over {16\pi ^2}}
\Big [-\sum c_i^{\prime}g_i^2+3{\bf DD^{\dagger }}+{\bf UU^{\dagger }}
+{\bf Tr}[3{\bf DD^{\dagger }}+{\bf EE^{\dagger }}]\Big ]{\bf D}\;,
\label{dDdt}
\end{eqnarray}
\begin{eqnarray}
{{d{\bf E}}\over {dt}}&=&{1\over {16\pi ^2}}
\Big [-\sum c_i^{\prime \prime}g_i^2+3{\bf EE^{\dagger }}
+{\bf Tr}[3{\bf DD^{\dagger }}+{\bf EE^{\dagger }}]\Big ]{\bf E}\;,
\label{dEdt}
\end{eqnarray}
where
\begin{equation}
b_i=({33\over 5},1,-3) \;, \qquad
c_i=({13\over 15},3,{16\over 3}) \;, \qquad
c_i^{\prime}=({7\over 15},3,{16\over 3}) \;, \qquad
c_i^{\prime \prime}=({9\over 5},3,0) \;. 
\end{equation}

In the most general case shown above,
the evolution equations involve matrices. For example
the Yukawa couplings form three-by-three Yukawa matrices: ${\bf U}$ for the 
up-type quarks, ${\bf D}$ for the 
down-type quarks, and ${\bf E}$ for the 
charged leptons. Similarly the soft-supersymmetry breaking parameters form the 
matrices ${\bf M}^2_{Q_L}$, ${\bf M}^2_{U_R}$, ${\bf M}^2_{D_R}$,
${\bf M}^2_{L_L}$, and ${\bf M}^2_{E_R}$ giving masses to the scalar 
supersymmetric particles. Finally there are in general 
matrices for the trilinear soft-supersymmetry breaking ``A-terms'':
${\bf A_U}$, ${\bf A_D}$, and ${\bf A_E}$. 
Since soft SUSY breaking trilinear scalar interactions are given by
corresponding terms in the superpotential 
(with superfields set to scalar components) times an $A$-parameter, 
it turns out to be useful to define the combinations 
${\bf U_A}_{ij}\equiv {\bf A_U}_{ij}{\bf U}_{ij}$, etc. 
in the matrix version of the RG equations. 
Then the  evolution of the soft-supersymmetry parameters (with our convention
for signs) is given by the
following 
renormalization group equations\cite{bbo2,bbmr}
%\begin{mathletters}
\begin{eqnarray}
{{dM_i}\over {dt}}&=&{2\over {16\pi ^2}}b_ig_i^2M_i\;, \\
{{d{\bf U_A}}\over {dt}}&=&{1 \over {16\pi ^2}}
\Bigg [-\Big ({13\over 15}g_1^2+3g_2^2+{16\over 3}g_3^2\Big ){\bf U_A}
+2\Big ({13\over 15}g_1^2{M_1}+3g_2^2{M_2}
+{16\over 3}g_3^2{M_3}\Big ){\bf U}\nonumber \\
&&+\Big \{\Big [4({\bf U_AU^{\dagger }U})
+6{\bf Tr}({\bf U_AU^{\dagger }}){\bf U}\Big ]
+\Big [5({\bf UU^{\dagger }U_A})+3{\bf Tr}({\bf UU^{\dagger }}){\bf U_A}\Big ]
\nonumber \\
&&+2({\bf D_AD^{\dagger }U})+({\bf DD^{\dagger }U_A})\Big \}\Bigg ]\;, \\
{{d{\bf D_A}}\over {dt}}&=&{1 \over {16\pi ^2}}
\Bigg [-\Big ({7\over 15}g_1^2+3g_2^2+{16\over 3}g_3^2\Big ){\bf D_A}
+2\Big ({7\over 15}g_1^2{M_1}+3g_2^2{M_2}
+{16\over 3}g_3^2{M_3}\Big ){\bf D}\nonumber \\
&&+\Big \{\Big [4({\bf D_AD^{\dagger }D})
+6{\bf Tr}({\bf D_AD^{\dagger }}){\bf D}\Big ]
+\Big [5({\bf DD^{\dagger }D_A})+3{\bf Tr}({\bf DD^{\dagger }}){\bf D_A}\Big ]
\nonumber \\
&&+2({\bf U_AU^{\dagger }D})+({\bf UU^{\dagger }D_A})
+2{\bf Tr}({\bf E_AE^{\dagger }}){\bf D}+{\bf Tr}({\bf EE^{\dagger }})
{\bf D_A}
\Big \}\Bigg ]\;, \\
{{d{\bf E_A}}\over {dt}}&=&{1 \over {16\pi ^2}}
\Bigg [-\Big (3g_1^2+3g_2^2\Big ){\bf E_A}
+2\Big (3g_1^2{M_1}+3g_2^2{M_2}
\Big ){\bf E}\nonumber \\
&&+\Big \{\Big [4({\bf E_AE^{\dagger }E})
+2{\bf Tr}({\bf E_AE^{\dagger }}){\bf E}\Big ]
+\Big [5({\bf EE^{\dagger }E_A})+{\bf Tr}({\bf EE^{\dagger }}){\bf E_A}\Big ]
\nonumber \\
&&+6({\bf D_AD^{\dagger }E})+3({\bf DD^{\dagger }E_A})\Big \}\Bigg ]\;, \\
{{dB}\over {dt}}&=&{2\over {16\pi ^2}}\Big ({3\over 5}g_1^2M_1+3g_2^2M_2
+{\bf Tr}(3{\bf UU_A}+3{\bf DD_A}+{\bf EE_A})\Big )\;, \\
{{d\mu }\over {dt}}&=&{\mu \over {16\pi ^2}}\Big (-{3\over 5}g_1^2-3g_2^2
+{\bf Tr}(3{\bf UU^{\dagger }}+3{\bf DD^{\dagger }}
+{\bf EE^{\dagger }})\Big )\;, \\
{{dm_{H_1}^2}\over {dt}}&=&{2 \over {16\pi ^2}}
\Big (-{3\over 5}g_1^2M_1^2-3g_2^2M_2^2\nonumber \\
&&+3{\bf Tr}({\bf D}({\bf M}_{Q_L}^2+{\bf M}_{D_R}^2){\bf D^{\dagger}}
+m_{H_1}^2{\bf DD^{\dagger }}+{\bf D_AD_A^{\dagger }})\nonumber \\
&&+{\bf Tr}({\bf E}({\bf M}_{L_L}^2+{\bf M}_{E_R}^2){\bf E^{\dagger}}
+m_{H_1}^2{\bf EE^{\dagger }}+{\bf E_AE_A^{\dagger }})\Big )\;, \\
{{dm_{H_2}^2}\over {dt}}&=&{2 \over {16\pi ^2}}
\Big (-{3\over 5}g_1^2M_1^2-3g_2^2M_2^2\nonumber \\
&&+3{\bf Tr}({\bf U}({\bf M}_{Q_L}^2+{\bf M}_{U_R}^2){\bf U^{\dagger}}
+m_{H_2}^2{\bf UU^{\dagger }}+{\bf U_AU_A^{\dagger }})\Big )\;, \\
{{d{\bf M}_{Q_L}^2}\over {dt}}&=&{2 \over {16\pi ^2}}
\Big (-{1\over 15}g_1^2M_1^2-3g_2^2M_2^2-{16\over 3}g_3^2M_3^2\nonumber \\
&&+{1\over 2}[{\bf UU^{\dagger }}{\bf M}_{Q_L}^2+{\bf M}_{Q_L}^2
{\bf UU^{\dagger }}+2({\bf U}{\bf M}_{U_R}^2{\bf U^{\dagger }}
+m_{H_2}^2{\bf UU^{\dagger }}+{\bf U_A^{}U_A^{\dagger }})]\nonumber \\
&&+{1\over 2}[{\bf DD^{\dagger }}{\bf M}_{Q_L}^2+{\bf M}_{Q_L}^2
{\bf DD^{\dagger }}+2({\bf D}{\bf M}_{D_R}^2{\bf D^{\dagger }}
+m_{H_1}^2{\bf DD^{\dagger }}+{\bf D_A^{}D_A^{\dagger }})]\Big )\;, \\ 
{{d{\bf M}_{U_R}^2}\over {dt}}&=&{2 \over {16\pi ^2}}
\Big (-{16\over 15}g_1^2M_1^2-{16\over 3}g_3^2M_3^2\nonumber \\
&&+[{\bf U^{\dagger }U}{\bf M}_{U_R}^2+{\bf M}_{U_R}^2
{\bf U^{\dagger }U}+2({\bf U^{\dagger }}{\bf M}_{Q_L}^2{\bf U}
+m_{H_2}^2{\bf U^{\dagger }U}+{\bf U_A^{\dagger }U_A^{}})]\Big )\;, \\
{{d{\bf M}_{D_R}^2}\over {dt}}&=&{2 \over {16\pi ^2}}
\Big (-{4\over 15}g_1^2M_1^2-{16\over 3}g_3^2M_3^2\nonumber \\
&&+[{\bf D^{\dagger }D}{\bf M}_{D_R}^2+{\bf M}_{D_R}^2
{\bf D^{\dagger }D}+2({\bf D^{\dagger }}{\bf M}_{Q_L}^2{\bf D}
+m_{H_1}^2{\bf D^{\dagger }D}+{\bf D_A^{\dagger }D_A^{}})]\Big )\;, \\
{{d{\bf M}_{L_L}^2}\over {dt}}&=&{2 \over {16\pi ^2}}
\Big (-{3\over 5}g_1^2M_1^2-3g_2^2M_2^2\nonumber \\
&&+{1\over 2}[{\bf EE^{\dagger }}{\bf M}_{L_L}^2+{\bf M}_{L_L}^2
{\bf EE^{\dagger }}+2({\bf E}{\bf M}_{E_R}^2{\bf E^{\dagger }}
+m_{H_1}^2{\bf EE^{\dagger }}+{\bf E_A^{}E_A^{\dagger }})]\Big )\;, 
\label{ll} \\
{{d{\bf M}_{E_R}^2}\over {dt}}&=&{2 \over {16\pi ^2}}
\Big (-{12\over 5}g_1^2M_1^2\nonumber \\
&&+[{\bf E^{\dagger }E}{\bf M}_{E_R}^2+{\bf M}_{E_R}^2
{\bf E^{\dagger }E}+2({\bf E^{\dagger }}{\bf M}_{L_L}^2{\bf E}
+m_{H_1}^2{\bf E^{\dagger }E}+{\bf E_A^{\dagger }E_A^{}})]\Big )\;, \label{er}
\end{eqnarray}
%\end{mathletters}
%
The above set of differential equations govern the evolution 
of the gauge and Yukawa couplings and the soft supersymmetry
breaking parameters in the MSSM. 
The two-loop generalizations of these RG equations 
can be found in Ref.~\cite{yamada,mv,jj,rge-all}.
Recently advances have been made in the derivation of the RG equations 
for soft supersymmetry breaking parameters. 
The RG equations for these soft parameters can be derived 
in a simple and systematic way from the same anomalous dimensions 
that are required in the derivation of the RG equations for the gauge and Yukawa 
couplings\cite{soft}. This is in effect a much more straightforward way
to derive the soft RG equations, 
and the procedure is applicable to all orders in perturbation theory. 
The RG equations shown above are easily derived using this method.  

In the exact supersymmetric limit 
the soft supersymmetry breaking parameters vanish, 
and only the gauge and Yukawa couplings remain. 
Even when supersymmetry is broken softly 
by the addition of explicit symmetry breaking terms, 
the RG equations for the gauge and Yukawa couplings depend only on each other, 
not on the soft (dimensionful) supersymmetry breaking parameters.
Thus, at the one loop order, the evolution of the gauge and Yukawa couplings 
is independent of the details of how supersymmetry is broken. 
Of course, various thresholds (specified by sparticle masses near the weak scale) 
depend on the soft supersymmetry breaking parameters, 
so detailed predictions of the gauge and Yukawa couplings
requires a careful treatment of these thresholds and 
the matching of the running couplings across them.
These effects are formally of two loop order. 
In practice the matching procedure can be a complicated problem 
if one performs it in its full generality\cite{pierce}.
Since the supersymmetric spectrum remains uncertain, 
there will remain uncertainties at some level in our predictions 
at the electroweak scale based on a given model 
at the grand unified scale ($M_{\rm GUT}$). 
Furthermore there are thresholds at the grand unified scale 
which generally depend on the details of the grand unified theory. 
These GUT threshold effects translate into 
corrections to the boundary conditions at the high energy scale.
Threshold corrections arise when one integrates out heavy degrees of freedom 
to create an effective theory without the heavy particles. 
The threshold corrections correspond to matching of 
the two theories across the threshold. 
The most common method of implementing threshold corrections is to alter 
the RG equations at the mass scale of the heavy particles being integrated out. 
A consistent treatment for calculations involving the $n$-loop RG equations 
requires threshold corrections
to be calculated to the $(n-1)$-order. For example, a calculation using the 
three-loop gauge coupling $\beta$-functions for the MSSM\cite{fjj}, would 
require threshold corrections at the two-loop level, and at the present time
the threshold corrections have not been calculated to this order.
 
\subsection{Mass Spectra at the Weak Scale}

The mSUGRA model is defined in terms of only five parameters 
as boundary conditions at the high energy scale\footnote{
The scale at which these parameters are specified is an uncertainty. 
It is frequently chosen to be the grand unified scale ($M_{\rm GUT}$).} 
\begin{eqnarray}
m_0\;, \quad m_{1/2}\;, \quad A_0\;, \quad B_0\;, \quad {\rm and} \quad \mu_0.   
\end{eqnarray}
The first and second parameters give the boundary condition for the masses
of the supersymmetric spin-$0$ and spin-$1/2$ particles respectively. $A_0$ 
fixes a universal value for the trilinear couplings at the high energy scale. 
For phenomenologically viable choices of parameters, 
the large top quark Yukawa coupling causes $m_{H_2}^2$ becomes negative 
so that electroweak symmetry is spontaneously broken. 
At the electroweak scale, the value of $\mu^2$ is then fixed 
to yield the experimental value of $M_Z^2$, from minimizing 
the effective potential at the electroweak scale with respect to the Higgs VEVs 
$v_{1,2} = \langle H_{1,2}\rangle$ [Eq.~(\ref{treemin1})]. 
The discrete choice for the sign of the supersymmetric Higgs mass parameter 
$\mu$ [${\rm sign}(\mu)$] is not fixed by the condition 
of radiative electroweak symmetry breaking [see below Eq.~(\ref{treemin2})].
The bilinear parameter $B$ can be expressed as a function 
in terms Higgs masses and $\tan\beta \equiv v_2/v_1$. 
It is customary to trade the parameter $B$ for $\tan\beta$ [Eq.~(\ref{treemin2})]. 
As a result, the hybrid parameter set,
\begin{equation}
m_0\;, \quad m_{1/2}\;, \quad A_0\;, \quad \tan\beta\;, 
\quad {\rm and}\; \quad {\rm sign}(\mu) \; ,
\end{equation}
is commonly used to specify the mSUGRA model 
with radiative breaking of the electroweak symmetry.

Several lessons can be learned from the above structure of the RG equations.
Firstly to the one-loop order in the RG equations the gauge couplings and 
the gaugino masses evolve together, i.e. 
\begin{eqnarray}
&&M_i(Q)={{g_i^2(Q)}\over {g_i^2(Q_0)}}m_{1/2}\;.\label{mi} 
\end{eqnarray}
The gaugino fields have a common mass $m_{1/2}$ 
near the scale that the gauge couplings unify.  
At any scale, the gaugino masses have the following approximate relations 
\begin{eqnarray}
{{M_1}\over {g_1^2}}\simeq {{M_2}\over {g_2^2}}\simeq {{M_3}\over {g_3^2}}\;.
\end{eqnarray}
In particular, at the electroweak scale this relation gives 
a definite prediction for the gaugino mass parameters ($M_i$) 
derived from a common boundary condition $m_{1/2}$ at the high energy scale.
A detailed analysis of the corrections to this renormalization group invariant
from two-loop contributions to the RG equations, threshold corrections and
possible sources from outside the mSUGRA paradigm can be found in 
Ref.~\cite{kribs}.

Analytical expressions can be obtained for the squark and slepton
mass parameters when
the corresponding Yukawa couplings are negligible (i.e. for the first two 
generations). For a universal scalar mass
$m_0^{}$ and gaugino mass $m_{1/2}^{}$ 
at the GUT scale, the expressions are \cite{il} 
\begin{eqnarray}
m_{\tilde {f}}^2&=&m_0^2+\sum _{i=1}^3f_im_{1/2}^2+(T_{3,\tilde{f}}
-e_{\tilde{f}}\sin ^2\theta _w)M_Z^2\cos 
2\beta \;, \label{fi} 
\end{eqnarray}
for the squark and slepton masses, where the $f_i$ are (positive) constants 
that depend on the evolution of the gauge couplings 
\begin{eqnarray}
f_i&=&{{d_i({f})}\over {b_i}}\left [1-{1\over 
{\left ({1-{{\alpha _G^{}}\over {2\pi }}b_it}\right )^2}}\right ]\;.
\end{eqnarray}
There is a contribution $f_i$ from each interaction of the 
$SU(3)\times SU(2)\times U(1)$ of the Standard Model. 
The last term in Eq.~(\ref{fi}) is the $D$-term contribution to the scalar 
masses, and $T_{3,\tilde{f}}$ is the SU(2) quantum number and 
$e_{\tilde{f}}$ is the electromagnetic charge of the sfermion.
The $b_i$ are given in section 1. 
The $d_i({f})$ is ${{N^2-1}\over {N}}$ for fundamental representations 
(zero for singlet representations) of $SU(N)$, 
and ${3\over {10}}Y^2$ for $U(1)_Y^{}$.
The squark mass spectrum of the third generation is more complicated 
for two reasons: (1) the effects of the third generation Yukawa couplings 
need not be negligible, and (2) there can be substantial mixing 
between the left and right top squark fields 
(and left and right bottom squark fields for large $\tan\beta$) 
so that the mass eigenstates are liner combinations 
of the left and right sfermion fields. 
Notice in particular that colored scalar supersymmetric particles 
will get the largest masses because of their QCD interactions. 
The universal boundary conditions in mSUGRA at the GUT scale
yields correlations between the masses in the supersymmetric spectrum through
Eqs.~(\ref{mi}) and (\ref{fi}).

The sleptons have only EW quantum numbers, and the lepton Yukawa couplings
are small for $\tan\beta < 40$. 
The relations at the scale $M_{EW}$ which determine the mass eigenstates
are given by\footnote{
We have omitted to write the contribution from the last term in Eq.~(\ref{fi}).}
\begin{equation}
m_{L_{1,2}}^2 \simeq  m_{L_3}^2 \simeq m_0^2 + 0.5 m^2_{1/2} 
\,; \qquad
m_{E_{1,2}}^2 \simeq  m_{E_3}^2 \simeq m_0^2 + 0.15 m^2_{1/2}.
\label{eq:slep_mass}
\end{equation}
The only Yukawa coupling that might be important in the evolution of 
the slepton masses is the tau Yukawa coupling when $\tan\beta \ge 40$. 
In that case, the third generation slepton mass parameters 
also receive non--negligible contributions 
in their running which can modify these expressions.
If \m0~and \mhalf~are of the same order of magnitude, 
physical slepton masses are dominately given by \m0.  
The $\sneutrino$ mass is fixed by a sum rule
\begin{equation} 
m_{\sneutrino_{\ell}}^2 = m_{\slepton_L}^2 + M_W^2\cos 2\beta\;,
\end{equation}
which follows directly from $SU(2)$ gauge symmetry.

The squark mass parameters have a stronger dependence on 
the common gaugino mass $m_{1/2}$ because of color. 
For the first and second generation squarks,
the left-- and right--handed soft SUSY--breaking parameters at $M_{EW}$
are given approximately by
\begin{equation}
\label{eq:approxmQU}
m_{Q_{1,2}}^2 \simeq  m_0^2 + 6.3 m^2_{1/2} 
\,; \qquad
m_{U_{1,2}}^2 \simeq  m_{D_{1,2}}^2 \simeq m_0^2 + 5.8 m^2_{1/2}\,.
\end{equation}
In general, the squarks are heavier than the sleptons 
and the lightest neutralino and chargino.
The first and the second generation squark soft SUSY--breaking parameters 
have the same value for squarks with the same quantum numbers 
since the contributions from the Yukawa couplings is negligible. 

For the third--generation squarks, the large
top and bottom Yukawa couplings
play a crucial role in the RG equation evolution.
When the top quark Yukawa coupling $h_t$ at the grand unified scale is
sufficiently large, its low--energy value is independent of
its exact value at $M_{GUT}$ because the top quark Yukawa coupling 
has an infrared fixed point \cite{IRFP}.
With the definition $\displaystyle Y_t\equiv h_t^2 / (4\pi)$,
the infrared fixed point value of $Y_t$ at the scale $m_t$ is
$Y_t^{ir}\simeq 8\alpha_3/9$.
Within the one--loop approximation, the effects
of the top Yukawa coupling on the RG equation evolution can
be parameterized in terms of the ratio $\yratlo$.
For small and moderate values of $\tan\beta$,
the left-- and right--handed soft SUSY--breaking parameters which determine 
the stop and sbottom masses are then given by\,\cite{COPW-C,COPW2,Drees}
\begin{eqnarray}
m_{Q_3}^2 &\simeq& {m_0^2}\left(1-\frac{1}{2}\yratio\right) +
  m^2_{1/2}\left(6.3+\yratio\left(-\frac{7}{3} + \yratio\right)\right) 
\nonumber \\
m_{U_3}^2 &\simeq& {m_0^2}\left(1-\yratio\right) +
               m^2_{1/2}\left(5.8+\yratio\left(-\frac{14}{3}+
2\yratio\right)\right),
\label{eq:soft_3gen}
\end{eqnarray}
and $m_{D_3} \simeq  m_{D_{1,2}}$.
For large $\tan\beta$, assuming $t-b$ Yukawa coupling unification 
at high energies ($Y_b=Y_t$ at $M_{GUT}$, which is a generic
prediction of $SO(10)$ $GUT$ models), 
the expressions for the third generation
soft SUSY--breaking parameters 
are:\,\cite{COPW2}
\begin{eqnarray}
m_{Q_3}^2 &\simeq& {m_0^2}\left(1-\frac{6}{7}\yratio\right) +
m^2_{1/2}\left(6.3+\yratio\left(-4 + \frac{12}{7}\yratio\right)\right) 
\nonumber \\
m_{U_3}^2 \simeq m_{D_3}^2 &\simeq& {m_0^2}\left(1-\frac{6}{7}\yratio\right) + 
m^2_{1/2}\left(5.8-\yratio\left(-4+\frac{12}{7}\yratio\right)\right).
\label{eq:soft_3gen_tanb}
\end{eqnarray}
Contributions proportional to $A_0^2$ and $A_0 m_{1/2}$ with a 
prefactor proportional to $(1-\yratlo)$
are also present in Eqs.~(\ref{eq:soft_3gen}) and (\ref{eq:soft_3gen_tanb}).
For $m_t \simeq 175$ GeV, the value of 
the ratio $\yratlo$ varies from $3/4$ to $1$ depending on
$\tan\beta$, with $ \yratlo \to 1$ as $\tan\beta\to 1$, and 
$\yratlo \simeq $ 0.85 for $\tan\beta=40$.
The value of $A_t$ is governed by $m_{1/2}$,
and, for large values of
the top Yukawa coupling, depends weakly on its 
initial value and $\tan\beta$ \cite{COPW-C},
\begin{equation}
A_t \simeq \left(1-\yratio\right)A_0 - 2 m_{1/2}.
\end{equation}
The exact values of $A_b$ and $A_{\tau}$ are not important, since
the mixing in the stau and sbottom sectors is governed by the terms 
$m_b \mu\tan\beta$ and $m_\tau \mu\tan\beta$, respectively.
In SUGRA models,
the above relations between the mass parameters 
leads to the general prediction, $m_{\squark} \ge
0.85 M_{\gluino}$ (for the five lightest squarks and small or moderate
$\tan\beta$).

\subsection{The Higgs Sector}

The soft--SUSY breaking parameters in the Higgs sector also have
simple expressions.
For small and moderate $\tan\beta$, \cite{COPW-C}
\begin{eqnarray}
m_{H_1}^2 \simeq m_0^2 + 0.5 m^2_{1/2} \,, \qquad
m_{H_2}^2 \simeq {m_0^2}\left(1-\frac{3}{2}\yratio\right) +
m^2_{1/2}\left(0.5+\yratio\left(-7+3\yratio\right)\right).
\label{eq:higgs_parameters}
\end{eqnarray}

The Higgs mass parameters that enter into the Higgs potential in the mSUGRA
model also assume the boundary condition $m_0$ at the high energy scale,
and their subsequent evolution breaks the electroweak symmetry in a radiative
fashion.
The Higgs potential which 
at tree level is 
\begin{eqnarray}
V_0&=&(m_{H_1}^2+\mu ^2)|H_1|^2+(m_{H_2}^2+\mu^2)|H_2|^2
-m_3^2(\epsilon_{ij}{H_1}^i{H_2}^j+{\rm h.c.})\nonumber \\
&&+{1\over 8}(g^2+g^{\prime 2})\left [|H_1|^2-|H_2|^2\right ]^2
+{1\over 2}g^2|H_1^{i*}H_2^i|^2\;, \label{rge-tree}
\end{eqnarray}
where $m_{H_1}^{}$, $m_{H_2}^{}$, and 
$m_3^2=B\mu$ are soft-supersymmetry breaking parameters. 
The soft Higgs mass $m_1^2$ and $m_2^2$ parameters are defined as 
\begin{eqnarray}
m_1^2&=&m_{H_1}^2+\mu ^2\;, \\
m_2^2&=&m_{H_2}^2+\mu ^2\;.
\end{eqnarray}
Figure 1 shows a typical evolution of the soft-supersymmetry
breaking parameters. The characteristic
behavior exhibited by the mass parameters are typical
of renormalization group equation evolution. The colored particles 
are generally driven heavier at low $Q$
by the large strong gauge coupling. The 
Higgs mass parameter $m_2^2$ is usually driven negative, 
giving the electroweak 
symmetry breaking. 

\begin{figure}[h]
\epsfxsize=4in
\centering
\hspace*{0in}
\epsffile{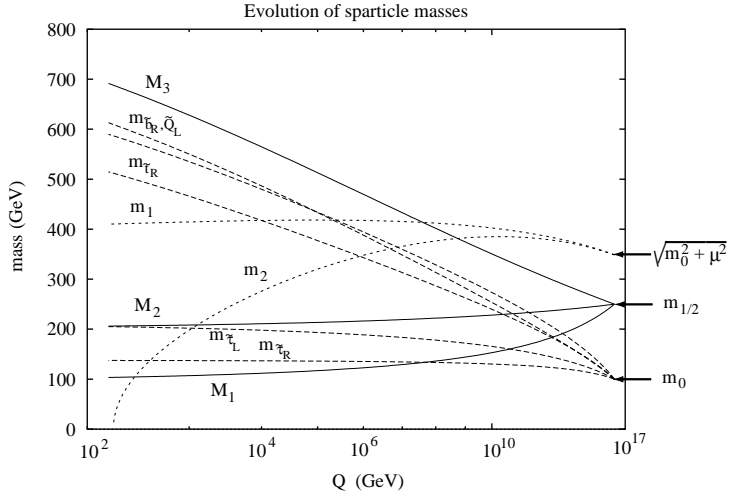}

\caption[]{An example (from Ref.~\cite{bbo2}) 
of the running of the 
soft-supersymmetry breaking parameters with $Q$. The particle which feel
the strong interaction are typically heavier than the non-colored ones, and
the Higgs mass parameter
$m_2^2$ is driven negative giving rise to breaking of the electroweak 
symmetry. }
\end{figure}
 
Minimizing the tree level Higgs potential $V_0$ one obtains 
the familiar tree-level conditions
\begin{eqnarray}
{1\over 2}M_Z^2&=&{{m_{H_1}^2-m_{H_2}^2\tan ^2\beta }
\over {\tan ^2\beta -1}}-\mu ^2 \;, \label{treemin1}
\end{eqnarray}
and
\begin{eqnarray}
B\mu &=&{1\over 2}(m_{H_1}^2+m_{H_2}^2+2\mu ^2)\sin 2\beta \;.
\label{treemin2}
\end{eqnarray}
Notice that the signs of $B$ and $\mu $ are not determined by
the minimization conditions
giving rise to two distinct cases mentioned above.

Substituting Eq.~(\ref{eq:higgs_parameters}) into Eq.~(\ref{treemin1}) yields
the result:
\begin{eqnarray}
\mu^2 + M_Z^2/2 = m_0^2 \left(1 + \left(\frac{3}{2} \yratio -1\right) \tan^2
\beta\right)\frac{1}{ \tan^2
\beta -1} + \nonumber \\
        m_{1/2}^2 \left(0.5 -
\left(0.5+\yratio\left(-7+3\yratio\right)\right)\tan^2 \beta\right)
\frac{1}{ \tan^2 \beta -1}.
\label{eq:EWSB}
\end{eqnarray}
Note in Eq.~(\ref{eq:higgs_parameters}) that $m_{H_2}^2<0$, which
is usually a sufficient condition to induce EWSB.
For large $\tan\beta$, the Higgs mass parameters
are more complicated.  In the limit of $t-b$ Yukawa unification,
they simplify to\,\cite{COPW2}
\begin{equation}
m_{H_1}^2 \simeq 
m_{H_2}^2 \simeq {m_0^2}\left(1-\frac{9}{7}\yratio\right) + m^2_{1/2}
\left(0.5+\yratio\left(-6+\frac{18}{7}\yratio\right)\right),
\end{equation}
and Eq.~(\ref{eq:EWSB}) must be modified accordingly.
All of these relations are only approximate:  the coefficients of
\mhalf~depend on the exact values of $\alpha_s$ and the scale of
the sparticle masses; the coefficients of \m0~and $A_0$ depend
mainly on $\tan\beta$. 

Given the mSUGRA boundary conditions, the RG equations give the values 
of the parameters in the Higgs potential $V_0$, and hence in the minimization
conditions in Eqs.~(\ref{treemin1}) and (\ref{treemin2}). The Higgs potential
can depend strongly on the scale $Q$ at which it is evaluated.
It was first argued in Ref.~\cite{grz} that this scale dependence
resulting from the RG equations can be compensated by including
the one-loop contributions to the
Higgs potential.
The one-loop effective potential is given by 
\begin{eqnarray}
V_1&=&V_0+\Delta V_1\;,
\end{eqnarray}
\begin{eqnarray}
\Delta V_1&=&{1\over {64\pi ^2}}{\rm Str}\left [{\cal M}^4
\left (\ln {{{\cal M}^2}\over {Q^2}}-{3\over 2}\right )\right ]\;, \label{V1}
\end{eqnarray}
is the one-loop contribution given in the dimensional reduction 
($\overline{DR}$) renormalization scheme\cite{dimred}. 
The supertrace is defined as
${\rm Str} f({\cal M}^2)=\sum _iC_i(-1)^{2s_i}(2s_i+1)f(m_i^2)$ where
$C_i$ is the color degrees of freedom and $s_i$ is the spin of the
$i^{th}$ particle.
To determine the minimum one must set the first derivatives of the 
effective potential to zero
\begin{eqnarray}
{{\partial V_1}\over {\partial \psi }}=
{{\partial V_0}\over {\partial \psi }}+
{1\over {32\pi ^2}}{\rm Str}\left [
{{\partial {\cal M}^2}\over {\partial \psi }}
{\cal M}^2
\left (\ln {{{\cal M}^2}\over {Q^2}}-1\right )
\right ]=0\;. \label{V1d1}
\end{eqnarray}
One way to incorporate these corrections into the one-loop 
generalizations of Eqs.~(\ref{treemin1})
and (\ref{treemin2}) is to set to zero the appropriate linear combinations of
Higgs boson tadpole diagrams\cite{bbo2}. 
The corrections to the one-loop minimization 
conditions can then be completely written in terms of analytical formulas.
Evaluated at the minimum of $V_1$,
tadpole contributions involve the coupling $\partial {\cal M}^2/\partial\psi$ 
and the usual integration factor 
${1\over {32\pi ^2}}{\cal M}^2\left (\ln {{{\cal M}^2}\over {Q^2}}-1\right )$;
setting tadpole contributions to zero is therefore equivalent to minimizing
the potential.

The Higgs masses are given at tree-level by
\begin{eqnarray}
M_A^2&=&m_{H_1}^2+m_{H_2}^2+2\mu^2\;, \\
M_{H,h}^2&=&{1\over 2}\left [M_Z^2+M_A^2\pm \sqrt{(M_Z^2+M_A^2)^2
-4M_Z^2M_A^2\cos ^22\beta}\right ]\;, \\
M_{H^\pm}^2&=&M_W^2+M_A^2\;.
\end{eqnarray}
The one-loop corrections in Eq.~(\ref{V1}) give corrections to the Higgs 
masses with the most important ones of order $m_t^4$\cite{CPeven}.

%
% Mass of the CP-odd Higgs pseudoscalar: $A^0$
%

In the MSSM, the Higgs potential is constrained by supersymmetry 
such that all tree-level Higgs boson masses and couplings 
are determined by just two independent parameters,  
commonly chosen to be the mass of the CP-odd pseudoscalar ($m_A$) 
and the ratio of vacuum expectation values (VEVs) of Higgs fields 
($\tan\beta \equiv v_2/v_1$). 

In the mSUGRA, mass of the CP-odd Higgs pseudoscalar ($m_A$) 
corresponding to the one loop effective potential has the following form:
\begin{eqnarray}
m^2_A      & = & (\tan\beta+\cot\beta) ( B\mu +\Delta^2_A ),   \nonumber \\
\Delta^2_A 
& = & \frac{3g^2}{32\pi^2}\frac{\mu A_t m^2_t}{M^2_W\sin^2\beta}
\left[ \frac{ f(m^2_{\tilde{t}_2})-f(m^2_{\tilde{t}_1}) }
     { m^2_{\tilde{t}_2} -m^2_{\tilde{t}_1} }\right] \nonumber \\
&   &+\frac{3g^2}{32\pi^2}\frac{\mu A_b m^2_b}{M^2_W\cos^2\beta}
\left[ \frac{ f(m^2_{\tilde{b}_2})-f(m^2_{\tilde{b}_1}) }
            { m^2_{\tilde{b}_2} -m^2_{\tilde{b}_1} } \right] \nonumber \\
&   &+\frac{ g^2}{32\pi^2}\frac{\mu A_\tau m^2_\tau}{M^2_W\cos^2\beta}
\left[ \frac{ f(m^2_{\tilde{\tau}_2})-f(m^2_{\tilde{\tau}_1}) }
            { m^2_{\tilde{\tau}_2} -m^2_{\tilde{\tau}_1} } \right] \, , 
\end{eqnarray}
where 
\begin{equation}
f(m^2) = m^2\left[ \ln(\frac{m^2}{Q^2})-1 \right] \, .
\end{equation}
 
We suggest that the Higgs-boson masses and couplings 
should be calculated with leading one loop corrections 
not only from the top Yukawa interactions 
but also from the bottom and the tau Yukawa interactions 
in the one-loop effective potential \cite{CPeven},
and should be evaluated at a scale close to 
$Q = \sqrt{m_{\tilde{t}_L}m_{\tilde{t}_R}}$ \cite{rge-Baer}. 
With this scale choice, 
the numerical value of the CP-odd Higgs boson mass ($m_A$) 
at large $\tan\beta$ \cite{rge-Baer,Madison} 
is relatively insensitive to the exact scale 
and the loop corrections to $M_A$ are small compared 
to the `tree level' contribution.
In addition, when this high scale is used, 
the RGE improved one-loop corrections approximately 
reproduce the dominant two loop perturvative calculation 
of the mass of the lighter CP-even Higgs scalar ($m_h$).  
The numerical values of $m_h$ calculated at this scale 
can be very close to the results of Ref.~\cite{Two-Loop} 
where somewhat different scales higher than $M_Z$ 
have been adopted in evaluating the effective potential.

%-----------------------------------------------------------------------
%   REFERENCES
%-----------------------------------------------------------------------

%-----------------------------------------------------------------------
%   END DOCUMENT
%-----------------------------------------------------------------------

%% file: Zhang/wgnew.tex
\def\gsim{\raise.3ex\hbox{$>$\kern-.75em\lower1ex\hbox{$\sim$}}}
\def\lsim{\raise.3ex\hbox{$<$\kern-.75em\lower1ex\hbox{$\sim$}}}

\section{Radiative Corrections to Particle Masses}
%\footnote{Subgroup members: J.A. Bagger, C. Kao, K.T. Matchev,
%D.M. Pierce, and R.-J. Zhang}

\subsection{Introduction}

In this section, we present approximations
for the one-loop supersymmetric radiative corrections
to particle masses in the MSSM.  These formulae have
been tested in the mSUGRA model and are typically good to
$1-2\%$.  The calculations
are performed in the dimensional reduction scheme
where the supersymmetric Ward-Takahashi identities
are ensured to at least the two-loop
order. The divergences are removed by the (modified)
minimal subtraction procedure, which we shall call
the $\overline{\rm DR}$ scheme \cite{drscheme}.

The full set of one-loop corrections
and the approximations presented here
can be found in Ref.~\cite{BMPZ}. We have changed some of
the notation to conform with the SUGRA Working Group conventions.
For example, we adopt the Working Group
convention on the sign of $\mu$-parameter, and relabel $\tilde t_1$
to be the lighter top squark, etc.
In some regions of the parameter space, these corrections can be 
important. We suggest them be included in event
simulation programs such as the ISAJET \cite{z-isajet} and the
SPYTHIA \cite{spythia}, which currently have only leading 
logarithmic contribution in RG evolutions.

We express the one-loop integrals for the self-energy diagrams
in terms of the standard
Passarino-Veltman functions \cite{PVfunc}, which are defined as
\begin{eqnarray}
&& A(m) = -i 16\pi^2 Q^{4-D}\int{d^Dk\over (2\pi)^D}{1\over k^2-m^2}\\[2mm]
&& B_0(p,m_1,m_2) = -i 16\pi^2 Q^{4-D}
	\int{d^Dk\over (2\pi)^D}
	{1\over\left[k^2-m_1^2\right]\left[(k-p)^2-m_2^2\right]}\\[2mm]
&& p_\mu B_1(p,m_1,m_2) = -i 16\pi^2 Q^{4-D}
	\int{d^Dk\over (2\pi)^D}
	{k_\mu\over\left[k^2-m_1^2\right]\left[(k-p)^2-m_2^2\right]}\\[2mm]
&& g_{\mu\nu}B_{22}(p,m_1,m_2) + p_\mu p_\nu B_{21}(p,m_1,m_2) %\nonumber\\
 = -i 16\pi^2 Q^{4-D}
	\int{d^Dk\over (2\pi)^D}
	{k_\mu k_\nu\over\left[k^2-m_1^2\right]\left[(k-p)^2-m_2^2\right]}
\end{eqnarray}
where $Q$ is the renormalization scale.
In certain limits, these functions have simple forms.
For example, we shall take
\begin{equation}
B_{0}(0,m_1,m_2) = - \ln\left(M^2\over Q^2\right) + 1 + {m^2\over
m^2-M^2}\ln\left(M^2\over m^2\right)\ ,
\label{b0(0)}
\end{equation}
\begin{equation}
B_{1}(0,m_1,m_2) = {1\over2}\left[- \ln\left(M^2\over Q^2\right)
+ {1\over2} + {1\over 1-x} + {\ln x \over (1-x)^2} - \theta(1-x)\ln x
\right]\ ,
\label{b1(0)}
\end{equation}
where $M=\max(m_1,m_2),$ $m=\min(m_1,m_2),$ $x=m_2^2/m_1^2$, and
\begin{equation}
B_1(p,0,m) = -{1\over2}\ln\left({M^2\over Q^2 }\right) + 1 - {1\over
2y} \left[ 1+{(y-1)^2\over y}\ln |y-1| \right] +
{1\over2}\theta(y-1)\ln y\ ,
\label{b1_p0m2}
\end{equation}
where $M = \max(p^2,m^2)$ and $y=p^2/m^2$.

These formulae can also be used in an effective field theory approach
to the particle masses.  When the SUSY particles are heavy compared
to the weak scale, it is appropriate to decouple them at an effective
SUSY scale, $M_{\rm SUSY}$.  Below $M_{\rm SUSY}$, the couplings
should then be run in an effective theory which contains just the
light SM particles.  This is, in fact, the approach used by many groups
\cite{RGrunning}.  Of course, in this approach, it is still necessary
to include some enhanced finite threshold corrections.
These corrections can be extracted from the following formulae by
omitting the terms which contain logarithms of the
renormalization scale.

\subsection{Quark and lepton masses}

The quark and lepton masses contain SM and SUSY radiative corrections.
In some cases, they can be important or even dominant. The following
approximations contain the most important contributions.

\subsubsection{Top quark mass}

The precise calculation of the running top quark mass has both
theoretical and experimental significance. For example, radiative
electroweak symmetry breaking and the
radiative correction to the lightest
Higgs boson mass both depend on the value of the running top quark
mass, $m_t(Q)$.  We found the following formula approximates the
full one-loop radiative correction to about $\pm 1\%$,
\begin{equation}
M_t = m_t(Q)
\left[ 1 +\left({\Delta m_t \over m_t}\right)^{\rm QCD}
+\left({\Delta m_t \over m_t}\right)^{\rm SUSY}\right]
\label{mtphys}\ ,
\end{equation}
where
\begin{equation}
\left({\Delta m_t \over m_t}\right)^{\rm QCD} =
{\alpha_s(Q)\over 3\pi} \left[3\ln\left({Q^2\over M_t^2}\right)+5\right]
\label{mtcorr}\ .
\end{equation}
and \cite{BMPZ,donini}
\begin{eqnarray}
\left({\Delta m_t\over m_t}\right)^{\rm SUSY} &=& -\
{\alpha_s(Q)\over 3\pi}\Bigg\{ B_{1}(m_t,m_{\tilde g},m_{\tilde t_1}) +
B_{1}(m_t,m_{\tilde g},m_{\tilde t_2}) \nonumber\\ 
&& - {2m_{\tilde g}(A_t-\mu\cot\beta)\over m_{\tilde t_1}^2-m_{\tilde t_2}^2}
\Bigg[B_{0}(m_t,m_{\tilde g},m_{\tilde t_1}) - B_{0}(m_t,m_{\tilde
g},m_{\tilde t_2})\Bigg] \Bigg\}\ .\nonumber \\
\label{mtgluino}
\end{eqnarray}
The SUSY QCD contribution (\ref{mtgluino}) can be as large as the SM
QCD  contribution (\ref{mtcorr}) for TeV-scale gluino and squark masses.

\subsubsection{Bottom quark mass}

Corrections to the bottom quark mass in the MSSM have received much
attention because they can contain significant enhanced supersymmetric
contributions in the region of large $\tan\beta$ \cite{bottomcorr}.
The best way to proceed is to first find the running bottom quark mass
at $M_Z$ in the SM.  This is related to running bottom quark mass at
$M_b$ by the SM RG equations.  At the two-loop level, one finds \cite{Gray}
\begin{equation}
m_b(M_b)^{\rm SM}=M_b\left[1+{5\alpha_s(M_b)\over3\pi}+12.4\left(
{\alpha_s(M_b)\over\pi}\right)^2\right]^{-1}\ .
\end{equation}
The running bottom quark mass at $M_Z$ in the MSSM is then
\begin{equation}
m_b(M_Z)^{\rm MSSM} = m_b(M_Z)^{\rm SM} \left[1 - \left( {\Delta
m_b \over m_b}\right)^{\rm SUSY}\right]\ .
\end{equation}
where
\begin{eqnarray}
\left({\Delta m_b\over m_b}\right)^{\rm SUSY} &=&
-\ {\alpha_s(M_Z)\over 3\pi}\Bigg\{ B_{1}(0,m_{\tilde g},m_{\tilde b_1}) +
B_{1}(0,m_{\tilde g},m_{\tilde b_2}) \nonumber\\ &-&
{2m_{\tilde g}(A_t-\mu\tan\beta)\over m_{\tilde b_1}^2-m_{\tilde b_2}^2}
\Bigg[B_{0}(0,m_{\tilde g},m_{\tilde b_1}) - B_{0}(0,m_{\tilde
g},m_{\tilde b_2})\Bigg] \Bigg\}\nonumber\\
&-&{\lambda_t^2\over16\pi^2}{\mu\ (A_t\tan\beta-\mu) \over m_{{\tilde
t}_1}^2-m_{{\tilde t}_2}^2} \Bigg[B_0(0,\mu,m_{{\tilde
t}_1})-B_0(0,\mu,m_{{\tilde t}_2})\Bigg] \nonumber \\
&-&
{g^2\over16\pi^2}\Bigg\{ {\mu M_2\tan\beta \over
\mu^2-M_2^2}\Bigg[c_t^2 B_0(0,M_2,m_{{\tilde t}_1}) + s_t^2
B_0(0,M_2,m_{{\tilde t}_2}) \Bigg]
+(\mu\leftrightarrow M_2)
\Bigg\}\ ,
\end{eqnarray}
where $s_t = \sin\theta_{\tilde t}$ and $c_t=\cos\theta_{\tilde t}$.
The angle $\theta_{\tilde t}$ is the
mixing angle of the top squark mass-squared matrix, and satisfies
\begin{equation}
\sin 2\theta_{\tilde t} = {2 m_t (A_t -\mu\cot\beta)\over m^2_{{\tilde t}_1}
-m^2_{{\tilde t}_2}}\ .
\end{equation}

\subsubsection{Tau lepton mass}

The SUSY corrections to the tau lepton mass are also enhanced in the large
$\tan\beta$ region.  To find the correction, we proceed as before, and
first use the SM RG equations and radiative correction to find the running
tau lepton mass, $m_\tau(M_Z)^{\rm SM}$, at the scale $M_Z$.  This is
related to the running tau lepton mass in the MSSM as follows,
\begin{equation}
m_\tau(M_Z)^{\rm MSSM} = m_\tau(M_Z)^{\rm SM} \left[1 - \left( {\Delta
m_\tau \over m_\tau}\right)^{\rm SUSY}\right]\ .
\end{equation}
where
\begin{equation}
\left({\Delta m_{\tau}\over m_{\tau}}\right)^{\rm SUSY} = -\ {g^2\over16\pi^2}
{\mu M_2 \tan\beta \over \mu^2 - M_2^2}\Bigg[
B_0(0,M_2,m_{\tilde\nu_{\tau}}) - B_0(0,\mu,m_{\tilde\nu_{\tau}})
\Bigg]\ .
\end{equation}

\subsection{Supersymmetric particle masses}

\subsubsection{Gluino mass}

The gluino mass corrections arise from
gluon/gluino and quark/squark loops \cite{PP1,donini}.
The corrections can be rather
large, so we include them in a way which automatically incorporates
the one-loop renormalization group resummation,
\begin{equation}
m_{\tilde g}\ =\ M_3(Q) \left[1 - {3\alpha_s(Q)\over4\pi}
\left(5+3\ln\left({Q^2\over M_3^2}\right)
-4B_{1}(M_3,0,M_{\widetilde Q_1})\right)
\right]^{-1}\ .
\end{equation}
We have approximated all squark masses
by a common mass $M_{\widetilde Q_1}$, which is the soft mass of the first
generation of left-handed squarks.

\subsubsection{Neutralino and chargino masses}

The complete set of corrections to the neutralino and chargino masses
is very involved \cite{PP1,neucharcorr,BMPZ}.
In the following we will present approximation
formulae which typically work to better than $2\%$.
We shall assume $|\mu|\,>\,M_1,M_2, M_Z,$
which is correct in the mSUGRA and minimal gauge-mediated models,
with radiative breaking of the electroweak symmetry.
This approximation leads to an error of order
$(\alpha/4\pi)M^2_Z/\mu^2$ in the masses.

The one-loop radiatively corrected neutralino/chargino masses are
determined from their tree-level running masses by
\begin{equation}
M_i = M_i(Q)\biggl[ 1 + \left({\Delta M_i\over M_i}\right) \biggr] \ ,
\end{equation}
where $i$ labels the different species of neutralinos and charginos.

For $i=\widetilde\chi^0_1$, we find
\begin{eqnarray}
\left({\Delta M_i \over M_i}\right)
&=& {g'^2 \over 32 \pi^2}
\biggl\{11 \theta_{M_1 M_{\widetilde Q_1}}
+ 9 \theta_{M_1 M_{\widetilde L_1}} +
\theta_{M_1 \mu M_Z} - 2B_1(0,\mu,m_A) \nonumber \\
&& - {2 \mu\over M_1} \sin(2\beta)
\biggl(B_0(0,\mu,0)-B_0(0,\mu,m_A)\,\biggr) -
{23 \over 2}\biggr\}\ ,
\label{M1b}
\end{eqnarray}
where $\theta_{m_1 \ldots m_2} \equiv
\ln(M^2/Q^2)$ with $M^2 = {\rm max}(m_1^2,...,m_2^2)$.
We have approximated the tree-level mass of
$\widetilde\chi^0_1$ by $M_1$,
and the squark and slepton masses by the first generation
left-handed soft masses $M_{\widetilde Q_1}$ and $M_{\widetilde L_1}$.

For $i=\widetilde\chi^0_2$ and $\widetilde\chi^+_1$, we find
\begin{eqnarray}
\left({\Delta M_i \over M_i}\right) &=& {g^2 \over 32 \pi^2}
\bigg\{9 \theta_{M_2M_{\widetilde Q_1}} + 3 \theta_{M_2M_{\widetilde L_1}} +
\theta_{M_2\mu M_Z} - 12 \theta_{M_2M_W}
 - \theta(M_W-M_2) \left(
12.56 {M_2 \over M_W} - 14.80 \right) \nonumber\\
&& + \theta(M_2-M_W) \left[4.32 \ln \left({M_2 \over M_W} -
0.8\right) + 9.20\right] \\
&& - 2B_1(0,\mu,m_A)  - {2 \mu\over
M_2} \sin(2\beta) \bigg(B_0(0,\mu,0) - B_0(0,\mu,m_A) \bigg)
 - {15 \over 2}\biggr\} \nonumber\ ,
\end{eqnarray}
where we have approximated the tree-level masses of
$\widetilde\chi^0_2$ and $\widetilde\chi^+_1$ by $M_2$.

For $i=\widetilde\chi^0_3,\ \widetilde\chi^0_4$
and $\widetilde\chi^+_{2}$, we find
\begin{eqnarray}
\left({\Delta M_i \over M_i}\right) &=& -\ {3\over 32 \pi^2}
\bigg[(\lambda_b^2 + \lambda_t^2) B_1(\mu,0,M_{\widetilde Q_3})
+ \lambda_t^2 B_1(\mu,0,M_{\widetilde U_3})\nonumber\\
&&\qquad\qquad + \lambda_b^2 B_1(\mu,0,M_{\widetilde D_3})\bigg] \nonumber\\ 
&&\ + {3g^2 \over 64\pi^2} \bigg[ {1\over 2} \theta_{\mu M_2M_Z}
 -  3 \theta_{\mu M_Z} - B_1(\mu,0,m_A)  + 4\bigg]\ ,
\end{eqnarray}
where we have approximated the tree-level masses of
$\widetilde\chi^0_{3,4}$ and $\widetilde\chi^+_2$ by $|\mu|$,
and $M_{{\widetilde Q_3},{\widetilde U_3},{\widetilde D_3}}$
denote the third generation soft squark masses.

\subsubsection{Squark masses}

The dominant corrections to the first two generation
and the bottom squark masses come from SUSY QCD \cite{12squarkcorr}.
For these cases, the one-loop radiative corrections are
given by
\begin{equation}
m_{\tilde q}^2 = m_{\tilde q}^2(Q) \left[1 +
{2\alpha_s(Q) \over 3\pi}
\bigg\{1 + 3x + (x-1)^2\ln|x-1| - x^2\ln x\ + 2x\ln\left({Q^2\over
m_{\tilde q}^2}\right)\bigg\}\right]\ ,
\label{QCDcorr}
\end{equation}
where $x=m_{\tilde g}^2/m_{\tilde q}^2$.

The radiative corrections to the top squark mass need a more careful
treatment \cite{donini,BMPZ} because the off-diagonal element of
the top squark mass-squared matrix is proportional to the top quark
mass.
The full one-loop formulae for the top squark mass
are very involved. We present here approximations which
work for the cases of light and heavy top squarks. To make
the approximations as good as possible, we use a matrix
formalism in which the top squark mass-squared matrix has the form:
\begin{equation}
{\cal M}_{\tilde t}^2 = {\cal M}_{\tilde t}^2(Q)\ + \pmatrix{
\Delta M^2_{LL} & \Delta M^2_{LR} \cr \Delta M^2_{LR} & \Delta
M^2_{RR}}\ ,\label{top sq app}
\end{equation}
where the $\Delta M^2$ entries are as follows:
\begin{eqnarray}
\Delta M^2_{LL} & =&  {2\alpha_s \over 3\pi} \bigg\{ 2 m_{{\tilde
t}_1}^2 \left[ s_t^2 B_1(m_{{\tilde t}_1},m_{{\tilde t}_2},0)  +
c_t^2 B_1(m_{{\tilde t}_1},m_{{\tilde t}_1},0) \right]\nonumber\\ &&
\qquad + A_0(m_{\tilde g}) + A_0(m_t)  - ( m_{{\tilde t}_1}^2 -
m_{\tilde g}^2 - m_t^2) B_0(0,m_{\tilde g},m_t) \bigg\} \nonumber \\
&&-{1\over16\pi^2} \bigg[ \lambda_t^2 c_t^2 A_0(m_{{\tilde t}_2}) +
\lambda_b^2 A_0(m_{\tilde b}) \nonumber \\
&& \qquad - 2 (\lambda_t^2 +
\lambda_b^2) A_0(\mu)  + (\lambda_t^2 c_{\beta}^2 + \lambda_b^2
s_{\beta}^2) A_0(m_A) \bigg] \nonumber \\
&&- {\lambda_t^2\over32\pi^2}
\bigg[\Lambda(\theta_t,\beta) B_0(0,m_{{\tilde t}_2},m_A) +
\Lambda(\theta_t-{\pi\over2},\beta)B_0(0,0,m_A) \nonumber \\ & &
\qquad + \Lambda(\theta_t,\beta-{\pi\over2}) B_0(0,m_{{\tilde
t}_2},0) + \Lambda(\theta_t-{\pi\over2},\beta-{\pi\over2})
B_0(m_{{\tilde t}_1},m_{{\tilde t}_1},m_Z) \bigg] \nonumber \\
&&-{1\over16\pi^2} \bigg[ \left( \lambda_t^2 m_t^2 c_\beta^2  +
\lambda_b^2 (\mu c_\beta + A_b s_\beta)^2 \right) B_0(0,m_{\tilde
b},m_A) \nonumber \\ && \qquad +  \left( \lambda_t^2 m_t^2 s_\beta^2
+ \lambda_b^2 (\mu s_\beta - A_b c_\beta)^2 \right) B_0(0,m_{\tilde
b},0) \bigg] \\
\Delta M^2_{LR} &=&\ {2\alpha_s\over3\pi} c_ts_t \left[
(m_{{\tilde t}_1}^2 + m_{{\tilde t}_2}^2) B_0(m_{{\tilde t}_1},
m_{{\tilde t}_2},0) + 2 m_{{\tilde t}_1}^2
B_0(m_{{\tilde t}_1},m_{{\tilde t}_1},0) \right]
\nonumber \\ &&-{4\alpha_3\over 3\pi} m_t m_{\tilde g}
B_0(0,m_t,m_{\tilde g}) + {3\lambda_t^2\over16\pi^2} c_t s_t
A_0(m_{{\tilde t}_2}) \nonumber \\
&&- {\lambda_t^2\over32\pi^2} \bigg[
\Omega( \theta_t,\beta) B_0(0,m_{{\tilde t}_2},m_A) +
\Omega(-\theta_t,\beta) B_0(0,0,m_A) \nonumber \\
&& \qquad +\Omega( \theta_t,{\pi\over2}+\beta)B_0(0,m_{{\tilde t}_2},0) +
\Omega(-\theta_t,{\pi\over2}+\beta)B_0(m_{{\tilde t}_1},
m_{{\tilde t}_1},M_Z) \bigg] \nonumber \\
&&-{1\over16\pi^2} \bigg[ \biggl(
\lambda_t^2 m_t c_\beta (\mu s_\beta + A_t c_\beta) + \lambda_b^2
m_t s_\beta (\mu c_\beta + A_b s_\beta) \biggr)
B_0(0,m_{\tilde b},m_A)\nonumber\\
&& \qquad - \lambda_t^2 m_t s_\beta (\mu c_\beta
 - A_t s_\beta) B_0(0,m_{\tilde b} ,0) \bigg] \\
\Delta M^2_{RR} &=&  {2\alpha_s \over3\pi}\bigg\{ 2 m_{{\tilde
 t}_1}^2 \left[ c_t^2 B_1(m_{{\tilde t}_1},m_{{\tilde t}_2},0)  +
s_t^2 B_1(m_{{\tilde t}_1},m_{{\tilde t}_1},0) \right] \nonumber \\
&& \qquad + A_0(m_{\tilde g}) + A_0(m_t) - ( m_{{\tilde t}_1}^2 -
 m_{\tilde g}^2 - m_t^2) B_0(0,m_{\tilde g},m_t) \bigg\} \nonumber \\
&&-{\lambda_t^2\over16\pi^2} \left[ s_t^2 A_0(m_{{\tilde t}_2}) +
A_0(m_{\tilde b}) - 4 A_0(\mu) + 2 c_{\beta}^2 A_0(m_A) \right]
 \nonumber \\
&&- {\lambda_t^2\over32\pi^2} \bigg[
 \Lambda({\pi\over2}-\theta_t,\beta) B_0(0,m_{{\tilde t}_2},m_A)  +
\Lambda(-\theta_t,\beta)B_0(0,0,m_A) \nonumber \\
& & \qquad +
\Lambda({\pi\over2}-\theta_t,\beta-{\pi\over2})B_0(0,m_{{\tilde
 t}_2},0)  + \Lambda(-\theta_t,\beta-{\pi\over2}) B_0(m_{{\tilde
 t}_1},m_{{\tilde t}_1},m_Z) \bigg] \nonumber \\
&&- {1\over16\pi^2}
 \bigg[ \left( \lambda_b^2 m_t^2 s_\beta^2 +\lambda_t^2 (\mu s_\beta +
 A_t c_\beta)^2 \right) B_0(0,m_{\tilde b},m_A) \nonumber \\
&& \qquad + \lambda_t^2 (\mu c_\beta - A_t s_\beta)^2 B_0(0,m_{\tilde b},0)
 \bigg] \ ,
\label{stop self}
\end{eqnarray}
with $s_\beta=\sin\beta$ and $c_\beta=\cos\beta$.
We have defined the two functions
\begin{eqnarray}
\Lambda(\theta_t,\beta) & = & \left( 2 m_t \cos \beta \sin\theta_t -
(\mu \sin \beta + A_t \cos \beta ) \cos\theta_t\right)^2 \nonumber \\
&&\qquad + (\mu \sin \beta + A_t \cos \beta )^2 \cos^2\theta_t \\
\Omega(\theta_t,\beta) & = & - 2 m_t^2 \cos^2 \beta \sin 2\theta_t
+ 2 m_t \cos\beta (\mu \sin \beta + A_t \cos \beta ) \ .
\end{eqnarray}
The eigenvalues of the mass-squared matrix (\ref{top sq app}) determine
the one-loop radiatively corrected top squark masses.

\subsubsection{Slepton masses}

The corrections to the slepton masses are small in the mSUGRA model.
We find that the corrections to the electron or muon slepton masses
are typically in the range $\pm(1-2)\%$, as are
the (predominantly) left- and right-handed tau
slepton corrections. In the gauge-mediated model, tau slepton
can receive appreciable corrections at large $\tan\beta$.

\subsubsection{Higgs boson masses}

The Higgs sector of the MSSM is composed of five physical Higgs
bosons. Generically, one of them ($h^0$) is light while the other four
($H^0, A^0$ and ${H^\pm}$) are heavy.  In the decoupling
limit\footnote{In the MSSM, the decoupling limit
occurs for $\tan\beta\gsim 1.5$ and $m_A\gsim 180$ GeV.} where
$m_A\rightarrow\infty$, the lightest Higgs boson $h^0$ becomes
indistinguishable from that of the SM.

The radiative corrections to the lightest CP-even Higgs-boson mass
are extremely important and have been extensively studied by many
groups \cite{lighthiggscorr,2loopcorr}.
The most important contributions come
from an incomplete cancellation of the quark and squark loops.
Interested readers should consult summaries by the LEP2 Higgs
boson working group and by Haber \cite{lephiggs}
for the one-loop formulae and some further refinements.
The corrections to the other heavy Higgs boson masses,
$m_A$, $m_{H^\pm}$ and $m_H$,
are typically less than 1\%~\cite{heavyhiggscorr}.

\subsection{Summary}

Radiative corrections are very important for a precise
determination of the SUSY parameters in the mSUGRA and gauge
mediated models.
We have surveyed the radiative corrections
to the SM and SUSY particle masses and presented a series of
useful approximations. These formulae include one-loop leading
logarithmic corrections as well as potentially large finite
contributions.

%% file: Feng/feng-new.tex
%%%%%%%%%%%%%%%%%%%%%%%%%%%%%%%%%%%%%%%%%%%%%%%%%%%%%%%%%%%%%%%%%%%%%%%%
\newcommand{\rem}[1]{{\bf #1}}
\newcommand{\susyU}{\widetilde{U}}
\def\ifb{\hbox{ fb}^{-1}}
\def\gev{\hbox{ GeV}}
\def\tev{\hbox{ TeV}}

\section{Radiative Corrections to Couplings and Superoblique Parameters}

\subsection{Soft vs.\ Hard SUSY-breaking}

If supersymmetry (SUSY) were an exact symmetry of nature, the
properties of supersymmetric particles would be completely determined
by the properties of their standard model partners.  For example,
their masses would satisfy the relations
\begin{equation}
\label{soft}
m_{\tilde{f}} = m_f \ ,
\end{equation}
where $m_{\tilde{f}}$ and $m_f$ are the masses of supersymmetric
particles and their standard model partners, respectively.
As is well known, relations between {\em
dimensionful\/} parameters such as Eq.~(\ref{soft}) are broken via the introduction of soft
SUSY-breaking parameters.  These soft terms are free parameters of
SUSY theories, and their {\em a priori} arbitrariness is responsible
for the lack of predictability of SUSY for collider experiments.

In addition to these relations, however, SUSY also predicts the
equivalence of {\em dimensionless\/} couplings.  By dimensional
arguments, these identities cannot be broken by soft SUSY-breaking
parameters at tree level and are therefore known as ``hard SUSY
relations''~\cite{Hikasa}.  For example, supersymmetry implies
\begin{equation}
\label{hard}
g_i=h_i \ ,
\end{equation}
where $g_i$ are the standard model gauge couplings, $h_i$ are their
supersymmetric analogues, the gaugino-fermion-sfermion couplings, and
the subscript $i=1,2,3$ refers to the U(1), SU(2), and SU(3) gauge
groups, respectively.  In contrast to other predictions, such as the
universality of scalar or gaugino masses, hard SUSY relations are
independent of the SUSY-breaking mechanism, and are therefore valid in
all supersymmetric theories.  The verification of hard SUSY relations
provides the only model-independent method of quantitatively
confirming that newly-discovered particles are indeed
superpartners~\cite{FMPT}.  At lepton colliders, hard SUSY relations
have been demonstrated to be verifiable at the percent level for many
scenarios through a variety of processes~\cite{FMPT,NFT,gex,NPY,KNPY},
as will be discussed below. (At hadron colliders, the possible
scenarios are far more numerous and much less well-studied. Such hadron collider studies are harder to do because several slepton signals may be involved and the backgrounds are more complex.)

\subsection{Radiative Corrections to Couplings}

The relations of Eq.~(\ref{hard}) are valid at tree-level.  However,
in the presence of soft SUSY-breaking, radiative corrections violate
Eq.~(\ref{hard}) and lead to ``hard SUSY-breaking.''  The radiative
corrections may be most easily understood in the language of
renormalization group equations.  Consider a simplified model with
some scalar superparticles at a heavy mass scale $M$ and all other
particles at a light mass scale $m$, typically taken to be the weak
scale.  Above the scale $M$, SUSY is unbroken, and we have $h_i=g_i$.
Below $M$, where the heavy superpartners decouple, light fermion loops
still renormalize the gauge boson wavefunction (and thus, $g_i$) but
heavy sfermion loops and sfermion-fermion loops decouple from gauge
boson and gaugino wavefunction renormalization,
respectively~\cite{C}. (Gauge loops still renormalize both
wavefunctions in the non-Abelian case.)  Since not all loops involving
the scalar and fermion fields of each supermultiplet decouple
simultaneously, supersymmetry is broken in the gauge sector, and
therefore the gauge couplings $g_i$ and gaugino couplings $h_i$ evolve
differently between the scales $M$ and $m$.  The splitting between
$h_i$ and $g_i$ at the light scale is given by
\begin{equation}
\susyU_{i} \equiv \frac{h_{i}(m)}{g_{i}(m)} - 1
\approx \frac{g_{i}^{2}(m)}{16\pi^{2}}\Delta b_i
\ln \frac{M}{m} \ ,
\label{delta}
\end{equation}
where $i=1,2,3$ denotes the gauge group as before, and $\Delta b_i$ is
the one-loop $\beta$-function coefficient contribution from all light
particles whose superpartners are heavy.\footnote{The contribution of
a particle $j$ with spin $S^j$ to the $\beta$-function coefficient
$\Delta b_{i}$ is $b_i^j = N_i^j a^j T_i^j$, where $N_i^j$ is the
appropriate multiplicity; $a^j = \frac{1}{3}, \frac{2}{3},
-\frac{11}{3}$ for $S^j = 0, \frac{1}{2}, 1$, respectively; and $T_i^j
= 0, \frac{1}{2}, N$, or $\frac{3}{5}Y^2$ for a singlet, a particle in
the fundamental representation of $SU(N)$, a particle in the adjoint
representation of $SU(N)$, or, for $i = 1$, a particle with
hypercharge $Y=2(Q-I_3)$, respectively.} The coupling splittings may also
receive contributions from split exotic supermultiplets, such as the
messenger supermultiplets of gauge-mediated SUSY-breaking
theories~\cite{gth,KRS}.

Equation (\ref{delta}) has a number of interesting properties.  In
particular, although the splittings between $h_i$ and $g_i$ are
one-loop effects, they can be greatly enhanced for large heavy sectors
(large $\Delta b_i$) and large mass splittings (large $M/m$).  The
corrections of Eq.~(\ref{delta}) are just a subset of all one-loop
corrections.  However, because of these enhancements, these splittings
are in many cases the dominant effects, as has been demonstrated
through complete one-loop calculations~\cite{KNPY}.  Note that the
fact that they grow with $\ln\frac{M}{m}$ implies that these effects are
{\em non-decoupling}, that is, they are in principle sensitive to
particles with arbitrarily high mass, in contrast to many other
observables.  Note also that the coupling $h_i$ is more asymptotically
free than $g_i$, $h_i(m) > g_i(m)$, and so the parameters $\susyU_i$
are always positive (at the leading log level).

\subsection{Superoblique Parameters}

The hard SUSY-breaking corrections have a beautiful analogy to the
oblique corrections of the electroweak sector of the standard model:

\begin{itemize}
\item In the standard model, SU(2) multiplets with custodial
SU(2)-breaking masses, such as the $(t,b)$ multiplet, renormalize the
propagators of the $(W,Z)$ vector multiplet differently, leading to
explicit custodial SU(2)-breaking in the vector multiplet at the
quantum level, and introducing non-decoupling effects that grow with
the mass splitting.
\item In supersymmetric models, supermultiplets with soft
SUSY-breaking masses, such as the $(\tilde{f},f)$ supermultiplet,
renormalize the propagators of the $(\hbox{gauge boson},
\hbox{gaugino})$ vector supermultiplet differently, leading to
explicit SUSY-breaking in the vector supermultiplet at the quantum
level, and introducing non-decoupling effects that grow with the mass
splitting.
\end{itemize}
On the strength of this analogy, hard SUSY-breaking corrections have
been dubbed ``super-oblique corrections'' in Ref.~\cite{gth}, and the
parameters $\susyU_i$, which are most analogous to the wavefunction
renormalization oblique parameter $U$~\cite{PT}, are called
``superoblique parameters.''  Further details concerning this analogy
may be found in Refs.~\cite{gth,NPY,KRS}.

\subsection{Experimental Implications of Superoblique Parameters}

Superoblique parameters have relevance for experimental studies in
many ways.  Among these are the following:

\begin{itemize}

\item Superoblique corrections influence superparticle production
rates and branching ratios, which depend on the couplings $h_i$ and
therefore the superoblique parameters $\susyU_i$.

\item As noted above, the verification of hard SUSY relations (with
their superoblique corrections) allows us to quantitatively determine
whether newly-discovered particles are in fact supersymmetric.  If
these particles are determined to be supersymmetric, verifications of
hard SUSY relations will play an important part in establishing the
underlying supersymmetry of the couplings.

\item Perhaps most exciting, if new particles are determined to be
supersymmetric, small but unexplained violations of hard SUSY
relations will provide unambiguous evidence for as-yet-undiscovered
SUSY particles.  If portions of the SUSY spectrum are beyond the reach
of colliders, the non-decoupling superoblique parameters may provide
indirect experimental evidence for superparticles with arbitrarily
high masses (if they belong to highly split supermultiplets).
\end{itemize}

With regard to the last point, it is worth noting that fine-tuning
arguments place stringent upper bounds only on sfermions that couple
strongly to the Higgs boson, namely, the 3rd generation
squarks~\cite{finetune,211}.  The squarks and sleptons of the 1st and
2nd generations may have masses far beyond the reach of the Tevatron
and even the LHC, and in fact, such massive squarks and sleptons
provide an elegant solution to the SUSY flavor problem by naturally
suppressing dangerous contributions to, for example, $K^0 - \bar{K}^0$
mixing and $\mu \to e\gamma$~\cite{211,212}.  In such models, mass
hierarchies of order $M/m \sim 40 - 200$ are possible.

The numerical values of the superoblique parameters $\susyU_i$ for two
representative models are given in Table~\ref{table:t1}.  We find that
fractional splittings of several percent are possible, depending
logarithmically on the mass hierarchy $M/m$.

\begin{table}
\caption{The superoblique parameters $\susyU_{i}$ in two
representative models: ``2--1 Models,'' with all first and second
generation sfermions at the heavy scale $M$, and ``Heavy QCD Models,''
with all squarks and gluinos at the heavy scale.}
\centering\leavevmode
\begin{tabular}{cccc}
   & $\susyU_1$ & $\susyU_2$ & $\susyU_3$ \\
\hline
2--1 Models &
$0.35\% \times \ln \frac{M}{m}$ &
$0.71\% \times \ln \frac{M}{m}$ &
$2.5\%  \times \ln \frac{M}{m}$ \\
Heavy QCD Models &
$0.29\% \times \ln \frac{M}{m}$ &
$0.80\% \times \ln \frac{M}{m}$ &
--- 
\end{tabular}
\label{table:t1}
\end{table}

\subsection{Measurements at Colliders}
\medskip

The experimental observables that are dependent on superoblique
parameters have been exhaustively categorized in Ref.~\cite{gex} for
both lepton and hadron colliders.  The most promising observables at
colliders are cross sections and branching ratios involving gauginos,
and several of these possibilities have been examined in detailed
studies.  The results are, of course, highly dependent on the
underlying SUSY parameters realized in nature, but we present a brief
synopsis below.

\begin{itemize}
\item {\em Measurements of $\susyU_1$.} Selectron pair production at
electron colliders includes a contribution from $t$-channel gaugino
exchange.  In particular, the reaction $e^+ e^- \to \tilde{e}_R^+
\tilde{e}_R^-$ depends on the $\tilde{B}$-$e$-$\tilde{e}$ coupling
$h_1$, and has been studied in Ref.~\cite{NFT}.  Under the assumption
that the selectrons decay through $\tilde{e} \to \tilde{B} e$, the
selectron and gaugino masses may be measured through kinematic
endpoints.  Combining this information with measurements of the
differential cross section, $\susyU_1$ may be determined to ${\cal O}
(1\%)$ with $20\ifb$ of data at $\sqrt{s} = 500 \gev$.

This high precision measurement may be further improved by considering
the process $e^- e^- \to \tilde{e}_R^- \tilde{e}_R^-$.  This process
is made possible by the Majorana nature of gauginos.  Relative to the
$e^+ e^-$ process, this reaction benefits from large statistics for
typical SUSY parameters and extremely low backgrounds, especially if
the electron beams are right-polarized.  Depending on experimental
systematic errors, determinations of $\susyU_1$ at the level of
$0.3\%$ may be possible with integrated luminosities of
$50\ifb$~\cite{gex}.

\item {\em Measurements of $\susyU_2$.} Chargino pair production has a
dependence on $\susyU_2$ at lepton colliders through the $\tilde{\nu}$
exchange amplitude.  This process was first studied as a way to verify
hard SUSY relations in Ref.~\cite{FMPT}.  In Ref.~\cite{gex},
estimates of 2--3\% uncertainties for $\susyU_2$ were obtained from
pair production of 172 GeV charginos with $\sqrt{s} = 400 - 500 \gev$.
In Ref.~\cite{KNPY}, these estimates were found to be conservative in the
sense that these measurements are improved in most other regions of
parameter space.

The process $e^+ e^- \to \tilde{\nu}_e \tilde{\nu}_e^*$ also depends
on $\susyU_2$ through the $t$-channel chargino exchange amplitude.
With a data sample of $100 \ifb$, $\susyU_2$ may be determined to
$\sim 0.6\%$~\cite{NPY}.

\item {\em Measurements of $\susyU_3$.} Finally, the superoblique
parameter $\susyU_3$ may be measured through processes involving
squarks.  Squark pair production cross sections at lepton colliders
are independent of superoblique corrections, but three-body production
processes, such as $\tilde{t} t \tilde{g}$ and $\tilde{b} b \tilde{g}$
have been suggested as a probe~\cite{gex,KRS}.

Squark branching ratios may also sensitive to superoblique corrections
if there are two or more competing modes.  In Ref.~\cite{gex},
parameters were studied in which the two decays $\tilde{b}_L \to b
\tilde{g}$ and $\tilde{b}_L \to b \tilde{W}$ were open.  For parameters
where the gluino decay is suppressed by phase space, these modes may
be competitive, and measurements of the branching ratios yield
constraints on $\susyU_3$.  For example, for $m_{\tilde{b}_L} = 300
\gev$, $\tilde{b}_L$ pair production at a $\sqrt{s} = 1\tev$ collider
with integrated luminosity $200\ifb$ yields measurements of $\susyU_3$
at or below the 5\% level for $10 \gev \alt m_{\tilde{b}_L} -
m_{\tilde{g}} \alt 100 \gev$. While these measurements are typically
numerically less stringent than those discussed above, the SU(3)
superoblique correction is typically large and so merits attention.
In addition, the possibility of studying squark branching ratios is
one that has particular promise at hadron colliders.

\end{itemize}

%\begin{references}

%% file: Pierce/susy-corrs.tex
\renewcommand{\gsim}{\lower.7ex\hbox{$\;\stackrel{\textstyle>}{\sim}\;$}}
\renewcommand{\lsim}{\lower.7ex\hbox{$\;\stackrel{\textstyle<}{\sim}\;$}}

\def\npb#1 #2 #3 #4 {Nucl.~Phys. {\bf B#1}, #2 (#3)#4 }
\def\plb#1 #2 #3 #4 {Phys.~Lett. {\bf B#1}, #2 (#3)#4 }
\def\prd#1 #2 #3 #4 {Phys.~Rev.  {\bf D#1}, #2 (#3)#4 }
\def\prl#1 #2 #3 #4 {Phys.~Rev.~Lett. {\bf #1}, #2 (#3)#4 }
\def\zpc#1 #2 #3 #4 {Zeit.~Phys. {\bf C#1}, #2 (#3)#4 }

\section{Supersymmetric Corrections to Standard Model Processes}

\subsection{Introduction}

The upgraded Tevatron collider will provide a wealth of new
information. The data and analyses will shed new light on many aspects
of the Standard Model, and may even include direct discoveries of
physics beyond the Standard Model. For example, if low energy
supersymmetry exists, supersymmetric particles might be discovered
through their direct production and decay processes. Alternatively,
new physics scenarios might be indirectly observed, constrained or
excluded via virtual corrections to Standard Model processes. In this
report we will discuss the implications of virtual supersymmetric
corrections to various Standard Model processes in light of the
expected measurements at the upgraded Tevatron collider. In the next
section we discuss the supersymmetric radiative corrections to the top
quark pair production cross section. In Section~\ref{single top} we
discuss single top quark production, and in Section~\ref{tgc} we
discuss di-boson production. We summarize the report in the last
section.

%-----------------------------------------------
% 2 Top quark pair production
%-----------------------------------------------
\subsection{Top quark pair production}

The upgraded Tevatron, with center-of-mass energy 2 TeV and total
integrated luminosity 2 (10) fb$^{-1}$, will produce approximately
$10^4$ ($7\times10^4$) top quark pairs. This should yield a
measurement of the cross-section to a precision of about 12\% (5.5\%)
\cite{TeV2000}.

The supersymmetric QCD (SQCD) corrections to the top quark pair
production cross section were first considered in Refs.~\cite{LHYH,KLNR}.Those papers did not include the full set of corrections. The
calculation was completed for the case of degenerate squark masses by
including the gluon self energy and box diagram contributions in
Ref.~\cite{Alam}. A relative sign between the two box diagram
contributions was corrected in Ref.~\cite{ZS}, and the overall sign
of the box diagram contribution was corrected in
Ref.~\cite{DW}\footnote{This was confirmed for this working group by
Z.~Sullivan.}.  The SQCD correction to the cross section depends
crucially on the box diagram contribution. With the correct sign the
box diagram contribution largely cancels against the vertex
correction, so that the total SQCD correction is less than 7\% with a
200 GeV gluino mass and a 100 GeV degenerate squark mass \cite{DW}. If
both the gluino and squark masses are above 200 GeV, the total SQCD
correction varies between 0 and $-3\%$. Such small corrections will
not be discernible at the Tevatron upgrade. Below the anomalous
threshold ($m_{\tilde{g}}^2+m_{\tilde{t}}^2 < m_t^2$), the SQCD
correction is greater than $+15\%$~\cite{ZSERR}.  However, the current
lower bounds on the gluino and top squark masses are close to ruling
out this region.  If there is a large hierarchy between the two top
squark masses, and the top squark mixing is large, the correction can
be as large as $\pm7\%$. Such a correction could be observed at the
upgraded Tevatron with sufficient luminosity, if the standard
theoretical uncertainty (from scale uncertainty, parton distribution
functions, etc.) is not too large.

The SUSY electroweak-like (SEW) corrections to the top quark pair
production cross-section have also been computed \cite{KLNR,ttmssm}.
As one would expect, these corrections are typically small, of order a
few percent. The SEW corrections can be enhanced in a couple of
ways. If one of the top squarks is light (i.e. below 100 GeV), the
light chargino is Higgsino-like, and $\tan\beta$ is very small and
very large (e.g. $\tan\beta=0.7$\footnote{Values of $\tan\beta$ below
1 do not correspond to solutions of the electroweak symmetry breaking
conditions. Also, if $\tan\beta<1.2$ the top Yukawa coupling hits a
Landau pole below the GUT scale. Additionally, in the MSSM $\tan\beta$
less than 1.5 has been (preliminarily) excluded based on Higgs
searches at LEP 2 \cite{delphi}.} and 50) these corrections can reach
about $-5.8\%$ and $-4.3\%$, respectively, at the Tevatron collider.
Alternatively, the SEW corrections can be enhanced if the resonant
condition $m_t=m_{\tilde t}+m_{\tilde\chi^0}$ is close to being
met. In this case the SEW correction can be as large as $-30\%$
($-15\%$) with $\tan\beta=0.7 (\gsim1.4)$. If the MSSM parameters are
such that such a large enhancement occurs, the correction will be
observable at the Tevatron upgrade, even with 2 fb$^{-1}$ integrated
luminosity. It should be noted, however, that in these cases direct
production of the light top squark will likely be observable as well.

MSSM loop-induced parity violating asymmetries in the strong
production of polarized top quark pairs has also been investigated,
both at the Tevatron \cite{ZS,DW,ttsm,Chung,LOYY,Doreen} and the LHC
\cite{Doreen}.  In addition to the small SM contribution \cite{ttsm}
induced by diagrams involving the $Z$, the $W^+$, and the Goldstone
boson $G^+$, the parity violating asymmetries in the MSSM are
generated by diagrams involving the charged Higgs boson ($H^\pm$), the
charginos ($\tilde\chi^{\pm}_i$) and the neutralinos
($\tilde\chi^0_i$).  Since QCD preserves parity, parity violating
asymmetries in the strong production of left- and right-handed top
quark pairs can only arise beyond leading order in perturbation
theory.  We consider
\begin{itemize}
\item
the differential left-right asymmetry
\begin{equation}\label{eq:thirteen}
\delta {\cal A}_{LR}(M_{t\bar t})= \frac{d\sigma_{{+\frac{1}{2},-\frac{1}{2}}}/d M_{t\bar t}-d\sigma_{{-\frac{1}{2},+\frac{1}{2}}}
/ d M_{t\bar t}}{d\sigma_{{+\frac{1}{2},-\frac{1}{2}}}/ d M_{t\bar t}
+d\sigma_{{-\frac{1}{2},+\frac{1}{2}}}/ d M_{t\bar t}} \; ,
\end{equation}
with $d\sigma_{\lambda_t,\lambda{\bar t}} / d M_{t\bar t}$ denoting
the invariant mass distribution of the polarized top quark pair,
\item
the integrated left-right asymmetry
\begin{equation}\label{eq:fourteen}
{\cal A}_{LR}= \frac{\sigma_{{+\frac{1}{2},-\frac{1}{2}}}
-\sigma_{{-\frac{1}{2},+\frac{1}{2}}}}
{\sigma_{{+\frac{1}{2},-\frac{1}{2}}}+\sigma_{{-\frac{1}{2},+\frac{1}{2}}}}~,
\end{equation}
\item
and the integrated left-right asymmetry assuming that the
antitop quark polarization is not measured in the experiment
\begin{equation}\label{eq:fifteen}
{\cal A} = \frac{\sigma_{{+\frac{1}{2},-\frac{1}{2}}}
+\sigma_{{+\frac{1}{2},+\frac{1}{2}}}
-(\sigma_{{-\frac{1}{2},+\frac{1}{2}}}
+\sigma_{{-\frac{1}{2},-\frac{1}{2}}})}{\sigma} \; ,
\end{equation}
where $\sigma(S)=\sum_{\lambda_t,\lambda_{\bar t}=\pm 1/2}\sigma_{\lambda_t, \lambda_{\bar t}}$is the total unpolarized $t\bar t$ production cross section.
\end{itemize}
At the upgraded Tevatron the parity violating effects within the MSSM
result in differential asymmetries $\delta{\cal A}_{LR}$ as large as a
few percent at large $t\bar t$ invariant mass, and integrated
left-right asymmetries of up to $|{\cal A}_{LR}|\approx 2.2\%$ and
$|{\cal A}| \approx 1.5\%$.
It is clear that any MSSM induced parity violation will be very
difficult, if not impossible, to measure at the Tevatron. If the
luminosity can be upgraded to ${\cal L}=100 \mbox { fb}^{-1}$,
polarization asymmetries of $|{\cal A}_{LR}| \gsim 0.7 \%, |{\cal A}|
\gsim 0.5 \%$ might be visible at the upgraded Tevatron \cite{Doreen}.
Also, preliminary studies of parity-violating effects induced by SQCD
one-loop contributions to strong $t\bar t$ production \cite{ZS} show
the possibility of a small enhancement to the SEW induced asymmetries
\cite{DW}.

%-----------------------------------------------
% 3 Single top production}
%-----------------------------------------------
\subsection{Single top production}
\label{single top}

At the Tevatron upgrade, single top production will provide a
significant source of top quarks. The single-top-quark cross section
at Run 2 will be 3.32~pb \cite{SSW1,SW}, which is about 50\% of the
top-quark pair production cross section.  Hence, one could hope to
measure the cross section accurately enough to be sensitive to
radiative corrections. With 1 fb$^{-1}$ of integrated luminosity, the
cross section will be measured to about 23--26\% \cite{TeV2000,SSW2}.
With 10~fb$^{-1}$ a 10--12\% measurement should be possible for
$s$-channel production, while theoretical uncertainties limit a
measurement of single-top-quark production via $W$-gluon fusion to
16\% \cite{TeV2000,SSW2}.

The SQCD corrections to single top production at the Tevatron have
been calculated in Ref.~\cite{LOYZ}. They found the corrections to the
cross section are at most a few per cent if the squark and gluino
masses are above 200 GeV. Hence, the SQCD corrections to this process
will not be observable at the Tevatron.

The SEW corrections to the single top production cross section have
also been calculated \cite{LOY}. Just as in the case of top quark pair
production, the SEW corrections are typically a few per cent or less,
but enhancements are possible. For example, if the lightest chargino
is Higgsino-like and $\tan\beta<1$\footnote{See previous footnote.}
the SEW corrections can be larger than 10\%. Also, if a resonant
condition is close to being met ($m_t\simeq m_{\tilde
t}+m_{\tilde\chi^0}$) the correction can be as large as $-7\%$.  If
the Tevatron detectors collect a very high luminosity (\gsim10
fb$^{-1}$) and the SUSY parameters happen to be just right, the SEW
corrections to the single top production process could be observable
at the Tevatron upgrade. In this situation it is likely that direct
production of the light top squark will also be observed.

The single top production cross section has also been considered in
$R$-parity violating extensions of the MSSM \cite{RPV single
top}. After taking into account the current bounds on the $R$-parity
violating couplings, significant bounds on certain combinations of
$R$-parity violating couplings can be obtained at the Tevatron
upgrade.

%-----------------------------------------------
% 4 Gauge boson production
%-----------------------------------------------
\subsection{Gauge boson production}
\label{tgc}

Di-boson production serves as a direct probe of anomalous gauge boson
couplings. At the Tevatron upgrade, constraints on anomalous gauge
boson couplings will be obtained by studying $W^+W^-$, $WZ$,
$W\gamma$, and $Z\gamma$ production processes.  With 1 fb$^{-1}$
luminosity, the expected limits on the various anomalous couplings
range from 0.3\% to ${\cal O}(1)$ \cite{TeV2000}.

The supersymmetric corrections to di-boson production at the Tevatron
have not been calculated. In Ref.~\cite{AKM} the supersymmetric
corrections to the $WW\gamma$ and $WWZ$ form factors are considered in
the context of $e^+e^-$ collider experiments. Hence, the corrections
to the form factors are presented at fixed incoming $Z$ or $\gamma$
momenta. In $p\overline p$ collisions, the form factors are probed
over a large range of incoming momenta. Additionally, different
orientations of the vertex will be probed, and the parts of the box
diagram contributions which must added to the vertex corrections for
gauge invariance (the ``pinched boxes'') are also different. All in
all, no direct translation of the results of Ref.~\cite{AKM} to the
case of hadronic colliders is possible. Nonetheless, we expect the
size of the supersymmetric correction to the triple gauge coupling
vertices in $p\overline p$ experiments to be roughly equal in
magnitude to that found in Ref.~\cite{AKM}. In that reference, a
careful scan over parameter space was carried out in order to find the
maximal possible correction in the MSSM. The largest correction to the
$WW\gamma$ vertex $\Delta\kappa_\gamma$ was found to be about 1.7\%
(this includes the SM contribution). Anomalous contributions to this
vertex are expected to be measured at about the 40\% (20\%) level at
the Tevatron, with 1 (10) fb$^{-1}$ luminosity \cite{TeV2000}. The
corrections in the MSSM are in this case about a factor of 10 or 20
too small to be observed at the Tevatron upgrade. In the $WWZ$ vertex
case, the maximal supersymmetric corrections to $\Delta\kappa_Z$ are
about 0.8\% \cite{AKM}. Again, this is too small by a huge factor
(e.g. 20 or 50) to be observable at the Tevatron upgrade.

Other studies corroborate these expectations. In Ref.~\cite{ALPS} at
some reference points in SUSY parameter space, the corrections to the
same $WW\gamma$ and $WWZ$ vertices are given at $\sqrt{s}=190$ GeV,
and are smaller than the maximal values quoted above. Also,
corrections to other $WWZ$ and $WW\gamma$ form factors $\lambda_Z$ and
$\lambda_\gamma$ are given. At $\sqrt{s}=190$ GeV the supersymmetric
corrections are ${\cal O}(0.1\%)$. They are expected to be measured to
about 5 or 10\% at the Tevatron upgrade, with 10 fb$^{-1}$
luminosity. Hence, we expect the supersymmetric corrections to these
couplings will be a factor of 50 or 100 times too small to be
observed. The supersymmetric corrections to the $WW\gamma$ couplings
have also been studied at zero momentum and found to be quite small
\cite{Lahanas}. If anomalous triple gauge boson couplings are observed
at the Tevatron upgrade, it will be a remarkable discovery, but
supersymmetry will not be a candidate for its origin.

%-----------------------------------------------
% 5 Conclusions
%-----------------------------------------------
\subsection{Conclusions}

In this report we have briefly reviewed the prospects for constraining
or indirectly obtaining evidence for supersymmetry by measuring
certain standard model processes at the upgraded Tevatron collider. We
first considered top quark pair production. The SQCD corrections
typically cancel to a large extent. The SUSY electroweak-like
corrections are also typically small, but can be enhanced to
observable levels in special regions of SUSY parameter space.  MSSM
induced parity violation in top-quark pair production was also
considered. The typical 1 to 2\% asymmetries will be very difficult,
if not impossible, to observe at the Tevatron upgrade.  We next
considered single top quark production. The conclusions are much the
same as in the top-quark pair production case. Again, the SQCD
corrections largely cancel, so that the largest corrections are due to
electroweak interactions, in resonance regions of parameter space
(i.e. $m_t\simeq m_{\tilde t}+m_{\tilde\chi^0}$). With large
luminosity, the particular region of parameter space with greatly
enhanced corrections could be observable at the Tevatron upgrade. In
both the single top and top pair production processes, it appears that
if the corrections are large enough to be observable, direct detection
of supersymmetric partners will also be possible (e.g. top squark pair
production). We lastly considered supersymmetric corrections to
di-boson production processes. Here the couplings will not be measured
accurately enough at the Tevatron to provide constraints on
supersymmetric models. The supersymmetric corrections are typically
one or two orders of magnitude too small to be observed.

A more promising approach to constrain supersymmetry by its virtual
effects is to consider the supersymmetric corrections to electroweak
precision observables \cite{ew} and other low energy observables, such
as the anomalous muon magnetic moment \cite{muon} and $b\rightarrow
s\gamma$ \cite{bsg,ew}. Many of the observables in such an analysis
are measured to very high precision, and this gives rise to enhanced
sensitivity to virtual effects. The upgraded Tevatron will greatly
improve this program by providing new precision data, for example
measurements of $M_W$, $m_t$ and $\alpha_s$. Also, corrections to the
high $p_T$ jet production cross section could provide some constraint
on (or an indication of) heavy squarks \cite{jet}.

%-----------------------------------------------
% References
%-----------------------------------------------

%% file: Spira/paper.tex
%***********************************************************************

\renewcommand{\st}{\tilde{t}}
\newcommand{\stb}{\bar{\tilde{t}}}
\renewcommand{\sq}{\tilde{q}}
\newcommand{\sqb}{\bar{\tilde{q}}}
\newcommand{\gl}{\tilde{g}}
\newcommand{\gau}{\tilde{\chi}}
\newcommand{\MS}{\mbox{$\overline{\rm MS}$}}
\newcommand{\MSSM}{\mbox{$\MSSM}$}
\newcommand{\SUSY}{\mbox{${\cal SUSY}$}}
\newcommand{\TeV}{\mbox{Te$\!$V}}
\newcommand{\GeV}{\mbox{Ge$\!$V}}
\newcommand{\tgb}{\mbox{$\tan\beta$}}
\newcommand{\nn}{\noindent}

 \renewcommand{\zp}[3]{{Z.\ Phys.} {\bf #1} (19#2) #3}
 \renewcommand{\np}[3]{{Nucl.\ Phys.} {\bf #1} (19#2)~#3}
 \renewcommand{\pl}[3]{{Phys.\ Lett.} {\bf #1} (19#2) #3}
 \newcommand{\pr}[3]{{Phys.\ Rev.} {\bf #1} (19#2) #3}
 \renewcommand{\prl}[3]{{Phys.\ Rev. Lett.} {\bf #1} (19#2) #3}
 \renewcommand{\prep}[3]{{\sl Phys. Rep.} {\bf #1} (19#2) #3}
 \renewcommand{\fp}[3]{{\sl Fortschr. Phys.} {\bf #1} (19#2) #3}
 \renewcommand{\nc}[3]{{\sl Nuovo Cimento} {\bf #1} (19#2) #3}
 \renewcommand{\ijmp}[3]{{\sl Int. J. Mod. Phys.} {\bf #1} (19#2) #3}
 \renewcommand{\ptp}[3]{{\sl Prog. Theo. Phys.} {\bf #1} (19#2) #3}
 \renewcommand{\sjnp}[3]{{\sl Sov. J. Nucl. Phys.} {\bf #1} (19#2) #3}
 \renewcommand{\cpc}[3]{{\sl Comp. Phys. Commun.} {\bf #1} (19#2) #3}
 \renewcommand{\mpl}[3]{{\sl Mod. Phys. Lett.} {\bf #1} (19#2) #3}
 \newcommand{\cmp}[3]{{\sl Commun. Math. Phys.} {\bf #1} (19#2) #3}
 \renewcommand{\jmp}[3]{{\sl J. Math. Phys.} {\bf #1} (19#2) #3}
 \renewcommand{\nim}[3]{{\sl Nucl. Instr. Meth.} {\bf #1} (19#2) #3}
 \newcommand{\el}[3]{{\sl Europhysics Letters} {\bf #1} (19#2) #3}
 \renewcommand{\ap}[3]{{\sl Ann. of Phys.} {\bf #1} (19#2) #3}
 \renewcommand{\jetp}[3]{{\sl JETP} {\bf #1} (19#2) #3}
 \newcommand{\jetpl}[3]{{\sl JETP Lett.} {\bf #1} (19#2) #3}
 \newcommand{\acpp}[3]{{\sl Acta Physica Polonica} {\bf #1} (19#2) #3}
 \newcommand{\vj}[4]{{\sl #1~}{\bf #2} (19#3) #4}
 \newcommand{\ej}[3]{{\bf #1} (19#2) #3}
 \newcommand{\vjs}[2]{{\sl #1~}{\bf #2}}
 \newcommand{\hep}[1]{{hep--ph/}{#1}}
 \newcommand{\desy}[1]{{DESY-Report~}{#1}}

%***********************************************************************

\section{Next to Leading Order SUSY Cross Sections}

\subsection{Introduction} 
%        ============ 

%% this paragraph used to be the abstract
The calculation of the next-to-leading order SUSY-QCD 
corrections to the production of squarks, gluinos and gauginos at the
Tevatron is reviewed. The NLO corrections stabilize the theoretical
predictions of the various production cross sections significantly and
lead to sizeable enhancements of the most relevant cross sections for
scales near the average mass of the produced massive particles. We
discuss the phenomenological consequences of the results on present
and future experimental analyses.

The search for supersymmetric particles is among the most important
endeavors of present and future high energy physics.  At the upgraded
$p\bar p$ collider Tevatron, the searches for squarks and gluinos, as
well as for the weakly interacting charginos and neutralinos, will
cover a wide range of the MSSM parameter space~\cite{CCRFM-97}.

%The novel colored particles, squarks and gluinos, and the weakly
%interacting gauginos can be searched for at the upgraded Tevatron, a
% with a c.m.\ energy of 2 TeV.  Until now the search
%at the Tevatron has set the most stringent bounds on the colored SUSY
%particle masses.  At the 95\% CL, gluinos have to be heavier than
%about 180 GeV, while squarks with masses below about 180 GeV have been
%excluded for gluino masses below $\sim 300$ GeV \cite{bounds}.  Stops,
%the scalar superpartners of the top quark, have been excluded in a
%significant part of the MSSM parameter space with mass less than about
%80 GeV by the LEP and Tevatron experiments \cite{bounds}.  Finally
%charginos with masses below about 90 GeV have been excluded by the LEP
%experiments, while the present search at the Tevatron is sensitive to
%chargino masses of about 60--80 GeV with a strong dependence on the
%specific model \cite{bounds}. Due to the negative search at LEP2 the
%lightest neutralino $\tilde \chi_1^0$ has to be heavier than about 30
%GeV in the context of SUGRA models \cite{bounds}. In the
%$R$-parity-conserving MSSM, supersymmetric particles can only be
%produced in pairs.  All supersymmetric particles will decay to the
%lightest supersymmetric particle (LSP), which is most likely to be a
%neutralino, stable thanks to conserved $R$-parity.  Thus the final
%signatures for the production of supersymmetric particles will mainly
%be jets, charged leptons and missing transverse energy, which is
%carried away by neutrinos and the invisible neutral LSP.

The cross sections for the production of SUSY particles in hadron
collisions have been calculated at the Born level already quite some
time ago \cite{LO}. Only recently have the theoretical predictions
been improved by calculations of the next-to-leading order SUSY-QCD
corrections \cite{sqgl,gaunlo}.  The higher-order corrections in
general increase the production cross section compared to the
predictions at the Born level and thereby improve experimental mass
bounds and exclusion limits. Moreover, by reducing the dependence of
the cross section on spurious parameters, {\it i.e.} the
renormalization and factorization scales, the cross sections in NLO
are under much better theoretical control than the leading-order
estimates.

The paper is organized as follows. In Section 2 we shall review the
calculation of the next-to-leading order SUSY-QCD corrections
\cite{sqgl,gaunlo}, by using the case of $\sq \sqb$ production as an
example. The NLO results for the production of squarks and gluinos are
presented in Section~3. We first focus on the scalar partners of the
five light quark flavors, which are assumed to be mass degenerate.
The discussion of final-state stop particles, with potentially large
mass splitting and mixing effects, is presented in Section 4.  In
Section 5 we discuss the NLO cross sections for the production of
charginos and neutralinos. We conclude the paper with a summary of the
relevant MSSM particle production cross sections at the upgraded
Tevatron, including next-to-leading order SUSY-QCD
corrections.\footnote{The MSSM Higgs sector will not be discussed
here, see instead Ref.\cite{MS-98}.}

\subsection{SUSY-QCD corrections}
%        ====================
The evaluation of the SUSY-QCD corrections consists of two pieces, the
virtual corrections, generated by virtual particle exchanges, and the
real corrections, which originate from real-gluon radiation as well as
from processes with an additional massless (anti)quark in the final
state.

\subsubsection{Virtual corrections}
%           ===================
The one-loop virtual corrections, {\em i.e.}\ the interference of the
Born matrix element with the one-loop amplitudes, are built up by
gluon, gluino, quark and squark exchange contributions (see
Fig.~\ref{fg:virt}a). We have adopted the fermion flow
prescription~\cite{fermion} for the calculation of matrix elements
including Majorana particles. The evaluation of the one-loop
contributions has been performed in dimensional regularization,
leading to the extraction of ultraviolet, infrared and collinear
singularities as poles in $\epsilon = (4-n)/2$. For the chiral
$\gamma_5$ coupling we have used the naive scheme, which is well
justified in the present analysis at the one-loop level.\footnote{We
have explicitly checked that the results obtained with a consistent
$\gamma_5$ scheme are identical to those obtained with the naive
scheme.}  After summing all virtual corrections no quadratic
divergences are left over, in accordance with the general property of
supersymmetric theories. The renormalization of the ultraviolet
divergences has been performed by identifying the squark and gluino
masses with their pole masses, and defining the strong coupling in the
$\overline{\rm MS}$ scheme including five light flavors in the
corresponding $\beta$ function. The massive particles, {\em i.e.}\
squarks, gluinos and top quarks, have been decoupled by subtracting
their contribution at vanishing momentum transfer \cite{decouple}. In
dimensional regularization, there is a mismatch between the gluonic
degrees of freedom (d.o.f. = $n-2$) and those of the gluino (d.o.f. =
$2$), so that SUSY is explicitly broken. In order to restore SUSY in
the physical observables in the massless limit, an additional finite
counter-term is required for the renormalization of the novel $\sq \gl
\bar q$ vertex. These counter-terms have been shown to render
dimensional regularization consistent with supersymmetry~\cite{count}.

\subsubsection{Real corrections} 
\begin{figure}[ht] \begin{center}\leavevmode
\epsfig{file=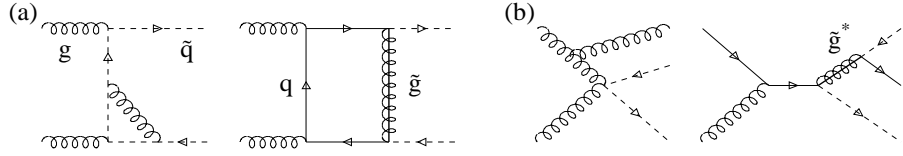,angle=0,width=12cm}
\caption[]{\label{fg:virt} \label{fg:real} 
  \it Typical diagrams of the virtual (a) and real (b) corrections.}
\end{center} \end{figure}

%           ================ 
The real corrections are generated by real-gluon radiation off all
colored particles and by final states with an additional massless
(anti)quark, obtained from interchanging the final state gluon with a
light quark in the initial state (see Fig.~\ref{fg:virt}b).  The
phase-space integration of the final-state particles has been
performed in $n=4-2\epsilon$ dimensions, leading to the extraction of
infrared and collinear singularities as poles in $\epsilon$.  After
evaluating all angular integrals and adding the virtual and real
corrections, the infrared singularities cancel.  The left-over
collinear singularities are universal and are absorbed in the
renormalization of the parton densities at next-to-leading order. We
have defined the parton densities in the conventional $\overline{\rm
MS}$ scheme including five light flavors, {\em i.e.}\ the squark,
gluino and top quark contributions are not included in the mass
factorization. We finally obtain an ultraviolet, infrared and
collinear finite partonic cross section.

There is, however, an additional class of physical singularities,
which have to be regularized. In the second diagram of
Fig.~\ref{fg:real}b, an intermediate $\sq \gl^*$ state is produced,
before the (off-shell) gluino splits into a $q\sqb$ pair. If the
gluino mass is larger than the common squark mass, and the partonic
c.m.\ energy is larger than the sum of the squark and gluino masses,
the intermediate gluino can be produced on its mass-shell. Thus the
real corrections to $\sq \sqb$ production contain a contribution of
$\sq \gl$ production. The residue of this part corresponds to $\sq
\gl$ production with the subsequent gluino decay $\gl \to \sqb q$,
which is already contained in the leading order cross section of $\sq
\gl$ pair production, including all final-state cascade decays.  This
term has been subtracted in order to derive a well-defined production
cross section. Analogous subtractions emerge in all reactions: if the
gluino mass is larger than the squark mass, the contributions from
$\gl \to \sq \bar q, \sqb q$ have to be subtracted, and in the reverse
case the contributions of squark decays into gluinos have to
subtracted.\smallskip

\subsection{Results}
%        =======
In the following, we will present numerical results for SUSY particle
production cross sections at the upgraded Tevatron ($\sqrt{s}=2$~TeV),
including SUSY-QCD corrections.  The hadronic cross sections are
obtained from the partonic cross sections by convolution with the
corresponding parton densities. We have adopted the CTEQ4L/M parton
densities \cite{CTEQ} for the numerical results presented below. The
uncertainty due to different parametrizations of the parton densities
in NLO is less than $\sim 15$~\%. The average final state particle
mass is used as the central value of the renormalization and
factorization scales and the top quark mass is set to $m_t =
175$~GeV. The $K$-factor is defined as $K = \sigma_{NLO} /
\sigma_{LO}$, with all quantities ($\alpha_s(\mu_R)$, parton densities, 
parton cross section) calculated consistently in lowest and in
next-to-leading order.

\subsubsection{Production of Squarks and Gluinos}
\begin{figure}[ht] \begin{center}\leavevmode
\epsfig{file=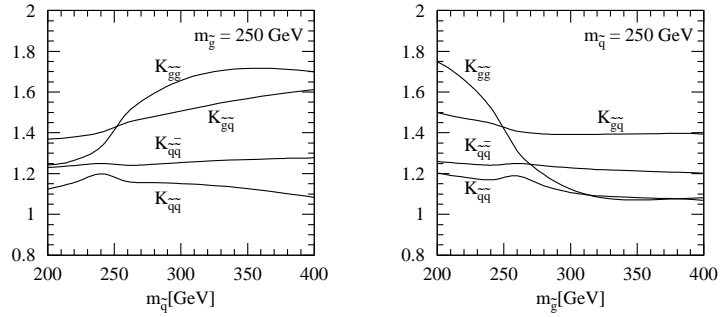,angle=270,width=10cm}
\caption[]{\label{fg:kfac} 
  \it $K$-factors of the squark and gluino
      production cross sections at $Q=m$.}
\end{center} \end{figure}

%           =================================
Squarks and gluinos can be produced in different combinations via
$p\bar p \to \sq \sqb, \sq \sq, \sq \gl, \gl \gl$. We first focus on
the five light-flavored squarks, taken to be mass degenerate. At the
central renormalization and factorization scale $Q=m$, where $m$
denotes the average mass of the final-state squarks/gluinos, the
SUSY-QCD corrections are large and positive, increasing the total
cross sections in general by 10--90\% \cite{sqgl}. This is shown in
Fig.~\ref{fg:kfac}, where the $K$-factors are presented as a function
of the corresponding SUSY particle mass. The inclusion of SUSY-QCD
corrections leads to an increase of the lower bounds on the squark and
gluino masses by 10--30 GeV with respect to the leading-order
analysis.

\begin{figure}[ht] \begin{center}\leavevmode
\epsfig{file=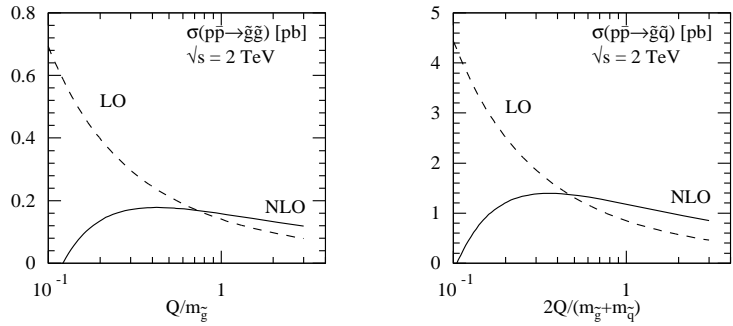,angle=270,width=10cm}
\caption[]{\label{fg:scale} 
  \it Scale dependence of the total squark and gluino production 
      cross sections for 
      $m_{\sq}=250$~GeV and $m_{\gl}=300$~GeV.}
\end{center} \end{figure}

The residual renormalization/factorization scale dependence in leading
and next-to-leading order is presented in Fig.~\ref{fg:scale}. The
inclusion of the next-to-leading order corrections reduces the scale
dependence by a factor 3--4 relative to the lowest order and reaches a
typical level of $\sim 15\%$, when varying the scale from $Q = 2m$ to
$Q = m/2$. This may serve as an estimate of the remaining theoretical
uncertainty due to uncalculated higher-order terms.

Finally, we have evaluated the QCD-corrected single-particle exclusive
transverse-momentum and rapidity distributions for all different
processes.  As can be inferred from Fig.~\ref{fg:pty}, the
modification of the normalized distributions in next-to-leading
compared to leading order is less than about 15\% for the
transverse-mo\-men\-tum distributions and even smaller for the
rapidity distributions.  It is thus a sufficient approximation to
rescale the leading order distributions uniformly by the $K$-factors
of the total cross sections.

\subsubsection{Stop Pair Production} 
%           ==================== 

Stop production has to be considered separately since the strong
Yukawa coupling between top/stop and Higgs fields gives rise to
potentially large mixing effects and mass splitting. At leading-order
in the strong coupling constant $\alpha_s$, only diagonal pairs of
stop quarks can be produced in hadronic collisions, $p\bar p \to
{\st}_1\stb_1/{\st}_2\stb_2$. In contrast to the production of
light-flavor squarks, the leading-order $t$-channel gluino exchange
diagram is absent for stop production via $q\bar{q}$ initial states,
since top quarks are not included in the parton densities. The
leading-order stop cross section is thus in general significantly
smaller than the leading order cross section for producing
light-flavor squarks, where the threshold behavior is dominated by
$t$-channel gluino exchange. Mixed ${\st}_1 {\st}_2$ pair production can
safely be neglected since it can proceed only via one-loop $\alpha_s$
or tree level $G_F$ amplitudes and is suppressed by several orders of
magnitude \cite{stops}.  The evaluation of the QCD corrections
proceeds along the same lines as in the case of squarks and
gluinos.\footnote{The results obtained for the case of stop production
can also be used to predict the sbottom pair cross section at NLO
including mixing and mass splitting.} The strong coupling and the
parton densities have been defined in the \MS~scheme with five light
flavors contributing to their scale dependences, while the stop masses
are renormalized on-shell.

The magnitude of the SUSY-QCD corrections is illustrated by the
$K$-factors at the central scale $Q=m_{\st}$ in Fig.~\ref{fg:kst}. In
the mass range relevant for the searches at the Tevatron, the SUSY-QCD
corrections are positive and reach a level of 30 to 45\% if the $gg$
initial state dominates. If, in contrast, the $q\bar{q}$ initial state
dominates, the corrections are small.  The relatively large mass
dependence of the $K$-factor for stop production at the Tevatron can
therefore be attributed to the fact that the $gg$ initial state is
important for small $m_{\tilde{t}}$, whereas the $q\bar{q}$ initial
state dominates for large $m_{\tilde{t}}$.

In complete analogy to the squark/gluino case, the scale dependence of
the stop cross section is strongly reduced, to about 15\% at
next-to-leading order in the interval $m_{\st}/2<Q<2m_{\st}$. The
virtual corrections at the NLO level depend on the stop mixing angle,
the squark and gluino masses, and on the mass of the second stop
particle. It turns out, however, that these dependences are very weak
for canonical SUSY masses and can safely be neglected, as can be
inferred from the light-stop production cross section in
Fig.~\ref{fg:kst}. On the other hand, internal particles with masses
smaller than the external particle mass, e.g. a light stop state
propagating in the loops for heavy stop production, will contribute to
the cross section.  This feature explains the small but noticeable
difference between the ${\st}_1$ and ${\st}_2$ $K$-factors at $m_{\st} =
300$~GeV shown in Fig.~\ref{fg:kst}.

The next-to-leading order transverse-momentum and rapidity
distributions are presented in Fig.~\ref{fg:tpty}. While the shape of
the rapidity distribution is almost identical at leading and
next-to-leading order, the transverse momentum carried away by hard
gluon radiation in higher orders softens the NLO transverse momentum
distribution considerably.

\subsubsection{Chargino and Neutralino Production} 
%           ================================== 
At leading order, the production cross sections for chargino and
neutralino final states depend on several MSSM parameters, {\em i.e.}\
$M_1, M_2, \mu$ and $\tgb$ \cite{LO}. The cross sections are sizeable
for chargino/neutralino masses below about 100 GeV at the upgraded
Tevatron. Due to the strong sensitivity to the MSSM parameters, the
extracted bounds on the chargino and neutralino masses depend on the
specific region in the MSSM parameter space \cite{CCRFM-97}.  The
outline of the determination of the QCD corrections is analogous to
the previous cases of squarks, gluinos and stops. The resonance
contributions due to $gq \to \gau_i \sq$ with $\sq \to q \gau_j$ have
to be subtracted in order to avoid double counting of the associated
production of electroweak gauginos and strongly interacting squarks.
The parton densities have been defined with five light flavors
contributing to their scale evolution in the \MS~scheme, while the
$t$-channel squark masses have been renormalized on-shell.\footnote{
The next-to-leading order SUSY-QCD corrections to slepton pair
production can be trivially obtained from the corresponding results
for chargino/neutralino production. Numerically, the SUSY-QCD
corrections for slepton production agree with the pure QCD
corrections~\cite{sleptons}, provided the squark and gluino mass are
not chosen smaller than the final state slepton mass.}

At the average mass scale, the QCD corrections enhance the production
cross sections of charginos and neutralinos typically by about
10--35\% (see Fig.~\ref{fg:kgausc}), depending in detail on the final
state and the choice of MSSM parameters. The leading order scale
dependence is reduced to about 10\% at next-to-leading order (see
Fig.~\ref{fg:kgausc}), which implies a significant stabilization of
the theoretical prediction for the production cross sections
\cite{gaunlo}.

The individual leading order contributions of the $s$-channel gauge
boson and the $t,u$-channel squark exchange are presented in
Fig.~\ref{fg:gauind}. For neutralino pair production the
$(t+u)$-channel contributions are by far dominating, while the
$s$-channel and interference terms are suppressed. Since the
$\gau^0_{1,2}$ states are predominantly gaugino-like, this reflects
the absence of a purely neutral trilinear gauge boson coupling in the
Standard Model. Contrary to that the $s$-channel of $\gau^\pm_1
\gau^0_2$ production is dominant and the $(t+u)$-channel term is
suppressed for large squark masses.  However, the interference turns
out to be sizeable.

\subsection{Conclusions} 
%        =========== 
We have reviewed the status of SUSY particle production at the
upgraded Tevatron at next-to-leading order in supersymmetric QCD. A
collection of relevant sparticle production cross sections is shown in
Fig.~\ref{fg:sum_tev}.  The higher-order corrections at the average
mass scale of the massive final-state particles significantly increase
the production cross section compared to the predictions at the Born
level. Experimental mass bounds are therefore shifted upwards.
Moreover, the theoretical uncertainties due to variation of
renormalization/factorization scales are strongly reduced to a level
of typically $\sim 15$~\%, so that the cross sections in
next-to-leading order SUSY-QCD are under much better theoretical
control than the leading order estimates. The NLO results for total
cross sections and differential distributions are available in the
form of the computer code PROSPINO \cite{prospino}. \\

%\noindent {\bf Acknowledgements} \\ We would like to thank P.M.\
%Zerwas for his collaboration and encouragement. Moreover, we are
%grateful to R.\ H\"opker and M.\ Klasen for their contribution to
%different parts of this work.

\clearpage

{\renewcommand{\topfraction}{1.0}   
\renewcommand{\bottomfraction}{1.0}
\renewcommand{\textfraction}{0.0}  

%% 2nd set of 3:
\begin{figure}[t] \begin{center}\leavevmode
\epsfig{file=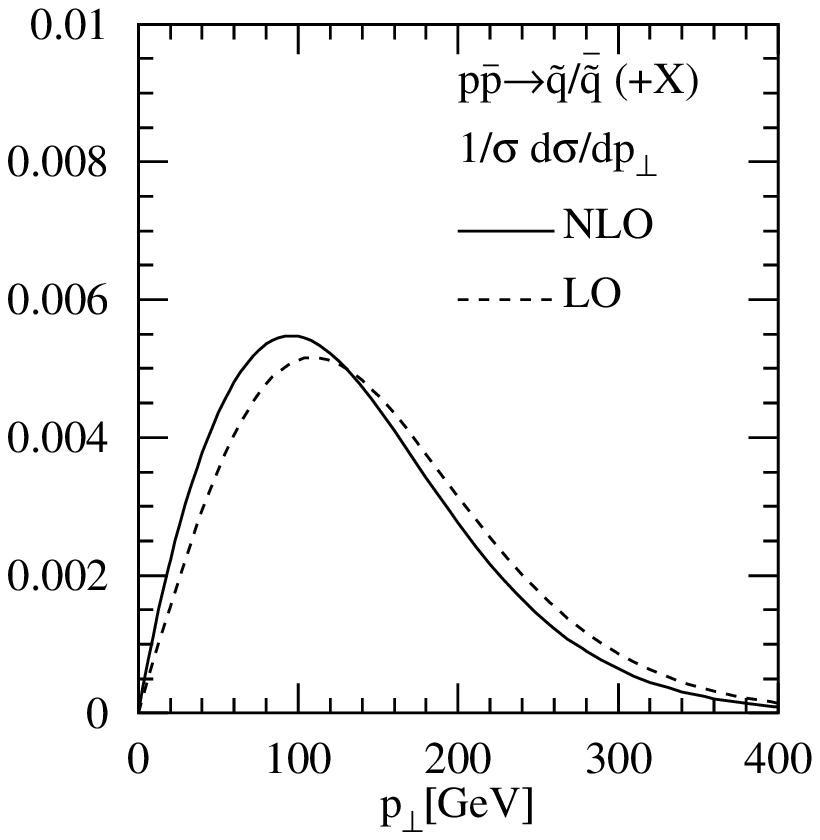,angle=270,width=5cm} \hspace{0.5cm}
\epsfig{file=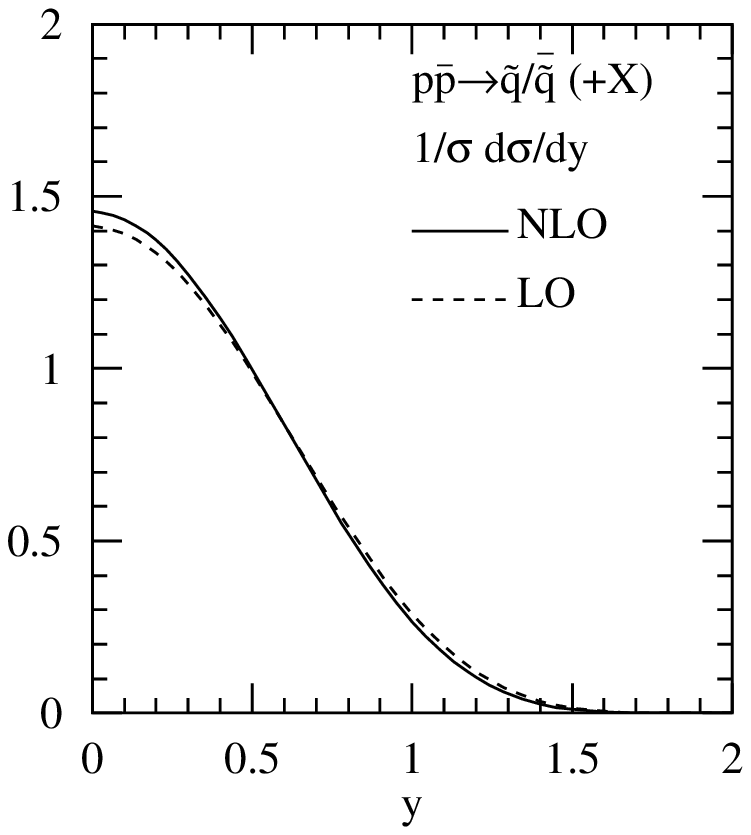,angle=270,width=5cm}
\caption[]{\label{fg:pty} 
  \it Normalized transverse-momentum and rapidity
      distributions for squark production at $Q=m$.
      Mass parameters: $m_{\sq}=250$~GeV and $m_{\gl}=300$~GeV.}
\end{center} \end{figure}

\begin{figure}[h] 
\begin{center}\leavevmode
\epsfig{file=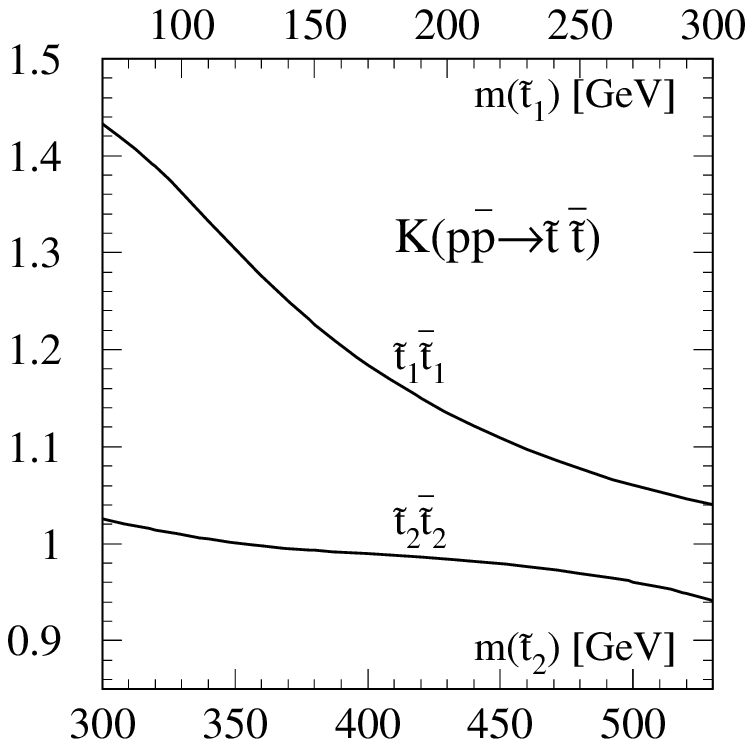,angle=0,width=4.9cm} \hspace{0.5cm}
\epsfig{file=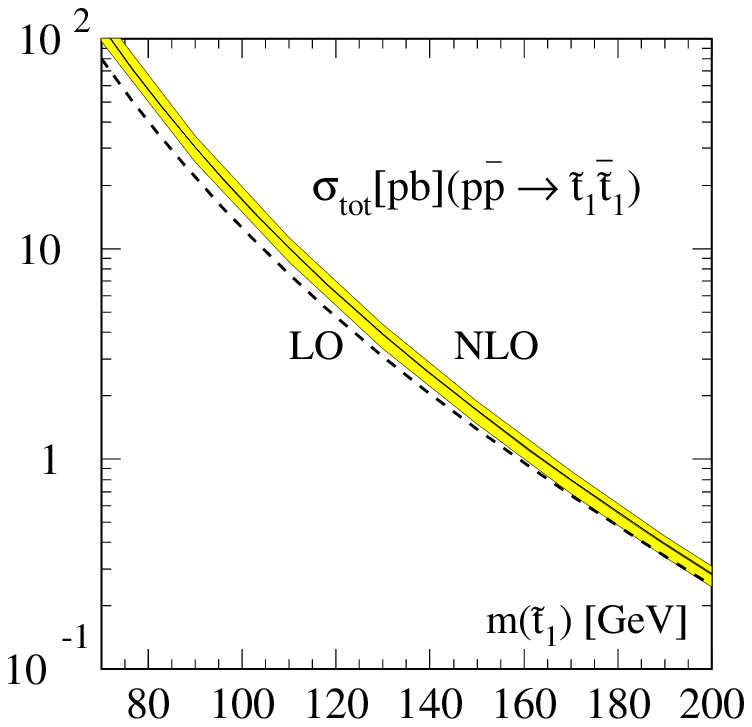,angle=0,width=4.9cm}
\caption[]{\label{fg:kst} 
  \it Left: $K$-factor of the stop production cross sections at
      $Q=m_{\st}$ as a function of the stop masses [top/bottom scale].  
      Right: Production cross sections for the light stop
      state.  The thickness of the NLO curve represents the dependence
      of the cross sections on the stop mixing angle and the gluino
      and squark masses.  The shaded band indicates the theoretical
      uncertainty due to the scale dependence [$m/2<Q<2m$].}
\end{center} \end{figure}

%%fig 10
\begin{figure}[b] 
\begin{center}\leavevmode
\epsfig{file=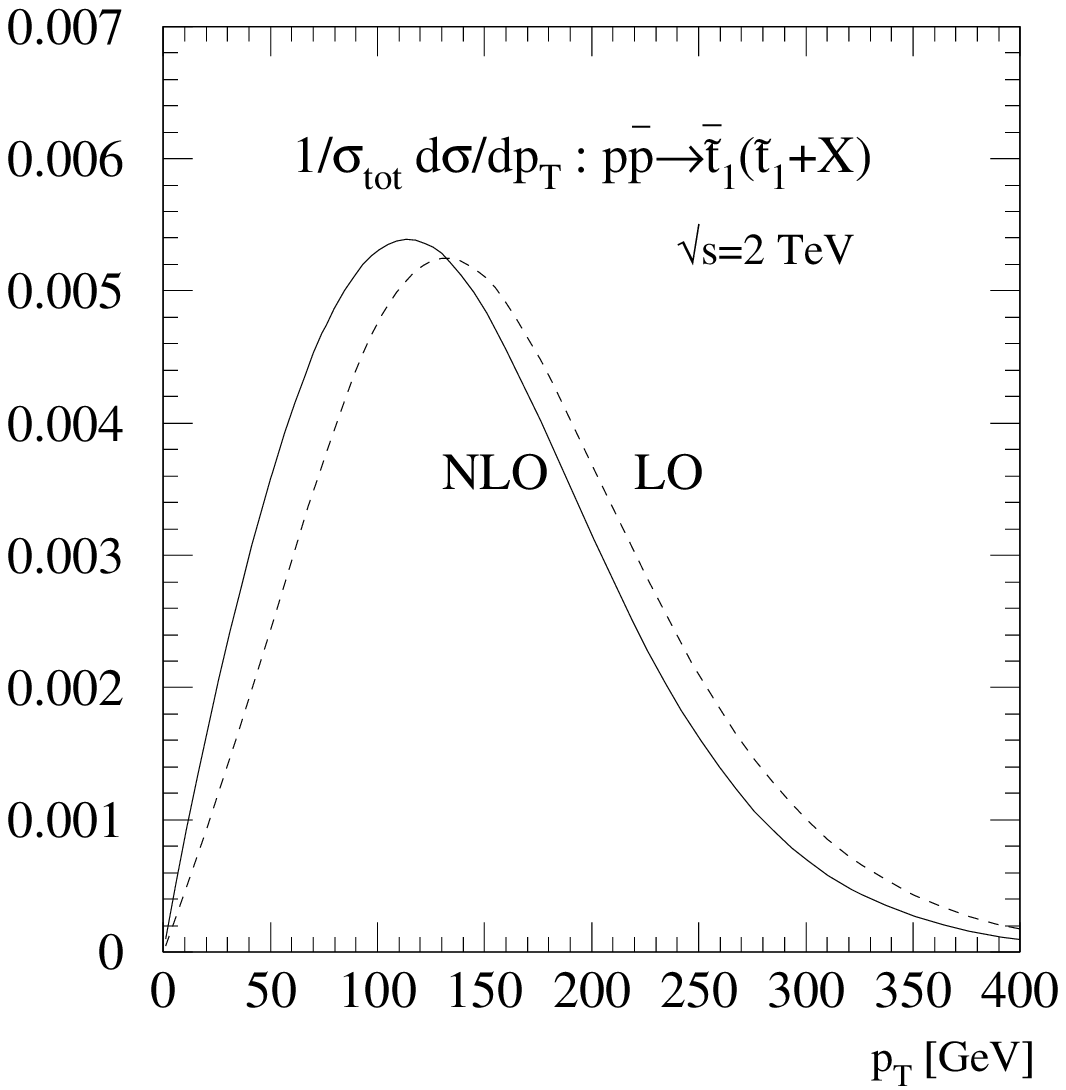,angle=0,width=4.8cm} \hspace{0.5cm}
\epsfig{file=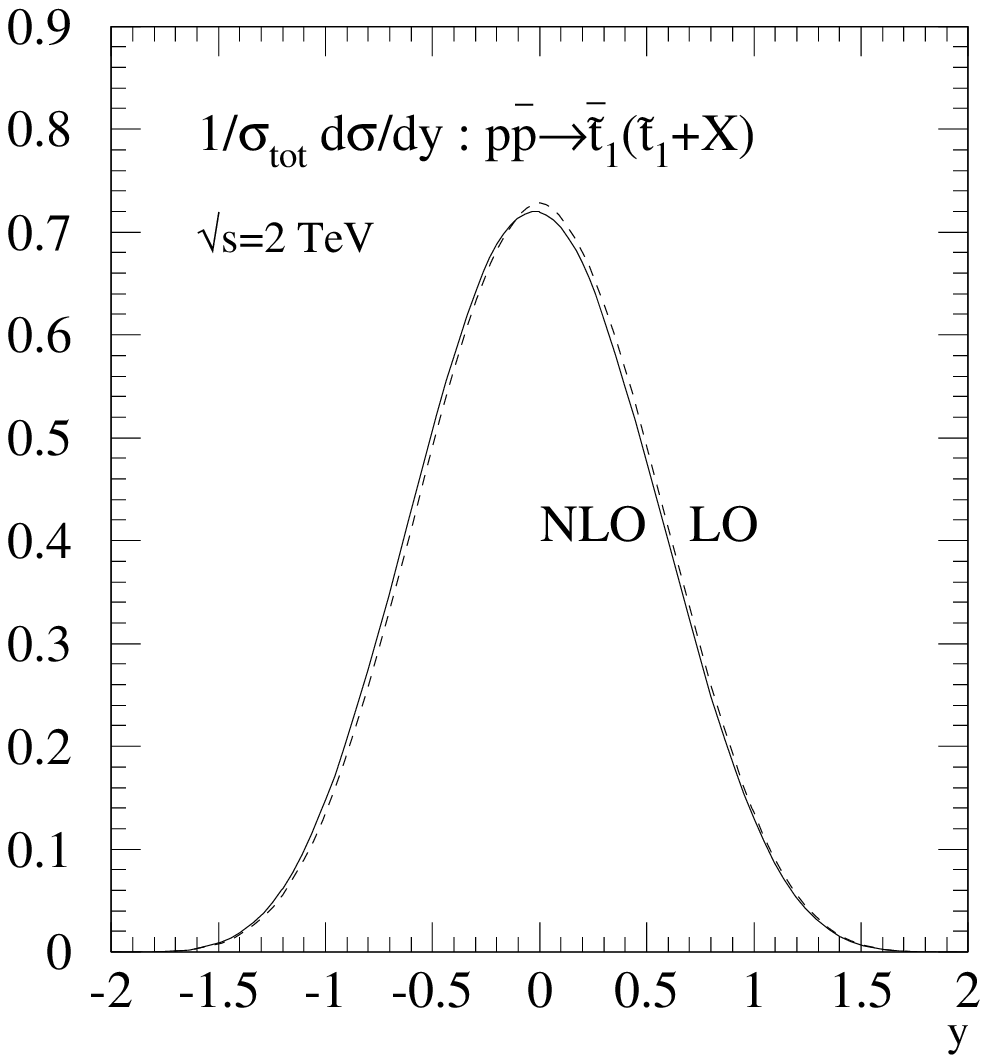,angle=0,width=4.8cm}
\caption[]{\label{fg:tpty} 
  \it Normalized transverse-momentum and rapidity
      distributions for stop production at $Q=m_{\st}=200$~GeV.}
\end{center} \end{figure}

% last set of 3:
\begin{figure}[t] \begin{center} \leavevmode
\epsfig{file=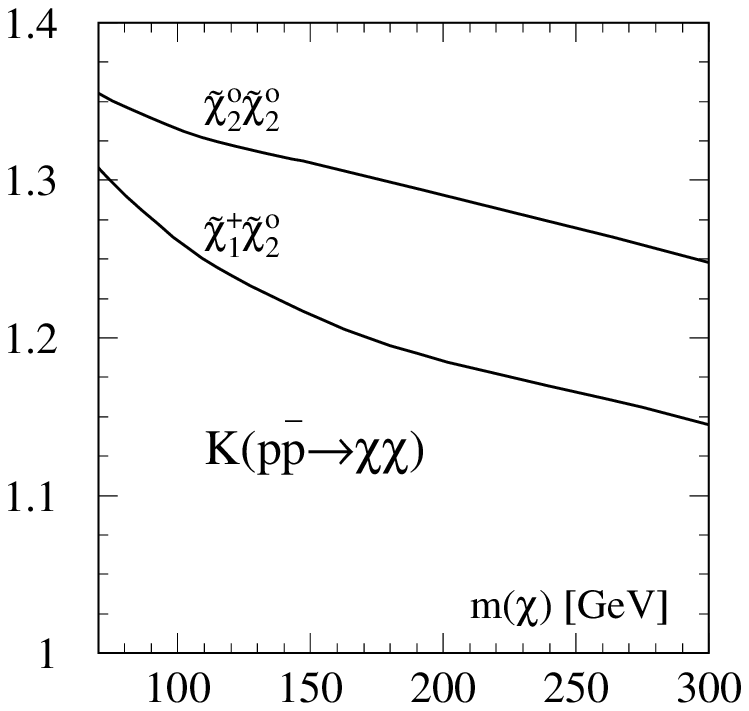,angle=0,width=5cm} \hspace{0.5cm}
\epsfig{file=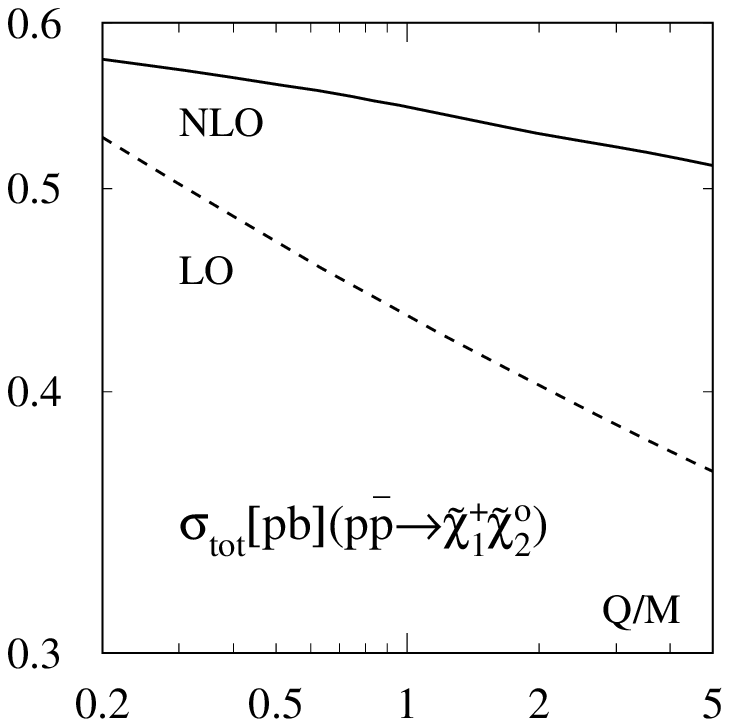,angle=0,width=5cm}
\caption[]{\label{fg:kgausc} 
  \it Left: $K$-factor of the $\tilde{\chi}_2^{0}\tilde{\chi}_2^{0}$
  and $\tilde{\chi}_1^{+}\tilde{\chi}_2^{0}$ production cross sections
  at the central scale $Q=m$. Right: Scale dependence of the
  $\tilde{\chi}_1^{+}\tilde{\chi}_2^{0}$ cross section.  SUGRA
  parameters: $m_0 = 100$ GeV, $A_0 = 300$ GeV, $\tgb = 4$, $\mu>0$.
  [The sign convention used in this section differs 
   from the ISAJET convention by a factor of $(-1)$ for $A_0$]}
\end{center} \end{figure}

\begin{figure}[h] \begin{center} \leavevmode
\epsfig{file=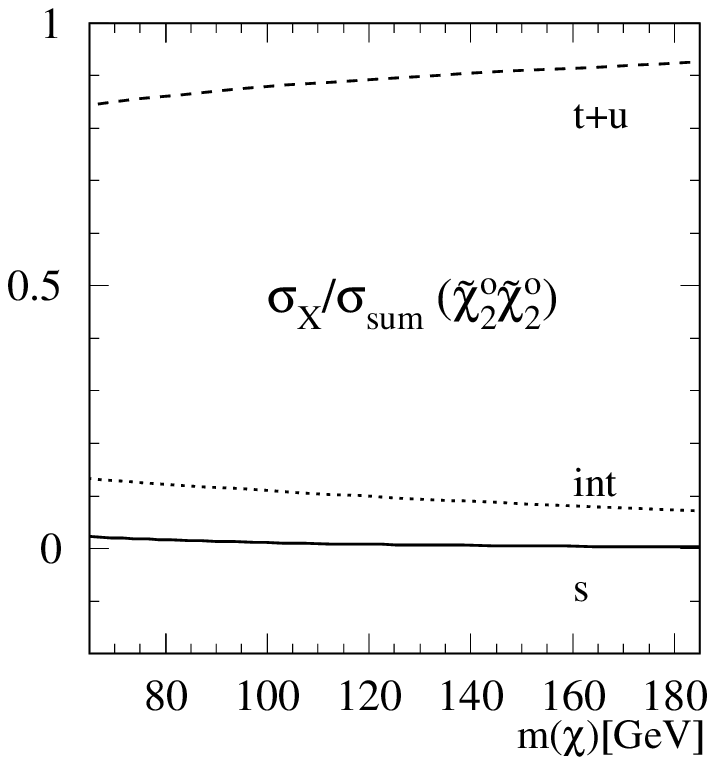,angle=0,width=5cm} \hspace{0.5cm}
\epsfig{file=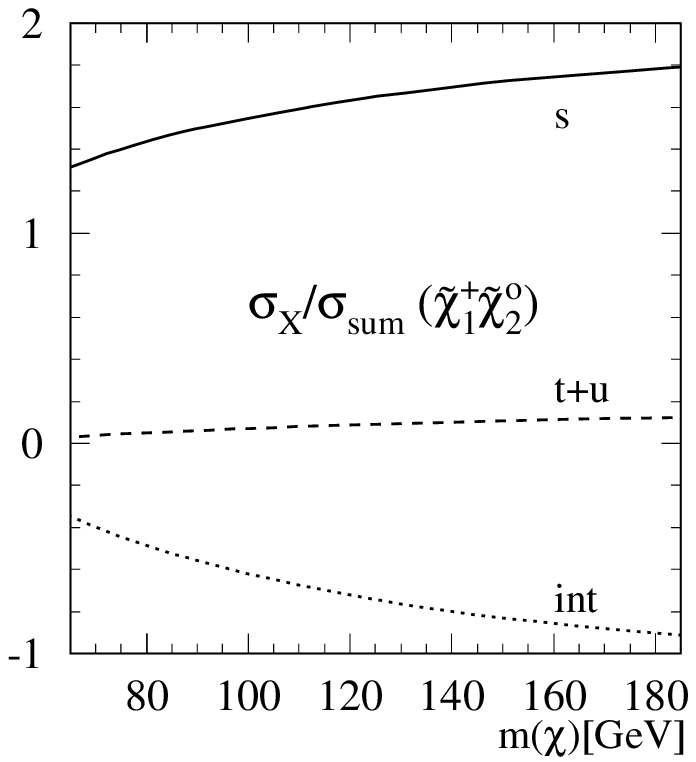,angle=0,width=5cm}
\caption[]{\label{fg:gauind} 
  \it The $s$-channel, $(t+u)$-channel and interference contributions 
  to the 
  leading order $\tilde{\chi}_2^{0}\tilde{\chi}_2^{0}$ and 
  $\tilde{\chi}_1^{+}\tilde{\chi}_2^{0}$ production cross section.
  SUGRA parameters as in Fig.~\ref{fg:kgausc}}
\end{center} \end{figure}

\begin{figure}[b] \begin{center} \leavevmode
\epsfig{file=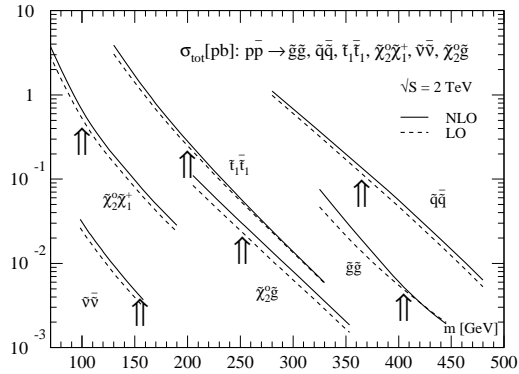,angle=0,width=7cm}
\caption[]{\label{fg:sum_tev} 
  \it The NLO production cross sections included in PROSPINO as
      a function of the final state particle mass; the arrows indicate
      the SUGRA inspired scenario: $m_{1/2}=150$~GeV, $m_0 = 100$~GeV,
      $A_0 = 300$~GeV, $\tgb = 4$, $\mu > 0$.  All cross sections are
      given at the average mass scale of the massive final-state
      particles.}
\end{center} \end{figure}
}

%% file: E.Berger/fnalrep.tex
%%%%%%%%%%%%%%%%%%%%%%%%%%%%%%%%%%%%%%%%%%%%%%%%%%%
\def\tell{\tilde{e}}
\def\tnu{\tilde{\nu}}

\section{Analysis of Next to Leading Order SUSY Production Cross Sections}

%%%%%%%%%%%%%% Begin Section I %%%%%%%%%%%%%%%%%%%%%%%%%%%%%%%%%%%%%%%%%
\subsection{Perturbative QCD Results}
\label{sec:1}

The possibility of supersymmetry (SUSY) at the electroweak scale and 
the ongoing search for the Standard Model (SM) Higgs boson constitute two 
major related aspects of the motivation for the Tevatron upgrade currently 
under construction at Fermilab.  The increase in the center-of-mass energy to 
2~TeV and the luminosity to an expected 2 fb$^{-1}$, together with detector
improvements, should permit discovery or exclusion of supersymmetric partners 
of the standard model particles up to much higher masses than at 
present~\cite{torah}. 

Estimates of the production cross sections for pairs of supersymmetric 
particles may be computed analytically from fixed-order 
quantum chromodynamics (QCD) perturbation theory.  Calculations that include 
contributions through next-to-leading order (NLO) in QCD have been performed 
for the production of squarks and gluinos \cite{Hopker}, top squark pairs 
\cite{Plehn}, slepton pairs \cite{Baer,Klasen}, gaugino pairs \cite{Klasen}, 
and the associated production of gauginos and gluinos \cite{BKT}. The cross 
sections can be calculated as functions of the sparticle masses and mixing 
parameters. 

In a recent paper \cite{Berger}, Berger, Klasen, and Tait provide numerical 
predictions at next-to-leading order for the production of squark-antisquark, 
squark-squark, gluino-gluino, squark-gluino, and top squark - antitop squark 
pairs in 
proton-antiproton collisions at the hadronic center-of-mass energy 2~TeV.  
These calculations are based on the analysis of Refs. \cite{Hopker,Plehn}, 
and the CTEQ4M parametrization \cite{CTEQ4} of parton densities.   The hard 
scale dependence of the cross section at leading order (LO) in perturbative 
QCD is reduced at NLO but not absent.  An estimate of the theoretical 
uncertainty at NLO is approximately $\pm 15$~\% about a central value.  
The central value is obtained 
with the hard scale chosen to be equal to the average of the masses of the 
produced sparticles, and the band of uncertainty is determined from a 
variation of the hard scale from half to twice this average mass.  The 
next-to-leading order contributions increase the production cross sections 
by 50~\% and more from their LO values.  For example, in the case of 
squark-antisquark production the next-to-leading order cross section lies 
above the leading order cross section by 59~\%.  This increase translates into 
a shift in the lower limit of the produced squark mass of 19~GeV.  The cross 
sections for squark-antisquark production, gluino pair production, and the 
associated production of squarks and gluinos of equal mass are of similar 
magnitude, whereas the squark pair production and top squark-antitop squark 
production cross sections are smaller by about an order of 
magnitude \cite{Hopker,Plehn}.

The cross sections reported in Ref.~\cite{Berger} are for inclusive yields, 
integrated over all transverse momenta and rapidities.  In the search for 
supersymmetric states, a selection on transverse momentum will normally be 
applied in order to improve the signal to background conditions.  The 
theoretical analysis can also be done with similar selections.  A tabulation 
of cross sections for various squark and gluino masses is available upon 
request from the authors of Ref. \cite {Berger}.

Next-to-leading order calculations of the production of neutralino pairs, 
chargino pairs, and neutralino-chargino pairs ar-mon the way to completion \cite{Klasen}, but final numerical predictions are 
not yet available for general use.  

The strongly interacting squarks and gluinos may also be produced singly 
in association with charginos and neutralinos.  Leading-order production cross 
sections for the associated production of a chargino plus a squark or gluino 
and of a neutralino plus a squark or gluino are published~\cite{assoclo}, and 
a next-to-leading order calculation of associated production of a gaugino plus 
a gluino is also now published~\cite{BKT}.  

\begin{figure}[h]
  \vspace*{-10mm}
\centering\leavevmode
  \epsfig{file=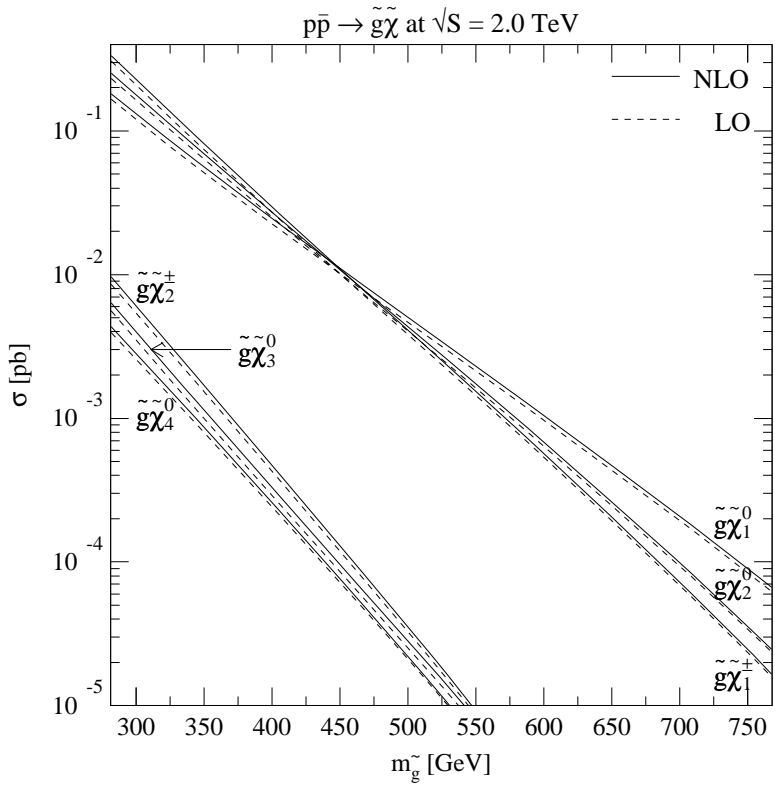,width=12cm}
  \vspace*{-4mm}

 \caption{Total hadronic cross sections for the associated production
  of gluinos and gauginos at Run II of the Tevatron.  NLO results are shown 
  as solid curves, and LO results as dashed curves.  The SUGRA scenario 
  parameter $m_{1/2}$ is varied $\in [100;400]$ GeV, and the cross sections 
  are shown as a function of the physical gluino mass. The chargino 
  cross sections are summed over positive and negative chargino charges.}
\label{elb-xsec}
\end{figure}

Berger, Klasen, and Tait~\cite{BKT} compute total cross sections for all the 
gaugino-gluino production reactions $\tilde g \tilde{\chi}^0_{(1-4)}$ and 
$\tilde g \tilde{\chi}^{\rm \pm}_{(1-2)}$ in next-to-leading order SUSY-QCD.  
For numerical results, they select an illustrative mSUGRA scheme in which the 
GUT scale common scalar mass $m_0 = 100 $ GeV, the common gaugino mass 
$m_{1/2} =150 $ GeV, the trilinear coupling $A_0 = 300 $ GeV, 
$\rm{tan}(\beta) = 4$ and $\rm{sgn}(\mu) = +$.  (The sign convention for 
$A_0$ is opposite to that in the ISASUGRA code).  They convolute the NLO hard 
partonic cross sections with the CTEQ4M parametrization \cite{CTEQ4} of parton 
densities, and present physical cross sections as a function of the 
$\tilde g$ mass or of the average mass 
$m = (m_{\tilde \chi} + m_{\tilde g})/2$.  For $p\bar p$ collisions at 
$\sqrt{S}=2 $ TeV the cross sections at $m_{\tilde g} = 300 $ GeV range from 
${\cal O}( 0.2 \,{\rm pb} )$ for the $\tilde {\chi}^0_2$ and the 
$\tilde {\chi}^{\rm \pm}_1$ to ${\cal O}( 0.2 \times 10^{-3}{\rm pb} )$ for the 
$\tilde {\chi}^0_4$.  The $\tilde g \tilde{\chi}^0_{(1,2)}$ and 
$\tilde g \tilde{\chi}^{\rm \pm}_1$ cross sections are of hadronic size 
despite the fact that the overall coupling strength is 
${\cal O}(\alpha_{EW}\alpha_s)$, not ${\cal O}(\alpha_s^2)$.  The 
masses of the $\tilde{\chi}^0_{(1,2)}$ and 
$\tilde{\chi}^{\rm \pm}_1$ ar- significantly smaller in a typical 
mSUGRA scenario than those of the squarks and gluinos.  The phase space 
and the parton luminosity ar- therefore greater for associated production of a 
gluino and a gaugino than for a pair of squarks or gluinos, and the smaller 
coupling strength is compensated.  
The next-to-leading-order cross sections are enhanced by 
typically 5\% to 15\% relative to the leading order values.  The theoretical 
uncertainty resulting from variations of the factorization/renormalization 
scale is approximately $\pm 10\%$ at NLO for the 
$\tilde {\chi}^0_2$ and the $\tilde {\chi}^{\rm \pm}_1$, a factor of 2 
smaller than the LO variation.  Shown in Fig.~\ref{elb-xsec} are the predicted cross 
sections as a function of $m_{\tilde g}$.  

Baer, Harris, and Reno~\cite{Baer} compute total cross sections for all the 
slepton pair production reactions $\tell_L\tnu_L$, $\tell_L\bar{\tell_L}$,
$\tell_R\bar{\tell_R}$ and $\tnu_L\bar{\tnu}_L$ in next-to-leading
order QCD.  The analytic calculations are very similar to the QCD corrections
to the Standard Model massive lepton-pair production (Drell-Yan) process.
Numerical results are based on the CTEQ4M parametrization \cite{CTEQ4} of 
parton densities.   For $p\bar p$ collisions at $\sqrt{S}=2 $ TeV, the cross 
sections range from ${\cal O}( 1{\rm pb} )$ at $m_{\rm slepton}=50~$GeV 
to ${\cal O}( 10^{-3}{\rm pb} )$ at $m_{\rm slepton}=200~$GeV.  The 
next-to-leading-order cross sections are enhanced by typically 35\% to 40\% 
relative to the leading order values.  The theoretical uncertainty resulting
from variations in the hard scattering scale and parton distribution
functions is approximately $\pm 15\%$.
In the mSUGRA model, slepton pair production is most important for small
values of the parameter $m_0$.
The next-to-leading order enhancements of slepton pair cross sections 
at Tevatron energies can push predictions for leptonic SUSY signals to 
higher values than typically
quoted in the literature in these regions of model parameter space.

For current expectations of the hierarchy of masses and cross sections, 
consult Ref.~\cite{torah}.  

%%%%%%%%%%%%%% Begin Section II %%%%%%%%%%%%%%%%%%%%%%%%%%%%%%%%%%%%%%%%%
\subsection{Monte Carlo Methods}
\label{sec:2}

Experimental searches for supersymmetry rely heavily on Monte Carlo 
simulations of cross sections and event topologies. Two Monte Carlo generators 
in common use for hadron-hadron collisions include SUSY processes; they 
are ISAJET \cite{Paige}
and SPYTHIA \cite{Sjostrand,Mrenna}.  Both the Monte Carlo approach and the 
fixed order pQCD approach have different
advantages and limitations.  Next-to-leading order perturbative calculations 
depend on very few parameters, e.g., the renormalization and factorization 
scales, and the dependence of the production cross sections on these 
parameters is reduced significantly in NLO with respect to LO. 
Therefore, the normalization of the cross section can be calculated quite 
reliably if one includes the NLO contributions.  On the other hand, the 
existing next-to-leading order calculations provide predictions only for 
fully inclusive quantities, e.g., a differential cross section for 
production of a squark or a gluino, after integration over all other particles 
and variables in the final state.  In addition, they do not include sparticle 
decays. This approach does not allow for event shape studies nor for 
experimental selections on missing energy or other variables associated with 
the produced sparticles or their decay products that are crucial if one wants 
to enhance the SUSY signal in the face of substantial backgrounds from 
Standard Model processes.

The natural strength of Monte Carlo simulations consists in the fact that
they generate event configurations that resemble those observed in experimental 
detectors.  Through their parton showers, these generators include, in the 
collinear approximation, contributions from all orders of perturbation theory.  
In addition, they incorporate phenomenological hadronization models, a 
simulation of particle decays, the possibility to implement experimental cuts, 
and event analysis tools.  However, the hard-scattering matrix elements in 
these generators are accurate only to leading order in QCD, and, owing to 
the rather complex nature of infra-red singularity cancellation in higher 
orders of perturbation theory, it remains a difficult challenge to incorporate 
the full structure of NLO contributions successfully in Monte Carlo simulations.
The limitation to leading-order hard-scattering matrix elements leads to 
large uncertainties in the normalization of the cross section.  The parton 
shower and hadronization models rely on tunable parameters, another source of 
uncertainties.

In Ref. \cite{Berger} a method is suggested to improve the accuracy of the 
normalization of cross sections computed through Monte Carlo simulations.
In this approach, the renormalization and factorization (hard) scale in the 
Monte Carlo LO calculation is chosen in such a way that the normalization 
of the Monte Carlo LO calculation agrees with that of the NLO 
perturbative calculation.  The scale choice depends on which partonic 
subprocess one is considering and on the kinematics.  This choice of  
the hard scale will affect both the hard matrix element {\em and} the 
initial-state and final-state parton shower radiation.   On the other hand, 
an alternative rescaling of the cross section by an overall $K$-factor will 
have no bearing on the parton shower radiation.  A reduction in the hard scale 
leads generally to less evolution and less QCD radiation, and vice-versa, in 
the initial- and final-state showering.  A change of the hard scale will be 
reflected in the normalization of the cross section as well as in the event 
shape.  Investigations are underway to determine how significant the changes 
are in computed final state momentum distributions.  

%%%%%%%%%%%%%% Begin References %%%%%%%%%%%%%%%%%%%%%%%%%%%%%%%%%%%%%%%%

%%%%%%%%%%%%%% End of References %%%%%%%%%%%%%%%%%%%%%%%%%%%%%%%%%%%%%%%

%%%%%%%%%%%%%% End of Figures %%%%%%%%%%%%%%%%%%%%%%%%%%%%%%%%%%%%%%%%%%

%% file: BaerHan/runii.tex
%%%%%%%%%%%%%%%%%%%%%%%%%%%%%%%%%%%%%%%%%%%%%%%%%%%%%%%%%%%%%%%%%%%%%%
%
%  This is a LaTeX file!
%
%%%%%%%%%%%%%%%%%%%%%%%%%%%%%%%%%%%%%%%%%%%%%%%%%%%%%%%%%%%%%%%%%%%%%%

\renewcommand{\topfraction}{1.0}    % to help placement of floats
\renewcommand{\bottomfraction}{1.0}
\renewcommand{\textfraction}{0.0}

\def\eslt{\not\!\!{E_T}}
\def\to{\rightarrow}
\def\te{\tilde e}
\def\tl{\tilde l}
\def\tb{\tilde b}
\def\tf{\tilde f}
\def\td{\tilde d}
\def\tst{\tilde t}
\def\ttau{\tilde \tau}
\def\tmu{\tilde \mu}
\def\tg{\tilde g}
\def\tnu{\tilde\nu}
\def\tell{\tilde\ell}
\def\tq{\tilde q}
%\def\tw{\widetilde W}
%\def\tz{\widetilde Z}
%% Change of notation:  (the easy way)
\def\tw{\widetilde\chi^\pm}
\def\tz{\widetilde\chi^0}

\section{Sparticle Cross Sections and Branching Fractions}

\subsection{Sparticle production cross sections}

If the mSUGRA model is correct, then there may be a wide variety of
new sparticles that can be produced at the Tevatron collider. What is 
especially important for determining the signal events is the relative 
sparticle production rates at different points in model parameter
space. A simplifying feature of the mSUGRA model is that the gaugino 
mass unification condition is approximately true throughout
parameter space:
\begin{eqnarray}
 {M_1\over \alpha_1} ={M_2\over \alpha_2} ={M_3\over \alpha_3}\,,
\end{eqnarray}
where $\alpha_1 = {5\over3}{g'^2\over4\pi}$ and $g'$ is the Standard Model U(1) gauge coupling.
In particular, this relation holds at the weak scale, so that if we know
any of the weak scale gaugino masses $M_i$, then we can calculate the others.

Another simplifying feature of mSUGRA models derived from the radiative
electroweak symmetry breaking (REWSB) constraint is that over most of 
parameter space, the magnitude of the $\mu$ parameter
\begin{eqnarray}
|\mu |\gg M_1,\ M_2,
\end{eqnarray}
so that the two lightest neutralinos and lightest chargino 
are gaugino-like and the two heaviest
neutralinos and heavier chargino are higgsino-like. The exception to this 
statement occurs at parameter space values near the regions where the
REWSB mechanism almost breaks down, and at large $\tan\beta$; in both of these regions, 
$|\mu |$ can become quite small.

\begin{figure}[t]
\centering\leavevmode
\begin{minipage}{5.5cm}
\psfig{figure=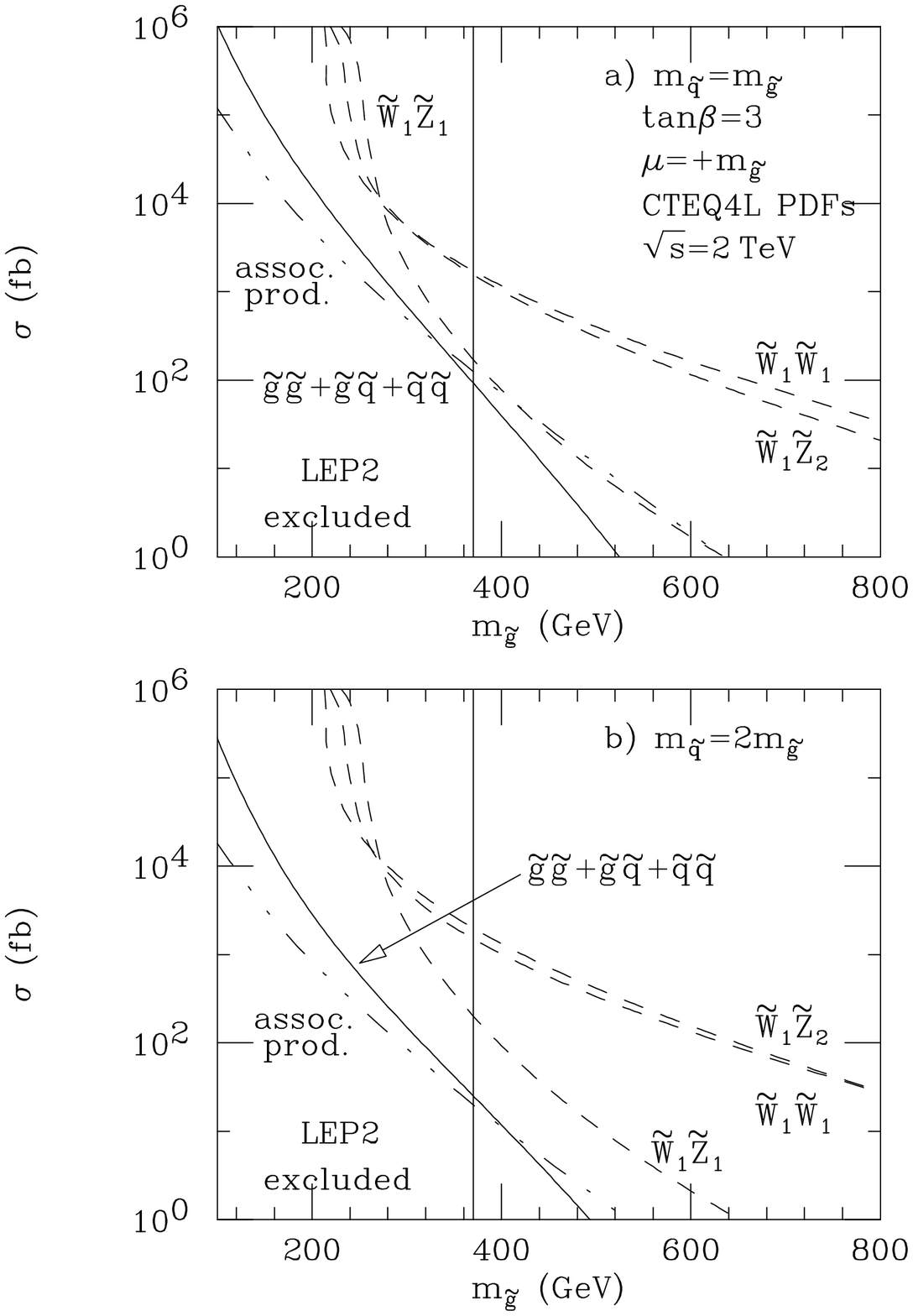,width=5.5cm}
\caption{Sparticle production cross sections as a function of $m_{\tilde g}$
for $\mu=+m_{\tilde g}$ and $\tan\beta =3$.\label{FIG1}}
\end{minipage}
%\end{figure}
\hspace{1cm}
%\begin{figure}[htb]
\begin{minipage}{5.5cm}
\psfig{figure=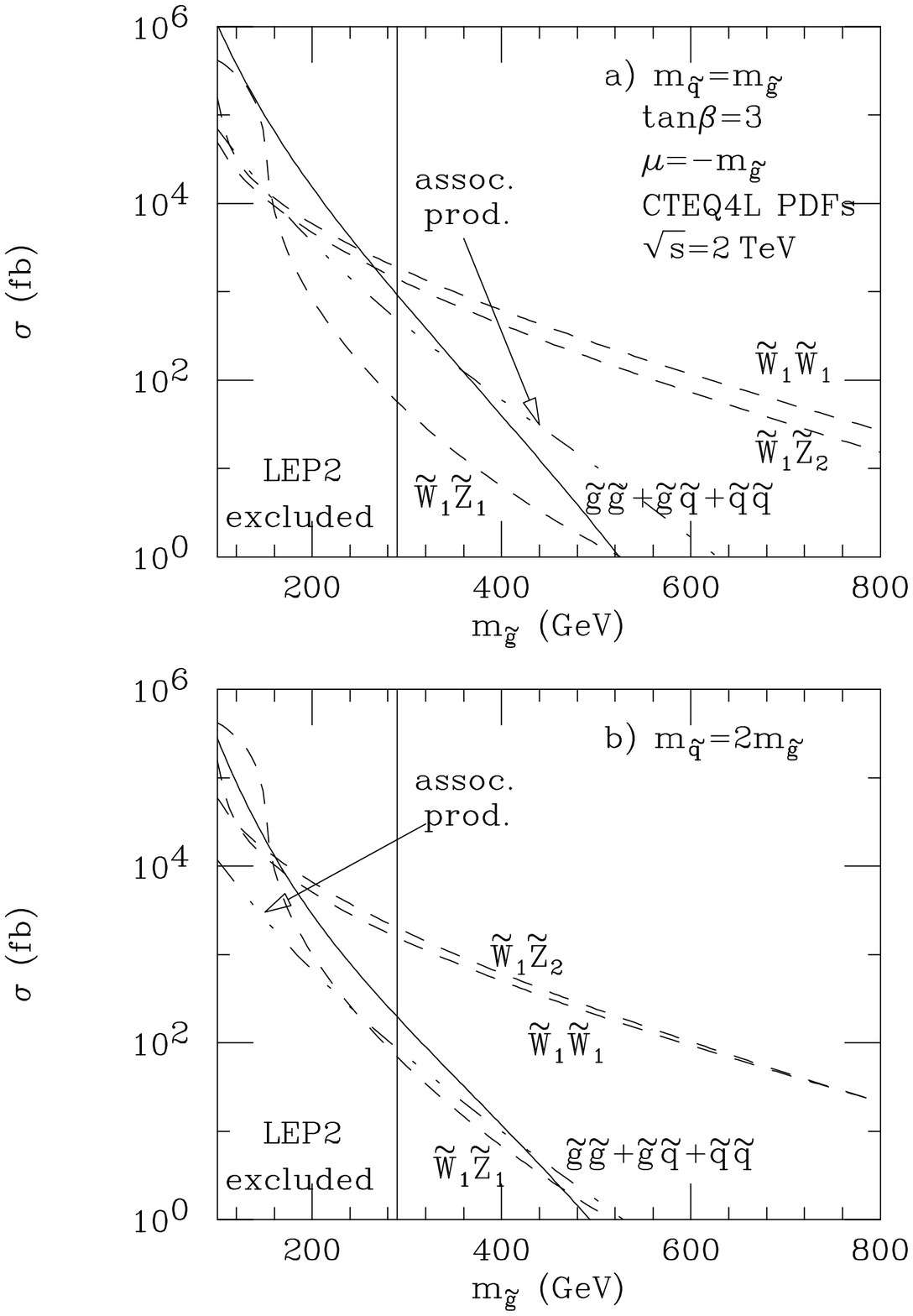,width=5.5cm}
\caption{Sparticle production cross sections as a function of $m_{\tilde g}$
for $\mu=-m_{\tilde g}$ and $\tan\beta =3$.\label{FIG2}}
\end{minipage}
%\end{figure}
%
\vskip1cm

%\begin{figure}[h]
\centering\leavevmode
\psfig{figure=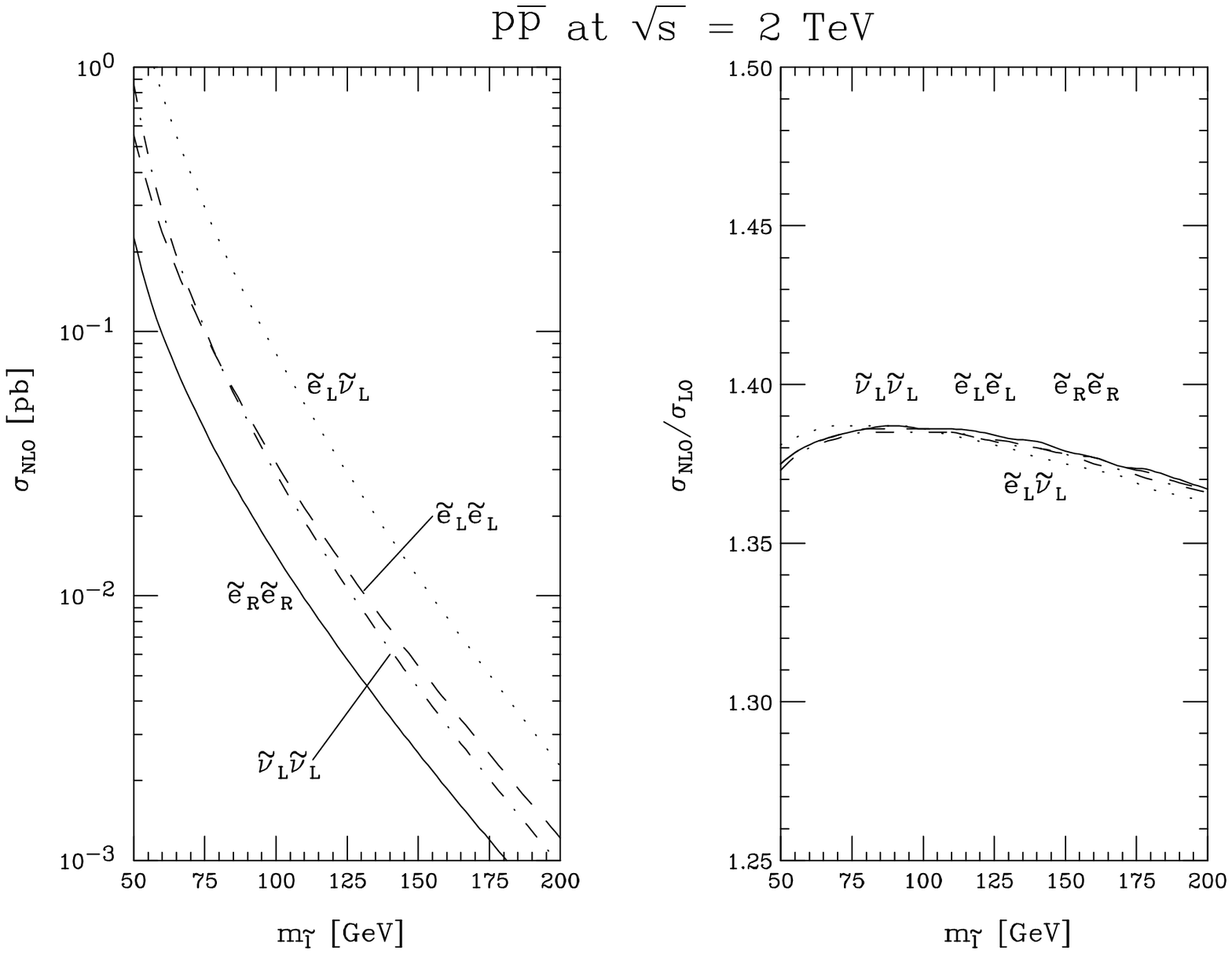,width=9cm}
\caption{{\it a}) Slepton production cross sections at NLO 
as a function of $m_{\tilde\ell}$; {\it b}) the ratio
${\sigma_{NLO}/ \sigma_{LO}}$ as a function of $m_{\tilde\ell}$.\label{FIG3}}
\end{figure}

Given the above inputs, it is possible to plot typical sparticle 
pair production cross sections expected at the Tevatron collider\cite{prod}.
In Fig. \ref{FIG1}, we plot\cite{bh-bckt} 
sparticle cross sections as a function of
$m_{\tilde g}$ assuming five generations of degenerate squark masses,
for {\it a}) $m_{\tq}=m_{\tg}$ and {\it b}) $m_{\tq}=2m_{\tg}$, 
assuming $\mu=+m_{\tg}$, and $\tan\beta =3$. In this plot, $m_{\tg}$
is the pole gluino mass. The region to the left of the vertical line
is excluded by the LEP2 chargino mass limit $m_{\tw_1}>90$ GeV.
Thus, it can be seen that direct chargino pair production $\tw_1\tw_1$
and $\tw_1\tz_2$ production dominate over the strongly produced $\tg\tg$,
$\tg\tq$ and $\tq\tq$ cross sections over essentially all
of parameter space for which $|\mu |\gg M_1,\ M_2$. The reaction 
$\tw_1\tz_1$ is also sub-dominant, while the squark or gluino plus
chargino or neutralino associated production reactions (summed over all
sparticle types and labelled by dash-dot-dot curves) are also relatively
suppressed. These qualitative features hold for both frames shown.

Similar results are shown in Fig. \ref{FIG2} for $\mu =-m_{\tg}$.
In this case, for low values of $m_{\tg}$ with $m_{\tq}\simeq m_{\tg}$, 
the strong production 
cross sections can be dominant, but only in parameter space regions 
now already excluded by LEP2.

The implications of these results are that in the mSUGRA model, 
$\tw_1\tw_1$ and $\tw_1\tz_2$ will likely be the dominant production 
cross sections over much of parameter space, and their signals may well
lead to the maximal reach of Tevatron experiments for SUSY particles.
For instance, the $\tw_1\tz_2\to 3\ell+\eslt$ signal can yield
a Run 2 reach of Tevatron experiments to values of $m_{\tg}\simeq 600$ 
GeV\cite{bcpt}.

If the parameter $m_0$ is very small, then slepton and sneutrino 
masses may be relatively small and their production 
cross sections may be important. These cross sections are shown at 
next-to-leading order\cite{bhr} in Fig. \ref{FIG3}{\it a}), while the
ratio ${\sigma_{NLO}/ \sigma_{LO}}$ is shown in Fig. \ref{FIG3}{\it b}).

\subsection{Sparticle branching fractions}

Once sparticles are produced, they typically decay through a cascade
of decays with lighter sparticles until it terminates with the production of the lightest SUSY particle, usually assumed to be absolutely stable.
Sparticle branching fractions are in general complicated functions of
SUSY model parameter space\cite{decay}. 
These branching fractions are embedded in 
several event generator programs
(such as ISAJET\cite{bh-isajet}) and PYTHIA\cite{bh-pythia}), 
and separate programs exist
that can output complete lists of
sparticle branching fractions. From the previous section, however, it is
clear that $\tw_1$ and $\tz_2$ production cross sections are dominant,
so their branching fractions will be the most relevant for SUSY searches
at the Tevatron within the mSUGRA framework.

If the parameter $m_{1/2}$ is small, then the light chargino $\tw_1$ 
usually decays via a 3-body mode via virtual $W$, $\tq_L$, $\tell_L$
or $H^\pm$ exchange. Over much of paramater space, the $W^*$
exchange dominates, so that the $\tw_1\to\tz_1 f\bar{f'}$ branching
fraction is similar to the $W\to f\bar{f'}$ one. Exceptions can occur if
$m_0$ is very small, in which case light virtual sleptons can enhance 
the leptonic decays, or if $\tan\beta$ is large so that large $\tau$
Yukawa coupling effects enhance decays to $\tau$ leptons\cite{bcdpt}. 
If $m_{1/2}$
becomes large, then ultimately $\tw_1$ becomes so heavy that the
decays  $\tw_1\to\tz_1 W$ (or even decays to sfermions) become accessible.

\begin{figure}[h]
\centering\leavevmode
\psfig{figure=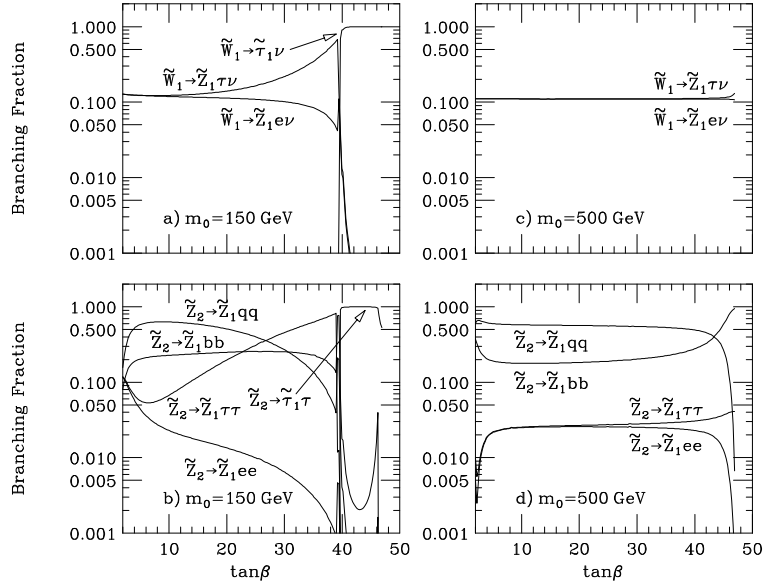,width=10cm}
\caption[]{Chargino ($\widetilde\chi^\pm$) and neutralino ($\widetilde\chi^0$) branching fractions versus $\tan\beta$. In {\it a}) and {\it
b}), we take the parameters ($m_0,m_{1/2},A_0)=(150,150,0)$ GeV while
in {\it c}) and {\it d}) we take ($m_0,m_{1/2},A_0)=(150,500,0)$
GeV. In all frames, $\mu >0$ and $m_t=170$ GeV. The discontinuities
are an artifact of the narrow width approximation. In ISAJET,
widths for three body and two body
decays are separately computed: the transition is, of course, smooth
since the virtual
particle smoothly goes on-shell.}
\label{FIG4}
\end{figure}

If $m_{1/2}$ is small, then the neutralino $\tz_2$ can decay via 3-body
modes $\tz_2\to\tz_1f\bar{f}$ where $f=\ell,\ \nu$ or $q$'s. These relative 
branching fractions are very model dependent. For instance, for moderate $m_0$
and $\mu <0$, the $\tz_2$ leptonic decays are enhanced, yielding a large 
Tevatron reach for SUSY via $\tz_2\to \tz_1\ell\bar{\ell}$\cite{bh-bckt,bcpt}. 
However, for 
the opposite sign of $\mu$, the leptonic decays can be suppressed, and there 
is no reach for $3\ell$ events by the Tevatron beyond the parameter 
space limits established by LEP2\cite{bh-bckt,bcpt}. 
As $m_{1/2}$ increases, the decays
$\tz_2\to \tz_1 h$ or $\tz_2\to \tz_1 Z$ can open up; the former decay is 
known as a ``spoiler mode'', since when it opens it can dominate the 
branching fractions and destroy the clean $3\ell$ signal from $\tw_1\tz_2$
production. If $m_0$ becomes small, then $\tz_2$ might be able to decay to
real or virtual sleptons and/or sneutrinos, and leptonic decays are again 
enhanced. As $\tan\beta$ becomes large, $b$-quark and $\tau$-lepton Yukawa
couplings can become important, and if $m_0$ is low enough, decays to 3rd 
generation fermions can be enhanced. The decay rate of $\tw_1$ and $\tz_2$
to various fermions and sfermions versus $\tan\beta$ is shown in 
Fig. \ref{FIG4} for small and large values of $m_0$. In these plots,
the enhancement of decays to third generation fermions for low $m_0$
is evident.

%

%%%%%%%%%%%%%%%%%%%%%%%%%%%%%%
%  Bibliography
%%%%%%%%%%%%%%%%%%%%%%%%%%%%%%

%% file: Falk/lepbnds.tex
\def\ga{\mathrel{\raise.3ex\hbox{$>$\kern-.75em\lower1ex\hbox{$\sim$}}}}
\def\la{\mathrel{\raise.3ex\hbox{$<$\kern-.75em\lower1ex\hbox{$\sim$}}}}
\def\gev{{\rm \, Ge\kern-0.125em V}}
\def\tev{{\rm \, Te\kern-0.125em V}}
\def\mchi{m_{\chi}}
\def\ohsq{\Omega_{\chi} h^2}
\def\m12{m_{1\!/2}}
\newcommand{\chp}{\tilde{\chi}^+}
\newcommand{\chm}{\tilde{\chi}^-}
\newcommand{\nt}{\tilde{\chi}^0}

%\begin{document}
%\baselineskip=18pt 
\section{LEP Constraints on mSUGRA}

Searches for supersymmetry in $e^+e^-$ collisions at LEP have already
probed significant regions of the mSUGRA parameter space.  Recent runs
at 133, 161, 172, 183 and 189 GeV center-of-mass energy have excluded
much of the region at low gaugino mass, and future runs at $\sim 192-200\gev$ 
will further extend the experimentally forbidden
region, or discover supersymmetry.   In this section we summarize those of the LEP bounds which
most strongly constrain the mSUGRA parameter space.  It includes bounds from the
run at 183 GeV centre-of-mass energy, and preliminary results from the run at 189 GeV can be found
on the web pages for the individual LEP experiments \cite{urls} and
from the LEP-SUSY Working Group home page \cite{lepsusywg}. 

Supersymmetric particle searches at LEP include searches for charginos
and neutralinos\cite{alcn183,decn183,l3cn183,opcn183}, sleptons\cite{alsl183,desl183,l3sl183,opsl183}, stops and
sbottoms\cite{desl183,tb183}, and Higgs bosons.  Constraints from the
Higgs searches will be discussed in detail in another section.   
The combined  constraints are presented as exclusion plots in the
$\{\m12,m_0\}$ plane, at fixed $\tan\beta$ and $A_0$.  The
outer envelope of the excluded regions in mSUGRA is generally set by the
chargino and slepton searches alone, though for large $A_0$ the stop bounds
can become important as well, as the lighter stop mass squared can be
driven negative at large $A_0$.  The stop mass constraints will be
summarized later in this secton.

Chargino pairs are produced via s-channel photon or $Z^0$ exchange
and t-channel sneutrino exchange.  In most of the mSUGRA parameters
space, the chargino production rate is large enough, and the chargino-neutralino
mass difference sufficient, so that the
chargino mass bound effectively saturates the kinematic limit.
However, destructive interference between the sneutrino and gauge
boson exchange channels can reduce the bounds for sneutrino masses
slightly above the chargino mass.  For sneutrino masses just below the
chargino mass, the produced charginos can decay into a sneutrino and
soft lepton, and if the mass difference between the sneutrino and
chargino is sufficiently small, the soft leptons will remain
undetected.

\begin{figure}[h]
\centering\leavevmode
%\hspace*{-0.2in}
\epsfig{file=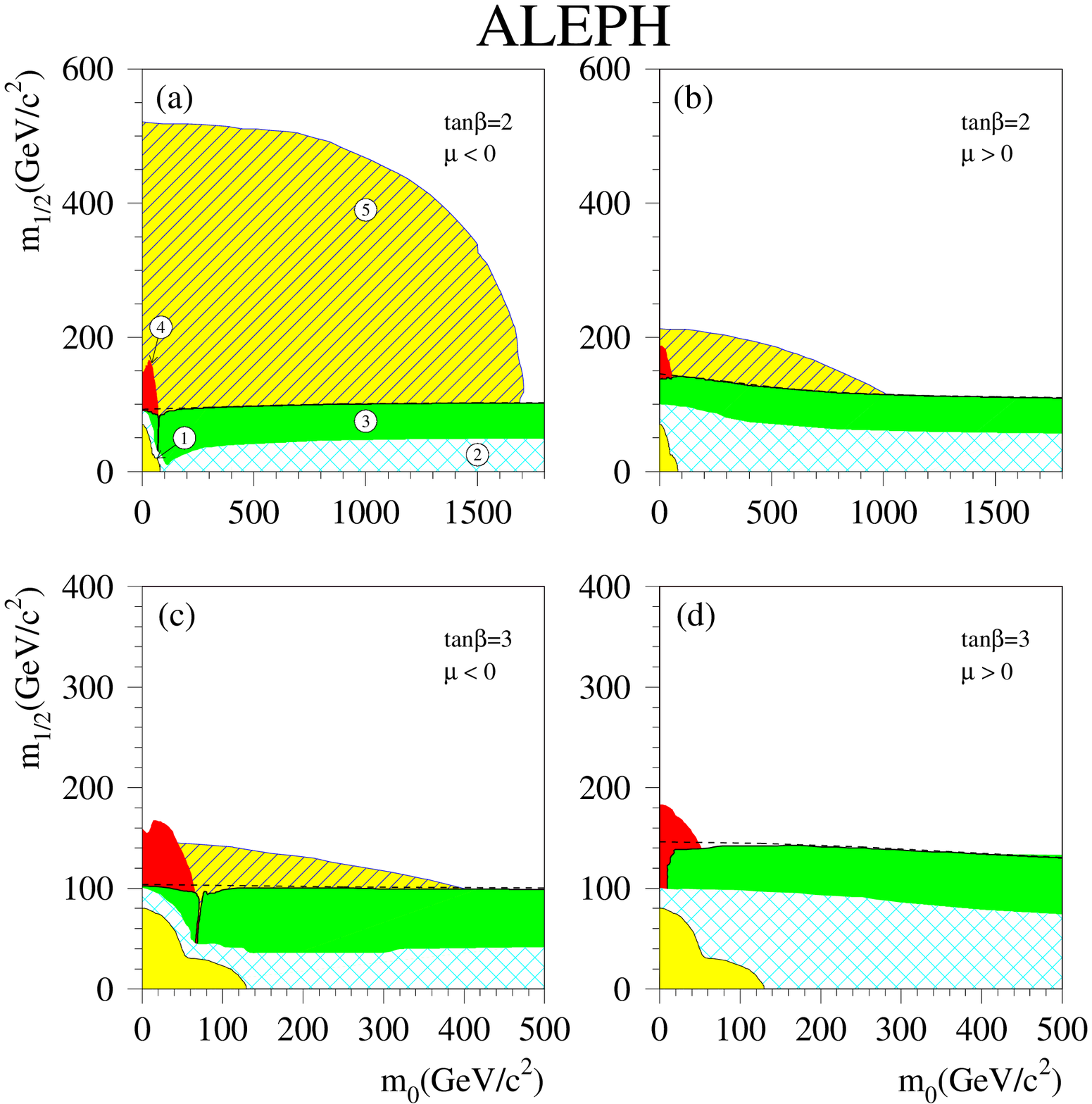,width=5in} 
\caption{Combined chargino and selectron bounds from LEP 183, 
for $\tan\beta=2$ and 3, and for $A_0=0$.  Region 1 is theoretically forbidden.
The other regions are excluded by the $Z$ width measurement at LEP1 (2),
chargino (3) and slepton (4) searches, and by Higgs boson searches (5).  The dashed lines 
represent the kinematic limit for direct chargino searches. \label{fig:23}}
\end{figure}

\begin{figure}
\vskip-1in
%\hspace*{-0.2in}
\centering\leavevmode
\epsfig{file=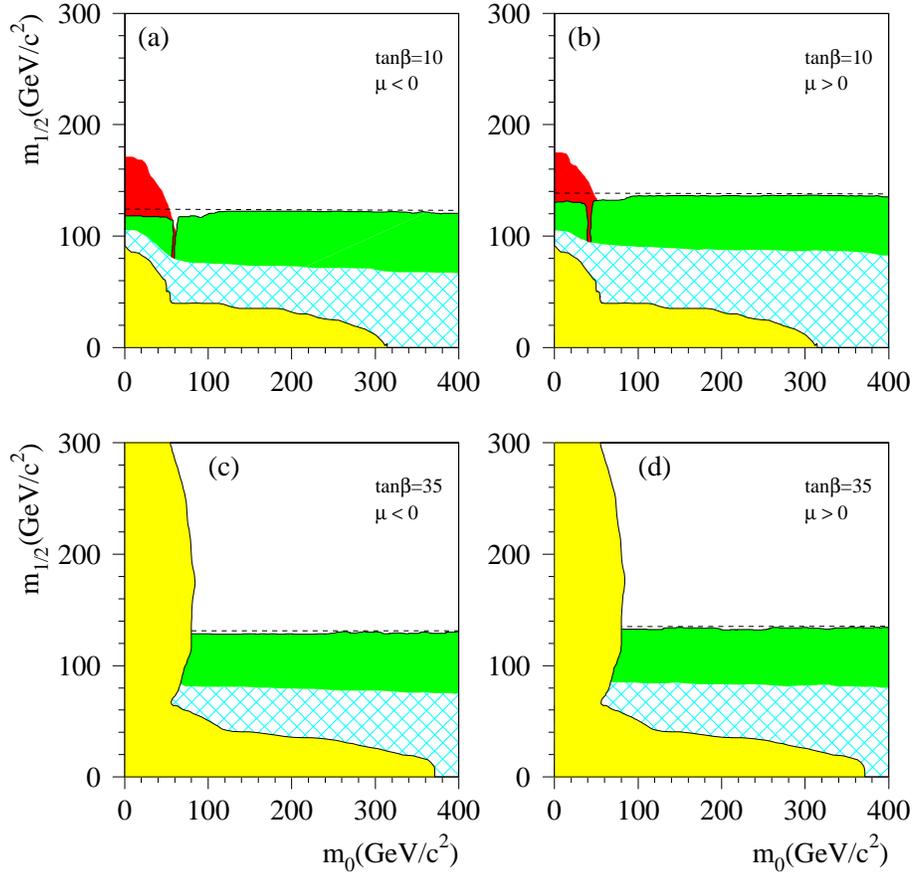,width=5in} 
\caption{Combined chargino and selectron bounds from LEP 183, 
for $\tan\beta=10$ and 35, and for $A_0$. The regions are defined as in Fig.~\protect{\ref{fig:23}}\label{fig:1035}}
\end{figure}

Fortunately, much of the latter loophole is covered by the selectron
searches.  Slepton pairs of all generations are produced via s-channel
photon or $Z^0$ exchange and t-channel neutralino exchange (for
selectrons).  However, limits on smuon and stau production do not
significantly strengthen the bounds obtained from selectron searches
alone.  Fig.~\ref{fig:23} and Fig.~\ref{fig:1035} display the combined selectron and
chargino bounds \cite{alcn} in the $\{\m12,m_0\}$ plane for $\tan\beta=2, 3, 10$, and 35, for
$\mu<0$ and $\mu>0$.  Here $A_0=0$, but the constraints are too not sensitive to $A_0$.
The chargino bound is strongest at large $m_0$, where the
sneutrino mass is large.  As $m_0$ is decreased, the chargino
constraint is observed to become weaker as destructive interference
between the s-channel and t-channel processes becomes important, and
near $m_0\sim 50-75\gev$ the sneutrino mass drops below the chargino mass,
and the chargino bound retreats.  For $\mu>0$, the
discontinuity in the chargino bound is covered completely by the
slepton bound.  The bounds in Figs.~\ref{fig:23} and \ref{fig:1035} are computed neglecting 
stau mixing.  Stau mixing changes the overlap between the regions which yield a
    kinematically accessible stau and those giving a light selectron, 
    and at large $\tan\beta$ it can open up small gaps in the 
chargino bounds, where the stau-neutralino mass difference is small ($<5$ GeV)
\cite{susywg}, and where the chargino can decay invisibly via chargino
$\rightarrow$ stau + neutrino followed by stau $\rightarrow$ neutralino + tau.
At low $\tan\beta$  these configurations are covered by selectron searches.

\begin{figure}
\centering\leavevmode
%\begin{minipage}{6.0cm}
%\hspace*{-1in}
\mbox{\epsfig{file=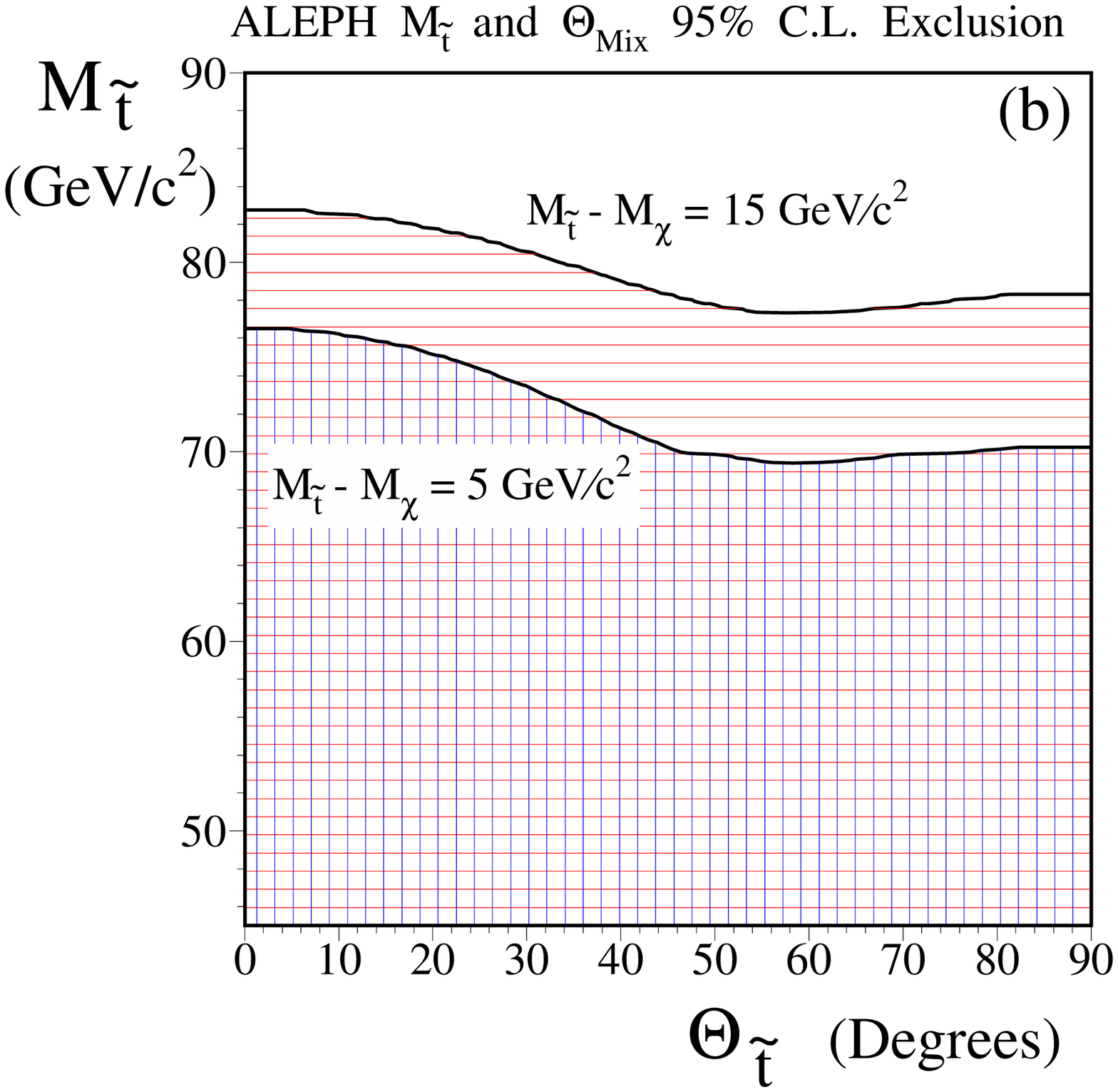,height=3in} 
%\end{minipage}
\hspace*{0.1in}
%\begin{minipage}{6.0cm}
\epsfig{file=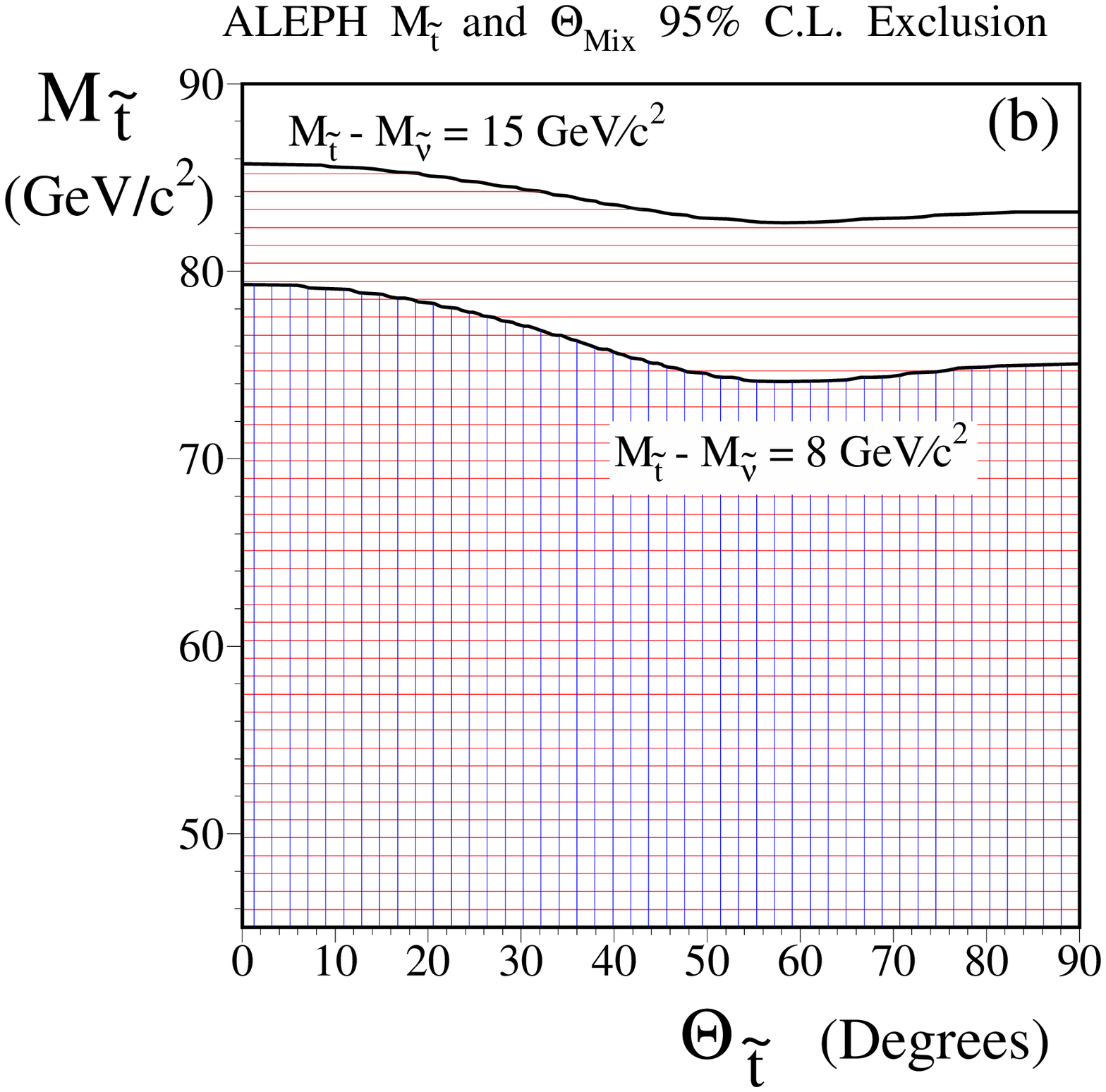,height=3in}} 
%\end{minipage}\hfill
\caption{Lower bound on the stop mass as a function of 
$\tan\beta$, for a) $\tilde t\rightarrow c\nt$ and
b) $\tilde t\rightarrow b l \tilde\nu$. \label{fig:mstops}}
\end{figure}

For sufficiently large $A_0$, the mass of the lighter stop tends to be
driven negative, so separate bounds on the mass of the lightest stop
can provide important constraints in this region.  The dominant
constraints come from searches for $\tilde t\rightarrow c\nt$ and
$\tilde t\rightarrow b l \tilde\nu$~\cite{tb183}, and the corresponding
limits are displayed in  Fig.\ref{fig:mstops}.  Bounds on the stop
mass depend on both the stop-neutralino (or stop-sneutrino) mass
difference and on the mixing angle between left and right stops.
However, unless the mass of the stop is less
than $10\gev$ above both the mass of the neutralino {\it and}
sneutrino, an absolute lower bound of $74\gev$ on the mass of the
lightest stop can be obtained~\cite{tb183}, independent of the stop
mixing angle.  Since the stop mixing angle is fixed as a function of
the mSUGRA parameters, this bound may be improved by a few $\gev$ in
any specific case.  

\begin{figure}
\vskip-1.in
\centering\leavevmode
\epsfig{file=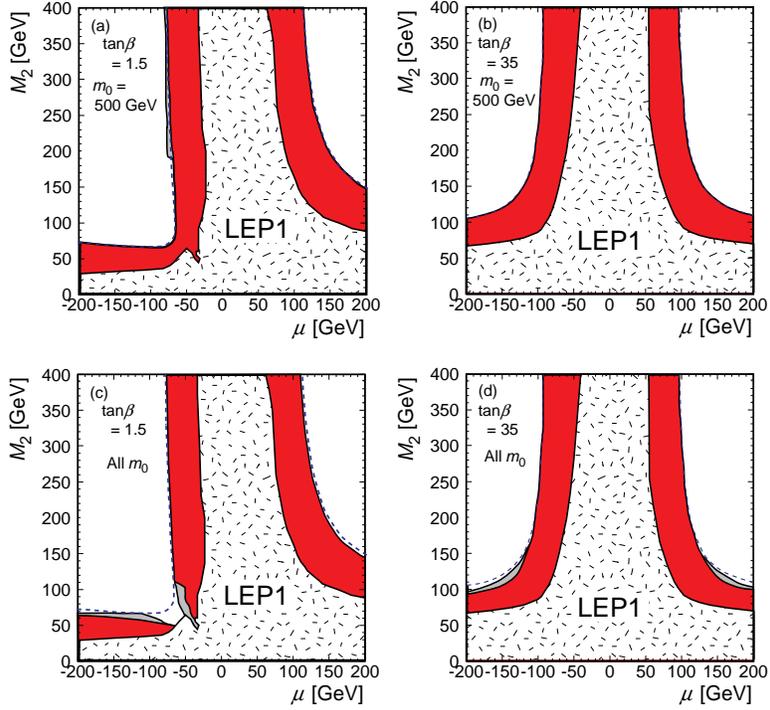,width=4in} 
\caption[]{Exclusion regions at 95\% C.L. in the ($M_2$,$\mu$) plane 
  with $m_0 \geq 500$~GeV for (a) $\tan \beta=1.5$ and for (b) $\tan
  \beta=35$.  Exclusion regions valid for all $m_0$ for (c) $\tan
  \beta=1.5$ and for (d) $\tan \beta=35$.  The speckled areas show the
  LEP1 excluded regions and the dark shaded areas show the additional
  exclusion region using the data from $\sqrt{s}=181$--184~GeV.  The
  kinematical boundaries for $\chp_1 \chm_1$ production are shown by
  the dashed curves.  The light shaded region in (a) extending beyond
  the kinematical boundary of the $\chp_1 \chm_1$ production is
  obtained due to the interpretation of the results from the direct
  neutralino searches.  The light shaded regions elsewhere show the
  additional exclusion regions due to direct neutralino searches and
  other OPAL search results (see \protect{\cite{opcn183}}).
\label{fig:muM2}}
\end{figure}

\begin{figure}
\vskip-1.0in
\centering\leavevmode
\epsfig{file=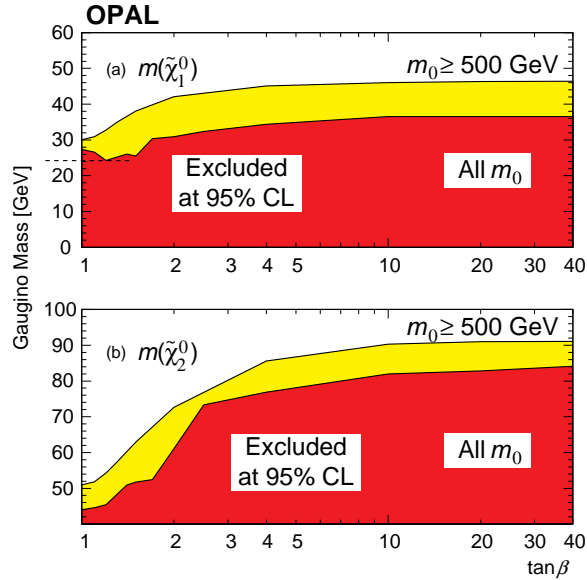,width=3in} 
\caption[]{The 95\% C.L.  mass limit on
(a) the lightest neutralino $\nt_1$ and
(b) the second-lightest neutralino $\nt_2$ as a function of 
$\tan\beta$ for $m_0 \ge 500$~GeV.  The mass limit on
$\nt_2$ is for the additional requirement of $m_{\tilde\chi^0_2}- m_{\tilde\chi^0_1}> 10$~GeV.
The exclusion region for $m_0 \ge 500$~GeV is shown
by the light shaded area and the excluded region valid
for all $m_0$ values by the dark shaded area.
\label{fig:mchi}}
\end{figure}

In the general MSSM, searches for associated neutralino $\nt_1\nt_2$
production through \mbox{s-channel} $Z^0$ or t-channel selectron
exchange can play a r\^ole in limiting the parameter space.
Fig.~\ref{fig:muM2} displays the region of $\{\mu,M_2\}$ parameter
space excluded by the combined set of LEP MSSM constraints
\cite{opcn183} for two representative values of $\tan\beta$, taking a
common mass parameter $m_0$ for the sfermions.  One can also extract
from the combined constraints a lower bound on the neutralino mass as
a function of $\tan\beta$, and this is displayed for the MSSM \cite{opcn183} in
Fig.~\ref{fig:mchi}.

%--------------------------------------------------

%\end{document}

%% file: Kamon-cdf-do/1e2.tex
\renewcommand{\topfraction}{1.0}    % to help placement of floats
\renewcommand{\bottomfraction}{1.0}
\renewcommand{\textfraction}{0.0}

\newcommand{\schitwozero }{\mbox{$\tilde{\chi}_{2}^{0}$}}
\newcommand{ \schionepm }{\mbox{$\tilde{\chi}_{1}^{\pm}$}}
\newcommand{ \lsp}    {\mbox{$\tilde{\chi}_{1}^{0}$}}
\newcommand{\mets}{\mbox{${E\!\!\!\!/_T}$}}
\newcommand{ \schionezero }{\mbox{$\tilde{\chi}_{1}^{0}$}}
\renewcommand{\pt}{\mbox{$p_{T}$}}
\newcommand{\et}{\mbox{$E_T$}}
\newcommand{ \chione }{\mbox{$\tilde{\chi}_{1}^{\pm}$}}
\newcommand{ \chitwo }{\mbox{$\tilde{\chi}_{2}^0$}}
\newcommand{\mgev}{\mbox{$\;{\rm GeV}/c^2$}}
\renewcommand{\gluino} {\mbox{$\tilde{g}$}}
\renewcommand{\squark} {\mbox{$\tilde{q}$}}
\newcommand{\pgev} {\mbox{$\;{\rm GeV}/c$}}
\newcommand{\intlum}{\mbox{${ \int {\cal L} \; dt}$}}
\newcommand{\invpb}{\mbox{$\;{\rm pb}^{-1}$}}
\def \gtsim    {\relax\ifmmode{\mathrel{\mathpalette\oversim >}}
                  \else{$\mathrel{\mathpalette\oversim >}$}\fi}
\def \ltsim    {\relax\ifmmode{\mathrel{\mathpalette\oversim <}}
                  \else{$\mathrel{\mathpalette\oversim <}$}\fi}
\def\oversim#1#2{\lower4pt\vbox{\baselineskip0pt \lineskip1.5pt
            \ialign{$\mathsurround=0pt#1\hfil##\hfil$\crcr#2\crcr\sim\crcr}}}
\newcommand{ \sstop}    {\mbox{$\tilde{t}$}}
\renewcommand{\sbottom}    {\mbox{$\tilde{b}$}}
\newcommand{ \stopone}    {\mbox{$\tilde{t}_{1}$}}
\def \dk       {\relax\ifmmode{\rightarrow}\else{$\rightarrow$}\fi}
\def \sp       {\relax\ifmmode{\;}\else{$\;$}\fi}
\newcommand{\gevcc}{\mbox{$\;{\rm GeV}/c^2$}}

\def\Journal#1#2#3#4{{#1} {\bf #2}, #3 (#4)}
\def \PRL      {Phys. Rev. Lett.~}
\def \PR       {Phys. Rev.}
\def \PRD      {Phys. Rev. D}
\def \PL       {Phys. Lett.~}
\def \PLB      {Phys. Lett. B}
\def \ZPC      {Z. Phys. C}	% - Particles and Fields}
\def \NPB      {Nucl. Phys. B}
\def \PR       {Phys. Rep.~}
\def \INC      {Il Nuovo Cimento}
\def \NIM      {Nucl. Instrum. Methods}
\def \NIMA     {Nucl. Instrum. Methods Phys. Res. Sect. A}
\def \etal     {\relax\ifmmode{et \; al.}\else{$et \; al.$}\fi}
\newcommand{\Dzero}{\mbox{D\O}}
\newcommand{\DZERO}{\Dzero\ Collaboration}
\newcommand{ \lplm    }{\mbox{$\ell^{+} \ell^{-}$}}
\newcommand{ \squarkb} {\mbox{$\bar{\tilde{q}}$}}
\newcommand{\epem}{\mbox{$e^+e^-$}}
\newcommand{ \snu}     {\mbox{$\tilde{\nu}$}}
\newcommand{ \seleR} {\mbox{$\tilde{e}_{R}$}}
\newcommand{ \seleL} {\mbox{$\tilde{e}_{L}$}}
\newcommand{ \stoptwo}    {\mbox{$\tilde{t}_{2}$}}
\newcommand{ \schionep }{\mbox{$\tilde{\chi}_{1}^{+}$}}
\newcommand{\bbb}{\mbox{$b\overline{b}$}}
\newcommand{ \chionem   }{\mbox{$\tilde{\chi}_{1}^{-}$}}
\newcommand{ \sbottomone}    {\mbox{$\tilde{b}_{1}$}}
\newcommand{\ppb}{\mbox{$p\overline{p}$}}
\newcommand{ \sbottomoneb}    {\mbox{$\bar{\tilde{b}}_{1}$}}
\newcommand{\ifm}[1]{\relax\ifmmode #1\else $#1$\fi}
\newcommand{\abseta}{\ifm{|\eta|}}
\newcommand{\ISAJET}{{\sc isajet}}
\newcommand{ \chizero}{\mbox{$\tilde{\chi}_{1}^0$}}
\newcommand{\met}{\mbox{${E\!\!\!\!/_T}$}}
\newcommand{\ttb}{\mbox{$t\overline{t}$}}
\renewcommand{\ppbar}{\mbox{$p\overline{p}$}}
\newcommand{\sqsq}   {\mbox{$\squark\squark$}}
\newcommand{ \sqsqb}   {\mbox{$\squark\squarkb$}}
\newcommand{\ipb}{\mbox{${\rm pb}^{-1}$}}
\newcommand{\tanb}{\ifm{\tan\beta}}
\renewcommand{\slepton} {\mbox{$\tilde{\ell}$}}
\newcommand{\gt}{\ifm{>}}
\newcommand{\lt}{\ifm{<}}
\renewcommand{\ttbar}{\mbox{$t\overline{t}$}}
\newcommand{\VECBOS}{{\sc vecbos}}
\newcommand{\HERWIG}{{\sc herwig}}
\newcommand{\Ht}{\ifm{H_T}}
\newcommand{\azero}{\ifm{A_0}}
\newcommand{\mzero}{\ifm{m_0}}
\renewcommand{\mhalf}{\ifm{m_{1/2}}}
\newcommand{\SPYTHIA}{{\sc spythia}}
\def \svxp {SVX$^{\prime}$}
\newcommand{\zb}{\ifm{Z}}
\newcommand{ \stoponeb}    {\mbox{$\bar{\tilde{t}}_{1}$}}
\newcommand{\degr}{\mbox{$^{\circ}$}}
\newcommand{ \schionem }{\mbox{$\tilde{\chi}_{1}^{-}$}}
\renewcommand{\gev}  {\mbox{${\rm GeV}$}}
\def \mc {\multicolumn}
\newcommand{\MET}{\mbox{${E\!\!\!\!/_T}$}}

%%%%%
\newlength{\pushupfigure}
\setlength{\pushupfigure}{-55.5pt}
\def \epsfin_v1#1#2{
        \vspace{\pushupfigure}
        \center
        \leavevmode
        \epsfxsize=#1
        \epsffile[20 143 575.75 698.75]{#2}
}
%%%%%

\section{SUGRA Limits in Run I }

\subsection{Introduction}

There has been a great effort in searches for SUSY particles by CDF and D\O\
using data samples
in $p \overline{p}$ collisions at $\sqrt{s}$ = 1.8 TeV 
at the Fermilab Tevatron.
The data were collected 
during the 1992-93 (Run Ia) and 1994-95 (Run Ib) runs.

With the assumption of 
the gaugino unification provided by supergravity (SUGRA) \cite{basicSUGRA},
the Minimal Supersymmetric Standard Model (MSSM) framework \cite{SUSYreview}
has a simple mass relation between gluino ($\gluino$),
chargino ($\schionepm$) and neutralinos
($\schitwozero$ and $\lsp$):
$(1/3 \sim 1/4) \; M(\gluino) \approx M(\schionepm) \approx 
M(\schitwozero) \approx 2~M(\lsp)$.
We assume the lightest neutralino ($\lsp$) 
is the lightest supersymmetric particle 
(LSP) and stable.  
The LSPs do not interact with the detector and 
therefore result in missing transverse energy (\mets). 

One canonical SUSY signature is trilepton events, which would come from 
chargino-neutralino  ($\schionepm \schitwozero$) 
pair production with  subsequent
leptonic decays ($\schionepm \rightarrow \ell^{\pm} \nu \lsp$
and $\schitwozero \rightarrow \lplm \lsp$) \cite{susy_trilepton}. 
Another signature is squark and gluino ($\squark$ and
$\gluino$) production followed by decays into the LSP and jets.
Yet another
is based on the Majorana nature of the gluino, which allows the gluino to
decay into a chargino of either sign. One consequently searches for
like-sign (LS) dilepton pairs from cases where a pair of 
gluinos have decayed into charginos of the same sign
\cite{susy_dilepton}.

Table \ref{tab:susy_br_example_sugra} shows
decay branching ratios 
of chargino, neutralino, gluino, and squarks, for
four minimal SUGRA (mSUGRA) cases:
(1)~$M(\squark) \gg M(\gluino)$;
(2)~$M(\squark) \simeq M(\gluino) + 7\ \mgev$;
(3)~$M(\squark) \simeq M(\gluino) -7\ \mgev$;
(4)~$M(\squark)/M(\gluino) \simeq 0.9$.
Other mSUGRA parameters are fixed as
	$\tan \beta = 2$, $\mu < 0$ and $A_0 = 0$.
These mSUGRA points can be characterized as:
\begin{itemize}
\item [{\bf 1:}]
	The gluino pair production is dominant, and
	the decays of $\schionepm$ and $\schitwozero$ are key.
	For example, $Br(\gluino \to e^{\pm} + X)$ = 4.9\% and
	$Br(\gluino \to e^{+}e^{-} + X)$ = 2.6\%.
	We should access to this point via
	trilepton, $\mets + jets$, and LS dilepton + $\mets + jets$ channels.
\item [{\bf 2:}] 
	The $\tilde{g}\tilde{g}$,
	$\squark\squarkb$ and 
	$\tilde{g}\tilde{q}$ production is dominant.
	The key decay is $\gluino \to \sbottom b$, because
	$M(\sbottom) - M(\gluino) > M(b)$ and 
	$M(\sbottom) < M(\squark)$.
	We have $Br(\gluino \to e^{\pm} + X)$ = 0.6\% and
	$Br(\gluino \to e^{+}e^{-} + X)$ = 17.2\%.
	Leptons are also expected from light squark decays.
	We should access to this point via
	trilepton, $\mets + jets$, and $2\ell + \mets + jets$ channels.
\item [{\bf 3:}]
	Similar to Case 2.
        $Br(\gluino \to e^{\pm} + X)$ = 0.7\% and
        $Br(\gluino \to e^{+}e^{-} + X)$ = 16.5\%.
	This point is
	close to D\O's 95\% C.L. exclusion contour
	based on $\mets +$ jets and
	$ee + \mets +$ jets analyses (see Section
	\ref{sec:d0_glsq_limit_run1}).
\item [{\bf 4:}]
	$\schitwozero \to \epem \lsp$ is suppressed, because
	$\snu_L$ (43 \mgev) is lighter than $\seleR$ (49 \mgev)
	and $\seleL$ (75 \mgev).
	The electron from $\seleR \to e \lsp$ ($Br = 100\%$) will be
	softer, so that the LS dilepton + $\mets$ channel
	will be important for a search for
	a direct production of $\schionepm\schitwozero$.
        Note $Br(\gluino \to e^{\pm} + X)$ = 6.3\% and
        $Br(\gluino \to e^{+}e^{-} + X)$ = 0.7\%.
\end{itemize}

Given the large mass for the top quark, Yukawa interactions should drive
one of the top squark (stop) masses to a value much lower than that of the
other squarks. There are three alternative decay modes possible for the
lighter stop ($\stopone$) if it is lighter than the top quark \cite{lstop}:
(i) $\stopone \to c\ \lsp$ 
	if the charginos, sleptons, sneutrinos are heavier than the stop; or
(ii) $\stopone \to b\ \schionepm$
		if the chargino is lighter than the stop; or
(iii) $\stopone \to \ell \snu b \; (\slepton \nu b)$
	if sneutrinos (or sleptons) are light enough.

The effect of the Yukawa interactions on the bottom squark (sbottom)
is much smaller, because the bottom quark mass is
much smaller than the top quark mass.
A large splitting between mass eigenstates can still occur
when the values of $\tan\beta$ are large. 
Thus, one of the sbottoms
($\sbottomone$) can be lighter than other squarks. 
We could study a direct production of sbottom pair 
($\ppb \to \sbottomone\sbottomoneb + X$)
and an indirect production via $\gluino \to \bar{b}\ \sbottomone$.
The decay modes could be (i) $\sbottomone \to b\ \lsp$ and 
(ii) $\sbottomone \to b\ \schitwozero$ \cite{lsbot}.

We summarize the current results of searches for SUSY particles
in the MSSM or SUGRA frameworks from CDF and D\O\ experiments 
using the data 
taken in 1992-95 (Run Ia+Ib; corresponding to
approximately 110 \invpb). 
It should be noted that results based on the data in 1992-93 
(Run Ia; 20 \invpb) can be found in elsewhere \cite{RIPS}
and are not covered in this paper.

\break

\begin{table}[ht]
\caption{Example of decay branching ratios 
	of chargino, neutralino, gluino, and squarks
	for four reference points in the mSUGRA model (\ISAJET\ version 7.20):
	(1) $M(\squark) \gg M(\gluino)$;
	(2) $M(\squark) \simeq M(\gluino) + 7\ \mgev$;
	(3) $M(\squark) \simeq M(\gluino) -7\ \mgev$;
	(4) $M(\squark)/M(\gluino) \simeq 0.9$.
	Here $m_0$ ($m_{1/2}$) is a common scalar (gaugino) mass.
	Other mSUGRA parameters are fixed as
	$\tan \beta = 2$, $\mu < 0$ and $A_0 = 0$.}
\label{tab:susy_br_example_sugra}
\begin{center}
\begin{tabular}{ l c  c  c c}
%\hline
{\bf mSUGRA:} $(m_0, m_{1/2}) \to$ & 
			(500, 90) & (170, 90) & (140, 90) & (~~0, 90)\\
\hline
\hline
{\bf Mass (\mgev)}	&	&	& 	& \\
$\schionepm$/$\schitwozero$ & 85.1/85.5 & 89.2/89.9 & 89.5/90.1 & 90.0/90.1\\
$\schionezero$             & 39.7       & 40.4      & 40.4	& 40.6 \\
$\gluino$                  & 292        & 275       & 273	& 264 \\
$\squark$                  & 543	& 283       & 267       & 231 \\
$\sbottom_1$/$\sbottom_2$  & 426/541    & 251/281   & 242/264	& 219/227 \\
$\stopone$/$\stoptwo$      & 297/458    & 247/309   & 241/304	& 225/296 \\
\hline
\hline
{\bf Decay Branching Ratio (\%)}    &	&	& & \\
$\schionep \to \schionezero e^+ \nu_e$  & 11.0 & 11.5 & 12.3 & 27.8 ($\snu e$)\\
$\schitwozero \to \schionezero e^+ e^-$ & ~6.4 & 19.1 & 21.3 & 
		~3.8 ($\tilde{e}e$) \\
\hline
$\gluino \to \schitwozero q \bar{q}, \; \schitwozero \bbb$ 
					& 41.3 & ~0.0 & ~0.0 & ~0.0 \\
$\gluino \to \schionep \bar{u} d, \; 
		\schionep \bar{c} s, \; + c.c.$ & 44.9 & ~5.6 & ~0.0 & ~0.0 \\
$\gluino \to \squark \bar{q}$ 			& ~0.0 & ~0.0 & 17.3 & 70.7 \\
$\gluino \to \sbottom_1 b$ 			& ~0.0 & 90.0 & 77.5 & 18.4 \\
$\gluino \to \sbottom_2 b$ 			& ~0.0 & ~0.0 & ~5.2 & 10.9 \\
\hline
$\tilde{u}_L \to \schitwozero u$ 		& ~7.1 & 27.1 & 27.3 & 25.7 \\
$\tilde{u}_L \to \schionep d$ 			& 15.3 & 65.0 & 66.8 & 67.2 \\ 
$\tilde{u}_L \to \gluino\ u$ 			& 76.9 & ~2.5 & ~0.0 & ~0.0 \\
\hline
$\tilde{d}_L \to \chitwo d$ 			& ~7.7 & 33.9 & 34.9 & 35.1 \\
$\tilde{d}_L \to \chionem u$ 			& 14.8 & 61.9 & 63.3 & 62.1 \\
$\tilde{d}_L \to \gluino\ d$ 			& 77.3 & ~2.8 & ~0.0 & ~0.0 \\
\hline
$\tilde{u}_R \to \chitwo  u$ 			& ~0.0 & ~1.6 & ~1.8 & ~1.6 \\
$\tilde{u}_R \to \gluino\ u$ 			& 94.8 & 11.1 & ~0.0 & ~0.0 \\
\hline
$\tilde{d}_R \to \chitwo  d$ 			& ~0.0 & ~1.1 & ~1.8 & ~1.6 \\
$\tilde{d}_R \to \gluino\ d$ 			& 98.6 & 33.3 & ~0.0 & ~0.0 \\
\hline
$\tilde{b}_1 \to \chitwo b$ 			& 12.6 & 99.8 & 99.7 & 98.4 \\
$\tilde{b}_1 \to \chionem t$ 			& 16.0 & ~0.0 & ~0.0 & ~0.0 \\
$\tilde{b}_1 \to \gluino\ b$ 			& 71.2 & ~0.0 & ~0.0 & ~0.0 \\
\hline
$\tilde{b}_2 \to \chitwo  b$ 			& ~0.0 & ~1.7 & ~1.7 & ~1.6 \\
$\tilde{b}_2 \to \gluino\ b$ 			& 98.6 & ~2.5 & ~0.0 & ~0.0 \\
\hline
$\tilde{t}_1 \to \schionezero t$ 		& 79.5 & 25.5 & 16.7 & ~6.1 \\
$\tilde{t}_1 \to \schitwozero t$ 		& ~6.2 & ~0.0 & ~0.0 & ~6.8 \\
$\tilde{t}_1 \to \schionep b$ 			& 14.3 & 74.5 & 74.5 & 55.3 \\
%\hline
%\hline
%$\gluino \to e^\pm + X$				& ~4.9 & ~0.6 & ~0.7 & ~6.3 \\
%$\gluino \to e^+ e^- + X$			& ~2.6 & 17.2 & 16.5 & ~0.7 \\
\hline
\end{tabular}
\end{center}
\end{table}

\subsection{CDF and D\O\ Detectors}

\subsubsection{CDF}

The CDF detector is a general purpose detector described 
in detail elsewhere \cite{cdf1det,cdftop94}.
The inner most part of CDF, the silicon vertex detector (\svxp),
allows a precise measurement of  a track's impact parameter with respect to
the primary vertex in the plane transverse to the beam direction.
The position of the primary vertex along the beam direction is determined by
a time projection chamber. The momenta of the charged particles are measured
in the central drift chamber which is located inside a 1.4 T superconducting
solenoidal magnet.
Outside the drift chamber there is a calorimeter, which is organized into
electromagnetic and hadronic components, with projective towers covering 
the pseudo-rapidity range $\abseta < 4.2$. The muon system is located outside
the calorimeter and covers the range $\abseta < 1.0$.

\subsubsection{D\O}

The \Dzero\ detector is a large multipurpose detector used to
measure charged leptons, photons, and jets.  The hermeticity of the detector
allows for a good measurement of the missing transverse energy.  Moving
radially from the beamline the \Dzero\ detector consists of a non-magnetic
central tracking system, a compact uranium/liquid-argon sampling
calorimeter, and a muon spectrometer~\cite{d01det}.

\subsection{Search for 
	$\schionepm\schitwozero$ Using Trilepton Events}
\label{sec:sugra_trilepton_run1}	

It has long been suggested that
one of the most promising channels for the discovery of SUSY at a hadron
collider is three isolated charged leptons 
plus missing energy, 
arising from $\schionepm\schitwozero$ pair
production with subsequent leptonic decays
($\schionepm \rightarrow \ell \nu \schionezero$ and 
$\schitwozero \rightarrow \ell^+ \ell^- \schionezero$) \cite{susy_trilepton}.

D\O\ and CDF search for direct production of $\schionepm\schitwozero$ 
in the trilepton channels ($e^+e^-e$, $e^+e^-\mu$,
$e\mu^+\mu^-$ and $\mu^+\mu^-\mu$). 
The detailed analyses of the D\O\ and CDF searches are described 
elsewhere \cite{d0_tril,cdf_tril}.
Table \ref{tab:cdf_d0_tril} is a summary of
the trilepton selections at D\O\ and CDF.
It should be noted that the $\pt$ cuts in the D\O\ analysis
depend on the triggers used, so that we only indicate
approximate values.
Both analyses find no events in their selection cuts,
which are consistent with expectation from SM background events.

\begin{table}[ht]
\caption{Summary of D\O\ and CDF trilepton analyses in Run I.
The \pt\ cuts in the D\O\ analysis depend on the trilepton channel
($eee$, $ee\mu$, $e\mu\mu$ or $\mu\mu\mu$).
\mets, \pt, and $M$ are given in GeV, \pgev, and \mgev, respectively.
Experimentally, electrons (muons) are selected by $\et$ ($\pt$).
We simply refer $\pt$ for both electron and muon.}
\label{tab:cdf_d0_tril}
\begin{center}
\begin{tabular}{l c c }

Experiment 	& D\O\ \cite{d0_tril} & CDF \cite{cdf_tril} \\
$\intlum$	& 107 \invpb	& 106 \invpb \\
\hline
\hline
{\bf Primary cuts:}   &	&	\\
$p_T$ for $\ell_1$, $\ell_2$, $\ell_3$ & 
	$\gtsim\ 10$, $\gtsim\ 10$, $> 5$ & $> 11$, $> 5$, $> 5$\\
$\abseta$ for $e_1$, $e_2$, $e_3$  & 
	$< 3.0$, $< 3.0$, $< 3.0$ & $< 1.1$, $< 2.4$, $< 2.4$ \\
$\abseta$ for $\mu_1$, $\mu_2$, $\mu_3$ & 
	$< 1.0$, $< 1.0$, $< 1.0$ & $< 0.6$, $< 1.0$, $< 1.0$ \\
Isolation cut for $\ell$ & yes &  yes \\
$\Delta \phi(\ell, \ell)$ cut & yes & yes \\
$Z$ veto	& $81$-$101$ (for $eee$) & $76$-$106$ \\
$M(\ell, \ell)$ veto & $<5$ (for $e\mu\mu$ and $\mu\mu\mu$) & 
			$J/\psi$(2.9-3.1), $\Upsilon$(9-11) \\
$\mets$       & $>$ 10-15 & $>15$ \\
\hline
{\bf Results:} & & \\
$N_{obs}$	& 0	& 0	\\
$N_{BG}$ & $1.5\pm0.5$ & $1.2\pm0.2$ \\
\hline
\end{tabular}
\end{center}
\end{table}

Table \ref{tab:simssm_vs_sugra} shows a summary of parameters in
a SUGRA-inspired MSSM framework \cite{tannenbaum_thesis,RIPS}
in the CDF analysis and an mSUGRA model in the D\O\ analysis.
In the mSUGRA model, 
the universality of $m_0$ 
automatically suppresses unwanted flavor changing neutral currents (FCNC).
The existence of the superparticles below 1 TeV generally leads to
large FCNC, especially in $K^{0}$-$\bar{K}^{0}$ oscillations.
To avoid this requires the squark masses at least in
the 1$^{st}\,$ and 2$^{nd}\,$ generations to be highly degenerate.
Thus, five squark masses are approximately degenerate
and one of stops could be lighter (or heavier) than the other squarks.
The MSSM framework,
in which five squarks are degenerate,
is adopted in the CDF search (and the stop mass is a free parameter).
In the framework, stops are set heavier than the other
squarks and $A_t = \mu / \tan\beta$ is chosen to remove the mixing
between $\sstop_L$ and $\sstop_R$. 
Slepton and sneutrino masses are related to squark and gluino masses
        as inspired by supergravity models \cite{rge_slepton}:
\begin{eqnarray}
%%%        \left.
        \begin{array}{lll}
        M(\tilde{\ell}_{L})^{2} & = &
        M(\squark)^{2} - 0.73 M_(\gluino)^{2} - 0.27 M(Z)^{2} \cos 2 \beta
                \\
        M(\tilde{\ell}_{R})^{2} & = &
        M(\squark)^{2} - 0.78 M(\gluino)^{2} - 0.23 M(Z)^{2} \cos 2 \beta
                \\
        M(\tilde{\nu}_{L})^{2} & = &
        M(\squark)^{2} - 0.73 M(\gluino)^{2} + 0.5 M(Z)^{2} \cos 2 \beta
        \end{array}
%%%        \right\}  
	&  & \nonumber
        %\label{eq:RG_Eqs}
\end{eqnarray}
The equations, which are designed to be simplified approximation of the mSUGRA mass sepctrum,
require $M(\squark) \; \gtsim \; 0.9 M(\gluino)$.
If the gluinos and squarks are rather close in mass,
the sleptons can be considerably lighter than squarks.
One major difference from mSUGRA models is that we do not assume the unification
of Higgs masses. Thus, we set $\mu$ (the Higgsino mixing parameter) free. 
This $\mu$ parameter also controls the mixing of bino, wino and higgsinos,
so that it is very important parameter to study.

\begin{table}
\caption{Comparion of two scenarios used in CDF and D\O\ trilepton analyses
in Run I: SUGRA-inspired MSSM vs minimal SUGRA.}
\label{tab:simssm_vs_sugra}
\begin{center}
\begin{tabular}{ l l l }
%\hline
        &  SUGRA-inspired MSSM    & minimal SUGRA \\
	&	(CDF)		& (D\O)	\\
\hline
\hline
Inputs          & $M(\gluino)$                  & $m_{1/2}$     \\
        & $M(\squark)$                  & $m_{0}$       \\
        & \hspace{.2in} [= $M(\sstop_L)$ = $M(\sstop_R)$ &              \\
        & \hspace{.2in} $\;$= $M(\sbottom_L)$ = $M(\sbottom_R)$] &      \\
        & 
                $\begin{array}{l}
                A_t  =  \mu/\tan\beta \\
                A_b  =  \mu \tan\beta
                \end{array}
                \left.\vbox to 3ex{\vfil}\right\}$  
                & $A_0$ \\        
        & $\tan\beta$                   & $\tan\beta$   \\
        & $\mu$                         & sign of $\mu$    \\
        & $M(H_A) =$ 500 \mgev          & n/a \\
\hline
\hline
Gaugino unification & Yes & Yes \\
\hline
Degeneracy of light squarks & Yes &
                        Approximately yes (if $\tan\beta \ltsim 10$) \\
\hspace{.5in} Squark mass       & Input & from $m_0$ and $m_{1/2}$      \\
\hspace{.5in} Stop mass ($A_t$) & $A_t = \mu/\tan\beta$ & 
                                        from $A_0$, $m_0$ and $m_{1/2}$ \\
                & $\stopone$ is set heavy. & $\stopone$ can be lighter. \\
\hline
Slepton mass:   &               & \\
\hspace{.5in} [$M(\squark) \gtsim\ 0.9 M(\gluino)$] & 
                                from $M(\squark)$ and $M(\gluino)$ &
                                        from $m_0$ and $m_{1/2}$ \\
%% \hspace{.5in} [$M(\squark) < M(\gluino)$] & 350 \mgev & n/a \\
\hline
Unification of Higgs masses &   No      & Yes \\
                & $\mu$ is a free parameter. & $| \mu |$ is fixed. \\
                & $h$ is SM Higgs-like. & $h$ is SM Higgs-like. \\
                & $H^0, H_A, H^{\pm}$ are set heavy. & $H^0, H_A, H^{\pm}$
                                                                are heavy. \\
\hline
\end{tabular}
\end{center}
\end{table}

\begin{figure}[ht]
\vspace{-0.2in}
%	\centering\leavevmode
%	\psfig{figure=cdf_trilepton_fig1.ps,height=2.9in}
%
\epsfin_v1{3in}{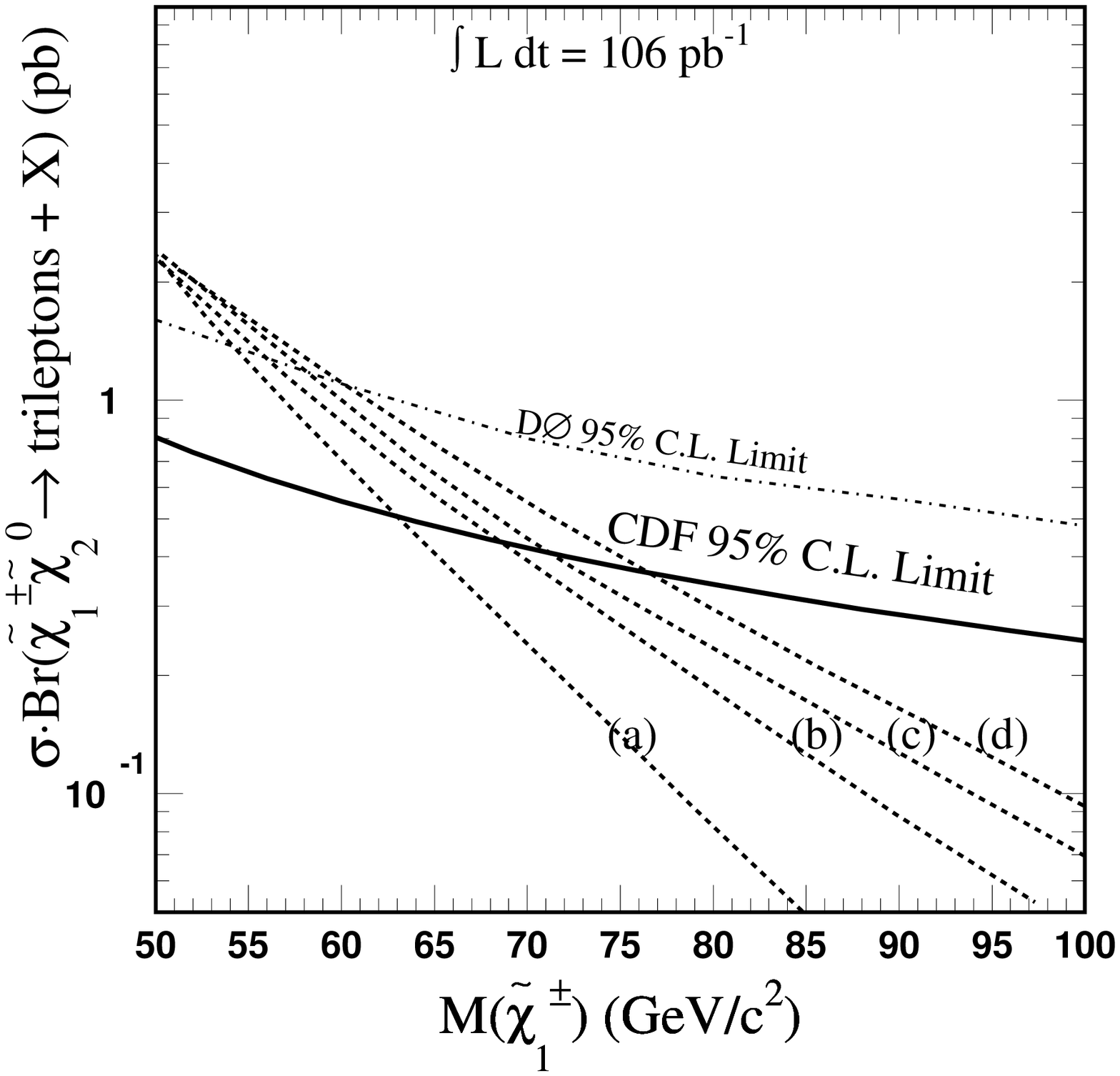}
\vspace{0.7in}
\caption{D\O\ \protect\cite{d0_tril} and CDF \protect\cite{cdf_tril} 
 95\% C.L. upper limits 
on 
$\sigma\cdot Br(\schionepm\schitwozero\dk3\ell + X)$ as a function of
$M(\schionepm)$,
compared to predictions at four SUGRA-inspired MSSM points 
at $\mu=-400$ \mgev, tan $\beta$ = 2 for
$M(\squark)/M(\gluino)$ = (a) 2.0, (b) 1.5, (c) 1.2 and (d) 1.0. 
Here
$Br(\schionepm\schitwozero \rightarrow 3\ell+X)$ =
$Br(eee) + Br(ee\mu) + Br(e\mu\mu) + 
Br(\mu\mu\mu)$.\label{fig:run1_cdf_3l_limit}}
\vspace{.5in}
%\end{figure}

%% fig12
%\begin{figure}
%\vspace{-1.in}
%\centering\leavevmode
%%%    \parbox{7.0in}{
%	\psfig{figure=sugra_run1_trilepton.ps,height=6.3in}
%%%}
%%%\end{center}
\epsfin_v1{3in}{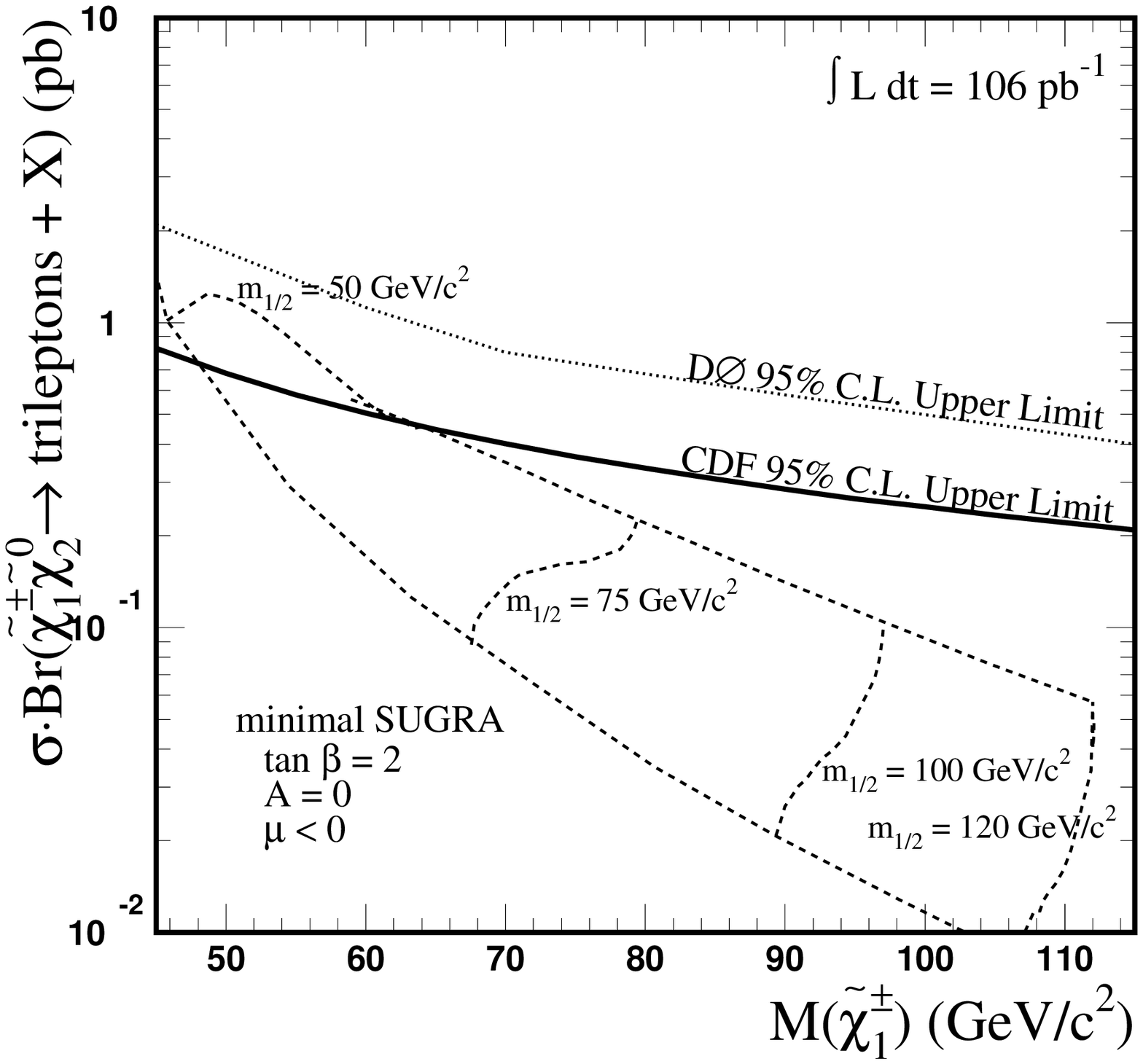}
\vspace{0.1in}
\caption{D\O\ \protect\cite{d0_tril} and CDF \protect\cite{cdf_tril} 
95\% C.L. upper limits on
$\sigma \cdot Br(\schionepm\schitwozero \rightarrow 3\ell+X)$, 
compared to the mSUGRA prediction.
Upper and lower dashed curves represent 
the predictions at $m_0$ = 100 \mgev\ and 
2500 \mgev,
varying $m_{1/2}$ from 50 \mgev to 120 \mgev.
}
\label{fig:sugra_run1_trilepton}
\end{figure}

Figure~\ref{fig:run1_cdf_3l_limit} shows the CDF and D\O\ upper limits
on $\sigma\cdot Br$ at 95\% confidence level (C.L.), where
$Br(\chione\chitwo \rightarrow 3\ell+X) \equiv$ 
$Br(eee) + Br(ee\mu) + Br(e\mu\mu) + Br(\mu\mu\mu)$.
The D\O\ upper limit on $\sigma \cdot Br$ in Ref.~\cite{d0_tril} is given for 
a single trilepton mode and therefore,
to compare with the CDF result, 
the D\O\ limit has been scaled up by a factor of 4.
We also overlay predictions at four representative points at
$\mu = -400$ \mgev\ and $\tan\beta = 2$ for 
the MSSM scenario.
The lower limit on $M(\schionepm)$ is maximized for 
$M(\squark) = M(\gluino)$.
CDF also studies the lower limits on $M(\schionepm)$ as a function of
$\mu$ at $\tan\beta = 2$ and $M(\squark) = M(\gluino)$.
The strongest limit is $M(\schionepm) >$~81.5~\mgev\sp 
and $M(\schitwozero) >$ 82.2 \mgev\sp at
$\mu = -600$ \mgev.
Limits on $M(\schionepm)$ 
at other $\mu$ values of $-1000$ $-800$, $-400$ and $-200$ \mgev\ are 
78.5, 81.0, 76.5, and 72.5 \mgev, respectively.

In SUGRA models,
$|\mu|$ is determined by demanding
the correct radiative electroweak symmetry breaking.
Using \ISAJET\ \cite{tk-isajet}, we find a mSUGRA parameter point 
corresponding to a MSSM point of
$M(\schionepm) \simeq\ 80\ \mgev$, 
$M(\squark) \simeq M(\gluino)$, and $\tan\beta = 2$:
($m_0$, $m_{1/2}$) = (130 \mgev, 75 \mgev) 
at $\mu < 0$ and $A_0 = 0$.
The value of $\mu$ is $-178$ \mgev, which is
roughly a region with $| \mu | \approx 200\ \mgev$.
The limits obtained on $M(\schionepm)$ in mSUGRA
are weaker than the limits in the MSSM as shown in
Fig.~\ref{fig:sugra_run1_trilepton}.
The strongest limit on $M(\schionepm)$ in mSUGRA is 62 \mgev\
\cite{tannenbaum_thesis},
corresponding to
$m_0$ = 160 \mgev, 
$m_{1/2}$ = 50 \mgev,
$\mu = - 158\ \mgev$ 
for $\tan\beta = 2$ and $A_0 = 0$
where
$M(\gluino)$ = 166 \mgev, $M(\squark)$ = 206 \mgev,
$M(\lsp)$ = 24 \mgev,
$Br(\schionepm \to e^\pm \nu \lsp)$ = 10.9\% and
$Br(\schitwozero \to e^+ e^- \lsp)$ = 18.7\%.

\subsection{Search for 
	Gluinos/Squarks Using $\mets + jets + X$ Events}
\label{sec:sugra_met_jets_run1}	

Pairs of gluino-gluino ($\tilde{g}\tilde{g}$), 
squark-squark ($\squark\squark, \squark\squarkb$), and 
gluino-squark ($\tilde{g}\tilde{q}$) would be produced via the strong
interaction.
Depending on the relative masses of gluinos and squarks, production of 
$\tilde{g}\tilde{g}$, $\squark\squarkb$, or
$\tilde{g}\tilde{q}$ may predominate. 
Here, degeneracy of five of the squarks ($\tilde{u}$, $\tilde{d}$,
$\tilde{s}$, $\tilde{c}$, $\tilde{b}$) is assumed to use
the next-to-leading order (NLO) calculation of the 
$\gluino\gluino$, $\squark \bar{\squark}$, and $\gluino \squark$ production
cross sections at Tevatron \cite{susy_nlo_xsec1}.

%Figure~\ref{fig:prod_xsec_glgl_sqsq_glsq_tevatron} shows
%the next-to-leading order (NLO) calculation of the 
%$\gluino\gluino$, $\squark \bar{\squark}$, and $\gluino \squark$ production
%cross sections at Tevatron \cite{susy_nlo_xsec1}.
%Here, degeneracy of five of the squarks ($\tilde{u}$, $\tilde{d}$,
%$\tilde{s}$, $\tilde{c}$, $\tilde{b}$) is assumed.

%% figure 13 used to be called here

Direct decays of each $\tilde{q}$ and
$\tilde{g}$ to quark(s) + $\chizero$ result in one and two quark jets
respectively, while
cascade decays through charginos and neutralinos may result in two or more
additional jets.  
Events therefore always contain  two jets, and most
should contain three or more jets, in addition to missing energy.
As a complement to the classic $\mets$+multijets analysis 
in the search for $\gluino\gluino$ production,
a LS dilepton approach has been proposed
to maximize the experimental sensitivity 
\cite{susy_dilepton}.

%%%%% fig.13
%\begin{figure}[t]
%\centering\leavevmode
%%%%%    \parbox{8.0in}{
%\psfig{figure=prod_xsec_glgl_sqsq_glsq_tevatron.ps,height=7.0in}
%%%%%}
%
%\vspace{-1.0in}
%\caption{
%Total cross-section for the Tevatron ($\protect\sqrt{s}$ = 1.8 TeV).
%NLO (solid): GRV94 parton densities, with scale $Q$ = $m$;
%compared with LO (dashed): EHLQ parton densities, at the scale $Q$ =
%$\protect\sqrt{s}$.
%	Taken from Ref. \protect\cite{susy_nlo_xsec1}. }
%\label{fig:prod_xsec_glgl_sqsq_glsq_tevatron}
%\end{figure}

\subsubsection{CDF Search}
\label{sec:cdf_glsq_limit_run1}

%%\subsubsection*{Dilepton, Jets and Missing Transverse Energy }
\medskip\noindent\underline{Dilepton, Jets and Missing Transverse Energy }

The search for 
$\gluino \gluino$, $\gluino \squark$, $\squark\squarkb$, and $\squark\squark$
production in the 
dilepton ($e^\pm e^\pm$, $e^\pm \mu^\pm$, $\mu^\pm \mu^\pm$) + jets + \met\ 
channel 
was first undertaken by CDF in Run Ia \cite{cdf1a_2ljmet}.
Since the production cross section times branching ratio is small,
we searched for the dilepton signature without the LS dilepton requirement.

We update the analysis 
with 106 \invpb\ of the Run Ia+Ib data \cite{cdf1_2ljmet}.
It should be noted we assume
degeneracy of five of the squarks in the MSSM framework 
(see Table \ref{tab:simssm_vs_sugra}) for both analyses.
The differences in the two analyses are summarized in
Table~\ref{tab:cdf1_2l_1a_vs_1b}.
No LS dilepton candidate events survive this selection, 
while 19 opposite sign (OS) dilepton events are retained.

The principal  SM backgrounds to the LS dilepton signature are events
from (i)~Drell-Yan process, (ii)~diboson production, 
(iii)~$b\bar{b} / c \bar{c}$ production,
and (iv)~$\ttb$ production.
The yield for each process is estimated by using \ISAJET\ and
the CDF detector simulation program.
We correct the \ISAJET\ cross section to 
the CDF measurements or to NLO calculations.
The expected number of
background events from SM processes is obtained to be
$0.55 \pm 0.25\ ({\rm stat}) \pm 0.06\ ({\rm sys})$ for LS events
and $14.1 \pm 1.3\ ({\rm stat}) \pm 3.1\ ({\rm sys})$ for OS events
\cite{cdf1_2ljmet}.

The sources of systematic uncertainty on the kinematic acceptances
for these analyses include 
initial and final state gluon radiation,
uncertainty on the integrated luminosity,
lepton identification,
Monte Carlo statistics,
jet energy scale,
and uncertainty on the trigger efficiency.
The total uncertainty on the kinematic acceptance
ranges from 16\% to 29\%
for various SUSY parameter points.

\begin{table}[ht]
\caption{SUSY dilepton analyses in Run Ia and Ia+Ib at CDF.}
\label{tab:cdf1_2l_1a_vs_1b}
\begin{center}
\begin{tabular}{ l l l }
%\hline
        &      Run Ia \cite{cdf1a_2ljmet} & Run Ia+Ib \cite{cdf1_2ljmet} \\
\hline
\hline
$p_T(\ell_1)$   & $>$12 \pgev   & $>$11 \pgev \\
$| \eta(\ell_1) |$& $<$1.1 ($e$), 0.6 ($\mu$)   & $<$1.1 ($e$), 0.6 ($\mu$) \\
$p_T(\ell_2)$   & $>$11 \pgev   & $>$~5 \pgev \\
$| \eta(\ell_2) |$& $<$2.4 ($e$), 1.0 ($\mu$)   & $<$1.1 ($e$), 1.0 ($\mu$) \\
$\Delta \phi(\ell_1 \ell_2)$ cut & Yes  & No \\
$\pt (\ell_1 \ell_2)$ cut & Yes  & No \\
$\Delta R_j^{cone}$ & 0.7       & 0.4  \\
$E_T(j)$        & $>$15 GeV     & $>$15 GeV \\
$| \eta(j) |$   & $<$2.4        & $<$2.4 \\
$N(j)$          & $\geq$2 (at least one central jet)    & $\geq$2       \\
$\Delta \phi(j_1 \mets)$ cut    & Yes   & No \\
$\mets$         & $>$25 GeV     & $>$25 GeV \\
Charges of dilepton & OS+LS & LS \\
\hline
{\bf Results:} &	& \\
$N_{obs}$  & 1 ($\mu^+\mu-$)	&	0 \\
$N_{BG}$   & $2.39 \pm 0.63 ^{+ 0.77}_{- 0.42}$ & $0.55 \pm 0.25 \pm 0.06$ \\
\hline
\end{tabular}
\end{center}
\end{table}

We set limits on the cross section times branching ratio for two cases:
(i) $M(\squark) \gg M(\gluino)$ where we fix $M(\squark) = 500 \mgev$, and
(ii) $M(\squark) \simeq M(\gluino)$.
In each case, we exclude:
\begin{eqnarray}
\sigma (\ppbar \rightarrow \gluino\gluino+\gluino\squark+\sqsqb+\sqsq) 
\cdot Br(\gluino\gluino+\gluino\squark+\sqsqb+\sqsq
\rightarrow \ell_{1}\ \ell_{2} + X) & \geq  &
\frac{N_{95\%}}{\epsilon_{tot} \cdot
\int {\cal L} \; dt}, \nonumber
\end{eqnarray}
where  $N_{95\%}$ is the
Poisson 95\% C.L. upper limit for observing 
zero events combined with a Gaussian distribution for the 
systematic uncertainty: $N_{95\%} =$ 3.1 $\sim$ 3.3 events 
for various SUSY points.
The acceptance, $\epsilon_{tot}$, 
is the product of the kinematic and geometric acceptance,
the efficiency of identifying LS dilepton with LS and two jets, and
the trigger efficiency for dileptons.
We calculate the event acceptance using 
\ISAJET\ version 7.20 with CTEQ3L parton distribution functions and 
the CDF detector simulation program.  
The integrated luminosity is $\int {\cal L} \; dt = 106 \pm 4$ \ipb.

\begin{figure}[ht]
\vspace{.5in}  
%%\centering\leavevmode
%%epsfxsize=3.5in\epsfysize=3in\epsffile{xsecbf_2l_limit_band_th.ps}
\epsfin_v1{3.5in}{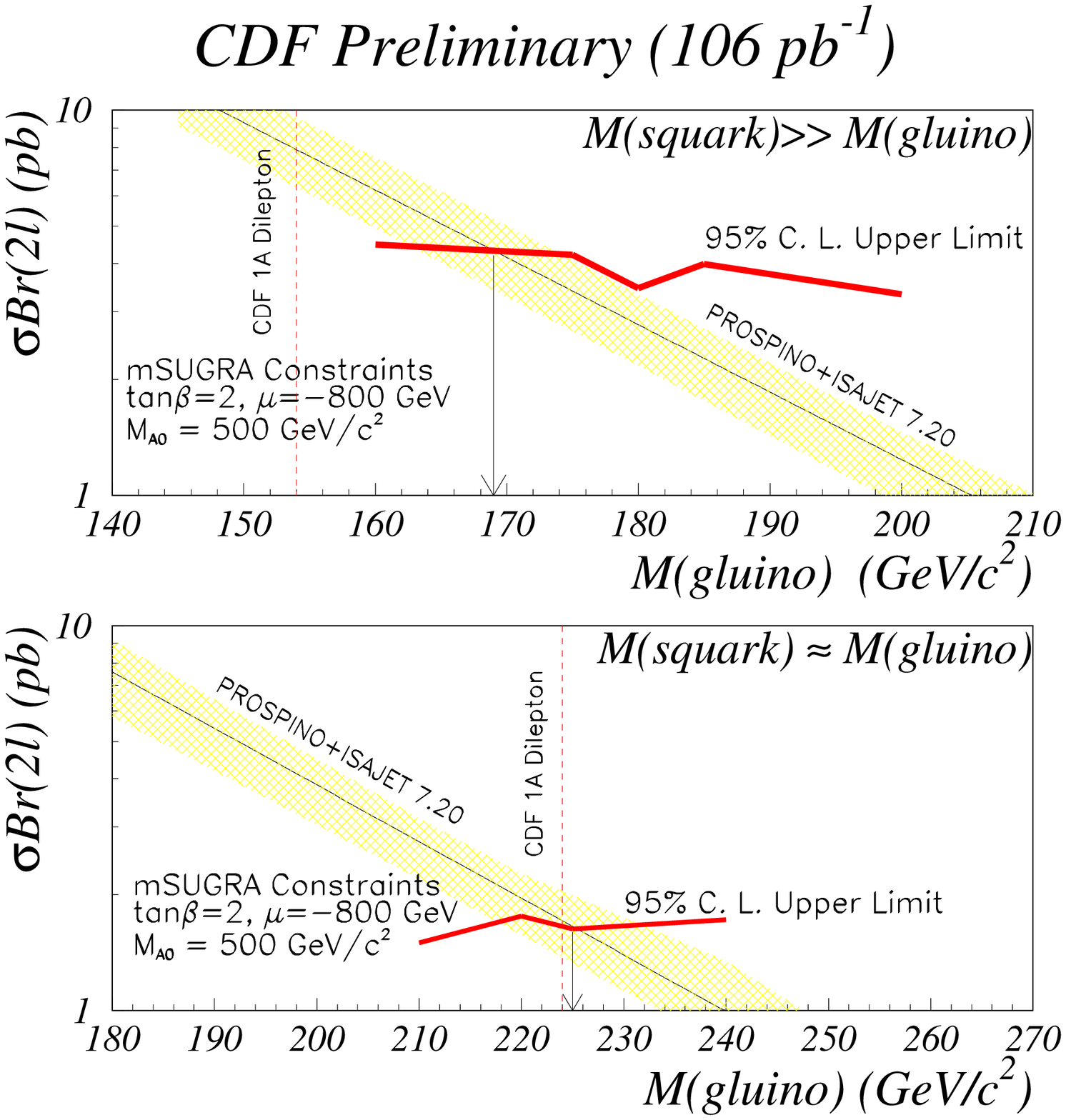}
\caption{CDF 95\% C.L. upper limits \protect\cite{cdf1_2ljmet}
on $\sigma\cdot Br$ as a function of $M(\gluino)$,
compared to NLO predictions \protect\cite{prospino-k}
in the MSSM framework
at $\tanb$ = 2 and $\mu=-800$ \mgev\ for two cases:
(i)~$M(\squark) \gg M(\gluino)$ and 
(ii)~$M(\squark) \simeq M(\gluino)$.
The nominal value of the cross section 
for $\gluino\gluino$, $\gluino\squark$, $\sqsqb$, and $\sqsq$ production 
is calculated with CTEQ3M parton distribution function at $Q^2$ = $M^2$.
The band shows the theoretical uncertainty in the calculation
varying $Q^2$ between $(2M)^2$ and $(0.5 M)^2$.}
\label{fig:cdf1_2l2jmet_limit}
\end{figure}

Figure \ref{fig:cdf1_2l2jmet_limit} shows the 95\% C.L. upper limits
on $\sigma \cdot Br$ compared with the NLO calculation \cite{prospino-k}.
For our nominal choice of $Q^2 = M^2$,
the lower gluino mass limit at 95\% C.L. is 
169 \mgev\sp for $M(\squark) \gg M(\gluino)$
and 225 \mgev\sp for $M(\squark) = M(\gluino)$
at $\tan \beta = 2$ and $\mu = -800$ \mgev\sp \cite{cdf1_2ljmet}.
We also indicate the corresponding limits 
from the Run Ia analysis \cite{cdf1a_2ljmet}.
The LS requirement is effective in selecting
$\gluino\gluino$ production in the case of $M(\squark) \gg M(\gluino)$,
which is explained by the Majorana nature of the gluino.

The CDF Run Ia+Ib analysis in mSUGRA framework
is also in progress.
However, we deduce the mSUGRA limits at $\tan\beta = 2$ based on
the following points:
\begin{itemize}
\item Since the stop pair production is not included 
	in the Run Ia+Ib analysis,
	the results are insensitive to the choice of $A_0$;
\item There is a weak $\mu$ dependence in the dilepton branching ratio
	between $-800\ \mgev$ and $-200\ \mgev$, if
	$M(\slepton) > M(\schionepm)$ where
	a principal decay mode is $\schionepm \to \ell^\pm \nu \lsp$.
	The $\mu$ value of $-450$ to $-200\ \mgev$ is roughly what 
	mSUGRA requires in the parameter space we explore.
\end{itemize}
Then, one can find the corresponding mSUGRA points for $\tanb = 2$,
$\mu < 0$ and $A_0 = 0$ as shown in Table \ref{tab:cdf1_2ljmet_sugra}.

\begin{table}[h]
\caption{Approximate estimate of\break mSUGRA
 (\ISAJET\ version 7.20) points
	corresponding to results of the CDF LS dilepton analysis
	for 
	(i) $M(\squark) \gg M(\gluino)$ and
	(ii) $M(\squark) \simeq M(\gluino)$.
        Other mSUGRA parameters are fixed as
	$\tan \beta = 2$, $\mu < 0$ and $A_0 = 0$.
	SUSY masses and SUGRA parameters are given in \mgev.}
\label{tab:cdf1_2ljmet_sugra}
\begin{center} \tabcolsep=2em
\begin{tabular}{ c r r  }
%\hline
\noalign{\smallskip}
{\bf Mass}	 &  (i) & (ii) \\
$\squark$       & 502   & 223  \\
$\gluino$       & 169   & 225  \\
$\schionepm$ 	& 50   & 78  \\
\hline
\hline
\noalign{\smallskip}
{\bf mSUGRA}  &   &   \\
$m_0$      & 490 &  120 \\
$m_{1/2}$  & 47  &  73 \\
$\mu$		& $-437$ & $-169$ \\
\hline
\end{tabular}
\end{center}
\end{table}

\subsubsection{D\O\ Searches}
\label{sec:d0_glsq_limit_run1}

%%\subsubsection*{Jets and Missing Transverse Energy }
\medskip\noindent\underline{Jets and Missing Transverse Energy}

The D\O\ search for 
$\gluino\gluino$, $\squark \bar{\squark}$, and $\gluino \squark$ production
in the jets + \met\ channel is based
on 72~\ipb\ of data~\cite{d0jmet1b,d0jmet1a} taken during the 1994-95 run.
The cross section values are determined using the NLO
calculation in Ref.~\cite{prospino-k} 
by assuming degeneracy of five of the squarks; 
top squarks are excluded in this analysis.

\begin{figure}[ht]
\centerline{\psfig{figure=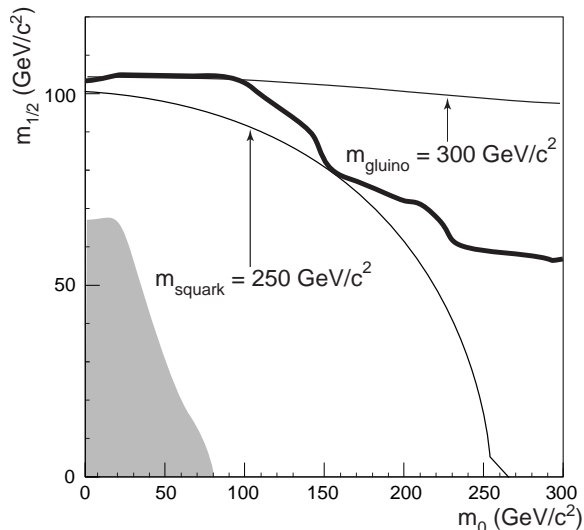,width=3.0in}}
\caption{D\O\ 95\% C.L. exclusion contour (heavy line)
in the \mhalf-\mzero\ plane
($\tanb = 2$, $\azero = 0$, and $\mu\lt 0$) for the jets + \met\ search.
The thin lines are contours of constant gluino or squark mass.
The dark shaded area is where mSUGRA does not predict
electroweak symmetry breaking or the sneutrino is the LSP.}
\label{fig:d0jmet}
\end{figure}

The analysis requires three jets with \et\gt\
25~GeV, where the leading jet has \et\gt\ 115~GeV and \abseta\lt 1.1.
The jet angle, with respect to the hard scattering vertex, is confirmed
using the tracks associated with the leading jet; this is because an
average event is likely to have more than one interaction and choosing the
incorrect vertex will lead to significant spurious \met.  
The \met\ is
required to be greater than 75~GeV to stay above the trigger threshold. 
The jet energy has a large resolution ($\sigma = 0.8 \sqrt{\pt}$) thus events
where the \met\ is correlated with the jets in azimuth are removed.
To remove events that are likely background, any candidates with isolated
electrons or muons are removed from the sample.  This cut reduces
backgrounds due to vector boson production in association with jets and
\ttbar\ production. The vector boson backgrounds are simulated with
\VECBOS~\cite{vecbos} and the \ttbar\ background with \HERWIG~\cite{herwig}. 
The multijet background is determined from a data set taken 
without a \met\ term in the trigger.  
The \met\ distribution is fit and extrapolated into the region of interest
for this analysis, {\it i.e.} greater than 75~GeV.  
The cuts on \met\ and \Ht, where \Ht\ is the scalar sum of 
the non-leading jets, are optimized for significance at each point 
in parameter space.  At the lowest
\met\ and \Ht\ point, \met\gt\ 75~GeV and \Ht\gt\ 100~GeV, the number of
events observed is 15 with an expected background of $8.3\pm 3.4$.  The
probability to observe more than 15 events, given this background estimation,
is 9.2\%.  We interpret this result as an
exclusion contour in the \mzero\ and \mhalf\ plane for 
$\tanb = 2$, $\azero = 0$, and $\mu\lt 0$ (see Fig.~\ref{fig:d0jmet}). 
At low \mzero, gluinos
with masses less than 300~\mgev\ are excluded.  At the \mzero\ and \mhalf\
point where the squark and gluino masses are equal this analysis excludes
masses less than 260~\mgev.

%%\subsubsection*{Dileptons, Jets, and Missing Transverse Energy}
\medskip\noindent\underline{Dileptons, Jets, and Missing Transverse Energy}

The D\O\ search for 
$\gluino\gluino$, $\squark \bar{\squark}$, and $\gluino \squark$ production
in the dielectron + jets + \met\ channel is based on
108~\ipb~\cite{d0dilep} taken during the 1992-95 run.
Top squarks are excluded in this analysis.
The cross section values are determined using the LO
calculation from Ref.~\cite{spythia-k}.

\begin{figure}
\centerline{\psfig{figure=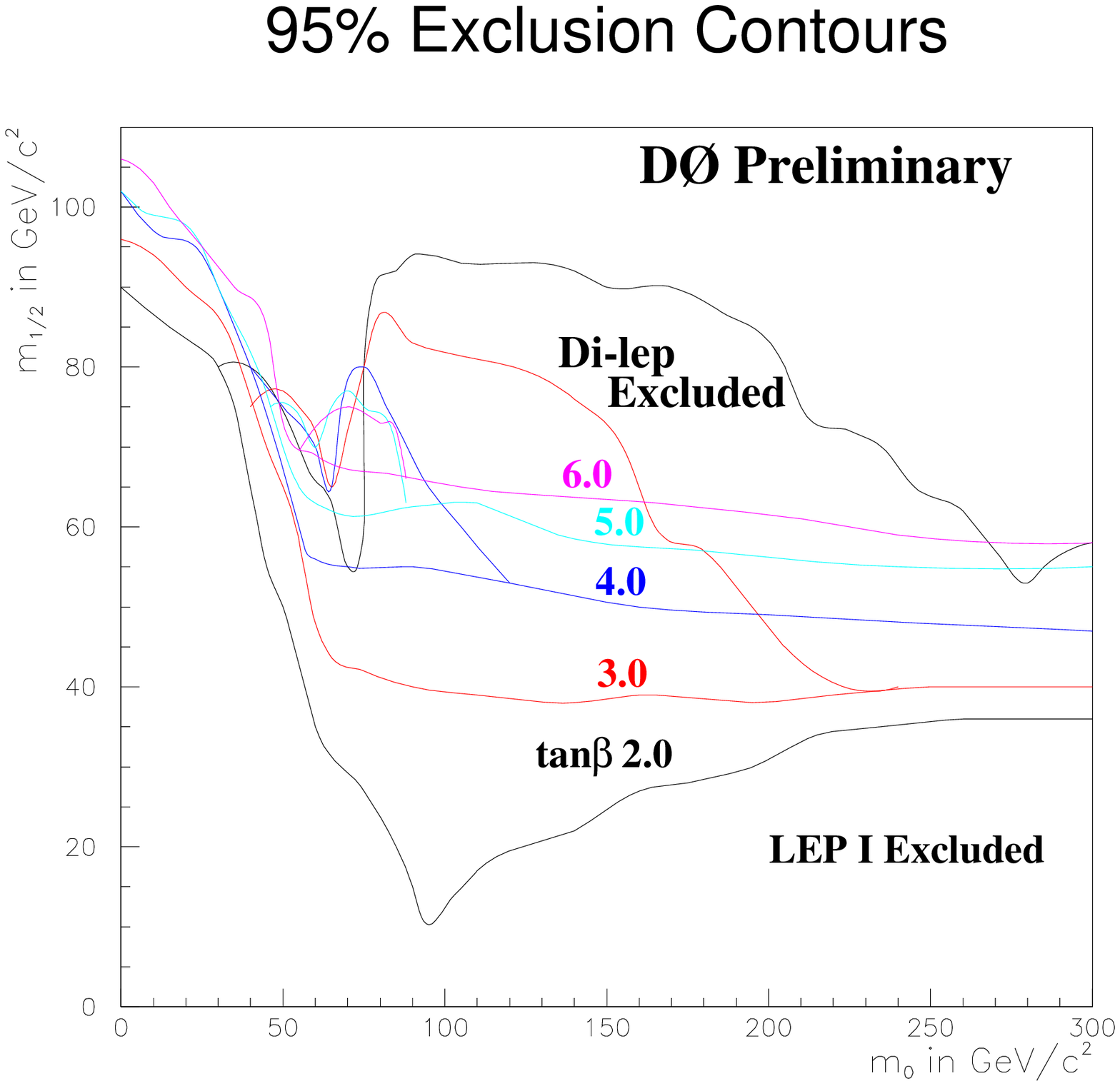,width=3.0in}}
\caption{D\O\ 95\% C.L. exclusion contour in the \mhalf-\mzero\ plane
($\azero = 0$ and $\mu\lt 0$) for the dielectron + jets + \met\ search.
The values of \tanb\ vary from 2 to 6, where the strictest limits 
come from $\tanb = 2$ and decrease as \tanb\
 increases. Also shown are the limits from LEP I for the 
same \tanb\ values.}
\label{fig:d0dilep}
\end{figure}

The dilepton modes used are electron-electron ($ee$),
electron-muon ($e\mu$), and muon-muon ($\mu\mu$).  
In order to optimize the ratio of signal to
background several kinematic thresholds for the leptons, jets, and \met\ are
employed.  The $ee$ channel requires two electrons with $\et\gt 17$ and
15~GeV.  The $e\mu$ requires an electron with $\et\gt 17$~GeV and a muon
with $\pt\gt$ 4, 7, or 10~\pgev.
The $\mu\mu$ channel requires two muons with $\pt\gt 20$ and 10~\pgev.
Each dilepton channel also requires two jets and \met.  The jet cuts are
also variable with $\et\gt 20$ or 45~GeV.  The \met\ is required to be
greater than 20, 30, or 40~GeV.  In addition, events with a dilepton
invariant mass near the \zb\ mass are sometimes rejected.  At different
points in \mzero-\mhalf\ space a fast Monte Carlo, interfaced with
\SPYTHIA~\cite{spythia-k}, is used to predict which set of thresholds on the
$ee, e\mu$, and $\mu\mu$ channels give the greatest significance.  The
set of cuts at each point are then applied to the data and compared to the
expected background.  The maximum significance is taken from $ee$, $e\mu$,
and $\mu\mu$ channels separately, from the 
combination of any two-channels, or three-channels.  
The backgrounds are due to $W$ boson, $Z$ boson, \ttbar, and QCD.  The $W$
boson, $Z$ boson, and \ttbar\ backgrounds are included in the fast Monte
Carlo with the cross sections normalized to the \Dzero\ measured values.
The QCD and vector boson background is taken from 
data. No excess of events over background
is observed and the result is presented as a limit 
in the \mhalf-\mzero\ plane 
for  $\azero = 0$, $\mu\lt 0$, and at $\tanb = 2, 3, 4, 5$, and 6.
(see Fig.~\ref{fig:d0dilep}).  
For equal \squark\ and \gluino\ masses a
limit of 270~GeV is obtained.  
In Fig.~\ref{fig:d0dilep} the mass limits
are reduced because as \tanb\ increases the branching ratio into
first and second generation leptons decreases.

\subsection{Search for Sbottom and Stop Quarks}
\label{sec:stopsbot_limit_run1}

At the Tevatron, the third generation scalar quarks are produced in
pairs via $gg$ and $q\overline{q}$ fusion. 
The leading order terms in
the production cross sections depend only on the scalar quark
masses. In the NLO terms, the largest
theoretical uncertainty is due to the QCD $\mu$ scale. 
The dominant SUSY corrections depend on other scalar quark masses 
and are small ($\sim 1\%$).  
For example, a third generation scalar quark 
($\tilde{Q}_1$ = $\sbottomone$ or $\stopone$) with a
mass of 110 \mgev\ would have a production cross section of
\mbox{7.4$\pm$1.1~pb}, using the mass scale $\mu= M(\tilde{Q}_1)$ for the
central value and $\mu = M(\tilde{Q}_1) / 2$ and 
$\mu = 2 M(\tilde{Q}_1)$ for the uncertainties~\cite{susy_nlo_xsec2}.

\subsubsection{Search for Sbottom and Stop Quarks in \met + Jets Channel}

D\O\ and CDF search for direct production of 
$\sbottomone\sbottomoneb$ and
$\stopone\stoponeb$ in the \met\ + jets channel:
\begin{itemize}
\item 
For the sbottom search we consider 
the $\tilde{b}_1 \rightarrow b\tilde{\chi}_{1}^{0}$ decay mode. 
Since this is a
tree-level decay it dominates over most of the parameter space if
$\tilde{\chi}_{2}^{0}$ is heavier than $\sbottomone$.
\item 
In the absence of
flavor changing neutral currents the 
$\stopone \rightarrow c \schionep$ decay proceeds via a one-loop
diagram and will become dominant when the tree-level decay
$\tilde{t}_1 \rightarrow b\tilde{\chi}_{1}^{+}$ is kinematically
forbidden.
\end{itemize}
Thus, $Br(\sbottomone \to b \lsp) = 100\%$ and
$Br(\stopone \to c \lsp) = 100\%$ are assumed.
Both analyses use the cross sections for
$\sbottomone\sbottomone$ and $\stopone\stopone$ production
calculated at NLO \cite{susy_nlo_xsec2}
in setting the lower limits on the sbottom and stop masses.

%\subsubsection*{D\O\ Search for Bottom Squark}
\medskip\noindent\underline{D\O\ Search for Sbottom Quark}

D\O~has searched for sbottom quarks in four channels \cite{sbottom_d0},
which are combined to set limits on the production of sbottom quarks. 
The first required a \met+jets topology, while the other three
channels impose the additional requirement 
that at least one jet have an associated muon, which
was used to tag $b$ quark decay.  For all channels, the presence of
significant \met~was used to identify the non-interacting LSPs. Backgrounds
arose from events where neutrinos produced significant \met; for example, in
$W$ plus multijet events where $W\rightarrow \ell\nu$.

Events for the \met+jets channel were collected using a trigger which
required \met$>35$ GeV. During offline analysis, events with two jets
($E_T^{\rm jet}>30$
GeV), \met $>40$ GeV, and no isolated electrons or muons were selected. To
eliminate QCD backgrounds, additional cuts were made on the angles between
the two jets, and between jets and the \met\ direction. Data with an
integrated luminosity of 7.4 pb$^{-1}$ yielded three events satisfying the
selection criteria with and expected background of 
$3.5\pm 1.2$~\cite{stop_d0}.

The muon channel used a combination of three triggers.  The triggers had the
following requirements: a trigger with two low-$p_T$ muons ($p_T^\mu >3.0$
GeV/c), a trigger with a single low-$p_T$ muon plus a jet with $E_T>10$ GeV,
or a trigger which required a high-$p_T$ muon ($p_T^\mu >15$ GeV/c) plus a
jet with $E_T>15$ GeV. Integrated luminosities of 60.1 \invpb,
19.5 \invpb, and 92.4 \invpb, respectively were collected using the three
muon triggers.
Either two muons, each associated with its own jet, or a single
jet-associated muon was required. If the event had only one muon-jet
combination, an additional jet with $E_T>25$ GeV was required. To remove QCD
backgrounds, events were selected with \met~$>35$ GeV and an azimuthal
angular separation between the \met\ direction and the nearest jet
greater than 0.7 radians.
For the single muon channels, backgrounds from $W$ boson decays were removed
by cuts on the muon-jet correlation; and top backgrounds were minimized by
cuts on the scalar sum of jet $E_T$.  Following all selection criteria, two
events remained in the data.

Combining the four channels gave five events observed in the data with a
total background estimated to be $6.0\pm 1.3$ events (1.4 $t\bar{t}$, 4.0
$W$ boson, and 0.6 $Z$ boson). 

We set limits on the cross section by
combining the acceptances and integrated luminositiees for the different
channels.  For any given $M(\sbottomone)$, we determined the value of
$M(\lsp)$ where our 95\% C.L. limit intersected the theoretical cross
section, and excluded the region where the cross section was greater than
our limit.  We used the program {\sc prospino}~\cite{prospino-k} 
to calculate the
theoretical sbottom  pair production cross section as a function of
$M(\sbottomone)$.

%\subsubsection*{CDF Searches for Stop and Sbottom Quarks}
\medskip\noindent\underline{CDF Searches for Stop and Sbottom Quarks}

The CDF analysis begins with a sample of events collected using a trigger
which required uncorrected missing transverse energy \mets\ \gt\ 35
GeV, corresponding to 88.0 $\pm$ 3.6 \invpb\ \cite{cdf1_stopsbot}.

We select events with 2 or 3 jets which have uncorrected transverse
energy \et\ \gt\ 15 GeV and $\abseta < 2$, requiring that there are no
other jets in the events with \et\ \gt\ 7 GeV and $\abseta < 3.6$.
Jets are found from calorimeter information using a fixed cone algorithm
with a cone radius of 0.4 in $\eta$-$\phi$ space.  The \mets\ cut is
increased beyond the trigger threshold to 40 GeV.  To reduce the
contribution from processes where missing energy comes from jet energy
mismeasurement we require that the \mets\ direction is neither
parallel to any jet ($j$) nor anti-parallel to the leading \et\ jet:
$min \Delta\phi(\mets,j) > 45 \degr$, 
$\Delta\phi(\mets,j_1) < 165 \degr$, where
the jet indices are ordered by decreasing \et.  A further reduction in
the QCD multijet background is made by requiring that $45 \degr <
\Delta\phi(j_1,j_2) < 165 \degr$.  We reject events with one or more
identified electrons (muons) with \et\ (\pt) \gt\ 10 GeV (\pgev).

After applying all these requirements, the data sample (called the
{\it pretagged sample}) contains 396 events. The largest source of
background in the sample is the production of $W$+jets, where the $W$
boson decays to a neutrino and an electron or muon that is not
identified, or a tau which decays hadronically.  The sample also
contains QCD multijet events where the large \mets\ is due to jet
energy mismeasurement.

The \svxp\ information is used to tag heavy flavor jets.  We associate
tracks to a jet by requiring that the track is within a cone of 0.4 in
$\eta$-$\phi$ space around the jet axis.  We require tracks to have
\pt\ \gt 1.0 \pgev, positive impact parameter, and a good \svxp\ hit
pattern.  A good \svxp\ hit pattern consists of 3 or 4 hits in the
\svxp\ detector with no hits shared by other tracks.  The impact
parameter of a track is positive if its projection on the jet axis is
positive, and negative otherwise.

For each track the probability that the track comes from the primary
vertex is determined taking into account the impact parameter
resolution function.  The resolution function is measured from the
negative impact parameter signature distribution in the data, which
does not carry lifetime information.  We call this probability the {\it
track probability}.  By construction, the distribution is flat for
tracks originating from the primary vertex. For tracks from a
secondary vertex, the distribution peaks near zero.  The joint
probability for tracks associated to a jet is called the {\it jet
probability} (${\cal P}_{jet}$) \cite{cdf_mu_bjet}.  
The jet probability is flat for a
primary jet data sample by construction.  
For bottom and charm jets, the jet probability peaks near zero.

We select events for the stop search analysis by requiring that the
event have at least one jet with ${\cal P}_{jet}$ \lt\ 0.05.  
This rejects 97\%
of the background while its efficiency for signal events is 25\% .
For the sbottom search analysis, we require the event to have at least
one jet with ${\cal P}_{jet}$  \lt\ 0.01.  
This rejects 99\% of the background
while retaining 45\% of the sbottom events.

In the sbottom (stop) analysis, we observe 5 (11) events in data,
which is consistent with $5.8 \pm 1.8$ ($14.5 \pm 4.2$) events
expected from the Standard Model processes such as $W$ and $Z$
production, $\ttb$ production, diboson production, and QCD events.

The sources of systematic uncertainty on the expected signal,
which are common to both sbottom and stop analyses,
are
(i) theoretical uncertainty on the NLO cross section for 
squark production,
(ii) initial and final state gluon radiation,
(iii) uncertainty on the efficiency for the heavy flavor tagging,
(iv) jet energy scale,
(v) uncertainty on the trigger efficiency,
(vi) uncertainty on the degradation of the signal efficiency 
for events with multiple primary vertices,
(vii) uncertainty on the integrated luminosity
and
(viii) MC statistics.
The total systematic uncertainty ranges from 31\% to 36\%
for the mass range 30 \mgev\ to 150 \mgev.
Using a background-subtracted method \cite{bs_method}, we find an
exclusion region in the $M(\lsp)$-$M(\tilde{Q}_1)$ plane at 95\%
C.L. limit.

%\subsubsection*{CDF and D\O\ Exclusion Regions}
\medskip\goodbreak\noindent\underline{CDF and D\O\ Exclusion Regions}

\begin{figure}[ht]
\vspace{-0.6in}
\epsfin_v1{0.5\textwidth}{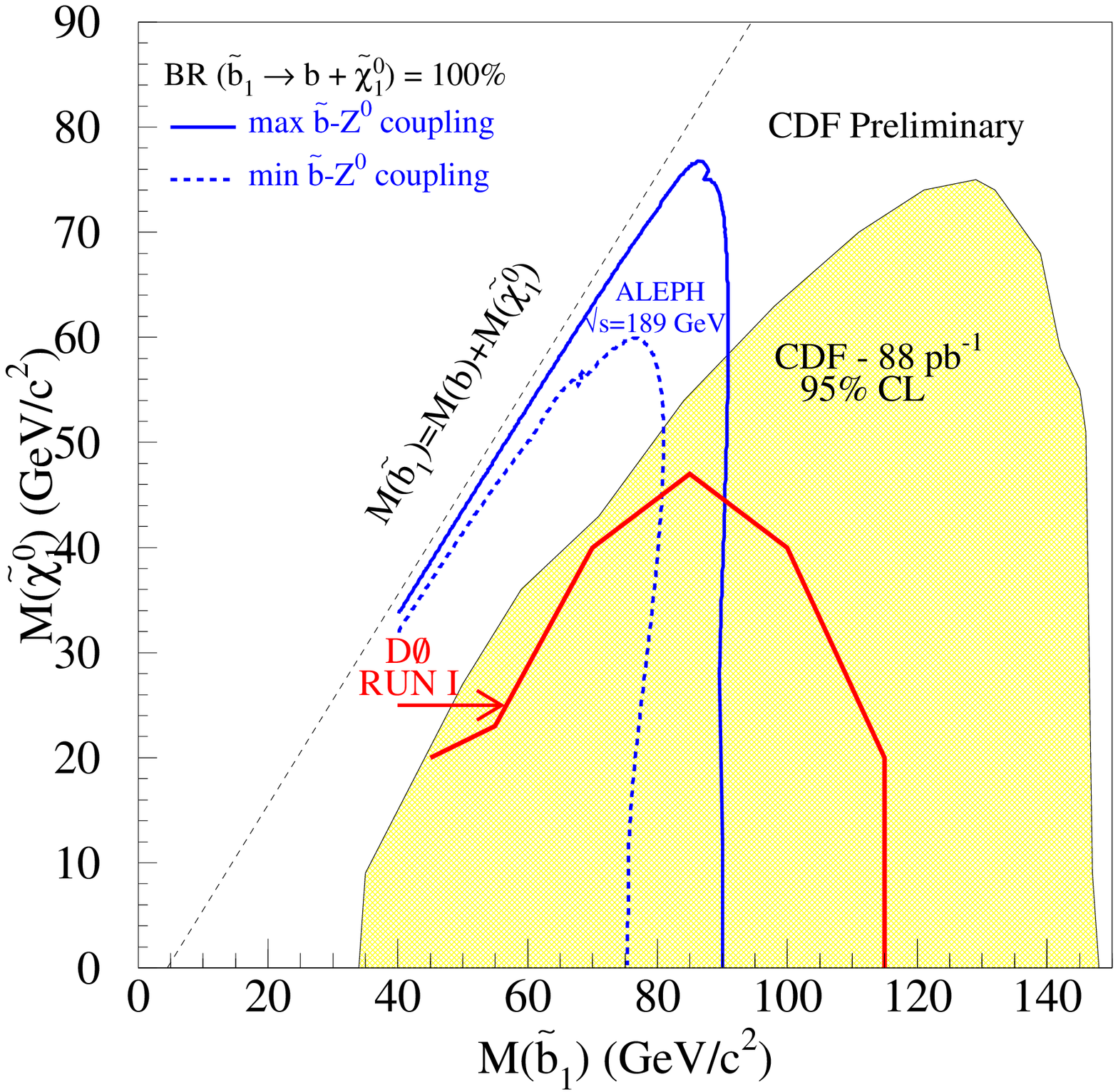}
\vspace{0.8in}
\caption{CDF and D\O\ 95\%  C.L. exclusion contours
in the $M(\lsp)$-$M(\tilde{b}_{1})$ plane for
$\sbottomone \to b \lsp$.
Also shown are the exclusion contours from
the ALEPH experiment at LEP for minimal and maximal
couplings \protect\cite{stopsbot_lep189}.}
\label{fig:run1_sbotlim}
\end{figure}

The excluded regions from CDF and D\O\ 
in the $M(\lsp)$-$M(\sbottomone)$ plane are shown 
in Fig. \ref{fig:run1_sbotlim}.  
Also plotted are the latest results from ALEPH \cite{stopsbot_lep189}.
The maximum $M(\sbottomone)$ excluded is 148 \mgev\ for $M(\lsp) = 0\ \mgev$
by CDF and 115 \mgev\ for $M(\lsp) < 20\ \mgev$ by D\O.

\begin{figure}[ht]
\vspace{-0.6in}
\epsfin_v1{0.5\textwidth}{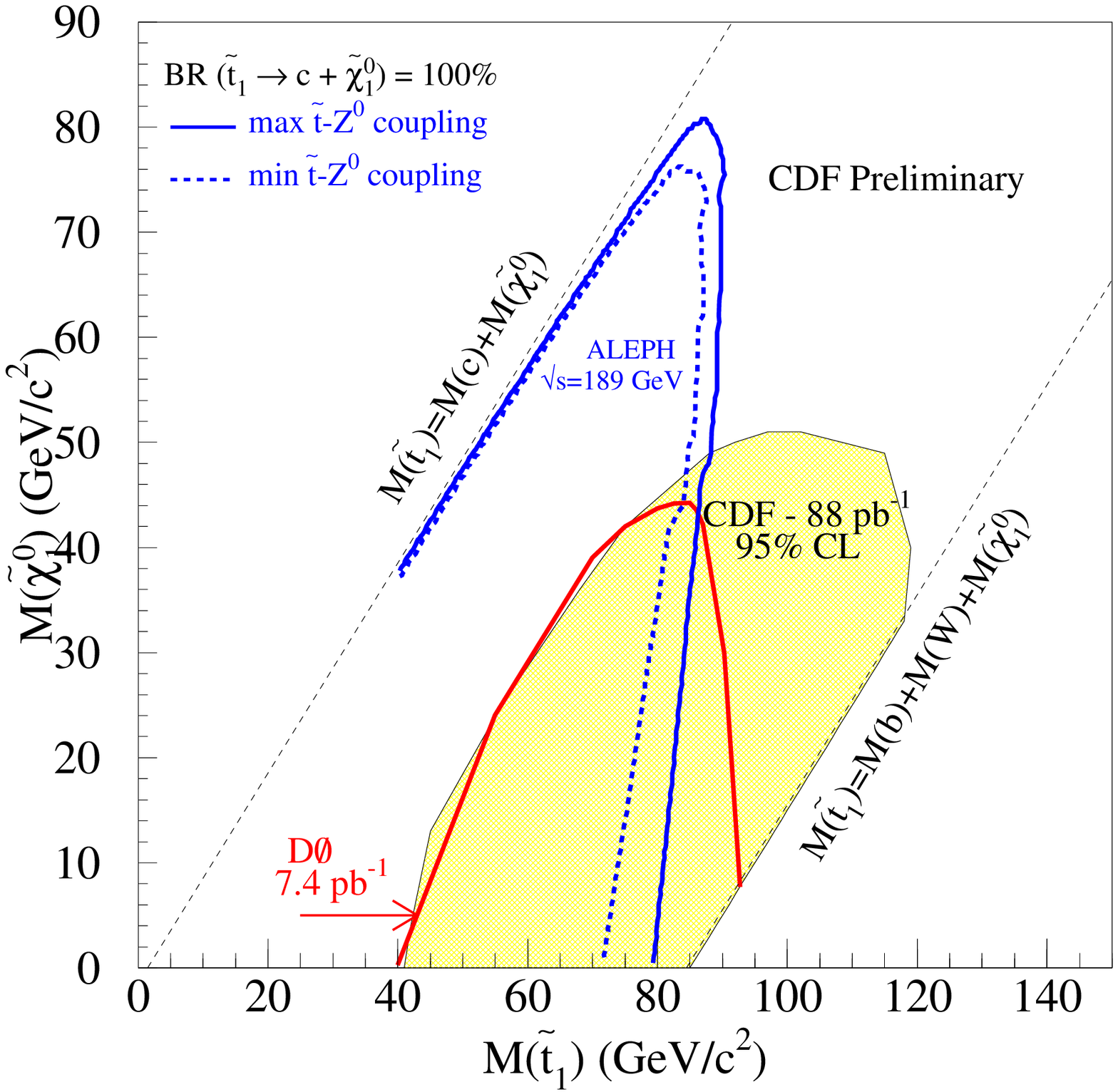}
\vspace{0.8in}
\caption{CDF and D\O\ 95\%  C.L. exclusion contours
in the $M(\lsp)$-$M(\tilde t)$ plane for
$\stopone \to c \lsp$.
Also shown are the exclusion contours from
the ALEPH experiment at LEP for minimal and maximal
couplings \protect\cite{stopsbot_lep189}.}
\label{fig:run1_stoplim}
\end{figure}

Figure \ref{fig:run1_stoplim} shows the exclusion region for the stop
analyses from CDF and D\O, compared to the ALEPH result
\cite{stopsbot_lep189}.  
The D\O\ results are based on 7.4 \invpb\ \cite{stop_d0}.  
The maximum $M(\stopone)$ excluded is 119 \mgev\ for
$M(\lsp) = 40\ \mgev$ by CDF.  The maximum value of the neutralino mass
is 51 \mgev\ which corresponds to the stop mass of 102 \mgev.

One can investigate the results within mSUGRA framework \cite{b-dutta}.
For the sbottom case,
the \tanb\ value has to be larger to generate
light sbottom mass.
There is little mSUGRA parameter space to satisfy 
$M(\schitwozero) \approx M(\schionepm) > M(\sbottomone)$.
For example, an MSSM point of
$M(\sbottomone) \approx 100\ \mgev$ and $M(\lsp) \approx 50\ \mgev$
is excluded in the CDF analysis.
The corresponding mSUGRA point is
$m_0 = 134\ \mgev, m_{1/2} = 130\ \mgev, \tanb = 45, \mu > 0, 
A_0 = -655\ \mgev$, which generates
$M(\sbottomone) = 98\ \mgev$, $M(\lsp) = 53\ \mgev$, and 
$M(\schionepm) = 98\ \mgev$.
However, the problem is 
that the electroweak symmetry does not break radiatively.
In contrast with the sbottom case, 
the light stop mass can be generated in lower \tanb\ values.
One of such mSUGRA points is
$m_0 = 130\ \mgev, m_{1/2} = 125\ \mgev, \tanb = 7, \mu > 0, 
A_0 = -535\ \mgev$, which generates
$M(\stopone) = 88\ \mgev$, $M(\lsp) = 49\ \mgev$, and 
$M(\schionepm) = 89\ \mgev$.

\subsubsection{Search for Stop Quark in Lepton+\met+jets Channel}

The CDF search for evidence of direct production of  $\stopone\stoponeb$ in
the lepton+\met+jets channel is based on 88 \invpb of inclusive lepton ($e$
and $\mu$) data \cite{cdf1_stop_ljmet,cdf1_stop_ljmet2}.  
Two separate \stopone~decay channels
were investigated.  
In the first channel, we look for 
$\tilde{t}_1 \rightarrow b \tilde{\chi}_1^+$ 
(with a branching ratio of 100\%).
We then require
one of the charginos, which decay via a virtual $W^\pm$, to decay as 
$\tilde{\chi}_1^+ \rightarrow e^+ \nu ~\tilde{\chi}_1^0$
or $\mu^+  \nu ~ \tilde{\chi}_1^0$ with an assumed branching ratio of 11\%
for each lepton type.  For models where $\tilde{t}_1 \rightarrow b
\tilde{\chi}_1^+$ is not kinematically allowed, we considered a second decay
scenario in which $\stopone~\rightarrow b l^+\tilde{\nu}$, where
$\tilde{\nu}$ is a scalar neutrino and each $l=e,~\mu,~\tau$ has a branching
ratio of 33.3\%.  In our two scenarios, either the $\tilde{\chi}_1^0$ or the
$\tilde{\nu}$ is the LSP.  For either process the signature is an isolated
lepton, missing transverse energy, \met~from the LSP's, and two jets from
the $b$ quarks.

The data for this analysis was obtained by requiring (i)  an electron with
\mbox{$E_T\ge$ 10 \pgev} or muon with \mbox{$p_T\ge$ 10 \pgev} originating
from  the primary vertex and passing lepton identification cuts, (ii)
$\MET\ge$ 25 GeV, and (iii) at least two jets, one with $E_T \ge$ 12 GeV and
the second with $E_T\ge$ 8 GeV.  The lepton identification cuts were
identical to those used in previous CDF analyses \cite{cdftop94}.  
%%For electron identification, the electron was required to be in the central
%%electromagnetic calorimeter, to have lateral and longitudinal shower profiles
%%consistent with those of an electron, have less than 5\% of its energy
%%deposited in the hadronic calorimeter, and be well-matched to a track from
%%the CTC.  A muon was required to have tracks in the inner and outer central
%%muon chambers which were well-matched to a track from the CTC.  
We also 
required the leptons to pass an isolation cut in which the calorimeter $E_T$
in a cone of $\Delta R =0.4$
%%\equiv \sqrt{(\Delta\eta)^2 + (\Delta\phi)^2}
around the lepton was less than 2 GeV (excluding the lepton $E_T$).

We used the SVX$^\prime$ to identify secondary vertices from $b$ quark decays
and selected events with at least one secondary vertex.   The tagging
algorithm is described in Ref.~\cite{cdftop94} with improvements given in
Ref.~\cite{cdftop95}.   
We reduced the $Z/\gamma\rightarrow l^+l^-$  background
in our sample by removing events with either two isolated, opposite-sign
leptons or an isolated lepton that reconstructs an invariant mass $\ge$ 50
\mgev\ with a second CTC track with no lepton identification requirements.
Finally, we reduced the background from $b \overline{b}$ events and events
with hadrons misidentified as leptons (fake leptons) by requiring that the
$\Delta\phi$ between the \MET~and the nearer of the two highest-$E_T$ jets be
$\ge$ 0.5 rad.  This reduces fake \MET~which is the result of jet energy
mismeasurement.  The number of events remaining in our sample after all cuts
is 81.

%%Signal and background selection cut efficiencies were estimated using a
%%variety of Monte Carlo generators followed by a CDF detector simulation.
Signal event samples were created using \ISAJET\ version 7.20 \cite{tk-isajet}.  
The supersymmetric particle masses used in signal
simulation were: $M(\tilde{\chi}_1^\pm)=$ 90 \mgev, 
$M(\tilde{\chi}_1^0)=$ 40 \mgev, and $M(\tilde{\nu}) \ge$ 40 \mgev.  
%%The signal selection efficiency increases with $m_{\stopone}$, reaching a
%%plateau as event energies advance from cut thresholds.  This plateau is near
%%6\% for the $\stopone~\rightarrow b l^+ \tilde{\nu}$ decay scenario for the
%%$m_{\tilde{\nu}}$ considered, but is lower for the $\tilde{t}_1 \rightarrow
%%b \tilde{\chi}_1^+$ decay scenario for the $m_{\tilde{\chi}_1^\pm}$ and
%%$m_{\tilde{\chi}_1^0}$ considered due to the branching ratio of the forced
%%$\tilde{\chi}_1^\pm$ decay.  
The sources of uncertainty for signal selection efficiency are
(i) the $b$-jet tagging efficiency,
(ii) the trigger efficiencies,
(iii) the integrated luminosity,
(iv) initial- and final-state radiation,
(v) Monte Carlo statistics, 
(vi)  the parton distribution function,
(vii) the corrections to jet energies from the underlying event,
(viii) the jet energy scale, and
(ix) the lepton identification and isolation efficiencies. 
The effects of some of these sources
vary with $M(\stopone)$, but none contribute more than 10\% to the
overall uncertainty, which is less than 16\%
for all $M(\stopone)$ considered.

%%Standard Model backgrounds come from any process that can produce two or more
%%jets and either real or fake leptons and real or fake $\MET$.  This includes
%%heavy flavor quark production, vector boson production with two or more
%%accompanying jets, and inclusive jet production with real or fake leptons.
The number of background events from heavy flavor quark production were
predicted using  measured or calculated cross sections and selection
efficiencies determined from Monte Carlo.  
Top-pair and single-top
production were simulated using \HERWIG\ version 5.6 \cite{herwig}
with
$\sigma_{t \bar{t}}$  = 5.1 $\pm$ 1.6 ~{\rm  pb} \cite{top_prod}
for $M(t)$ = 175 \mgev.
The value of $\sigma_{t \bar{b}}$ from $W$-gluon
fusion for $M(t)$ = 175 \mgev\ from a NLO calculation
is 1.70 $\pm$ 0.15 ~{\rm pb} \cite{stelzer}.  Vector boson samples were
generated using \VECBOS ~version 3.03 \cite{vecbos}.  
Drell-Yan, $b \bar{b}$,
and $c \bar{c}$ samples were generated with \ISAJET\ version 7.06 and
normalized to independent CDF data samples.

To determine the number of events with fake leptons in our sample, we used a
data sample passing all our selection cuts with the exceptions of a  
non-overlapping  \MET~requirement ($15\le \MET \le 20$ GeV) and no
requirement on $\Delta\phi$(\MET, nearer jet).  The number of fake lepton
events was normalized to this data sample, which contained negligible signal,
after other backgrounds were subtracted.  The number of fake lepton events
was then extrapolated to the signal region using cut efficiencies determined
from an independent fake-lepton event sample.

% The complete list of backgrounds and the number of expected events
% remaining after all cuts is given in Table \ref{after}.   
%The only significant backgrounds remaining 
The significant backgrounds are $t\overline{t}$, $b \overline{b}$, 
$W^\pm \rightarrow l^+\nu~+\ge$2 jets, and fake lepton events.
The number of data events agrees well with the expected background.
%
% Table 1 here
%

To determine the number of potential signal events in this final data sample,
we performed extended, unbinned likelihood fits for each \stopone~mass
considered for both decay scenarios.  The likelihood fits compared the shapes
of distributions of the signal and background and included Gaussian terms
tying the fit background levels to their predicted levels.  The fit
parameters were the number of signal events, the number of $t\overline{t}$
events, the number of $b \overline{b}$ plus fake lepton events, and the
number of vector boson events (represented in the fit by the $W^\pm
\rightarrow l^+\nu~+\ge$2 jets distributions).  
%We used the Kolmogorov statistic applied to the simulated distributions of
%signal and combined backgrounds to determine the most sensitive kinematic
%distributions to use in the fit.  
%The kinematic distributions evaluated include lepton $p_T$, lepton
%momentum transverse to the nearer of the two highest-$E_T$ jets, $H_T$ (the
%scalar sum of  lepton $E_T$, \MET, and jet $E_T$ for all jets with $E_T\ge$ 8
%GeV), jet multiplicity, and $\Delta \phi(j_1,j_2)$.

\begin{figure}
\vspace{-0.3in}
\epsfin_v1{0.5\textwidth}{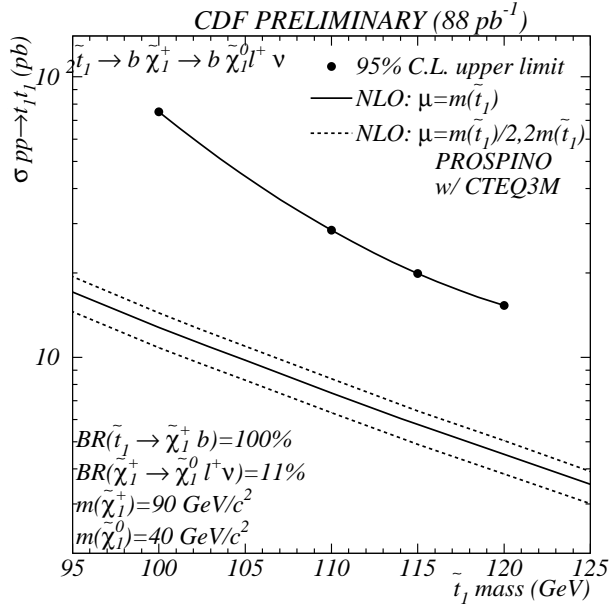}
\vspace{0.8in}
\caption{The points represent the CDF 95\% C.L. cross section limit as
  a function of \stopone~mass in the model where the decay $\tilde{t}_1
  \rightarrow b \tilde{\chi}_1^\pm$ has a branching ratio of 100\%
  \protect\cite{cdf1_stop_ljmet}.
  The masses in the model were $M(\tilde{\chi}_1^\pm)=$ 90 \mgev\ and
  $M(\tilde{\chi}_1^0)=$ 40 \mgev. The line without markers represents the
  NLO theoretical prediction for $\sigma_{\stopone\stoponeb}$ using the
  renormalization scale $\mu= M(\stopone)$ and parton distribution function
  CTEQ3M.  The dashed lines represent the NLO cross section for
  $\mu=M(\stopone)/2$ and $\mu=2 M(\stopone)$. } 
\label{fig:cdf1_stop_ljmet}
\end{figure}

For the $\tilde{t}_1 \rightarrow b \tilde{\chi}_1^+$ decay, sensitivity to
signal was greatest for a two-dimensional  fit to the combined probability
distributions for $H_T$ and $\Delta \phi(j_1,j_2)$.  Fit results at all
masses were consistent with zero signal events.  
%% The result from the two-dimensional fit to data for $\tilde{t}_1
%% \rightarrow b \tilde{\chi}_1^\pm$ with $m_{\stopone} =115$ \mgev,
%% $m_{\tilde{\chi}_1^\pm}$ = 90 \mgev, and $m_{\tilde{\chi}_1^0}$ = 40 \mgev\
%% is shown in Fig.~\ref{fig:afterfit}.   
The 95\% C.L. limits on $\sigma_{\stopone\stoponeb}$ for this decay are shown
in Fig.~\ref{fig:cdf1_stop_ljmet} 
as a function of $M(\stopone)$.
The NLO theoretical prediction for $\sigma_{\stopone\stoponeb}$ using the
renormalization scale $\mu= M(\stopone)$ and parton distribution function
CTEQ3M is shown in Fig.\ \ref{fig:cdf1_stop_ljmet} for comparison
\cite{susy_nlo_xsec2}.

\begin{figure}
\vspace{-0.3in}
\epsfin_v1{0.5\textwidth}{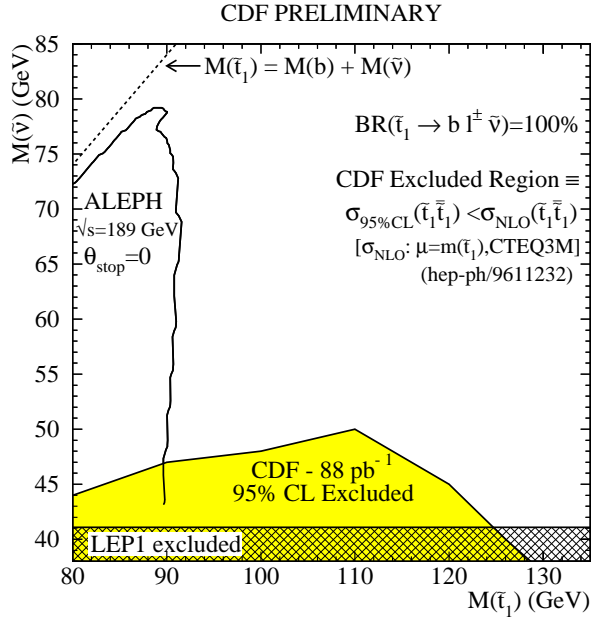}
\vspace{0.8in}
\caption{CDF 95\% C.L. excluded region 
  in the $M(\tilde{\nu})$-$M(\stopone)$ plane 
  when the $\stopone~\rightarrow b e^+ \tilde{\nu}$,
  $\stopone~\rightarrow b \mu^+ \tilde{\nu}$, and $\stopone~ \rightarrow b
  \tau^+ \tilde{\nu}$ branching ratios are assumed to be 33.3\%
  \protect\cite{cdf1_stop_ljmet2}.
  We define
  the exclusion region as that region of supersymmetric parameter space for
  which the 95\% C.L. limit on $\sigma_{\stopone\stoponeb}$ is less than the
  NLO theoretical prediction ($\mu= M(\stopone)$).  
  The LEP1 limit on $M(\tilde{\nu})$ and ALEPH excluded region in the
  $M(\tilde{\nu})$-$M(\stopone)$ plane are also shown
  \protect\cite{aleph}.  The ALEPH excluded region corresponds to the case in
  which the \stopone~decouples from the $Z^0$ ($\theta_{\stopone}=$ 0.98
  rad).}   
\label{fig:cdf1_stop_ljmet_3bod_excl}
\end{figure}

For the $\stopone~\rightarrow b l^+\tilde{\nu}$ decay scenario, sensitivity
to signal was greatest for a fit to the $H_T$ distribution.  Again, all fit
results were consistent with zero signal events.  The 95\% C.L. limits on
$\sigma_{\stopone\stoponeb}$ for the $\stopone~\rightarrow b l^+ \tilde{\nu}$
decay were generated for $M(\tilde{\nu})$ between 40 and 50 \mgev, and the
resulting region in the $M(\tilde{\nu})$-$M(\stopone)$ plane 
for which the limit on $\sigma_{\stopone\stoponeb}$ is less than the NLO
prediction ($\mu = M(\stopone)$) is shown in
Fig.~\ref{fig:cdf1_stop_ljmet_3bod_excl}.

\subsection{Summary}

The CDF and D\O\ collaborations have actively carried out
analyses of various SUSY signatures in the MSSM or mSUGRA framework.
We describe the most up-to-date analyses.
A summary of the SUSY mass limits is given in Table \ref{tab:susy_mass_run1}.
There are several analyses which are in progress and
are not included in this paper.
The status of
the SUSY analyses using the Run Ia+Ib data is summarized
in Table \ref{tab:susy_summary_run1}.

\begin{table}[ht]
\vspace{0.2in}
\caption{Summary of 95\% C.L. lower mass limits on SUSY particles
in mSUGRA or MSSM framework. 
The strongest limits from CDF and D\O\ are listed. }
\label{tab:susy_mass_run1}
\begin{center}
\begin{tabular}{l l r r c c c c c }
%\hline
Decay & \multicolumn{3}{c}{Mass Limit (\mgev)} & 
		$m_{1/2}$ & $m_0$ & $\tanb$ & 
		$sgn(\mu)$ & $A_0$ \\
\hline \hline
{\bf mSUGRA} &&& &  & & & &  \\
$\schionepm\schitwozero\to 3\ell+X$ & $\schionep$ & 
		~62 & (106 \invpb) & 50 & 160  & 2 & $-1$ & 0 \\
$\gluino\gluino,\gluino\squark,\squark\squarkb\to jets+\met$ & $\gluino$   & 
		300 &(~72 \invpb) & 105 & 0   & 2 & $-1$ & 0 \\
$\gluino\gluino,\gluino\squark,\squark\squarkb\to 
		2\ell+jets+\met$ & $\gluino$     	    & 
		270 &(108 \invpb) & 89  & 150 & 2 & $-1$ & 0 \\
$\gluino\gluino,\gluino\squark,\squark\squarkb\to 2\ell+jets+\met$ & $\gluino$ 
	 (heavy $\tilde{q}$) & 169 &(106 \invpb) & 50  & 500 & 2 & $-1$ & 0 \\
$\sbottomone\to b + \lsp$ 
	& $\sbottomone$ & --- &(~88 \invpb)  &  
	\multicolumn{5}{c}{not available in mSUGRA} \\
$\stopone\to c + \lsp$ & $\stopone$  &   ~90 &(~88 \invpb)  &  
	125 & 130 & 7 & +1 & $-535$ \\
\hline
{\bf MSSM} &&& &  & & & &  \\
$\sbottomone\to b + \lsp (40\ \mgev)$ 
	& $\sbottomone$ & 140 &(~88 \invpb)  &  & & & &  \\
$\stopone\to c + \lsp (40\ \mgev)$ & $\stopone$  &   119 &(~88 \invpb)  &  
	& & & & \\
$\stopone\to b + l + \lsp (40\ \mgev)$ & $\stopone$  & 125 &(~88 \invpb)  &  
	& & & & \\
\hline
\end{tabular}
\end{center}
\end{table}

\begin{table}
\caption{Status (as of June 1999)
	of SUSY analyses in MSSM or SUGRA framework at CDF and D\O\
	using the Run Ib (or Run Ia+Ib) data. }
\label{tab:susy_summary_run1}
\begin{center}
\begin{tabular}{ c  l  c c   }
%\hline 
Prod. & \multicolumn{1}{c}{SUGRA Signature} & CDF & D\O \\
\hline
\hline
 & $(\ell^\pm \nu \; \chizero) \; 
	(\ell^+ \ell^- \; \chizero)$ & \cite{cdf_tril} & \cite{d0_tril} \\
$\schionepm  \schitwozero$  & $(\ell^\pm \nu \; \chizero) \;
	 (\tau^+ \tau^- \; \chizero)$ & Progress & \\ 
 & $(\tau^\pm \nu \; \chizero) \; 
	(\ell^+ \ell^- \; \chizero)$ & Progress & \\
\hline
\mc{4}{l}{\sp} \\
%\hline
$\gluino \gluino, \squark \gluino, \squark \bar{\squark}$
	& \met$+\ge$3,4 jets &  Progress & \cite{d0jmet1b} \\ 
%\hline
\mc{4}{l}{\sp} \\
\hline
$\gluino \gluino$ & $(q \bar{q}^{\prime} \chione) \; 
		(q \bar{q}^{\prime} \chione)
           \rightarrow \ell^\pm \ell^\pm$ + jets + $\MET$
				& \cite{cdf1_2ljmet} & \cite{d0dilep} \\ 
	& $(\bar{b} \sbottom) \; 
		(\bar{b} \sbottom) \to
	   (\bar{b} b \schionezero) (\bar{b} b \schionezero)$
					& Progress &  \\ 
\hline
\mc{4}{l}{\sp} \\
%\hline
$\sbottomone\sbottomone$ & $(b \schionezero) \; 
			(\bar{b} \schionezero)$   & 
		\cite{cdf1_stopsbot} & \cite{sbottom_d0} \\
%\hline
\mc{4}{l}{\sp} \\
\hline
	& $(c \schionezero) \; 
		(\bar{c} \schionezero)$  & \cite{cdf1_stopsbot} & \\ 
$\stopone\stopone$         & $(b \schionep) \; 
			(\bar{b} \schionem) \to
                        (b \ell^+ \nu \schionezero) \;
                        (\bar{b} q \bar{q^\prime} \schionezero)$ & 
			\cite{cdf1_stop_ljmet} & \\ 
	& $(b \ell \snu) \; 
		(\bar{b} \ell \snu) \to 
		(b \ell \nu \schionezero) \; (\bar{b} \ell \nu \schionezero)$
		 & \cite{cdf1_stop_ljmet2} & \\ 
	& $(b \schionep)\; 
			(\bar{b} \schionem) \to
                        (b \ell^+ \nu \schionezero) \;
                        (\bar{b} \ell^- \nu \schionezero)$  & 
			Progress & \\ 
\hline
\end{tabular}
\end{center}
\end{table}

The experience from Run I analyses will greatly help
us to design new triggers 
for previously inaccessible channels,
particularly those involving $\tau$'s and heavy flavor.
This will increase the quality of the Run II searches.

%% file: Szalap/final.tex
\section{Constraints on the mSUGRA Parameter Space from Electroweak Precision Data}

%\author{G.C.~Cho, K.~Hagiwara, C.~Kao and 
%R.~Szalapski\footnote{subgroup convener}}

%%%%%%% ABSTRACT %%%%%%% ABSTRACT %%%%%%% ABSTRACT %%%%%%% ABSTRACT %%%%%

We place constraints on the parameter space of 
the mSUGRA model by studying the loop-level contributions of SUSY particles to
electroweak precision observables.  In general the Higgs bosons and the 
superpartner particles of SUSY models contribute to electroweak observables 
through universal propagator corrections as well as process-specific vertex and
box diagrams.  However, due to the bound on the mass of the lightest chargino, 
$m_{{\tilde{\chi}}_1^\pm} > 91$~GeV, we find that the process-dependent 
contributions to four-fermion amplitudes are negligibly small.  Hence, the full
analysis may be reduced to an analysis of the propagator corrections, and in 
some regions of parameter space the constraints from the $b\rightarrow s\gamma$
process are quite important.  The propagator corrections are dominated by the 
contributions of the scalar fermions, and we summarize the results in the 
Peskin-Takeuchi $S$--$T$ plane and the contributions to the $W$-boson mass, 
$m_W$.  We then present the results in the mSUGRA $m_0$--$m_{1/2}$ plane and 
find that our analysis of the propagator corrections provides constraints in 
the small-$m_0$--small-$m_{1/2}$ region, precisely the region of interest for 
collider phenomenology.  In some regions of parameter space, especially for 
$\mu < 0$ and large $\tan\beta$, the constrained region is enlarged 
considerably by including the process $b\rightarrow s\gamma$.  The work 
presented here is part of a  larger collaborative effort, and results will be 
presented more completely elsewhere.\cite{chks99}  

%%%%%% SECTION 2 %%%%%% SECTION 2 %%%%%% SECTION 2 %%%%%% SECTION 2 %%%%%

The loop-level contributions of supersymmetric (SUSY) particles to electroweak
observables have been extensively discussed in the 
literature\cite{susy_loop1,susy_loop2,susy_loop3,susy_loop4}.  In particular,
processes with four external light fermions have been studied including 
observables  which are sensitive to the $Zbb$ coupling.  The branching 
fraction ${\rm Br}(B\rightarrow X_s\gamma)$ is sensitive to 
SUSY effects in some regions of parameter space\cite{bsg-theory,more-bsg}.
The relationship between $m_W$ and $m_Z$ will provide stronger constraints
as the measurement of $m_W$ improves.

The complete one-loop corrections to four-fermion amplitudes include the 
universal propagator corrections as well as the process-dependent vertex 
and box corrections.  However, when the extra Higgs bosons and the superpartner
particles become sufficiently massive, it is necessary to retain only the 
leading propagator corrections\cite{susy_loop_heavy}, and these contributions 
may be summarized in terms of the $S$, $T$ and $U$ parameters of Peskin and 
Takeuchi\cite{stu} or some other triplet of parameters\cite{others}.  The 
recent bounds\cite{susy_bound} on the mass of the lightest chargino,  
$m_{\tilde{\chi}^\pm_1} > 91$~GeV, and on the mass of the lighter scalar-top 
quark, $m_{\tilde{t}_1} > 80$~GeV, imply a sufficiently massive spectrum such 
that the process-dependent vertex and box contributions may be safely 
neglected.  In the context of the mSUGRA model, the chargino mass bound alone 
is sufficient to reach this conclusion.

In our analysis we adopt, in the notation of Hagiwara {\em et al.}\cite{hhkm}, 
a form factor, $g_L^b$, to describe corrections to the $Zbb$ vertex as well 
as the $S$ and $T$ parameters, which include corrections to the gauge-boson 
propagators.  We find that it is more convenient to drop the $U$ 
parameter in favor of the directly measured $W$-boson mass.  We first obtain 
constraints from the electroweak data on the four parameters $\Delta g_L^b$, 
$\Delta S$, $\Delta T$ and $\Delta m_W$, which measure deviations from their 
corresponding SM reference values calculated at $m_t = 175$~GeV and 
$m_H = 100$~GeV.  We then calculate the contributions to these parameters 
and to the $B\rightarrow X_s\gamma$ decay width from the superpartner and 
Higgs particles to obtain constraints on the mSUGRA parameters.

The electroweak data through 1998 including the LEP 
and SLC experiments as well as low-energy neutral-current experiments may 
be summarized as 
%%%
\begin{equation}
\left.
\begin{array}{ccc}
\Delta S - 24.2 \Delta g_L^b & = & -0.114 \pm 0.14 \\
\Delta T - 42.9 \Delta g_L^b & = & -0.215 \pm 0.14 
\end{array}
\right\}\makebox[0.4cm]{}
\rho_{\rm corr} = 0.77 \;,
\end{equation}
%%%
where $\rho_{\rm corr}$ denotes the correlation between the two one-sigma 
errors.  Because the correlation is strong, we present our results in the 
$\Delta S^\prime$--$\Delta T^\prime$ plane where $\Delta S^\prime = \Delta 
S - 24.2 \Delta g_L^b$ and $\Delta T^\prime = \Delta T - 42.9 \Delta 
g_L^b$.  
Note that $m_W$ is not correlated with $\Delta S^\prime$ and $\Delta T^\prime$,
and hence it may be treated separately.  Averaging the LEP2 and Tevatron 
measurements of the $W$-boson mass,  $m_W = 80.375 \pm 0.064$~GeV.
The deviation of the data from the SM reference value for the $W$-boson mass is
%%%
\begin{equation}
\Delta m_W = -0.027 \pm 0.064 \rm GeV \;. 
\end{equation}
%%%
For the measurement of the branching fraction for the process $b\rightarrow
s\gamma$ we use
%%%
\begin{equation}
{\rm Br}(B\rightarrow X_s\gamma) = 
3.11 \pm 0.80 \pm 0.72 \times 10^{-4}\;,
\end{equation}
%%%
from the ALEPH\cite{bsg-aleph} collaboration.  Results from the more recent 
CLEO measurement\cite{bsg-cleo} will be reported elsewhere\cite{chks99}.

%%%%%% SECTION 3 %%%%%% SECTION 3 %%%%%% SECTION 3 %%%%%% SECTION 3 %%%%%

\begin{figure}[h]
\centering
\leavevmode\psfig{figure=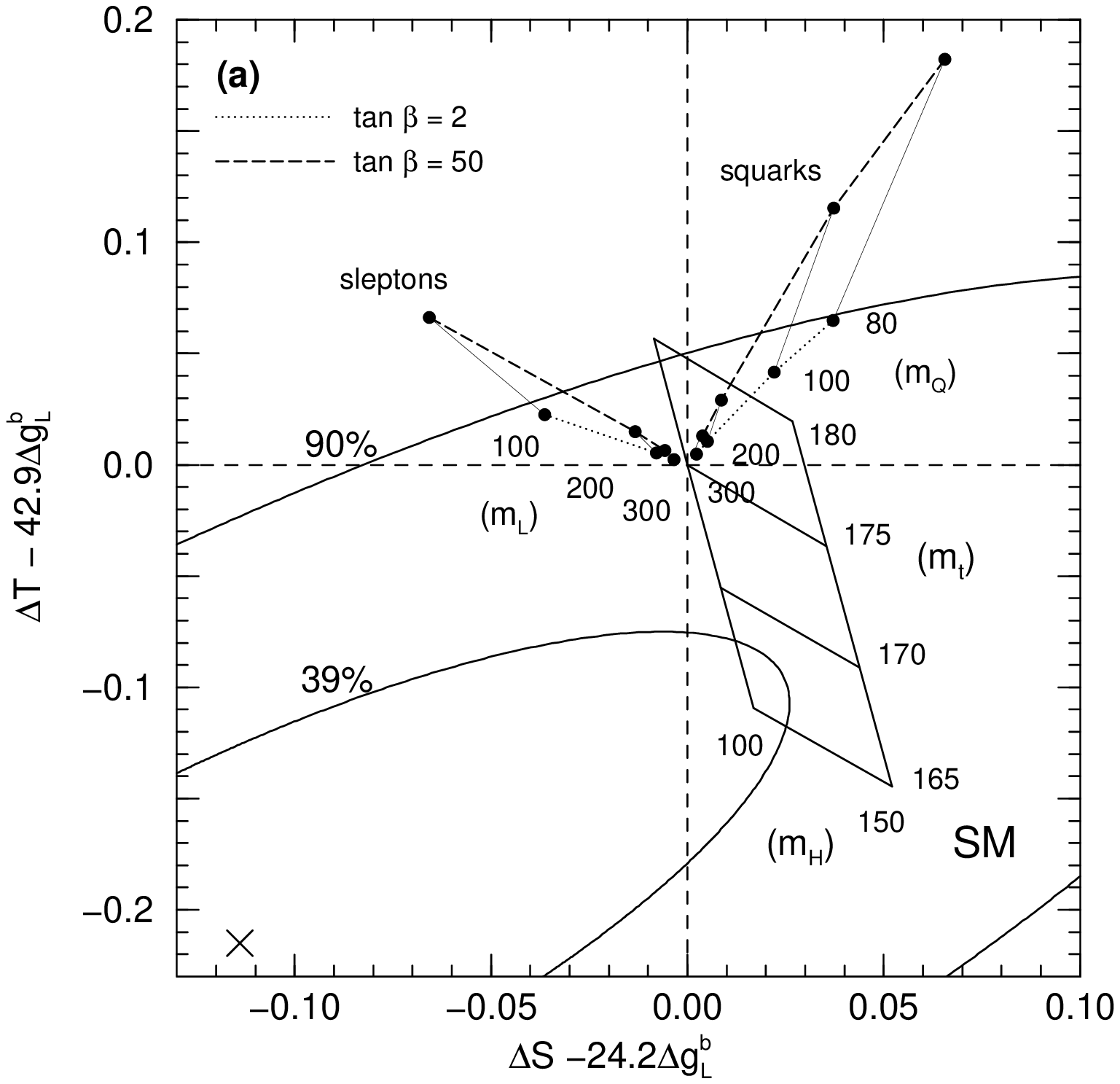,width=6.5cm,silent=0}
\hspace*{0.5cm}
\psfig{figure=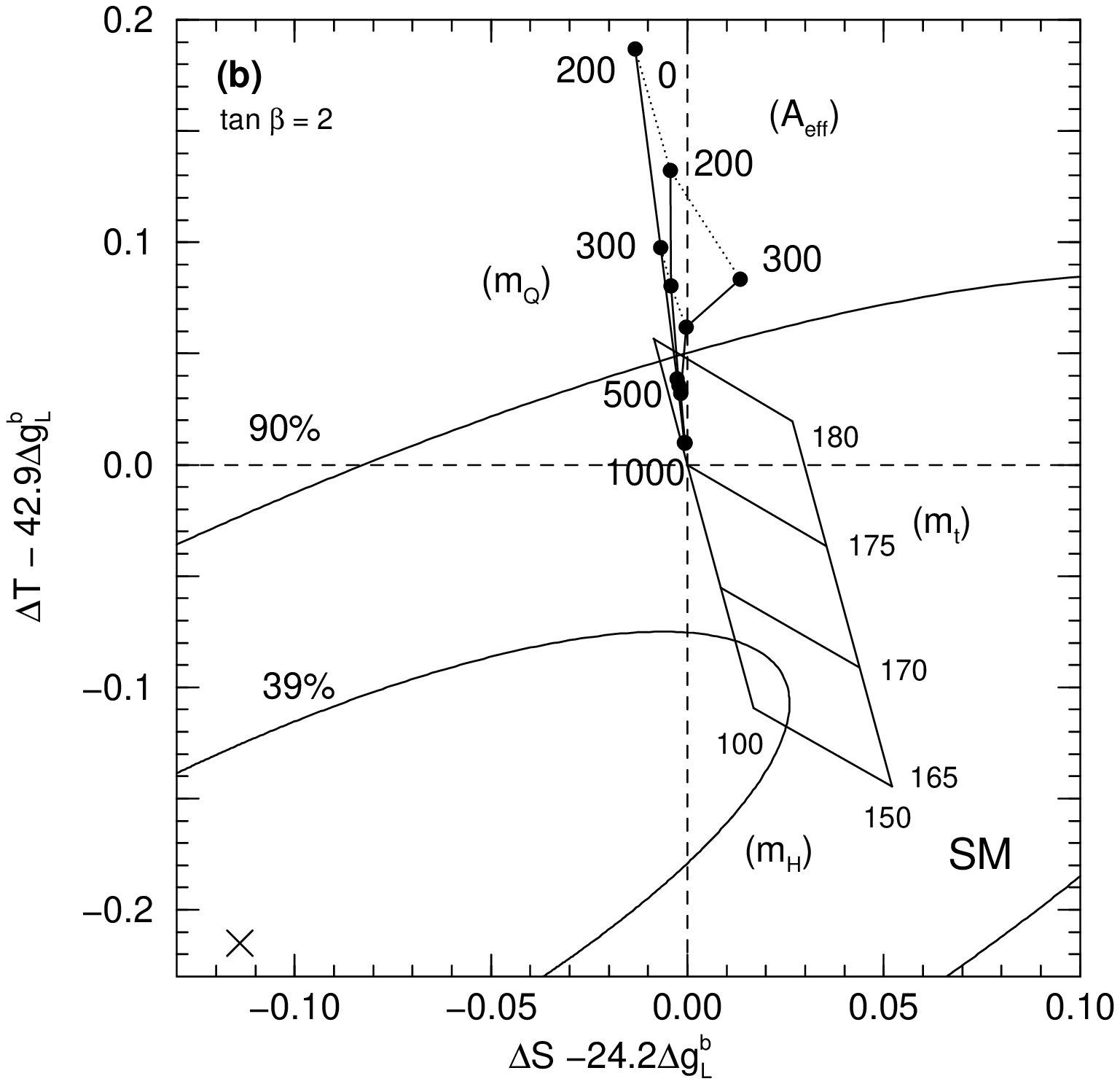,width=6.5cm,silent=0}

\caption{(a) shows the sfermion contributions for the first two families,
and (b) shows the stop-sbottom contributions.  Details are given in the text.}
\label{fig_s-t-plane}
\end{figure}

The SUSY contributions to $\Delta S^\prime$, $\Delta T^\prime$ and 
$\Delta m_W$ are dominated by the contributions of the sfermions.
Hence, we begin with a discussion of the sfermion contributions.
%%%
%%%
In Figure~\ref{fig_s-t-plane}(a) and (b) the `$\times$' marks the location of 
the best fit to the experimental data in the 
$\Delta S^\prime - \Delta T^\prime$
plane, and the ellipses show the 39\% (one-sigma) and 90\% confidence-level 
(CL) contours as
indicated.  A grid has been included which shows the SM predictions for 
$\Delta S^\prime$ and $\Delta T^\prime$ as a function of $m_t$ and $m_H$.
We choose the point where $m_t = 175$~GeV and $m_H = 100$~GeV as our reference
point, {\em i.e.} $\Delta S^\prime = \Delta T^\prime = 0$, and the dashed-line
axes are drawn through this point. The same point serves as the SUSY prediction
in the limit of very large masses for the non-SM particles and when the 
lightest SUSY Higgs particle behaves like the SM Higgs boson.

Figure~\ref{fig_s-t-plane}(a) includes the contributions of the sfermions of 
the first two generations with the squark and slepton contributions 
shown separately.  The contribution of a sfermion loop to the $S$ parameter is
proportional to the hypercharge of the sfermion.  Since $Y = \frac{1}{6}$ for 
the squarks and $Y = -\frac{1}{2}$ for the sleptons, we see that the 
squarks increase $\Delta S^\prime$ while the sleptons decrease 
$\Delta S^\prime$.  Dotted contours are used to show the case where 
$\tan\beta = 2$ while dashed contours are used to show the $\tan\beta = 50$ 
case.  For the slepton contributions, we show the cases where the explicit 
soft-SUSY-breaking slepton-doublet mass parameter has the nonzero values 
$m_L = 100, 200$ and 300~GeV.  Contours of equal $m_L$, but varying 
$\tan\beta$, are drawn using thin solid lines.  Similarly, we consider the 
squark contributions where the explicit soft-SUSY-breaking squark-doublet mass 
parameter has the values $m_Q = 80, 100, 200$ and 300~GeV; contours of 
constant $m_Q$, but varying $\tan\beta$, are indicated by the thin solid lines.
While the contributions to $\Delta S^\prime$ tend to cancel between the 
squark and slepton sectors, the contributions to $\Delta T^\prime$ always 
add constructively, and for light sfermions lead to an unacceptably large
deviation from the SM prediction and the experimental measurement of  
$\Delta T^\prime$.

The large mass of the top quark leads to large left-right mixing of the 
top squarks, and to a lesser degree the mass of the bottom quark leads to 
left-right mixing of the bottom squarks.  For this reason the third-family
sfermions require a separate discussion, and we summarize the stop--sbottom
contributions in Figure~\ref{fig_s-t-plane}(b).  In the mass matrix for the 
stop squarks it is the off-diagonal element $-m_t A_{\rm eff}^t$ where 
$A_{\rm eff}^t = A_t + \mu\cot\beta$ that determines the level of left-right 
mixing, while in the sbottom-squark mass matrix the off-diagonal element 
$-m_b A_{\rm eff}^b$ where $A_{\rm eff}^b = A_b + \mu\tan\beta$ determines
the degree of mixing.  We plot our results for $A_{\rm eff}^t = A_{\rm eff}^b
= A_{\rm eff}$ showing contours of constant $A_{\rm eff}$ by the dashed lines
and lines of constant $m_Q$ by the dotted lines.  In 
Figure~\ref{fig_s-t-plane}(a) we saw that, with a value as small as 
$m_Q = 80$~GeV, the contributions of the squarks of the first two generations 
to $\Delta T^\prime$ are still fairly small, while for the third family a 
value of $m_Q = 300$~GeV already produces an unacceptable result for 
reasonable  values of $A_{\rm eff}$.  It may be tempting to abandon
universality of the soft-SUSY-breaking parameters and consider cases with a
relatively small value of $m_Q$ for the first two families and a much larger 
value to decouple the third family.  While this is possible in principle, 
caution is required to avoid large flavor-changing neutral currents.  In the 
context of the mSUGRA model we will, of course, use the soft-SUSY-breaking 
parameters which are obtained from the common mass parameters at the GUT scale.
We also note that large values of $A_{\rm eff}$ tend to produce smaller 
$\Delta T^\prime$ but larger $\Delta S^\prime$.  We have shown only the case 
$\tan\beta = 2$ since we find similar results for large $\tan\beta$.

%%%%%% SECTION 4 %%%%%% SECTION 4 %%%%%% SECTION 4 %%%%%% SECTION 4 %%%%%

\begin{figure}[htbp]
\begin{center}
\leavevmode\psfig{figure=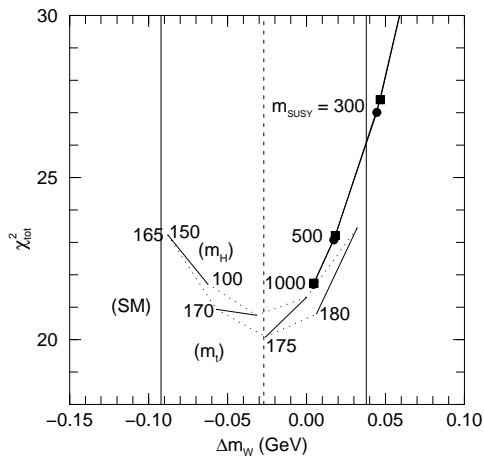,width=6.5cm,silent=0}
\end{center}
\caption{The sfermion contributions in the $\chi^2_{\rm tot} - \Delta m_W$
plane where $\chi^2_{\rm tot}$ refers to the total $\chi^2$ coming from the 
simultaneous fitting of $\Delta S^\prime$, $\Delta T^\prime$ and $\Delta m_W$.}
\label{fig_chi2-mw-plane}
\end{figure}

Figure~\ref{fig_chi2-mw-plane} shows the sfermion contributions to the 
$W$-boson mass.  We include a grid that shows the SM prediction for 
$\Delta m_W$ as a function of $m_H$ and $m_t$.  Along the upper dotted contour 
$m_H = 100$~GeV, while the lower dotted contour corresponds to $m_H = 150$~GeV.
Points of equal $m_t$ are connected by the solid line segments.  The vertical 
dashed line represents the world average for the central value of the $m_W$ 
measurement with the one-sigma errors represented by the vertical solid lines. 
For simplicity we set the explicit soft-SUSY-breaking squark-doublet, 
squark-singlet, slepton-doublet and slepton-singlet mass parameters to a 
common value, $m_{\rm SUSY}$.  We then plot the total chi-squared from the 
simultaneous fitting of $\Delta S^\prime$, $\Delta T^\prime$ and $\Delta m_W$, 
{\em i.e.} $\chi^2_{\rm tot}$, {\em versus} $\Delta m_W$ for $\tan\beta = 2$ 
(represented by the squares) and $\tan\beta = 50$ (represented by the circles).
For $m_{\rm SUSY} = 1000$~GeV the $\tan\beta = 2$ and $\tan\beta = 50$ points 
are nearly indistinguishable.  We note that the contributions of the SUSY 
particles always increase $m_W$.  However, a value of $m_{\rm SUSY} = 300$~GeV 
leads to only a one-sigma discrepancy with the data.  Hence, at the current 
time, the measurement of the $W$-boson mass provides only a minor constraint. 

%%%%%% SECTION 5 %%%%%% SECTION 5 %%%%%% SECTION 5 %%%%%% SECTION 5 %%%%%

Although the Higgs bosons, the charginos and the neutralinos also contribute 
to $\Delta S^\prime$, $\Delta T^\prime$ and $\Delta m_W$, in the mSUGRA model
the contributions are small compared to the sfermion contributions.  Hence, 
even though we include these contributions in the numerical analysis, we 
do not show the Higgs-boson, chargino and neutralino figures that correspond 
to Figure~\ref{fig_s-t-plane} and Figure~\ref{fig_chi2-mw-plane}.

%%%%%% SECTION 6 %%%%%% SECTION 6 %%%%%% SECTION 6 %%%%%% SECTION 6 %%%%%

%%%
\begin{figure}[t]
\centering\leavevmode
\mbox{\psfig{figure=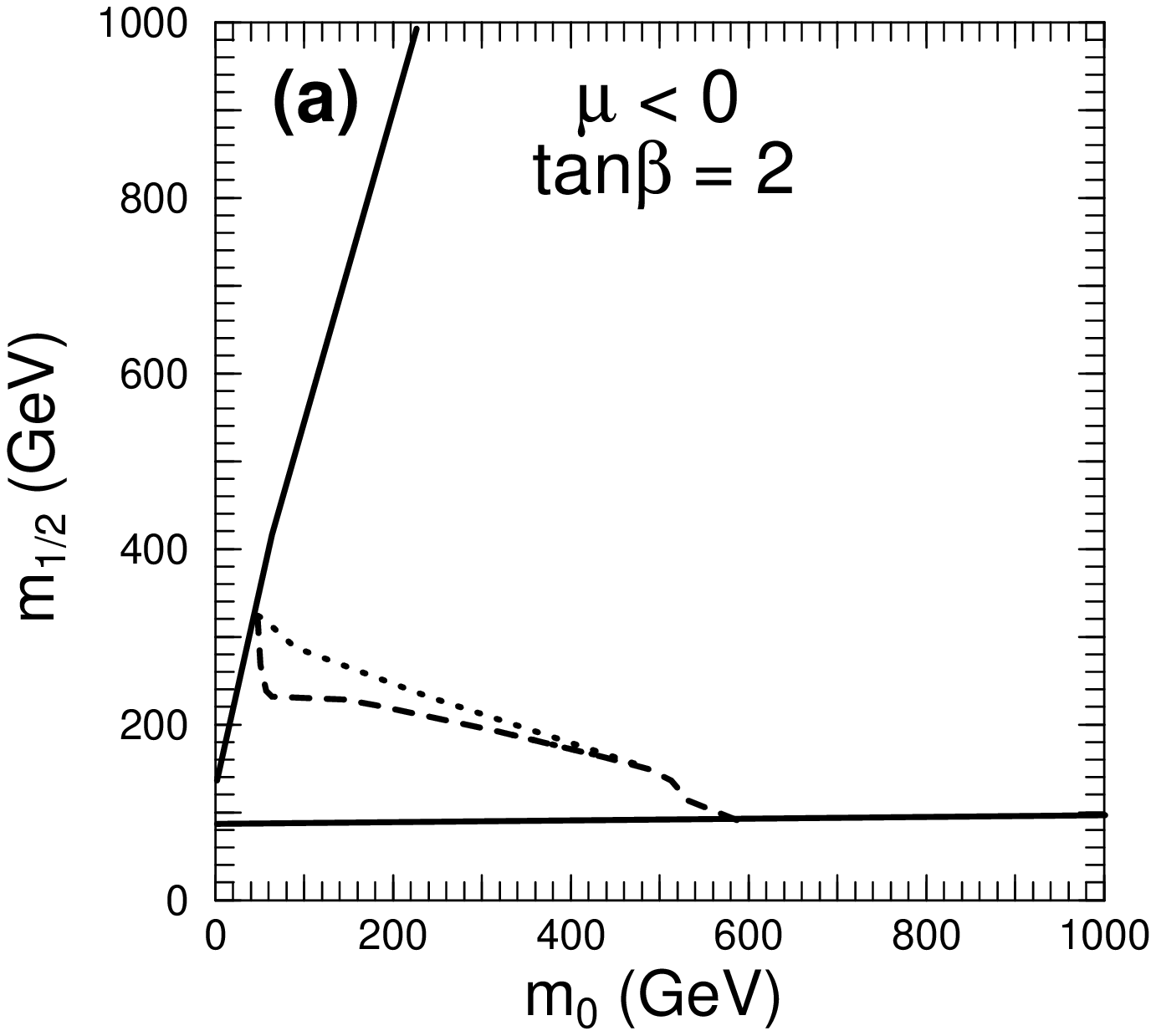,width=5.5cm,silent=0}
%\hspace*{0.2cm}
\psfig{figure=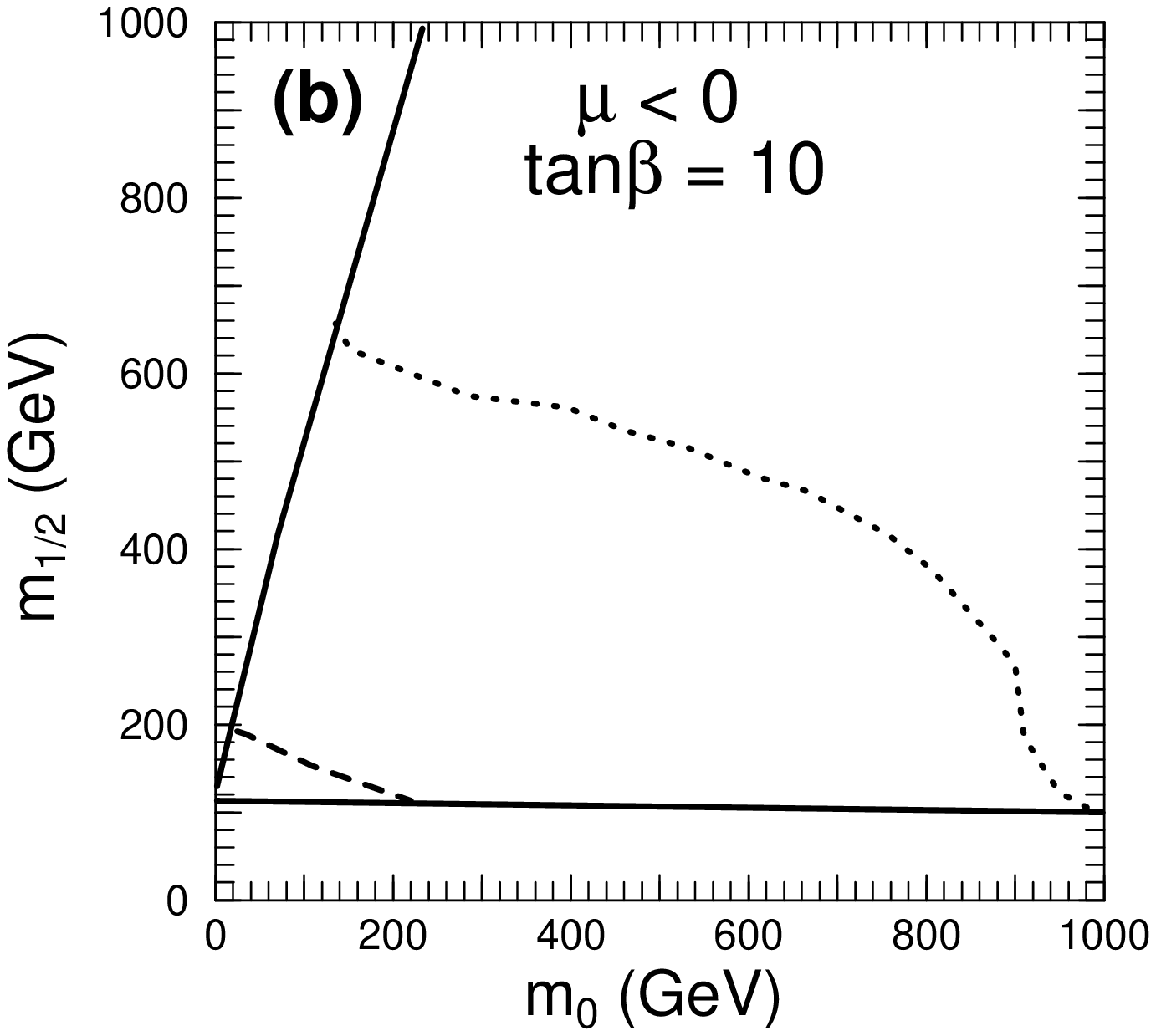,width=5.5cm,silent=0}
%\hspace*{0.2cm}
\psfig{figure=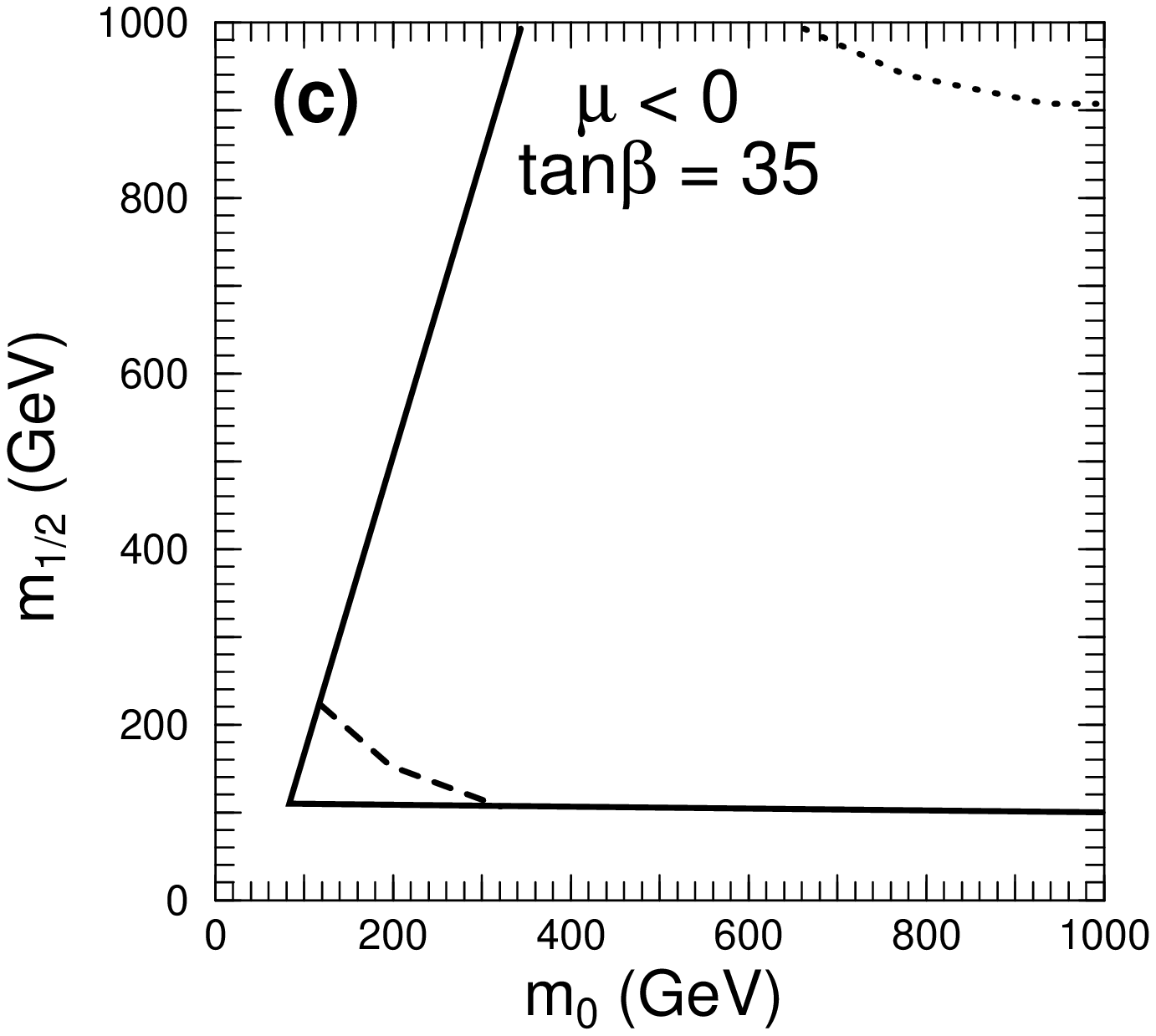,width=5.5cm,silent=0}}

\vspace*{0.5cm}
\centering\leavevmode
\mbox{\psfig{figure=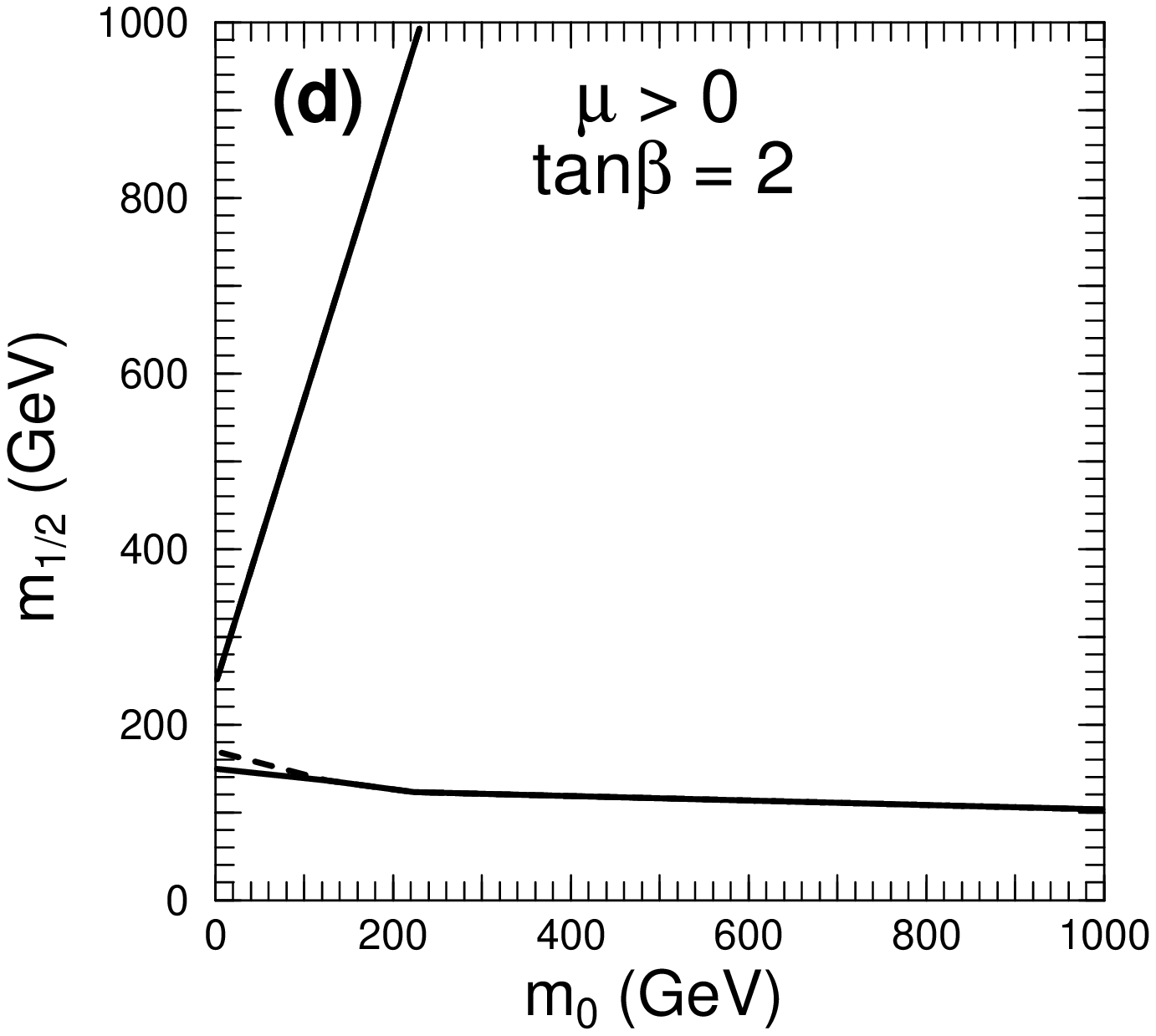,width=5.5cm,silent=0}
%\hspace*{0.2cm}
\psfig{figure=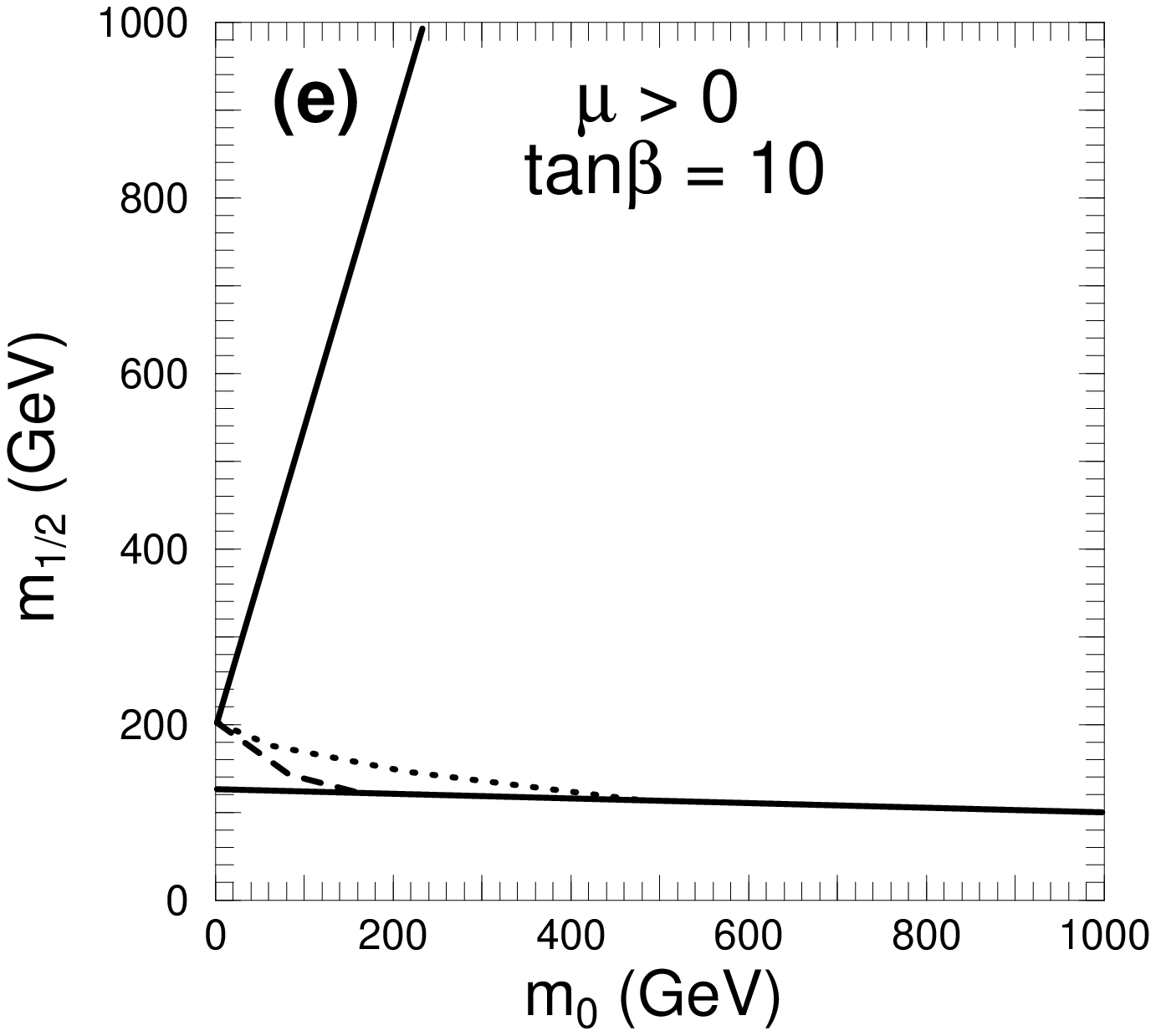,width=5.5cm,silent=0}
%\hspace*{0.2cm}
\psfig{figure=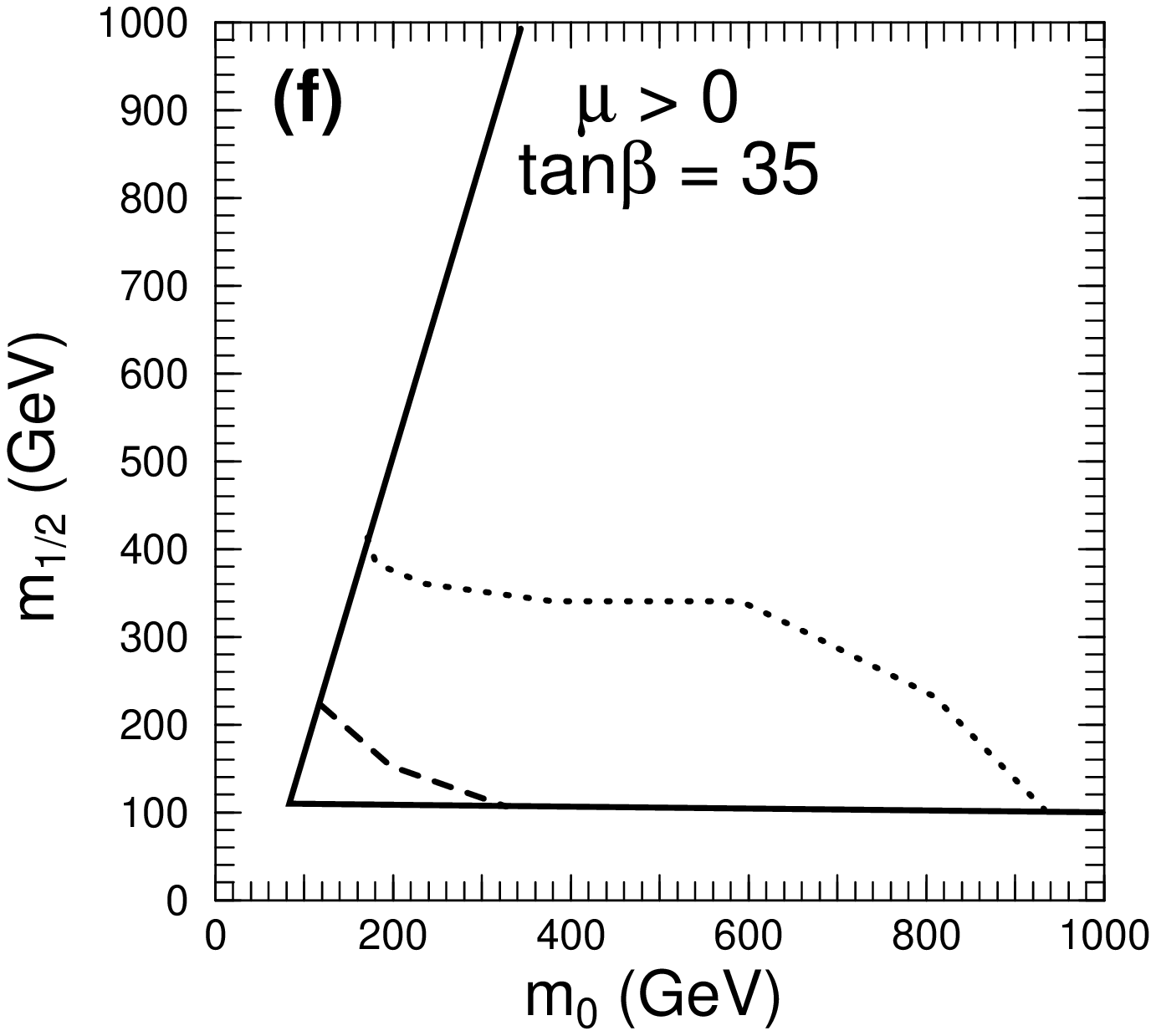,width=5.5cm,silent=0}}

\caption{Favored regions in the mSUGRA $m_0$--$m_{1/2}$  plane lie in the 
region which is above and to the right of all drawn contours.  Further
explanation is provided in the text.}
\label{fig_m0-m1/2}
\end{figure}
%%%
Next we discuss Figures~\ref{fig_m0-m1/2}(a)--(f).  In each of these figures 
the values for $\tan\beta$ and ${\rm sign}(\mu)$ are held to the constant
values indicated.  We allow
$A_0$ to vary in the range $-$500~GeV $ < A_0 < $ 500~GeV, and we scan the 
$m_0$--$m_{{1/2}}$ plane between 0~GeV and 1~TeV.  For each point in the
five-dimensional parameter space of unification-scale input parameters we 
employ the mSUGRA RGE portion of ISAJET\cite{isajet7.40} to determine the 
RGE evolution to the electroweak scale.  We then verify whether that point is 
either excluded or allowed according to the following tests:
%%%
\begin{enumerate}
\item{Verify that the obtained particle spectrum is physical, that the correct 
vacuum for electroweak symmetry breaking is obtained and that the lightest 
superpartner particle is a neutralino, {\em i.e.} $\widetilde{\chi}^0_1$.  
This leads to a disallowed region in the upper left corner of each of the 
figures extending to the solid line with positive slope.\label{test-phys}}
\item{Verify that the chargino mass bound, $m_{\tilde{\chi}^\pm_1} > 
91$~GeV, is satisfied.  We find that region below the horizontal solid line is 
excluded.\label{test-mass}}
\item{Calculate $\Delta S^\prime$, $\Delta T^\prime$ and $\Delta m_W$ and 
check $\chi^2_{\rm tot}$.  Points which are disallowed at the 95\%~CL extend
the disallowed region in the $m_0$--$m_{{1/2}}$ plane from the solid
contour to the dashed contour.\label{test-stu}}
\item{Calculate the contribution to ${\rm Br}(B\rightarrow X_s\gamma)$.
Points which are disallowed at the 95\%~CL extend the disallowed region of the 
$m_0$--$m_{{1/2}}$ plane from the dashed contour up to the dotted 
contour.\label{test-bsg}}
\end{enumerate}
%%%
The portion of the $m_0$--$m_{{1/2}}$ plane which is above and to 
the right of all the contours is deemed the `favored' region for the mSUGRA 
model.  The portion of the $m_0$--$m_{{1/2}}$ plane which is excluded by 
Test~\ref{test-mass}, the chargino mass bound, is significant.  Once this has 
been taken into account, Test~\ref{test-stu} excludes a corner of the remaining
$m_0$--$m_{{1/2}}$ plane corresponding to small values of $m_0$ and 
$m_{{1/2}}$.  This region is fairly large in Figure~\ref{fig_m0-m1/2}(a)
while it is barely observable in Figure~\ref{fig_m0-m1/2}(d).  When 
${\rm sign}(\mu) < 0$, when $\tan\beta$ is large, and especially when both of these
conditions are true Test~\ref{test-bsg} excludes a significant region of the 
parameter space.  In Figure~\ref{fig_m0-m1/2}(c) all but a tiny portion of the
figure has been disallowed.  Our excluded regions from Test~\ref{test-bsg} 
are larger than those of Ref.~\cite{more-bsg} due to a different treatment of 
strong corrections.

%%%%%% SECTION 7 %%%%%% SECTION 7 %%%%%% SECTION 7 %%%%%% SECTION 7 %%%%%

In conclusion, the direct constraints which come from the nonobservation of 
the lightest chargino at LEP2 have important consequences.  First of all, the 
process dependent vertex and box corrections to four-fermion amplitudes become
negligibly small, and as a result the analysis of electroweak data has been 
simplified and has become more transparent.  After taking into account the 
chargino mass bound the $Z$-pole data, the low-energy neutral-current data and 
the measurement of the $W$-boson mass exclude only a small portion of the 
$m_0$--$m_{{1/2}}$ plane.  However, this is still significant because 
the excluded region is where $m_0$ and $m_{{1/2}}$ are small, precisely
the region of interest for collider studies, and especially relevant for the 
Tevatron.  We find that the excluded region is largest for smaller $\tan\beta$
with ${\rm sign}(\mu) < 0$.  For ${\rm sign}(\mu) < 0$ or $\tan\beta$ large,
a significant portion of the $m_0$--$m_{{1/2}}$ plane is excluded by the 
${\rm Br}(B\rightarrow X_s\gamma)$ measurement, and the constraint becomes 
very severe when both of these conditions are met.

%%%%% ACKNOWLEDGEMENTS %%%%% ACKNOWLEDGEMENTS %%%%% ACKNOWLEDGEMENTS %%%%
%% deleted
%%%%%%%% SECTION  %%%%%%%%%%% BIBLIOGRAPHY  %%%%%%%%%%% SECTION  %%%%%%%%

%% file: Falk-cosmo/cosmonew.tex
\def\ga{\mathrel{\raise.3ex\hbox{$>$\kern-.75em\lower1ex\hbox{$\sim$}}}}
\def\la{\mathrel{\raise.3ex\hbox{$<$\kern-.75em\lower1ex\hbox{$\sim$}}}}
\def\gev{{\rm \, Ge\kern-0.125em V}}
\def\tev{{\rm \, Te\kern-0.125em V}}
\def\mchi{m_{\chi}}
\def\ohsq{\Omega_{\chi} h^2}
\def\m12{m_{1\!/2}}
\def\st{{\widetilde \tau}_{\scriptscriptstyle\rm R}}

\section{Cosmological Constraints}

An appealing feature of the mSUGRA model  is that  it naturally
provides a dark matter candidate, since the LSP typically has a
large and cosmologically interesting relic density\cite{rd}.   However,  this
same feature makes mSUGRA susceptible to cosmological constraints.  

In particular, the age of the universe provides an observational upper
limit on the neutralino relic density $\ohsq$.  Specifically, the
constraints $t_0 > 12{\rm \, G\kern-0.125em yr}$ and
$\Omega_{\chi}\le1$ together require that $\ohsq < 0.3$.  A non-zero
cosmological constant does not loosen this bound, provided that
$\Omega_{\chi} + \Omega_{\Lambda} \le 1$.  

\begin{figure}[t]
\vspace*{-.8in}

\centering\leavevmode
\epsfig{file=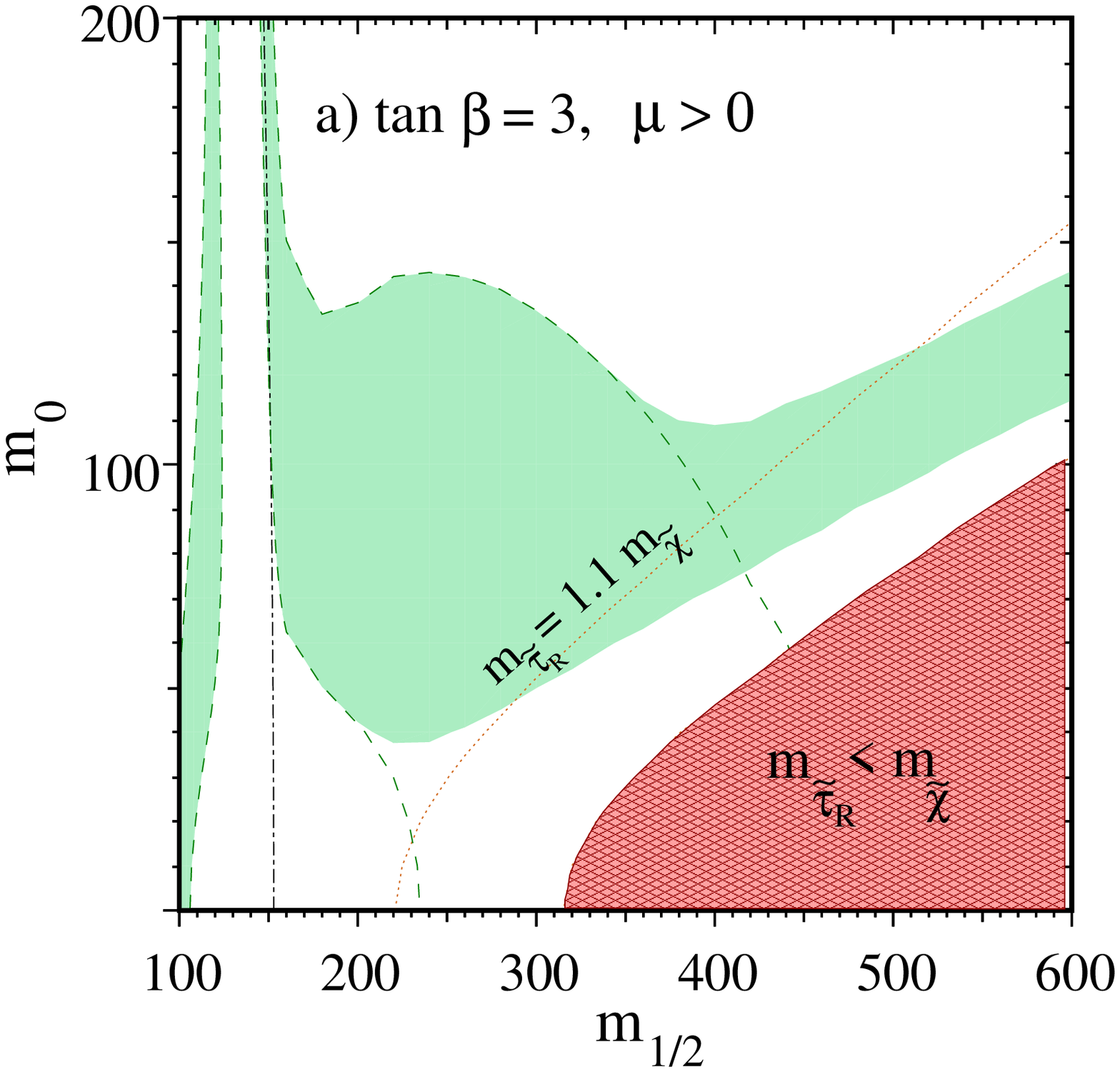,height=3.5in}
%\vspace*{-1.5in}
%\centering\leavevmode
\hspace*{.25in}
\epsfig{file=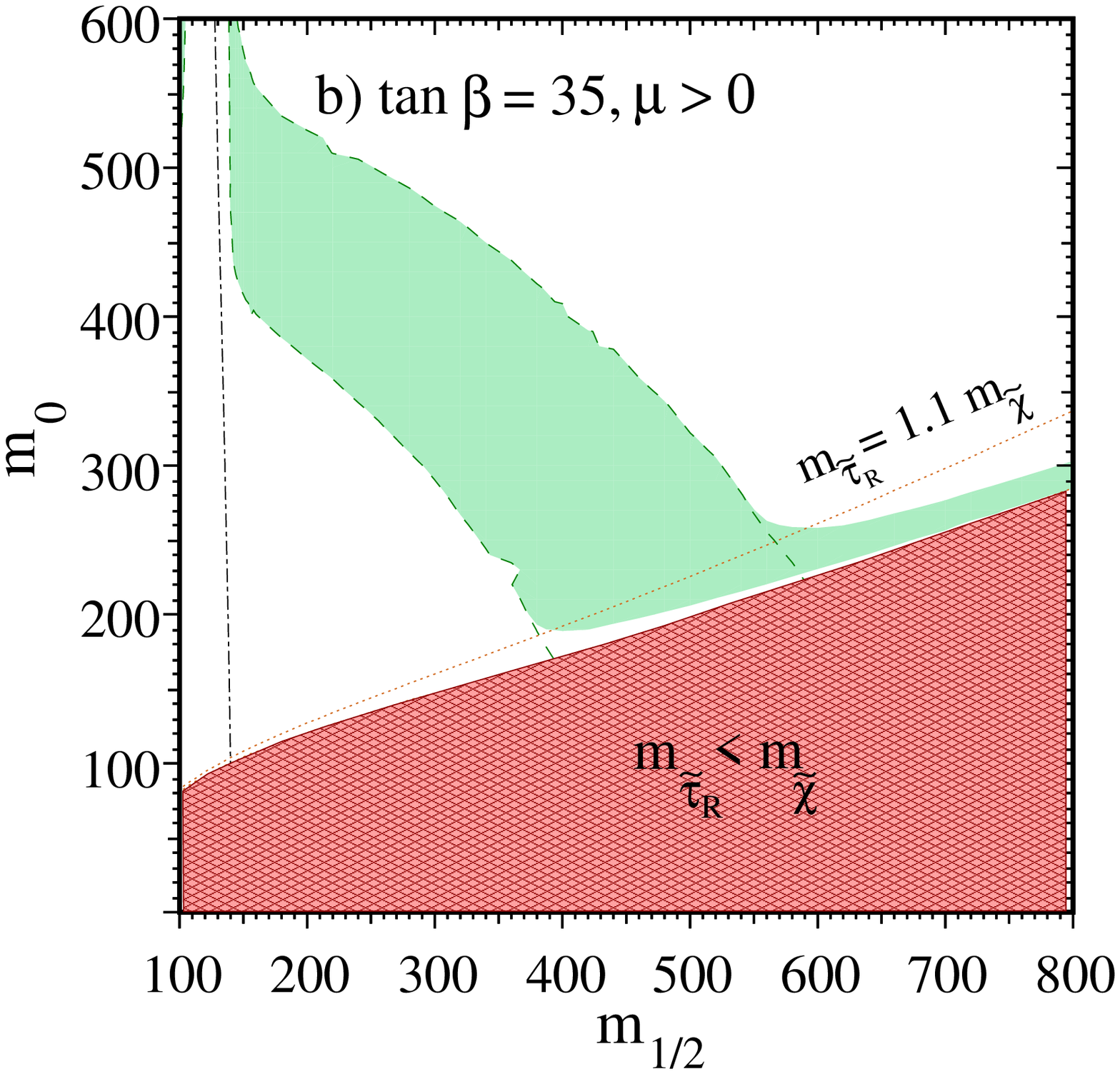,height=3.5in}
\vspace*{-0.35in}
\caption{
  The light-shaded area is the cosmologically preferred 
region with \mbox{$0.1\leq\ohsq\leq 0.3$}, for a)~$\tan\beta=3$ and b)~$\tan\beta=35$.   
The dashed line shows the location  of the
cosmologically preferred region  if one ignores the light sleptons.  
In the dark shaded
regions, the LSP is the ${\tilde
\tau}_R$, leading to an unacceptable abundance
of charged dark matter.  Also shown as a dotted line is the
contour $m_{\st}=1.1 \,\mchi$. The
  dot-dashed lines are contours of $m_{\chi^\pm}=91\gev$.   { \label{fig:rd2}}
    }

\end{figure}

The relic abundance of neutralinos is inversely proportional to the
thermallly averaged neutralino annihilation cross-section
$\langle\sigma_{\rm ann} v\rangle$.  
For gaugino-type neutralinos, typically 
annihilation is primarily into fermion pairs via sfermion exchange.
The age constraint then translates into an upper bound on the sfermion
masses, and hence on $m_0$ and $\m12$.  There is, however, a stripe
cutting through the mSUGRA parameter space where the mass of the
neutralino is close to the mass of the right-handed stau, and in this region
coannihilation between the neutralino and $\tilde\tau_R$ (and possibly the other
$\tilde\ell_R$) can greatly reduce the relic density \cite{efo}. 
Fig.~(\ref{fig:rd2}a) displays
the neutralino relic density in the $\{\m12,m_0\}$
plane for $\tan\beta=3$, with $\mu>0$ and $A_0=0$.  The shaded area is
the cosmologically preferred region with $0.1\le\ohsq\le0.3$; the
region above the shaded area is excluded by the constraint $t_0>
12{\rm \, G\kern-0.125em yr}$.   The dashed lines show the corresponding
preferred region if one ignores $\chi-\tilde\ell_R$ coannihilation.
The shape of the shaded region is
insensitive to $\tan\beta$, for small to moderate $\tan\beta$, and 
results are similar for $\mu<0$ and for other moderate
values of $A_0$, although very large $A_0$ can affect the position of the 
$m_{\st}=\mchi$ contour.

Although the lightest neutralino is predominantly gaugino everywhere
in this figure, it does have a small Higgsino admixture.  Annihilation
through a $Z^0$ or Higgs in the s-channel is then possible, and if
the mass of the neutralino is close to half the $Z^0$ or Higgs mass,
this process is enhanced and can dominate the sfermion exchange; this
dramatically reduces the relic abundance, so that $\ohsq$ is
sufficiently low, independent of the sfermion mass, and hence $m_0$.
The Higgs and $Z^0$ poles are visible in Fig.~(\ref{fig:rd2}a) at
values of $\m12$ between $120\gev\la\m12\la 150\gev$.  However, the
pole region is excluded by the LEP chargino searches, which have
constrained $m_{\chi^\pm}>91\gev$.  The Higgs pole does move to the
right slowly for larger values of $\tan\beta$, exposing a sliver of
the pole region for $\tan\beta>4$.  The upcoming runs at LEP2 will
substantially close this loophole, though some narrow regions near the pole will
survive at moderate to large $\tan\beta$.   It is a general constraint, then,
that except for in the close vicinity of a pole or near $m_{\st}=\mchi$, 
one is restricted to $m_0<150\gev$ and $\m12<450\gev$ for
$\tan\beta\la 7$.   Below the contour $m_{\st}=\mchi$, there is an
unacceptable abundance of charged dark matter, and the
$m_{\st}=\mchi$ and $\ohsq=0.3$ contours cross at 
$\m12\sim1500\gev$.    Therefore there are overall cosmological constraints
$m_0<350\gev$ and $\m12<1500\gev$ at low to moderate $\tan\beta$.

While the constraint $\ohsq<0.3$ is a genuine observational bound, the
lower limit on $\ohsq$ comes from the desire that SUSY provide some or
all of the dark matter in galaxies and on larger scales where the
presence of additional  missing mass is inferred.  The lower limit
$\ohsq>0.1$ represents a ``cosmologically interesting'' relic
abundance and also indicates how $\ohsq$ varies as $m_0$ is decreased.
However, the actual amount of neutralino dark matter is quite unknown,
may well be zero, and so doesn't constitute a true bound in the same
sense as the upper limit.

For large $\tan\beta$, the Higgsino admixture in LSP is no longer
negligible for the purposes of computing the relic density.
The presence of additional annihilation channels, e.g. through the pseudoscalar
Higgs,  allows for smaller annihilation rates through sfermions and
hence permits larger sfermion masses.  Fig.~(\ref{fig:rd2}b) displays
the cosmologically preferred region for $\tan\beta=35$, with $\mu>0$.  Note the
increased scales in both $\m12$ and $m_0$ over Fig.~(\ref{fig:rd2}a).  For values of
$\tan\beta\ga 40$,  the presence of annihilation poles into the
pseudo-scalar Higgs and heavy Higgs can dramatically reduce the
neutralino relic abundance~\cite{bk}.

As a first step away from the parameter restrictions of mSUGRA, one
may relax the Higgs soft mass$^2$ unification constraint, so that
$\mu$ and $m_A$ are no longer determined by the conditions of
electroweak symmetry breaking.  This allows for $M_2\gg\mu$, so that
the lightest neutralino is Higgsino-like.  In this case the neutralino
can annihilate through $Z^0$ and Higgs exchange, and annihilation is
typically quite efficient.  Additionally, in the pure Higgsino region,
the mass of the lightest neutralino is close to the masses of the
next-to-lightest neutralino and lightest chargino, and the
coannihilation of the LSP with the NLSP and NNLSP can play an
important r\^ole in further reducing the neutralino relic abundance.
Searches for chargino and associated neutralino production at LEP2
have already excluded the regions of parameter space where the relic
density of light Higgsinos (i.e. with masses $\la 500\gev$)
 is greater than 0.3~\cite{efgos}.  In the pure gaugino
regime $\mu\gg M_2$, the bounds are similar to the mSUGRA case.

As one wanders farther from the constraints of mSUGRA, the neutralino
relic density will depend on new CP-violating phases in $\mu$ and
$A$\cite{fos}, the conditions of gaugino and scalar mass
unification~\cite{an,dnry}, etc., and the relic
density will then have to be considered separately for each case.

%--------------------------------------------------

%% file: Arnowitt-cdm/cdm.tex
\section{Cold Dark Matter Searches}
%\noindent \it Arnowitt, Baer, Barger, Falk, Kao, Li, Nath}

\subsection{Introduction}

Supergravity GUT models with $R$-parity invariance possess a cold dark matter candidate: the lightest neutralino ($\chi_1^0$) \cite{1}. Relic density  calculations indicate that the predicted  amount of relic $\chi_1^0$ is in accord with astronomical measurements of cold dark matter, i.e. $0.05\le \Omega_{\chi_1^0}h^2 \le 0.30$, for a significant part of the SUSY parameter space for mSUGRA models. Measurements of rotation curves of a large number of spiral galaxies (including the Milky Way) indicate that their halos are composed mostly of dark matter. For the Milky Way, the local density of dark matter (DM) is estimated to be $\rho_{DM}= 0.3\ GeV/cm^3$. (This number may be in error by a factor of two, e.g.\ if the halo is flattened it would be larger, while if part of the halo is baryonic machos, the SUSY component would be smaller). These dark matter particles are incident on the solar system with a velocity of $v\simeq 300\ km/s$, and hence with a flux of $\simeq 10^5\ (100\ GeV/m_{\chi_1^0}) cm^{-2}s^{-1}$. Astronomical measurements essentially observe only the gravitational interactions of the dark matter. However, with such a flux it may be possible to observe dark matter more directly through electroweak interactions using detectors on the Earth. Techniques that have been proposed for observing neutralino DM fall into three categories:
\begin{enumerate}
\item Direct detection of incident $\chi_1^0$ by their scattering off quarks in a nuclear target.
\item Indirect detection of $\chi_1^0$ which accumulate in the center of the Sun or Earth and annihilate producing $\nu_{\mu}$.
\item Indirect detection from annihilation of halo $\chi_1^0$ into final states of antiprotons, positrons and gamma rays.
\end{enumerate}

\subsection{Direct Detection of Dark Matter}
\label{directdm}

Neutralinos may scatter off quarks via $s$-channel squark poles and $t$-channel $Z$ and Higgs poles. The effective Lagrangian that governs this interaction has the form ${\cal L}= {\cal L}^{SD}+{\cal L}^{SI}$ \cite{2} where the spin dependent (SD) part has the form
% eq1
\begin{equation} {\cal L}^{SD}= (\bar{\chi}_1\gamma^{\mu}\gamma^5\chi_1) [\bar{q}\gamma^{\mu}(A_LP_L+A_RP_R)q]
\end{equation}
($P_{L,R}= {1\over 2}(1\mp\gamma^5)$, $q(x)=$ quark field) and the spin independent (SI) part has the form
% eq2
\begin{equation} {\cal L}^{SI}= (\bar{\chi}_1\chi_1) (\bar{q}Cm_qq)
\end{equation}
The coefficients $A_{L,R}$ come from the $Z$ pole and squark pole, while $C$ arises from the neutral $h$ and $H$ Higgs exchanges and the squark pole. Even if $m_H^2\gg m_h^2$ (as is common in mSUGRA for low and intermediate $\tan \beta$), the heavy Higgs can make a significant contribution to the $d$-quark part of $C$. The reason for this is that for $d$-quarks the $H$ contribution relative to the $h$ contribution is ($m_h^2/m_H^2 \tan \alpha$) where $\alpha$ is the rotation angle that diagonalizes the $h-H$ mass matrix. For a wide range of parameters one finds (including loop corrections) $\tan\alpha  = O(1/10)$ which can overcome the smallness of $m_h^2/m_H^2$. Further, for large $\tan \beta$, $m_H$ need not be very large. To obtain the total nuclear cross section, one must add up the scattering from each quark in a nucleon, and each nucleon in the nucleus. The SI scattering adds coherently giving a scattering amplitude proportional to $M_N$, the nuclear mass, while the SD scattering is incoherent. The total detector scattering event rate then has the form
% eq3
\begin{equation}
R= [R_{SI}+R_{SD}] \biggl [ {\rho_{\chi_1^0}\over 0.3\rm\ GeV cm^{-3}}
\biggr ] \biggl [ {v_{\chi_1^0}\over 320\rm\ km/s} \biggr ] \rm {events\over kg\ d} \label{cdm-eq3}
\end{equation}
where
% eq4
\begin{equation} R_{SI}= {16m_{\chi_1^0} M_N^3M_Z^4\over [M_N+m_{\chi_1^0}]^2} |A_{SI}|^2; \quad R_{SD}= {16m_{\chi_1^0} M_N \over [M_N+m_{\chi_1^0}]^2} J(J+1)|A_{SD}|^2 
\end{equation}
where $A_{SI,SD}$ are the SI, SD amplitudes and $J$ is the spin of the nuclear target. We note that for $M_N$ large $R_{SI}\sim M_N$ (due to the coherent nature of SI scattering) while $R_{SD}\sim 1/M_N$ and so for heavy targets the spin independent scattering dominates. This is true even for relatively light  nuclei unless the nucleus has very large spin interactions (e.g. $F$) and even there, $R_{SD}$ dominates only when both $R_{SD}$ and $R_{SI}$ are small. The above calculations of $R$ contain a number of uncertainties. Thus one needs to know the quark content of the nucleon and there are experimental uncertainties, particularly in the strange quark contribution. In addition, the nucleons bind in the nucleus, and assumptions about nuclear form factors must be made. Along with the uncertainty in the value of $\rho_{\chi_1^0}$ mentioned above, predictions of Eq.(\ref{cdm-eq3}) are probably uncertain to within a factor of three. A large number of different terrestial dark matter detectors have been built or are being considered \cite{3}. These include cryogenic detectors (with recoil and ionization signals) based on Ge, Al$_2$O$_3$, LiF, Sn, Si, superconducting granule detectors, and scintilation detectors based NaI, Xe, CaF and A. (The last type of detector has the possibility of becoming quite large with nuclear targets of 100\ kg or more.)

\begin{figure}[h]
\centering\leavevmode
\psfig{figure=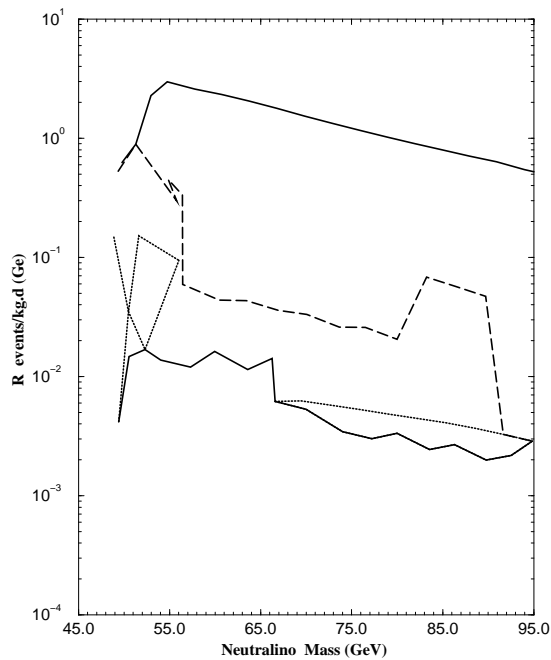,width=3.2in}
\vspace*{-.2in}

\caption{Maximum and minimum event rates for scattering of $\chi_1^0$ by a
$Ge$ target with relic density constraint $0.05 \stackrel{<}{\sim} \Omega_{\chi_1^0}h^2 \stackrel{<}{\sim} 0.30$ without $p$ decay constraint
(solid), with constraint $\tau (p\rightarrow \bar{\nu} K) > 5.5 \times 10^{32}
yr$ (dashed and dotted)~[4]. \protect\label{cdm-f1}}
\end{figure}

Figure~\ref{cdm-f1} shows the maximum and minimum event
rates expected for a Ge detector (solid curve) for the mSUGRA model when the SUSY parameter space of $m_0$, $m_{\tilde{g}}\le 1\ TeV$, $|A_t/m_0|\le 7$, $2\le \tan\beta \le 25$ is scanned. One sees that event rates may vary by a factor of $10^3$ over this space, the high event rates coming from large $\tan\beta$ and the low rates from small $\tan\beta$. The maximum event rate curve is governed by a somewhat complex play of phenomena. Thus in the relic density analysis, the $h$ and $Z$ $s$-channel poles are a dominant contribution in the region $40\ GeV \stackrel{<}{\sim} m_{\chi_1^0} \stackrel{<}{\sim}55\ \rm GeV$ and this gives rise to rapid annihilation which can be compensated for by $m_0$ becoming large so that too much annihilation does not occur (i.e. so that $\Omega_{\chi_1^0}h^2\ge 0.05$).  As one moves to higher $m_{\chi_1^0}$ in the relic density analysis, the $t$-channel sfermion contributions become dominant, and $m_0$ must remain  small (i.e. $m_0 \stackrel{<}{\sim} 200\rm\ GeV$) to get sufficient annihilation (i.e. so that $\Omega_{\chi_1^0}h^2\le 0.3)$.  The rise in the maximum event rate curve of Fig.\ref{cdm-f1} as $m_{\chi_1^0}$ increases toward 60\ GeV reflects this reduction in $m_0$, since then the squark mass is reduced and the $\chi_1^0$-quark scattering through the $s$-channel squark pole is increased.  The fall off of the event rate for $m_{\chi_1^0}\stackrel{>}{\sim} 60\rm\ GeV$ is due to the decreasing $\chi_1^0$-quark cross section as $m_{\chi_1^0}$ increases. Current detectors hope to obtain a sensitivity of $R\ge 10^{-2}\rm\ events/kg\ d$ which will impinge on the part of the SUSY parameter space with large $\tan\beta$. The large $\tan\beta$ domain is the more difficult one for accelerator discovery of SUSY, since the tri-lepton signal is weakened due to increased $\tau$ and $b$ final state events \cite{5,barkao2}. Thus dark matter searches and accelerator searches are to some extent complementary. However, the detection of soft leptons from $Z$ decays restores the trilepton search coverage of the large $\tan\beta$ region\cite{barkao2}

\subsection{Indirect Detection of Dark Matter}

Dark matter particles in the halo of the galaxy can annihilate into ordinary matter. Three possible final state particles that might be observable are $\bar{p}$, $e^+$ and $\gamma$'s \cite{1}. For the $\bar{p}$ case, the primary cosmic ray background is reduced by considering the energy region $E_{\bar{p}}\lsim 1\ GeV$, and it is conceivable that a signal of this type is detectable. However, there is still a considerable uncertainty concerning backgrounds. The Majorana nature of the $\chi_1^0$ prevents an annihilation into a two body $e^++e^-$ final state, and hence, as in the $\bar{p}$ case one expects a continuum spectrum that is difficult to separate from background. However, a heavy neutralino might annihilate into a $W^++W^-$ pair followed by $W^+\rightarrow e^++\nu_e$. One expects here a peak at $E_{e^+} \simeq {1\over 2} m_{\chi_1^0}$ in the positron distribution. There are again background uncertainties, though this represents a possible signal  for $\chi_1^0$ dark matter. The cleanest signal is the two body annihilation 
% eq5
\begin{equation} \chi_1^0 + \chi_1^0  \rightarrow \gamma + \gamma; \quad E_{\gamma} \cong  m_{\chi_1^0}
\end{equation}
since there are no astrophysical backgrounds with discrete photon energy. Thus an observation of this process would be a clean signal for halo $\chi_1^0$ dark matter. However, since the process proceeds through a loop and is order $\alpha^4$, the annihilation cross section is small, and one would likely require $\rho_{\chi_1^0}$ to be a factor of 10 larger than it is currently thought to be for this process to be observable. Such large values of $\rho_{\chi_1^0}$ might possibly be true in the center of the galaxy or if dark matter clusters in the halo, though again, astrophysical uncertainties make such posibilities unclear. An alternate way of indirectly observing halo $\chi_1^0$ arises from the fact that $\chi_1^0$ incident on the sun or earth may be gravitationally captured. From subsequent scattering they will sink to the center and accumulate there. Annihilation can then occur into ordinary matter resulting in neutrinos at the end of a cascade decay chain:
% eq6
\begin{equation}
\chi_1^0 + \chi_1^0  \rightarrow \nu_{\mu} + X \label{cdm-eq6}
\end{equation}
The $\nu_{\mu}$ can escape from the center of the sun or earth and be detected by terrestial  neutrino telescopes. The clearest signals come from upward going muons arising from $\nu_{\mu}\rightarrow \mu$ conversion in the rock of the earth. In general, over the age of the solar system, the capture rate and annihilation rate of neutralinos come into equilibrium for the sun and in at least part of the parameter space also for the earth. Thus the number of $\nu_{\mu}$ observed will be scaled by $\rho_{\chi_1^0}$. Further, process (\ref{cdm-eq6}) gives rise to a spectrum of $\nu_{\mu}$ with energies of about $E_{\nu_{\mu}}\approx {1\over 3} m_{\chi_1^0}$. Thus the neutrinos from $\chi_1^0$ annihilation can be distinguished from atmospheric neutrinos and other solar neutrinos. Neutrino telescopes being built are NESTOR and AMANDA with sizes of about $(10^3\mbox{--}10^4)\rm\,m^2$ with a proposed expansion of AMANDA to $10^6\rm\,m^2\ (ICE^3)$. For the coherent SI interactions (which are generally the largest interactions) one has roughly that a neutrino telescope of $(10^5\mbox{--}10^6)\rm\,m^2$ corresponds to a direct Ge detector of 1\ k \cite{6} (though the neutrino telescope may become more sensitive for very large $\tan\beta$ and non-universal soft breaking \cite{7}). For the smaller SD interaction, a neutrino telescope of $10^3\rm\ m^2$ roughly corresponds to a 1\ kg\ Ge detector.

\subsection{What Can Dark Matter Tell Accelerator Physics}

Current accelerator searches have begun to significantly restrict the mSUGRA parameter space. Thus the $b\rightarrow s+\gamma$ branching ratio combined with the nearness of the top quark mass to the Landau pole has eliminated most of the $\mu <0$ part of the parameter space and almost all of the $A_t>0$ region (in ISAJET notation). Direct searches at the Tevatron imply  $m_{\tilde{g}}\stackrel{>}{\sim} 250\ GeV$, and LEP189 data implies $m_{\chi_1^{\pm}}> 95\rm\ GeV$. (The latter, by scaling, implies $m_{\tilde{g}}\stackrel{>}{\sim} 350\rm\ GeV$ for most of the allowed parameter space.) If some dark matter detector were to discover $\chi_1^0$ dark matter, it would determine at least two things: the mass of the $\chi_1^0$ (from recoil  energy) and the event rate. As an example of what this might imply, the DAMA NaI experiment has recently suggested that its data may be showing a seasonal variation in event rates \cite{8}. Assuming this effect is real, a fit to the data gave $m_{\chi_1^0}\simeq 60\rm\ GeV$. From the sensitivity of the apparatus, the mSUGRA event rate must be close to the maximum event rate curve of Fig.\ref{cdm-f1}, which implies that $\tan\beta$ is relatively large, i.e. $\tan\beta\approx$ 10--20. By scaling one expects then $m_{\chi_1^{\pm}}\simeq 100\rm\ GeV$ and $m_{\tilde{g}}\simeq 400\rm\ GeV$. Relic density constraints imply $m_0\stackrel{<}{\sim} (150\mbox{--}200)\rm\ GeV$ in this region, implying that for the first two generations $m_{\tilde{q}} \simeq m_{\tilde{g}}$ and $m_h\simeq 110\rm\ GeV$. The above results imply that the $\chi_1^{\pm}$ may not be accessible to LEP. However, the Run~IIb (with $20\rm\ fb^{-1}$ of luminosity) should be able to see the light Higgs and the $\chi_1^{\pm}$, while the gluino would be at the edge of observability. Further, with most of the SUSY parameters determined, most of the mSUGRA mass spectrum not accessible to the Tevatron would be predicted for the LHC. While the above NaI detector data has, of course, not been confirmed, and if dark matter is discovered in some other region of the parameter space it may be harder to extract the SUSY parameters and masses, the example illustrates the impact dark matter detection could have on accelerator physics.

\subsection{Proton decay}

\begin{figure}
\centering\leavevmode
\psfig{figure=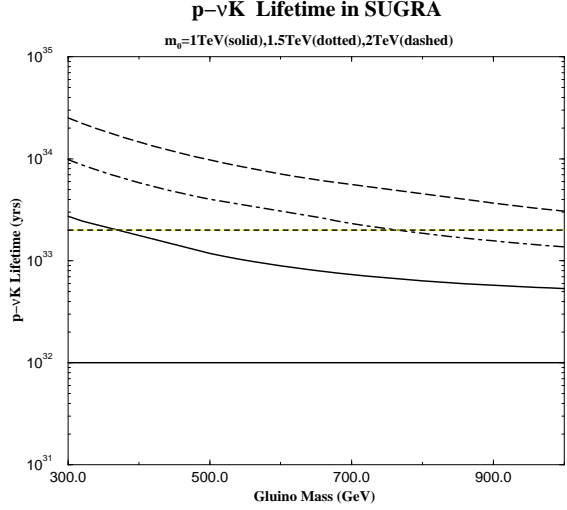,width=3.5in,angle=-90}

\caption{Maximum $\tau (p\rightarrow \bar\nu K)$ for minimal $SU(5)$ mSUGRA with $m_0\le 1\rm\ TeV$ (solid), $m_0\le 1.5\ TeV$ (dot-dashed), $m_0\le 2\ TeV$ (dashed) [4]. The solid horizontal line is the Kamiokande experimental bound, and the dashed horizontal line the expected sensitivity of Super K. [4]. The current Super K bound is $\tau_p>5.5\times 10^{32}yr$ at 90\% C.L. [11]. \protect\label{cdm-f2}}
\end{figure}

Grand unified models where both quarks and leptons reside in the same representation, generally give rise to proton decay. SUSY models with $R$-parity invariance forbid the dimension four operators which would yield a much too rapid proton decay, but generally allow for dimension five operators arising due to the exchange of color triplet Higgsinos, $\tilde{H}_3$. It is possible to suppress this operator by assuming that more than one pair of $3+\bar{3}$ color triplets exist, and with only a moderate amount of fine tuning, one can increase the lifetime by a factor of 10 in this way. Thus predictions of $p$-decay are more model dependent than other mSUGRA predictions. We consider first the mSUGRA $SU(5)$ model where there is only one pair of color triplets embedded in the $SU(5)$ $5+\bar{5}$ representations. The dominant decay mode is then usually $p\rightarrow \bar{\nu}K^+$ with a decay rate given by \cite{9}
% eq7
\begin{equation}
\Gamma (p\rightarrow \bar{\nu}K^+) = \sum_{i=,e,\mu,\tau} \Gamma (p\rightarrow \bar{\nu}_iK^+) = \biggl ( {\beta_p \over M_{H_3}} \biggr )^2  |A|^2|B|^2C \label{cdm-eq7}
\end{equation}
where $M_{H_3}$ is the Higgs triplet mass, $\beta_p$ is the matrix element of the three quark operator between the vacuum and the proton state (lattice gauge calculations give $\beta_p=5.6\times 10^{-3}\rm\,GeV^3$ \cite{10}). $A$ is a factor depending on quark masses and CKM matrix elements, $B$ contains the dressing loop integral involving SUSY masses, and $C$ contains the chiral current algebra factors that convert quarks into mesons and baryons. There are a number of uncertainties in inputs in the above result, e.g. the value of $\beta_p$, CKM elements etc., and one estimates that the prediction of $(M_{H_3})^2\Gamma$ has an uncertainty of perhaps a factor of two or three. In the following we assume $M_{H_3}\le 10M_G$. For mSUGRA $SU(5)$, the second generation contribution to Eq.(\ref{cdm-eq7}) dominates, and one roughly finds that
\begin{equation}
\Gamma \approx const \biggl ( {m_{\tilde{g}}\over m_0^2} \tan\beta \biggr )^2 \label{cdm-eq8}
\end{equation}
Equation (\ref{cdm-eq8}) shows the effects that bounds on the proton lifetime have on the SUSY parameter space, since large $m_{\tilde{g}}$, large $\tan\beta$ and small $m_0$ will destabilize the proton. These effects can be seen in Fig.\ref{cdm-f2} where the maximum $p\rightarrow \bar{\nu} K$ lifetime is plotted vs.\ $m_{\tilde{g}}$ (as one scans the parameter space with $\tan\beta \le 25$) \cite{4}. The current data does not yet constrain the model, though Super Kamiokande will be able to test the model completely for $m_0\le 1\ TeV$ (solid curve). For $m_0 > 1\ TeV$, the curves lie higher and there will still be regions not tested by Super Kamiokande.

\begin{figure}
\centering\leavevmode
\psfig{figure=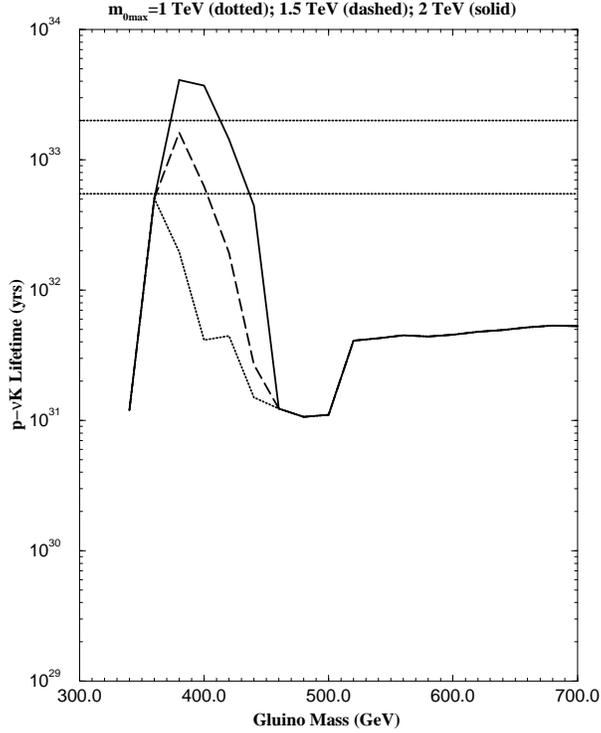,width=3.5in}

\caption{Maximum $\tau (p\rightarrow \bar{\nu}K)$ for minimal $SU(5)$ mSUGRA with the relic density constraint $0.05 \leq \Omega_{\chi_1^0}h^2 \leq 0.30$ with $m_0 \leq 1\rm TeV$ (dotted) and $m_0\leq 1.5\rm\ TeV$ (dashed), and $m_0 \leq 2 TeV$ (solid) [4]. Horizontal lines are the current and expected future Super Kamiokande bounds. \protect\label{cdm-f3}}
\end{figure}

It is interesting now to combine the above results with the constraints arising from the dark matter analyses. In the relic density analysis, there are roughly two separate domains. As discussed above, for the region $m_{\tilde{g}}\stackrel{<}{\sim} 450\rm\ GeV$ ($m_{\chi_1^0}\stackrel{<}{\sim} 55\rm\ GeV$) annihilation of $\chi_1^0$ in the early universe can proceed sufficiently rapidly through $s$-channel $Z$ and $h$ poles so that $m_0$ can be quite large and still the relic density obeys the astronomical bounds of $0.05\le \Omega_{\chi_1^0}h^2 \le 0.30$. Thus in this region, as can be seen from Eq.(\ref{cdm-eq8}), the maximum $p$ lifetime can be quite large. In the region $m_{\tilde{g}}\stackrel{>}{\sim} 450\rm\ GeV$, the annihilation of $\chi_1^0$ proceeds mainly through $t$-channel sfermion poles, and to get sufficient annihilation these must be light. Here one finds that $m_0\stackrel{<}{\sim} 150\rm\ GeV$, reducing the maximum $p$ lifetime by a factor of between 10 and 30. This effect can be seen in Fig.\ref{cdm-f3} where the dotted curve shows the maximum $p$ lifetime when the relic density constraint is imposed for $m_0\le 1\rm\ TeV$. We see now that current $p$ decay data ($\tau_p > 5.5\times10^{32}\rm\, yr$ [11]) already excludes the domain $m_{\tilde{g}}\stackrel{>}{\sim} 350\rm\ GeV$, and even if the $m_0$ bound is raised to $m_0 \leq 2 TeV$ (dashed curve) the excluded region is $m_{\tilde{g}}\stackrel{>}{\sim} 450\rm\ GeV$. Since the current LEP189 bound of $m_{\chi_1^{\pm}} > 95\rm\ GeV$ already excludes $m_{\tilde{g}}\stackrel{<}{\sim} 350\rm\ GeV$, most of the parameter space of this model has been eliminated when the relic density constraints are included. One can consider non-minimal extensions of the $SU(5)$ model. The inclusion of Yukawa textures so that the low energy limit of the theory gives rise to correct quark-lepton mass ratios raises the $p$ lifetime by a factor of about 3--5. Combined with the above mentioned uncertainty in the calculations of $\Gamma (p \rightarrow \bar{\nu} K)$ of a factor of 2--3, one might have a total increase in the theoretical curves of 5--15. The above exclusion region of $m_{\tilde{g}}\stackrel{>}{\sim} 450\rm\ GeV$ would then be reached when Super Kamiokande obtains a sensitivity of about $8 \times 10^{32}\rm\, yr$.

The above considerations hold for other GUT groups that contain an $SU(5)$ subgroup and have matter embedded in that $SU(5)$ in the usual way. A notable exception is $SO(10)$ where $t-b$ Yukawa unification requires $\tan\beta$ to be very large, i.e. $\tan\beta = 56$. From Eq.(\ref{cdm-eq8}) one sees that this requirement will significantly reduce the $p$ lifetime. (In $SU(5)$, current data generally requires $\tan\beta < 10$.) Further analyses is required to see which $SO(10)$ models with conventional proton decay will remain viable. Equation~(\ref{cdm-eq8}) shows that the maximum $p$ lifetime occurs mainly from the low $\tan\beta$ and high $m_0$ part of the parameter space. As discussed in Sec.~\ref{directdm}, the dark matter detector event rates increase with $\tan\beta$ and decrease with $m_0$, the maximum event rates generally occuring at high $\tan\beta$ and low $m_0$. One expects, therefore, a reduction of event rates if one imposes the constraints from $p$ lifetime bounds. This can be seen in Fig.\ref{cdm-f1}. The dotted and dashed curves give the maximum and minimum event rates when the proton decay constraint [11] is imposed. [In the dotted curve, we have multiplied the conventional result [9] for $\tau (p \rightarrow \bar{\nu} K)$ by 20 to account for the above mentioned corrections for Yukawa textures and other uncertainities.
(Note in this case the region $56{\rm\ GeV} \leq m_{\chi^0_1} \leq 67\rm\ GeV$ is forbidden.) The dashed curve multiplies the conventional result by 100 corresponding to an additional decay suppression that could arise from more than one pair of superheavy Higgs color triplets.] Proton decay in $SU(5)$-type models can always be suppressed by chosing a complicated Higgs sector. Predictions are thus model dependent. If one considers the conventional models where this tuning is not done, then the combined effects of relic density constraints and $p$ decay bounds generally imply that the gluino is relatively light i.e. $m_{\tilde{g}}\stackrel{<}{\sim} \rm\ 450 GeV$. The Run~IIb is expected to be sensitive to gluinos with mass up to 400\ GeV in direct searches, and indirectly from the discovery potential of the $\chi_1^{\pm}$ to gluinos of mass greater than $500\rm\ GeV$. Thus in such models gluinos will be in a range that may be accessible to the Tevatron.

%\pagebreak

%% file: Abel/run-II.tex
\section{Charge and Colour Breaking Constraints}

%       macros
\newcommand{\hepph}[1]{{hep-ph/#1 }}
\newcommand{\phrd}[3]{{{\it Phys.~Rev.}~{\bf D#1} (#3) #2}}
\renewcommand{\plb}[3]{{{\it Phys.~Lett.}~{\bf B#1} (#3) #2}}
\renewcommand{\prd}[3]{{{\it Phys.~Rev.}~{\bf D#1} (#3) #2}}
\renewcommand{\prl}[3]{{{\it Phys.~Rev.~Lett}~{\bf #1} (#3) #2}}
\renewcommand{\npb}[3]{{{\it Nucl.~Phys.}~{\bf B#1} (#3) #2}}
\renewcommand{\ptp}[3]{{{\it Prog.~Theor.~Phys.}~{\bf #1} (#3) #2}}
\renewcommand{\rpp}[3]{{{\it Rept.~Prog.~Phys.}~{\bf #1} (#3) #2}}
\newcommand{\leqsim}{\,\raisebox{-0.6ex}{$\buildrel < \over \sim$}\,}
\newcommand{\geqsim}{\,\raisebox{-0.6ex}{$\buildrel > \over \sim$}\,}
\renewcommand{\be}{\begin{equation}}
\renewcommand{\ee}{\end{equation}}
\newcommand{\ba}{\begin{eqnarray}}
\newcommand{\ea}{\end{eqnarray}}
\renewcommand{\etal}{\mbox{\em et al}}
\newcommand{\ie}{\mbox{\em i.e.~}}
\newcommand{\eg}{\mbox{\em e.g.~}}
\newcommand{\cf}{\mbox{\em c.f.~}}
\renewcommand{\nn}{\nonumber}
\newcommand{\dif}{\mbox{d}}
\def\gev{\,{\rm GeV }}
\def\tev{\,{\rm TeV}}
\def\mcha{m_{\chi^\pm}}
\def\dd{\mbox{d}}
\def\mchi{m_{\tilde\chi}}
\def\m12{m_{1/2}}
\def\tb{\tan\beta}
\def\ohsq{\Omega_\chi h^2}
\newcommand{\smallfrac}[2]{\frac{\mbox{\small #1}}{\mbox{\small #2}}}

Vacuum stability bounds are 
a particularly important issue for supersymmetric models 
because of the large number of scalars, 
any of which can get a vacuum expectation value, possibly breaking 
charge and/or colour. Here we shall give a short discussion of
these bounds concentrating on two questions; 
\begin{itemize}
\item 
How do the bounds restrict the parameter space in the mSUGRA model? 
\item
How do they restrict models with non-universal supersymmetry breaking?
\end{itemize}

\subsection{Introduction and the mSUGRA Model}

Insisting that the physical vacuum be stable results in a set 
of constraints on the possible supersymmetry breaking parameters 
which are known as Charge and Colour Breaking (CCB) 
bounds~\cite{ccb1,casas1,baer,as98a,as98b,casas2,af98}. 
There are different schools of thought regarding the precise
cosmological meaning of these bounds (see for example Ref.\cite{af98} 
for a recent discussion). Here we take the point of view 
that unphysical minima 
lower than the physical vacuum should be avoided. 
There are two important kinds of bounds which result; 
$D$-flat directions which develop a minimum due to large trilinear 
supersymmetry breaking terms; $D$ {\em and} $F$ flat directions which 
correspond to a combination of gauge invariants involving $H_2$. 
The first kind of flat directions give a familiar 
set of constraints on the trilinear couplings which is typically of the form 
\be 
A_t^2 \leqsim 3 (m_{H_U}^2 + m_{t_R}^2 + m_{t_L}^2 ),
\ee
where the notation is conventional. These constraints turn out to 
be very weak. A more sophisticated treatment (in which the 
constraint is optimised by finding the deepest direction close to 
the $D$ flat direction) results
in a set of `generalised' CCB bounds which are much stronger~\cite{casas1}. 
By far the most severe bounds, however, come from the directions which are 
$F$ and $D$ flat and we shall concentrate on these\footnote{These
bounds are sometimes referred to as Unbounded From Below although 
we shall avoid this terminology since it can be confusing. Typically the 
directions are not unbounded from below but develop a radiative minimum.}.

$F$ and $D$ flat directions develop a 
minimum because of the negative mass-squared 
term, $m_{H_U}^2$, of the Higgs which couples to the top-quark.
Whilst the latter is a great success for driving electroweak symmetry breaking,
it can generate an undesirable minimum radiatively. 
These directions are
constructed from conjunctions of $L_i H_2$ plus any one of the following 
gauge invariants~\cite{as98b}, 
\be 
LLE\mbox{, }
LQD\mbox{, }
QULE\mbox{, }
QUQD\mbox{, }
QQQLLLE.
\ee
Absence of CCB minima along the first two directions is usually 
enough to guarantee their absence along the rest~\cite{as98b}.
As an example consider the $L_iL_3E_3$, $L_iH_2$ direction, which 
corresponds to the choice of VEVs,
\ba
\label{komkom}
h_2^0             &=& -a^2 \mu/h_{E_{33}} \nonumber \\
\tilde{e}_{L_3}=\tilde{e}_{R_3} &=& a \mu/h_{E_{33}} \nonumber \\
\tilde{\nu}_i   &=& a  \sqrt{1+a^2} \mu/h_{E_{33}},
\ea 
where $a$ parameterizes the distance along the flat direction.
The potential along this direction is $F$ and $D$-flat and
depends only on the soft supersymmetry breaking terms;
\be
\label{softv}
V=\frac{\mu ^2}{h_{D_{33}}^2} a^2 (a^2 (m_{H_U}^2+m_{L_{ii}}^2) + 
m_{L_{ii}}^2+m_{E_{33}}^2+m_{L_{33}}^2 ).
\ee 
The first term dominates at large VEVs when $a\gg 1$ and 
it is this that generates the dangerous CCB minimum with
a VEV which is typically a few orders of magnitude larger 
than the weak scale~\cite{baer}.
There is an `optimum' direction very close to this one which
is slightly deeper~\cite{casas1,baer}. 
In order to minimise the one-loop corrections, the mass squared parameters 
in Eq.(\ref{softv}) should be evaluated at either the scale of the 
running top quark mass, $h_t H_U^0$, or at $M_{susy}$, whichever 
is the larger. 

\begin{figure}
%\vspace*{-1.3in}
%\begin{minipage}{4.0cm}
%\hspace*{-1in}
\centerline{\epsfig{file=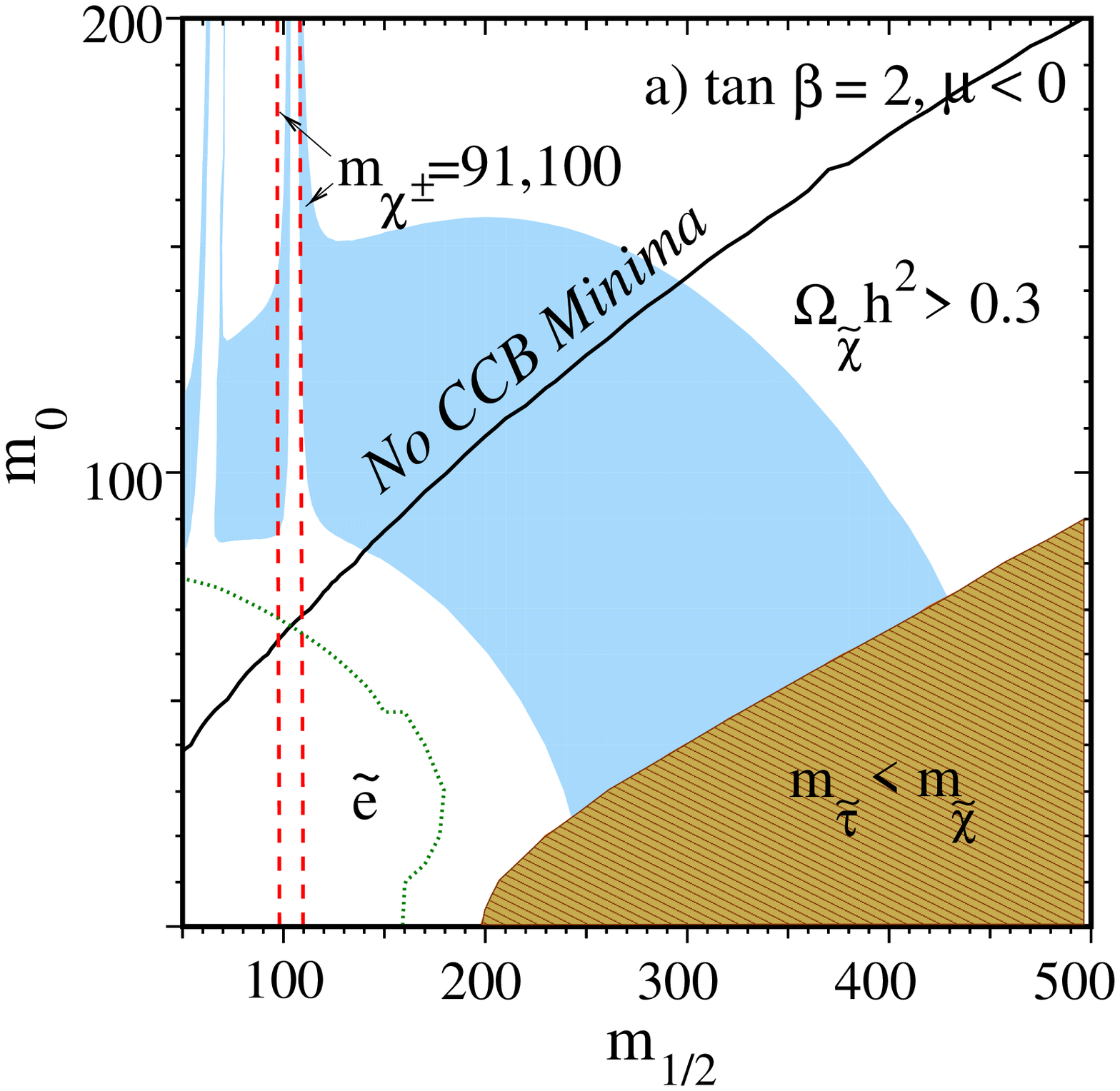,height=3.2in} 
%\end{minipage}
\hspace*{-.5in}
%\begin{minipage}{4.0cm}
\epsfig{file=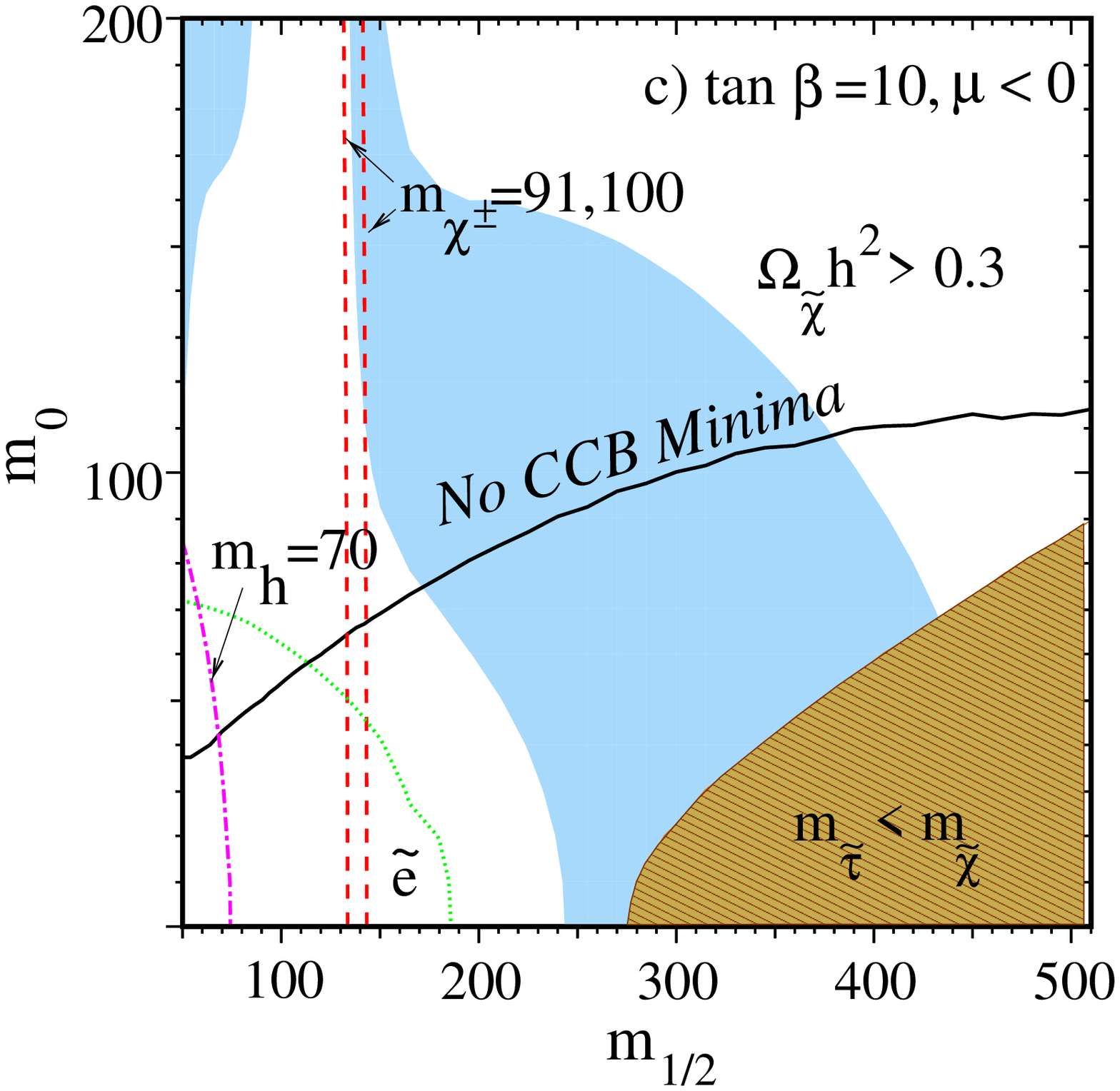,height=3.2in} 
%\end{minipage}
\hspace*{-.5in}
%\begin{minipage}{4.0cm}
\epsfig{file=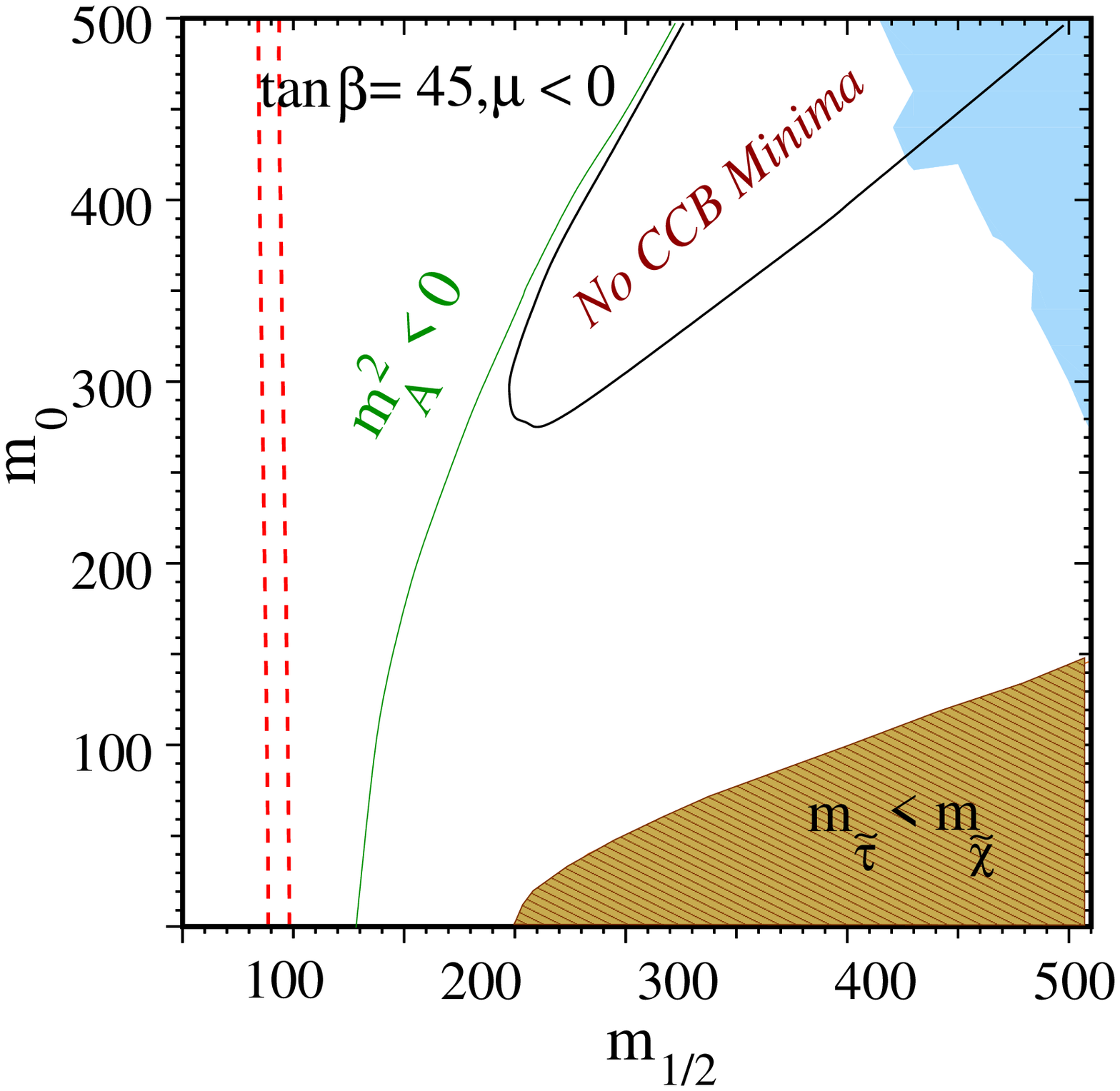,height=3.2in} }
%\end{minipage}
\vspace{-.25in}

\caption{\label{fig:bnds} Experimental, dark matter and CCB 
Constraints on the mSUGRA model.}
\end{figure}

For a given choice of supersymmetry breaking parameters it is now 
straightforward to determine whether the potential has a minimum 
which is lower than the physical one. The results for mSUGRA are 
shown in Figure~\ref{fig:bnds} where we plot the regions in the ($m_0,\m12$) plane 
in which there are CCB minima, for three different values of $\tan\beta $
and $\mu < 0$. We choose a universal trilinear coupling of 
$A_0=\m12 $ for the low $\tan\beta $ diagrams, since this is the value 
which minimises the regions with a 
CCB minimum~\cite{baer,as98a,as98b}. The $\tan\beta =45 $ diagram
has $A_0=0$.
The light shading denotes the cosmolgically preferred region with $0.1<\Omega_\chi h^2 < 0.3$, where the upper bound originates from a lower limit of 12~Gyr on the age of the universe. 
Thus we can see that, in mSUGRA, CCB minima when combined with dark matter 
constraints and experimental constraints are 
extremely restrictive. 
For example, we find that the Higgs mass lower bound, together with the cosmological constraint, implies that the mSUGRA electroweak vacuum cannot be stable for values of $\tan\beta<2.3 $ for $\mu<0$. In addition, at high $\tan\beta $, large regions of parameter space become excluded by trilinear constraints and by 
the requirement of correct electroweak symmetry breaking. 

\subsection{Non-universal supersymmetry breaking}

Although mSUGRA continues to provide a good working model of physics
beyond the Standard Model, at the time of writing there is much
interest in models, derived from both perturbative and
non-perturbative string theory, in which the supersymmetry breaking is
non-universal.  In specific cases, where there are few parameters, it
is possible to continue using the numerical techniques outlined
above. Indeed, recently, a variety of string derived models were shown
to also suffer from severe CCB problems~\cite{casas2}. For theories
with a larger number of independent parameters however, it becomes
increasingly difficult to pursue this option. In more general models
progress can be made at low $\tan\beta$ if some accuracy is
sacrificed~\cite{as98a,as98b}. In Ref.\cite{as98b} general expressions
were derived which are valid in the one-loop approximation and in the
(low $\tan\beta $) approximation that the bottom and tau Yukawas can
be neglected. Defining $\tilde{m}=m/\m12 $, CCB minima are absent when
the following bound is satisfied.
\be
\label{arp3}
\left.(2 \tilde{m}_{L_{ii}}^2 + \tilde{m}_{H_U}^2-\tilde{m}^2_{U_{33}}-
\tilde{m}^2_{Q_{33}})\right|_0
\geqsim 
f(\tilde{B}|_0)+(\rho_p-1)\left( g(\tilde{B}|_0)+3 \tilde{M}^2|_0
-\rho_p (1- {\tilde{A}_{U_{33}}}|_0)^2 \right),
\ee
where 
\ba 
\tilde{B}|_0&=&\left.(\tilde{m}_{L_{ii}}^2+\tilde{m}_{L_{33}}^2+
\tilde{m}_{E_{33}}^2)\right|_0\nn\\
\tilde{M}|_0&=&\left.(\tilde{m}_{H_U}^2+\tilde{m}_{Q_{33}}^2+
\tilde{m}_{U_{33}}^2)\right|_0\nn\\
f(x)&=& 1.43-0.16 x +0.02 x^2 \nn\\
g(x) &=& 2.94 -0.20 x + 0.02 x^2 \nn\\
\frac{1}{\rho_p}& =& 1+3.17 (\sin^2\beta -\sin^2 \beta^{QFP})
\ea
and where $\tan\beta^{QFP}\approx 1.6 $. Note that, at the low $\tan\beta$
fixed point, the constraint is on the single combination 
$\left.(2 \tilde{m}_{L_{ii}}^2 + \tilde{m}_{H_U}^2-\tilde{m}^2_{U_{33}}-
\tilde{m}^2_{Q_{33}})\right|_0$ of GUT scale parameters.
In the above, $f(x)$ and $g(x)$ are fits to the bounds for $\mu = 200\gev$ 
which are accurate for $x\geqsim 0$, and are valid for the MSSM with 
{\em any} pattern of supersymmetry breaking. 
Compared to the numerically 
determined bounds in mSUGRA, they are a slight overestimate at low 
$\tan\beta$ (by $ 5-10 \%$) and an underestimate at higher $\tan\beta $ 
as the bottom quark Yukawa becomes increasingly important in the 
running~\cite{af98}. 
The main inaccuracy comes from the poor determination of 
$\tan\beta^{QFP} $ and of 
course from the dependence on $\mu $ 
(which enters the bounds only logarithmically). 
For example, if we take $\mu = 500 \gev $ we find 
\be
f(x) = 1.20-0.14 x +0.02 x^2.
\ee
Although only approximate, these expressions are accurate enough to 
indicate whether a model is likely 
to suffer from severe CCB problems. For example it is found that 
certain versions of Horava-Witten M-theory in which supersymmetry 
breaking comes from bulk moduli fields (see Ref.\cite{choi} and 
references therein) are always unstable~\cite{as98b,casas2}. 
Eqn.(\ref{arp3}) also indicates that patterns of supersymmetry breaking 
with large slepton mass-squareds at the GUT scale are likely to be 
safe from CCB minima. However large slepton mass-squareds at the GUT scale 
also lead to large relic densities so that, in this case, the dark matter 
bounds and CCB bounds are negatively correlated. 
Note that, for the above expressions, universal 
gaugino masses were assumed at the GUT scale. The case of non-universal 
gaugino masses was examined in Ref.\cite{casas2}. In particular, in that
work it was found that small $M_3$ and large $M_1$ and $M_2$ avoids 
dangerous minima. This is in accord with Eq.(\ref{arp3}) since 
the renormalisation of $m_{H_U}^2$ is dominated by $M_3$ whereas the 
renormalisation of $m_L^2$ is dominated by $M_1$ and $M_2$.

Minor modifications to the superpotential of 
the MSSM may also lift the 
dangerous minimum whilst giving phenomenology which is (currently) 
indistinguishable from the MSSM. The simplest option 
is to break $R$-parity by additional (but undetectable) 
lepton number violating operators. 
Of course, in this case the constraints on $\Omega h^2$ no longer apply, since the LSP can decay.
Alternatively, models in which electroweak symmetry breaking is driven by 
a singlet superfield rather than a $\mu$-parameter have a very different 
potential at large field values, whereas at low values they typically 
mimic the MSSM (see for example Ref.\cite{savoy}).

%% file: Bhrlik/bsg.tex
% Note: this was the file that was modified by Toby
%
%%%%%%%%%%%%%%%%%%%%%%%%%%%%%%%%%%%%%%%%%%%%%%%%%%%%%%%%%%%%%%%%%%%%%%%%%%%%%%%
\def\doublespaced{\baselineskip=\normalbaselineskip\multiply
    \baselineskip by 200\divide\baselineskip by 100}
\def\singleandabitspaced{\baselineskip=\normalbaselineskip\multiply
    \baselineskip by 120\divide\baselineskip by 100}
\def\singlespaced{\baselineskip=\normalbaselineskip}
\newcommand{ \centeron }[2]{{\setbox0=\hbox{#1}\setbox1=\hbox{#2}\ifdim
                             \wd1>\wd0\kern.5\wd1\kern-.5\wd0\fi \copy0
                             \kern-.5\wd0\kern-.5\wd1\copy1\ifdim\wd0>\wd1
                             \kern.5\wd0\kern-.5\wd1\fi}}
\newcommand{ \ltap }{\>\centeron{\raise.35ex\hbox{$<$}}
                     {\lower.65ex\hbox{$\sim$}}\>}
\newcommand{ \gtap }{\>\centeron{\raise.35ex\hbox{$>$}}
                     {\lower.65ex\hbox{$\sim$}}\>}
\def\mol{Mol}
\def\etmiss{E\llap/_T}
\def\eslt{E\llap/_T}
\def\esl{E\llap/}
\def\msl{m\llap/}
\def\to{\rightarrow}
\def\te{\tilde e}
\def\tmu{\tilde\mu}
\def\ttau{\tilde\tau}
\def\tl{\tilde\ell}
\def\ttau{\tilde \tau}
\def\tg{\tilde g}
\def\tnu{\tilde\nu}
\def\tell{\tilde\ell}
\def\tq{\tilde q}
\def\tu{\tilde u}
\def\tc{\tilde c}
\def\tb{\tilde b}
\def\tst{\tilde t}
\def\tt{\tilde t}
\def\tw{\widetilde W}
\def\tz{\widetilde Z}

\hyphenation{mssm}
%\def\ds{\displaystyle}
%\def\ts{${\strut\atop\strut}$}
%
%\include{psfig}
%

%%%%%%%%%%%%%%%%%% MAIN TEXT %%%%%%%%%%%%%%%%%%%%%%%%%%%%%%%%%%%%%%%%%%%%%%%

\section{$\lowercase{b\to s}\gamma$ Constraints}

An important experimental check of mSUGRA is provided by the measurement of the 
$b\to s\gamma$ decay rate. Weak scale SUSY particles contribute to the one loop 
decay amplitude, and their presence can significantly modify the Standard Model 
result.

The best experimental value available for the inclusive $B\to X_s \gamma$ 
branching ratio as measured by the CLEO collaboration $B(B\to X_s 
\gamma)=(3.15\pm0.54)\times 10^{-4}$\cite{bsg-alex} has to be compared to the 
next-to-leading order Standard Model prediction including non-perturbative 
corrections $B(B\to X_s) \gamma=(3.38\pm 0.33)\times 10^{-4}$ \cite{bsg-misiak}. 
SUSY contributions can correct the theoretical value in both directions, and a 
requirement that the branching ratio remain restricted within the experimental 
95\% confidence level band $1\times 10^{-4}< B(B\to X_s\gamma )<4.2\times 
10^{-4}$ translates into constraints on the mSUGRA parameter space. All the 
next-to-leading order calculation ingredients for SUSY models are already 
available, with the exception of the exact two-loop matching conditions between the 
full theory and the effective theory. In this case, an approximate procedure 
effectively decoupling the heavy particles in the loop at different scales can 
be used for the dominating SUSY contributions \cite{bb1}.  

In the MSSM, the SUSY contribution involves the charged Higgs-top quark loop, 
chargino-squark loops, gluino-squark loops and neutralino-squark loops
\cite{hp,masiero}. The first are most important, while the gluino and neutralino 
loops can be safely neglected within the mSUGRA framework. The charged Higgs 
loop contribution is always of the same sign as the SM contribution and adds 
constructively with the SM loop. The chargino-squark loops, on the other hand, 
can interfere with the SM contribution either constructively or destructively 
depending on the sign of $\mu$. This feature, together with the fact that the 
magnitude of the total chargino-squark contribution grows with $\tan\beta$,
characterizes constraints imposed on the mSUGRA parameter space by $b\to 
s\gamma$.

\begin{figure}[h]              
\centering
\epsfxsize=5.0in                % the width; height is automatically scaled
\hspace*{0in}
\epsffile{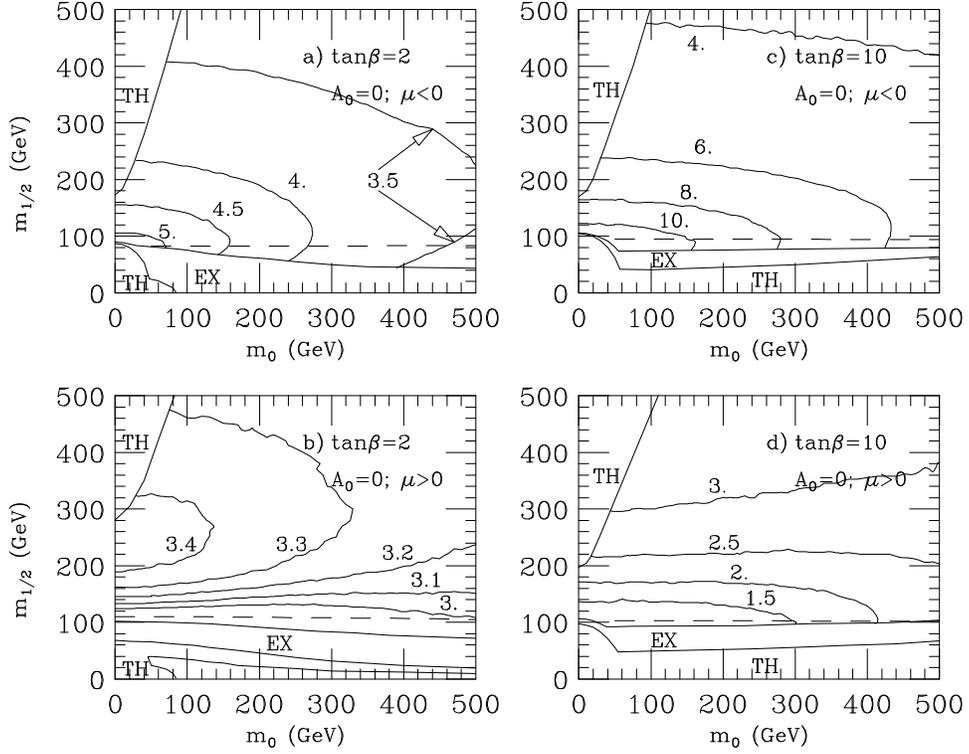}
\bigskip

\caption{Plot of contours of constant branching ratio $B(b\to s\gamma )$ in 
the $m_0\ vs.\ m_{1/2}$
plane, where $A_0=0$ and $m_t=175$ GeV. 
Each contour should be multiplied
by $10^{-4}$. 
The regions labelled by TH (EX) are excluded by theoretical (experimental)
considerations. The dashed line corresponds to the LEP2 limit
of $m_{\tw_1}>80$ GeV for a gaugino-like chargino.}
\label{bsg-fig1}
\end{figure}

For $\tan\beta \ltap 5$, the chargino squark contribution is small enough to 
permit large allowed regions for both signs of $\mu$, with larger branching 
ratios arising for the case of $\mu <0$.  With $\tan\beta$ increasing, the allowed 
region in the $\mu<0$ case narrows, requiring heavy charginos to suppress 
$\tan\beta$ enhancement in the chargino-squark loops.  In the $\mu>0$ case ,
cancellation between the chargino loops and the charged Higgs loop makes it 
possible to obtain branching ratios in some cases very close to the CLEO mean 
value and certainly consistent with the experimental limits over the whole 
range of parameters. As an example, Fig.~\ref{bsg-fig1} displays contours of constant 
branching ratio $B(b\to s \gamma)$ in the $m_0$ vs $m_{1/2}$ plane for 
$\tan\beta =2$ and $10$, and both signs of $\mu$.

In the large $\tan\beta$ parameter region with $\tan\beta \gtap 35$, the 
$\mu>0$ branch of the parameter set is also constrained by  $b\to s \gamma$,
since a very large chargino-squark contribution drives the branching ratio 
below the lower experimental value of $1\times 10^{-4}$. The $\mu<0$ case 
requires extremely heavy superpartners ($m_{\tg}\gtap 1500\ \rm{GeV}$, 
$m_{\tq}\gtap 1600\ \rm{GeV}$) in order to satisfy experimental constraints on 
$B(b\to s \gamma)$ and thus imposes quite severe constraints on models with Yukawa 
coupling unification \cite{ltb}.

Additional dependence on $A_0$ does not change the general picture as determined 
by the values of $ \tan\beta$ and sign of $\mu$ but needs to be considered for
a detailed analysis of mSUGRA models, since it can change the branching ratio by 
up to several tens of percent in either direction.

An interesting question arises in connection with CP-violating phases in 
SUGRA models. Traditionally, $\mu$ and $A_0$ have been considered to be real in all 
analyses concerned with $b\to s \gamma$ decay rates in this scenario. It was 
recently shown \cite{bsg-nath} that both $\mu$ and $A_0$ can be complex without 
violating experimental limits on the neutron and electron dipole moments. Since 
the sign of $\mu$  plays such a crucial role in the branching ratio 
calculation, a completely general analysis including the complex phase of $\mu$
is necessary here to fully describe $b\to s \gamma$ constraints on the MSSM.

%%%%%%%%%%%%%%%%%%%%% REFERENCES %%%%%%%%%%%%%%%%%%%%%%%%%%%%%%%%%%%%%%%%%%%%%%

%

%% file: Blazek/g_2wrkshp.tex
\def\bsgam{b\rightarrow s\gamma}
\def\as{\alpha_s}
\def\be{\begin{equation}}
\def\ee{\end{equation}}
\def\bea{\begin{eqnarray}}
\def\eea{\end{eqnarray}}

\def\sign{{\sf sign}}

\def\ul{\underline}
\def\g2{$g_\mu-2$}

\section{$\lowercase{g}-2$ for Muons in an MSSM Analysis
with SUGRA Induced SUSY Breaking}

%\section{Introduction}
%
The anomalous magnetic moment of the muon is potentially a significant constraint for any
extension of the Standard Model (SM).
Currently, the measured value \cite{a_mu_exp} is
$a_\mu^{exp} \equiv \left(\frac{g_\mu - 2}{2}\right)^{exp} = 
                        (11\,659\,230 \pm 84) \times 10^{-10}$.
%\label{a_exp}
%\ee
On the other hand, the SM prediction (\cite{a_mu_th} and references therein) yields 
$a_\mu^{SM} = (11\,659\,159.6 \pm 6.7) \times 10^{-10}$
%\ee
where the QED, electroweak and hadronic contributions are sumed and the errors are combined
in quadrature. That constrains the contributions from new physics beyond the SM to fit within
the
\be
    -70 \times 10^{-10} < \delta a_\mu < +210 \times 10^{-10}
\label{da_now}
\ee
window, at 90\% C.L..  The width of this window is dominated by the experimental
uncertainty. That, however, will change after the 1998-1999 run of the E821 
experiment at BNL 
which is expected to reach the accuracy $\pm 10 \times 10^{-10}$. If
the observed central value then becomes exactly equal to the SM prediction, 
the new constraint will be 
\be
    -20 \times 10^{-10} < \delta a_\mu < +20 \times 10^{-10}
\label{da_99}
\ee
at 90\% C.L., with the window for new physics narrowed by a factor of 7.

In the MSSM, there are significant contributions from new physics due to the chargino-sneutrino 
and neutralino-smuon loops \cite{a_mu_MSSM}. Some of these contributions are proportional
to tan$\beta$. The effect can be traced back to the diagrams with the chirality flip inside 
the loop (or in one of the vertices) as opposed to the diagrams where the chirality
flip takes place in the external muon leg. There are no similarly enhanced terms in the SM, 
where the chirality can only be flipped in the external muon leg. Thus for tan$\beta \gg 1$
we expect that the SUSY contribution scales versus the SM electroweak contribution 
$ \delta a_\mu^{EW} $ as
\be
 \delta a_\mu^{SUSY} \simeq   \delta a_\mu^{EW} \: \left( \frac{M_W          }{\tilde{m}} \right)^2\, \tan\!\beta
                     \simeq  15 \times 10^{-10} \: \left( \frac{100 {\rm GeV}}{\tilde{m}} \right)^2\, \tan\!\beta,
\label{eq:compEW}
\ee
where $\tilde{m}$ stands for the heaviest sparticle mass in the loop (see also \cite{a_mu_MSSM}).

%\section{Numerical Analysis}
%
Our numerical analysis of the SUSY sector has been completely based on the top down 
approach. Three cases with low, medium, and large value of tan$\beta$ were analyzed.
For low and medium tan$\beta$ ($2$ and $20$), the minimal set of the initial SUSY parameters 
read 
$M_{1/2},\; m_0,\; A_0,\; \mbox{\rm sign}\mu,\; \mbox{\rm and tan} \beta$,
%\label{4params}
%\ee
with the universal gaugino mass, scalar mass, and trilinear coupling
introduced at $M_{GUT}$. The Yukawa couplings of the third generation
fermions were free to vary independently on each other. 
For tan$\beta={\cal O}(50)$, in order to reduce
severe fine-tuning required for the correct electroweak symmetry breaking, the scalar 
Higgs masses were allowed to deviate from $m_0$. The third generation yukawas at $M_{GUT}$ 
were strictly set equal to each other and we used the results of the global 
analysis of model 4c, an SO(10) model with minimal number of effective operators 
leading to realistic Yukawa matrices \cite{lr,bcrw}. In fact, the SO(10)-based 
equality $\lambda_t=\lambda_b=\lambda_\tau$
is the main reason why such a large tan$\beta$ is attractive.
The rest of the analysis was then the same for each of the three cases. 
Particular values of $m_0$ and $M_{1/2}$ were picked up, while the rest of the initial parameters
at the unification scale varied in the optimization procedure. Using the 2(1)-loop RGEs 
for the dimensionless (dimensionful) couplings the theory was renormalized down to 
low energies where the radiative electroweak symmetry 
breaking was checked at one loop as in \cite{bcrw} and one-loop SUSY threshold corrections
to fermion masses were calculated. (The latter is of particular importance for
$m_b$ which receives significant corrections if tan$\beta$ is large.) 
Also, the experimental constraints imposed by the observed branching ratio $BR(\bsgam)$ 
and by direct sparticle searches were taken into account.
$ \delta a_\mu^{SUSY} $ was evaluated 
for those values of the initial parameters which gave the lowest $\chi^2$ calculated 
out of the ten low energy observables $M_Z$, $M_W$, $\rho_{new}$, $\alpha_s(M_Z)$, $\alpha$,
$G_\mu$, $M_t$, $m_b(M_b)$, $M_\tau$, and $BR(\bsgam)$ \cite{bcrw,br}.

\begin{figure}[b]
\centering\leavevmode
\begin{minipage}{2.7in}
\epsfysize=2.7in
\epsffile{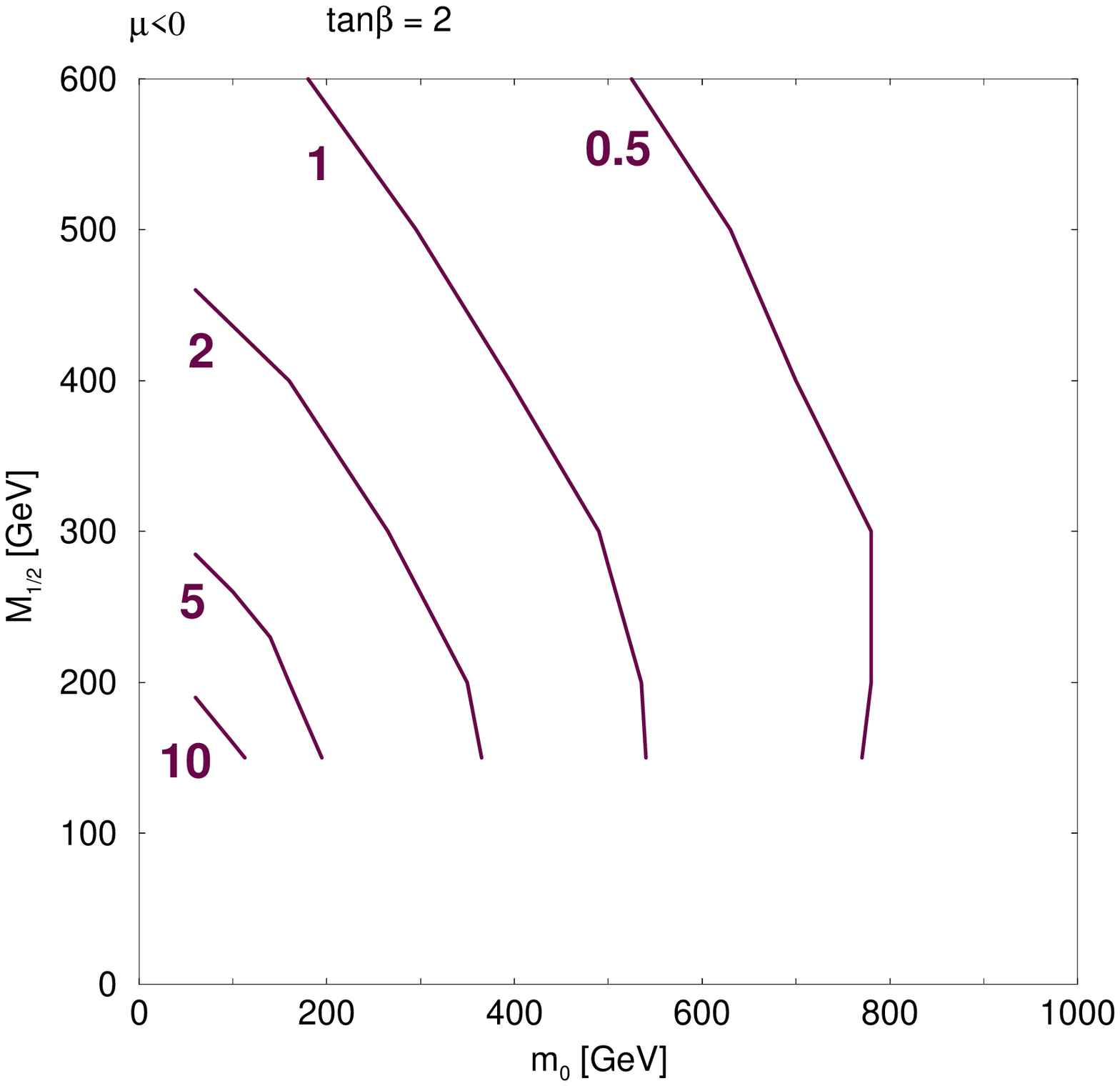}

\caption[]{
Contour lines of constant $\delta a_\mu^{SUSY}\times 10^{10}$ in the MSSM analysis
with fixed tan$\beta=2$. 
\label{famu_tb2_SUSY}}
\end{minipage}
%\end{figure}
\hspace{.5in}
%\begin{figure}
\centering\leavevmode
\begin{minipage}{2.7in}
\epsfysize=2.7in
\epsffile{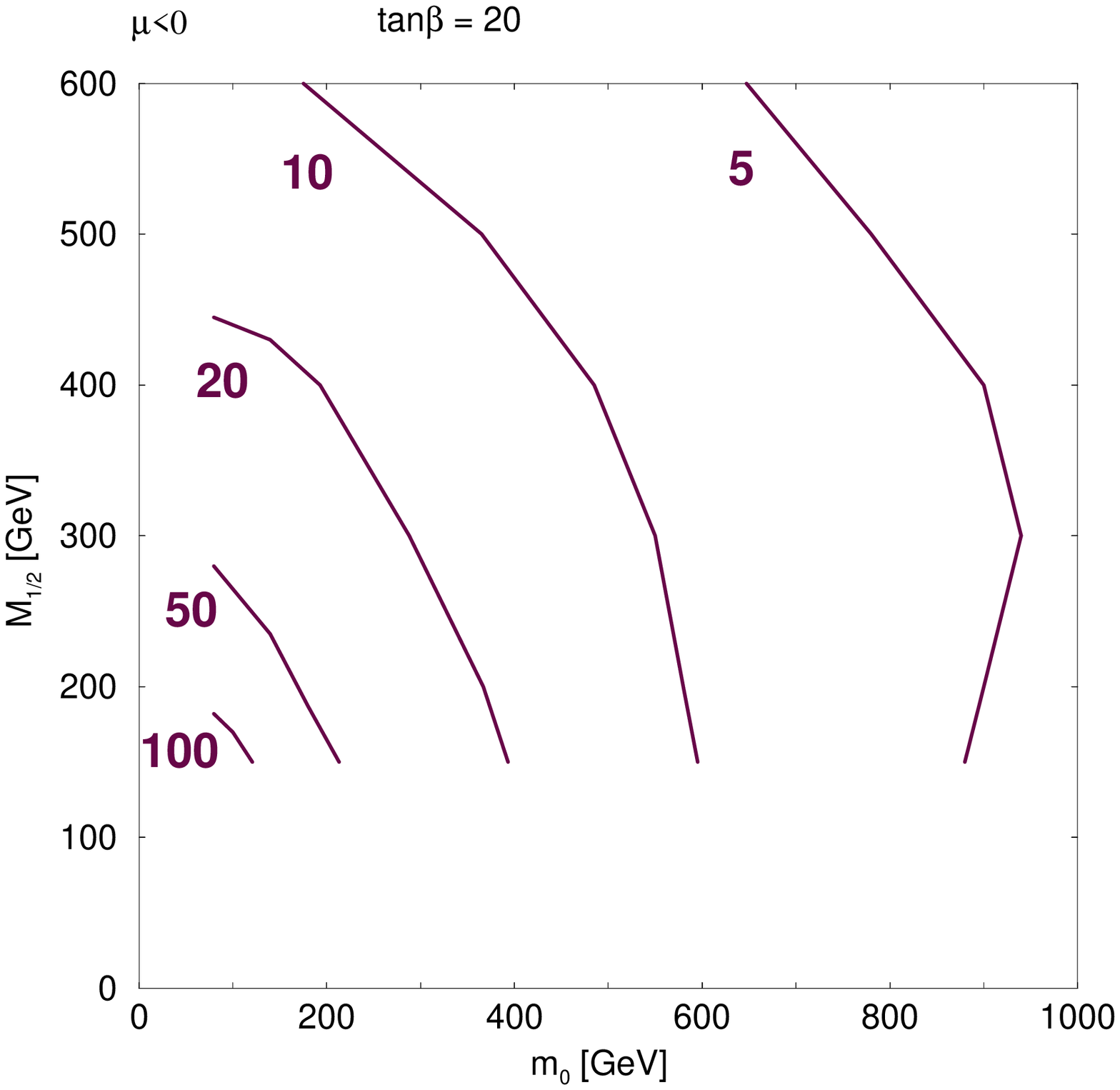}

\caption[]{
Contour lines of constant $\delta a_\mu^{SUSY}\times 10^{10}$ in the MSSM analysis
with fixed tan$\beta=20$.
\label{famu_tb20_SUSY}}
\end{minipage}
\end{figure}

%\section{Results and Conclusions}
%
The results for tan$\beta=2$ and $20$ are shown in figures \ref{famu_tb2_SUSY}--\ref{famu_tb20_SUSY}. 
The $ \delta a_\mu^{SUSY}\times 10^{10} $ contour lines are bound from below
by the limit on the neutralino mass (a limit $m_{\chi^0} > 55$GeV was imposed),  
and from above by the stau mass  ($m_{\tilde{\tau}} > 60$GeV).
The main observation is that
the present limits on $ \delta a_\mu^{SUSY}$, eq.(\ref{da_now}) are far from excluding
any region of the parameter space which is left allowed by other experiments. 
Note the simple pattern suggested by figures \ref{famu_tb2_SUSY} and \ref{famu_tb20_SUSY}
shows how well the approximate relation (\ref{eq:compEW}) works. 
Overlapping the figures the
contour lines marked as 10, 5, 2, 1, and 0.5   in figure \ref{famu_tb2_SUSY}  are 
roughly on top of the contour lines marked as 100, 50, 20, 10, and 5 in figure 
\ref{famu_tb20_SUSY}. The pattern seems to indicate that for tan$\beta=50$ even
the present data (eq.(\ref{da_now})) already start constraining the MSSM analysis.

\begin{figure}
\begin{minipage}{2.75in}
\centering\leavevmode
\epsfxsize=2.75in
\epsffile{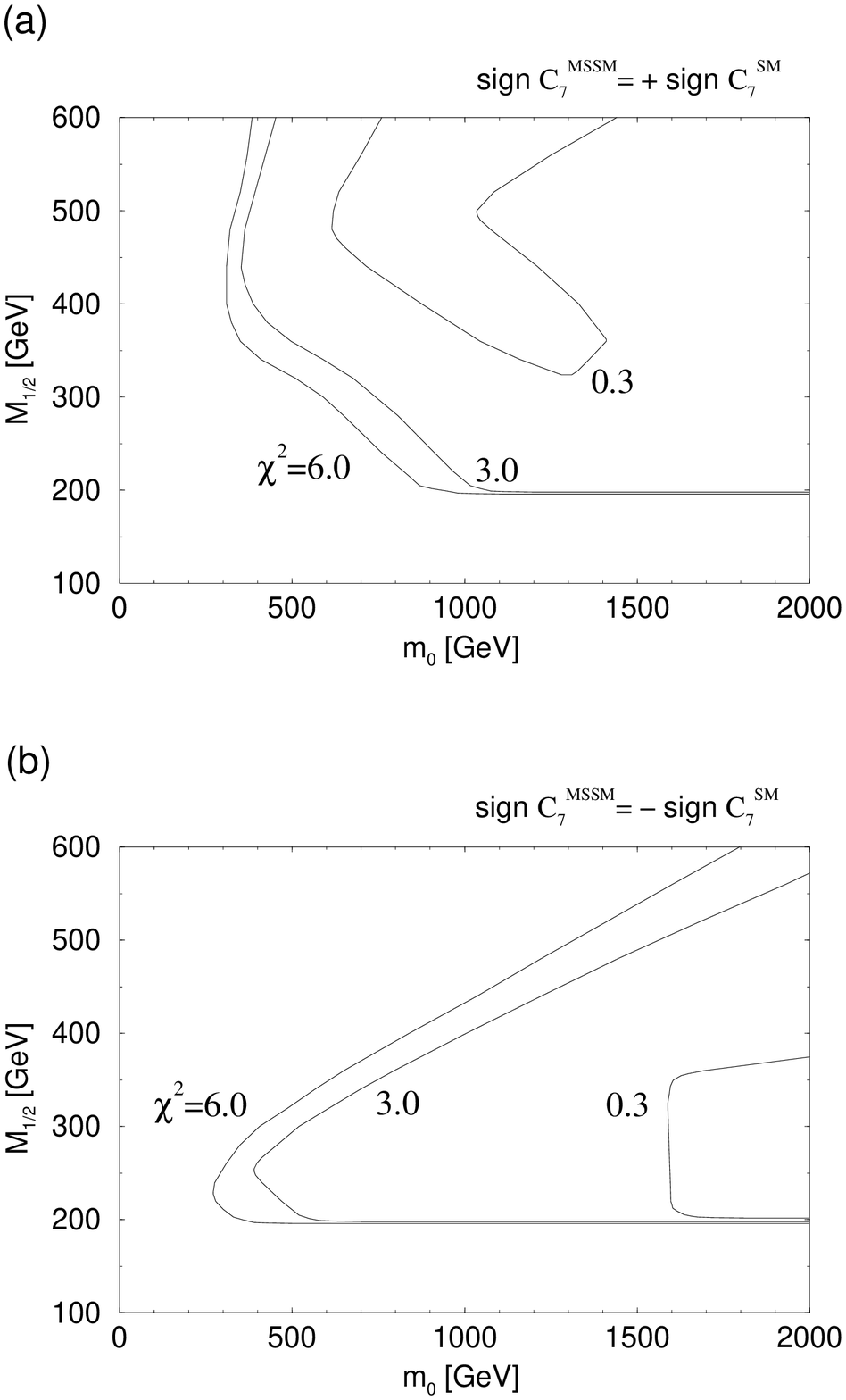}
\caption[]{
Best fit $\chi^2$ contour plots in the SO(10) model analysis, 
with the Wilson coefficient $C_7^{MSSM}$ of 
({\em a}) the same ({\em b}) the opposite sign as compared to $C_7^{SM}$. 
As indicated, the contour lines correspond
to $\chi^2=6,\,3$, and $0.3$ per $3\; d.o.f.$, respectively.
tan$\beta$ varies between 50 and~55.
\label{f_F4c110}}
\end{minipage}
%\end{figure}
\hspace{.5in}
%\begin{figure}[b]
\centering\leavevmode
\begin{minipage}{2.75in}
\epsfxsize=2.75in
\epsffile{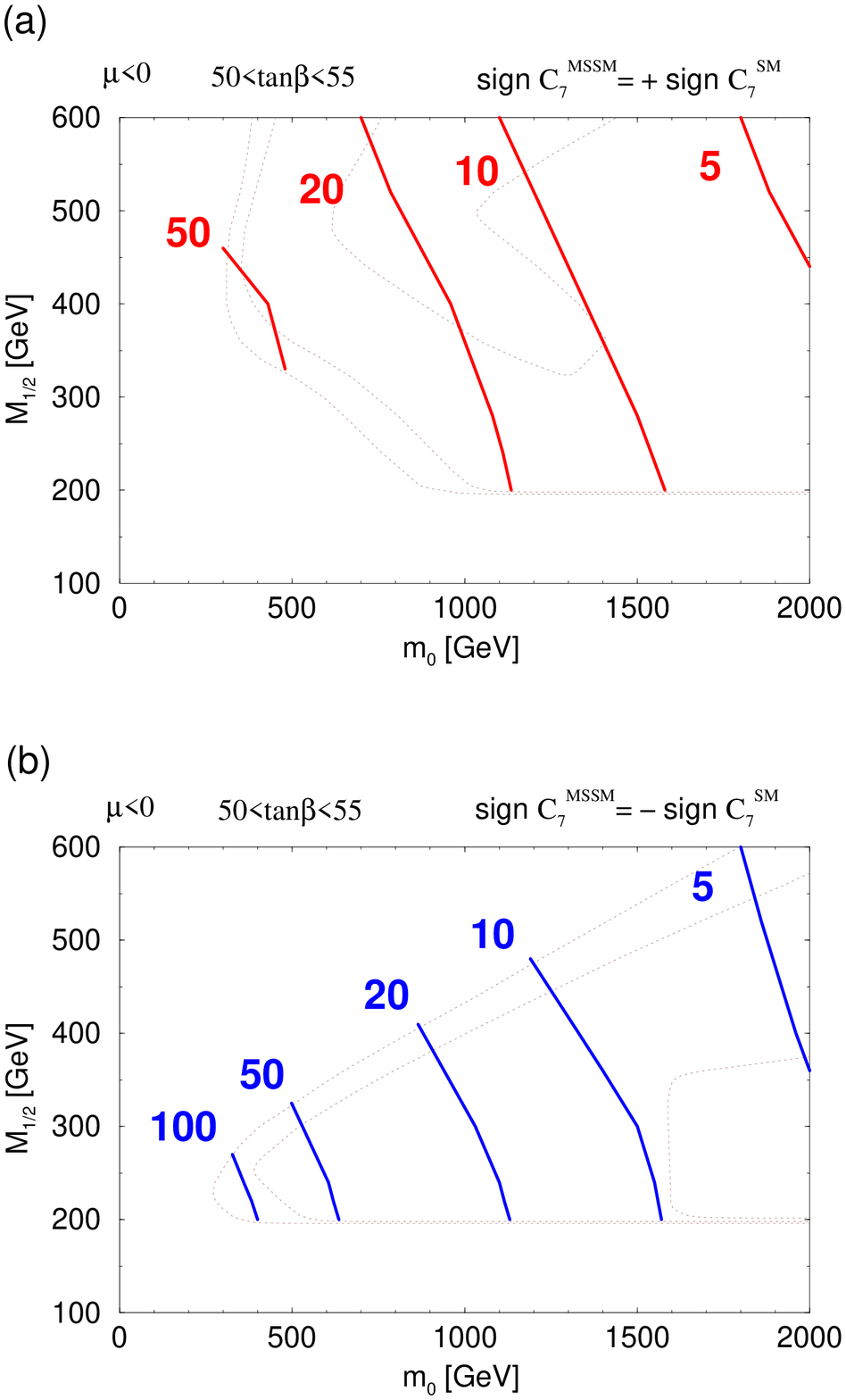}
\caption[]{
Contour lines of constant $\delta a_\mu^{SUSY}\times 10^{10}$ in the 
SO(10) model analysis, 
with the Wilson coefficient $C_7^{MSSM}$ of ({\em a}) the
same ({\em b}) the opposite sign as compared to $C_7^{SM}$. 
For better reference, the $\chi^2$ contour lines 
of figures \ref{f_F4c110}a and \ref{f_F4c110}b
are shown in the background of (a) and (b), respectively, as dotted lines.
\label{famu_F4c}}
\end{minipage}
\end{figure}

The analysis of the case tan$\beta\approx 50$ is qualitatively different.
The global analysis yields two distinct fits \cite{br}, see 
figs. \ref{f_F4c110}a--b, which differ by the sign of the Wilson coefficient 
$C_7^{MSSM}$. ($C_7$ determines the $\bsgam$ decay amplitude.)
In this regime, the sign of $C_7$ can be the same, or the opposite, as in the SM, 
due to the fact that the chargino contribution is enhanced 
by tan$\beta$ compared to the SM. (Thus, the flipped sign of $C_7$ 
cannot be obtained for low tan$\beta$.) 
For large tan$\beta$ however, the fit with the flipped sign is equally good as the fit with the sign 
unchanged, see figure \ref{famu_F4c}. Neglecting the fine-tuning issue, they just differ in the range 
of the allowed parameter space \cite{br}. This difference then affects
 $\delta a_\mu^{SUSY}$ which is more significant in
the fit with the flipped sign of $C_7^{MSSM}$ than in the fit where the signs are the same,
see figs. \ref{famu_F4c}a and \ref{famu_F4c}b.
The most important lesson is that the MSSM contribution to the muon anomalous magnetic moment
again stays within the currently allowed range, eq.(\ref{da_now}), 
assuming all other particle physics constraints are taken into account --- in 
contradiction to the naive extrapolation from figures \ref{famu_tb2_SUSY}-\ref{famu_tb20_SUSY}. 
This is mainly because the allowed SUSY parameter range is further reduced for large 
tan$\beta$ by strong constraints on the $b$ quark mass and the branching ratio $BR(\bsgam)$.
For $\mu > 0$, 
the chargino contribution to $C_7$ changes sign too, which leads to unacceptable
values of $BR(\bsgam)$ for medium and large tan$\beta$. For low tan$\beta$, 
$\delta a_\mu^{SUSY}$ is small and will be hard to observe, regardless
of the sign of $\mu$ \cite{a_mu_MSSM}.

The E821 BNL experiment, however, turns the muon anomalous magnetic moment 
into a powerful MSSM constraint. Figures 
\ref{famu_tb2_SUSY}, \ref{famu_tb20_SUSY}, and \ref{famu_F4c} show that
it will be a major constraint for large and medium tan$\beta$ in the region $m_0 < 400-500$GeV.
Of course, it may drastically affect the MSSM analysis with any value of tan$\beta$ 
if the observed central value turns out to be below (or well above) the SM prediction. 
For such an outcome,
the muon anomalous magnetic moment will actually become a dominant constraint for the MSSM analysis
with universal SUGRA mediated SUSY breaking terms.

%% file: Bhrlik/cp-new.tex
% new version from Toby 9/7/99
% this is the version that was modfied by toby
%
%%%%%%%%%%%%%%%%%%%%%%%%%%%%%%%%%%%%%%%%%%%%%%%%%%%%%%%%%%%%%%%%%%%%%%%%%%%%%%%
\def\doublespaced{\baselineskip=\normalbaselineskip\multiply
    \baselineskip by 200\divide\baselineskip by 100}
\def\singleandabitspaced{\baselineskip=\normalbaselineskip\multiply
    \baselineskip by 120\divide\baselineskip by 100}
\def\singlespaced{\baselineskip=\normalbaselineskip}
\def\gev{{\rm \, Ge\kern-0.125em V}}
\def\m12{m_{1\!/2}}

\def\mol{Mol}
\def\etmiss{E\llap/_T}
\def\eslt{E\llap/_T}
\def\esl{E\llap/}
\def\msl{m\llap/}
\def\to{\rightarrow}
\def\te{\tilde e}
\def\tmu{\tilde\mu}
\def\ttau{\tilde\tau}
\def\tl{\tilde\ell}
\def\ttau{\tilde \tau}
\def\tg{\tilde g}
\def\tnu{\tilde\nu}
\def\tell{\tilde\ell}
\def\tq{\tilde q}
\def\tu{\tilde u}
\def\tc{\tilde c}
\def\tb{\tilde b}
\def\tst{\tilde t}
\def\tt{\tilde t}
\def\tw{\widetilde W}
\def\tz{\widetilde Z}

\hyphenation{mssm}
%\def\ds{\displaystyle}
%\def\ts{${\strut\atop\strut}$}
%
%\include{psfig}
%

%%%%%%%%%%%%%%%%%% MAIN TEXT %%%%%%%%%%%%%%%%%%%%%%%%%%%%%%%%%%%%%%%%%%%%%%%

\section{Effects of CP-violating Phases}

In the minimal supergravity model, all parameters are assumed to be real and
no CP-violating effects can originate in the superpartner sector of the
Lagrangian. Once this assumption is relaxed, $m_{1/2}$, $A_0$ and $\mu$
can become complex, while $\tan\beta$ is still set to be real by a convenient
redefinition of the Higgs fields in order to prevent spontaneous CP violation.
On the other hand, R-symmetry relates the phases of the parameters and
only two of the phases are physically independent. Usually, $\varphi_{\mu_0}$
and $\varphi_{A_0}$ are chosen to supplement the set of GUT scale parameters
defining a non-minimal SUGRA model.

It is important to realize that $\varphi_{\mu}$ does not run as parameters are 
evolved from the GUT scale down to the electroweak scale while the phase of $A$ 
does run and is strongly affected by the magnitude of the gaugino mass terms in 
the RGE's for $A$.
The effect of the $\varphi_{A}$ phases increases with increasing
values of $A_0$. Furthermore, the effects of $\varphi_{A}$
enter through the L-R off-diagonal mixing term in the squark and slepton mass
matrix $m_f (A_f e^{-i\varphi_{A_f}}-\mu e^{i\varphi_{\mu}} f(\beta))$, where 
$f(\beta))=\tan\beta$ (or $\cot\beta$) for $T_3=-1/2$ ($+1/2$). This means that
the phase of $A$ will be irrelevant if $|A_f| \ll |\mu| f(\beta)$.

\begin{figure}[h]
 % "t" for "top"
\vspace*{-1.0in}
\centering
\epsfxsize=4in
\hspace*{0in}
\epsffile{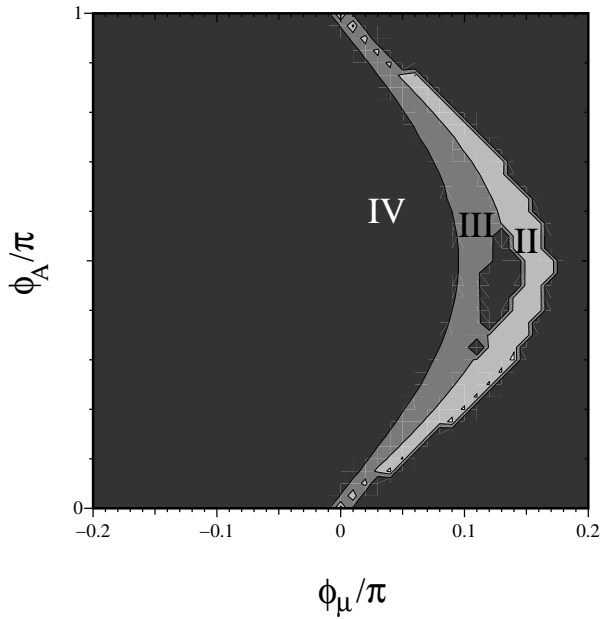}
\vspace*{-0.5in}
\caption{Plot of the required values of $\m12$ to satisfy
the electron and neutron EDM constraints for $\tan\beta=2, A_0=1000\gev$.
 The zones labelled ``II'',
``III'', and ``IV'' correspond to $200\gev<\m12^{\rm min}<300\gev$,
$300\gev<\m12^{\rm min}<450\gev$, and $\m12^{\rm min}>450\gev$,
respectively. }
\label{cp-fig1}
\end{figure}

Experimental upper limits for electron and neutron electric dipole moments
impose severe constraints on the values of $\varphi_{A_0}$ and
$\varphi_{\mu_0}$ if they are considered separately, requiring, e.g.,  
$\varphi_{\mu_0} \lsim 0.01$ for $m_{1/2} \lsim 1$~TeV.
Recently, it has been pointed out \cite{nath} that cancellations
between different contributions to the electric dipole moments can permit
large values of these phases.
As an example of the effect of cancellations
between various contributions to the neutron and electron EDMs,
Fig.~\ref{cp-fig1} displays for $\tan\beta=2, A_0=1000\gev$ the
minimum value of $\m12$ in mSUGRA required to bring both the neutron and
electron EDMs below their experimental limits (there is an identical
region with $\varphi_A\rightarrow-\varphi_A,
\varphi_\mu\rightarrow-\varphi_\mu$) \cite{olive}.  The cancellations persist  
for a limited
range in $\m12$, particularly at larger $\varphi_\mu$, and the allowed regions
are between 20 and $40 \gev$ wide in $\m12$ for $\varphi_\mu/\pi >
0.14$.  Larger values of $A_0$ permit larger $\varphi_\mu$, but for a
smaller range in $\m12$, while smaller $A_0$ requires smaller
$\varphi_\mu$.  Note that $\varphi_{\mu_0}>0.17 \pi$ in this example would  
require a heavy
sparticle spectrum. The maximum $\varphi_\mu$ scales roughly like
$1/\tan\beta$.  The maximum value of $\varphi_{A_0}$ is
virtually unrestricted even for relatively light spectra in the sense that
for any value of $\varphi_{A_0}$ there exists such a value of $\varphi_{\mu_0}$
and a range of $m_{1/2}$ as above so that the electric dipole moment  
constraints are satisfied.

The implications of the phases are significant from the point of view of
reconstructing the parameters of the Lagrangian from experimental data
\cite{cp-bk}. For instance, in the chargino sector the presence of phases makes it
impossible to correctly determine all parameters just by measuring the chargino
masses and production cross sections as the number of parameters exceeds the
number of observables. For this to be possible a measurement at
an $e^+e^-$ collider with a polarized beam would be necessary. Also the Higgs 
sector is affected by the phases and the mass of the light Higgs as well as
its production cross section can vary substantially with the variation of the 
phases.

Another important issue is the impact of the phases on the lightest
neutralino relic density and direct detection rates \cite{cp-bk,gor}.
The annihilation cross
section of the lightest neutralino depends on $\varphi_{\mu}$ and as a result 
the relic density depends on the allowed range of
$\varphi_{\mu}$. Although the lightest
neutralino in SUGRA is an almost pure bino, the Higgsino admixture is very
important for the direct detection rate calculation since the heavy Higgs
exchange contribution dominates the scalar cross section. The rate can then
vary even by a factor of two or more within the SUGRA models depending on
the value of $\varphi_{\mu}$.

Minimal SUGRA is a model with a
simplifying set of assumptions, and when one steps away from
those assumptions, by relaxing the sfermion mass constraints, for
example, or by including complex phases, some of the analyses will
change (see for example \cite{cgss}, for an analysis of the effect of phases on the
trilepton signature). Therefore one should cautiously interpret  simulations,
efficiencies or analyses of data based on that model, until we know that
its assumptions are confirmed experimentally.
It should be noted here that when all the phases are included in the analysis
of the electric dipole moment limits, none of them are required to be small
\cite{bgk}, so they should all be measured once there is data.

%%%%%%%%%%%%%%%%%%%%% REFERENCES %%%%%%%%%%%%%%%%%%%%%%%%%%%%%%%%%%%%%%%%%%%%%%
%

%

%% file: Cheng/r2report.tex
\newcommand{\lae}{\begin{array}{c}\,\sim\vspace{-21pt}\\< \end{array}}
\newcommand{\gae}{\begin{array}{c}\,\sim\vspace{-21pt}\\> \end{array}}

\section{Flavor Violation}

If the soft squark and slepton masses are non-universal,
then in general they do not have to be diagonal in
the same basis as the quark and lepton mass matrices. 
In that case, they can induce flavor-changing effects.
The low energy flavor-changing processes, such as 
$K-\bar{K}$ mixing, $\mu \to e\gamma$, provide some
probes and constraints on the non-degeneracies and mixings
among different generation sfermions. On the other hand,
the flavor-changing effects can also show up when the 
sfermions are directly produced. The superGIM mechanism
suppresses the flavor-changing effects only when $\Delta m
\lsim \Gamma$ for real sfermions, where $\Delta m$ is the mass 
difference between sfermions of different generations and $\Gamma$ is the decay width of the sfermions, in constrast
with $\Delta m < m$ for virtual sfermions in the low energy 
processes\cite{ACFH}. Therefore, for $\Gamma \lsim \Delta m
\ll m$, direct production could be a powerful probe of the
flavor mixings in the sfermion sector.

However, detecting squark flavor mixings is difficult because
it is hard to disentangle them from the quark flavor mixings which
are already present in the Standard Model. In addition, only
heavy quarks can be tagged. On the other hand, lepton flavor
is conserved in the Standard Model in the absence of neutrino 
masses. Even in the
presence of neutrino masses and mixings, the induced 
flavor-violating effects for the charged leptons in the rare decay
processes or at colliders are too small to be observed, due
to the tiny neutrino masses. Therefore, any observed lepton
flavor violation in these processes implies new flavor-violating
physics beyond the Standard Model.

Lepton flavor violation from slepton oscillations at 
lepton colliders have been studied in Ref.~\cite{ACFH,LFV}.
It is found that the direct slepton production often provides
a more powerful probe than low energy processes, such as
$\mu \to e\gamma$, in the $\Delta m-\sin 2\theta$ plane. The
mixing angle $\sin \theta$ between sfermions of different generations can be probed down to $10^{-1}-10^{-2}$,
which is quite interesting in comparison with the CKM matrix
elements.

At the upgraded Tevatron, the sleptons are predominantly pair
produced through Drell-Yan processes, mediated by a $W, \;
\gamma,$ or $Z$ in the $s$-channel. The production cross sections
and the possibility of detection were studied in 
Ref.~\cite{BCPT,BHR}. The conclusion is that the detection of
sleptons at the Tevatron is very difficult, unless sleptons are 
very light. As a result, for lepton flavor violation to have
any chance to be observed at the Tevatron, sleptons must be
light and have very large mixings among them.

%\looseness=-1
To get a more quantitative idea of the reach at the upgraded
Tevatron, we consider the following MSSM parameters,
$m_{\tilde{l}_R}=100$~GeV, $m_{\tilde{l}_L}=110$~GeV,
$M_1=60$~GeV, $M_2=120$~GeV, $\mu=-300$~GeV, $\tan \beta =2(30)$;
these parameters do not satisfy the minimal supergravity boundary
condition. Lighter sleptons can be 
produced at LEP II, which will do a better job in probing
lepton flavor violation anyway. For heavier sleptons, the
cross section will be too small for the sleptons to be observed.
For this choice of parameters, both left and right handed sleptons
decay directly to (lepton+LSP), so the signal for slepton
production is $l\bar{l} \not\!\! E_T$. If two leptons are of
different flavors, it is a signal for flavor violation.
If the chargino decay channels for the left handed sleptons
are open, the branching fractions of the signal will be reduced.
The major background to the
dilepton signature comes from $W$ pair production. Other backgrounds
from $t\bar{t}$ production, from $Z\to \tau\bar{\tau}$, and from
SUSY processes such as chargino pair production are also calculated in
Ref.~\cite{BCPT}, and they are much smaller. The two leptons are required 
to be energetic and isolated to veto the Standard Model backgrounds
from $W$ pair, $\tau \bar{\tau}$, and top quark pair production.
The cuts imposed in Ref.~\cite{BCPT} are $p_T(l) > 15$~GeV, missing
tranverse energy $\not\!\! E_T > 20$~GeV, no
jet, and transverse opening angle $30^\circ < \Delta \phi(l\bar{l})
\leq 150^\circ$. From Ref.~\cite{BCPT}, including the next-to-leading
(NLO) order QCD corrections\cite{BHR}, the production cross sections
for a single generation of charged sleptons in the flavor-conserving
case are $\sigma_{0L}(\tilde{e}_L \tilde{e}_L) \sim 21$~fb,
$\sigma_{0R}(\tilde{e}_R \tilde{e}_R) \sim 14$~fb, for the above
slepton mass parameters. The major background from $W$ pair
production is about $\sigma_B \sim 47$~fb (including NLO QCD corrections)
after cuts. For a total integrated luminosity $L=\int {\cal L} dt$
and signal efficiency $\epsilon$, the $3\sigma$ significance requires
\begin{equation}
\sigma(\tilde{l}\tilde{l}') > \frac{3}{\epsilon}
\sqrt{\frac{\sigma_B}{L}} \approx 14 \left(\frac{30\%}{\epsilon} \right)
\sqrt{\frac{25 {\rm fb}^{-1}}{L}} {\rm fb} .
\end{equation}

\begin{figure}[h]
\centerline{\psfig{file=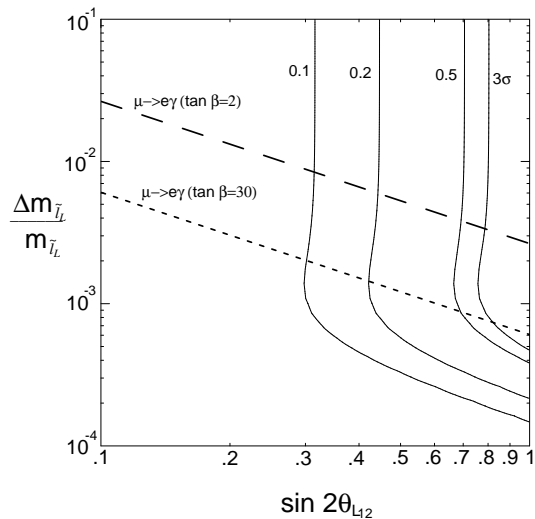,width=3in}}
\caption[]{The constant contours of the ratio of the flavor-violating slepton
production cross section $\sigma_{L}
(\tilde{e}_{L}^{\pm} \tilde{\mu}_{L}^{\mp})$ to $\sigma_{0L}$,
The MSSM parameters are chosen to be $m_{\tilde{l}_R}=100$~GeV,
$m_{\tilde{l}_L}=110$~GeV, $M_1=60$~GeV, $M_2=120$~GeV,
$\mu=-300$~GeV, $\tan\beta=2(30)$. The $3\sigma$ contour assumes
the integrated luminosity $L=25 {\rm fb}^{-1}$ and signal efficiency
$\epsilon=30\%$. The current constraints from $B(\mu \to e\gamma)$
for $\tan\beta=2$ and $30$ are also shown; the regions above the
dashed lines are ruled out. \label{cheng-f1}}
\end{figure}

\begin{figure}[h]
\centerline{\psfig{file=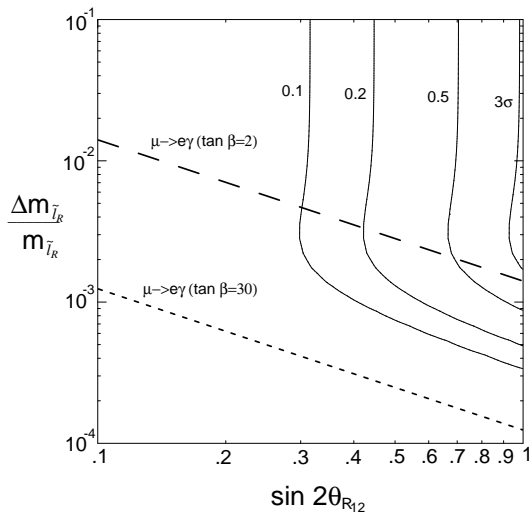,width=3in}}
\caption[]{Same as Figure \ref{cheng-f1}, except for right handed sleptons. \label{cheng-f2}}
\end{figure}

In Figures \ref{cheng-f1} and \ref{cheng-f2} we plot the contours of constant ratios of 
flavor-violating slepton production cross sections $\sigma_{L(R)}
(\tilde{e}_{L(R)}^{\pm} \tilde{\mu}_{L(R)}^{\mp})$ to 
$\sigma_{0L(R)}$
(we assume 2 generation mixing for simplicity) in the 
$\sin 2\theta_{L(R)} - \Delta m_{\tilde{l}_{L(R)}}$ plane.
The $3\sigma$ contours for $\epsilon=30\%$ and $L=25\mbox{ fb}^{-1}$
are also shown. In addition, we superimpose the current constraint
of $B(\mu \to e\gamma)< 4.9\times 10^{-11}$ for $\tan \beta =2$
(dashed line) and 30 (dotted line). The areas above these lines are ruled out.
We can see that even for an integrated luminosity as high as
$L=25\mbox{ fb}^{-1}$, lepton
flavor violation be observed only for a limit range of $\Delta m$
in small $\tan \beta$ and nearly maximal mixing. Therefore, we conclude that
probing SUSY flavor violation is not very promising at the
upgraded Tevatron. However, if flavor violation is seen at the
upgraded Tevatron, it will point to a very specific and interesting
region of the parameter space.

%% file: Keung/cp-top.tex
\section{3rd Generation Scalar Quarks and
Electric Dipole Moments in Supersymmetric Theories}

%\begin{abstract}
CP violation  in the trilinear couplings   of the Higgs bosons  to the
scalar-top or the scalar-bottom quarks may lead  to large loop effects
of CP noninvariance in the  Higgs sector of the minimal supersymmetric
standard model (MSSM).    These third-generation Yukawa  couplings may
directly be constrained by electric dipole moments induced by two-loop
Barr-Zee-type   graphs. Our  analysis   shows  that large  $\tan\beta$
scenarios with $\tan\beta >  40$, $\mu, A >  0.5$~TeV and large CP-odd
phases are highly disfavoured.
%\end{abstract}

Supersymmetric   (SUSY)   theories  including  the  MSSM  \cite{GD,NS}
generally require some  degree of fine  tuning in order to account for
the small flavour-changing neutral currents (FCNC) involving the first
two families of quarks, and the  absence of any electric dipole moment
(EDM) of   the   neutron  and  electron  \cite{BS}.    Presently,  the
experimental limits  on the neutron  EDM $d_n$  and electron EDM $d_e$
are   very  stringent \cite{PDG},  i.e.\ $|d_n|  <  10^{-25}\ e$cm and
$|d_e|  < 10^{-26}\ e$cm.   Therefore, solutions  to  the CP  and FCNC
problems are mainly  based  on  suppressing the contributions  of  the
first two families  of   scalar  quarks  without affecting much    the
paramater space of the third generation.

In addition to the usual CP-odd phase in the Cabbibo-Kobayashi-Maskawa
(CKM) matrix and the strong  QCD phase $\theta$, the MSSM supplemented
by the universality condition at a unification  scale predicts two new
CP-violating phases \cite{EFN}.  Specifically, two of the four complex
parameters $\{  \mu ,\ B,\ m_\lambda ,\   A \}$ are field-independent,
where $\mu$    is the  mixing   parameter of    the  two  Higgs chiral
superfields    in  the superpotential,    $B\mu$  is the soft bilinear
Higgs-mixing mass, $m_\lambda$ represents the gaugino mass, and $A_f =
A$ is the universal soft trilinear coupling of the Higgs fields to the
scalar fermions $\tilde{f}$.  These can be chosen to be $\mu$ and $A$.
In such a weak basis, a qualitative estimate of the combined effect of
all one-loop  contributions to the electron and  $u$-, $d$- quark EDMs
may be given by \cite{EFN,KO,FOS,IN}
\begin{equation}
  \label{estim}
\Big( \frac{d_f}{e}\Big)^{\rm 1-loop}\ \sim\ 10^{-25}\ {\rm cm}\times
\frac{\{ {\rm Im}\ \mu ,\ {\rm Im}\  A_f \} }
{\max (M_{\tilde{f}}, m_{\lambda} )}\
\Big(\, \frac{1\ {\rm TeV}}{\max (M_{\tilde{f}}, m_{\lambda} )}\,\Big)^2
\Big(\, \frac{m_f}{10\ {\rm MeV}}\,\Big)\ .
\end{equation}
Eq.\ (\ref{estim}) leads to the known conclusion \cite{KO} that in the
MSSM, large CP-violating phases are  only possible if scalar quarks of
the   first two families or   gauginos have masses in   the TeV range.
However,   there is also another   interesting scheme  to suppress the
one-loop EDM contributions and have  most  of the SUSY particles  much
below  the TeV    scale.    One may    require that ${\rm    arg}(\mu)
\stackrel{\displaystyle <}{\sim }  10^{-2}$, which is also favoured by
dark-matter constraints \cite{FOS},  and assume  an hierarchic pattern
for  $A_f$'s, {\em e.g.},  $[A] = {\rm  diag} (\epsilon,\epsilon,1) A$
with $\epsilon  \stackrel{\displaystyle  <}{\sim  }  10^{-3}\,  \mu/A$
\cite{CKP}.    In this scheme, $A_\tau=A_t=A_b=A$  are  the only large
trilinear couplings in the theory   with CP-violating phases of  order
unity.  We should  note that  there is  also a  significant three-loop
contribution to  neutron EDM  through Weinberg's three-gluon  operator
\cite{cp-SW,DDLPD}.  However, these  effects scale as $1/m_{\tilde{g}}^3$
and are  therefore well below the  experimental  neutron EDM  bound if
gluinos are heavier than about  400 GeV \cite{IN}.  Allowing for large
CP-violating trilinear couplings of the Higgs bosons to the scalar-top
and scalar-bottom  quarks is   a rather interesting   phenomenological
scenario, since such Yukawa couplings can lead  by themselves to large
loop  effects of CP noninvariance  in  the Higgs  sector  of the MSSM.
More details may be found in \cite{APLB}.

\begin{figure}

\begin{center}
\begin{picture}(400,150)(0,0)
\SetWidth{0.8}

\ArrowLine(10,30)(45,60)\Text(30,35)[lb]{$f_L$}
\DashLine(45,60)(68,85){5}\Text(55,75)[rb]{$a$}
\Photon(92,85)(115,60){3}{3}\Text(105,80)[l]{$\gamma,g$}
\ArrowLine(45,60)(115,60)\Text(80,51)[l]{$f_R$}
\ArrowLine(115,60)(150,30)\Text(118,35)[lb]{$f_R$}
\Photon(80,120)(80,150){3}{3}\Text(86,140)[l]{$\gamma,g$}
\DashArrowArc(80,100)(20,0,360){3}\Text(55,100)[r]{$\tilde{q}_1,\tilde{q}_2$}

\Text(75,20)[]{\bf (a)}

\ArrowLine(210,30)(245,60)\Text(230,35)[lb]{$f_L$}
\DashLine(245,60)(268,85){5}\Text(255,75)[rb]{$a$}
\Photon(301,100)(315,60){3}{4}\Text(315,80)[l]{$\gamma,g$}
\Photon(301,100)(315,130){3}{3}\Text(292,130)[l]{$\gamma,g$}
\ArrowLine(245,60)(315,60)\Text(280,51)[l]{$f_R$}
\ArrowLine(315,60)(350,30)\Text(318,35)[lb]{$f_R$}
\DashArrowArc(280,100)(20,0,360){3}
\Text(255,100)[r]{$\tilde{q}_1,\tilde{q}_2$}

\Text(275,20)[]{\bf (b)}

\end{picture}
\end{center}

\caption{Two-loop contribution to EDM and CEDM in supersymmetric
  theories (mirror graphs are not displayed.)}\label{f1}
\end{figure}
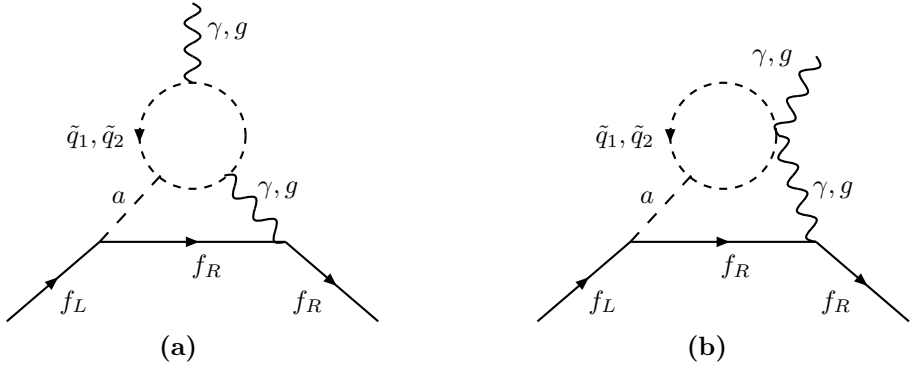

Most recently, we  have  found  \cite{CKP} that  CP-violating   Yukawa
couplings involving the third   generation of scalar quarks  can  also
give rise to EDMs of electron  and neutron at  the observable level by
virtue of the  two-loop  Barr-Zee-type graphs \cite{BZ,CKY}   shown in
Fig.\ \ref{f1}.  The two-loop  EDM effects are  rather enhanced in  the large
$\tan\beta$ regime.   However,  we should stress  that  apart from the
MSSM, these novel EDM  contributions are present in any supersymmetric
theory.  In particular,  even  if the   first two  families of  scalar
quarks are arranged so as to give small  effects on EDMs, the two-loop
Barr-Zee-type graphs may even dominate  by several orders of magnitude
over   all other one-,  two-   and three-loop contributions  discussed
extensively            in        the     existing           literature
\cite{EFN,KO,FOS,IN,DDLPD,TKNO}. The  SUSY  scenario we  have in  mind
contains a large CP-violating phase in the third  family $A_\tau = A_t
= A_b   =  A$, and CP violation   is  induced through  the interaction
Lagrangian having the generic form
\begin{equation}
  \label{Lcp}
{\cal L}_{\rm CP}\ =\ -\, \xi_f v\, a\, (\tilde{f}_1^* \tilde{f}_1\, -\,
\tilde{f}_2^* \tilde{f}_2)\
+\ \frac{ig_w m_f}{2M_W}\, R_f\, a\,\bar{f}\gamma_5 f\, ,
\end{equation}
where  $A$ is the would-be CP-odd  Higgs boson, $M_W = {1\over2}g_w v$
is the $W$-boson mass, $\tilde{f}_{1,2}$  are the two mass-eigenstates
of  scalar quarks  of third  family, $R_b=\tan\beta$, $R_t=\cot\beta$, 
and $\xi_f$  is a model-dependent
CP-violating parameter. In the MSSM,  only $\tilde{t}$ and $\tilde{b}$
are  expected   to  give   the largest  contributions,   as  the these
quantities  $\xi_t$  and $\xi_b$ are  Yukawa-coupling  enhanced,  viz.
\cite{APLB}
\begin{equation}
  \label{xiq}
\xi_t\ =\ \frac{\sin 2\theta_t m_t {\rm Im} ( \mu
  e^{i\delta_t})}{\sin^2\beta\ v^2}\,,\qquad
\xi_b\ =\ \frac{\sin 2\theta_b m_b {\rm Im} ( A_b
  e^{-i\delta_b})}{\sin\beta\, \cos\beta\, v^2}\, ,
\end{equation}
where  $\delta_q =   {\rm arg}  (A_q + R_q   \mu^*)$,  and $\theta_t$,
$\theta_b$ are the mixing angles  between weak and mass eigenstates of
$\tilde{t}$ and $\tilde{b}$, respectively.
%******************************************************************
%%%Figure 1
%******************************************************************

In  the   calculation of  the  $\tilde{t}$-  and  $\tilde{b}$-mediated
two-loop graphs   shown in  Fig.\ \ref{f1},   we  have neglected  subdominant
contributions emanating from  analogous Barr-Zee-type graphs where the
photon is replaced by  a $Z$ boson, since  the  vectorial part  of the
$Z$-boson-mediated interaction is  suppressed relative to the photonic
one  by a factor $(1 -  4\sin^2\theta_w )/4 \approx 2.4\ 10^{-2}$ with
$\cos\theta_w =  M_W/M_Z$ for the electron case,  and is  smaller by a
factor  $1/4$  for the $u$  and  $d$  quarks.   Taking  the above into
consideration, we calculate the EDM of a light fermion $f$
\begin{equation}
  \label{EDMf}
\left( \frac{d_f}{e}\right)^{\tilde{q}}_{\rm EW} \ =\ Q_f\,
\frac{3\alpha_{\rm em}}{32\pi^3}\, \frac{R_f\, m_f}{M^2_a}
 \sum_{q = t,b}\
\xi_q\, Q^2_q\,\left[\,
F\left({M^2_{\tilde{q}_1} \over M^2_a}\right)
 -
F\left({M^2_{\tilde{q}_2} \over M^2_a}\right) \right]\, ,
\end{equation}
where  $\alpha_{\rm  em}  = e^2/(4\pi)$  is   the electromagnetic fine
structure constant and all  kinematic parameters are evaluated  at the
electroweak  (EW)  scale. In Eq.\  (\ref{EDMf}),  $F(z)$ is a two-loop
function given by
\begin{equation}
  \label{Fz}
F(z)\ = \ \int_0^{1} dx\ \frac{x(1-x)}{z - x(1-x)}\
\ln \Big[\,\frac{x(1-x)}{z}\,\Big]\ ,
\end{equation}
which  may  also be  expressed  in  terms  of dilogarithmic  functions
\cite{CKP}; $F(z\ll 1) = \ln z + 2$; $F(z\gg 1)  = - {1\over6}(\ln z +
{5\over3} )/z$.   Furthermore,  the  two-loop   Barr-Zee  type  graphs
depicted in Fig.\ \ref{f1} also give rise  to $u$- and $d$- chromo-EDMs, {\em
  i.e.},
\begin{equation}
  \label{CEDMf}
\left( \frac{d^C_f}{g_s}\right)^{\tilde{q}}_{\rm EW} =\
\frac{\alpha_s}{64\pi^3}\, \frac{R_f\, m_f}{M^2_a}\ \sum_{q = t,b}\
\xi_q\, \left[
F\left( {M^2_{\tilde{q}_1} \over M^2_a} \right) -
F\left( {M^2_{\tilde{q}_2} \over M^2_a} \right)
       \right]\, .
\end{equation}
Using the valence quark  model, we can  now  estimate the neutron  EDM
$d_n$ induced by   $d_u$ and $d_d$  at  the hadronic scale   $\Lambda$
including QCD renormalization effects \cite{CKY}. The neutron EDM is
given by
\begin{equation}
  \label{dne}
\frac{d_n}{e}\ \approx\
\left(\frac{g_s (M_Z )}{g_s (\Lambda )}\right)^{32\over23}
\left[\, \frac{4}{3}\,\left( \frac{d_d}{e}\right)_\Lambda
 - \frac{1}{3}\,\left( \frac{d_u}{e}\right)_\Lambda\, \right]\, .
\end{equation}
Note that in Eq.\ (\ref{dne}), all the anomalous dimension factors are
explicitly  separated   out   from quantities  $(d_d/e)_\Lambda$   and
$(d_u/e)_\Lambda$ which are simply given by Eq.\ (\ref{EDMf}) with the
running  couplings  and the   running masses of   $u$-  and $d$-quarks
evaluated   at the  low-energy    hadronic   scale  $\Lambda$.  We
take $m_u (\Lambda ) = 7$  MeV, $m_d (\Lambda ) = 10$
MeV, $\alpha_s (M_Z) = 0.12$, and $g_s (\Lambda )/(4\pi) = 1/\sqrt{6}$
\cite{cp-SW}. Likewise, the light-quark  CEDMs $d_u^C$ and $d^C_d$ give
rise to a neutron EDM
\begin{equation}
  \label{cdne}
\frac{d^C_n}{e}\ \approx\
\left(\frac{g_s (M_Z )}{g_s (\Lambda)}\right)^{74\over23}
\left[\,  \frac{4}{9} \left( \frac{d^C_d}{g_s}\right)_\Lambda
      + \frac{2}{9} \left( \frac{d^C_u}{g_s}\right)_\Lambda
      \,\right]\, ,
\end{equation}
where quantities  in the last bracket are  given by Eq.\ (\ref{CEDMf})
with the strong  coupling constant  $g_s$  and the $u$-  and $d$-quark
masses evaluated at the scale $\Lambda$.

\begin{figure}[h]
\centering\leavevmode
  \epsfxsize=11.5cm
    \epsffile{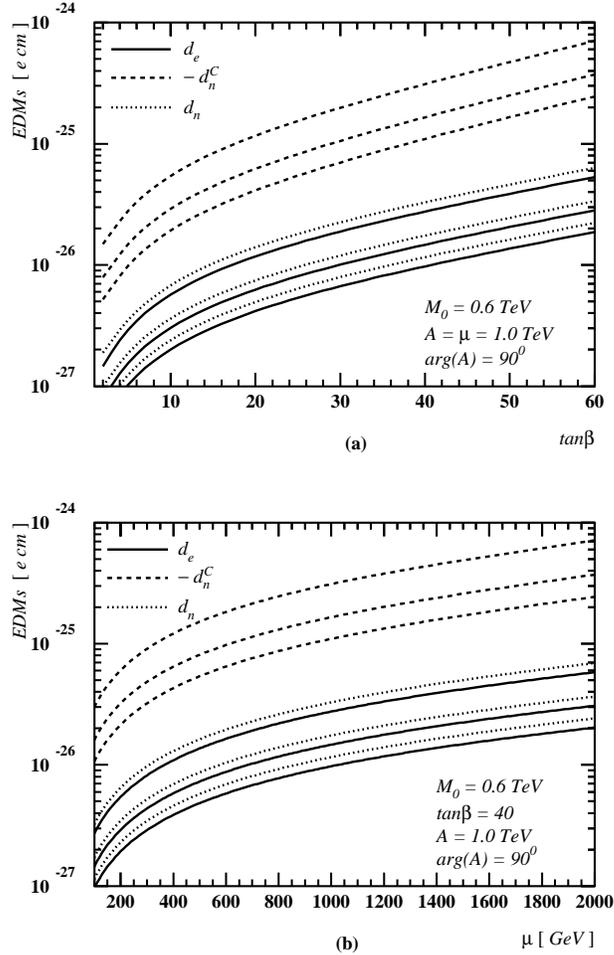}

\caption{Numerical estimates of EDMs. Lines of the same type
from the upper to the lower one correspond to $M_a = 100,\ 300,\
500$ GeV, respectively.}\label{f2}
\end{figure}

Figure \ref{f2} shows the dependence  of the EDMs  $d_e$ (solid line), $d^C_n$
(dashed line),  and $d_n$ (dotted line)  on $\tan\beta$ and $\mu$, for
three different masses of the would-be CP-odd  Higgs boson $A$, $M_A =
100, 300, 500$  GeV.  Since  the coupling  of   the $a$ boson  to  the
down-family  fermions   such as the  electron   and  $d$ quark depends
significantly on $\tan\beta$,  we find a  substantial increase of $d_n$
and $d_e$ in the large $\tan\beta$ domain (see Fig.\ \ref{f2}(a)).  As can be
seen  from  Fig.\ \ref{f2}(b),   EDMs   also depend   on  $\mu$  through  the
$a\tilde{f}^*\tilde{f}$   coupling in  Eq.\  (\ref{Lcp}).  {}From  our
numerical analysis, we  can exclude large $\tan\beta$ scenarios,  {\em
i.e.}, $40< \tan\beta < 60$ with $\mu,\ A > 0.5$  TeV, $M_a \le 0.5$
TeV, and large  CP  phases.  Nevertheless, the situation  is different
for     low    $\tan\beta$    scenarios,    {\em   e.g.}\   $\tan\beta
\stackrel{\displaystyle <}{\sim} 20$, where the two-loop Barr-Zee-type
contribution to EDMs is not   very restrictive for natural values   of
parameters   in   the MSSM.   Finally,  EDMs   display  a  mild linear
dependence  on  the mass of the  $a$  boson for the  range of physical
interest, $0.1< M_a \stackrel{\displaystyle <}{\sim} 1$ TeV.

%******************************************************************
%%%Figure 2
%******************************************************************

In summary, we   have found that   the  SUSY version of the   two-loop
Barr-Zee  mechanism (see also Fig.\ \ref{f1})   induces observable effects on
the EDMs of electron and  neutron. In this way,  we  were able to  put
strict limits  on  the CP-violating  parameters related   to the third
generation scalar    quarks;  especially   the  limits  become    very
significant for large $\tan\beta$ scenarios \cite{CKP}.  Obviously, these
novel constraints will have an important impact on possible effects of
CP   violation at   collider and  other    low-energy observables,  on
dark-matter detection rates and searches, and  on baryogenesis in SUSY
theories.

%% file: Kamon-runii-param/runii-new.tex
\renewcommand{\schitwozero }{\mbox{$\tilde{\chi}_{2}^{0}$}}
\renewcommand{\schionepm }{\mbox{$\tilde{\chi}_{1}^{\pm}$}}
\renewcommand{\lsp}    {\mbox{$\tilde{\chi}_{1}^{0}$}}
\renewcommand{\mets}{\mbox{${E\!\!\!\!/_T}$}}
\renewcommand{\schionezero }{\mbox{$\tilde{\chi}_{1}^{0}$}}
\renewcommand{\pt}{\mbox{$p_{T}$}}
\renewcommand{\et}{\mbox{$E_T$}}
\renewcommand{\chione }{\mbox{$\tilde{\chi}_{1}^{\pm}$}}
\renewcommand{\chitwo }{\mbox{$\tilde{\chi}_{2}^0$}}
\renewcommand{\mgev}{\mbox{$\;{\rm GeV}/c^2$}}
\renewcommand{\gluino} {\mbox{$\tilde{g}$}}
\renewcommand{\squark} {\mbox{$\tilde{q}$}}
\renewcommand{\pgev} {\mbox{$\;{\rm GeV}/c$}}
\renewcommand{\intlum}{\mbox{${ \int {\cal L} \; dt}$}}
\renewcommand{\invpb}{\mbox{$\;{\rm pb}^{-1}$}}
\def \gtsim    {\relax\ifmmode{\mathrel{\mathpalette\oversim >}}
                  \else{$\mathrel{\mathpalette\oversim >}$}\fi}
\def \ltsim    {\relax\ifmmode{\mathrel{\mathpalette\oversim <}}
                  \else{$\mathrel{\mathpalette\oversim <}$}\fi}
\def\oversim#1#2{\lower4pt\vbox{\baselineskip0pt \lineskip1.5pt
            \ialign{$\mathsurround=0pt#1\hfil##\hfil$\crcr#2\crcr\sim\crcr}}}
\renewcommand{\sstop}    {\mbox{$\tilde{t}$}}
\renewcommand{\sbottom}    {\mbox{$\tilde{b}$}}
\renewcommand{\stopone}    {\mbox{$\tilde{t}_{1}$}}
\def \dk       {\relax\ifmmode{\rightarrow}\else{$\rightarrow$}\fi}
\def \sp       {\relax\ifmmode{\;}\else{$\;$}\fi}
\renewcommand{\gevcc}{\mbox{$\;{\rm GeV}/c^2$}}

\def\Journal#1#2#3#4{{#1} {\bf #2}, #3 (#4)}
\def \PRL      {Phys. Rev. Lett.~}
\def \PR       {Phys. Rev.}
\def \PRD      {Phys. Rev. D}
\def \PL       {Phys. Lett.~}
\def \PLB      {Phys. Lett. B}
\def \ZPC      {Z. Phys. C}	% - Particles and Fields}
\def \NPB      {Nucl. Phys. B}
\def \PR       {Phys. Rep.~}
\def \INC      {Il Nuovo Cimento}
\def \NIM      {Nucl. Instrum. Methods}
\def \NIMA     {Nucl. Instrum. Methods Phys. Res. Sect. A}
\def \etal     {\relax\ifmmode{et \; al.}\else{$et \; al.$}\fi}
\renewcommand{\Dzero}{\mbox{D\O}}
\renewcommand{\DZERO}{\Dzero\ Collaboration}
\renewcommand{\lplm    }{\mbox{$\ell^{+} \ell^{-}$}}
\renewcommand{\squarkb} {\mbox{$\bar{\tilde{q}}$}}
\renewcommand{\epem}{\mbox{$e^+e^-$}}
\renewcommand{\snu}     {\mbox{$\tilde{\nu}$}}
\renewcommand{\seleR} {\mbox{$\tilde{e}_{R}$}}
\renewcommand{\seleL} {\mbox{$\tilde{e}_{L}$}}
\renewcommand{\stoptwo}    {\mbox{$\tilde{t}_{2}$}}
\renewcommand{\schionep }{\mbox{$\tilde{\chi}_{1}^{+}$}}
\renewcommand{\bbb}{\mbox{$b\overline{b}$}}
\renewcommand{\chionem   }{\mbox{$\tilde{\chi}_{1}^{-}$}}
\renewcommand{\sbottomone}    {\mbox{$\tilde{b}_{1}$}}
\renewcommand{\ppb}{\mbox{$p\overline{p}$}}
\renewcommand{\sbottomoneb}    {\mbox{$\bar{\tilde{b}}_{1}$}}
\renewcommand{\ifm}[1]{\relax\ifmmode #1\else $#1$\fi}
\renewcommand{\abseta}{\ifm{|\eta|}}
\renewcommand{\ISAJET}{{\sc isajet}}
\renewcommand{\chizero}{\mbox{$\tilde{\chi}_{1}^0$}}
\renewcommand{\met}{\mbox{${E\!\!\!\!/_T}$}}
\renewcommand{\ttb}{\mbox{$t\overline{t}$}}
\renewcommand{\ppbar}{\mbox{$p\overline{p}$}}
\renewcommand{\sqsq}   {\mbox{$\squark\squark$}}
\renewcommand{\sqsqb}   {\mbox{$\squark\squarkb$}}
\renewcommand{\ipb}{\mbox{${\rm pb}^{-1}$}}
\renewcommand{\tanb}{\ifm{\tan\beta}}
\renewcommand{\slepton} {\mbox{$\tilde{\ell}$}}
\renewcommand{\gt}{\ifm{>}}
\renewcommand{\lt}{\ifm{<}}
\renewcommand{\ttbar}{\mbox{$t\overline{t}$}}
\renewcommand{\VECBOS}{{\sc vecbos}}
\renewcommand{\HERWIG}{{\sc herwig}}
\renewcommand{\Ht}{\ifm{H_T}}
\renewcommand{\azero}{\ifm{A_0}}
\renewcommand{\mzero}{\ifm{m_0}}
\renewcommand{\mhalf}{\ifm{m_{1/2}}}
\renewcommand{\SPYTHIA}{{\sc spythia}}
\def \svxp {SVX$^{\prime}$}
\renewcommand{\zb}{\ifm{Z}}
\renewcommand{\stoponeb}    {\mbox{$\bar{\tilde{t}}_{1}$}}
\renewcommand{\degr}{\mbox{$^{\circ}$}}
\renewcommand{\schionem }{\mbox{$\tilde{\chi}_{1}^{-}$}}
\renewcommand{\gev}  {\mbox{${\rm GeV}$}}
\def \mc {\multicolumn}
\renewcommand{\MET}{\mbox{${E\!\!\!\!/_T}$}}

%%%%%
%\newlength{\pushupfigure}
\setlength{\pushupfigure}{-55.5pt}
\def \epsfin_v1#1#2{
        \vspace{\pushupfigure}
        \center
        \leavevmode
        \epsfxsize=#1
        \epsffile[20 143 575.75 698.75]{#2}
}
%%%%%

\def\invfb{\mbox{fb$^{-1}$}}

\section{Run II Parameters}

The Tevatron is currently being upgraded to provide
a peak luminosity of $2 \times 10^{32}$ cm$^{-2}$s$^{-1}$
(from $2 \times 10^{31}$ cm$^{-2}$s$^{-1}$ in Run I)
at $\sqrt{s}$ = 2 TeV (from 1.8 TeV).
The next run, called Run II, is expected to start in 2000 and
to accumulate 2 \invfb\ of data.

Both CDF \cite{cdf2_tdr} and D\O\ \cite{d02_tdr} detectors 
are also being upgraded for Run II, and will be
are capable of operating under Run II conditions.
%%Figures \ref{fig:cdf2} and \ref{fig:d02} are 
%%upgraded CDF and D\O\ detectors.
A major upgraded detector subsystem is the tracking system, including a
silicon microstrip chamber and outer tracker.
In CDF, the outer tracker is a remake of the Run I drift chamber,
but  adapted to the more demanding conditions of Run II.
In D\O, the tracker is based on a new tracking technique using
scintillation fibers with a 2 T solenoidal magnet.
Another key element of the CDF and D\O\ upgrades is their
sophisticated trigger systems for much higher interaction rate
in Run II.
Parameters related to identification of particles
at both detectors are briefly summarized in Table \ref{tab:run2_para}.
It should be noted that:
\begin{itemize}
\item The $\eta$ coverage for leptons
	is such that we can determine the momentum of a track
	(and, its charge).
	For instance, the CDF detector can identify the electron
	candidate down to
	$\eta \approx 3$ if the momentum measurement is not required.
	The coverage of the lepton trigger
	is determined by the available hardware trigger.
\item The secondary vertex trigger is available in Run II.
	Here, the trigger efficiency is assumed to be
	the same as the efficiency of the offline analysis.
\item Identification (ID) of a charm quark jet is based on
	a technique called jet probability, which
	reconstructs a probability that the ensemble of tracks
	in a jet is consistent with being from the primary vertex.
	We assume the ID efficiency in Run II is the same as in Run I.
%%\item There is a design of ID for high $p_T$ $\tau$ object
%%	($i.e.,$ $\gtsim$ 20 \pgev) from $W^\pm$ and $H^\pm$ events.
%%	The efficiency is reasonably high.
%%	For SUSY events, we need to identify them with
%%	$p_T \gtsim$ 10 \pgev.
%%	This work is under way.
\end{itemize}

%%
%% CDF Run-II Detector
%%
%%\begin{figure}[ht]
%%\centering\epsfig{file=cdf2.eps,width=0.45\linewidth,angle=270}
%5\caption{CDF-II detector.}
%%\label{fig:cdf2}
%%\end{figure}

%%\begin{figure}[h]
%% \centering\leavevmode
%%\psfig{figure=cdf2_elev.ps,height=4.0in}
%%\caption{Elevation view of the CDF-II detector.}
%%\label{fig:cdf2}
%%\end{figure}

%%
%% D0 Run-II Detector (new)
%%
%%\begin{figure}[ht]
%%\centering\epsfig{file=d02.eps,width=0.45\linewidth,angle=0}
%%\caption{Upgraded D0  detector.}
%%\label{fig:d02}
%%\end{figure}
%%

\begin{table}[ht]
\caption{Summary of CDF and D\O\ parameters for particle identification
(ID) expected in Run II.
The number in each parenthesis is
the $\eta$ coverage for a trigger leg.}
\label{tab:run2_para}
\begin{center}
\begin{tabular}{l l  c c  c c  l }
%%\hline
	&  & \multicolumn{2}{c}{CDF-II} &
		\multicolumn{2}{c}{D\O-II} & Comments \\
\hline
\hline
\multicolumn{2}{l }{$|\eta|$ coverage} &  &	&  &	& \\
 &$\gamma$ & $<2.0$ & ($<1.1$)	&  $<1.5$ & ($<1.5$) & \\
 &$e$ 	 & $<2.0$ & ($<1.1$)	&  $<3.0$ & ($<1.5$) & \\
 &$\mu$ & $<2.0$ & ($<1.1$) 	&  $<2.0$ & ($<1.7$) & \\
 &$\tau$ & $<2.0$ & ($<1.1$)	&  $<2.0$ & ($<1.7$) & \\
 &$j$	 & $<3.0$ &	&  $<3.0$ & & \\
 &$b$	 & $<2.0$ &     &  $<2.0$ & & \\
 &$c$	 & $<2.0$ &     &  $<2.0$ & & \\
\hline
\multicolumn{2}{l }{Efficiency (\%)} & & & & \\
 & $b$ ID	  & \multicolumn{2}{c}{50\%} &  \multicolumn{2}{c}{50\%} &  
for \ttb \\
 & $c$ ID	  & \multicolumn{2}{c}{50\%} &  \multicolumn{2}{c}{50\%} & 	\\
%%%%% & $\tau$ Trigger & ? 		&  ?		& 	\\
 & $\tau$ ID      & \multicolumn{2}{c}{40\%} &  \multicolumn{2}{c}{40\%} &  
for $W \to \tau \nu$ \\
\hline
\end{tabular}
\end{center}
\end{table}

%% file: HanTata/file.tex
%
%%%%%%%%%%%%%%%%%%%%%%%%%%%%%%%%%%%%%%%%%%%%%%%%%%%%%%%%%%%%%%%%%%%%%%%%%%%%%%%
\def\pT{p_T^{\phantom{7}}}
\def\MW{M_W^{\phantom{7}}}
\def\ET{E_T^{\phantom{7}}}
\def\bh{\bar h}
\def\lm{\,{\rm lm}}
\def\lo{\lambda_1}                                              
\def\lt{\lambda_2}
\def\pslt{p\llap/_T}
\def\eslt{E\llap/_T}
\def\to{\rightarrow}
\def\Re{{\cal R \mskip-4mu \lower.1ex \hbox{\it e}}\,}
\def\Im{{\cal I \mskip-5mu \lower.1ex \hbox{\it m}}\,}
\def\SU{SU(2)$\times$U(1)$_Y$}
\def\te{\tilde e}
\def\tmu{\tilde \mu}
\def\tl{\tilde l}
\def\ttau{\tilde \tau}
\def\tg{\tilde g}
\def\tga{\tilde \gamma}
\def\tnu{\tilde\nu}
\def\tell{\tilde\ell}
\def\tq{\tilde q}
\def\tt{\tilde t}
\def\tw{\tilde \chi^{\pm}}
\def\twb{\tilde \chi^{\mp}}
\def\tz{\tilde \chi^0}
\def\cmsec{{\rm cm^{-2}s^{-1}}}
\def\sgn{\mathop{\rm sgn}}
\hyphenation{mssm}
\def\ds{\displaystyle}
\def\ts{${\strut\atop\strut}$}

\section{Review of Previous Studies on Accessible Regions and an Estimation of What Needs to be Calculated}

Squarks and gluinos, if kinematically acessible, should be copiously
produced in hadronic collisions. Once produced, these rapidly decay into
primary jets and charginos and neutralinos. The charginos and
neutralinos further decay until the cascade terminates in the lightest
neutralino, which we assume is the LSP. Events with hard jets (from the
primary decay) together with $\eslt$ from the escaping LSPs, and
possibly leptons, from the secondary decays of charginos and neutralinos
would signal gluino and/or squark production at the Tevatron. If squarks
and gluinos are too heavy to be produced at the Tevatron, the reactions
$p\bar{p} \to \tw_1\tz_2$, and $p\bar{p} \to \tw_1\twb_1$ are the
dominant SUSY processes at the Tevatron. The subsequent leptonic decays
of charginos and neutralinos can lead to clean, {\it i.e.} jet-free
multilepton plus $\eslt$ events which have very low Standard Model (SM)
backgrounds. Since charginos and neutralinos are expected to be
lighter than gluinos in models where gaugino masses unify near the GUT
scale, the clean multilepton signals potentially offer the largest
reach, provided a sufficient integrated luminosity is accumulated, and the
leptonic decays of $\tw_1$ and $\tz_2$ are not suppressed.

\subsection{Jetty Channels}

Gluino and squark production is signalled by $n-jet + m-leptons +\eslt$
events. Such events can also come from chargino and neutralino
production and associated production. Several groups
\cite{ht-baer,mrenna,comp,tev33,ltanb} have studied these signals and obtained
an assessment of the SUSY reach. It is difficult to make direct
comparisons as the analyses differ from one another. Moreover, some
studies \cite{mrenna} focus on the signal SUSY reaction by reaction,
while others~\cite{comp,ltanb} compute the signal by considering
simultaneously all SUSY processes that can contribute to the particular
event topology. In the latter case, one has to adopt a particular framework
(usually mSUGRA) for the analysis. 

Multijet + $\eslt$ events (without leptons) form the classic SUSY
signature. In Ref.~\cite{mrenna,tev33} it has been argued from a
simulation of $\eslt$ events from gluino pair production that
experiments at the Tevatron Main Injector should be sensitive to a
gluino mass of about 390~GeV, assuming an integrated luminosity of
2~fb$^{-1}$. We will refer to this as the Main Injector (MI) data
sample. In contrast to this, a different analysis \cite{comp} of
the $\eslt$ signal within the mSUGRA framework finds a sensitivity up to
$m_{1/2} = 150$~GeV (or $m_{\tg}$ somewhat over 400~GeV), if $m_0$ is not
very large. It should be remembered that this analysis includes all
production processes, in particular $\tq\tq$ and $\tg\tq$ as well as
chargino and neutralino production, which can contribute substantially
to the signal (the latter if $m_{1/2}$ is large). The claimed reach also
varies with other model parameters. It should be kept
in mind that the analyses of Baer et al.\ have {\it not} been optimized for
this channel.

Cascade decays of squarks and gluinos may also result in jetty multilepton
events which have been studied in Ref. \cite{comp}. The most touted of
these are the same-sign (SS) dilepton and the trilepton (3$\ell$)
channels as these have low SM backgrounds. Several things about these
signals are worth noting.
\begin{itemize}
\item The branching fraction for leptonic decays of charginos, and
especially neutralinos, is sensitive to model parameters. For instance,
the leptonic branching fraction of $\tz_2$ may be enhanced if sleptons
are significantly lighter than squarks, and if the neutralino coupling
to $Z$ is suppressed by mixing angle factors~\cite{ht-baer,mrenna,lepbf}. 
There are
significant ranges of mSUGRA parameters where this is the case. On the
other hand, there are equally significant parameter regions where the
leptonic decay of $\tz_2$ is strongly suppressed. In other words,
multilepton signals are sensitive to model parameters.

\item When $\tan\beta$ is large, third generation sfermions are
significantly lighter than their siblings of the first two
generations. While this enhances~\cite{ltanb} decays to taus, chargino
and neutralino decays to $e$ and $\mu$, the multilepton
signals we have been discussing, become strongly suppressed. We will return to
this below.

\item The LEP lower bound on the chargino mass has crept up to $\sim$90~GeV
since the analyses of the jetty multilepton signals were
carried out. The analysis of Ref.~\cite{comp} then shows that the SS
dilepton signal is not expected to be observable at the MI, for model
parameters that respect the LEP chargino bound. There could, however,
still be observable signals in other multilepton channels. Indeed, there
are some regions of parameter space not accessible via the $\eslt$
search, but for which there would be an observable signal via 
the $3\ell$ channel --- it should, however, be kept in mind that the
$\eslt$ analysis was not completely optimized in this paper \cite{comp}. For small values of $m_0$ where sleptons are light so that chargino and neutralino
decays to leptons are strongly enhanced, it may be possible to probe
$m_{1/2}$ as large as 200~GeV at the MI. On the other hand, there are
other ranges of parameters for which there is no observable signal even 
if charginos are at their LEP limit.
\end{itemize}

\subsection{Clean Multilepton Channels: The Low $\tan\beta$ Case}

As we have discussed, the clean trilepton channel from $\tw_1\tz_2$
production potentially offers the
greatest reach at luminosity upgrades of the Tevatron, and has,
therefore, received the maximum
attention~\cite{ht-baer,mrenna,comp,tev33,ht-snow}. 
These analyses have been performed within the mSUGRA type framework, for
low to medium values of $\tan\beta$ for which the bottom and tau Yukawa
couplings are not very large. 
It should be clear from our earlier discussion that the reach is
model-dependent, and even within mSUGRA, highly sensitive to model
parameters. Again, detailed direct comparison is difficult, but the
analyses seem to be in qualitative agreement. Ref.~\cite{mrenna,tev33}
find a maximum reach of $\sim 200$~GeV for the chargino mass at the MI,
which compares reasonably with the maximum value of $m_{1/2} \sim
250$~GeV (for $\mu > 0$) obtained in Ref.~\cite{comp}. Thus for
favourable ranges of parameters (where leptonic decays of $\tw_1$ and
$\tz_2$ are enhanced), the MI reach substantially exceeds that of LEP2. 
Also, the observation of this signal may provide the first direct
evidence of the neutralino whose production cross section in $e^+e^-$
colliders may be very suppressed.
The important
thing is that this channel offers a chance for SUSY discovery for ranges
of parameters not accessible via the jetty signatures. It is
complementary to the $\eslt$ and other jetty searches in that there are
regions of parameters where there is a signal in the jetty channels but
none in the clean 3$\ell$ channel. Unfortunately though, the
non-observation of any signal in this channel will not allow us to infer
any lower bound on $m_{\tw_1}$.

\subsection{The Large $\tan\beta$ Scenario}

For very large values of the mSUGRA parameter $\tan\beta$, decays of
gluinos to third generation quarks, and of charginos and neutralinos (too
heavy to decay into real $W$, $Z$ or Higgs bosons) to $b$-quarks and 
$\tau$ leptons can be very much enhanced, thereby suppressing their
decays to $e$ and $\mu$. In the extreme case, the decays $\tw_1 \to
\ttau_1\nu$ and $\tz_2 \to \ttau_1\tau$ may be the only allowed two body decay
modes of charginos and neutralinos --- these would then essentially
always decay to taus. It is clear that for large $\tan\beta$, both jetty
as well as clean multilepton signals would be strongly suppressed and
the Tevatron reach via these channels greatly reduced. On the other
hand, there is the hope that there might be observable signals in new
channels involving $b$-jets or $\tau$ leptons. 

\begin{figure}[t]
\centering\leavevmode
\epsfxsize=4.5in\epsffile{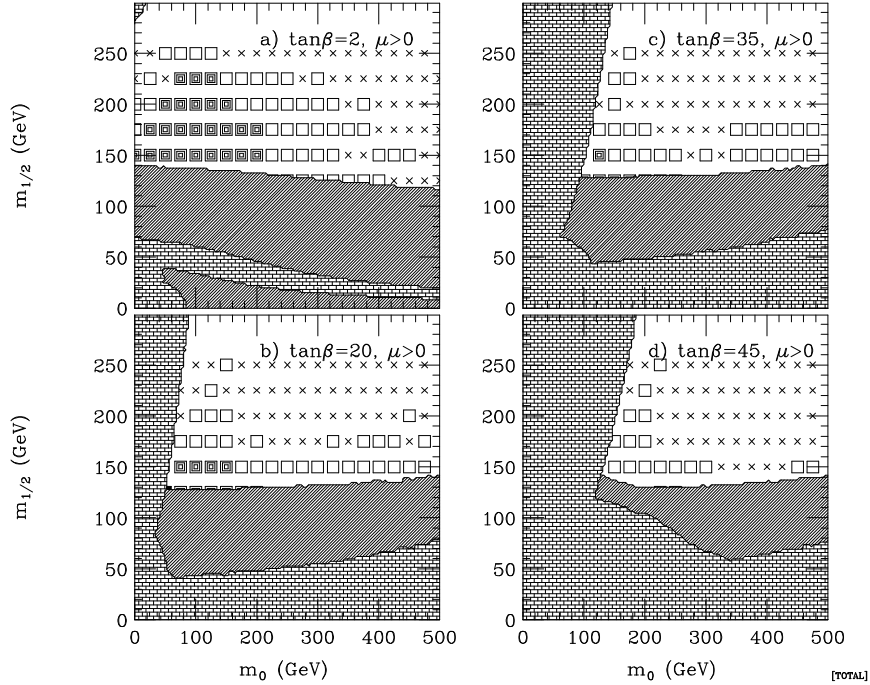}
\caption[]{The combined SUSY reach of the Main Injector (2~fb$^{-1}$) [grey squares]
and Run~IIb (25~fb$^{-1}$) [hollow squares] within the mSUGRA framework. The
bricked (hatched) regions are excluded by theoretical (experimental)
constraints. The $\eslt$, $eslt$+tagged $b$, clean 3$\ell$ and clean
3$\ell$ plus identified $\tau$ channels establish this plot, though for
some parameter points, other signals may also be observable.\label{HT-total}}
\end{figure}

The situation has been recently analysed in Ref.~\cite{ltanb}.  Here, it
was assumed that it would be possible to tag central $b$-jets with an
efficiency of 50\%, and that it would be possible to identify
hadronically decaying taus as narrow jets, if their visible $E_T \geq
15$~GeV. With otherwise the same analysis cuts that they had used in
earlier low $\tan\beta$ analyses \cite{comp}, the region of the
$m_0-m_{1/2}$ plane that can be probed at the Tevatron MI and with an
integrated luminosity of 25~fb$^{-1}$ via any one of several channels is
shown by grey and open squares in Fig.~\ref{HT-total}. The bricked (hatched)
portions are already excluded by theoretical (experimental) constraints.
The $\eslt$, $\eslt$ + tagged $b$, the clean $3\ell$ and clean $3\ell$ with
tagged $\tau$ channels establish the entire plot~\cite{ltanb}, though
confirmatory signals may also be present in other channels.
We see that the reach of the Tevatron 
monotonically decreases with increasing values of $\tan\beta$, until for
$\tan\beta=45$ there is no reach in any of the channels examined,
including the new channels with $b$-jets or $\tau$ leptons. Moreover,
there are parameter regions just beyond the current LEP2
bound for which this analysis finds no observable signal, even with an
integrated luminosity of 25~fb$^{-1}$.

The lack of an observable signal was traced in Ref.~\cite{ltanb}
to the fact that the
secondary leptons and jets from tau decay were usually too soft to
pass the acceptance and trigger criteria. Indeed these authors pointed out that
it would be worthwhile to investigate whether it was possible to relax
these cuts without introducing unacceptably large backgrounds or losing
the capability to trigger on the events.

The Wisconsin Group \cite{ht-bkl} noted that Ref.~\cite{ltanb} had
required the $p_T$ of the leptons in clean 3$\ell$ events exceed (20,
15, 10)~GeV. In the earlier low $\tan\beta$ study~\cite{comp}, this was
to ensure that physics backgrounds from heavy flavour as well as
instrumental backgrounds would not obscure the signal, the cross section
for which is only ${\cal O}$(few fb) at the discovery
limit. Ref.~\cite{ht-bkl} showed that by loosening the lepton cuts to (12,
5, 5)~GeV, they were able to include secondary leptons from tau decays
in the signal, thereby increasing the parameter range where the signal
might be observable. Although they did not veto events with jets and
did not explicitly compute the worrisome backgrounds mentioned above, it
appears \cite{kamon} that these are under control. A more complete analysis \cite{barkao}
that delineates the mSUGRA parameter region where the trilepton signal
with soft cuts might be observable has recently been completed.

The Wisconsin analysis notwithstanding, it would be of considerable
interest to be able to identify taus in SUSY events. Indeed if it were
possible to conclusively establish a significant excess of taus relative
to $e$ and $\mu$ in the SUSY event sample, it would point to the third
generation of sparticles being different from the first two, either by
virtue of large Yukawa couplings, or simply because $m_{\ttau_1} <
m_{\te_R}$ and $m_{\tmu_R}$. Such a determination could serve as an
important stepping stone for clarifying the nature of the underlying
physics. The possibility of identifying and even triggering on
hadronically decaying $\tau$ leptons with visible $p_T < 15$~GeV is also
under active investigation. 

We conclude by again highlighting some of the issues that would be worthy
of further investigation.
\begin{itemize}
\item The classic $\eslt$ channel remains an important search channel
for SUSY, and it is worthwhile to optimize the cuts and maximize the
range that can be explored this way. 

\item The region of the $m_0-m_{1/2}$ plane that can be explored using
the soft cuts suggested by the Wisconsin Group~\cite{ht-bkl} needs to be
delineated. This is especially important for large $\tan\beta$. A
documentation of the physics and non-physics backgrounds as a
function of these cuts would also be useful.

\item It would be interesting to examine whether it is possible to
reduce the cut on the visible $p_T$ of the hadronically decaying taus
(perhaps by focussing on single prong events for which the QCD
background is smaller) without being overwhelmed by SM backgrounds. An
efficient tau trigger may also prove very useful, since one would then
not have to rely on hard $e$ and $\mu$ to trigger the events.

\item Improved $b$-jet identification may also prove useful if
$\tan\beta$ is large. Already in the analysis of Ref.~\cite{ltanb}, the
$\eslt$ sample with tagged $b$-jets extends the parameter space region 
over what can be probed via other channels. Moreover, observations of
excess of $b$ and $\tau$ signals could serve as a pointer to large
$\tan\beta$ (at least within the mSUGRA model), though it is not clear
whether this could be established solely with the Tevatron data.
\end{itemize}

%%%%%%%%%%%%%%%%%%%%% REFERENCES %%%%%%%%%%%%%%%%%%%%%%%%%%%%%%%%%%%%%%%%%%%%%%
%

%
%\newpage
%%%%%%%%%%%%%%%%%%%%%%%%%%%%%%%%%%%%%%%%%%%%%%%%%%%%%%%%%%%%%%%%%%%%%

%% file: Baer-4D/sims.tex
\def\notE{\ \hbox{{$E$}\kern-.60em\hbox{/}}}
\def\to{\rightarrow}
\def\te{\tilde e}
\def\tx{\tilde\chi}
\def\tl{\tilde l}
\def\tb{\tilde b}
\def\tf{\tilde f}
\def\td{\tilde d}
\def\tst{\tilde t}
\def\ttau{\tilde \tau}
\def\tmu{\tilde \mu}
\def\tg{\tilde g}
\def\tnu{\tilde\nu}
\def\tell{\tilde\ell}
\def\tq{\tilde q}
\def\tw{\tilde\chi^\pm}
\def\tz{\tilde\chi^0}

%\begin{document}
\section{Results of five mSUGRA case study points}

%\author{H. Baer, V. Barger, C. Kao, P. Mercadante, \\
%S. Mrenna, P. Quintana, F. Paige, X. Tata and Y. Wang}

\subsection{Five SUGRA points for detailed analyses}

Four points in the mSUGRA parameter space were chosen for detailed
analyses of signals and backgrounds.
A fifth point was chosen with non-universal soft SUSY breaking
Higgs masses at the GUT scale.
The parameter values for the five case studies are listed in  
Table~\ref{casestudy-tab1}
along with some of the more relevant sparticle masses.
ISAJET 7.44 \cite{case-isajet} was used to generate the sparticle masses,
branching fractions, production cross sections and simulated events 
with decay matrix elements incorporated in the computation of the
momenta of leptons from chargino and neutralino decays.

\subsection{SUSY channel contributions to signals}

The total cross section for production of {\it all} sparticle types
is listed in Table~\ref{casestudy-tab1} in $fb$. 
We also list the percentage of cross section 
for various relevant sparticle production mechanisms.
We note the following features of each case study point.
\begin{itemize}

\item {\bf Case 1:} The mSUGRA parameters for this point lie in the
cosmologically favored region of parameter space, and give rise to a
reasonable relic density of neutralinos. The dominant production mechanisms
at the Tevatron are $\tw_1\tilde\chi_1^\mp$ and $\tw_1\tz_2$ production.
For this case,
$\tz_2\to \ell\tell_R$ at $\sim 100\%$, so a large rate for clean trilepton
events is expected.

\item {\bf Case 2:} This parameter space point occurs with a large value
of $\tan\beta =35$ so that $\tw_1\to\ttau_1\nu_\tau$ and $\tz_2\to\ttau_1\tau$
occur at $\sim 100\%$. The dominant production cross section is again
$\tw_1\tilde\chi_1^\mp$ and $\tw_1\tz_2$ production. Here, we anticipate
that an inclusive
trilepton signal can be extracted with relatively soft lepton $p_T$ cuts,
since the detected leptons typically come from $\tau$ decays.
Events containing a mixture of $e$'s, $\mu$'s and hadronically
identified $\tau$'s should also be present.

\item {\bf Case 3:}
This parameter space point also occurs at large $\tan\beta$,
but the $A_0$ parameter was chosen so that relatively light
$\tst_1$, $\tb_1$ and $\ttau_1$ are generated.
$\tw_1\to\ttau_1\nu_\tau$ and $\tz_2\to\ttau_1\tau$
occur again at $\sim 100\%$, so that the trileptons should occur
at a similar rate as in case 2. However, the rather large $\tst_1\tst_1$
production cross section may yield an observable $\tst_1$ signal.
$\tst_1\to b\tw_1$ at $\sim 100\%$, but since $\tw_1\to\ttau_1\nu_\tau$,
hard leptons are not generated in the $\tst_1$ cascade decay.

\item {\bf Case 4:} This parameter space choice has
dominant $\tw_1\tilde\chi_1^\mp$, $\tw_1\tz_2$ and $\tst_1\tst_1$ production. 
It could contain a sample of high $p_T$ trilepton events, but also
one may search for $\tst_1\tst_1$ production where $\tst_1\to b\tw_1$
with $\tw_1\to \ell\nu_\ell\tz_1$.

\item {\bf Case 5:} This point was chosen to have rather large GUT scale
Higgs masses, so that scalar universality is broken. The $\mu$ parameter
is relatively small so that the lower lying charginos and neutralinos have
a substantial higgsino component. In this case, $\tw_1\tilde\chi_1^\mp$,
$\tw_1\tz_2$ and $\tw_1\tz_3$ all occur at large rates.
$\tz_2\to ee\tz_1$ occurs with a 3\% branching ratio, but $\tz_3\to\ttau_1\tau$
at $\sim 100\%$. This case may lead to both clean, hard trileptons, but
also contain a soft trilepton component from $\tw_1\tz_3$ production.

\end{itemize}

\subsection{Results of simulations}

For the five case study points above, we have performed detailed
simulations of signal and background for the trilepton signal
\cite{case-Trilepton1,case-Trilepton2,case-Trilepton3,case-CDF,case-D0,%
case-Madison1,case-Madison2,case-Matchev,case-BDPQT} 
which is expected to be one of the most important signal channels for 
mSUGRA at Tevatron Run 2. In our studies, we used the toy detector 
simulation package ISAPLT, assuming calorimetry between $-4<|\eta |<4$, 
with an array of calorimeter cells of size 
$\Delta\eta\times\Delta\phi =0.1\times 0.2618$. Electromagnetic 
energy resolution was taken as $0.15/\sqrt{E}$ and hadronic
calorimeter resolution was taken to be $0.7/\sqrt{E}$. Jets were
coalesced in towers of $\Delta R =0.7$ and were called a jet if
$E_T(j)>15$ GeV, using the jet finding algorithm GETJET.
Leptons ($e$'s or $\mu$'s) were taken to be isolated if the hadronic $E_T$
in a cone about the lepton of $\Delta R=0.4$ was less than 2 GeV.

Three sets of acceptance cuts were studied for SUSY signals and backgrounds.
A relatively hard set of cuts chosen to maximize the
reach for $\tw_1\tz_2\to 3\ell$ in $m_{1/2}$ were taken from
Ref. \cite{case-BKT} where the focus was on the signal for low values of
$\tan\beta$. In the present study, the isolation requirement is
slightly different from Ref. \cite{case-BKT}, however; these cuts are listed
in column 3 of Table~\ref{casestudy-tab2}. 
The CDF \cite{case-CDF} and the D0 \cite{case-D0} collaborations, 
and the authors of Ref. \cite{case-Trilepton3} 
have used relatively softer cuts in their analysis. 
These soft cuts were advocated in Refs. \cite{case-Madison1,case-Madison2} 
as being more effective in 
eliciting signal from background, especially for large $\tan\beta$, where
many of the signal leptons arise as secondaries
from $\tau$ decay, and are quite soft.
These cuts are listed in column 2 of Table~\ref{casestudy-tab2}.
Note that these cuts lack a
jet veto, so that the signal will be {\it inclusive},
containing both clean and jetty
trilepton events. Finally, we also examined the SUSY signal and
background with soft cuts plus a jet veto in addition (clean trileptons).

%------------------------
%	Table I
%------------------------
\begin{table}[htb]
\begin{center}
%\begin{minipage}[b]{3in}
\caption{Parameter space choices, sparticle masses and
total signal cross sections for the five chosen case studies of the mSUGRA
group. We also list the fractional contribution to the signal from
various subprocesses.
We take $m_t=175$ GeV. \label{casestudy-tab1}}

\bigskip
\begin{tabular}{|l|c|c|c|c|c|}
\hline
case & $(1)$ & $(2)$ & $(3)$ & $(4)$ & $(5)$ \\
\hline
$m_0$ & 100 & 140 & 200 & 250 & 150 \\
$m_{1/2}$ & 200 & 175 & 140 & 150 & 300 \\
$A_0$ & 0 & 0 & -500 & -600 & 0 \\
$\tan\beta$ & 3 & 35 & 35 & 3 & 30 \\
$m_{H_1},m_{H_2}$ & -- & -- & -- & -- & 500,500 \\
$A_t$ & -359 & -326 & -374 & -387 & -543 \\
$A_b$ & -575 & -433 & -626 & -945 & -742 \\
$m_{\tg}$ & 508 & 455 & 375 & 403 & 734 \\
$m_{\tq}$ & 450 & 410 & 370 & 415 & 650 \\
$m_{\tst_1}$ & 306 & 297 & 153 & 134 & 440 \\
$m_{\tst_2}$ & 502 & 448 & 392 & 448 & 626 \\
$m_{\tb_1}$  & 418 & 329 & 213 & 346 & 566 \\
$m_{\tb_2}$  & 441 & 408 & 342 & 413 & 592 \\
$m_{\tw_1}$ & 141 & 126 & 106 & 109 & 100 \\
$m_{\tz_1}$ & 76 & 69 & 56 & 57 & 80 \\
$m_{\tz_2}$ & 143 & 127 & 107 & 111 & 124 \\
$m_{\tz_3}$ & 316 & 252 & 296 & 373 & 141 \\
$m_{\tell_R}$ & 132 & 162 & 212 & 260 & 195 \\
$m_{\tell_L}$ & 180 & 194 & 229 & 275 & 266 \\
$m_{\ttau_1}$ & 131 & 104 & 88 & 257 & 132 \\
$m_{A}$     & 372 & 206 & 185 & 479 & 443 \\
$m_{H}$     & 376 & 206 & 185 & 481 & 444 \\
$m_{h}$     &  99 & 110 & 112 & 104 & 115 \\
$m_{H^\pm}$ & 380 & 224 & 205 & 485 & 452 \\
$\mu$ & 312 & 241 & 286 & 369 & -110 \\
$\sigma_{tot.}(fb)    $ & 404 & 653 & 2712 & 3692 & 1393 \\
$\tg ,\tq (\%)        $ & 4.3 & 6.6 & 50.4 & 66.2 & 0.01 \\
$\tg\tx ,\tq\tx (\%)  $ & 2.4 & 3.6 & 2.9  & 1.2  & 0.01 \\
$\tx\tx (\%)          $ & 85.0& 85  & 45.7 & 32.6 & 99.5 \\
$\tell\tell (\%)      $ & 8.3 & 4.7 & 1.0  & 0.04 & 0.4  \\
$\tst_1\tst_1 (\%)    $ & 1.8 & 1.5 & 41   & 65   & 0.01 \\
$\tw_1\tz_2 (\%)      $ & 43.8& 45  & 26.5 & 18   & 16.7 \\
$\tw_1\tilde\chi_1^\mp (\%)      $ & 33.5& 33  & 17.6 & 13   & 24.6 \\
\hline
\end{tabular}
%\end{minipage}
\end{center}
\end{table}

%------------------------
%	Table II
%------------------------
\begin{table}[htb]
\begin{center}
%\begin{minipage}[b]{2.2in}
\caption{Hard and soft cuts for Tevatron SUSY trilepton results.  
\label{casestudy-tab2}}
\bigskip
\begin{tabular}{|l|c|c|c|}
\hline
Cut & Soft A & Soft B & Hard \\
\hline
$p_T(\ell_1,\ell_2,\ell_3)$ 
                     & $>$11,7,5 GeV & $>$11,7,5 GeV & $>$20,15,10 GeV \\
$|\eta (\ell_{1,2/3})|$     & $<$1.0,2.0 & $<$1.0,2.0 & $<$1.0,2.0 \\
$ISO_{\Delta R=0.4}$ & $<$2 GeV & $<$2 GeV & $<$2 GeV \\
$\notE_T$ & $>$25 GeV & $>$25 GeV & $>$25 GeV \\
Require $M_{\ell\bar{\ell}}\; (\gamma-{\rm veto})$ 
                    & $>$ 18 GeV & $<$ 20 GeV & $<$12 GeV \\
Require $M_{\ell\bar{\ell}}\; (Z-{\rm veto})$ 
                    & $<$ 75 GeV & $<$ 81 GeV & $<$81 GeV \\
Reject $M_T(\ell,\notE_T)\; (W-{\rm veto})$ 
                    & 65--85 GeV & 60--85 GeV & 60--85 GeV \\
\hline
\end{tabular}
%\end{minipage}
\end{center}
\end{table}

The dominant SM backgrounds are listed in Table~\ref{casestudy-tab3} 
for the three sets of cuts.
After suitable cuts, there are two major sources of the SM background
\cite{case-Trilepton1,case-Trilepton2,case-Trilepton3,%
case-Madison1,case-Madison2,case-Matchev,case-BDPQT,case-Chanowitz,case-Ellis}:
(i) $q\bar{q} \to W^* Z^*, W^* \gamma^* \to \ell\nu \ell\bar{\ell}$
or $\ell'\nu' \ell\bar{\ell}$ ($\ell = e$ or $\mu$)
with one or both gauge bosons being virtual\footnote{
If it is not specified,
$W^*$ and $Z^*$ represent real or virtual gauge bosons,
while $\gamma^*$ is a virtual photon.}, and
(ii) $q\bar{q} \to W^* Z^*, W^* \gamma^* \to \ell\nu\tau\bar{\tau}$
or $\tau\nu \ell\bar{\ell}$ and subsequent $\tau$ leptonic decays.
We have employed the programs MADGRAPH \cite{case-Madgraph} 
and HELAS \cite{case-Helas} to calculate
the cross section of $p\bar{p} \to 3l +\notE_T +X$ 
via four subprocesses (i) $q\bar{q}'\to e^+\nu_e\mu^+\mu^-$, 
(ii) $q\bar{q}'\to e^+\nu_e e^+ e^-$, 
(iii) $q\bar{q}'\to e^+\nu_e \tau^+ \tau^-$, and 
(iV) $q\bar{q}'\to \tau^+\nu_\tau e^+ e^-$, 
including contributions from intermediate states with $W^*Z^*$, 
$W^*\gamma^*$, and other diagrams. 
We have also evaluated contributions from 
$q\bar{q} \to \tau^+\tau^- e^+ e^-$ 
via $Z^*Z^*$, $Z^*\gamma^*$, and $\gamma^*\gamma^*$, with one leptonic 
and one hadronic tau decays.
We use ISAJET to calculate the background from $t\bar{t}$.
We also ran $Z+jets$ and $W+jets$ background jobs. 
For these latter two cases, 
no events passed any of the cuts out of $5\times 10^5$ and $10^6$ events
generated, respectively. These correspond to backgrounds of less than
0.3 and 4 fb, respectively. In runs of $10^8$ $W+jets$ events
with somewhat different cuts,
some $3\ell$ events could be generated leading to sizable backgrounds;
these sources always had $b\to c\ell\nu$ followed by $c\to s\ell\nu$,
so that these sources of background could be removed by imposing
an angular separation cut between the isolated leptons, giving a background
consistent with zero.
We list in Table~\ref{casestudy-tab3} 
the total background rate as well as 
the signal rates necessary to achieve a 99\% C.L. and a $5\sigma$ signal
with 2 and 30 fb$^{-1}$ of integrated luminosity, respectively.
At Run II with 2 fb$^{-1}$ integrated luminosity,
we expect about three event per experiment with soft cuts
from the background cross section of 1.60 fb (Soft A) or 1.30 (Soft B).
The signal cross section must yield
a minimum of five signal events for discovery.
The Poisson probability for the SM background to fluctuate
to this level is less than $0.6\%$.

%------------------------
%	Table III
%------------------------
\begin{table}[htb]
\begin{center}
\caption{SM backgrounds (fb) for hard and soft cuts for Tevatron
SUSY trilepton signals. \label{casestudy-tab3}}
\bigskip
\begin{tabular}{|l|c|c|c|}
\hline
BG & soft A & soft B & hard \\
\hline
$\ell'\nu'\ell\bar{\ell}$ 
           & $0.60\pm 0.003$ & $0.45\pm 0.003$ & $0.19\pm 0.001$ \\
$\ell \nu \ell\bar{\ell}$ 
           & $0.30\pm 0.004$ & $0.20\pm 0.004$ & $0.09\pm 0.002$ \\
$\ell\nu\tau\bar{\tau}$ 
           & $0.41\pm 0.008$ & $0.36\pm 0.008$ & $0.22\pm 0.005$ \\
$\tau\nu\ell\bar{\ell}$ 
           & $0.13\pm 0.009$ & $0.13\pm 0.008$ & $0.06\pm 0.005$ \\
$\ell\ell\tau\bar{\tau}$ 
           & $0.06\pm 0.001$ & $0.06\pm 0.001$ & $0.04\pm 0.001$ \\
$t\bar{t}$ & $0.06\pm 0.003$ & $0.06\pm 0.004$ & $0.003\pm 0.003$ \\
$total$    & $1.56$ & $1.26$ & $0.60$ \\
$99\%\,{\rm C.L.} (2\ fb^{-1})$ & $2.5$ & $2.5$ & $2.0$ \\
$5\sigma (30\ fb^{-1})$ & $1.14$ & $1.01$ & $0.71$ \\
$3\sigma (30\ fb^{-1})$ & $0.68$ & $0.61$ & $0.42$ \\
\hline
\end{tabular}
\end{center}
\end{table}

The SUSY signal rates for $3\ell$ events are listed in
Table~\ref{casestudy-tab4} for the five parameter space choices, and the
three sets of cuts.  Using soft inclusive cuts, cases 1, 3, and 4 should
be detectable with just 2 fb$^{-1}$ at Run 2, and all five cases should
be visible with 30 fb$^{-1}$.  Implementing a jet veto can reduce the
signal from between a factor of 2 to a factor of 4, so that only a
fraction of events extracted with the soft inclusive cuts are without
jets. Using soft cuts plus a jet veto, only point 1 is visible at 2
fb$^{-1}$, while cases 1, 3, 4 and 5 are visible with 30
fb$^{-1}$. Using hard cuts, again only case 1 is visible at 2 fb$^{-1}$,
while just the low $\tan\beta$ cases 1 and 4 are observable with 30
fb$^{-1}$.

%------------------------
%	Table IV
%------------------------
\begin{table}[htb]
\begin{center}
\caption{SUSY $3\ell$ signal (fb) for hard and soft cuts at the Tevatron.  
\label{casestudy-tab4}}
\bigskip
\begin{tabular}{|l|c|c|c|c|}
\hline
Case & Soft A & Soft B & Soft A+Jet Veto & Hard \\
\hline
(1) & $8.41\pm 0.13$ & $7.39\pm 0.12$ & $4.50\pm 0.10$ & $3.78\pm 0.09$ \\
(2) & $0.97\pm 0.06$ & $0.93\pm 0.06$ & $0.26\pm 0.06$ & $0.47\pm 0.04$ \\
(3) & $1.18\pm 0.13$ & $1.08\pm 0.12$ & $0.23\pm 0.06$ & $0.32\pm 0.07$ \\
(4) & $2.97\pm 0.16$ & $2.72\pm 0.23$ & $1.29\pm 0.16$ & $1.52\pm 0.17$ \\
(5) & $0.73\pm 0.07$ & $0.63\pm 0.07$ & $0.28\pm 0.05$ & $0.39\pm 0.06$ \\
\hline				   
\end{tabular}
\end{center}
\end{table}

\subsection{Endpoint reconstruction studies}

The same flavor, opposite sign dilepton invariant mass distribution
can be a sensitive guide to neutralino and slepton masses.
In the figure shown, we show the dilepton mass distribution after
soft inclusive cuts
for each of the case studies. The final histogram includes the contribution
from the $WZ$ and $t\bar t$ backgrounds. The background is folded into the
histograms for each case study.

For case 1, the trilepton signal is large and a clear mass endpoint should
be visible even at Run 2 integrated luminosity values.
In this case, $\tz_2\to \ell\bar{\ell}\tz_1$ via a real $\tell_R$, so
an endpoint is expected at
\begin{eqnarray*}
m_{\ell\bar{\ell}}^{max}=m_{\tz_2}\sqrt{1-{m_{\tell_R}^2\over m_{\tz_2}^2}}
\sqrt{1-{m_{\tz_1}^2\over m_{\tell_R}^2}}\simeq 45\ {\rm GeV},
\end{eqnarray*}
and is clearly visible in the plot \cite{case-Madison2,case-BDPQT}.

For case 2, there should be a similar edge -- but in the $m(\tau\bar{\tau})$
distribution -- at 54.5 GeV. Dileptons from the subsequent $\tau$ leptonic
decays should also respect this bound, 
but with a softened mass distribution.
This situation is shown in the figure. The statistical sample will be
quite limited even with 30 fb$^{-1}$ of integrated luminosity since
less than 30 signal events will make up the plot. Slepton, sneutrino and
heavier neutralino production also contributes to this plot. As a
result, the
edge in the $m(\ell\bar{\ell})$ distribution is washed out, making extraction
of information on neutralino or slepton masses very difficult.
It might be interesting to examine the possibility of constructing the mass
edge using identified $\tau$'s; since this depends on detector
capabilities, we have not done so.

Case 3 also involves $\tz_2\to\tau\ttau_1$ with a branching fraction of  
100\%, so that $m(\tau\bar{\tau})$ should be bounded by 47 GeV. 
In spite of the fact that several SUSY sources contribute to the
trilepton signal, 
the dilepton  
mass
reconstruction again happens to respects this bound. As in case 2, the
dilepton mass endpoint is washed out so extraction of precision mass
information will be difficult.

In case 4, $\tz_2\to \ell\bar{\ell}\tz_1$ via virtual particles,
so we expect $m(\ell\bar{\ell})$ to be bounded by
$m_{\tz_2}-m_{\tz_1}=54$ GeV. There is no sharp edge and the signal cross
section is small.

Finally, in case 5, dileptons can occur from $\tz_2$ via virtual
sparticles and $\tz_3$ decay via a real $\ttau_1$.
The $m(\ell\bar{\ell})$ distribution shown in the
figure exhibits a mass edge at $m_{\tz_2}-m_{\tz_1}=44$ GeV.
The decay $\tz_3\to\tau\ttau_1$ will likewise have a $m(\tau\bar{\tau})$
edge at 39 GeV with a correspondingly softer dilepton mass
distribution.
Note that with 30 fb$^{-1}$ of integrated luminosity, 
$< 25$ signal events will be used to create this plot, 
so the statistical sample will be very limited.
Extraction of masses which is further complicated by the fact that
$\tw_1\tz_3$ production contributes significantly will be difficult. 
\begin{figure}[htb]
\centering\leavevmode
\psfig{figure=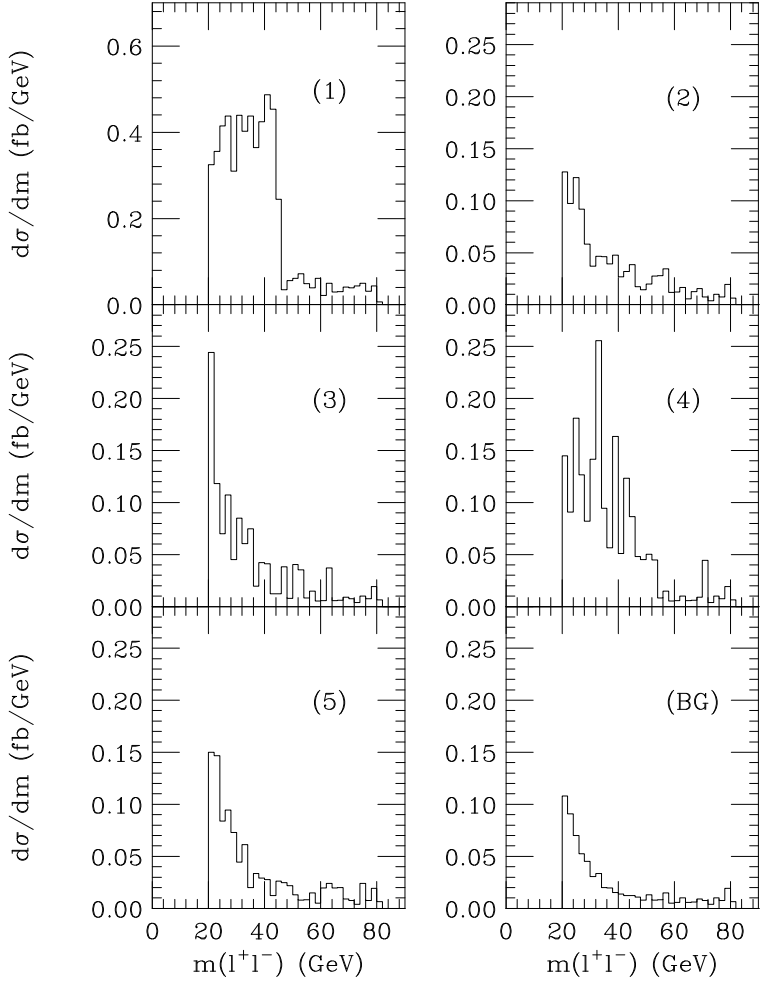,width=10cm}
\bigskip
\caption{Opposite sign, same flavor dilepton mass reconstruction for
the five case study points and the $WZ$ plus $t\bar t$ background.
The background is included in the histogram for each case study.}
\label{4D-FIG1}
\end{figure}

\subsection{$\notE_T +jets$ analysis}

The classic signature for SUSY at hadron colliders is the appearance of
multijet events accompanied by large $\notE_T$. To see if the
$\notE_T +$jets signal is seeable, we use the cuts of
Mrenna {\it et al.} \cite{case-Trilepton3}. These consist of requiring:
\begin{itemize}
\item $\notE_T > 75$ GeV,
\item veto isolated leptons within $|\eta|<4$ if $p_T(\ell )>15$ GeV,
\item transverse sphericity $S_T>0.2$,
\item $\Delta\phi (j,\notE_T )>0.5$ and
\item $E_T(j_1)+E_T(j_2)+\notE_T >300$ GeV.
\end{itemize}
Using ISAJET 7.44, the backgrounds are listed 
in Table~\ref{casestudy-tab5} along with the
rate needed for a $5\sigma$ signal at 2 and 30 fb$^{-1}$ of integrated
luminosity.\footnote{Our background levels differ substantially from those
quoted in Mrenna {\it et al.} \cite{case-Trilepton3} using PYTHIA,
where background estimates of
24, 11 and 5 fb are obtained from $t\bar{t}$, $W+jets$ and $Z+jets$,
respectively.}

The corresponding signal rates for the five case studies are listed in  
Table~\ref{casestudy-tab6}.
{}From the tables, we see that case 3 reaches nearly the $5\sigma$ level
for just 2 fb$^{-1}$ of integrated luminosity.
For case 3, $m_{\tg}\simeq m_{\tq}\sim 370$ GeV.
For 30 fb$^{-1}$,
case 3 should be clearly seeable, while case 2 and 4 just reach the
$5\sigma$ level\footnote{
With such a large background cross section, 
it might be difficult to establish a signal of 
$\notE_T +jets$ with a statistical significane of $S/\sqrt{B} = 5$ 
($S=$ number of signal events and $B$= number of  background evets), 
for ${\cal L} = 30$ fb$^{-1}$, 
because the ratio of $S/B$ is only $5\%$.}.

We conclude that it might be possible to probe gluinos and squarks up to
$\sim 350-375$~GeV if $m_{\tq} \simeq m_{\tg}$ at Run 2 and up to about
400~GeV with an integrated luminosity of 30~fb$^{-1}$.

%------------------------
%	Table V	
%------------------------
\begin{table}[htb]
\begin{center}
\begin{minipage}[b]{2in}
\caption{SM backgrounds (fb) for $\notE_T +$jets events for the Tevatron.  
\label{casestudy-tab5}}
\smallskip
\tabcolsep=1.5em
\begin{tabular}{|l|r|}
\hline
BG & rate (fb) \\
\hline
$t\bar{t}$ &  47.6 \\
$W+jets$   & 106.7 \\
$Z+jets$   & 139.6 \\
$total$    & 293.9 \\
$5\sigma ( 2\ fb^{-1})$ & 60.6 \\
$5\sigma (30\ fb^{-1})$ & 15.6 \\
\hline
\end{tabular}
\end{minipage}
%\end{center}
%\end{table}
\hspace{2.5cm}
%\begin{table}
%\begin{center}
\begin{minipage}[b]{2in}
\caption{SUSY signal (fb) for $\notE_T +$jets events for the Tevatron.  
\label{casestudy-tab6}}
\bigskip
\tabcolsep=2em
\begin{tabular}{|l|r|}
\hline
case & rate (fb) \\
\hline
(1) & $ 5.7\pm 0.1$ \\
(2) & $16.6\pm 0.2$ \\
(3) & $61.9\pm 0.9$ \\
(4) & $18.5\pm 0.6$ \\
(5) & $ 1.3\pm 0.2$ \\
\hline
\end{tabular}
\end{minipage}
\end{center}
\end{table}

%%%%%%%%%%%%%%%%%%%%%%%%%%%%%%
%  Bibliography
%%%%%%%%%%%%%%%%%%%%%%%%%%%%%%

%\end{document}

%% file: Eboli-4F/draft04.tex
%\documentstyle[12pt,epsfig]{report}
%\def\baselinestretch{1.2}

%\begin{document}

\hfuzz=12pt
\def\etm{E\llap/_T}

\section{Stops, Sbottoms, and Gluinos}

%%%%%%%%%%%%%%%%%%%%%%%%%%%%%%%%%%%%%%%%%%%%%%%%%%%%%%%%%%%%%%%%%%%%%%

\subsection{Stop pair production}

\begin{figure}[hb]
\vspace{-4ex}

\begin{center}
 \mbox{\epsfig{file=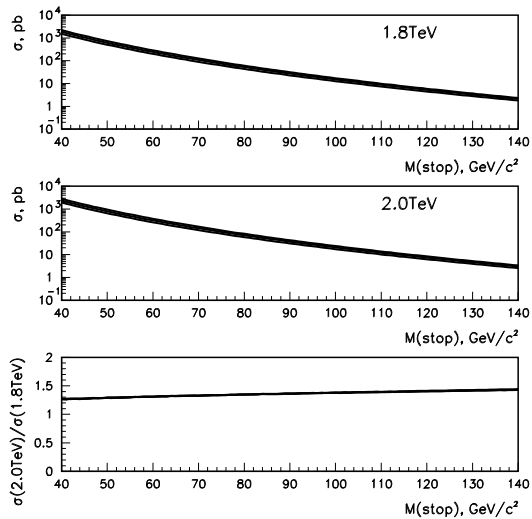,width=0.45\linewidth}}
\end{center}
\caption{Lightest stop  pair production cross section 
as a function of the stop mass at 1.8 TeV and 2.0 TeV center-of-mass
energies. Third plot shows the ratio of the first two. }
\label{st:pro}
\end{figure}

The stop pair production takes place through gluon-gluon and quark-quark
fusions, and consequently, the Born production cross section depends only on
the stop quark mass. The increase in the production cross section due to
variation of the center-of-mass energy from 1.8 to 2~TeV can be as large as 40\%, which is shown in Fig.\ \ref{st:pro} as a function of the stop mass. Moreover, the next-to-leading order SUSY corrections to the stop production have been evaluated \cite{xsec} and they can be as large as 40\%; see Fig.~\ref{fg:kst}(b).

The stop signals are determined by the available stop decay modes. Depending on the SUGRA particle spectrum, the lightest stop can decay at tree level into the
lightest chargino and a $b$-quark ($\tilde{t}_1 \to \tilde{\chi}^+_1 b$), or
into a sneutrino (slepton) accompanied by a charged lepton (neutrino) and a
$b$-quark ($\tilde{t}_1 \to b \ell \tilde{\nu}$ and $b \tilde{\ell} \nu$). If
these decay modes are not kinematically accessible, the stop decays via 1-loop
diagrams into a charm quark and the lightest neutralino ($\tilde{t}_1 \to c
\tilde{\chi}^0_1$). We exhibit in Table \ref{t1} the possible stop signatures
\cite{howie}.

%%%

\begin{table}[htb]
\caption{Possible signatures for stop pair production.}
\begin{center}
\begin{tabular}{||c|c|c||}
\hline
$\tilde{t}$ decay mode		& signature	& selection
\\ \hline \hline
$b \tilde{\chi}^+_1$		& 2 $b$-jets, 2 $W$'s, $\not{\!\!E_T}$
				& $b$-jet, jet, $\ell$, $\not{\!\!E_T}$
\\ \hline
$b\ell\tilde{\nu}$ or $b\nu\tilde{\ell}$ 
& 2 $b$-jets, 2 $\ell$, $\not{\!\!E_T}$	& 2 $\ell$, jet, $\not{\!\!E_T}$
\\ \hline
$c\tilde{\chi}^0_1$		& 2 $c$-jets, $\not{\!\!E_T}$
			        & 2 $c$'s,  $\not{\!\!E_T}$
\\ \hline 
\end{tabular}
\end{center}
%\vskip -24pt
\label{t1}
\end{table}

%%%%%%%%%%

\subsubsection{Reach in the $b \ell \not{\!\!E_T}$ topology }

If the chargino is lighter than the stop, then the main decay mode of
stops is $\tilde{t}_1 \to \tilde{\chi}^+_1 b$. Usually the chargino
decays into the lightest neutralino and a real or virtual $W$. In this
case the process exhibits two $W$'s and two $b$ quarks. Here we focus
on final states exhibiting a $b$-tagged jet, a second jet, a lepton
($e$ or $\mu$) and missing transverse energy $\not{\!\!E_T}$ . The
main standard model backgrounds are $b\bar{b}$ and $t\bar{t}$
production, as well as $W + 2 j$.  In order to enhance the signal
relative to the SM background, we imposed the following cuts:

\begin{itemize} \addtolength{\itemsep}{-2mm}

  \item the lepton must have $p_T > 10$ GeV;

    \item at least 2  jets with  $E_T > 12 $ (8) GeV;

      \item $\not{\!\!E_T} > 15$ GeV;

        \item there should be at least one $b$-tagged jet.

\end{itemize}

\begin{figure}[h]
\vspace{-4ex}

\begin{center}
\mbox{\epsfig{file=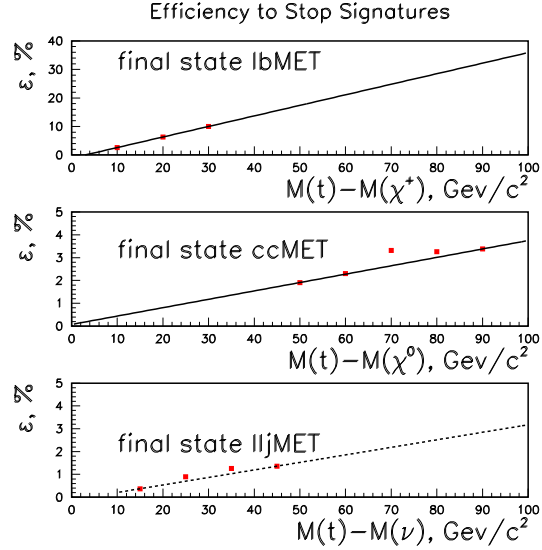,width=0.45\linewidth}}
\end{center}
\caption{Efficiencies for the stop signatures studied in this report.}
\label{eff}
\end{figure}

\begin{figure}[b]
\begin{center}
\mbox{\epsfig{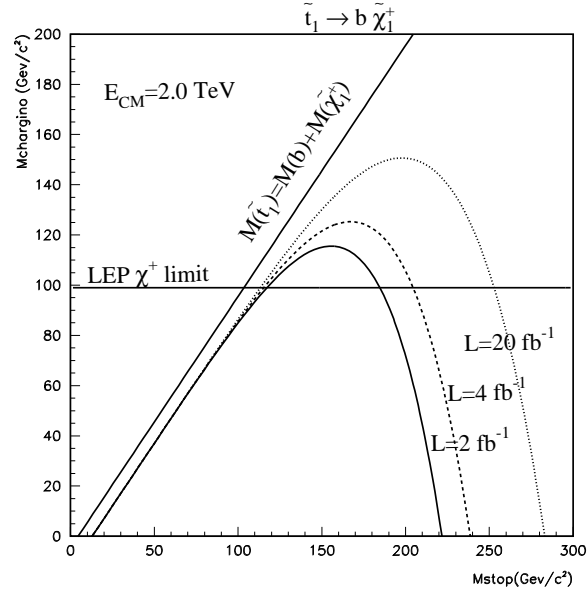}}
\end{center}
\caption{Sensitivity to searches for stop pair production in
$\tilde{t}_1 \to \tilde{\chi}^+_1 b$ channel.}
\label{sen:st}
\end{figure}

We present in Fig.\ \ref{eff} the detection efficiency for this
topology.  The $5\sigma$ sensitivity to searches for $\tilde{t}_1 \to
\tilde{\chi}^+_1 b$ is shown in Fig.\ \ref{sen:st} assuming that the
stop decays 100\% of the time into this channel. Clearly, the search
for this topology will allow us to unravel the existence of stops with
masses up to 185 GeV for an integrated luminosity of 2 fb$^{-1}$.

%%%%%%%%%%

\subsubsection{Reach in the $cc \not{\!\!E_T}$ topology}

The decay $\tilde{t}_1 \to c \tilde{\chi}^0_1$ gives rise to events presenting
two acolinear charm jets and missing $E_T$. 
In our analysis, we assumed
that this decay channel is dominant when it is kinematically allowed.

The main sources of SM backgrounds are the $W/Z$+jets production where
the vector boson decays into a $e$ or $\mu$ that is not identified or
into a $\tau$ which decays hadronically. There is also a small
contribution from QCD multiple jet production and $t\bar{t}$ pairs. In
order to extract the signal we required that

\begin{itemize}
  
\item the event must have 2 or 3 jets with $E_T > 15$ GeV and $45^\circ <
  \Delta \phi(j_1, j_2) < 165^\circ$;
  
\item $\not{\!\!E_T} > 40 $ GeV and is not in the direction of the jets,
{\em i.e.},\\
 $45^\circ < \Delta\phi( \not{\!\!E_T}, j) <165^\circ$;
\item at least one jet is tagged as a charm jet, 
using secondary vertex information.
\end{itemize}

\begin{figure}[hbt]
\begin{center}
\mbox{\epsfig{file=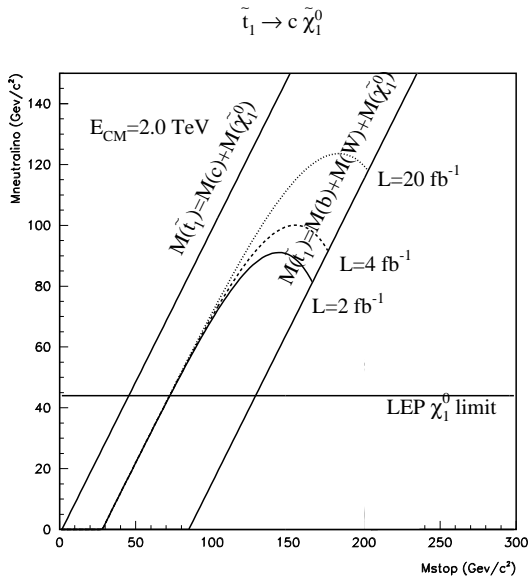,width=0.45\linewidth}}
\end{center}
\caption{Sensitivity to searches for $\tilde{t}_1 \to c \tilde{\chi}^0_1$
 for several integrated luminosities.}
\label{sen:cc}
\end{figure}

We exhibit in Fig.\ \ref{sen:cc} the expect $5\sigma$ sensitivity of
the RUN II for the search of $\tilde{t}_1 \to c \tilde{\chi}^0_1$.
As we can see, this
channel will allow a search for stops with masses up to 160 GeV,
depending on $m_{\tilde{\chi}^0_1}$, for an integrated luminosity of 2
fb$^{-1}$.  The gap between the kinematical limit and the
region of sensitivity is due to the low efficiency for this topology, as can
be seen from Fig.\ \ref{eff}, where the most stringent cut is on
$\etm$. In order to close this gap 
we can lower the missing $E_T$ cut to 25 GeV in the search for
$cc\not{\!\!E_T}$ topology  provided secondary vertex information can be used
at the trigger level.

%%%%%%%%%%

\subsubsection{Reach in the $j \ell^+ \ell^- \not{\!\!E_T}$ topology}

In the parameter space regions where the sleptons are light enough, it is
possible for stops to decay into $b \ell \tilde{\nu}$ or $b \tilde{\ell}
\nu$. In this scenario, the search for events presenting two leptons and a jet
becomes important. Moreover, in this case it is not necessary to require a
$b$-tagged jet. The main SM backgrounds for this topology are Drell-Yan and
the production of $b \bar{b}$, $t\bar{t}$, and $W$+ jets. In order to
select these events we required

\begin{itemize} \addtolength{\itemsep}{-2mm}

  \item two leptons with $p_T(\ell_1) > 8$ GeV and $p_T(\ell_2) > 5$ GeV;

    \item at least one jet with $E_T > 15$ GeV;
      
      \item $\not{\!\!E_T} > 20$ GeV.

\end{itemize}

\begin{figure}[hbt]
\begin{center}
\mbox{\epsfig{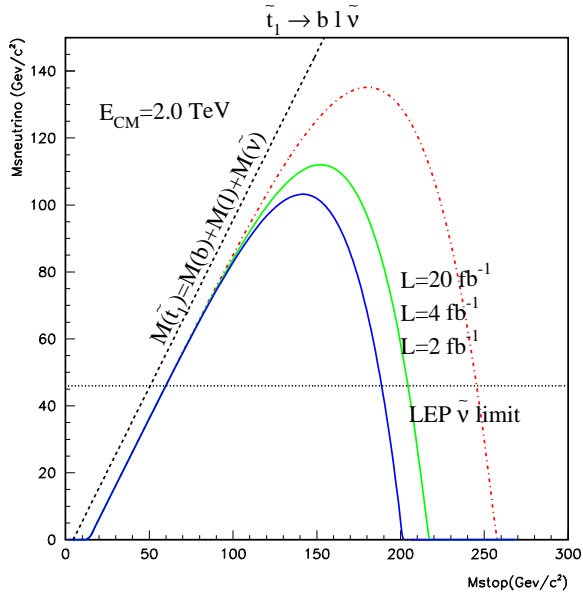}}
\end{center}
\caption{Sensitivity to searches for $\tilde{t}_1 \to b \ell
  \tilde{\nu}$ or $b \nu \tilde{\ell}$ for different integrated
  luminosities.}
\label{sen:ll}
\end{figure}

The reach of the stop search in the $j \ell^+ \ell^- \not{\!\!E_T}$
channel is presented in Fig.\ \ref{sen:ll} for several integrated
luminosities. Clearly this channel will be able to probe stop masses
up to 190 GeV for an integrated luminosity of 2 fb$^{-1}$.

%%%%%%%%%%%%%%%%%%%%%%%%%%%%%%%%%%%%%%%%%%%%%%%%%%%%%%%%%%%%%%%%%%%%%%

\subsection{Sbottom pair production}

It is possible that $\tilde{b}_1$, the lighter of the two $b$-squark
eigenstates, may be much lighter than other squarks. Moreover, in models (such
as mSUGRA) with universal squark masses at some high scale, it is reasonable
to suppose that $\tilde{b}_1 \approx \tilde{b}_L$ as long as the bottom Yukawa
coupling is small relative to its top counterpart.

The sbottom pair production signatures depend on the allowed sbottom
decay channels. In this analysis, we assume that the gluino is so heavy
that the decay $\tilde{b}_1 \to b\tilde{g}$ is kinematically
forbidden, and further, that the lightest neutralino
($\tilde{\chi}^0_1$) mass is such that the decay $\tilde{b}_1 \to
b\tilde{\chi}^0_1$ is always allowed.  The other possible two-body
decay modes are $\tilde{b}_1 \to b\tilde{\chi}^0_i$ and $\tilde{b}_1
\to t\tilde{\chi}^+_i$.  Given the available center-of-mass energy at the
Tevatron, it is sufficient to focus on the case where only the
neutralino decays of $\tilde{b}_1$ are accessible.

We use ISAJET 7.37~\cite{ISAJET} for our simulation, and we model the
experimental conditions at the Tevatron through the toy calorimeter
simulation package ISAPLT. We simulate calorimetry covering $-4<\eta
<4$ with cell size $\Delta\eta\times\Delta\phi =0.1\times 5^\circ$. We
take the hadronic (electromagnetic) energy resolution to be $50\%
/\sqrt{E}$ ($15\% /\sqrt{E}$).  Jets are defined as hadronic clusters
with $E_T > 15$~GeV within a cone with $\Delta R=\sqrt{\Delta\eta^2
+\Delta\phi^2} =0.7$. We require that $|\eta_j| \leq 3.5$.  Muons and
electrons are classified as isolated if they have $p_T>10$~GeV, $|\eta
(\ell )|<2$, and the visible activity within a cone of $R=0.3$ about
the lepton direction is less than 5~GeV.  For SVX tagged $b$-jets, we
require a jet (using the above jet requirement) to have in addition
$|\eta_j|<1$ and to contain a $B$-hadron with $p_T > 15$~GeV. Then the
jet is tagged as a $b$-jet with a 50\% efficiency.

%%%%%%

\subsubsection{The ${\tilde{b}_1} \to \tilde{\chi}^0_1 b$ case}

Here we assume that $m_{\tilde{\chi}^0_2} > m_{\tilde{b}_1}-m_b$ so that
$\tilde{b}_1 \to b\tilde{\chi}^0_1$ with a branching fraction of essentially
100\%. In this case, the signal naively consists of two $b$-jets recoiling
against $\etm$ from the two neutralinos that escape detection. The dominant SM
backgrounds come from $W + j$, $Z \to \nu\nu +j$, $Z \to \tau\tau +j$ and
$t\bar{t}$ production.  To enhance the signal relative to the SM background,
we impose the following requirements, hereafter referred to as the basic cuts:

\begin{enumerate}
\item  at least two jets with $p_T(j_1)>30$~GeV, $p_T(j_2)> 20$~GeV;
\item at least one jet in $|\eta_j|<1$;
\item $\etm > 50$~GeV;
\item $\Delta\phi(\vec{\etm},\vec{p}_{Tj})>30^\circ$;
\item for di-jet events only, $\Delta\phi(\vec{p}_{Tj1},\vec{p}_{Tj2})
<150^\circ$;
\item at least one SVX tagged B;
\item no isolated leptons ($e$ or $\mu$).
\end{enumerate}

The dominant SM background at 2 TeV is $t\bar{t}$ production; see Table
\ref{bg:sb}.  Because of the lepton veto, much of this background comes when
one of the tops decays into a tau lepton that decays hadronically, while the
other top decays completely hadronically. These events are, therefore, likely
to have large jet multiplicity, in contrast to the signal. We are, therefore,
led to impose the additional requirement,
\begin{enumerate}
\setcounter{enumi}{7}
\item $n_j =2,3\,,$
\end{enumerate}
designed to further reduce the top background with relatively modest
loss of signal. The corresponding background levels are shown in the
second row of Table \ref{bg:sb}. The entry +8 in the first column
denotes the cuts over and above the basic cuts 1-7. Indeed, we see
that the top background is reduced by a factor 5, and no longer
dominates.  Now the background comes mainly from $Z \to \nu\nu + j$
events.  Therefore we are led to impose, in addition to cuts 1-8, a
further requirement,
\begin{enumerate}
\setcounter{enumi}{8}
\item $\Delta\phi({j_1, j_2}) \geq 90^\circ$,
\end{enumerate} 
which significantly reduces the vector boson backgrounds, as can be seen from
the third row in Table \ref{bg:sb}.  For higher integrated luminosity, like
25~$fb^{-1}$, the high event rate makes it possible to require double
$b$-tagging,
\begin{enumerate}
\setcounter{enumi}{9}
\item $n_b \geq 2$,
\end{enumerate}
to greatly reduce the vector boson background.  Finally, we also considered
the cut,
\begin{enumerate}
\setcounter{enumi}{10}
  \item $m_{j_1, j_2} \leq 60$~GeV, where $j_1$ and $j_2$ are the two
  highest $p_T$ untagged jets in the event. If an event has less than
  two untagged jets, we retain it as part of the signal.
\end{enumerate}

%%%

\begin{table}[hbt]
\caption[]{Standard Model background cross sections in fb to the
$b$-squark signal after the basic cuts 1-7 described in the text, as
well as after additional cuts designed to further reduce
backgrounds. The ``plus entries'' in the first column refer to the cuts
in addition to the basic cuts; for instance, the last row has cuts 1-8
together with cut 11. We take $m_t=175$ GeV.}
\begin{center}
\begin{tabular}{|l|c|c|c|c|c|}
\hline
CUT &$W+j$ & $Z \to \nu\nu +j$ & $Z \to \tau\tau +j$ & $t\bar{t}$ & $Total$ \\ 
\hline
Basic & 65.5 & 92.4 &2.6  & 195 & 356 \\
$+8$ & 51.6 & 80.6 & 2.1 & 36.7 & 172 \\
$+8+9$ &21.9 & 26.2 & 0.9  & 28.0 & 77 \\
$+10$ & 5.6 & 7.2 & 0.1 & 37.5 & 50.4 \\
$+8+10$ & 3.8 & 6.6 & 0.1 & 6.7 & 17.2 \\
$+10+11$ &5.0 & 6.6 & 0 & 11.7 & 23.3 \\
\hline
\end{tabular}
\end{center} 
\bigskip
\label{bg:sb}
\end{table}

\begin{figure}[t]
\begin{center}
    \mbox{\epsfig{file=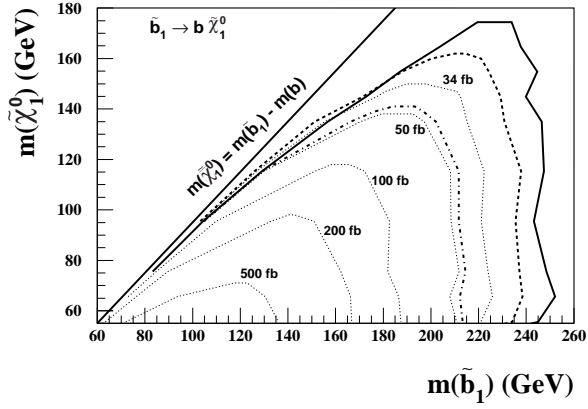,width=0.5\linewidth}}
\vspace{-5ex}
\end{center}
\caption{The region of the $m_{\tilde{b}_1}-m_{\tilde{\chi}^0_1}$ 
  plane that can be probed at a 2~TeV $p\bar{p}$ collider, assuming
  that $\tilde{b}_1 \to b\tilde{\chi}^0_1$ and that $\tilde{\chi}^0_1$
  escapes detection. The sbottom signal should be detectable with the
  observability criteria defined in the text in the region below the
  dot-dashed, dashed and solid contours for an integrated luminosity
  of 2~fb$^{-1}$, 10~fb$^{-1}$ and 25~fb$^{-1}$. Also shown in the
  figure are contours of constant signal cross section after cuts 1-8
  for the 2~fb$^{-1}$ case. The 34~fb contour marks the $0.2B$ level
  that we require as a minimum for the signal.  The diagonal line
  marks the boundary of the region beyond which $m_{\tilde{b}_1} >
  m_b+m_{\tilde{\chi}^0_1}$.  \label{re:sb}}

%\end{figure}

%\begin{figure}[h]
\begin{center}
    \mbox{\epsfig{file=Eboli-4F/sens_sbot.epsi,width=0.45\linewidth}}
\vspace{-2ex}
\end{center}
\caption{CDF discovery potential of bottom squarks in the channel
$\tilde{b}_1 \to b\tilde{\chi}^0_1$. \label{b:sens}}

\end{figure}

Turning to the signal, the sbottom production cross section is completely
determined by $m_{\tilde{b}_1}$ and $m_{\tilde{\chi}^0_1}$.  The $5\sigma$
reach of a 2~TeV $p\bar{p}$ collider \cite{bpt} is shown in Fig.\ \ref{re:sb}.
In addition, we also require the signal to exceed 20\% of the background.  The
diagonal solid line marks the boundary of the region where $m_{\tilde{b}_1}
\geq m_{\tilde{\chi}^0_1} + m_b$.  With our assumptions, we must be below this
line, since otherwise, $\tilde{b}_1$ would be the LSP.  The dot-dashed contour
shows the reach that should be attainable (using cuts 1-8) with an integrated
luminosity of 2~fb$^{-1}$. The dotted lines denote signal cross sections after
these cuts.  Preliminary CDF studies indicate that the attainable limits do
not change significantly when a full detector simulation is carried out
\cite{regi}; see Fig.\ \ref{b:sens}.

\renewcommand{\textfraction}{0.0}

%% \clearpage
%%%%%%

\subsubsection{The $m_{\tilde{b}_1} > m_b + m_{\tilde{\chi}^0_2}$ case}

The signal now depends on the branching fraction for the decay
$\tilde{b}_1 \to b \tilde{\chi}^0_2$, as well as the decay pattern of
$\tilde{\chi}^0_2$. In other words, the signal depends not only on the
$b$-squark mixing angle but also on the parameters of the neutralino
mass matrix.  To make our analysis tractable, we will assume that
$\tilde{b}_1 \approx \tilde{b}_L$.  We will further assume that
$|\mu|$ is much larger than electroweak gaugino masses. Assuming
gaugino mass unification, the two lighter neutralinos are
approximately the hypercharge gaugino and the $SU(2)$-gaugino, with
$m_{\tilde{\chi}^0_2} \approx m_{\tilde{\chi}^+_1} \approx
2m_{\tilde{\chi}^0_1}$. Note that these assumptions fix the branching
fraction for $\tilde{b}_1 \to b\tilde{\chi}^0_{1,2}$ decays in terms
of the sparticle masses. It is also worth pointing out that if
$\tilde{\chi}^0_2 \simeq SU(2)$-gaugino, it essentially decouples from
$\tilde{b}_R$, so that the maximum impact of the $\tilde{b}_1 \to
b\tilde{\chi}^0_2$ indeed occurs when $\tilde{b}_1 =\tilde{b}_L$.

Since we are interested in seeing how much the reach of the Tevatron may be
reduced from that shown in Fig.\ \ref{re:sb}, we consider extreme limits for
how $\tilde{\chi}^0_2$ might decay \cite{FN2}. If $\tilde{\chi}^0_2$
dominantly decays to leptons via $\tilde{\chi}^0_2 \to
\ell\bar{\ell}\tilde{\chi}^0_1$, sbottom pair production would result in
characteristic $bb+4\ell +\etm$ and $bb+2\ell + jets +\etm$ events for
which the background is small, and the corresponding reach,
presumably, larger than that in Fig.\ \ref{re:sb}. If
$\tilde{\chi}^0_2 \to b\bar{b}\tilde{\chi}^0_1$ it may be possible to
reduce the background (and hence increase the reach) by requiring two
(or more) $b$-tags. The worst case ``realistic'' scenario is when
$\tilde{\chi}^0_2$ decays into jets which are not amenable to any
tagging \cite{FN3}.  To simulate this situation, we have forced
$\tilde{\chi}^0_2$ to decay via $\tilde{\chi}^0_2 \to
u\bar{u}\tilde{\chi}^0_1$ and run these events through our simulation,
and once again obtained the reach for the three choices of integrated
luminosity in Fig.\ \ref{re:sb}.

\begin{figure}[hbt]
\vspace{-4ex}

\begin{center}
    \mbox{\epsfig{file=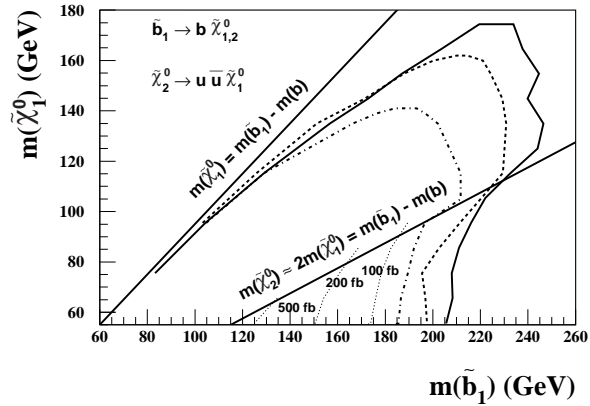,width=0.5\linewidth}}
\end{center}
\caption{The same as Fig.\ \protect\ref{re:sb} except that the decay
  $\tilde{b}_1 \to b\tilde{\chi}^0_2$ is also allowed when kinematically
  accessible below the lower diagonal line.  To illustrate the largest
  degradation of the reach in this scenario, we assume that $\tilde{\chi}^0_2$
  always decays via $\tilde{\chi}^0_2 \to u\bar{u} \tilde{\chi}^0_1$. The
  dotted lines are again contours of fixed cross section after the basic cuts
  1-7.  The dot-dashed contour that denotes our projection of the MI reach
  also corresponds to $S=0.2B$.  }
\label{re:sb2}
\end{figure}

The results of our analysis for this case are illustrated in Fig.\ 
\ref{re:sb2}.  The upper diagonal line is as in Fig.\ \ref{re:sb}, while the
lower line is where $m_{\tilde{b}_1} =2m_{\tilde{\chi}^0_1} +m_b \simeq
m_{\tilde{\chi}^0_2}+m_b$.  In our analysis, we have adjusted $A_b$ to cancel
the off diagonal term in the sbottom mass matrix in order to make $\tilde{b}_1
=\tilde{b}_L$. We have fixed $\mu = 500$~GeV and $\tan\beta =2$; this value of
$\mu$ is large enough for $\tilde{\chi}^0_1$ and $\tilde{\chi}^0_2$ to be
gauginos to a very good approximation. When the decay $\tilde{b}_1 \to
b\tilde{\chi}^0_2$ is inaccessible, the reach should be as given by our
analysis above; {\it i.e.}, the reach illustrated by the dot-dashed, dashed
and solid contours until just above this line, is identical to that in Fig.\ 
\ref{re:sb}.  The contours below this line show the extent to which the reach
might be reduced if the $\tilde{b}_1$ can also decay to $\tilde{\chi}^0_2$.
These contours in Fig.\ \ref{re:sb2} turn inwards just slightly above this
line precisely because the relation $m_{\tilde{\chi}^0_2} =
2m_{\tilde{\chi}^0_1}$ is slightly violated by our finite choice of $\mu$.

In summary, assuming that $\tilde{b}_1 \to b\tilde{\chi}^0_1$ and that
$\tilde{\chi}^0_1$ escapes detection, we have shown that it should be possible
for experiments at the MI to detect $b$-squark signals over SM backgrounds for
$m_{\tilde{b}_1} \leq 210$~GeV, even if the LSP is quite heavy. The capability
of tagging $b$-jets in the central region with high efficiency and purity is
crucial for this detection.  The reach may be somewhat degraded if sbottom can
also decay into $\tilde{\chi}^0_2$.  We have argued that in many models (with
$\tilde{\chi}^0_1$ as a stable LSP) this degradation is typically smaller than
30--40~GeV.

%%%%%%%%%%%%%%%%%%%%%%%%%%%%%%%%%%%%%%%%%%%%%%%%%%%%%%%%%%%%%%%%%%%%%%

%% AZZI SECTION REMOVED

%%%%%%%%%%%%%%%%%%%%%%%%%%%%%%%%%%%%%%%%%%%%%%%%%%%%%%%%%%%%%%%%%%%%%%

\subsection{Gluino production}

The gluino pair production takes place through gluon-gluon and
quark-quark fusions, and consequently, the production cross section
depends only on the gluino mass. However, the signals for gluino
production depend on properties of all sparticles lighter than these
since they can appear in the gluino cascade decays. Moreover, its
signal comes mixed with the ones due to the production of squark as
well as gluino squark pairs.

\subsubsection{$\etm +$ jets channel}\label{et0l}

\begin{figure}[hbt]
\vspace{-4ex}

\begin{center}
    \mbox{\epsfig{file=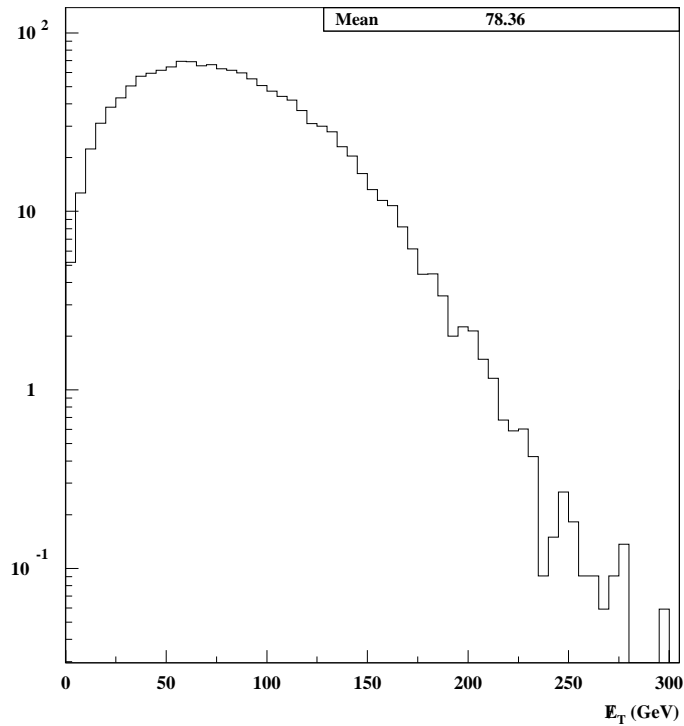,width=0.6\linewidth}}
\vspace{-4ex}
\end{center}
\caption{$\etm$ distribution for $\tan \beta =2$, $\mu < 0$, $A_0=0$, 
$m_0=100$ GeV, and $m_{1/2}=70$ GeV, which corresponds to
$m_{\tilde{q}} = 207$ GeV and $m_{\tilde{g}} = 215$ GeV.}
\label{glu:met}
\end{figure}

The canonical signature of supersymmetry is multi-jet events with a
large $\etm$ \cite{etj}; see Fig.\ \ref{glu:met}. This signal can
originate from many SUSY particle production like $\tilde{g}
\tilde{g}$ or $\tilde{q} \tilde{q}$. The main SM backgrounds for this 
channel are single and pair production of $W$'s and/or $Z$'s in
association with jets, $t \bar{t}$, and QCD events with $\etm$ due to
mismeasurement of the jets. In order to enhance the signal and
suppress the backgrounds we imposed the following cuts \cite{ncpt}:

\begin{itemize} \addtolength{\itemsep}{-1mm}

        \item jet multiplicity $n_{\rm jet} \ge 2$;
        
        \item transverse sphericity $S_T > 0.2$;

        \item $\etm > 40$ GeV;

        \item the missing energy does not point along a jet, {\em i.e.}
        $\Delta \phi ( \vec{\etm}, \vec{E}^j_T ) > 30^\circ$;

        \item $E_T(j_1)~,~E_T(j_2) > E_T^c$ and $\etm > E_T^c$ where
        the parameter $E_T^c$ was adjusted as described below;

        \item no isolated leptons.

\end{itemize}

%%%%

%%%%%

We considered a signal to be observable if, for a given integrated luminosity,
we have (i) at least 5 signal events, (ii) the statistical significance of the
signal exceeds $5\sigma$, and (iii) the signal is larger than 20\% of the
background. We checked the observability of the signal for $E_T^c=15$, 40, 60,
80, 100, 120, and 140 GeV and considered the signal to be observable if it is
so for any one of the values of $E_T^c$. In our analysis we used ISAJET to
generate the signal and backgrounds and assumed a toy calorimeter similar to
the one described in the sbottom analyses.

\begin{figure}[hbt]
\begin{center}
    \mbox{\epsfig{file=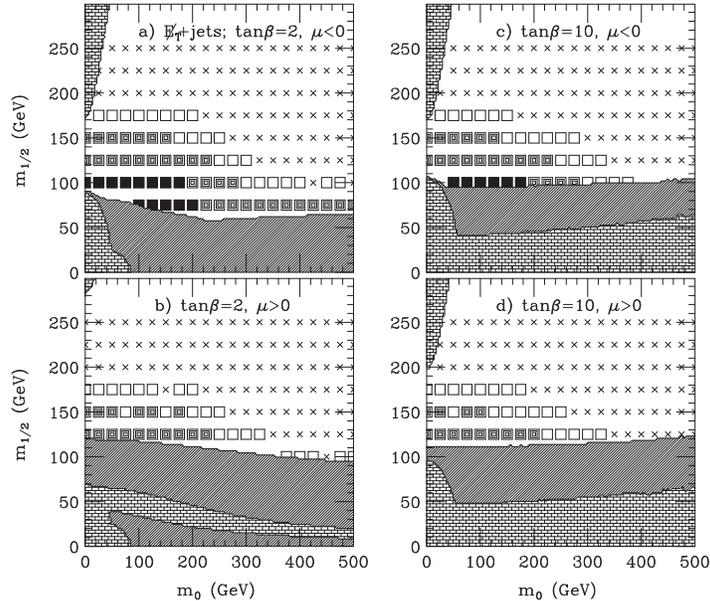,width=0.55\linewidth}}
\end{center}
\caption{Regions of the $m_0 \times m_{1/2}$ plane where the multi-jets
plus $\etm$ signal is observed at a 2 TeV $p \bar{p}$ collider. We
considered three values for the integrated luminosity: 100 pb$^{-1}$
(black squares), 2 fb$^{-1}$ (gray squares), and 25 fb$^{-1}$ (white
squares). The bricked regions are excluded by theoretical constraints
and the shaded regions are excluded by experiment. }
\label{glu:etmiss}
\end{figure}

In Fig.\ \ref{glu:etmiss}, we present the observable regions of the
plane $m_0 \times m_{1/2}$ according to the above criteria
\cite{ncpt}.  This figure was obtained for $\tan\beta =2$ (10) and
both signs of $\mu$ with $A_0$ being fixed to zero.  As we can see
from this figure, with a data sample of 2 fb$^{-1}$, the Tevatron
experiments should be able to probe $m_{1/2}$ up to 150 GeV,
corresponding to $m_{\tilde{g}} \simeq 400$ GeV, if $m_0 < 200 $ GeV.

%%%%

%%%%%

\subsubsection{Multijet plus leptons and $\etm$ channels}

The cascade decay of gluinos (or squarks) can also give rise to
leptons in addition to jets and missing $E_T$. We further classify the
events by their isolated lepton ($e$ or $\mu$) content as follows:

\begin{itemize} \addtolength{\itemsep}{-1mm}
  
\item $1\ell$ events with exactly one isolated lepton satisfying $E_T(\ell)>
  10$~GeV.  To reduce the background from $W$ production, we also required
  $M_T(\ell,\etm)> 100$~GeV;

 \item Opposite sign ($OS$) dilepton events with exactly two unlike
 sign isolated leptons, where we required $E_T(\ell_1)> 10$~GeV;

 \item Same sign ($SS$) dilepton events with exactly two same sign
 isolated leptons, again with $E_T(\ell_1)> 10$~GeV;

 \item $3\ell$ events, with exactly three isolated leptons with
 $E_T(\ell_1) > 10$~GeV. We veto events with
 $|M(\ell^+\ell^-)-M_Z|<8$~GeV.

\end{itemize}

\begin{figure}[hbt]
\begin{center}
    \mbox{\epsfig{file=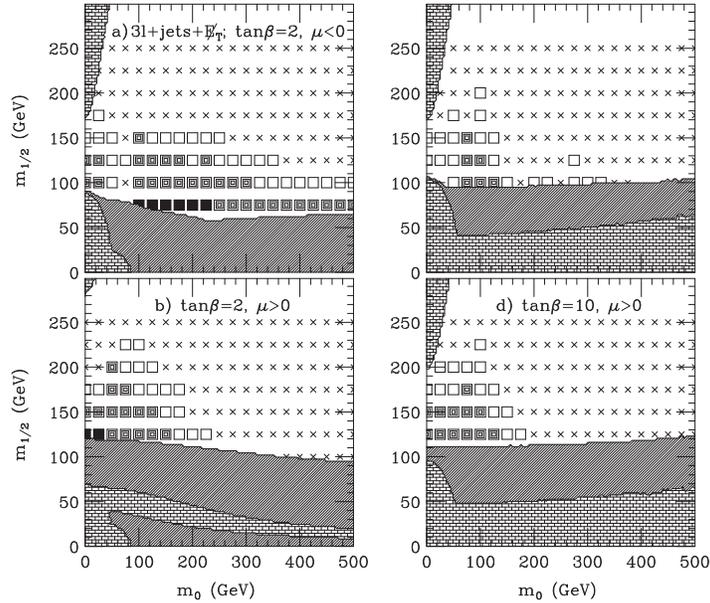,width=0.55\linewidth}}
\end{center}
\caption{The same as Fig.\ \protect\ref{glu:etmiss} except for the
multijet plus $3\ell$ and $\etm$ channel. }
\label{glu:3l}
\end{figure}

A detailed analysis of the above signatures \cite{ncpt} indicates that
the maximal reach is still obtained in the $\etm$ plus multiplet
channel, except for isolated values of SUGRA parameters where the
trilepton channel has a larger reach (see Fig.\ \ref{glu:3l}).

%%%%

%%%%%

\subsubsection{Large $\tan\beta$}

At large $\tan\beta$ the tau and bottom Yukawa couplings become
comparable to the electroweak gauge couplings and even to the top
Yukawa coupling. This has a significant impact \cite{tb} on the search
for supersymmetry at colliders. For large $\tan \beta$ the lightest tau slepton
and bottom squark can be considerably lighter than the corresponding
sleptons and squarks of the first two generations. Moreover, gluino,
chargino and neutralino decays to third generation particles are
significantly enhanced when $\tan\beta$ is large.

The phenomenological implications related to large values of
$\tan\beta$ are: the Tevatron signals in multilepton ($e$ and $\mu$)
channels are greatly reduced while there could be new signals
involving $b$-jets and $\tau$-leptons via which to search for SUSY
\cite{tb}. Furthermore, for very large $\tan\beta$ the greatest 
reach is attained in the multi-jet$+\etm$ signature. Figure~\ref{HT-total} (earlier in this report) shows the reach of the Tevatron upgrades for large and small $\tan\beta$, where a point is considered to be accessible if there is
a channel leading to a $5\sigma$ effect. In this analyses the same
cuts of the previous subsection were used \cite{tb}.  We can see from
this figure that the SUSY sensitivity of the Tevatron upgrades is
reduced as $\tan\beta$ increases.

%%%%

%\begin{figure}[hbt]
%\begin{center}
%    \mbox{\epsfig{file=Eboli-4F/newlastfig.eps,width=0.6\linewidth}}
%\end{center}
%\caption{A plot of points accessible to Tevatron MI and TeV33
%searches for mSUGRA via {\it all} of the signal channels 
%\protect\cite{tb}.  A $5\sigma$ signal above background is found for 
%some value of $E_T^c$ for the MI for gray squares, while white squares
%are accessible only at TeV33.  Points with a $\times$ symbol are
%inaccessible to MI and TeV33. }
%\label{glu:tb}
%\end{figure}

%%%%%

%%%%%%%%%%%%%%%%%%%%%%%%%%%%%%%%%%%%%%%%%%%%%%%%%%%%%%%%%%%%%%%%%%%%%%

%\end{document}

%% file: Kao/trilepton.tex
%=======================================================================
% Trilepton.tex for the mSUGRA group of Physics at RUN II 
% This manuscript is in LaTeX
% Version of August 25, 1999
%=======================================================================

%------------------------------------
% New Commands	
%------------------------------------

%\renewcommand{\alt}{\matrix{<\cr\noalign{\vskip-7pt}\sim\cr}}
%\renewcommand{\agt}{\matrix{>\cr\noalign{\vskip-7pt}\sim\cr}}
\renewcommand{\notE}{\ \hbox{{$E$}\kern-.60em\hbox{/}}}
\newcommand{\notp}{\ \hbox{{$p$}\kern-.43em\hbox{/}}}
\renewcommand{\tchi}{\tilde{\chi}}
\def\D0{\mbox{D\O }}

%=======================================================================
% TITLE PAGE
%=======================================================================

%-----------------------------------
%   Title
%-----------------------------------
\section{Trilepton Signal of Minimal Supergravity
at the Upgraded Tevatron}
 
%-----------------------------------
%   Authors
%-----------------------------------
%\author{
%R.~Arnowitt, H.~Baer, \underline{V.~Barger}, T.~Kamon, \underline{C.~Kao}, \\
%S.~Lammel, T.-J.~Li, P.~Mercadante, J.~Nachtman, \\ 
%B.~Tannenbaum, P.~Nath, X.~Tata and Y.~Wang}

%=======================================================================
%   BEGIN MAIN TEXT
%=======================================================================

%-----------------------------------------------------------------------
%   I. Introduction
%-----------------------------------------------------------------------
\subsection{Introduction}

%------------------------------------
% Trilepton Signature
%------------------------------------

In this report, we assess the prospects of discovering 
the trilepton signal along with missing transverse energy ($3\ell+\notE_T$) 
\cite{Wdecay,Trilepton1,Trilepton2,Trilepton3,BCDPT,%
Madison1,Madison2,Matchev1,Matchev2,Baer&Tata,D0-3l,CDF-3l} 
in the minimal supergravity model (mSUGRA) 
at the upgraded Tevatron with 2 TeV center of mass energy.  
We assume each of the CDF and the \D0 experiments 
will accumulate an integrated luminosity (${\cal L}$) of 2 fb$^{-1}$ 
at Run II. 
In addition, we consider a possible upgrade of the Tevatron luminosity 
to $10^{33}$ cm$^{-2}$ s$^{-1}$ at Run III and take the corresponding 
integrated luminosity to be ${\cal L} = 30$ fb$^{-1}$ \cite{Run3}.
The major source of this signal is associated production of 
the lightest chargino ($\tchi^\pm_1$) 
and the second lightest neutralino ($\tchi^0_2$) with decays to leptons. 

In the mSUGRA unified model, 
the sleptons ($\tilde{\ell}$), the lighter chargino ($\tchi^\pm_1$) 
and the lighter neutralinos ($\tchi^0_1,\tchi^0_2$) 
are typically less massive than gluinos and squarks.
Because of this, the $3\ell+\notE_T$ signal 
from associated production and decays of $\tchi^\pm_1 \tchi^0_2$ 
is one of the most promising channels for supersymmetric (SUSY) 
particle searches at the Tevatron.  
The background to this signal from processes in the Standard Model (SM) 
can be greatly reduced with suitable cuts.
In most of the mSUGRA parameter space, 
the weak-scale gaugino masses are related 
to the universal gaugino mass parameter $m_{1/2}$ by 
$m_{\tchi^0_1} \sim 0.4 m_{1/2}$ and 
$m_{\tchi^\pm_1} \sim m_{\tchi^0_2} \sim 0.8 m_{1/2}$. 
Consequently, this discovery channel could provide valuable information 
about the value of $m_{1/2}$.
We consider universal boundary conditions at $M_{\rm GUT}$
with a common gaugino mass $m_{1/2}$ and a common scalar mass $m_0$
to study the production cross section and decay branching fractions
of $\tchi^\pm_1$ and $\tchi^0_2$.
Non-universal boundary conditions among sfermion masses \cite{sfermions-3l}
or the gaugino masses \cite{gauginos-3l}
could change the production cross section and branching fractions
of the charginos and neutralinos.
For $m_{1/2} =$ 200 GeV and $\tan\beta \alt 25$,
a non-universality among sfermions may significantly enhance
the trilepton signal when 50 GeV $\alt m_0 \alt$ 130 GeV \cite{sfermions-3l}.

%-----------------------------------
% tan(b)
%-----------------------------------

The Yukawa couplings of the bottom quark ($b$) and the tau lepton ($\tau$) 
are proportional to $\sec\beta$ and are thus greatly enhanced 
when $\tan\beta$ is large.
In SUSY grand unified theories, 
the masses of the third generation sfermions 
are consequently very sensitive to the value of $\tan\beta$.
As $\tan\beta$ increases, the lighter tau slepton ($\tilde{\tau}_1$) 
and the lighter bottom squark ($\tilde{b}_1$) become lighter than 
charginos and neutralinos while other sleptons and squarks are heavy. 
Then, $\tchi^\pm_1$ and $\tchi^0_2$ can dominantly decay into 
final states with tau leptons via real $\tilde{\tau}_1$.

%------------------------------------------------
% tau tau tau, tau tau l, tau ll, lll
%------------------------------------------------

\begin{figure}[htb]
\centering\leavevmode
\epsfxsize=4in\epsffile{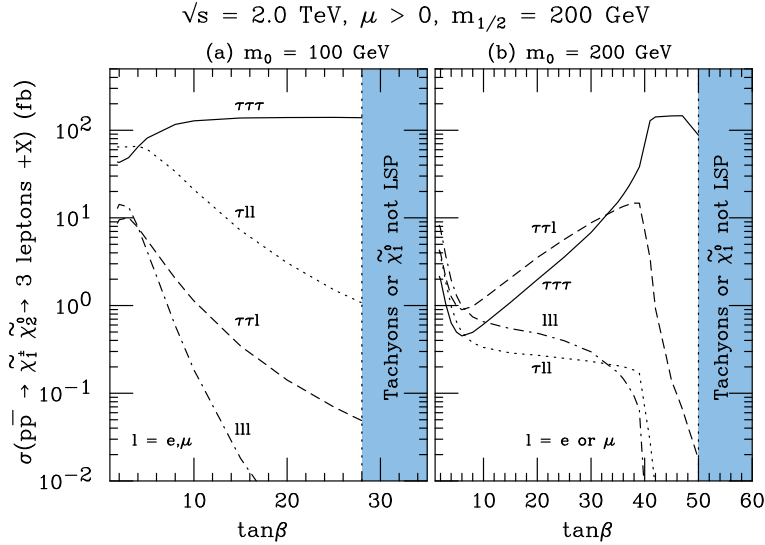}

\caption[]{
Cross section of 
$p\bar{p} \to \tchi^\pm_1 \tchi^0_2 \to 3\;{\rm leptons}\; +X$ 
without cuts at $\sqrt{s} = 2$ TeV versus $\tan\beta$, 
with $\mu > 0$, $m_{1/2} = 200$ GeV, $m_0 =$ 100 GeV 
for (a) $\tau\tau\tau$ (solid), (b) $\tau \tau \ell$ (dot-dash), 
(c) $\tau \ell \ell$ (dash) and (d) $\ell \ell \ell$ (dot), 
where $\ell = e$ or $\mu$. 
\label{fig:taus}}
\end{figure}

The relevance of $\tau$ leptons in the production 
and decays of $\tchi^\pm_1 \tchi^0_2$, are illustrated in Figure~\ref{fig:taus},
where the product of the cross section 
$\sigma(p\bar{p} \to \tchi^\pm_1 \tchi^0_2 +X)$ 
and the branching fraction 
$B(\tchi^\pm_1 \tchi^0_2 \to 3 \; {\rm leptons}\; +\notE_T)$ 
versus $\tan\beta$ is presented with $\mu > 0$, $m_{1/2} = 200$ GeV, 
$m_0 =$ 100 and 200 GeV, 
for four final states (a) $\tau\tau\tau$, (b) $\tau\tau\ell$, 
(c) $\tau\ell\ell$ and (d) $\ell\ell\ell$, where $\ell = e$ or $\mu$.
For $m_0 \alt$ 200 GeV and/or $\tan\beta \agt 40$, 
channels with at least one $\tau$ lepton are dominant.

%------------------------
% Figure 1
%------------------------

One way to detect $\tau$ leptons is through their one prong 
and three prong hadronic decays. 
The CDF and the \D0 collaborations are currently investigating 
the efficiencies for detecting these modes 
and for possibly implementing a $\tau$ trigger. 
Recently, it has been suggested that the $\tau$ leptons 
in the final state may be a promising way 
to search for $\tchi^\pm_1 \tchi^0_2$ production at the Tevatron 
if excellent $\tau$ identification becomes feasible 
\cite{BCDPT,Matchev1,Wells}.

Another way of exploiting the $\tau$ signals, 
that we consider in this report, 
is to detect the soft electrons and muons from leptonic $\tau$ decays 
by employing softer but realistic $p_T$ cuts on the leptons 
\cite{Madison1,Madison2}. 
We find that this can considerably improve the trilepton signal 
significance from $\tchi^\pm_1 \tchi^0_2$ production.

%------------------------------------------------------------------------
% II. Production Cross Section and Branching Fractions
%------------------------------------------------------------------------
\subsection{Cross Section and Branching Fractions}

%------------------------------------
% Production Rate 
%------------------------------------

In hadron collisions, associated production of chargino and neutralino 
occurs via quark-antiquark annihilation 
in the $s$-channel through a $W$ boson 
($q\bar{q'} \to W^{\pm} \to \tchi^\pm_1 \tchi^0_2$) and 
in the $t$ and $u$-channels through squark ($\tilde{q}$) exchanges.
If the squarks are light, 
a destructive interference between the $W$ boson and the squark exchange 
amplitudes can suppress the cross section by as much as $40\%$ 
compared to the $s$-channel contribution alone.
For squarks much heavier than the gauge bosons, 
the effect of negative interference is reduced 
and the $s$-channel $W$-resonance amplitude dominates.

\begin{figure}[t]
\centering\leavevmode
\epsfxsize=4in\epsffile{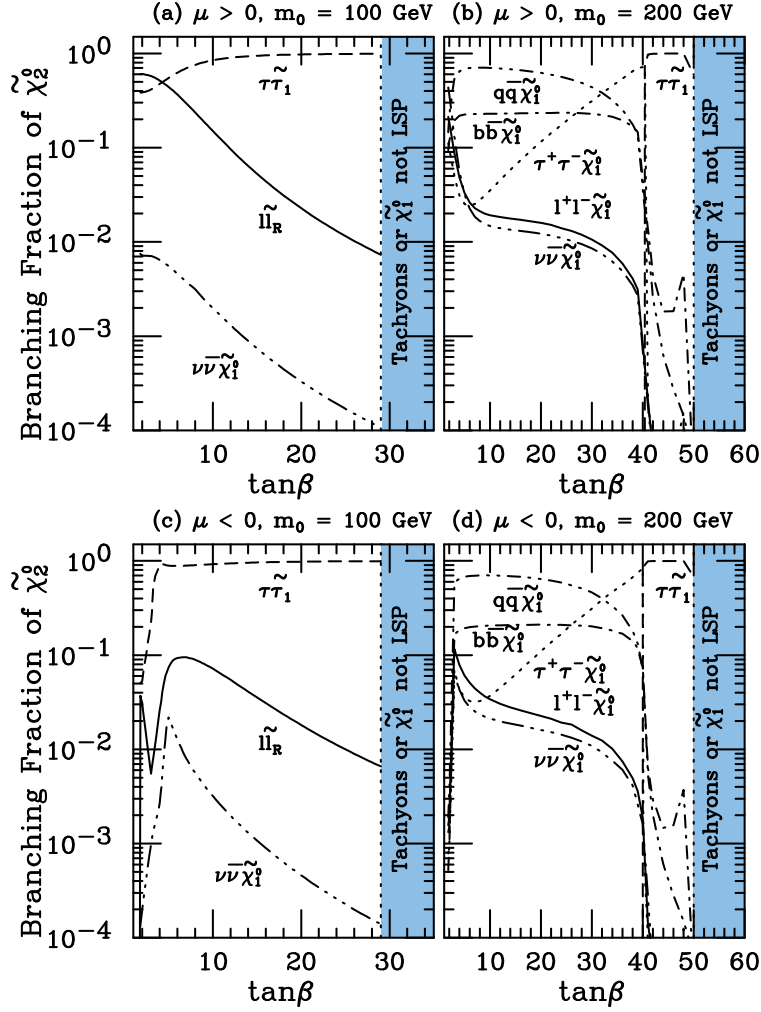}

\caption[]{
Branching fractions of $\tchi^0_2$ decays into various channels 
versus $\tan\beta$ with $m_{1/2} =$ 200 GeV, 
for $m_0 = 100$ GeV and 200 GeV. 
\label{fig:bfwz}
}\end{figure}

In Figure~\ref{fig:bfwz}, we present branching fractions of $\tchi^0_2$ 
versus $\tan\beta$ with $\mu > 0$ as well as $\mu < 0$ 
for $m_{1/2} = 200$ GeV and several values of $m_0$.\footnote{
In frames (b) and (d) of Figure~\ref{fig:bfwz}, 
the Higgs pseudoscalar mass ($m_A$) 
and the lighter Higgs scalar mass ($m_h$) 
are very sensitive to the value of $\tan\beta$. 
For $\tan\beta = 48$, we obtain $m_h \simeq m_A \simeq 103$ GeV. 
For $\tan\beta = 50$, we find that $m_h \simeq m_A \simeq 30$ GeV, 
which have already been excluded by LEP experiments.}
For $\tan\beta \alt 5$, 
the branching fractions are sensitive to the sign of $\mu$.

%------------------------
% Figure 2
%------------------------

%------------------------------------
% Branching Fractions
%------------------------------------

For $\mu > 0$ and $\tan\beta \sim 3$, we have found that: 
\begin{itemize}
\item For $m_0 \alt 100$ GeV, 
$\tchi^0_2$ decays dominantly to 
$\tilde{\ell}_R\ell$ and $\tilde{\tau}_1\tau$, 
and $\tchi^\pm_1$ decays into $\tilde{\tau}_1\nu$, 
\item For 120 GeV $\alt m_0 \alt$ 170 GeV,   
the $\tchi^\pm_1 \tchi^0_2 \to 3 \ell +\notE_T$ branching fraction 
is still significant due to light virtual sleptons.
\item For $m_0 \agt$ 180 GeV, 
$\tchi^\pm_1$ and $\tchi^0_2$ dominantly decay into  $q\bar{q}'\tchi^0_1$.
\end{itemize}

For $\mu < 0$ and $\tan\beta \sim 3$, we have found that: 
\begin{itemize}
\item For $m_0 \alt$ 140 GeV, 
$\tchi^0_2$ dominantly decays to $\tilde{\nu}_L\nu$, 
$\tilde{\ell}_R\ell$, and $\tilde{\tau}_1\tau$, 
and $\tchi^\pm_1$ decays into $\tilde{\nu}_L\ell$ and $\tilde{\tau}_1\nu$. 
\item For 140 GeV $\alt m_0 \alt$ 160 GeV,   
the $\tchi^\pm_1 \tchi^0_2 \to 3 \ell +\notE_T$ branching fraction 
is still significant due to light virtual sleptons.
\item For $m_0 \agt$ 170 GeV, 
$\tchi^\pm_1$ and $\tchi^0_2$ dominantly decay into  $q\bar{q}'\tchi^0_1$.
\end{itemize}

For $\mu > 0$ and $m_0 \sim $ 200 GeV, $\tchi^0_2$ dominantly decays 
(i) into $\tau\bar{\tau}\tchi^0_1$ for $25 \alt \tan\beta \alt 40$, 
(ii) into $\tau\tilde{\tau}_1$ for $\tan\beta \agt 40$.  
For $m_0 \alt 300$ GeV and $\tan\beta \agt 35$,
both $\tilde{\tau}_1$ and $\tilde{b}_1$ can be lighter than other sfermions, 
and $\tchi^\pm_1$ and $\tchi^0_2$ can decay dominantly into final states 
with $\tau$ or $b$ via virtual or real $\tilde{\tau}_1$ and $\tilde{b}_1$. 

%-----------------------------------------------------------------------
%   III. Discovery Potential at the Tevatron
%-----------------------------------------------------------------------
\subsection{Discovery Potential at the Tevatron}

The ISAJET 7.44 event generator program \cite{ISAJET-3l} 
with the parton distribution functions of CTEQ3L \cite{CTEQ-3l} 
is employed to calculate the $3\ell +\notE_T$ signal 
from all possible sources of SUSY particles. 
An energy resolution of $\frac{0.7}{\sqrt{E}}$ for the hadronic calorimeter 
and $\frac{0.15}{\sqrt{E}}$ for the electromagnetic calorimeter is assumed. 
Jets are defined to be hadron clusters with $E_T > 15$ GeV in a cone 
with $\Delta R \equiv \sqrt{\Delta\eta^2+\Delta\phi^2} = 0.7$. 
Leptons with $p_T > 5$ GeV and within $|\eta_{\ell}| < 2.5$  
are considered to be isolated if the hadronic scalar $E_T$ 
in a cone with $\Delta R = 0.4$ about the lepton is smaller than 2 GeV.

After suitable cuts, there are two major sources of the SM background 
\cite{Trilepton2,Trilepton3,Madison1,Madison2,Matchev1,Matchev2,Baer&Tata,%
Chanowitz,Ellis}: 
(i) $q\bar{q} \to W^* Z^*, W^* \gamma^* \to \ell\nu \ell\bar{\ell}$ 
or $\ell'\nu' \ell\bar{\ell}$ ($\ell = e$ or $\mu$) 
with one or both gauge bosons being virtual\footnote{
If it is not specified, 
$W^*$ and $Z^*$ represent real or virtual gauge bosons, 
while $\gamma^*$ is a virtual photon.}, and 
(ii) $q\bar{q} \to W^* Z^*, W^* \gamma^* \to \ell\nu\tau\bar{\tau}$ 
or $\tau\nu \ell\bar{\ell}$ and subsequent $\tau$ leptonic decays. 
We have employed the programs MADGRAPH \cite{Madgraph} 
and HELAS \cite{Helas} to calculate
the cross section of $p\bar{p} \to 3l +\notE_T +X$ 
via four subprocesses (i) $q\bar{q}'\to e^+\nu_e\mu^+\mu^-$, 
(ii) $q\bar{q}'\to e^+\nu_e e^+ e^-$, 
(iii) $q\bar{q}'\to e^+\nu_e \tau^+ \tau^-$, and 
(iv) $q\bar{q}'\to \tau^+\nu_\tau e^+ e^-$, 
including contributions from intermediate states with $W^*Z^*$, 
$W^*\gamma^*$, and other diagrams. We have also evaluated contributions 
from $q\bar{q} \to \tau^+\tau^- e^+ e^-$ via $Z^*Z^*$, $Z^*\gamma^*$, 
and $\gamma^*\gamma^*$, with one leptonic and one hadronic tau decays.
We use ISAJET to calculate the background from $t\bar{t}$.

We found that most $\ell_3$'s from the $\tchi^\pm_1 \tchi^0_2$ decays 
have relatively smaller $p_T$ than the $\ell_3$'s from the backgrounds. 
Therefore, it is very important to have a soft acceptance cut 
on $p_T(\ell_3)$ to retain the trilepton events 
from  $\tchi^\pm_1 \tchi^0_2$ decays. 
Our acceptance cuts are chosen to be consistent with the experimental cuts 
proposed for Run II \cite{Teruki,Jane-3l} at the Tevatron as follows: 
\begin{eqnarray}
p_T(\ell_1) & > & 11 \; {\rm GeV}, 
\;\; p_T(\ell_2) > 7 \; {\rm GeV}, 
\;\; p_T(\ell_3) > 5 \; {\rm GeV}, \nonumber \\
|\eta(\ell_1,\ell_2,\ell_3)| & < & 2.0, \nonumber \\
{\rm at \;\; least \;\; one}\;\; \ell \;\; {\rm with} \;\; 
p_T(\ell) & > & 11 \;{\rm GeV} \;\; {\rm and} 
\;\; |\eta(\ell)| < 1.0, \nonumber \\
\notE_T & > & 25 \;\; {\rm GeV}, \nonumber \\
|M_{\ell\ell}-M_Z| & \geq & 10 \;\; {\rm GeV} \nonumber \\
M_{\ell\ell}       & \geq & 12 \;\; {\rm GeV} 
\label{eq:Basic}
\end{eqnarray}
To further reduce the background from $W^* Z^*$ and $W^* \gamma^*$, 
we require that \cite{Madison2,Baer&Tata} 
\begin{eqnarray}
|M_{\ell\ell}-M_Z| & \geq & 15 \;\; {\rm GeV} \;\; (Z-{\rm veto}), 
                        \nonumber \\
M_{\ell\ell}       & \geq & 18 \;\; {\rm GeV} \;\; (\gamma-{\rm veto}), 
                        \nonumber \\
M_T(\ell',\notE_T)      & \leq & 65 \;\; {\rm GeV} \;\; {\rm or} \;\; 
M_T(\ell',\notE_T) \geq 85 \;\; {\rm GeV} \;\; (W-{\rm veto}),
\label{eq:cuts2}
\end{eqnarray}
where $M_{\ell\ell}$ is the invariant mass for 
any pair of leptons with the same flavor and opposite signs, 
and $M_T(\ell',\notE_T)$ is the transverse mass 
of the remaining lepton.

%------------ ------------ 
% W+Jets and Z+Jets
%------------ ------------ 
We have also checked backgrounds 
from the production of $Z+jets$ and $W+jets$.
Some $3\ell$ events could be generated from $W+jets$, 
leading to sizable backgrounds;
these sources always had $b\to c\ell\nu$ followed by $c\to s\ell\nu$,
so that these sources of background could be removed by imposing
an angular separation cut between the isolated leptons, 
giving a background consistent with zero.
This angular separation cut causes no signal loss.

The effect of cuts on the signal and background is demonstrated in Table~\ref{tab1-3l}.
The trileptons come from $\tchi^\pm_1\tchi^0_2$ production as well as 
additional SUSY particle sources that are discussed in the next section.
The cross sections of the signal with $m_{1/2} = 200$ GeV, $m_0 = 100$ GeV, 
and several values of $\tan\beta$, 
along with backgrounds from  
(i) $\ell'\nu'\ell\bar{\ell}$, 
(ii) $\ell \nu \ell\bar{\ell}$, 
(iii) $\ell \nu \tau\bar{\tau}$, 
(iv) $\tau \nu_\tau \ell\bar{\ell}$, 
(v) $\ell\bar{\ell}\tau\bar{\tau}$, 
and (vi) $t\bar{t}$.
We present cross sections with six sets of cuts: 
(a) Basic Cuts \cite{Madison1,Madison2}: cuts in Eq. (\ref{eq:Basic});
(b) Soft Cuts A1 \cite{Madison2}: 
cuts in Eqs. (\ref{eq:Basic}) and (\ref{eq:cuts2});
(c) Soft Cuts A2 \cite{Madison2}: the same cuts as soft cuts A1, 
except requiring 18 GeV $\leq M_{\ell\ell} \leq$ 75 GeV; 
(d) Soft Cuts B \cite{Baer&Tata}: cuts in Eq. (\ref{eq:Basic}), 
$20 \;\; {\rm GeV} \leq M_{\ell\ell} \leq 80 \;\; {\rm GeV}$, and 
$M_T(\ell',\notE_T) \leq 60 \;\; {\rm GeV}$ or  
$M_T(\ell',\notE_T) \geq 85 \;\; {\rm GeV}$; 
(e) Hard Cuts A \cite{Madison2}: the same cuts as soft cuts A1, 
except requiring $M_{\ell\ell} \geq 12 \;\; {\rm GeV}$, 
and $p_T(\ell_1,\ell_2,\ell_3) >$ 20, 15, and 10 GeV;  
(f) Hard Cuts B \cite{Baer&Tata}: the same cuts as in soft cuts B, 
except requiring 12 GeV $\leq M_{\ell\ell} \leq$ 80 GeV, 
and $p_T(\ell_1,\ell_2,\ell_3) >$ 20, 15, and 10 GeV. 
The reach with any set of soft cuts is qualitatively quite
similar. 
Hardening the cuts generally results in a reduced reach over
much of parameter space, but may possibly lead to an incremental
increase in the reach for the large values of $m_{1/2}$ and $m_0$. 
A more strict cut to require $M_{\ell\ell} < 75$ GeV as in soft cuts A2 
can further reduce the backgrounds from $\ell'\nu'\ell\bar{\ell}$
as well as $\ell\nu\ell\bar{\ell}$ 
with a slight reduction in the trilepton signal for most SUGRA
parameters and might slightly improve the statistical significance.
For brevity, we will present reach results with soft cuts A1 
in this report.

%------------------------------------
%   Table I
%------------------------------------

\begin{table}[htb]
\begin{center}
\caption[]{
The cross section of 
$p\bar{p} \to {\rm SUSY \;\; particles} \to 3\ell +X$ in fb
versus $\tan\beta$ for $m_{1/2} = 200$ GeV and $m_0 = 100$ GeV 
along with the trilepton cross sections of the SM backgrounds (BG)
and values of statistical significance 
($N_S = S/\sqrt{B}$, $S =$ number of signal events, 
and $B =$ number of background events)
for an integrated luminosity of ${\cal L} = 30$~fb$^{-1}$, 
at the upgraded Tevatron with six sets of cuts: 
(a) Basic Cuts: cuts in Eq. (\ref{eq:Basic});
(b) Soft Cuts A1: 
cuts in Eqs. (\ref{eq:Basic}) and (\ref{eq:cuts2});
(c) Soft Cuts A2: the same cuts as soft cuts A1, 
except requiring 18 GeV $\leq M_{\ell\ell} \leq$ 75 GeV; 
(d) Soft Cuts B: cuts in Eqs. (\ref{eq:Basic}), 
$20 \;\; {\rm GeV} \leq M_{\ell\ell} \leq 80 \;\; {\rm GeV}$, and 
$M_T(\ell',\notE_T) \leq 60 \;\; {\rm GeV}$ or
$M_T(\ell',\notE_T) \geq 85 \;\; {\rm GeV}$;
(e) Hard Cuts A: the same cuts as soft cuts A1, 
except requiring $M_{\ell\ell} \geq 12 \;\; {\rm GeV}$, 
and $p_T(\ell_1,\ell_2,\ell_3) >$ 20, 15, and 10 GeV.
(f) Hard Cuts B: the same cuts as soft cuts B, 
except requiring 12 GeV $\leq M_{\ell\ell} \leq$ 80 GeV; 
and $p_T(\ell_1,\ell_2,\ell_3) >$ 20, 15, and 10 GeV.
\label{tab1-3l}}

\medskip

\begin{tabular}{lccccccc}
$\tan\beta$ $\backslash$ Cuts 
& Basic & Soft A1 & Soft A2 & Soft B & Hard A & Hard B \\
\hline
3  &  12.8 & 8.82 & 8.41 & 7.37 & 4.04 & 3.44 \\
10 &  3.49 & 2.57 & 2.43 & 2.20 & 1.13 & 0.95 \\
20 &  1.18 & 0.90 & 0.79 & 0.74 & 0.34 & 0.26 \\
25 &  0.66 & 0.50 & 0.43 & 0.40 & 0.20 & 0.16 \\
\hline 
SM BG \\
\hline 
$\ell'\nu'\ell\bar{\ell}$ & 2.63 & 0.72 & 0.60 & 0.44 & 0.32  & 0.18 \\
$\ell \nu \ell\bar{\ell}$ & 2.09 & 0.41 & 0.30 & 0.19 & 0.20  & 0.09 \\
$\ell \nu \tau\bar{\tau}$ & 0.60 & 0.45 & 0.41 & 0.36 & 0.22  & 0.18 \\
$\tau \nu \ell\bar{\ell}$ & 0.37 & 0.20 & 0.13 & 0.12 & 0.11  & 0.06 \\
$\ell \ell\tau\bar{\tau}$ & 0.12 & 0.08 & 0.06 & 0.06 & 0.04  & 0.04 \\
$t\bar{t}$                & 0.14 & 0.11 & 0.06 & 0.06 & 0.009 & 0.005 \\
Total BG                  & 5.95 & 1.97 & 1.56 & 1.23 & 0.90  & 0.56 \\
\hline
$\tan\beta$ $\backslash$ $N_S$ \\
\hline
3  &  28.7 & 34.4 & 36.9 & 36.4 & 23.3 & 25.2 \\
10 &   7.8 & 10.0 & 10.7 & 10.8 &  6.5 &  7.0 \\
20 &   2.6 &  3.5 &  3.5 &  3.7 &  2.0 &  1.9 \\
25 &   1.5 &  1.9 &  1.9 &  2.0 &  1.2 &  1.1
\end{tabular}
\end{center}
\end{table}

At Run II with 2 fb$^{-1}$ integrated luminosity, 
we expect about 4 events per experiment 
from the background cross section of 1.97 fb.
The signal cross section ($\sigma_S$) must yield at least 6 signal events 
for observation at 99\% confidence level (C.L.).
The Poisson probability for the background to fluctuate 
to this level is less than $0.8\%$.
At Run III with ${\cal L} = 30$ fb$^{-1}$, 
we expect about 60 background events; 
a $5 \sigma$ signal would be 38 events corresponding to 
$\sigma_S = 1.28$ fb, 
and a $3 \sigma$ signal would be 23 events corresponding to 
$\sigma_S = 0.77$ fb. 

\begin{figure}[htb]
\centering\leavevmode
\epsfxsize=4in\epsffile{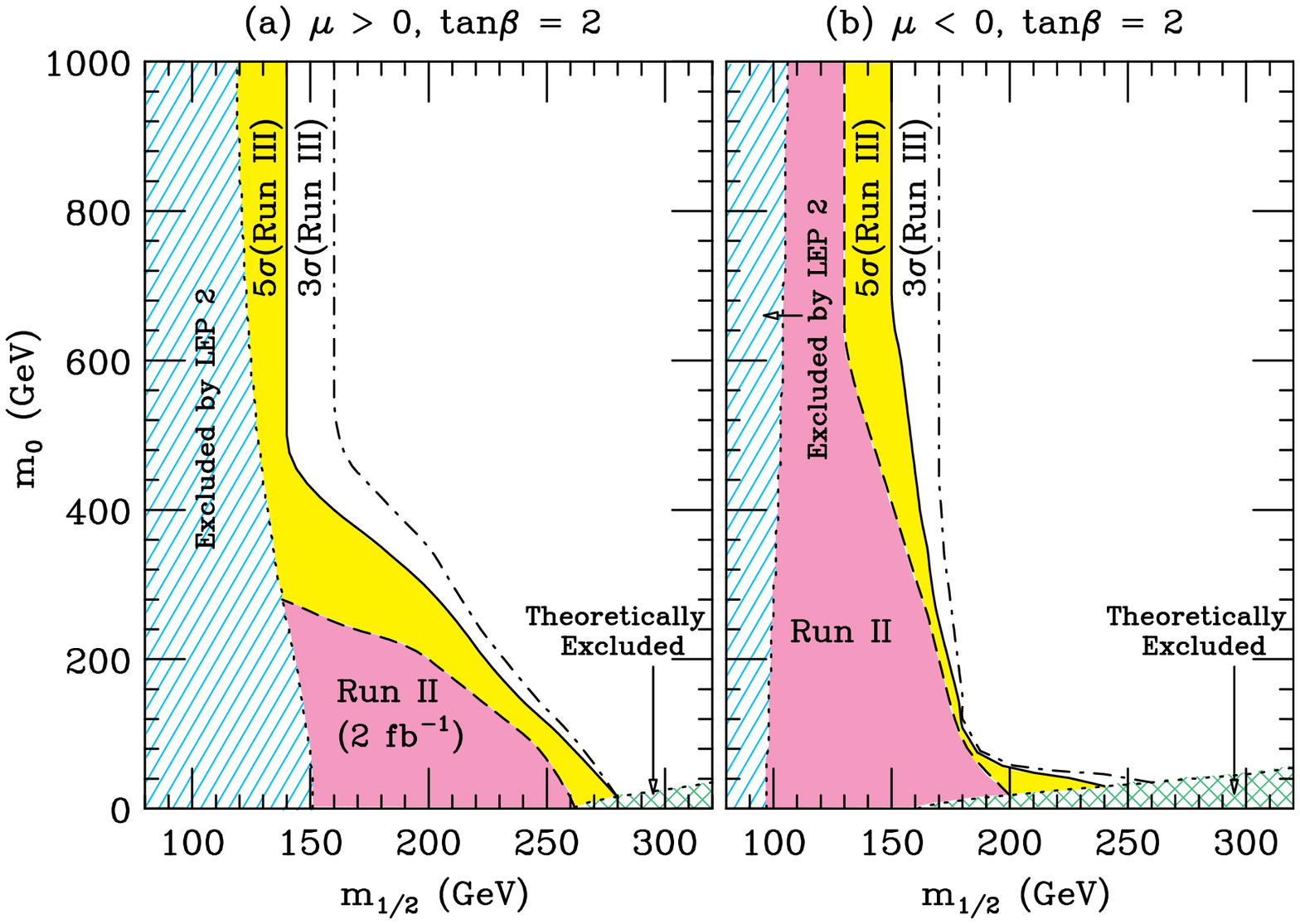}

\caption[]{
The contours of 99\% C.L. observation at Run II and 
5$\sigma$ discovery as well as 3$\sigma$ observation at Run III 
for $p\bar{p} \to {\rm SUSY \;\; particles} \to 3\ell +X$ 
with soft cuts A1, in the $(m_{1/2},m_0)$ plane, 
for $\tan\beta =$ 2, (a) $\mu > 0$ and (b) $\mu < 0$. 
\label{fig:contour2}
}\end{figure}

To assess the discovery potential of the upgraded Tevatron, 
we present the contours of 99\% C.L. observation at Run II 
and $5\sigma$ discovery as well as 3$\sigma$ observation at Run III 
in Figure~\ref{fig:contour2}, 
for $p\bar{p} \to {\rm SUSY \;\; particles} \to 3\ell +X$ 
at $\sqrt{s} = 2$ TeV with soft cuts A1 
[Eqs.~(\ref{eq:Basic},\ref{eq:cuts2})] in the $(m_{1/2},m_0)$ plane, 
for $\tan\beta =$ 2, $\mu > 0$ and $\mu < 0$. 
We include all SUSY sources for the trilepton signal. 
Also shown are the parts of the parameter space excluded by 
the theoretical requirements or 
the chargino search at LEP 2 ($m_{\tchi^\pm_1} \alt 95$ GeV) \cite{LEP2-3l}. 

%------------------------
% Figure 3
%------------------------

\begin{figure}[htb]
\vglue1cm
\centering\leavevmode
\epsfxsize=4in\epsffile{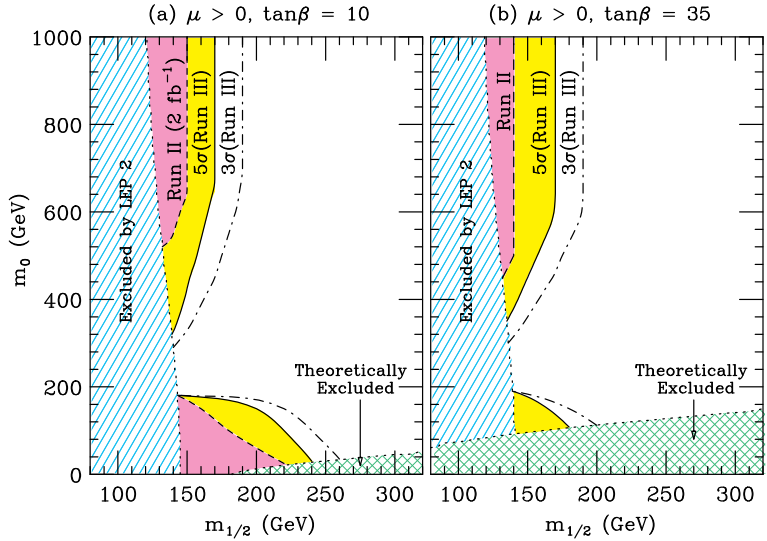}

\medskip
\caption[]{
The same as Figure~\ref{fig:contour2}, for $\mu > 0$, 
(a) $\tan\beta =$ 10 and (b) $\tan\beta = 35$.
\label{fig:contour10}
}\end{figure}

Figure~\ref{fig:contour10} shows the 99\% C.L. observation contour of Run II 
and the $5\sigma$ discovery contour 
as well as the 3$\sigma$ observation contour of Run III 
for $p\bar{p} \to {\rm SUSY \; particles} \to 3\ell +X$ 
at $\sqrt{s} = 2$ TeV with soft cuts A1, 
in the parameter space of $(m_{1/2},m_0)$, 
for $\mu > 0$, $\tan\beta = 10$ and $\tan\beta = 35$. 

%------------------------
% NEW PAGE
%------------------------
%\newpage

%------------------------
% Figure 4
%------------------------

\begin{figure}[htb]
\centering\leavevmode
\epsfxsize=4in\epsffile{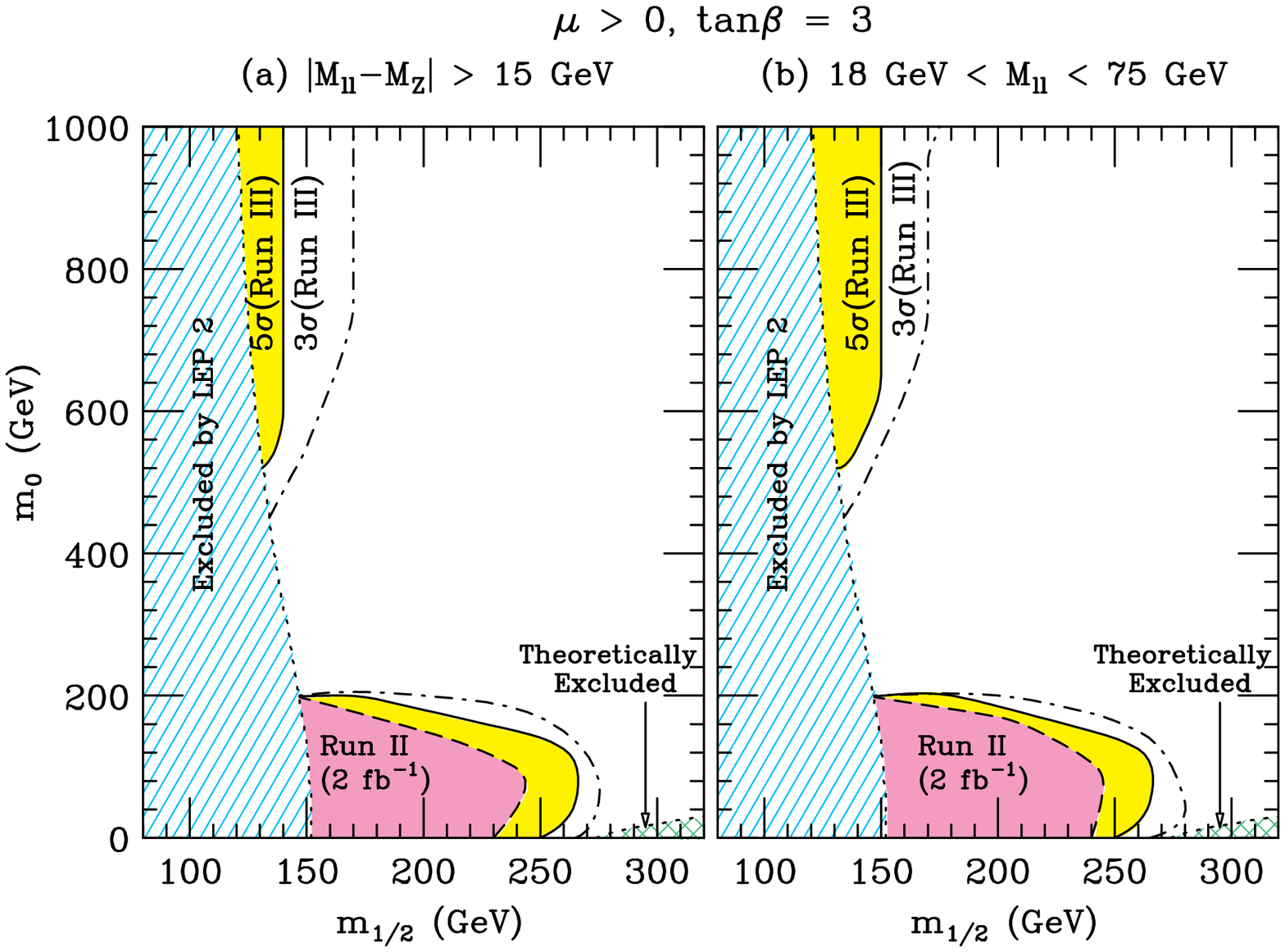}

\medskip
\caption[]{
The same as Figure~\ref{fig:contour2}, for $\mu > 0$ and $\tan\beta = 3$, 
with (a) soft cuts A1 ($|M_{\ell\ell}-M_Z| > 15$ GeV), 
and (b) soft cuts A2 ($M_{\ell\ell} < 75$ GeV).
\label{fig:contour3}
}\end{figure}

In Figure~\ref{fig:contour3}, we present the contours of 99\% C.L. observation at Run II 
and $5\sigma$ discovery as well as 3$\sigma$ observation at Run III 
for $p\bar{p} \to {\rm SUSY \; Particles} \to 3\ell +X$ 
in the $(m_{1/2},m_0)$ plane for $\tan\beta = 3$\footnote{
The $m_h$ is sensitive to the value of $\tan\beta$.
Taking $m_{1/2} = 200$ GeV, $m_0 = 100$ GeV, $A_0 = 0$ and $\mu > 0$,
we obtain $m_h = 89.5$ GeV for $\tan\beta = 2$ and
$m_h = 99.3$ GeV for $\tan\beta = 3$.} 
with soft cuts A1 and soft cuts A2. 

%------------------------
% Figure 5
%------------------------

%-----------------------------------------------------------------------
% VI. Conclusions
%-----------------------------------------------------------------------
\subsection{Conclusions}

In most of the mSUGRA parameter space,
$\tchi^\pm_1\tchi^0_2$ production is the dominant source of trileptons.
For $m_0 \alt 150$ GeV and $\tan\beta \agt 20$,
production of $\tilde{\ell}\tilde{\nu}$ and $\tilde{\ell}\tilde{\ell}$
can enhance the trilepton signal and may yield observable rates
at Run III in regions of parameter space that are otherwise inaccessible.
We summarize the contributions to trileptons 
from various channels for $\mu > 0$ in Table~\ref{tab2-3l}.

%------------------------------------
%   Table II
%------------------------------------

\begin{table}[htb]
\begin{center}
\caption[]{
The cross section of $p\bar{p} \to {\rm SUSY \;\; particles} \to\break 3\ell +X$ 
in fb versus $\tan\beta$ 
with contributions from various relevant SUSY channels 
at $\sqrt{s} = 2$ TeV with the acceptance cuts described in 
[Eqs. (\ref{eq:Basic}) and (\ref{eq:cuts2})] 
for $\mu > 0$, $m_{1/2} = 200$ GeV, $\tan\beta= 3, 10, 20$ 
and 35 (25 for $m_0 = 100$ GeV).  \label{tab2-3l}
}
\medskip
\begin{tabular}{crrrc} 
Channel $\backslash \tan\beta$ & 3 & 10 & 20 & 35(25) \\
\hline
(i) $m_0 = 100$ GeV & & & \\
Total               
  & 8.82 & 2.57 & 0.90 & 0.50 \\
$\tchi^\pm_1\tchi^0_2$
  & 7.16 & 1.74 & 0.40 & 0.13 \\
$\tilde{\ell}\tilde{\nu}$
  & 0.66 & 0.32 & 0.18 & 0.10  \\
$\tilde{\ell}\tilde{\ell}$
  & 0.33 & 0.16 & 0.15 & 0.13 \\
$\tchi^0_2\tchi^0_2$, $\tchi^0_2\tchi^0_3$, $\tchi^0_3\tchi^0_4$, 
  & 0.30 & 0.12 & 0.05 & 0.05 \\
$\tchi^\pm_1\tchi^0_{3,4}$, $\tchi^\pm_2\tchi^0_{3,4}$ 
$\tchi^\pm_1\tchi^\mp_2$, $\tchi^\pm_2\tchi^\mp_2$
  & 0.07 & 0.08 & 0.06 & 0.05 \\
$\tilde{g}\tchi^0_{2,3}$,
$\tilde{q}\tchi^0_{2,3}$,$\tilde{g}\tilde{g}$,
$\tilde{q}\tilde{q}$,$\tilde{\nu}\tilde{\nu}$
  & 0.30 & 0.15 & 0.06 & 0.04 \\
\hline
(ii) $m_0 = 200$ GeV & & & \\
Total               
  & 1.07 & 0.23 & 0.25 & 0.31 \\
$\tchi^\pm_1\tchi^0_2$
  & 0.98 & 0.16 & 0.17 & 0.19 \\
$\tilde{\ell}\tilde{\nu}$
  & 0.03 & 0.02 & 0.02 & 0.03  \\
$\tilde{\ell}\tilde{\ell}$
  & 0.01 & 0.01 & 0.01 & 0.01 \\
$\tchi^0_2\tchi^0_2$, $\tchi^0_2\tchi^0_3$, $\tchi^0_3\tchi^0_4$, 
  & 0.01 & 0.01 & 0.01 & 0.02 \\
$\tchi^\pm_1\tchi^0_{3,4}$, $\tchi^\pm_2\tchi^0_{3,4}$ 
$\tchi^\pm_1\tchi^\mp_2$, $\tchi^\pm_2\tchi^\mp_2$
  & 0.01 & 0.01 & 0.02 & 0.02 \\
$\tilde{g}\tchi^0_{2,3}$,
$\tilde{q}\tchi^0_{2,3}$,$\tilde{g}\tilde{g}$,
$\tilde{q}\tilde{q}$,$\tilde{\nu}\tilde{\nu}$
 &  0.03 & 0.02 & 0.02 & 0.04 \\
\hline
(iii) $m_0 = 500$ GeV & & &  \\
Total               
  & 0.28 & 0.48 & 0.46 & 0.42 \\
$\tchi^\pm_1\tchi^0_2$
  & 0.28 & 0.45 & 0.44 & 0.41 \\
$\tchi^0_2\tchi^0_2$, $\tchi^0_2\tchi^0_3$, $\tchi^0_3\tchi^0_4$, 
  & $-$  & 0.01  & $-$ & $-$  \\
$\tchi^\pm_1\tchi^0_{3,4}$, $\tchi^\pm_2\tchi^0_{3,4}$ 
$\tchi^\pm_1\tchi^\mp_2$, $\tchi^\pm_2\tchi^\mp_2$
  & $-$  & 0.01 & 0.01 & $-$  \\
$\tilde{g}\tchi^0_{2,3}$,
$\tilde{q}\tchi^0_{2,3}$,$\tilde{g}\tilde{g}$,
$\tilde{q}\tilde{q}$,$\tilde{\nu}\tilde{\nu}$
  & $-$  & 0.01 & 0.01 & 0.01 
\end{tabular}
\end{center}
\end{table}

In regions of the parameter space with $m_0 \alt 200$ GeV 
or $\tan\beta \agt 40$, 
the $\tchi^\pm_1$ and the $\tchi^0_2$ decay 
dominantly to final states with $\tau$ leptons. 
The subsequent leptonic decays of these $\tau$ leptons 
contribute importantly to the trilepton signal
from $\tchi^\pm_1 \tchi^0_2$ associated production.
With soft but realistic lepton $p_T$ acceptance cuts, 
these $\tau \to \ell$ contributions can significantly 
enhance the trilepton signal. 
The Tevatron trilepton searches are most sensitive to the region 
of mSUGRA parameter space with $m_0 \alt 100$ GeV and $\tan\beta \alt$ 10. 

The discovery potential of the upgraded Tevatron for $\mu > 0$ 
are summarized in the following:
\begin{itemize} \addtolength{\itemsep}{-2mm}
\item For $m_0 \sim 100$ GeV and $\tan\beta\sim 2$,
the trilepton signal should be detectable
at the Run II if $m_{1/2} \alt$ 240 GeV ($m_{\tchi^\pm_1} \alt 177$ GeV), and
at the Run III if $m_{1/2} \alt$ 260 GeV ($m_{\tchi^\pm_1} \alt 195$ GeV).
\item For $m_0 \sim 150$ GeV and $\tan\beta\sim 35$,
the trilepton signal should be detectable
at the Run III if $m_{1/2} \alt$ 170 GeV ($m_{\tchi^\pm_1} \alt 122$ GeV).
\item For $m_0 \agt 600$ GeV and $\tan\beta\sim 35$,
the trilepton signal should be detectable
at the Run III if $m_{1/2} \alt$ 170 GeV ($m_{\tchi^\pm_1} \alt 130$ GeV).
\end{itemize}

For 180 GeV $\alt m_0 \alt$ 350 GeV and $10 \alt \tan\beta \alt 35$, 
the $\tchi^0_2$ decays dominantly into $q\bar{q}\tchi^0_1$. 
In these regions it will be difficult to establish a trilepton signal. 
For $m_0 \agt$ 500 GeV and $m_{1/2}$ close to the reach of soft cuts, 
most trileptons from SUSY sources 
have relatively higher $p_T$ than those generated with lower $m_0$, 
and the statistical significance of the trilepton signal 
might be slightly improved by optimized harder cuts 
\cite{Matchev2,Baer&Tata}.
While there are regions of parameter space 
where it will be difficult to establish a trilepton signal because 
the leptonic decays of $\tchi^0_2$ is suppressed, 
the important point is that the experiments at the Tevatron 
may probe a substantial region not accessible at LEP 2.

%%% Supplementary paragraph from Matchev & Pierce
%% Their references are appended to the previous list of references

\def\gsim{\lower.7ex\hbox{$\;\stackrel{\textstyle>}{\sim}\;$}}
\def\lsim{\lower.7ex\hbox{$\;\stackrel{\textstyle<}{\sim}\;$}}
\def\t{\tilde }
\def\fiv{${\bf 5}+{\bf\overline5}$ }
\def\ten{${\bf 10}+{\bf\overline{10}}$ }
\def\met{\rlap{\,/}E_T}
\def\beq{\begin{equation}}
\def\eeq{\end{equation}}
\def\bear{\begin{eqnarray}}
\def\eear{\end{eqnarray}}

%%%%%%%%%%%%%%%%% JOURNALS %%%%%%%%%%%%%%%%%%%%%%%%
\def\ap  #1 #2 #3 #4 {Ann.~Phys.         {\bf  #1}, #2 (#3)#4 }
\def\aplb#1 #2 #3 #4 {Acta Phys.~Pol.    {\bf B#1}, #2 (#3)#4 }
\def\cpc #1 #2 #3 #4 {Comp.~Phys.~Comm.  {\bf  #1}, #2 (#3)#4 }
\def\jetp#1 #2 #3 #4 {JETP Lett.         {\bf  #1}, #2 (#3)#4 }
\def\npb #1 #2 #3 #4 {Nucl.~Phys.        {\bf B#1}, #2 (#3)#4 }
\def\mpla#1 #2 #3 #4 {Mod.~Phys.~Lett.   {\bf A#1}, #2 (#3)#4 }
\def\plb #1 #2 #3 #4 {Phys.~Lett.        {\bf B#1}, #2 (#3)#4 }
\def\pr  #1 #2 #3 #4 {Phys.~Rep.         {\bf  #1}, #2 (#3)#4 }
\def\prd #1 #2 #3 #4 {Phys.~Rev.         {\bf D#1}, #2 (#3)#4 }
\def\prl #1 #2 #3 #4 {Phys.~Rev.~Lett.   {\bf  #1}, #2 (#3)#4 }
\def\ptp #1 #2 #3 #4 {Prog.~Theor.~Phys. {\bf  #1}, #2 (#3)#4 }
\def\zpc #1 #2 #3 #4 {Zeit.~Phys.        {\bf C#1}, #2 (#3)#4 }
%%%%%%%%%%%%%%%%%%%%%%%%%%%%%%%%%%%%%%%%%%%%%%%%%%%

\subsection{Trilepton Analysis with Variable Cuts}

In this section we summarize the results from a trilepton analysis
with variable cuts \cite{MP3L-3l,talks-3l,MPPLB-3l}.  We perform a
detailed Monte Carlo simulation of signal and background
with a realistic detector simulation based on SHW v.2.2
\cite{SHW-a3l,TAUOLA-a3l,STDHEP-a3l}.
We used ISAJET \cite{ISAJET-3l} for simulation of the signal,
and PYTHIA \cite{PYTHIA-a3l} for all background determinations
except $WZ$. For $WZ$ we first use COMPHEP \cite{COMPHEP-a3l} 
to generate hard scattering events
at leading order, then we pipe those through
PYTHIA, adding showering and hadronization,
and finally, we run the resulting events through SHW.
We made several modifications in the SHW/TAUOLA package,
which are described in Refs. \cite{MP3L-3l,MPPLB-3l}.
In our analysis we use several alternative values for
the cut on a particular variable. 
1) Four $\met$ cuts: $\met>\{15,20,25\}$ GeV or no cut.
2) Six high-end invariant mass cuts for any pair of opposite sign,
same flavor leptons. The event is discarded if:
$|M_Z - m_{\ell^+\ell^-}|<\{10,15\}$ GeV;
or $m_{\ell^+\ell^-}>\{50,60,70,80\}$ GeV.
3) Eleven low-end invariant mass cuts for any pair of opposite sign,
same flavor leptons: $m^\gamma_{\ell^+\ell^-}<\{10,60\}$ GeV,
in 5 GeV increments.
4) Four azimuthal angle cuts on opposite sign, same flavor leptons:
two cuts on the difference of the azimuthal angle of the two highest
$p_T$ leptons, $|\Delta\varphi|<\{2.5,2.97\}$, one cut
$|\Delta\varphi|<2.5$ for {\em any pair} leptons, and no cut.
5) An optional jet veto (JV) on QCD jets in the event.
6) An optional cut on the the transverse mass $m_T$ of any
$\ell\nu$ pair which may originate from a $W$-boson:
$60<m_T(\ell,\nu)<85$ GeV. 
7) Five sets of lepton $p_T$ cuts:
$\{11,5,5\},~\{11,7,5\},~\{11,7,7\},~ \{11,11,11\}$ and $\{20,15,10\}$,
where the first four sets also require a
central lepton with $p_T>11$ GeV and $|\eta|<1.0$ or 1.5.
We then employ a parameter space
dependent cut optimization: at each point in SUSY parameter space, we
consider all possible combinations of cuts, and determine the best
combination by maximizing $S/\sqrt{B}$.

\begin{figure}[t]
\centerline{\psfig{figure=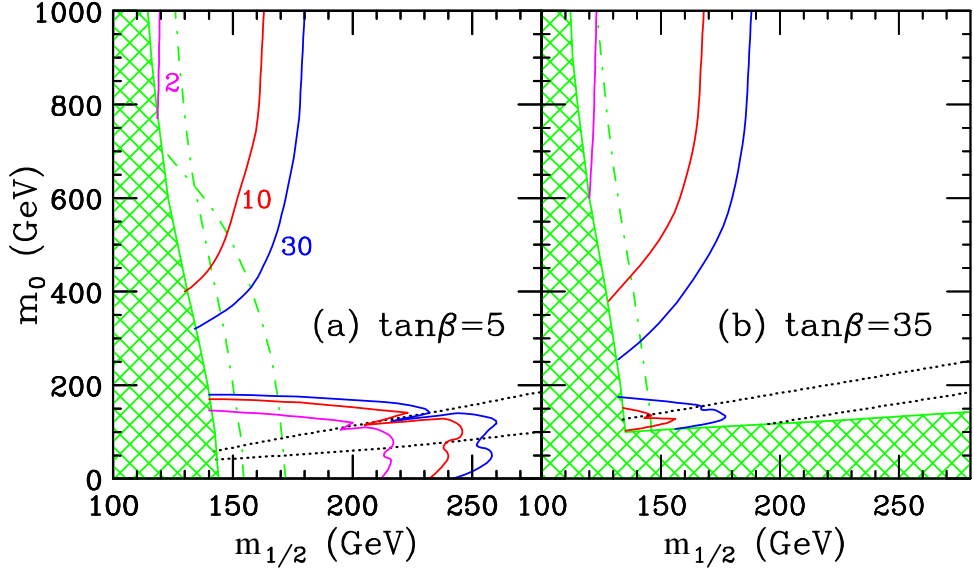,height=3.0in}}
\begin{center}
\parbox{5.5in}{
\caption[] {Tevatron reach in the trilepton channel in
the $m_0-m_{1/2}$ plane, for fixed values of $A_0=0$, $\mu>0$
and (a) $\tan\beta=5$, or (b) $\tan\beta=35$. Results are shown
for 2, 10 and 30 ${\rm fb}^{-1}$ total integrated luminosity.
\label{3l}}}
\end{center}
\end{figure}

We present our results for the Tevatron reach in the trilepton
channel in Figs.~\ref{3l}.
We require the observation of at least 5 signal events, and present
our results as $3\sigma$ exclusion contours in the $m_0-m_{1/2}$
plane, for two representative values of $\tan\beta$, 5 and 35. We fix
$\mu>0$ and $A_0=0$.  The cross-hatched region is excluded by current
limits on the superpartner masses. The dot-dashed lines correspond to
the projected LEP-II reach for the chargino and the lightest Higgs
masses. In Fig.~\ref{3l}a the left dotted line shows where
$m_{\tilde\nu_\tau}=m_{\tilde\chi_1^\pm}$ and the right dotted line
indicates $m_{\tilde\tau_1}=m_{\tilde\chi_1^\pm}$ (and
$m_{\tilde\tau}\simeq m_{\tilde\mu}\simeq m_{\tilde e}$). In Fig.~\ref{3l}b
the dotted lines show where $m_{\tilde e_R}=m_{\tilde\chi_1^\pm}$
(left) and $m_{\tilde\tau_1}=m_{\tilde\chi_1^\pm}$ (right).
We see that the trilepton channel provides for significant reach at
both small $m_0$ ($m_0\lsim 150$ GeV) and large $m_0$ ($m_0\gsim 400$
GeV). With only 2 fb$^{-1}$ the reach is quite
limited. 

\begin{figure}[!ht]
\centerline{\psfig{figure=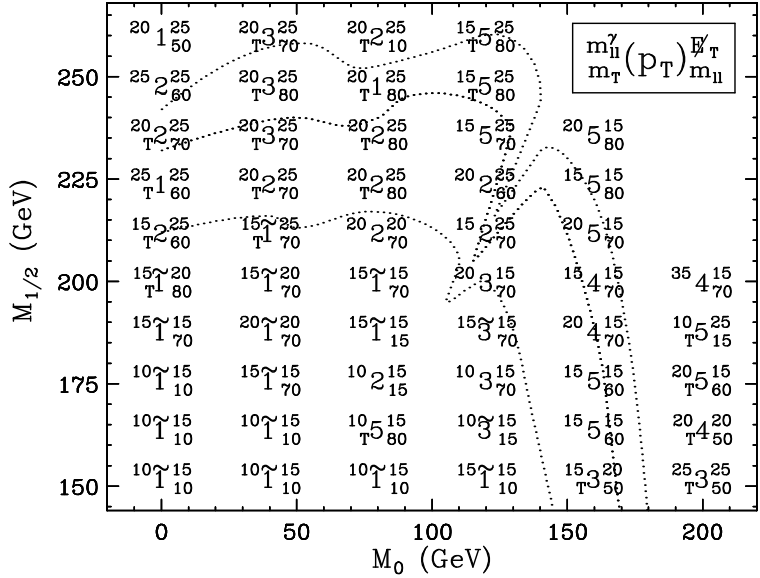,height=3.0in}}
\begin{center}
\parbox{5.5in}{
\caption[] {The optimal sets of trilepton cuts in the $m_0,~
m_{1/2}$ plane, for $\tan\beta=5$ and small $m_0$.  We show the
optimal low end dilepton mass cut $m^\gamma_{\ell^+\ell^-}$, missing
$E_T$ cut $\met$, high end dilepton mass cut $m_{\ell^+\ell^-}$,
transverse $\ell\nu$ mass cut and lepton $p_T$ cut (see text). The
dotted lines indicate the reach contours from Fig.~\ref{3l}. Note that the ranges of $m_0$ and $m_{1/2}$ are different from previous figures.
\label{bc5_m0_small}}}
\end{center}
\end{figure}

In Fig.~\ref{bc5_m0_small} 
we show the optimum cuts chosen in our optimization procedure,
in the $m_0$, $m_{1/2}$ plane, for $\tan\beta=5$,
in the small $m_0$ region.
We use the following notation to describe the set of cuts at each
point. The central symbol indicates the set of lepton $p_T$ cuts
with higher values corresponding to harder cuts.
The left superscript shows the
value (in GeV) of the low-end invariant mass cut
$m_{\ell^+\ell^-}^\gamma$. A left subscript ``T''
indicates that the cut on the transverse $\ell\nu$ mass was selected.
The right superscript shows the $\met$ cut: $\met>\{15,20,25\}$ GeV
(``15'',``20'',``25''), or no cut (no symbol).  A right subscript
denotes the high-end dilepton invariant mass cut:
$|m_{\ell^+\ell^-}-M_Z|>\{10,15\}$ GeV (``10'',``15'') or
$m_{\ell^+\ell^-}<\{50,60,70,80\}$ GeV
(``50'',``60'',``70'',``80''). And finally, a tilde over the central
symbol indicates that the luminosity limit came from requiring 5
signal events rather than $3\sigma$ exclusion.
We see from Fig.~\ref{bc5_m0_small} that in the regions where background
is an issue, the combination of the $m_T$ cut and a tighter low-end
dilepton mass cut $m^{\gamma}_{\ell^+\ell^-}\sim 20$ GeV is typically
preferred.  Notice that the transverse mass cut is never
enough by itself, i.e. whenever it is chosen, it is almost
always supplemented with a $m^{\gamma}_{\ell^+\ell^-}$
cut of 15 to 25 GeV (with the exception of
two points with high lepton $p_T$ cuts).  On the other hand, there are
significant regions where the low invariant mass cut
$m^{\gamma}_{\ell^+\ell^-}$ by itself is enough to kill the
background, and the transverse mass cut is not needed.

%-----------------------------------------------------------------------
%   REFERENCES
%-----------------------------------------------------------------------

%

%% file: Matchev-2l/dilep.tex
%% copied from Matchev/Pierce file rep2L.tex to format for report

\def\gsim{\lower.7ex\hbox{$\;\stackrel{\textstyle>}{\sim}\;$}}
\def\lsim{\lower.7ex\hbox{$\;\stackrel{\textstyle<}{\sim}\;$}}
\def\t{\tilde }
\def\fiv{${\bf 5}+{\bf\overline5}$ }
\def\ten{${\bf 10}+{\bf\overline{10}}$ }
\def\met{\rlap{\,/}E_T}
\def\beq{\begin{equation}}
\def\eeq{\end{equation}}
\def\bear{\begin{eqnarray}}
\def\eear{\end{eqnarray}}

%%%%%%%%%%%%%%%%% JOURNALS %%%%%%%%%%%%%%%%%%%%%%%%
\def\ap  #1 #2 #3 #4 {Ann.~Phys.         {\bf  #1}, #2 (#3)#4 }
\def\aplb#1 #2 #3 #4 {Acta Phys.~Pol.    {\bf B#1}, #2 (#3)#4 }
\def\cpc #1 #2 #3 #4 {Comp.~Phys.~Comm.  {\bf  #1}, #2 (#3)#4 }
\def\jetp#1 #2 #3 #4 {JETP Lett.         {\bf  #1}, #2 (#3)#4 }
\def\npb #1 #2 #3 #4 {Nucl.~Phys.        {\bf B#1}, #2 (#3)#4 }
\def\mpla#1 #2 #3 #4 {Mod.~Phys.~Lett.   {\bf A#1}, #2 (#3)#4 }
\def\plb #1 #2 #3 #4 {Phys.~Lett.        {\bf B#1}, #2 (#3)#4 }
\def\pr  #1 #2 #3 #4 {Phys.~Rep.         {\bf  #1}, #2 (#3)#4 }
\def\prd #1 #2 #3 #4 {Phys.~Rev.         {\bf D#1}, #2 (#3)#4 }
\def\prl #1 #2 #3 #4 {Phys.~Rev.~Lett.   {\bf  #1}, #2 (#3)#4 }
\def\ptp #1 #2 #3 #4 {Prog.~Theor.~Phys. {\bf  #1}, #2 (#3)#4 }
\def\zpc #1 #2 #3 #4 {Zeit.~Phys.        {\bf C#1}, #2 (#3)#4 }
%%%%%%%%%%%%%%%%%%%%%%%%%%%%%%%%%%%%%%%%%%%%%%%%%%%

\let\tt\relax

%===============END OF DEFINITIONS====================

\section{Like-Sign Dilepton Analysis}

%%%%%%%%%%%%%%%%%%%%%%%%%%%%%%%%%%%%%%%%%%%%%%%%%%%%%%%%%%%%%%%
%%  \subsection{Introduction}  NOT NEEDED - NO OTHER SUBSECTIONS

The inclusive like-sign dilepton (2L) channel has been
suggested \cite{JN-matchev} as an alternative to the trilepton (3L)
signature. By not requiring the odd-sign lepton in the
event, the signal acceptance goes up, and the backgrounds are
hopefully still under control.

In this section we summarize the results from a like-sign
dilepton analysis with variable cuts \cite{MP3L-2l,talks-2l,MPPLB}.
We perform a detailed Monte Carlo simulation of signal and
background with a realistic detector simulation based on
SHW v.2.2 \cite{SHW-matchev,TAUOLA,STDHEP}.
We used ISAJET \cite{ISAJET-matchev} for simulation of the signal,
and PYTHIA \cite{PYTHIA-matchev} for all background determinations
except $WZ$. For $WZ$ we first use COMPHEP \cite{COMPHEP} 
to generate hard scattering events
at leading order, then we pipe those through
PYTHIA, adding showering and hadronization,
and finally, we run the resulting events through SHW.
The resulting parton-level cross section
was integrated with the {\tt CTEQ4m} structure functions \cite{CTEQ-matchev}.
We have made several modifications in the SHW/TAUOLA package,
which are appropriate for our purposes:
1) We modified TAUOLA to account for the correct (on average)
polarization of tau leptons coming from decays of supersymmetric particles.
2) We extend the tracking coverage to $|\eta|<2.0$, which increases
the electron and muon acceptance, as is expected in Run II \cite{TDR}.
For muons with $1.5<|\eta|<2.0$, we apply the same fiducial efficiency
as for $1.0<|\eta|<1.5$. However,
we still require that tau jets are reconstructed only up to $|\eta|<1.5$.
3) We retain the existing electron isolation requirement and add a
muon isolation requirement $I<2$ GeV, where $I$ is the total
transverse energy contained in a cone of size $\Delta
R=\sqrt{\Delta\phi^2+\Delta\eta^2}=0.4$ around the muon.
4) We increase the jet cluster $E_T$ cut to 15 GeV and correct the
jet energy for muons. We also add a simple electron/photon rejection cut
$E_{em}/E_{had}<10$ to the jet reconstruction algorithm, where
$E_{em}$ ($E_{had}$) is the cluster energy from the
electromagnetic (hadronic) calorimeter.
5) We correct the calorimeter $\met$ for muons.
The addition of the muon isolation cut and the jet $E_{em}/E_{had}$
cut allows us to uniquely resolve the ambiguity arising in SHW v. 2.2,
when a lepton and a jet are very close.

We conservatively use leading order cross sections for all
processes, although NLO corrections should be incorporated in the
experimental Run II analyses. Since the $k$-factor is roughly the
same for both signal \cite{ino corr} and background
\cite{VV_corr,tt_corr,DY_corr}, we expect that the Tevatron
reach will be somewhat better once NLO corrections are
accounted for.

In our analysis we use several alternative values for
the cut on a particular variable. 
1) Four $\met$ cuts: $\met>\{15,20,25\}$ GeV or no cut.
2) Six high-end invariant mass cuts for any pair of opposite sign,
same flavor leptons. The event is discarded if:
$|M_Z - m_{\ell^+\ell^-}|<\{10,15\}$ GeV;
or $m_{\ell^+\ell^-}>\{50,60,70,80\}$ GeV.
3) Eleven low-end invariant mass cuts for any pair of opposite sign,
same flavor leptons: $m^\gamma_{\ell^+\ell^-}<\{10,60\}$ GeV,
in 5 GeV increments.
4) Four azimuthal angle cuts on opposite sign, same flavor leptons:
two cuts on the difference of the azimuthal angle of the two highest
$p_T$ leptons, $|\Delta\varphi|<\{2.5,2.97\}$, one cut
$|\Delta\varphi|<2.5$ for {\em any pair} leptons, and no cut.
5) An optional jet veto (JV) on QCD jets in the event.
6) An optional cut on the the transverse mass $m_T$ of any
$\ell\nu$ pair which may originate from a $W$-boson:
$60<m_T(\ell,\nu)<85$ GeV. 
7) Five sets of lepton $p_T$ cuts:
$\{11,9\},~\{11,11\},~\{13,13\},~ \{15,15\}$ and $\{20,20\}$.
We then employ a parameter space
dependent cut optimization: at each point in SUSY parameter space, we
consider all possible combinations of cuts, and determine the best
combination by maximizing $S/\sqrt{B}$.

We simulate the following background processes (with the generated
number of events in parentheses): $ZZ$ ($10^6$), $WZ$ ($10^6$), $WW$
($10^6$), $t\bar{t}$ ($10^6$), $Z+$jets ($8\cdot10^6$) and $W+$jets
($8\cdot10^6$).

A back-of-the-envelope comparison \cite{MP3L-2l} of the $WZ$ backgrounds
for the 3L and 2L channels reveals that vetoing the third
lepton is not a good idea, so in what follows we only consider the
{\em inclusive} 2L channel, just as in Ref.~\cite{JN-matchev}.
In this case, the signal acceptance is definitely increased.
Unfortunately, the corresponding increase in the background
is even larger and the 2L channel is competitive with the 3L on the basis
of $S/\sqrt{B}$ only if the lepton acceptance for the signal is less
than $1/\sqrt{7.3}\sim 37\%$ \cite{MP3L-2l}. However, for typical
values of the SUSY model parameters the lepton acceptance is much higher
and the 3L channel is preferred.

To make matters worse, the 2L channel suffers from a potentially large
new source of background: $W+$jet production where the jet fakes a
lepton.  Although the rate for a jet faking a lepton is quite small,
on the order of $10^{-4}$, the large $W+$jet cross section results in
a major background for the 2L channel. The
best way to estimate this background is from data, since Monte Carlo
simulations are not reliable for fakes. In our analysis we shall
follow the procedure of Ref.~\cite{JN-matchev}, where the rate for observing
an isolated track which would otherwise pass the lepton cuts was
measured in the Run I $Z+$jet event sample.  This rate was then
multiplied by the probability that, given an isolated track, it would
fake a lepton. This probability was measured in Run I minimum bias
events to be $\sim 1.5\%$, independent of $p_T$ \cite{JN-matchev}. In our
study we first simulate with Monte Carlo the $p_T$ distribution of
isolated tracks in $W$ and $Z$ production.
Then the 2L background cross section is
obtained by multiplying the cross section for isolated tracks by
the probability that an isolated track will fake a lepton. 
We normalize the isolated track rate to data. Using the measured 1.5\%
fake rate per isolated track, we find 1.5 fb of cross section when
running the simulation at $\sqrt{s}=1800$ GeV and using the set of
cuts from Ref.~\cite{JN-matchev}. This is half the cross section found in
Ref.~\cite{JN-matchev}. Hence, to match the data to PYTHIA/SHW we need to
double the isolated track rate obtained from Monte Carlo.

\begin{figure}[t]
\centerline{\psfig{figure=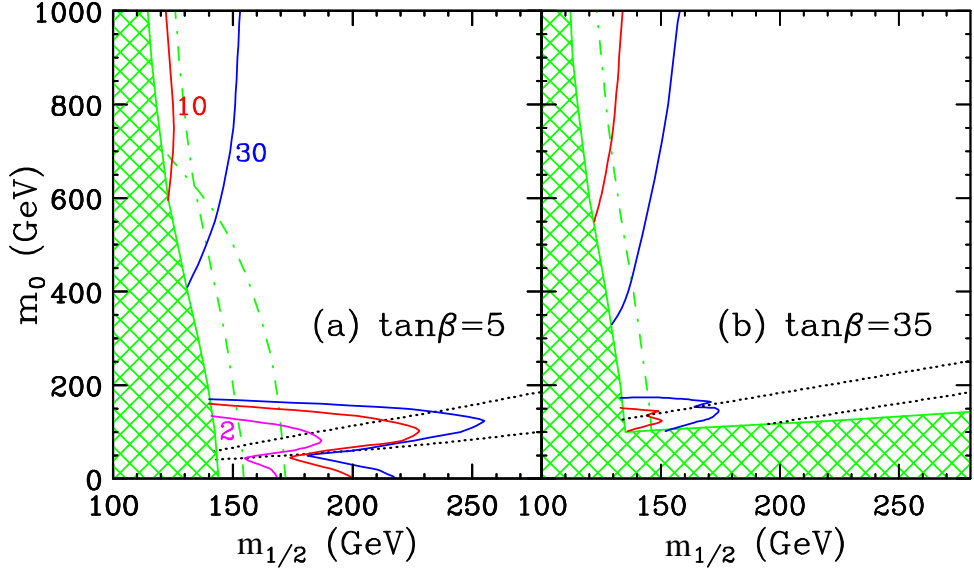,height=3.0in}}
\caption[] {Tevatron reach in the like-sign dilepton channel in
the $M_0-M_{1/2}$ plane, for fixed values of $A_0=0$, $\mu>0$
and (a) $\tan\beta=5$, or (b) $\tan\beta=35$. Results are shown
for 2, 10 and 30 ${\rm fb}^{-1}$ total integrated luminosity.
\label{2l}}
\end{figure}

We present our results for the Tevatron reach in the
like-sign dilepton channel in Fig.~\ref{2l}.
We require the observation of at least 5 signal events, and present
our results as $3\sigma$ exclusion contours in the $M_0-M_{1/2}$
plane, for two representative values of $\tan\beta$, 5 and 35. We fix
$\mu>0$ and $A_0=0$.  The cross-hatched region is excluded by current
limits on the superpartner masses. The dot-dashed lines correspond to
the projected LEP-II reach for the chargino and the lightest Higgs
masses. In Fig.~\ref{2l}a the left dotted line shows where
$m_{\tilde\nu_\tau}=m_{\tilde\chi_1^\pm}$ and the right dotted line
indicates $m_{\tilde\tau_1}=m_{\tilde\chi_1^\pm}$ (and
$m_{\tilde\tau}\simeq m_{\tilde\mu}\simeq m_{\tilde e}$).
In Figs.~\ref{2l}b
the dotted lines show where $m_{\tilde e_R}=m_{\tilde\chi_1^\pm}$
(left) and $m_{\tilde\tau_1}=m_{\tilde\chi_1^\pm}$ (right).

We see that the like-sign dilepton channel offers some reach at
both small and large $M_0$, for both small and large
values of $\tan\beta$. Based on the analysis presented here, the reach of the like-sign dilepton channel is not competitive with that of the clean trilepton channel\cite{3L-matchev}.

%% file: Nachtman/likesign.tex
%\global\firstfigfalse  % kludge to bypass foolishness of revtex to place
%\global\firsttabfalse  % figures and tables at end of paper

% Macros:

\def \chizero  {\tilde{\chi}_1^0}
\def \chitwo   {\tilde{\chi}_2^0}
\def \chione   {\tilde{\chi}_1^\pm}
\def \ET       {E_{T}}
\def \MET      {\mbox{$\raisebox{.3ex}{$\not$}\ET$}}
\def \ppb      {\mathrm{p\bar p}}
\def \lnu      {\ell\nu}
\def \pgev     {{\rm GeV}/{\it c}}
\def \bbb      {\mathrm{b\bar b}}
\def \ccb      {\mathrm{c\bar c}}
\def \ttb      {\mathrm{t\bar t}}
\def \ee       {\mathrm{e^+e^-}}
\def \mumu     {\mu^+\mu^-}
\def \PT       {P_{T}}
\def \mhalf    {m_{1/2}}
\def \pbinv    {{\rm pb}^{-1}}
\def \Gcsq     {{\rm GeV}/{\it c}^2}
\def \mzero    {m_{0}}
\def \tanb     {\tan\beta}

\let\mathrm\relax

\section{Study of a Like-Sign Dilepton Search for Chargino-Neutralino 
Production at CDF}

%\begin{abstract}
%We propose a like-sign dilepton search for chargino-neutralino 
%production in $\ppb$ collisions at $\sqrt{s} = 1.8$ TeV, which 
%complements the previously published trilepton search
%by the CDF detector using Fermilab Run I data.
%Monte Carlo predictions for the signal and background efficiencies
%indicate a significant increase in sensitivity to chargino-neutralino 
%production compared to the traditional trilepton analysis alone.
%\end{abstract}

\subsection{Introduction}

Previous searches for chargino-neutralino production at the Tevatron have 
focused primarily on signatures with three charged leptons (trileptons) plus 
missing  transverse energy ($\MET$)~\cite{nacht-trilep}.  In the Minimal 
Supersymmetric (SUSY) Standard Model, chargino-neutralino production occurs 
in proton-antiproton ($\ppb$) collisions via a virtual $W$ ($s$ channel) or a 
virtual squark ($t$ channel).  In a representative minimal Supergravity 
(SUGRA) model (parameters: 
$\mu < 0, \tanb = 2, \mathrm{A_{0}} = 0, \mzero = 200~\Gcsq$, 
$\mhalf = 90-140~\Gcsq$), we expect three-body chargino and neutralino decays through virtual 
bosons and sleptons in a chargino mass region of $80-130~\Gcsq$.  Conserving 
$R$-parity, these decays produce a distinct signature: trileptons plus
$\MET$ from a neutrino and the lightest supersymmetric
particle.  We demonstrate that the sensitivity to this signature can be
significantly increased by searching for events with two like-sign leptons.  
The Like-Sign Dilepton (LSD) search provides a strong rejection 
of Standard Model background through the like-sign requirement, and enhances 
the acceptance of the signal by requiring only two of the three leptons
produced in the chargino-neutralino decay.

\subsection{Like-Sign Dilepton Analysis}

Signal and most background processes were generated using ISAJET 7.20 and 
the CDF detector Monte Carlo simulation.  For the signal estimation, we used 
representative SUGRA parameters of 
 $\mu < 0, \tanb = 2, \mathrm{A_{0}} = 0, \mzero = 200~\Gcsq$, and
 $\mhalf = 90-140~\Gcsq$.  The relevant mass relations are  
 $\mathrm{M_{\chione}} \sim \mathrm{M_{\chitwo}} \sim 2\mathrm{M_{\chizero}}$, with
 $\mathrm{M_{\chione}}$ between $80-130~\Gcsq$.  The sleptons and sneutrinos
have masses between $200-220~\Gcsq$, so we generate only three-body
chargino and neutralino decays.

The LSD analysis begins with the selection of a pair of leptons ($ee, \mu\mu, 
e\mu$) with the same charge.  We then impose kinematic requirements 
on the selected events in order to
remove Standard Model and other non-SUSY backgrounds.  Our primary requirements
are minimum transverse momentum ($\PT > 11~\pgev$) for both leptons, 
and isolation, in which we remove events where at least one lepton
has excess transverse energy greater than 2 GeV in a cone of 0.4 radians 
around the lepton.  Monte Carlo simulations indicate that isolation 
removes heavy flavor ($\bbb,\ccb$) backgrounds most effectively.  
As the like-sign cut requires us to select both leptons from a $b$ or $c$ decay 
in such an event, and as semi-leptonic $b$ and $c$ decays produce 
leptons associated with jets, neither of the selected leptons will 
be isolated.  Isolation also reduces $\ttb$ background because at least one lepton from
the like-sign pair will be selected from a $b$ decay in such an event.  
The isolation cut, when applied to both like-sign leptons, 
reduces $\bbb$ and $\ccb$ backgrounds to a negligible level.

We remove diboson events 
through a $Z$-mass rejection, in which the combined mass of a third 
opposite-sign, same-flavor lepton selected by the analysis and either 
of the LSDs is between 80--100~$\Gcsq$, reducing $WZ$ and $ZZ$ backgrounds.  
We impose no requirement on $\MET$.  This leaves $WZ$ production as the 
dominant source of Standard Model background, as shown in Table~\ref{nacht-tab1}.

An important source of non-SUSY background estimated from data is events with 
one true lepton, such as W $\rightarrow \lnu$ + jets, and a ``fake'' 
lepton, {\it{i.e.}} an isolated track misidentified as a lepton.
This fake lepton, in combination with the true lepton from 
the W decay, can be selected as a signal event in this analysis.  In 
order to estimate this background, we first look at  $Z \rightarrow \ee$  
+ jets, which we assume provides a model for $W + \rm jets$ events.  
Removing the true leptons, we then measure the rate of underlying isolated 
tracks in the event.  Next we search minimum bias data, 
in which we assume there are no true leptons, to find the probability of an
isolated track to be misidentified as a lepton.  The probability
of misidentifying an isolated track as a lepton is 1.5$\%$ per track.
We multiply this probability by the isolated track rate from the $Z$ 
$\rightarrow \ee$ events, by the number of $W +\rm jets$ events 
expected~\cite{run1w}, and by a factor of 0.5 for the like-sign requirement.  
This ``fake'' rate drops rapidly with an increasing minimum $\PT$ 
requirement.  Optimization of the number of expected background events 
as a function of the $\PT$ requirement yields 0.3 events expected from 
$W + \rm jets$ in 100~$\pbinv$ of data.

\subsection{Results}

Applying the analysis requirements and normalizing the luminosity to
100~$\pbinv$, the expected background is a total of 0.56
events, as shown in Table~\ref{nacht-tab1}.  Drell-Yan and $W +\rm jets$ are the most 
significant non-SUSY backgrounds; $WZ$ production is the largest Standard
Model background. There is little background overlap of the 
trilepton and LSD analyses in the selected events based on Monte Carlo
studies.  Therefore, the backgrounds are treated as independent.  For 
the trilepton analysis, the expected background for the Run I luminosity
of 107~$\pbinv$ is 1.2 events~\cite{nacht-trilep}.  The total expected background 
for the combined LSD and trilepton analyses is 1.8 events.

\begin{table}

\caption{\label{nacht-tab1} Background estimates for the number of events
expected in 100~$\pbinv$ of data based on Monte Carlo (except for the 
$W +\rm jets$ data estimation).  The errors are one-sigma statistical errors.}

\begin{tabular}{|c|c|l|}

Process   &   Luminosity($\pbinv$)    & $\mathrm{N_{events}}$ expected \\ 
\tableline
\tableline
$WZ$ 	&	16,684		&  $0.11\pm 0.02$ \\ 
\tableline
$ZZ$      &	13,992		&  $0.01\pm 0.01$ \\ 
\tableline
$WW$      &        6,870		&  $0^{+0.02}_{-0}$ \\ 
\tableline
$\mathrm{t\bar{t}}$ & 5,558     &  $0^{+0.02}_{-0}$ \\ 
\tableline
Drell-Yan($\gamma^{*}/Z$)  &        1,728       &  $0.11^{+0.10}_{-0.06}$ \\ 
\tableline
$\mathrm{\bbb,\ccb}$   & 3,122     &   $0.03^{+0.04}_{-0.02}$ \\ 
\tableline
$W +\rm jets$  &  (from data)         &    0.30  \\ 
\tableline 
\tableline
Total     &                     &    0.56

\end{tabular}

\end{table}

\begin{figure}[ht]
	
\centerline{\epsfxsize=3.5in\epsfysize=3.6in \epsfbox{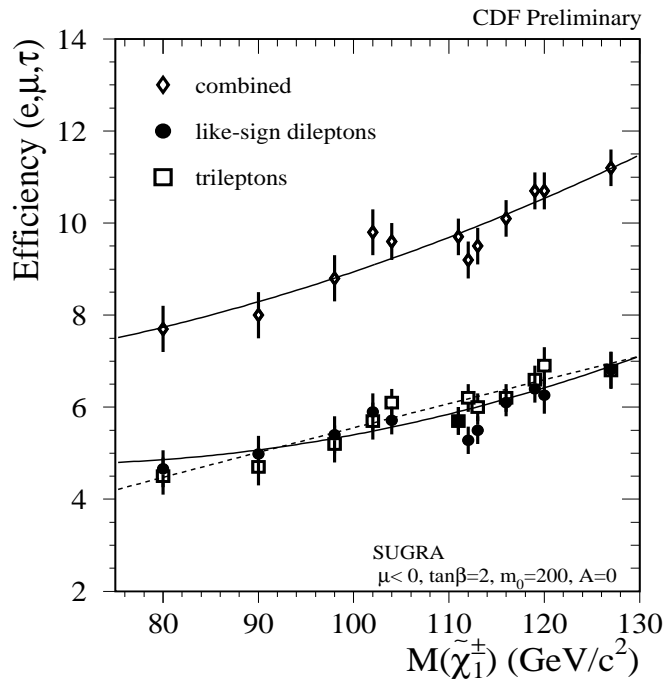}}   
\caption[]{
\label{nacht-eff}
The efficiency (as the percentage of events selected) for the 
trilepton, LSD, and combined analyses as a function of chargino mass.  
SUGRA parameters of 
 $\mu < 0, \tanb = 2, \mathrm{A_{0}} = 0, \mzero = 200~\Gcsq$, and
 $\mhalf = 90-140~\Gcsq$ were used to measure the efficiency.}

\end{figure}

Figure~\ref{nacht-eff} shows the efficiency versus
chargino mass for the trilepton analysis, the LSD analysis, and the combined
analyses, taking into account the signal overlap between the trilepton and LSD 
analyses.  These efficiencies are calculated for all three analyses as
number of selected events divided by total number of chargino-neutralino 
events where both sparticles decay leptonically, where a lepton
can be $e, \mu,$ or $\tau$.  All $\tau$ decays are included in this 
calculation, even though the analyses are only sensitive to the 
leptonic decays.

\begin{figure}[ht]

\centerline{\epsfxsize=3.5in \epsfbox{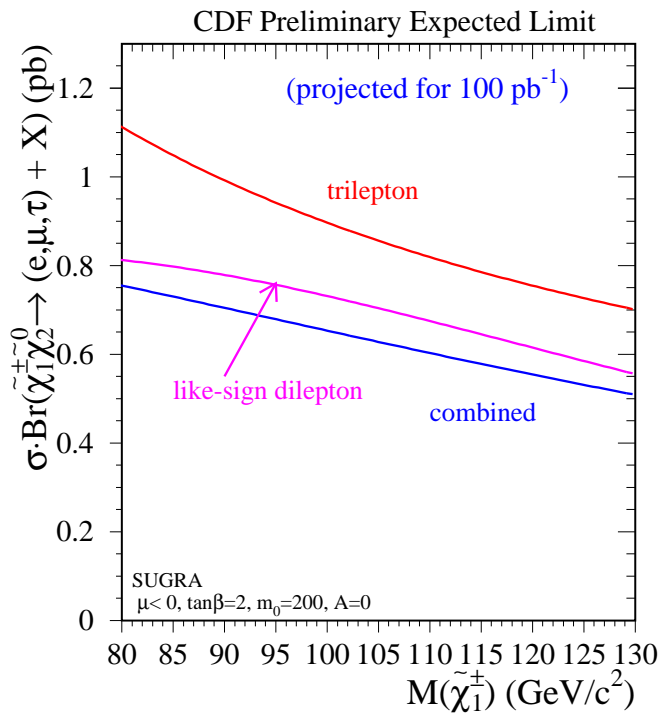}}   
\caption[]{
\label{explim}
Average expected limit on $\sigma\cdot\cal{B}$ as a function 
of $M_{\chione}$ for the LSD analysis, trilepton analysis, and the combination
of both analyses.  The parametrized efficiencies shown in 
Figure~\ref{nacht-eff} are used along with the expected background from
Monte Carlo in this calculation.}

\end{figure}

\break
Figure~\ref{explim} shows the average expected limit normalized to 
100~$\pbinv$ for the trilepton, LSD, and combined analyses.  These limits 
were calculated from the efficiencies in Figure~\ref{nacht-eff} and from the 
expected number of background events based on Monte Carlo.

\subsection{Conclusion}

This study indicates that a fully realized Like-Sign
Dilepton analysis will increase the sensitivity of searches 
for chargino-neutralino production with the CDF detector using existing data 
of $\ppb$ collisions at $\sqrt{s} = 1.8$ TeV.  
It has been shown that the sensitivity of the previously published  
trilepton analysis can be improved by combining it with this new
LSD signature search.  Significantly, the LSD search has fewer requirements 
than the trilepton analysis, {\it{e.g.}} the trilepton analysis requires 
$\MET > 15~\Gcsq$ whereas the LSD analysis has no $\MET$ requirement, 
making the Like-Sign Dilepton channel sensitive to a greater number of 
signatures.

%This research was supported by a grant (number DE-FG03-91ER40662) from the 
%U.S. Department of Energy.

%% file: Matchev-tau/tau.tex
\def\be{\begin{equation}}
\def\ee{\end{equation}}
\def\bear{\begin{eqnarray}}
\def\eear{\end{eqnarray}}
\def\gsim{\lower.7ex\hbox{$\;\stackrel{\textstyle>}{\sim}\;$}}
\def\lsim{\lower.7ex\hbox{$\;\stackrel{\textstyle<}{\sim}\;$}}
\def\dr{\mbox{\footnotesize{$\overline{\rm DR}$\ }}}
\def\GeV{{\rm GeV}}
\def\TeV{{\rm TeV}}
\def\met{\not\!\!\!E_T}
\def\msusy{M_{\rm SUSY}}

%%%%%%%%%%% journals
\def\cpc#1 #2 #3 #4 {Comp.~Phys.~Comm.  {\bf  #1}, #2 (#3)#4 }
\def\npb#1 #2 #3 #4 {Nucl.~Phys. {\bf B#1}, #2 (#3)#4 }
\def\plb#1 #2 #3 #4 {Phys.~Lett. {\bf B#1}, #2 (#3)#4 }
\def\prd#1 #2 #3 #4 {Phys.~Rev.  {\bf D#1}, #2 (#3)#4 }
\def\prl#1 #2 #3 #4 {Phys.~Rev.~Lett. {\bf #1}, #2 (#3)#4 }
\def\pr#1  #2 #3 #4 {Phys.~Rept. {\bf #1}, #2 (#3)#4 }
\def\mpl#1 #2 #3 #4 {Mod.~Phys.~Lett. {\bf A#1}, #2 (#3)#4 }
\def\zpc#1 #2 #3 #4 {Z.~Phys. {\bf C#1}, #2 (#3)#4 }
%
%===============END OF DEFINITIONS====================

\section{Signatures with Tau Jets}

%\author{J.Lykken, K.Matchev and D.Pierce}

%%%%%%%%%%%%%%%%%%%%%%%%%%%%%%%%%%%%%%%%%%%%%%%%%%%%%%%%%%%%%%%
\subsection{Introduction}

Searches for supersymmetry (SUSY) in Run I of the Tevatron
have been done exclusively in channels involving some
combination of leptons, jets, photons and missing transverse
energy ($\met$) \cite{Tevatron searches}. At the same time, several Run I
analyses have identified hadronic tau jets, e.g. in $W$-production
\cite{W to tau} and top decays \cite{top to tau}. 
Hadronically decaying taus have also been used to place limits on
a charged Higgs \cite{H+ to tau} and leptoquarks \cite{Leptoquark to tau}.
Since tau identification is expected to improve further in Run II,
this raises the question whether SUSY searches in channels involving
tau jets are feasible. 

SUSY signatures with tau leptons are very well motivated, since
they arise in a variety of models of low-energy supersymmetry, e.g.
gravity mediated (SUGRA) \cite{BCDPT_PRL,BCDPT_PRD,JW}
or the minimal gauge-mediated models \cite{JW,DN,BT}.
Here we present results from a study \cite{LM}
of all possible {\em experimental}
signatures with three identified objects
(leptons or tau jets) plus $\met$, and compare their reach
to the clean trilepton channel \cite{Run I 3L,3L,Barger,BCDPT_PRL}.
In evaluating the physics potential of the future Tevatron
runs in these new tau channels, it is important to
be aware not only of the physical backgrounds, but also of the 
experimental realities. Jets faking taus will comprise a significant
fraction of the background, and it is crucial to have a reliable
estimate of that rate, which requires a detailed Monte Carlo analysis.
We use PYTHIA \cite{PYTHIA} and TAUOLA \cite{tauola}
for event generation, and the SHW package \cite{SHW},
which provides a realistic detector simulation.

\subsection{Motivation}\label{sec:motivation}

The classic signature for supersymmetry at the Tevatron is
the clean $3\ell\,\met$ channel\footnote{See the
section on trilepton SUSY searches in this Report.}.
It arises in the decays of gaugino-like 
chargino-neutralino pairs $\tilde\chi^\pm_1 \tilde \chi^0_2 $. 
The reach is somewhat limited by the rather small leptonic
branching fractions of the chargino and neutralino. In the limit
of either heavy or mass degenerate squarks and sleptons,
the leptonic branching ratios are $W$-like
and $Z$-like, respectively. However, both
gravity-mediated and gauge-mediated models of SUSY breaking
allow the sleptons to be much lighter than the squarks,
thus enhancing the leptonic branching fractions of the gauginos.

There are several generic reasons as to why one may expect light sleptons
in the spectrum:
\begin{enumerate}
\item The slepton masses at the high-energy
(GUT or messenger) scale may be rather small to begin with.
This is typical for gauge-mediated models, since the sleptons are
colorless and do not receive large soft mass contributions $\sim \alpha_s$. 
This argument applies to all slepton flavors, including staus.
The minimal SUGRA models, on the other hand, predict light sleptons if
$M_0\ll M_{1/2}$. Various effects (non-flat Kahler metric,
RGE running above the GUT scale, D-terms from extra $U(1)$ gauge factors)
may induce nonuniversalities in the scalar masses at the GUT scale,
in which case the slepton-squark mass hierarchy can be affected.
In the absence of a specific model, we do not know which way the
splittings will go, but
as long as the soft scalar masses are small, the RGE running down to the
weak scale will naturally induce a splitting between the squarks
and sleptons, making the sleptons lighter.
\item The renormalization group equations for the scalar soft masses
contain terms proportional to Yukawa couplings, which tend to reduce 
the corresponding mass during the evolution down to low-energy scales.
This effect is significant for third generation scalars, and
for large values of $\tan\beta$ splits the staus from the first two
generation sleptons.
\item The mixing in the charged slepton mass matrix further reduces the
mass of the lightest eigenstate. The slepton mixing is enhanced at
large $\tan\beta$, since it is proportional
to $\mu m_\ell\tan\beta/m^2_{\tilde \ell}$, where 
$m_\ell$ ($m_{\tilde \ell}$) is the lepton (slepton) mass.
Notice that this effect again only applies to the staus,
since $m_\tau\gg m_{\mu,e}$.
\end{enumerate}
Due to these effects, or some peculiar scalar mass non-universality,
it may very well be that among all scalars, only the lightest sleptons
from each generation (or maybe just the lightest stau $\tilde\tau_1$)
are lighter than $\tilde\chi^\pm_1$ and $\tilde \chi^0_2$.
Indeed, in both SUGRA and minimal gauge mediated models
one readily finds regions of parameter space where either 
$m_{\tilde\chi^0_1}< m_{\tilde \tau_1}\sim m_{\tilde \mu_R}< m_{\tilde\chi^+_1}$
(typically at small $\tan\beta$) or
$m_{\tilde\chi^0_1}< m_{\tilde \tau_1}< m_{\tilde\chi^+_1}< m_{\tilde \mu_R}$
(at large $\tan\beta$). Depending on the particular model, and the values of the
parameters, the gaugino pair decay chain may then end up overwhelmingly
in {\em any one} of the four final states: $\ell\ell\ell$, 
$\ell\ell\tau$, $\ell\tau\tau$ or $\tau\tau\tau$. 

\subsubsection{Tau Jets}

In order to make a final decision as to which experimental signatures
are most promising, we have to factor in the tau branching ratios
to leptons and jets. About two-thirds of the subsequent tau decays are hadronic,
so it appears advantageous to consider signatures with tau jets
in the final state as alternatives to the clean trilepton signal\footnote{From
now on, we shall use the following terminology: a ``lepton'' ($\ell$) is either a
muon or an electron; a tau is a tau-lepton, which can later
decay either leptonically, or to a hadronic tau jet, which we
denote by $\tau_h$.}. The branching ratios for three
leptons or undecayed taus into a final state
containing leptons and tau jets is shown in Table~\ref{tau_branching}.
We see that the presence of taus in the underlying SUSY signal
always leads to an enhancement of the signatures with tau jets
in comparison to the clean trileptons. This disparity is most striking
for the case of $\tau\tau\tau$ decays, where
$BR(\tau\tau\tau\rightarrow \ell\ell\tau_h)/
 BR(\tau\tau\tau\rightarrow \ell\ell\ell)\sim 5.5$.

\begin{table*}[h]
\caption{ Branching ratios of the four possible
SUSY signals into the corresponding
experimental signatures involving final state leptons 
$l$ (electrons or muons) as well as identified
tau jets ($\tau_h$). \label{tau_branching}}
\begin{tabular}{||c||c|c|c|c||}\hline\hline
Experimental
  & \multicolumn{4}{c||}{Trilepton SUSY signal} \\
\cline{2-5}
signature
  & $\tau\tau\tau$ 
    & $\tau\tau\ell$
      & $\tau\ell\ell$
        & $\ell\ell\ell$  \\
\hline\hline
$\tau_h\tau_h\tau_h$
  & 0.268
    & ---
      & ---
        & ---  \\
\hline
$\ell\tau_h\tau_h$
  & 0.443
    & 0.416
      & ---
        & ---  \\
\hline
$\ell\ell\tau_h$
  & 0.244
    & 0.458
      & 0.645
        & ---  \\
\hline
$\ell\ell\ell$ 
  & 0.045
    & 0.126
      & 0.355
        & 1.00  \\
\hline
\end{tabular}
\end{table*}

\begin{figure}[h]
%\epsffile[-80 130 400 600]{pT_frac.ps}
\centering\leavevmode
\epsfxsize=3in\epsffile{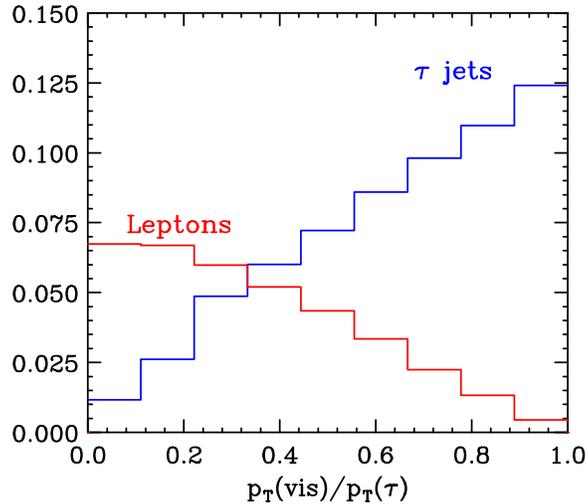}
\caption[]{Distribution of the $p_T$ fraction that the visible tau
decay products (charged leptons or tau jets) inherit from the tau parent.
\label{pT_frac}}
\end{figure}

An additional advantage of the tau jet channels is
that the leptons from tau decays are much softer than the tau jets.
In Fig.~\ref{pT_frac} we show the distribution of the
$p_T$ fraction carried away by the visible decay products from the tau decay.
We see that the lepton products are rather soft,
so the benefit of reducing the corresponding $p_T$ cuts is an issue
\cite{Barger}. 

It should be noted, however, that the tau jet
channels suffer from larger backgrounds than the clean trileptons.
The physical background (from {\em real} tau jets in the event)
is actually smaller, but a significant part of the background is due to
events containing narrow isolated QCD jets with the correct track
multiplicity, which can be misidentified as taus. 
The jetty signatures are also hurt by
the lower detector efficiency for tau jets than for leptons.
The main goal of this study, therefore, will be to see what is the
net effect of all these factors, on a channel by channel basis.
We shall present the results from our Monte Carlo analysis in
the next Section.

\subsubsection{A Challenging Scenario}

For our analysis we choose to examine one of the most challenging
scenarios for SUSY discovery at the Tevatron. 
We shall assume the typical large $\tan\beta$ mass hierarchy
$m_{\tilde\chi^0_1}< m_{\tilde \tau_1}<
m_{\tilde\chi^+_1}< m_{\tilde \mu_R}$. One then finds that
$BR(\tilde\chi^+_1\tilde\chi^0_2\rightarrow \tau\tau\tau+X)\simeq 100\%$
below $\tilde\chi^\pm_1\rightarrow W^\pm \tilde\chi^0_1$
and   $\tilde\chi^0_2\rightarrow Z\tilde\chi^0_1$ thresholds.
(The region of gaugino masses where the two-body 
decays to gauge bosons are open is irrelevant for the Tevatron,
as we shall see below.) 
In order to shy away from specific model dependence, we shall
conservatively ignore all SUSY production channels other than
$\tilde\chi^\pm_1\tilde \chi^0_2$ pair production.
Production of any other supersymmetric particles, which
can later decay into gauginos, will only enhance our signal.

The $p_T$ spectrum of the taus resulting from the chargino and
neutralino decays depends on the mass differences
$m_{\tilde\chi^+_1}-m_{\tilde\tau_1}$ and
$m_{\tilde\tau_1}-m_{\tilde\chi^0_1}$.
The larger they are, the harder the spectrum, and the better the
detector efficiency. However, as the mass difference gets large,
the $\tilde\chi^+_1$ and $\tilde\chi^0_2$ masses themselves
become large, so the production cross-section is severely suppressed.
Therefore, at the Tevatron we can only explore regions with favorable
mass ratios and at the same time small enough gaugino masses.
This suggests a choice of SUSY mass ratios for this study:
for definiteness we fix
$2m_{\tilde\chi^0_1}\sim (4/3)\ m_{\tilde \tau_1}
  \sim m_{\tilde\chi^+_1} (< m_{\tilde \mu_R})$
throughout the analysis, and vary the chargino mass.
The rest of the parameters (first two generation slepton,
heavier stau, tau sneutrino, squark, gluino, Higgs and higgsino masses)
have constant values corresponding to the SUGRA point $M_0=180$ GeV, 
$M_{1/2}=180$ GeV, $A_0=0$ GeV, $\tan\beta=44$ and $\mu>0$,
but we are not constrained to SUGRA models only.
Our analysis will apply equally to gauge-mediated models with
a long-lived neutralino NLSP, as long as the relevant gaugino and slepton
mass relations are similar. Note that our choice of heavy
first two generation sleptons is very conservative.
A more judicious choice of their masses, namely
$m_{\tilde \mu_R}<m_{\tilde\chi^+_1}$, would lead to a
larger fraction of trilepton events, and as a result, a higher reach.
Furthermore, the gauginos would then decay via two-body modes
to first generation sleptons, and the resulting lepton spectrum
would be much harder, leading to a higher lepton efficiency.
With our choice of the superpartner spectrum, all two-body gaugino
decays to first two generation sleptons are closed,
which diminishes the discovery reach of the trilepton channel.

Since the four experimental signatures in our analysis contain only
soft leptons and tau jets, an important issue is whether one can
develop efficient combinations of Level 1 and Level 2 triggers to
accumulate these data sets without squandering all of the available bandwidth.
We will not attempt to address this issue in detail here; instead we will
assume 100\% trigger efficiency for those signal events {\it which pass all
of our analysis and acceptance cuts}. We have nevertheless studied
the following set of triggers \cite{BCDPT_PRD}: 1) $\met>40$ GeV;
2) $p_T(\ell)>20$ GeV and 3) $p_T(\ell)>10$ GeV, $p_T({\rm jet})>15$ GeV
and $\met>15$ GeV; with pseudorapidity cuts
$|\eta(e)|<2.0$, $|\eta(\mu)|<1.5$ and $|\eta(jet)|<4.0$.
We found that they are efficient in picking out about 90 \% of
the signal events in the channels with at least one lepton (see below).
Dedicated low $p_T$ tau triggers for Run II,
which may be suitable for the new tau
jet channels, are now being considered by both CDF \cite{CDF tau trigger}
and D0 \cite{D0 tau trigger}.

\subsection{Analysis}\label{sec:analysis}

We used PYTHIA v6.115 and TAUOLA v2.5 for event generation.
We used the SHW v2.2 detector simulation package, which
simulates an average of the CDF and D0 Run II detector performance.
In SHW tau objects are defined as jets with $|\eta|<1.5$, net
charge $\pm 1$, one or three tracks in a $10$ degree
cone with no additional tracks in a $30$ degree
cone, $E_T>5$ GeV, $p_T>5$ GeV, plus an electron rejection cut.
SHW electrons are required to have $|\eta|<1.5$, $E_T>5$ GeV,
hadronic to electromagnetic energy
deposit ratio $R_{h/e}<0.125$, and satisfy standard isolation cuts.
Muon objects are required to have $|\eta|<1.5$, $E_T>3$ GeV
and are reconstructed using the expected Run II muon 
detection efficiencies. We use standard isolation
cuts for muons as well. Jets are required to have $|\eta|<4$ 
and $E_T>15$ GeV. In addition we have added jet energy correction
for muons and the rather loose jet id requirement $R_{h/e}>0.1$.
We have also modified the TAUOLA program in order to correctly
account for the chirality of tau leptons coming from SUSY decays.

The reconstruction algorithms in SHW already include some
basic cuts, so we can define a reconstruction efficiency
$\epsilon_{rec}$ for the various types of objects: electrons,
muons, tau jets etc. We find that as we vary the chargino mass
from 100 to 140 GeV the electron and tau jet reconstruction
efficiencies range from 42 to 49 \%, and from 29 to 36\%, correspondingly.
The lepton efficiency may seem surprisingly low, but this is because
a lot of our leptons are very soft and fail the $E_T$ cut.
The tau efficiency is in good agreement with the results from 
Ref.~\cite{CDF thesis} and \cite{Leslie}, once we account
for the different cuts used in those analyses.

The most important issue in these channels is the fake tau rate.
Several experimental analyses try to estimate it using Run I
data. Here we simulate the corresponding backgrounds to our
signal and use SHW to obtain the fake rate, thus
avoiding trigger bias \cite{CDF thesis}.  We find that the SHW
tau fake rate in $W$ production is roughly 1.5\%, and almost independent
of the tau $p_T$.

\subsubsection{Cuts}

As discussed earlier, we expect that the
reach in the classic $\ell\ell\ell\met$
channel will be quite suppressed, due to the softness of the
leptons. Therefore we apply the soft cuts proposed in
Refs.~\cite{Barger}. We require a central lepton with 
$p_T>11$ GeV and $|\eta|<1.0$, and in addition two more leptons with
$p_T(\ell_2)>7$ GeV and $p_T(\ell_3)>5$ GeV.
Leptons are required to be isolated: $I(\ell)<2$ GeV, where $I$ is
the total transverse energy contained in a cone of size
$\delta R=\sqrt{\Delta\phi^2+\Delta\eta^2}=0.4$ around the lepton. 
We impose a dilepton invariant mass cut for same flavor,
opposite sign leptons: $|m_{\ell\ell}-M_Z|>10$ GeV and $|m_{\ell\ell}|>11$ GeV.
Finally, we impose an optional veto on additional jets with
$p_T>15$ GeV, and require $\not\!\!\!E_T$ to be either
more than 20 GeV, or 25 GeV. This gives us a total of
four combinations of the $\not\!\!\!E_T$ cut and the jet
veto (shown in Table.~\ref{a-d}), which we apply for all
tau jet signatures later as well. 

\begin{table*}[h]
\caption{ Definition of the signal samples A-D.
\label{a-d}}
\begin{tabular}{||c||c|c||}\hline\hline
Sample & $\not\!\!\!E_T$ cut & Jet veto  \\ \hline\hline
A      & 20 GeV              & no        \\ \hline
B      & 25 GeV              & no        \\ \hline
C      & 20 GeV              & yes       \\ \hline
D      & 25 GeV              & yes       \\ \hline\hline
\end{tabular}
\end{table*}

For our $\ell\ell\tau_h\not\!\!E_T$ analysis we impose cuts similar to the
stop search analysis in the $\ell^+\ell^-j\!\met$ channel \cite{CDFanalysis}:
two isolated ($I(\ell)<2$ GeV) leptons with $p_T(\ell_1)>8$ GeV and
$p_T(\ell_2)>5$ GeV and one identified hadronic tau jet with
$p_T(\tau_h)>15$ GeV. Again, we impose
invariant mass cuts $|m_{\ell\ell}-m_Z|>10$ GeV and $|m_{\ell\ell}|>11$
for any same flavor, opposite sign dilepton pair.

A separate, very interesting signature arises if the two leptons
have the same sign, since the background is greatly suppressed.
In fact, we expect this background to be significantly smaller than the
trilepton background! Roughly one
third of the signal events in the general $\ell\ell\tau_h$ sample
are expected to have like-sign leptons.

For our $\ell\tau_h\tau_h\not\!\!E_T$ analysis we use some basic
identification cuts: two tau jets with $p_T>15$ GeV and  $p_T>10$ GeV
and one isolated lepton with $p_T(\ell)>7$ GeV.

Finally, for the $\tau_h\tau_h\tau_h\not\!\!E_T$ signature
we require three tau jets
with $p_T>15$ GeV, $p_T>10$ GeV
and $p_T(\tau_3)>8$ GeV, respectively.

\subsubsection{Signal}

\begin{figure}[t!]
\epsfxsize=3.5in
\centering\leavevmode
\epsffile{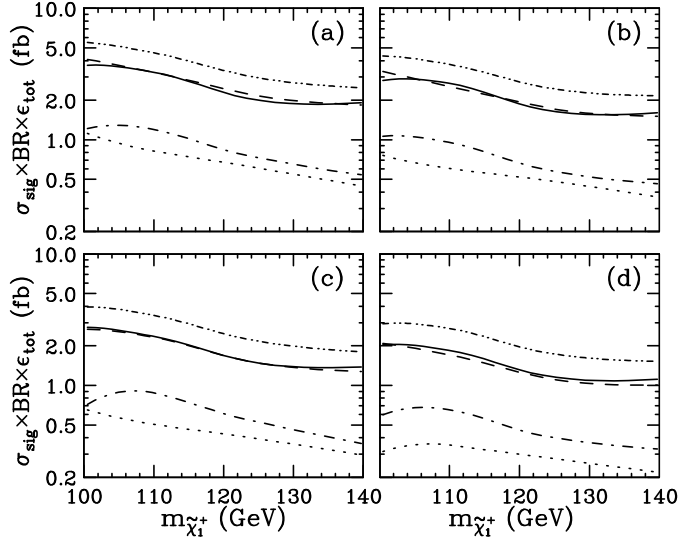}
\caption[]{ Signal cross-section times branching ratio after cuts for
the five channels discussed in the text:
$\ell\ell\ell\met$ (dotted),
$\ell\ell\tau_h\met$ (dashed),
$\ell^+ \ell^+ \tau_h\met$ (dot dashed),
$\ell\tau_h\tau_h\met$ (dot dot dashed) and
$\tau_h\tau_h\tau_h\met$ (solid); and for the four signal
samples from Table \ref{a-d}: a) A, b) B, c) C and d) D.
\label{sigeff}}
\end{figure}

One can get a good idea of the relative importance of the
different channels by looking at the corresponding signal samples
after the analysis cuts have been applied.
In Fig.~\ref{sigeff} we show the signal cross-sections
times the corresponding branching ratios times the total efficiency
$\epsilon_{tot}\equiv \epsilon_{rec}\epsilon_{cuts}$,
which accounts for both the detector acceptance and the
efficiency of the cuts (for each point we generated 100,000
signal events).
We see that the lines are roughly ordered according to the
branching ratios from Table~\ref{tau_branching}. This can be
understood as follows. The acceptance (which includes the
basic ID cuts in SHW) is higher for leptons than for $\tau$ jets.
Therefore, replacing a lepton with a tau jet in the
experimental signature costs us a factor of $\sim 1.5$ in acceptance,
due to the poorer reconstruction of tau jets, compared to leptons.
Later, however, the cuts tend to reduce the leptonic signal more than the
tau jet signal. This is mostly because the leptons are softer
than the tau jets.
Notice that we cannot improve the efficiency for leptons by further lowering
the cuts -- we are already using the most liberal cuts \cite{Barger}.
It turns out that these two effects mostly cancel each other, and the total
efficiency $\epsilon_{tot}$ is roughly the same for all channels.
Therefore the relative importance of each channel
will only depend on the tau branching ratios and the backgrounds. 
For example, in going from $\ell\ell\ell$ to $\ell\ell\tau_h$,
one wins a factor of 5.5 from the branching ratio. 
Therefore the background to $\ell\tau_h\tau_h\met$
must be at least $5.5^2\sim 30$ times larger in order
for the clean trilepton channel to be still preferred.

\subsubsection{Backgrounds}

We next turn to the discussion of the backgrounds involved.
We have simulated the following physics background processes:
$ZZ$, $WZ$, $WW$, $t\bar{t}$, $Z+jets$, and $W+jets$,
generating $4\times(10^6)$ and $2\times(10^7)$ events, respectively.
We list the results in Tables~\ref{BKND_A}-\ref{BKND_D},
where all errors are purely statistical.

A number of interesting observations can be made regarding these backgrounds:
\begin{enumerate}
\item $WZ$ is indeed the major source of background for the trilepton
channel. The majority of the background events
contain a leptonically decaying off-shell $Z$ and 
pass the invariant dilepton mass cut. The rest of the WZ background comes
from $Z\rightarrow\tau^+\tau^-\rightarrow \ell^+\ell^-\met$.
We find a total $WZ$ rate which is a factor of three higher than in
recent trilepton analyses prior to this workshop (see, e.g.
\cite{BCDPT_PRL}, \cite{BCDPT_PRD}, \cite{Barger}).
To simulate the diboson backgrounds, these previous estimates 
employed ISAJET, where the $W$ and $Z$ gauge bosons are always
generated exactly on their mass shell, and there is no finite-width
smearing effect \cite{MP_comp}, \cite{MP_PRD} \footnote{Since then,
the trilepton analysis has been redone independently by several
groups and the increase in the $WZ$ background has been confirmed
\cite{MP_comp,MP_PRD,MP_talks,MP_PLB,CE,BDPQT,BK4}. In addition, the
virtual photon contribution and the $Z-\gamma$ interference effect,
neither of which is modelled in either PYTHIA or ISAJET, have also
been included \cite{MP_talks,MP_PLB,CE,BDPQT,BK4}, which further increases
the background several times. This required new cuts for the trilepton
analysis, specifically designed to remove these additional contributions
\cite{MP_PLB,BDPQT}. See the trilepton analyses in this Report.}.
\item As we move to channels with more tau jets, the number
of background events with {\em real} tau jets decreases: first,
because of the smaller branching ratios of $W$ and $Z$ to taus;
and second, because the tau jets in $W$ and $Z$ decays are {\em softer}
than the leptons from $W$ and $Z$. This is to be contrasted with
the signal, where, conversely, the tau jets are harder than the leptons.
We also see, however, that the contribution from events with
fake taus (from hadronically decaying $W$'s and $Z$'s or from initial
and final state jet radiation) increases, and for the $3\tau$
channel events with fake taus are the dominant part of the $WZ$
background.
\item Notice that the $WZ$ background to the same-sign dilepton channel 
is smaller (by a factor of two) than for the trilepton channel. As expected,
it is also about a half of the total contribution to $\ell\ell\tau$
(recall that for the signal this ratio is only a third).
Indeed, one third of the events with opposite sign leptons
come from the $Z$-decay and are cut away by the dilepton mass cut.
\item Vetoing a fourth lepton in the event reduces the $ZZ$ background
to the trilepton channel only by 4--8  \%. The $ZZ$
trilepton background is due to one $Z$ decaying as $Z\rightarrow \tau\tau$,
thus providing the missing energy in the event, and the other $Z$
decaying to leptons: $Z\rightarrow \ell^+\ell^-$. Most of the events passing
the cuts contain an off-shell $Z/\gamma$ decaying leptonically\footnote{ISAJET
analyses would miss this component of the $ZZ$ background.},
and the third lepton coming from a leptonic tau.
But then it is 6 times more probable that the
second tau would decay hadronically and will not give a fourth lepton.
The rest of the $ZZ$ background events come from a regular
$Z\rightarrow \ell^+\ell^-$ decay, where one of the leptons is missed,
and the invariant mass cut does not apply. For those events, there
is obviously no fourth lepton.
\item The jet veto is very effective in reducing the $t\bar{t}$ background
for the first three channels. However, it also reduces the signal
(see Fig.~\ref{sigeff}).
\item In all channels, a higher $\not\!\!\!E_T$ cut did not
help to get rid of the major backgrounds. Indeed, $WZ$,
$t\bar{t}$ and/or $W+jets$ backgrounds tend to have a lot of missing energy,
due to the leptonic W-decays.
\item Our result for the $W+jets$ and $Z+jets$ backgrounds should be taken
with a grain of salt, in spite of the relatively small statistical errors.
Events with fake leptons are expected to comprise a major part of this
background, and SHW does not provide a realistic simulation of those. 
In fact, the most reliable way to estimate this background will
be from Run IIa data, e.g. by estimating the probability for an
isolated track from Drell-Yan events, and the lepton fake rate
per isolated track from minimum bias data \cite{JN,MP_PRD,MP_PLB}.
\item We have underestimated the total background
to the three-jet channel by considering only processes with at least
one real tau in the event. We expect sizable contributions from
pure QCD multijet events, or $Wj\rightarrow jjj$, where
{\em all} three tau jets are fake. 
\end{enumerate}

%%%%%%%%%%%%%%%%%%%%%%% CASE A
\begin{table*}[h]
\renewcommand{\arraystretch}{1.1}
\caption{ Results for the backgrounds in the various channels
in case A:\ $\not\!\!\!E_T>20$ GeV and no jet veto.
\label{BKND_A}}
\begin{tabular}{||c||c|c|c|c|c||}
\hline\hline
Case A:\ $\not\!\!\!E_T>20$  
  & \multicolumn{5}{c||}{Experimental signatures} \\ 
\cline{2-6}
  & $\ell\ell\ell\!\!\met$
    & $\ell\ell\tau_h\!\!\met$ 
      & $\ell^+\ell^+\tau_h\!\!\met$
        & $\ell\tau_h\tau_h\!\!\met$
          & $\tau_h\tau_h\tau_h\!\!\met$  \\ \hline\hline
%%%%%%%%%%%%%%%%%%%%%%%%%  LLL   %  LLT   % L+L+T  % LTT    % TTT    
$ZZ$                    & 0.196 $\pm$ 0.028   
                                 & 0.334 $\pm$ 0.036
                                          & 0.094 $\pm$ 0.019
                                                   & 0.181 $\pm$ 0.027
                                                            & 0.098 $\pm$ 0.020  \\ \hline
$WZ$     (fb)           & 1.058 $\pm$ 0.052   
                                 & 1.087 $\pm$ 0.053
                                          & 0.447 $\pm$ 0.034
                                                   & 1.006 $\pm$ 0.051
                                                            & 0.248 $\pm$ 0.025  \\ \hline
$WW$     (fb)           & ---   
                                 & 0.416 $\pm$ 0.061
                                          & ---
                                                   & 0.681 $\pm$ 0.078
                                                            & 0.177 $\pm$ 0.039  \\ \hline
$t\bar{t}$     (fb)     & 0.300 $\pm$ 0.057   
                                 & 1.543 $\pm$ 0.128
                                          & 0.139 $\pm$ 0.038
                                                   & 1.039 $\pm$ 0.105
                                                            & 0.161 $\pm$ 0.041  \\ \hline
$Z+jets$      (fb)      & 0.112 $\pm$ 0.079   
                                 & 7.34 $\pm$ 0.64
                                          & 0.168 $\pm$ 0.097
                                                   & 20.3 $\pm$ 1.1
                                                            & 17.9 $\pm$ 1.0  \\ \hline
$W+jets$      (fb)      & ---    & ---    & ---    & 37.2 $\pm$ 2.9
                                                            & 6.1 $\pm$ 1.2  \\ \hline\hline
$\sigma_{BG}^{\rm tot}$ 
(fb)                    & 1.67 $\pm$ 0.11   
                                 & 10.7  $\pm$ 0.7
                                          & 0.85 $\pm$ 0.11
                                                   & 60.4 $\pm$ 3.1
                                                            & 24.7 $\pm$ 1.6  \\ \hline\hline
\end{tabular}
\end{table*}
%
%
%
%%%%%%%%%%%%%%%%%%%%%%% CASE B
\begin{table*}[ht!]
\renewcommand{\arraystretch}{1.1}
\caption{ Results for the backgrounds in the various channels
in case B:\ $\not\!\!\!E_T>25$ GeV and no jet veto. \label{BKND_B}}
\begin{tabular}{||c||c|c|c|c|c||}
\hline\hline
Case B:\ $\not\!\!\!E_T>25$  
  & \multicolumn{5}{c||}{Experimental signatures} \\ 
\cline{2-6}
  & $\ell\ell\ell\!\!\met$
    & $\ell\ell\tau_h\!\!\met$ 
      & $\ell^+\ell^+\tau_h\!\!\met$
        & $\ell\tau_h\tau_h\!\!\met$
          & $\tau_h\tau_h\tau_h\!\!\met$  \\ \hline\hline
%%%%%%%%%%%%%%%%%%%%%%%%%  LLL   %  LLT   % L+L+T  % LTT    % TTT    
$ZZ$     (fb)           & 0.165 $\pm$ 0.025   
                                 & 0.271 $\pm$ 0.033
                                          & 0.090 $\pm$ 0.019
                                                   & 0.153 $\pm$ 0.024
                                                            & 0.086 $\pm$ 0.018  \\ \hline
$WZ$     (fb)           & 0.964 $\pm$ 0.050   
                                 & 1.001 $\pm$ 0.051
                                          & 0.423 $\pm$ 0.033
                                                   & 0.909 $\pm$ 0.049
                                                            & 0.204 $\pm$ 0.023  \\ \hline
$WW$     (fb)           & ---   
                                 & 0.380 $\pm$ 0.058
                                          & ---
                                                   & 0.602 $\pm$ 0.073
                                                            & 0.142 $\pm$ 0.036  \\ \hline
$t\bar{t}$     (fb)     & 0.300 $\pm$ 0.057   
                                 & 1.500 $\pm$ 0.127
                                          & 0.139 $\pm$ 0.038
                                                   & 0.996 $\pm$ 0.103
                                                            & 0.128 $\pm$ 0.037  \\ \hline
$Z+jets$      (fb)      & 0.056 $\pm$ 0.056   
                                 & 4.87 $\pm$ 0.52
                                          & 0.112 $\pm$ 0.079
                                                   & 13.61 $\pm$ 0.87
                                                            & 11.82 $\pm$ 0.81  \\ \hline
$W+jets$      (fb)      & ---    & ---    & ---    & 32.1 $\pm$ 2.7
                                                            & 5.5 $\pm$ 1.1  \\ \hline\hline
$\sigma_{BG}^{\rm tot}$ 
(fb)                    & 1.49  $\pm$ 0.10
                                 & 8.0  $\pm$ 0.5
                                          & 0.76 $\pm$ 0.09
                                                   & 48.4 $\pm$ 2.8
                                                            & 17.9 $\pm$ 1.4  \\ \hline\hline
\end{tabular}
\end{table*}
%
%
%
%%%%%%%%%%%%%%%%%%%%%%% CASE C
\begin{table*}[ht!]
\renewcommand{\arraystretch}{1.1}
\caption{ Results for the backgrounds in the various channels
in case C:\ $\not\!\!\!E_T>20$ GeV and no extra jets with $p_T>15$ GeV.
\label{BKND_C}}
\begin{tabular}{||c||c|c|c|c|c||}
\hline \hline
Case C:\ $\not\!\!\!E_T>20$  
  & \multicolumn{5}{c||}{Experimental signatures} \\ 
\cline{2-6}
jet veto
  & $\ell\ell\ell\!\!\met$
    & $\ell\ell\tau_h\!\!\met$ 
      & $\ell^+\ell^+\tau_h\!\!\met$
        & $\ell\tau_h\tau_h\!\!\met$
          & $\tau_h\tau_h\tau_h\!\!\met$  \\ \hline\hline
%%%%%%%%%%%%%%%%%%%%%%%%%  LLL   %  LLT   % L+L+T  % LTT    % TTT    
$ZZ$     (fb)           & 0.114 $\pm$ 0.021   
                                 & 0.220 $\pm$ 0.029
                                          & 0.071 $\pm$ 0.017
                                                   & 0.094 $\pm$ 0.019
                                                            & 0.031 $\pm$ 0.011  \\ \hline
$WZ$     (fb)           & 0.805 $\pm$ 0.046   
                                 & 0.828 $\pm$ 0.046
                                          & 0.347 $\pm$ 0.030
                                                   & 0.695 $\pm$ 0.043
                                                            & 0.136 $\pm$ 0.019  \\ \hline
$WW$     (fb)           & ---   
                                 & 0.301 $\pm$ 0.052
                                          & ---
                                                   & 0.354 $\pm$ 0.056
                                                            & 0.097 $\pm$ 0.029  \\ \hline
$t\bar{t}$     (fb)     & ---   
                                 & 0.086 $\pm$ 0.030
                                          & ---
                                                   & 0.032 $\pm$ 0.018
                                                            & ---   \\ \hline
$Z+jets$      (fb)      & 0.056 $\pm$ 0.056   
                                 & 4.93 $\pm$ 0.52
                                          & 0.056 $\pm$ 0.056
                                                   & 12.66 $\pm$ 0.84
                                                            & 10.36 $\pm$ 0.76  \\ \hline
$W+jets$      (fb)      & ---    & ---    & ---    & 25.8 $\pm$ 2.4
                                                            & 3.2 $\pm$ 0.9  \\ \hline\hline
$\sigma_{BG}^{\rm tot}$ 
(fb)                    & 0.97  $\pm$ 0.07
                                 & 6.4  $\pm$ 0.5
                                          & 0.47 $\pm$ 0.06
                                                   & 39.6 $\pm$ 2.5
                                                            & 13.8  $\pm$ 1.2  \\ \hline\hline
\end{tabular}
\end{table*}
%
%
%
%
%%%%%%%%%%%%%%%%%%%%%%% CASE D
\begin{table*}[b]
\renewcommand{\arraystretch}{1.1}
\caption{ Results for the backgrounds in the various channels
in case D:\ $\not\!\!\!E_T>25$ GeV and no extra jets with $p_T>15$ GeV.
\label{BKND_D}}
\begin{tabular}{||c||c|c|c|c|c||}
\hline\hline
Case D:\ $\not\!\!\!E_T>25$  
  & \multicolumn{5}{c||}{Experimental signatures} \\ 
\cline{2-6}
jet veto
  & $\ell\ell\ell\!\!\met$
    & $\ell\ell\tau_h\!\!\met$ 
      & $\ell^+\ell^+\tau_h\!\!\met$
        & $\ell\tau_h\tau_h\!\!\met$
          & $\tau_h\tau_h\tau_h\!\!\met$  \\ \hline\hline
%%%%%%%%%%%%%%%%%%%%%%%%%  LLL   %  LLT   % L+L+T  % LTT    % TTT    
$ZZ$     (fb)           & 0.098 $\pm$ 0.020   
                                 & 0.177 $\pm$ 0.026
                                          & 0.071 $\pm$ 0.017
                                                   & 0.075 $\pm$ 0.017
                                                            & 0.027 $\pm$ 0.010  \\ \hline
$WZ$     (fb)           & 0.732 $\pm$ 0.044   
                                 & 0.766 $\pm$ 0.045
                                          & 0.329 $\pm$ 0.029
                                                   & 0.622 $\pm$ 0.040
                                                            & 0.115 $\pm$ 0.017  \\ \hline
$WW$     (fb)           & ---   
                                 & 0.274 $\pm$ 0.049
                                          & ---
                                                   & 0.327 $\pm$ 0.054
                                                            & 0.071 $\pm$ 0.025  \\ \hline
$t\bar{t}$     (fb)     & ---   
                                 & 0.075 $\pm$ 0.028
                                          & ---
                                                   & 0.032 $\pm$ 0.018
                                                            & ---   \\ \hline
$Z+jets$      (fb)      & ---   
                                 & 3.25 $\pm$ 0.24
                                          & ---
                                                   & 7.62 $\pm$ 0.65
                                                            & 6.55 $\pm$ 0.61  \\ \hline
$W+jets$      (fb)      & ---    & ---    & ---    & 22.6 $\pm$ 2.3
                                                            & 3.0 $\pm$ 0.8  \\ \hline\hline
$\sigma_{BG}^{\rm tot}$ 
(fb)                    & 0.83  $\pm$ 0.05
                                 & 4.5  $\pm$ 0.3
                                          & 0.40 $\pm$ 0.03
                                                   & 31.3 $\pm$ 2.4
                                                            & 9.8  $\pm$ 1.0  \\ \hline\hline
\end{tabular}
\end{table*}

\subsubsection{Tevatron reach}

We are now ready to present our results for the Tevatron reach in Run II.
A $3\sigma$ exclusion limit would require
\begin{equation}
L\ =\ {9\sigma_{BG}\over 
\left(\sigma\times BR(\tilde\chi^+_1\tilde\chi^0_2\rightarrow X) \epsilon_{tot}\right)^2}.
\end{equation}
Notice that the limit depends linearly on the
background $\sigma_{BG}$ after cuts, but {\em quadratically} on
the signal branching ratios. This allows the jetty channels
to compete very successfully with the clean trilepton signature,
whose branching ratio is quite small.
\begin{figure}[t!]
\epsfxsize=3.5in
\centering\leavevmode
\epsffile{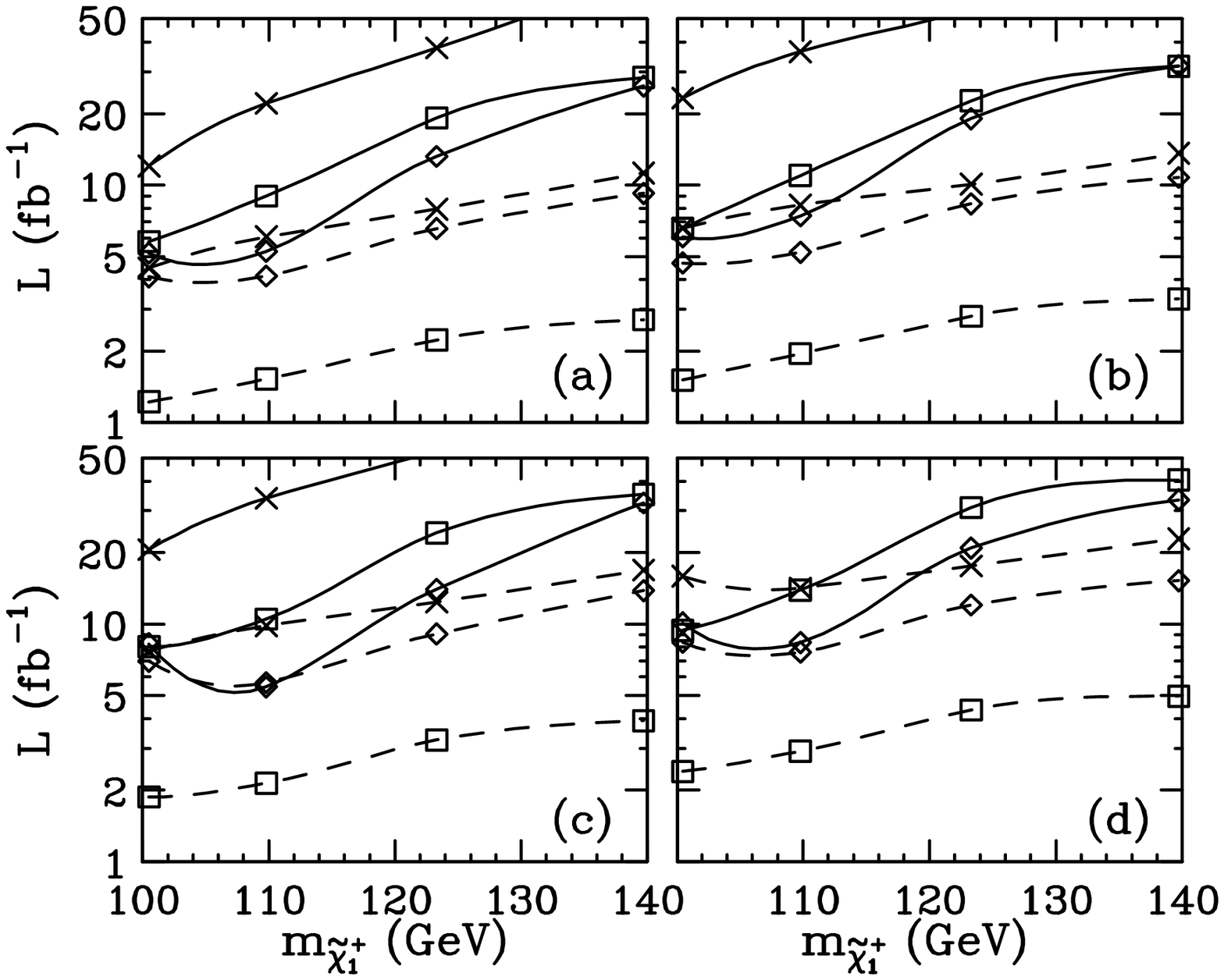}

\caption[]{ The total integrated luminosity $L$ needed
for a $3\sigma$ exclusion (solid lines) or observation of
5 signal events (dashed lines), as a function of the chargino
mass $m_{\tilde\chi^+_1}$, for (a) sample A; (b) sample B;
(c) sample C and (d) sample D, as defined in Table \ref{a-d};
and for the following channels: trileptons ($\times$),
dileptons plus a tau jet ($\Box$)
and like-sign dileptons plus a tau jet ($\Diamond$).
\label{reach}}
\end{figure}
In Fig.~\ref{reach} we show the Tevatron reach in the three
channels: trileptons ($\times$), dileptons plus a tau jet ($\Box$)
and like-sign dileptons plus a tau jet ($\Diamond$).
We see that the two channels with tau jets have a much better
sensitivity compared to the usual trilepton signature.
Assuming that efficient triggers can be implemented,
the Tevatron reach will start exceeding LEP II
limits as soon as Run IIa is completed and the two collaborations
have collected a total of $4\ {\rm fb}^{-1}$ of data.
Considering the intrinsic difficulty of the SUSY scenario
we are contemplating, the mass reach for Run IIb is quite
impressive. One should also keep in mind that
we did not attempt to optimize our cuts for the new channels.
For example, one could use angular correlation cuts to suppress Drell-Yan,
transverse $W$ mass cut to suppress $WZ$ \cite{BDPQT},
or (chargino) mass--dependent $p_T$ cuts for the leptons and tau jets
\cite{MP_PRD,MP_PLB},
to squeeze out some extra reach. In addition, the $ll\tau_h$ channel
can be explored at smaller values of $\tan\beta$ as well
\cite{BCDPT_PRD,Barger,MP_PRD,MP_PLB} (see also the section
on trileptons, dileptons and dileptons plus tau jets in this Report),
since the two-body chargino decays are preferentially to tau sleptons.
In that case, the clean trilepton channel still offers the best reach,
and a signal can be observed already in Run IIa. Then,
the tau channels will not only provide an important confirmation,
but also hint towards some probable values of the SUSY model parameters.

\subsection{Dilepton Plus Tau Jet Analysis with Variable Cuts}

Subsequently, the dilepton plus tau jet channel was
analysed again in Refs.~\cite{MP_PRD,MP_PLB}. The main improvements
were the inclusion of the $\gamma^\ast$ contribution of the $WZ$
background, and parameter space dependent cut optimization.
For details on the numerical procedure, see Refs.~\cite{MP_PRD,MP_PLB}
and the like-sign dilepton section in this report.

\begin{figure}[h]
\epsfxsize=3.5in
\centering\leavevmode
\epsffile{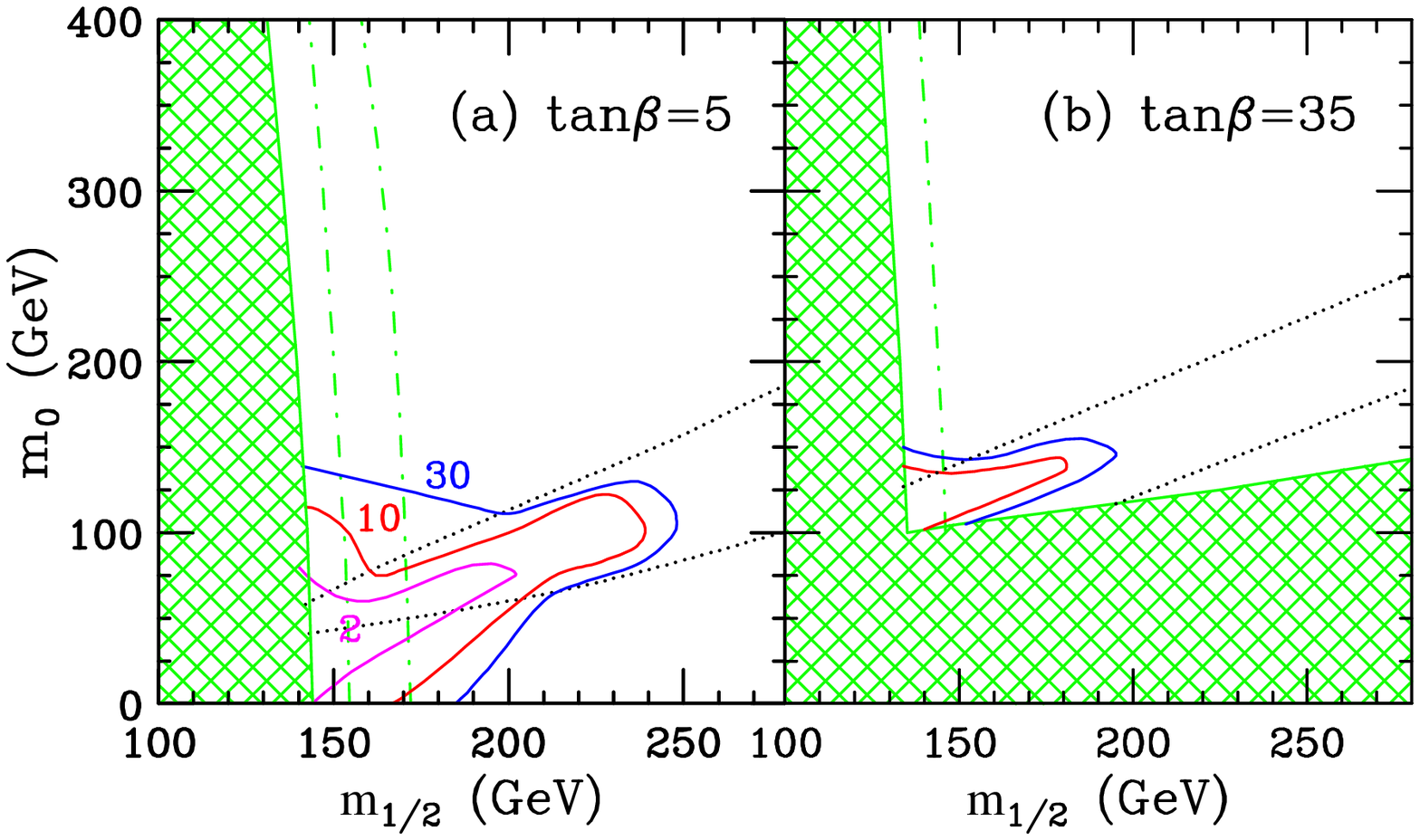}
\caption[]{ Tevatron reach in the dilepton plus tau jet
channel in
the $M_0-M_{1/2}$ plane, for fixed values of $A_0=0$, $\mu>0$
and (a) $\tan\beta=5$, or (b) $\tan\beta=35$. Results are shown
for 2, 10 and 30 ${\rm fb}^{-1}$ total integrated luminosity.
\label{2l1t}}
\end{figure}

The results for the Tevatron reach in the dilepton
plus tau jet channel are shown in Fig.~\ref{2l1t}.
We require the observation of at least 5 signal events, and present
the results as $3\sigma$ exclusion contours in the $M_0-M_{1/2}$
plane, for two representative values of $\tan\beta$, 5 and 35. We fix
$\mu>0$ and $A_0=0$.  The cross-hatched region is excluded by current
limits on the superpartner masses. The dot-dashed lines correspond to
the projected LEP-II reach for the chargino and the lightest Higgs
masses. In Fig.~\ref{2l1t}a the left dotted line shows where
$m_{\tilde\nu_\tau}=m_{\tilde\chi_1^\pm}$ and the right dotted line
indicates $m_{\tilde\tau_1}=m_{\tilde\chi_1^\pm}$ (and
$m_{\tilde\tau}\simeq m_{\tilde\mu}\simeq m_{\tilde e}$). In Fig.~\ref{2l1t}b
the dotted lines show where $m_{\tilde e_R}=m_{\tilde\chi_1^\pm}$
(left) and $m_{\tilde\tau_1}=m_{\tilde\chi_1^\pm}$ (right).
We see that the dilepton plus tau jet channel provides some 
reach, but only at small $M_0$. 

%% file: Baer-non-u/non-u.tex
%----------------------------------------------------------------
%%%%%%%%%%%%%%%%%%%%%%%%%%%%%%%%%%%%%%%%%%%%%%%%%%%%%%%%%%%%%%%%%%%%%%
%  This is a LaTeX file!
%  written by Howie Baer
%  modified 12/08/98
%%%%%%%%%%%%%%%%%%%%%%%%%%%%%%%%%%%%%%%%%%%%%%%%%%%%%%%%%%%%%%%%%%%%%%
%\usepackage{psfig}
\def\eslt{\not\!\!{E_T}}
\def\to{\rightarrow}
\def\Phat{\hat{\Phi}}
\def\te{\tilde e}
\def\tx{\tilde\chi}
\def\tl{\tilde l}
\def\tb{\tilde b}
\def\tf{\tilde f}
\def\td{\tilde d}
\def\tst{\tilde t}
\def\ttau{\tilde \tau}
\def\tmu{\tilde \mu}
\def\tg{\tilde g}
\def\tnu{\tilde\nu}
\def\tell{\tilde\ell}
\def\tq{\tilde q}
\def\tw{\tilde\chi^\pm}
\def\tz{\tilde\chi^0}

\section{Implications of Non-Universal Gaugino Masses
 for Tevatron SUSY Searches}

%\author{G. Anderson, H. Baer, C. H. Chen and X. Tata}

\subsection{Non-universality of soft SUSY breaking terms}

A crucial assumption underlying the mSUGRA framework 
is that the soft supersymmetry
breaking gaugino masses, scalar masses and trilinear ($A$) terms 
are universal at the unification scale.
This is, however, only an assumption--- guided by phenomenological
considerations such as the non-observation of neutral currents--- about the 
symmetries of physics at high scales. Without such a symmetry, 
quantum corrections to soft SUSY breaking terms can break their
degeneracy\cite{bagger} and universality fails. 
In fact, neither gauginos nor scalars
are required to have universal masses in supergravity models\cite{early}.
Indeed, there are many scenarios with {\it non-universal}
soft supersymmetry breaking terms at the $GUT$ scale. These include:
\begin{itemize}
\item models with universality at the Planck or string scales. RGE evolution 
between $M_{Planck}$ or $M_{String}$ and $M_{GUT}\simeq 2\times 10^{16}$ GeV
then leads to non-universality\cite{pp}.
\item Various superstring models can lead to non-universality in soft SUSY
breaking terms\cite{ibanez,cdg}.
\item If hidden sector fields whose vevs break SUSY 
live in higher dimensional GUT representations,
then specific patterns of non-universality can be achieved\cite{snowmass,nonu-abct}.
\item Large Yukawa couplings can lead to non-universality of mainly third
generation scalar masses\cite{hall}.
\end{itemize}
In this section, we investigate the consequences of the fourth item above
for SUSY searches at the Fermilab Tevatron collider and its upgrade options.

If one views the mSUGRA model in the context of $SU(5)$ grand
unification, then it can be assumed that SUSY breaking occurs in the
hidden sector via the development of a vacuum expectation value (vev) by
the auxiliary component of an $SU(5)$ singlet superfield
$\Phat$. However, $SU(5)$ SUGRA grand unified (GUT) models can also be
constructed\cite{snowmass,nonu-abct} in which SUSY breaking occurs via
an $F$-term that is {\it not} an $SU(5)$ singlet.  In this class of models,
gaugino masses are generated by a chiral superfield $\Phat$ that appears
linearly in the gauge kinetic function, and whose auxiliary $F$
component acquires an intermediate scale vev:
\begin{equation}
{\cal L}\sim \int d^2\theta \hat{W}^a\hat{W}^b
{\Phat_{ab}\over M_{\rm Planck}} + h.c.
\sim  {\langle F_{\Phi} \rangle_{ab}\over M_{\rm Planck}}
\lambda^a\lambda^b\, +\ldots .
\end{equation}
%where the $\lambda^{a,b}$ are the gaugino fields.

$F_{\Phi}$ belongs to an $SU(5)$ representation which
appears in the symmetric product of two adjoints:
\begin{equation}
({\bf 24}{\bf \times}
 {\bf 24})_{\rm symmetric}={\bf 1}\oplus {\bf 24} \oplus {\bf 75}
 \oplus {\bf 200}\,,
\label{irrreps}
\end{equation}
where only $\bf 1$ yields universal masses.
Only the component of $F_{\Phi}$ that is `neutral' with respect to
the SM gauge group should acquire a vev,
$\langle F_{\Phi} \rangle_{ab}=c_a\delta_{ab}$, with $c_a$
then determining the relative magnitude of
the gauginos masses at $M_{GUT}$.
The relations amongst the various GUT scale gaugino masses have been
worked out in Ref. \cite{snowmass}.
The relative $GUT$ scale
$SU(3)$, $SU(2)$ and $U(1)$ gaugino
masses $M_3$, $M_2$ and $M_1$ are listed in Table~\ref{masses}
along with the approximate masses after RGE evolution to $Q\sim M_Z$.
In principle, an arbitrary linear combination of the above
irreducible representations is also allowed. Here, we will
consider for simplicity the implications of each irreducible representation
separately.

The event generator ISAJET
7.37\cite{nonu-isajet} has been upgraded to accommodate SUGRA models 
with various non-universal
soft SUSY breaking terms. In this study, we use it to simulate models with
non-universal gaugino mass parameters at the scale $M_X$ assuming
universality of other parameters.
The model parameter space thus
corresponds to
\begin{equation}
m_0,\ M_3^0,\ A_0,\ \tan\beta\ {\rm and} \mathop{\rm sgn}(\mu),
\end{equation}
where $M_i^0$ is the $SU(i)$ gaugino mass at scale $Q=M_{GUT}$.  $M_2^0$
and $M_1^0$ can then be obtained in terms of $M_3^0$ as shown in
Table \ref{masses}.

\begin{table}
\caption{Relative gaugino masses at $M_{GUT}$ and $M_Z$
in  the four possible $F_{\Phi}$ irreducible representations.}
\begin{small}
\centering\leavevmode
\begin{tabular}{|c|ccc|ccc|}
\hline
\ & \multicolumn{3}{c|} {$M_{GUT}$} & \multicolumn{3}{c|}{$M_Z$} \cr
$F_{\Phi}$
& $M_3$ & $M_2$ & $M_1$
& $M_3$ & $M_2$ & $M_1$ \cr
\hline
${\bf 1}$   & $1$ &$\;\; 1$  &$\;\;1$   & $\sim \;6$ & $\sim \;\;2$ &
$\sim \;\;1$ \cr
${\bf 24}$  & $2$ &$-3$      & $-1$  & $\sim 12$ & $\sim -6$ &
$\sim -1$ \cr
 ${\bf 75}$  & $1$ & $\;\;3$  &$-5$      & $\sim \;6$ & $\sim \;\;6$ &
$\sim -5$ \cr
${\bf 200}$ & $1$ & $\;\; 2$ & $\;10$   & $\sim \;6$ & $\sim \;\;4$ &
$\sim \;10$ \cr
\hline
\end{tabular}
\end{small}
\smallskip
\label{masses}
\end{table}
%\newpage

\begin{figure}[htb]
\centering\leavevmode
\psfig{figure=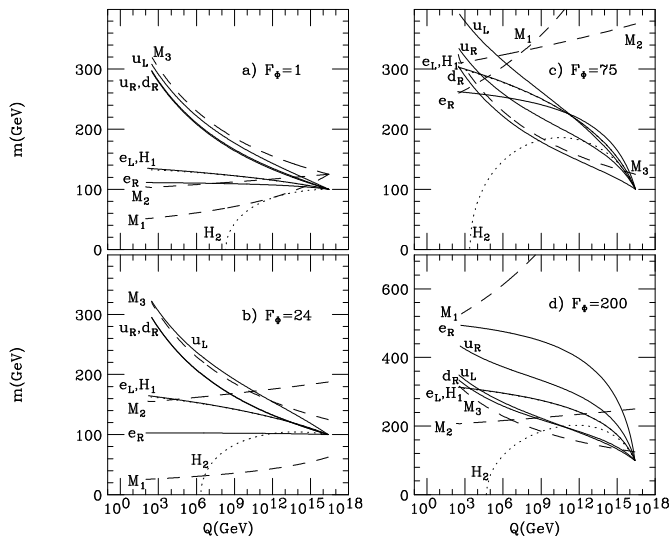,width=8.8cm}
\caption{
A plot of the evolution of soft SUSY breaking parameters versus
renormalization scale $Q$ from $M_{GUT}$ to $M_{weak}$ for SUGRA model
parameters $m_0=100$ GeV, $M_3^0=125$ GeV, $A_0=0$, $\tan\beta =5$
and $\mu >0$, for the {\it a}) $F_\Phi \sim{\bf 1}$,
{\it b}) $F_\Phi\sim{\bf 24}$, {\it c}) $F_\Phi\sim{\bf 75}$ and
{\it d}) $F_\Phi\sim{\bf 200}$ models.
}
\label{nonu-FIG1}
\end{figure}

We illustrate the evolution of the magnitude of
soft SUSY breaking masses versus scale $Q$
in Fig. \ref{nonu-FIG1} for the four model choices {\it a}) $F_\Phi\sim{\bf 1}$, {\it b}) $F_\Phi\sim {\bf 24}$, {\it c}) $F_\Phi\sim {\bf 75}$ and
{\it d}) $F_\Phi\sim {\bf 200}$.
We take $m_0=100$ GeV, $M_3^0=125$ GeV, $A_0=0$, $\tan\beta =5$ and
$\mu >0$. We take $m_t=175$ GeV.

The gaugino masses are denoted by dashed lines, while Higgs masses are
denoted by dotted lines and squark and slepton masses are denoted by
solid lines. For the usual mSUGRA case illustrated in
Fig. \ref{nonu-FIG1}{\it a}, the gaugino masses evolve from a common GUT
scale value. For the $F_\Phi \sim {\bf 24}$ model in frame {\it b}), the
splitting in GUT scale gaugino masses shown in Table~\ref{masses}
leads to a large mass gap between
$M_1$ and $M_2$ at the weak scale, and also a large mass gap between
left and right sfermions. In case {\it c}) for $F_\Phi \sim
{\bf 75}$, the large GUT scale splitting of gaugino masses leads to near
gaugino mass degeneracy at the weak scale, and also similar masses for
both squarks and sleptons. Finally, for case {\it d}) with
$F_\Phi\sim {\bf 200}$, the large GUT scale splitting leads to
$M_2,M_3 <M_1$ at the weak scale.
In addition, the large GUT scale values of $M_1$ and $M_2$ cause the weak
scale slepton masses to evolve to relatively high masses compared to the
$F_\Phi \sim {\bf 1}$ and ${\bf 24}$ models, so that
left sfermions are lighter than
right sfermions; this is in contrast to usual expectations from models
with universal gaugino masses. Note also that the different squarks can be 
significantly split. This has a considerable impact on gluino decays
when $m_{\tg}$ falls in between the masses of the different types of
squarks.

\subsection{Implications for Tevatron SUSY searches}

\begin{figure}[htb]
\centering\leavevmode
\psfig{figure=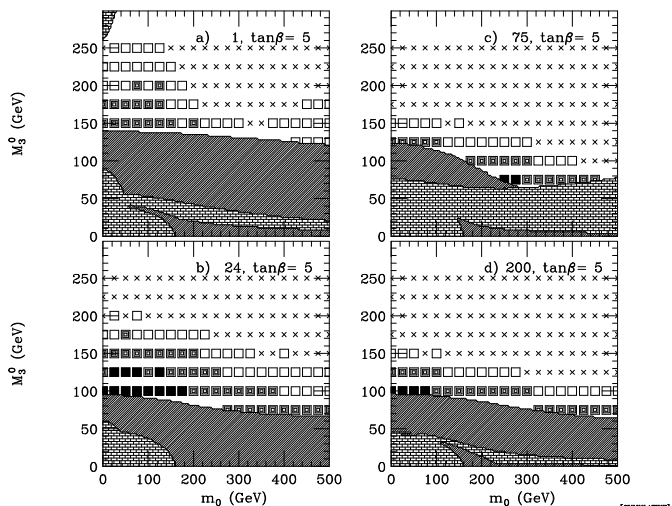,width=8.8cm}  % was 12
\caption{
A plot of parameter space points accessible to Fermilab Tevatron
collider experiments with integrated luminosity 0.1 fb$^{-1}$ (black squares),
2 fb$^{-1}$ (gray squares) and 25 fb$^{-1}$ (white squares) via
any of the SUSY search channels listed in the text.
Points are plotted in the
$m_0\ vs.\ M_3^0$ plane for
SUGRA model parameters $A_0=0$, $\tan\beta =5$ and $\mu >0$
for the {\it a}) $F_\Phi \sim {\bf 1}$,
{\it b}) $F_\Phi\sim {\bf 24}$, {\it c}) $F_\Phi\sim{\bf 75}$ and
{\it d}) $F_\Phi\sim{\bf 200}$ models.
}
\label{nonu-FIG19}
\end{figure}

We evaluate model parameter space points with regard to discovery of SUSY 
by the Tevatron collider. We investigated observability of SUSY in 
the jetty or clean channels labelled by lepton multiplicity as 
$J0L$, $J1L$, $JOS$,
$JSS$, $J3L$, $COS$, $C3L$ channels in Ref. \cite{nonu-abct}. 
We also examined the possibility that OS leptons come from a real $Z$ decay. 
We denote 
parameter space points seeable with $0.1$, $2$ or 25 fb$^{-1}$ of integrated
luminosity at a 5$\sigma$ level above SM backgrounds with a black square,
gray square or white square, respectively. Points not accessible in any of 
the above channels are denoted by a $\times$ symbol. Bricked regions are 
excluded theoretically, and shaded regions are excluded by SUSY searches at 
LEP2. Our main result is shown in Fig. \ref{nonu-FIG19}, where a parameter space
point is denoted by appropriate box if a SUSY signal is observable in {\it any}
of the above signal channels. We take $A_0=0$, $\tan\beta =5$ and $\mu >0$.
Frame {\it a}) shows the mSUGRA prediction for comparison purposes.
Almost all the accessible points shown in frame {\it a}) are seeable in the $C3L$
channel, although a few are also seeable in the $J0L$ channel.
These form the most important search channels for SUSY at the Tevatron
for the mSUGRA model.

In frame {\it b}), the results for the $F_\Phi\sim {\bf 24}$ model are shown. 
The large mass gap between the $\tz_2$ and $\tz_1$ leads to a large 
rate for hard isolated lepton production in cascade decay events.
Hence, for many of the parameter space points shown, a SUSY signal 
should be observable in several channels at once! In addition, the 
large mass gap between the $\tz_2$ and $\tz_1$ leads to a large rate for
$\tz_2\to\tz_1 Z$ decays. Thus, in this case one can search for SUSY in the
jets plus $Z\to \ell\bar{\ell}+\eslt$ channel\cite{btwz}, denoted JZ. 
In fact, for 
several of the parameter space points shown, a SUSY signal can be found
{\it only in this channel} at the Tevatron! 
%Since this channel isn't usually
%included in the list of experimental search channels, it is important 
%to emphasize it here. 
%The $JZ$ channel could well be indicative that nature 
%has chosen an $F_\Phi\sim {\bf 24}$ model 
%which leads to a large mass gap between 
%$\tz_2$ and $\tz_1$.
An observable signal in the $JZ$ channel could help zero in on the pattern of
SUSY breaking.

For the $F_\Phi\sim{\bf 75}$ and ${\bf 200}$ models shown in 
frames {\it c}) and {\it d}),
there is just a small mass gap between the lightest charginos and neutralinos,
which contain substantial higgsino components ($|\mu |$ is smaller than
or comparable to the gaugino masses). 
Hence, hard isolated leptons from
$\tw_1$ and $\tz_2$ decay are rare. The main channel for discovery in these
models is the $J0L$ channel, and the reach in $M_3^0$ is much more limited than
in the $F_\Phi\sim{\bf 1}$ or ${\bf 24}$ models.

%\section{Summary}
%
%If supersymmetry is broken by hidden sector fields transforming in higher
%dimensional representations of $SU(5)$, then non-universal gaugino
%masses can be generated with specific relationships for their relative
%magnitudes. In contrast to the mSUGRA case with exact GUT scale universality,
%the $F_\Phi\sim{\bf 24}$ model leads to Tevatron collider events with many
%hard isolated leptons. If this SUSY model is discovered at the Tevatron,
%then a signal will likely be visible in many channels at once. Of particular
%note is the {\it new} $JZ$ channel: for some parameter space points, SUSY 
%can only be discovered at the Tevatron in this channel. 
%The $F_\Phi\sim{\bf 75}$ and ${\bf 200}$ 
%models lead to near degeneracy amongst the
%lighter charginos and neutralinos so that few hard isolated leptons can get 
%produced. The Tevatron reach for SUSY is more limited in these models than
%in the mSUGRA or $F_\Phi\sim{\bf  24}$ 
%case. The dominant discovery channel is the
%$J0L$ channel, and only a few parameter space points can be observed via
%discovery channels involving hard isolated leptons.

\vspace*{-.25in}

%%%%%%%%%%%%%%%%%%%%%%%%%%%%%%
%  Bibliography
%%%%%%%%%%%%%%%%%%%%%%%%%%%%%%

%% file: Arnowitt-trilep/trilep-new.tex
%******************DEFINITIONS FOR FIGURES******************
% These should work on Mac's and Unix machines.  You need epsf.def.
% \input epsf.tex
\def\DESepsf(#1 width #2){\epsfxsize=#2 \epsfbox{#1}}
% Null macro in case the ones above don't work.
%\def \DESepsf(#1 width #2){\bf #1  here: just uncomment the macro.}
%******************END DEFINITIONS*************************

\section{Modified Trilepton Signal in Non-Universal Models}

At the Tevatron, the production cross section of  the  second
lightest neutralino and the lightest chargino ($\chi_1^{\pm},\chi^0_2$) is the largest among all the
supersymmetric particles. As a result, the final decay products of this
production mode is very crucial for the detection of SUSY. 

 In most of the allowed regions of the parameter space 
 (allowed by LEP II and the Tevatron
so far), the squarks are heavier than the sleptons and the gauginos. If the
squark masses are heavier and the slepton masses are lighter than these
gauginos, the chargino pair $\chi_1^{\pm}$ dominantly gives rise to final 
states of 2$l$+${\rlap/E}_T$ and 2$\tau$  +${\rlap/E}_T$  ($l$ is $e$,
$\mu$). Similarly the $\chi^{\pm}_1,\chi^0_2$ decay to final states consisting of  3 $\tau$, 3 $l$,
$l\tau\tau$,
$\tau ll$ and ${\rlap/E}_T$. (In the $\tau ll$ mode, the $\tau$ comes from the 
$\chi_1^{\pm}$ and the $ll$ comes from the  $\chi^0_2$.)
We will concentrate here on
the $\chi_1^{\pm},\chi^0_2$ productions. 

Detection 
efficiencies are not the same 
for all charged leptons. 
The $\tau$ detection efficiency for the Tevatron has not been  specified yet. The
$\tau$ can be detected hadronically
(`thin jet') or leptonically. If we depend on the leptonic modes (since we are
working in a machine with lots of jets), then the effective leptonic 
 cross section of the
final state with multi $\tau$'s becomes very small (due to the small leptonic branching ratio
of the $\tau$). So far the detection efficiency for the 3$l$ mode is the best.

The trilepton channel at the Tevatron has been analysed by many groups in various 
different supersymmetric models \cite{w1z2,gmsb-arnotri,arno-bk1}. These theoretical scenarios
range from the supergravity motivated models (SUGRA) to Gauge Mediated SUSY
breaking Models (GMSB). In absence of experimental evidence as well as a full 
understanding of the dynamics of SUSY breaking, no unique model  has as yet
emerged. Models where the different SUSY masses are related, are found to be more
compelling due to their predictivity. Among these, the SUGRA models with the
constraint that  all the scalar masses are same (and gaugino masses also the
same) at the grand unified theory scale $M_G$ (universal boundary conditions) is the most  popular
one. Recently, it has been shown that in this model, among all the 3 lepton plus ${\rlap/E}_T$
modes, only the   
3$\tau$+${\rlap/E}_T$ production cross section is large in
most of the parameter space where the sleptons are produced on shell \cite{arno-bk1}.
 For these results a universal soft breaking at  the GUT scale was  assumed. 
 Among the other modes, only the $\tau ll$ mode  becomes
comparable (also detectable) but only  for the very small region of $\tan\beta$ $\le 5$. 
The branching ratio (BR) in this mode
 decreases rapidly and at $\tan\beta$=10, for e.g. 
 $m_0=100$ GeV and $m_{1/2}=200$ GeV, it
  becomes about $1\over {10}$
of the BR of the 3$\tau$ mode (The  BR of the 3$\tau$ mode remains almost the same). 
The $\tau ll$ cross section becomes 21.2 fb. The situation worsens by decreasing $m_0$. For
example, at $m_0=30$ GeV (keeping the other parameters same) the cross section
for the $\tau l l$ mode 
becomes 11 fb (in this region the BR of the $\tau\tau l$ mode becomes large).   
Even when the sleptons become off shell, the branching ratio to 
the leptonic modes involving multi $\tau$'s are large.  The $\tau $ domination in
the signal (on shell or off shell case) is usually expected when $\tan\beta$ is large.  
But the domination seems to
persist even in the region of low $\tan\beta$.

The reason for this misfortune depends primarily on two factors. The lighter  stau mass is
lighter than the selectron mass. In the on shell case, the decay width
depends
on  $(\Delta m^2)^2$(where $\Delta m^2$ is the mass$^2$ difference between the
gaugino and the slepton), which  is larger for the modes involving $\tau$. So
even a very
small mass difference will be observed in the BR of the leptonic modes.  The other factor is
that the $\chi^0_2$
 is primarily a wino which has coupling to the left sleptons
only. But the left-handed selectrons are  heavier than both the right-handed
selectrons and  the $\chi^0_2$. On the other hand,  the lighter
$\tilde\tau_1$ is a mixture of $\tilde\tau_L$ and $\tilde\tau_R$ due to the 
large left right mixing $m_{\tau}\mu
\tan\beta$.
(We will assume $A=0$ at the GUT scale  throughout the analysis, but a non-zero value will not
change the conclusion). Consequently, the $\chi^0_2$
primarily would decay into the lighter
$\tilde\tau_1$ and a $\tau$.  Among these two
 factors, the first one has a larger impact. In addition the $\chi^{\pm}_1$ is a  mixture of charged wino and  Higgsino  giving rise to a dominating  $\tilde\tau_1\nu$ final state when lighter
stau is on shell. [However when the sneutrino mode contributes (either on shell or off shell)
 to the decay, the BR to the final
state involving $l\nu$ increases.]

We find that the situation changes in the most general case, i.e.\ where the
boundary condition at the GUT scale is not assumed to be universal. Non-universality at
the boundary can appear naturally and so it is reasonable to consider this
possibility. A general non-flat  Kahler metric (where the SUSY breaking field is coupled to
the observable fields with different couplings)  can induce non-universalities in
the scalar masses.   Since the Higgs sector is weakly constrained by the requirement
of FCNC suppression and  the third
generation is weakly coupled to the FCNC process, one  may assume that the third generation squark, slepton and
the Higgs masses  are non-universal at the GUT scale, while the first and 
second generation scalar masses and the gaugino masses are assumed to be
universal. (Non universalities in the gaugino sector can be also induced, but 
are small in most models, and so we assume these masses to be universal). 
Scalar mass non universalities can be also  generated from the running of the
RGE's from the Planck scale or string scale  to the GUT scale. In this case, 
 because of the quark-lepton
unification, not only the third generation squark masses, 
but the third generation slepton masses will also be different from the
other generation masses \cite{hb3}.  Finally, we mention that  non-universalities
can be generated from the  so-called D-terms arising from the rank reduction of
the groups which embed the SM as a subgroup, when the GUT group has  rank
higher than the SM.

Let us examine the parametrization of the  non-universalities. The Higgs soft
breaking masses are given by $ m_{H_1}^2=m_0^2(1+\delta_1)$;
$m_{H_2}^2=m_0^2(1+\delta_2)$.
The third generation fermion soft breaking masses are as follows:
$m_{q_L}^2=m_0^2(1+\delta_3)$; $m_{u_R}^2=m_0^2(1+\delta_4)$
$m_{e_R}^2=m_0^2(1+\delta_5)$;
$m_{d_R}^2=m_0^2(1+\delta_6)$;$m_{l_L}^2=m_0^2(1+\delta_{7})$, where
$q_L=(t_L,b_L)\,{\rm and}\, u_R=t_R$.
 The $\delta_i$ exhibit the amount of non-universality. If we use
a unifying group, where the fields belong to some represantaion of that group the
$\delta_i$ develop relations among themselves. For example, in the case of
the GUT group SU(5) the matter fields are embedded in $\bar 5$ and $10$
representations. The $\delta_i$ in the previous expressions have the following
relations:
\begin{equation}
\delta_3=\delta_4=\delta_5=\delta_{10};~~\delta_6=\delta_7=\delta_{\bar5}.
\end{equation} Any GUT group which has an SU(5) with the matter
fields in the
$10$ and $\bar 5$ representations will have the above pattern of
non-universalities. In the case of  $SO(10)$, if we demand a direct breaking  into
the SM and keep the $5+\bar 5$ Higgs in the same 10 of $SO(10)$, we get an
additional constraint, $\delta_5=\delta_2-\delta_1$. In this note we will use
$\delta_{1}$,$\delta_{2}$,
$\delta_{\bar5}$ and
$\delta_{10}$ to represent the non-universalities.  One need not be restricted in
this choice.

Using non-universality, we find that the BR of the
$\tau ll$ mode and the 3$l$ mode can become large and detectable for a wide range of
 parameter space. The new effects can
reduce the selectron masses and raise the lighter stau masses (and can make it even
heavier than the selectron masses). The BR to the selectron mode is then no longer suppressed. The
magnitude of $\mu$ can also be decreased, which increases  the lighter stau mass
compared to the lighter selectron mass. The nature of the $\chi^0_2$ can also change
with the change in the size of $\mu$ (the wino component can decrease and the bino
component can increase). However for the reasonable values of the deviations from
universality, this change is small.

Another important point to note is that, when we use the non-universal boundary
conditions one has the additional term 
$S\equiv \alpha_1{{3 Y_i}\over {10 \pi}}\sum_i(Y_i m_i^2)$ in the RGEs
of the fields. The contribution from this term is zero in the case of
universal boundary condition. In the non-universal case, this factor can decrease
the slepton mass and $\mu$ as well. Hence the $\tilde\tau_1$ mass does not decrease
as much as the selectron mass decreases. For example, in the universal case
for $m_0=100$ GeV, $tan\beta=10$ and  $m_{1/2}=200$ GeV, $\tilde
e_R,\,\tilde\tau_1,\,\chi^0_2 ,\,{\rm and }\, \chi_1^{\pm}$ masses and $\mu$ are 134,
126, 147, 146  and 284 GeV respectively. If we use non-universal boundary conditions $\delta_{1}=-0.5$
and $\delta_{2}=6$, the same
masses become 120, 115, 131, 127  and 203 GeV respectively.

\begin{figure}
\centerline{ \DESepsf(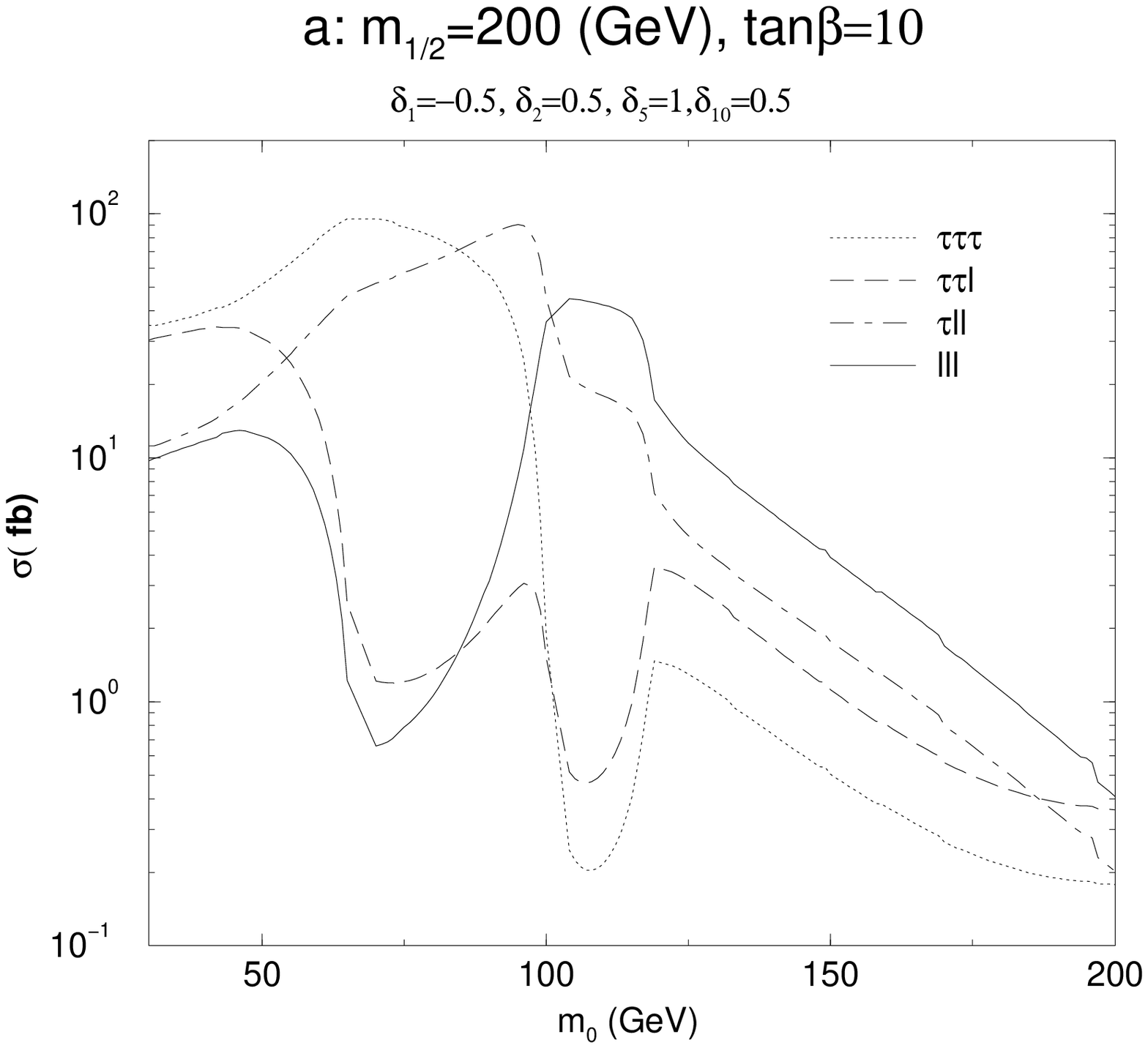 width 8 cm)
\qquad \DESepsf(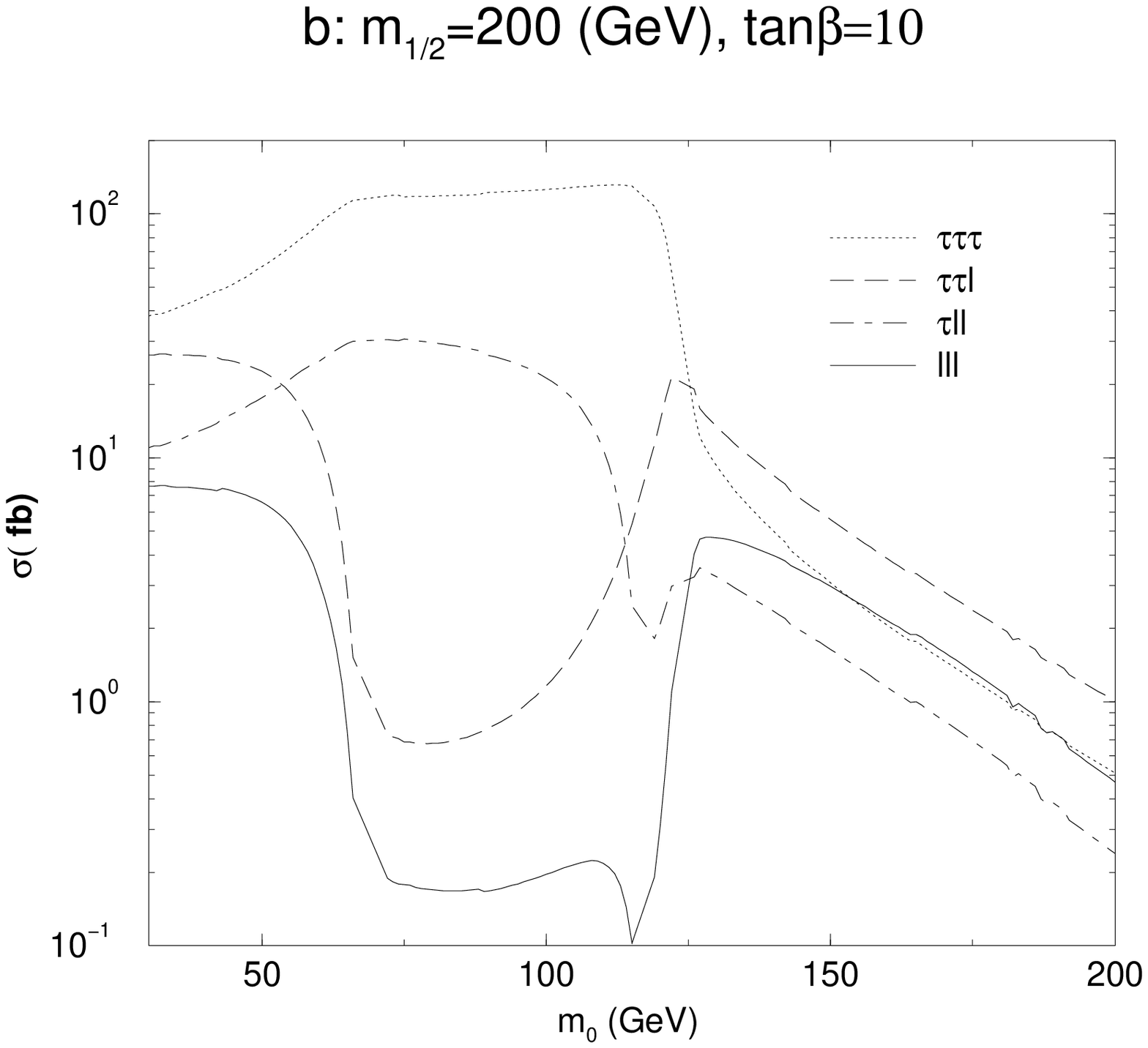 width 8 cm) }
\caption[]{The cross sections for all the leptonic modes are
shown. a)Non-universal case and b)universal case. \label{fig:arno-trilep1}}
\end{figure}

We now discuss the results.  To exhibit
the above ideas, we
have chosen a pattern of non-universality which obeys the $SU(5)$ group structure. 
(One need not be restricted
in this choice.) In the Fig.~\ref{fig:arno-trilep1}a, we have used $\delta_{1}=-0.5$, $\delta_{2}=0.5$,
 $\delta_{\bar 5}=1$ and $\delta _{{10}}=0.5$. 
These values of the $\delta_i$  also allows the simplest SO(10) 
breaking pattern at $M_G$ ($SO(10)\rightarrow SM$).
 
In Fig.~\ref{fig:arno-trilep1}a we plot  the production cross sections of the leptonic modes as function of $m_0$. The other
parameters are taken to be $\tan\beta=10$,
$m_{1/2}=200$ GeV and $\mu>0$. The sign convention we adopt is the same  as in
ref\cite{hk6}. In Fig.~\ref{fig:arno-trilep1}b we plot the same parameter space for the universal
boundary condition.
We observe the following:
\begin{itemize}
\item For $30\stackrel{<}{\sim} m_0 \stackrel{<}{\sim}70$ GeV, the 3$\tau$ and $\tau\tau l$ mode dominate 
initially in the non-universal case. But $\tau ll$ is not far behind and as $m_0$ increases the BR in
this mode increases. The reason for $\tau$ domination in this region (the $\tilde e_R$ and
$\tilde\tau_1$ masses
are almost same)  is
due to the wino nature of $\chi^0_2$. In this region the chargino can decay into
$l\tilde\nu_l$. As the sneutrino mass goes offshell toward the 
end of the region, the branching ratio of 3$l$ and
$\tau\tau l$ are reduced. In the universal case (Fig.~\ref{fig:arno-trilep1}b), the 3$\tau$ mode
dominates with the $\tau ll$ mode and the $\tau\tau l$ modes coming next.

\item For $70\stackrel{<}{\sim} m_0\stackrel{<}{\sim} 100$ GeV, the $\tau ll$ and 3$\tau$ mode dominate for the
non-universal case, and the
3$l$ mode starts becoming significant. The branching ratio of $\chi^0_2$ into $e$'s
 and $\mu$'s increases. Towards the end of the region, the sleptons (first $\tilde\tau_1$) becomes
 off shell. In the universal case the 3$\tau$ mode is dominant in this region.
 
\item For $100\stackrel{<}{\sim}  m_0\stackrel{<}{\sim}
 200$ GeV, the sleptons are mostly offshell. The 3$l$ is the
dominant decay mode in the non-universal case and next to that is the $\tau ll$ mode. The production cross
section decreases as we increase $m_0$. At the end of this region the 3$\tau$ and 
$\tau\tau l$ modes increase again due to the off shell Higgs contribution.
In the universal case the 3$\tau$ mode dominates initially. For $m_0\ge 130$,
the 3$l$  mode becomes equal to the 3$\tau$ mode, but these cross sections are
very reduced by that time and the $\tau\tau l$ mode dominates here. 
\end{itemize}

Using a dilepton trigger, if we use 5  events in RUN II as a bench mark for a SUSY
signal (corresponding to the cross section of 25 fb with 10$\%$ 
acceptance rate \cite{cdf-arnotri}), the inclusive $ll$+${\rlap/E}_T$ production allow us to scan 
$50\stackrel{<}{\sim}
m_0\stackrel{<}{\sim}130$. In the universal case we do not get 
5 events in this mode for any value of $m_0$. In the case of $\tau$l trigger, 
(assuming the same acceptance rate) the inclusive $\tau l$+${\rlap/E}_T$ production all
ow us to scan $30\stackrel{<}{\sim}
m_0\stackrel{<}{\sim}100$. 
Here again, in the universal case we do not find any $m_0$ value which gives
rise to 5 events.

\begin{figure}
\centerline{ \DESepsf(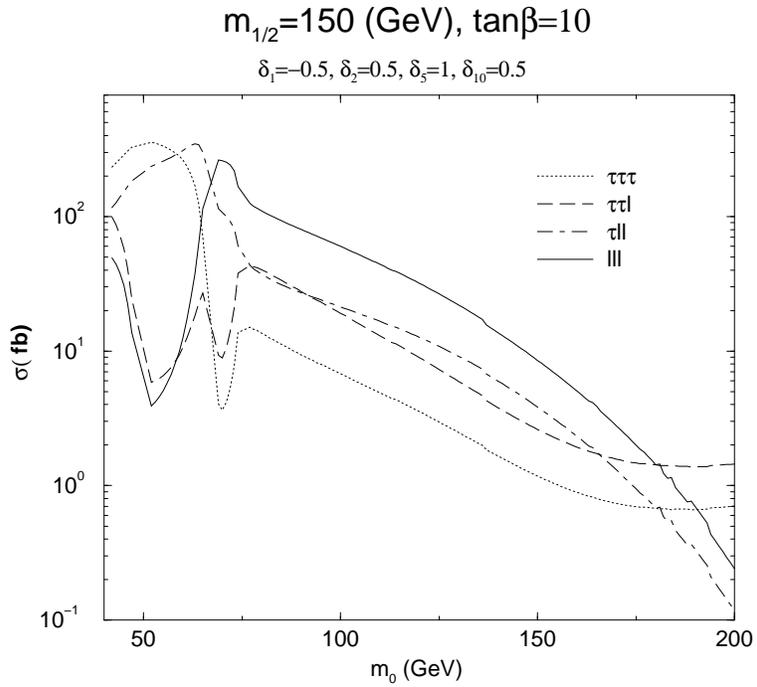 width 10 cm) }
\caption[]{The cross sections for all the leptonic modes are
shown in non-universal case. \label{fig:arno-trilep2}}
\end{figure}

If we reduce $m_{1/2}$, the BRs of the $\tau ll$ and the 3$l$ mode increase 
because of the increase in phase space. In Fig.~\ref{fig:arno-trilep2} we plot
the production cross sections of the leptonic modes as function of $m_0$ for
$m_{1/2}=150$ GeV and $\mu>0$.  The opposite effect would be observed
if $m_{1/2}$ is increased.

In the case of $\mu<0$, the
$\chi^{\pm}_1\chi^0_2$ production cross section are smaller due to the increased
chargino and  neutralino masses.  We find that this choice of the $\mu$ sign gives rise to
a scenario  analogous to what one could get by for $\mu>0$ increasing the value of
$m_{1/2}$. The contribution of the on-shell sneutrino for $m_0\stackrel{<}{\sim}
 80$ GeV ($m_{1/2}=200$ GeV)
enhances the  BR of the chargino into leptons. The 3$\tau$ channel BR
shrinks. 

In the universal case, 
 the trilepton final state originates mainly from the $\tau ll$ mode. Since one
of the leptons has to originate from a $\tau$, the production cross section of
the trilepton final state is mostly suppressed. On the other hand, in the  non-universal example considered here,  the trilepton signal where none of the
leptons is originating from a $\tau$, is dominant  when
$m_0\stackrel{>}{\sim}90$.  Thus the cross section for the trilepton channel is
higher.  For example, for
$m_0=$100 GeV, $m_{1/2}=$200 GeV and  $tan\beta=10$, in the universal case the
cross section is 14 fb. On the other hand, in the non-universal scenario, the
same parameters will generate $\sim$40 fb of cross section in the trilepton
channel, without any of the leptons arising from a
$\tau$. If we add the $\tau ll$ mode (with the $\tau$ decaying into a lepton)
the total  trilepton cross section becomes $\sim$ 60 fb. The main background for
the $\chi^{\pm}_1\chi^0_2$ process are WZ, WZ$^*$, , W$^*\gamma^*$
 production
 \cite {arno-mp,th,bk2}. In order to reduce the backgrounds, one can use cuts as suggested
in the
 Ref.\cite{bk2}. For example if one uses the soft cut (type A), the background becomes
1.97 fb. In Tables \ref{table:arno-trilep1} and \ref{table:arno-trilep2} we exhibit the cross sections with and without the
soft cuts for $m_{1/2}$=150 and 200 GeV and tan$\beta$=3 and 10. 
In the nonuniversal model we have chosen, the trilepton 
cross section appears to
be always bigger in the nonuniversal case compared to the universal one. The
cross section in the universal case goes down rapidly for larger values of
  $m_{1/2}$ and tan$\beta$. 
  
  Table \ref{table:arno-trilep3} shows the number of expected trilepton
  events (after soft cuts) for the universal and nonuniversal model for the
  SUSY parameters in Table \ref{table:arno-trilep1} and \ref{table:arno-trilep2}. $N_{\sigma}$ is the number of standard
  deviations that distinguish between the models. For RUN II (4 $fb^{-1}$) one
  can distinguish for this model, nonuniversality at better than 97$\%$ C.L.,
  while for the total RUN II/III (24 $fb^{-1}$) one would be statistically at
  $>5\sigma$, the discovery level. 
  In SUGRA models, recent LEP189 data
 already implies $m_{1/2}>$150 GeV \cite{bk3}. While there is at present no
 purely accelerator bound on tan$\beta$, the search for the Higgs at LEPII and
 the Tevatron RUNII/III \cite{wag} will give rise to lower bound on tan$\beta$,
 and help limit the SUSY parameter space.

\begin{figure}
\centerline{ \DESepsf(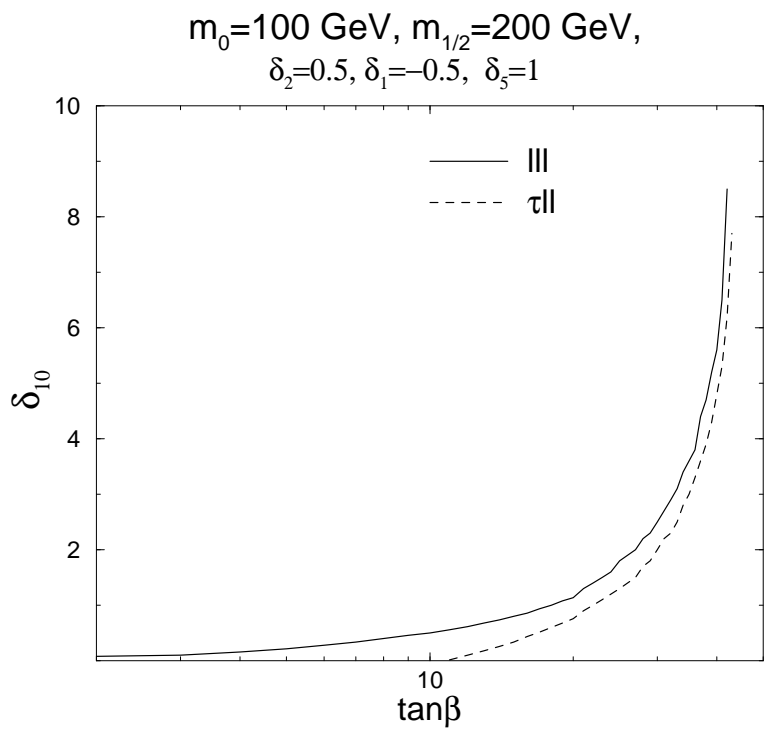 width 10 cm) }
\caption[]{The value of $\delta_{10}$ required as a fuction of $tan\beta$ so that the
production cross section for the 3$l$ and 
the $\tau ll$ modes be $35$fb.\label{fig:arno-trilep3}}
\end{figure}

In Fig.~\ref{fig:arno-trilep3} we show a parametric plot of the 3$l$ and $\tau ll$ with the production cross
sections fixed at 35 fb as functions of  $\delta_{10}$ and  $tan\beta$. Here we have used
$\delta_2=0.5$, $\delta_1=-0.5$, $\delta_5$=1, $m_0=100$ GeV,
$m_{1/2}=200$ GeV and $\mu>0$. 
Both $\delta _5$ and $\delta _{10}$ help to raise the $\tilde\tau_1$ mass. Since
$\delta_{10}$ affects the right stau mass, it has a larger impact in increasing the
$\tilde\tau_1$ mass and thereby decreasing the branching ratio of the
$\tau\tau\ l$ and the 3$\tau$ modes. The selectron mass is reduced by a small amount (a few GeV) through
the S term.  Hence an increment of $\delta_{10}$ will raise the BR of the 3$l$
and $\tau ll$ mode. It is evident from the figure that as 
$\tan\beta$ increases a larger
value of $\delta_{10}$ is needed. This is due to the fact that the $\tilde\tau_1$ mass
decreases as $\tan\beta$ increases. However, one sees that
for $\delta_{10}\leq 2$, the 3$l$ mode is large even for
$tan\beta\simeq 25$.  

To conclude, we have looked into the final states of the chargino-second
lightest neutralino production at Tevatron for $\sqrt s=2$ TeV. We have 
found that in the unified models with non universal boundary
conditions, the 3$l$ and $\tau l l$ final states can dominate over the
3$\tau$ or the $\tau\tau l$ modes for low and intermediate values of
$\tan\beta$. But in the models with universal boundary
conditions the 3$\tau$ mode dominates even for low values of $\tan\beta$.
Since the 3$l$ mode has by far the best detection efficiency, these
non universal boundary conditions
may be tested in RUN II.

%\Let us discuss the effects of these $\delta$ s. In the case of universal boundary
%\condition, $\tilde\tau_1$ and $\tilde e_R$ masses are 126.03 GeV and 133.87 GeV
%\respectively. $\mu$ is 284.37 GeV.  Let us choose  $\delta_2=0.5$,
%\$\delta_1=-0.5$, $\delta_{10}=0.5$ and $\delta_5=\delta_2-\delta_1$. $\delta_{10}$
%\raises the lighter stau mass and the reduction in the $\tilde e_R $ occurs through
%\S term.  
%\$\tilde\tau_1=$ mass becomes 143.83 GeV and $\tilde e_R$ becomes 131.87 GeV. The
%\large stau mass decreases the BR of $\tau\tau\tau$ mode.
%\If we assume $\delta_{10}$ to be 0, the stau mass decreases to 126.38 GeV,
%\$\mu$ increases.

\begin{table}[h]
\caption[]{The total trilepton cross section in the 
nonuniversal case for $m_0$=100 GeV.\label{table:arno-trilep1}}
\centering\leavevmode
\begin{tabular}{|c|c|c|c|c|}\hline tan$\beta$&$m_{1/2}$ (GeV)&without cut (fb)&
with soft cut (fb)\\\hline
3& 150& 139 & 31 \\\cline{2-4}
 & 200 &67&16\\\hline
10&150 & 70&16\\\cline{2-4}
&200&60&11\\\hline
\end{tabular}
\end{table}

\begin{table}[h]  
\caption[]{The total trilepton cross section in the universal case $m_0$=100 GeV. \label{table:arno-trilep2}}
\centering\leavevmode
\begin{tabular}{|c|c|c|c|c|c|}\hline tan$\beta$&$m_{1/2}$ (GeV)&without cut (fb)&
with soft cut (fb)\\\hline
3& 150& 104 & 23\\\cline{2-4}
 & 200 &41&9\\\hline
10&150 & 40&9\\\cline{2-4}
&200&14&2.6\\\hline
\end{tabular}
\end{table}

\begin{table}[h]
\caption[]{Numbers of events after soft cuts for universal case 
(N$_{univ}$) and nonuniversal case (N$_{nonuniv}$) for m$_0$=100 GeV.
(Background is 2 events/fb.) N$_{\sigma}$ is the number of std.\ between
universal and non-universal signal.\label{table:arno-trilep3}}
\centering\leavevmode
\begin{tabular}{|c|c|c|c c|c c|c c|c|c|}\hline &&
&$N_i$&&$N_i/\sqrt{B_i}$&&${\sigma_i}=\sqrt{N_i}$&&
&\\ L($fb^{-1}$)&tan$\beta$&$m_{1/2}$ (GeV)
&{univ}&{nonuniv}&{univ}&{nonuniv}&{univ}&{nonuniv}&${\sigma}=
\sqrt{\sigma_{univ}^2+\sigma_{non}^2}$&
$N_{\sigma}$\\\hline
4&3& 150& 92 & 124 &32.5 &43.8&9.6&11.1&14.7&2.2\\
(RUN II)& & 200& 36 & 64 &12.6&22.6&6.0&8.0&10.0&2.8\\\cline{2-11}
&10 & 150 &36&64&12.6&22.6&6.0&8.0&10.0&2.8\\& & 200
&10.4&44&3.7&15.6&3.2&6.6&7.4&4.6\\\hline
24&3& 150& 552 & 744&79.7&107.4&23.5&27.3&36.0&5.3\\
(RUN II/III)& & 200& 216 & 384&30.9&55.4&14.7&19.6&24.5&6.9\\\cline{2-11}
&10 & 150 &216 & 384&30.9&55.4&14.7&19.6&24.5&6.9\\& & 200
&62&264&9.1&38.2&7.9&16.2&18.1&11.2\\\hline
\end{tabular}
\end{table}

\clearpage

%% file: Tata-4E/tata-4E.tex
%%%%%%%%%%%%%%%%%%%%%% REVTEX FILE %%%%%%%%%%%%%%%%%%%%%%%%%%%%%%%%%%%%%%%%%%%
%\documentstyle[preprint,eqsecnum,aps,epsf]{revtex}	% preprint format
%\documentstyle[eqsecnum,aps]{revtex}			% PR format
%\newif\iftightenlines\tightenlinesfalse
%\tightenlines\tightenlinestrue

%\begin{document}
%

%%%%%%%%%%%%%%%%%%%%%%%%%%%%%%%%%%%%%%%%%%%%%%%%%%%%%%%%%%%%%%%%%%%%%%%%%%%%%%%
\def\pT{p_T^{\phantom{7}}}
\def\MW{M_W^{\phantom{7}}}
\def\ET{E_T^{\phantom{7}}}
\def\bh{\bar h}
\def\lm{\,{\rm lm}}
\def\lo{\lambda_1}                                              
\def\lt{\lambda_2}
\def\pslt{p\llap/_T}
\def\eslt{E\llap/_T}
\def\to{\rightarrow}
\def\Re{{\cal R \mskip-4mu \lower.1ex \hbox{\it e}}\,}
\def\Im{{\cal I \mskip-5mu \lower.1ex \hbox{\it m}}\,}
\def\SU{SU(2)$\times$U(1)$_Y$}
\def\te{\tilde e}
\def\tmu{\tilde \mu}
\def\tl{\tilde l}
\def\ttau{\tilde \tau}
\def\tg{\tilde g}
\def\tga{\tilde \gamma}
\def\tnu{\tilde\nu}
\def\tell{\tilde\ell}
\def\tq{\tilde q}
\def\tt{\tilde t}
\def\tw{\tilde \chi^{\pm}}
\def\twb{\tilde \chi^{\mp}}
\def\tz{\tilde \chi^0}
\def\cmsec{{\rm cm^{-2}s^{-1}}}
\def\sgn{\mathop{\rm sgn}}
\hyphenation{mssm}
\def\ds{\displaystyle}
\def\ts{${\strut\atop\strut}$}
%

%\medskip
%\pacs{PACS numbers: 14.80.Ly, 13.85.Qk, 11.30.Pb}
%{\tt$\backslash$\string pacs\{\}}

%\narrowtext

%%%%%%%%%%%%%%%%%% MAIN TEXT %%%%%%%%%%%%%%%%%%%%%%%%%%%%%%%%%%%%%%%%%%%%%%%
%\newpage

\section{Can SUSY Remain Hidden from Tevatron Searches?}
\label{hidden-susy}

We have seen that it is very possible that supersymmetry might first be
discovered in experiments at the Main Injector or its luminosity
upgrades, assuming that the mSUGRA framework provides a reasonable
approximation of how supersymmetry is realized. Moreover, SUSY might
then be detectable in several channels involving $\eslt$, jets, and hard
isolated leptons. Despite the fact that most SUSY analyses are done
within specific frameworks (usually mSUGRA or the minimal gauge-mediated
SUSY breaking model), it is often believed that sparticles will readily
be discovered at the Tevatron if their production is not kinematically
suppressed. It is worth remembering, however, that even in the mSUGRA model
there may be regions of parameter space where observable signals for
SUSY are strongly suppressed even though sparticles might be produced
with significant cross sections at the Tevatron.

Our purpose here is to point out SUSY scenarios in which
sparticles are kinematically accessible, but which may easily escape the
scrutiny of Tevatron experiments. This is not because these
``pessimistic scenarios'' are especially attractive or theoretically
compelling, but only to caution the reader of the diversity
of phenomenological possibilities. In view of our current lack of
understanding of the pattern of sparticle masses and mixing angles which
largely determine the size of SUSY signals at colliders, it is
imperative that we at least be aware of how changing underlying
assumptions about high scale physics
might alter our conclusions about the capabilities of
experimental facilities.

Events with $\eslt$, which form the canonical SUSY signal, arise because
the (neutral) LSP is stable in models where $R$-parity is
conserved. Signals with leptons come from leptonic decays of charginos,
neutralinos and sleptons which might be produced directly by electroweak
processes or via cascade decays of heavier gluinos and squarks produced
by strong interaction. Thus SUSY may remain hidden in those models where
({\it i})~the LSP is unstable and decays hadronically, ({\it ii})~ the
production of sleptons is inhibited, or alternatively, these dominantly
decay via lepton number violating couplings (on which there are strong
constraints), and ({\it iii})~decays of charginos and neutralinos into
hard, isolated leptons are suppressed --- this could be either for
kinematic reasons, or because mixing and mass patterns strongly favour
decays into hadrons, taus and/or neutrinos. This also happens in the
mSUGRA framework for some ranges of parameters.

\subsection{Hiding $\eslt$}

For the sake of discussion we will take the mSUGRA model as the starting
point, and consider how departures from it might serve to hide SUSY at
the Tevatron. Small baryon number violating superpotential interactions
can cause the LSP (often the lightest neutralino) to decay into three
quarks or three antiquarks. These interactions can be small enough that
they do not affect the usual calculation of sparticle masses, sparticle
production mechanisms or decays of sparticles other than the LSP, and
yet large enough that the LSP decays without any observable displaced
vertex. 

These hadronic decays of the LSP have two important effects. First, the $\eslt$
signal is obviously reduced. Second, the presence of the additional
hadronic activity from the decays of the two LSPs in each event makes it
difficult for the leptons from usual cascade decay sources to remain
isolated and reduces the multilepton cross sections.  Tevatron signals
in such a scenario have been analysed in Ref. \cite{RPV} assuming that
the LSP only decays via $\tz_1 \to cds,\bar{c}\bar{d}\bar{s}$, and the
jets cannot be flavour-tagged.  The main injector reach in the $\eslt$
channel was limited to $m_{\tg} \leq 200$~GeV even for $m_{\tg} \sim
m_{\tq}$. The greatest reach was found in the multilepton channels. It
was shown that (if $m_{\tq} \sim m_{\tg}$) gluinos as heavy as 350~GeV
could still be probed via the $3\ell$ channel. On the other hand, for
heavy squarks, the greatest reach was found in the same sign dilepton
channel but limited to $m_{\tg} \alt 200$~GeV.  It is worth noting that
in the mSUGRA model, $m_{\tg} \alt 200$~GeV usually implies that the
chargino should be within the kinematic reach of LEP 2.

Tevatron signals for $R$-parity violating decays are discussed in this Volume.

\subsection{Hiding Leptons}

Next, we examine whether it is possible to also hide the leptonic signals.
We outline several scenarios where this might happen.

\subsubsection{A Large R-violating coupling}

A simple possibility is to imagine\cite{baer-snowmass}
that the one $R$-parity violating coupling responsible for LSP decay is
large relative to electroweak gauge couplings.  In this case, it could
be that gluinos would dominantly decay via $\tg \to cds,
\bar{c}\bar{d}\bar{s}$ which is mediated by virtual squarks instead of
the usual cascade decays to charginos and neutralinos. Furthermore,
charginos and neutralinos might also dominantly decay via these
couplings. The only limit on this coupling that we are aware of in the
literature, $\lambda''_{cds} \leq 1.25$, comes from the
requirement~\cite{herbi} that the coupling satisfy the unitarity bound
up to the GUT scale, and is not an experimental limit. Thus, logically
speaking, the coupling may be even larger than this without any
violation of experimental constraints (though we may lose some of the
original motivation for weak scale SUSY if this is the case --- but this
is a headache for theorists only). In such a scenario, sparticles would
rapidly decay mostly into jets, and sparticle production would manifest
itself by an increase in the multi-jet cross section which has enormous
QCD backgrounds at hadron colliders. No one has analysed whether it
would be possible to extract this signal from the background. While this
may be possible --- {\it e.g.} by searching for $c$-tagged
hemispherically separated
events with the same invariant mass in the two hemispheres, or by
dividing the multijets into two parts by minimizing the mass difference
between them --- it is clear that it would be a formidable task to do so.
Such a scenario would obviously be much simpler to look for at $e^+e^-$
colliders where multijet cross sections from Standard Model sources are
much smaller.

\subsubsection{Non Universal Gaugino Masses}

While the scenario of the previous
paragraph modified the mSUGRA model by the introduction of one large
$R$-parity violating coupling (this would alter masses as well as
sparticle decay patterns from mSUGRA expectation, as well as allow for
single squark production), we may imagine that the assumption of a
universal gaugino mass (which leads to $m_{\tg} \sim 3m_{{\tilde W}} \sim
6m_{\tilde B}$) is altered. Indeed, it had been pointed out a long time
ago~\cite{ellis} that if there are non-canonical gauge kinetic terms (as
may well be the case in non-renormalizable theories such as
supergravity) and the gauge kinetic function develops a VEV which
spontaneously breaks the GUT gauge group, gaugino masses need not be
universal. 

The resulting phenomenology was reexamined at the last Snowmass summer
study~\cite{snow} for the SUSY $SU(5)$ GUT model. The VEV then has to
transform as the symmetric product of two $SU(5)$ adjoint
representations, {\it i.e.} according to the {\bf 1, 24, 75} or {\bf
200} dimensional representation of $SU(5)$. The singlet leads to a
universal gaugino mass, while in the other cases we obtain a different
pattern of gaugino masses. These GUT scale gaugino masses have then to
be evolved to the weak scale relevant for phenomenology, and then to be
substituted into the MSSM gaugino-higgsino mass matrices that determine
the chargino and neutralino eigenstates. It turns out that for the {\bf
75} case, the chargino and the lightest neutralino are degenerate to
within a GeV, with the second lightest neutralino just a few GeV
heavier. There is also considerable degeneracy for the {\bf 200}
case. As a result of this degeneracy, leptons from $\tw_1$ and $\tz_2$
decays are very soft and multilepton signals in such a scenario are
essentially unobservable~\cite{abct} at the Tevatron and the $\eslt$
channel offers the sole hope for SUSY discovery at the Tevatron. If we
now allow a small $R$-parity violating coupling that causes the LSP to
decay hadronically as we discussed above, it may well be impossible to
discover SUSY via searches at the Tevatron

Another variant which results in very soft leptons from chargino and
neutralino decays is the so-called O-II orbifold model~\cite{brignole},
also examined in Ref.~\cite{snow}. For a value of the model parameter
$\delta_{GS} \sim -4$, $m_{\tg} \sim m_{\tw_1} \sim m_{\tz_1}$ with
$\tz_2$ somewhat heavier, so that leptons from gluino cascade decays to
charginos are difficult to detect at hadron colliders.  The clean
trilepton signal from $\tw_1\tz_2$ production is also suppressed. There
is a significant SM background for jetty or clean opposite sign dilepton events
to contend with. This scenario has been analysed in
Ref.~\cite{chen}. Combining this with the $R$-parity violating hadronic
decay of the LSP may
again make SUSY events hard to detect at hadron
colliders.~\footnote{This has also been mentioned by J.~Gunion,
hep-ph/9810394 (1998).}

A different possibility is to imagine that coloured sparticles
accessible at the Tevatron are lighter than their colourless cousins. In
this case, their leptonic cascade decays are kinematically
suppressed. This possibility is realized in the medium light gluino
model of Raby and collaborators~\cite{raby} or in the O-II
model~\cite{brignole} 
with $\delta_{GS} \sim -3$. The gluino is the stable LSP. It is claimed
that allowing for non-perturbative contributions to the annihilation
cross section, the relic abundance can be small enough so as not to
violate the cosmological bounds. To understand how the gluino manifests
itself in the experimental apparatus, we have to understand how a
gluino-hadron interacts with matter. This has been studied in
detail by the authors of
Ref.\cite{BCG}, who examined several models for this interaction. These
models differ in the fraction of the gluino kinetic energy that is deposited
in the experimental
apparatus and the fraction that escapes (the mass always escapes,
of course), leading to $\eslt$. They conclude that a gluino with a mass
between $\sim 25-140$~GeV should have been detected at the Tevatron in
the $\eslt$ channel. If, however, the gluino rapidly decays to hadrons
via an $R$-violating coupling, detecting SUSY signals might prove very
difficult. 

In closing, it is instructive to note that in all the scenarios where
SUSY might remain hidden from Tevatron searches even when sparticles are
relatively light, we always appear to need two separate deviations from
the canonical mSUGRA framework --- one to suppress the $\eslt$ signal,
and a logically different mechanism to reduce cross sections for
isolated multilepton events.

%%%%%%%%%%%%%%%%%%%%% REFERENCES %%%%%%%%%%%%%%%%%%%%%%%%%%%%%%%%%%%%%%%%%%%%%%
%

%%%%%%%%%%%%%%%%%%%%%%%%%%%%%%%%%%%%%%%%%%%%%%%%%%%%%%%%%%%%%%%%%%%%%

%\end{document}

%% file: Tata-summary/summary.tex
\def\tG{\tilde G}
\def\ETC{E_T^c}
\def\eslt{\rlap/E_T}
\def\to{\rightarrow}
\def\te{\tilde e}
\def\tf{\tilde f}
\def \tlam{\tilde{\lambda}}
\def\tl{\tilde l}
\def\tb{\tilde b}
\def\tst{\tilde t}
\def\tt{\tilde t}
\def\ttau{\tilde \tau}
\def\tmu{\tilde \mu}
\def\tg{\tilde g}
\def\tga{\tilde \gamma}
\def\tnu{\tilde\nu}
\def\tell{\tilde\ell}
\def\tq{\tilde q}
\def\tw{\tilde{\chi}^{\pm}}
\def\tww{\tilde{\chi}_1^{\pm}\chi_1^{\mp}}
\def\tz{\tilde{\chi}^0}
\def\Rsl{\rlap /R}

\section{Summary: Sparticle Detection at Run II and Beyond}

Within the framework of the mSUGRA model (or any model where gaugino
masses are unified at some high scale), gluinos are much heavier than
charginos and neutralinos. Furthermore, renormalization effects tend to
make squarks even heavier. Thus for large enough gluino masses,
electroweak production of charginos and neutralinos becomes the most
important sparticle production mechanism at a high luminosity hadron
collider. Indeed we see from Figs.~\ref{FIG1} and \ref{FIG2} that $\tww$
and $\tw_1\tz_2$ production are the SUSY processes with the largest
cross sections at the Tevatron. QCD radiative corrections increase the
cross sections shown here by 10--35\%. $\tw_1\tz_2$ production followed
by their leptonic decays results in the clean trilepton signature for
SUSY. This signal~\cite{summ-trilep}, together with the $ jets + \eslt$
(possibly with leptons) signal, had been exhaustively examined even
before the Workshop, and together, these have generally been viewed to
be the main channels for SUSY search at the Tevatron. Just before the
start of the Workshop, it was pointed out~\cite{summ-bcdpt} that for
large values of $\tan\beta$, charginos and neutralinos preferentially
decay to third generation fermions (mostly taus), and possibly also
sfermions, as shown in Fig.~\ref{FIG4}. Thus cross sections for
multilepton signals, including the much touted trilepton signal, could
be much reduced relative to their expectation for low $\tan\beta$, and
the reach of the Tevatron correspondingly diminished. The pre-Workshop
projection of the Tevatron reach in the $m_0-m_{1/2}$ plane is
summarized in Fig.~\ref{HT-total}. We see from this that for $\tan\beta
\geq 35$, there is essentially no reach at the Main Injector beyond that
of LEP II. The reach of experiments at the Run~IIb upgrade is
significantly diminished relative to that for low
$\tan\beta$. Furthermore, because cross sections for multi $e$ and $\mu$
signals are reduced, $b$-jet and $\tau$-lepton tagging are necessary to
establish the reach in Fig.~\ref{HT-total} when $\tan\beta$ is large.

An important program for our Group revolved around efforts to identify
new signatures that would allow sparticle detection even for large
values of $\tan\beta$. The Wisconsin Group~\cite{wisc,wisc2} first
pointed out that by softening the cuts on the leptons, it may be
possible to detect the trilepton signal from $\tw_1\tz_2$ production
even if these decayed to taus which subsequently decay to $e$ or
$\mu$. Backgrounds to the soft lepton signals were carefully
reassessed. It was found \cite{wisc2,lm,ellis-summ} that $W^*Z^*$ and
$W^*\gamma^*$ production (where $\gamma^*$ is virtual and the $W$ and
$Z$ may be real or virtual) gave rise to trilepton events with a cross
section of ${\cal{O}}(10)$~fb even within the Standard Model.  While
this level is negligible for Run~I, new strategies \cite{wisc2,quint,mp}
had to be identified in order to maximize the reach for integrated
luminosities envisioned for Run~II and beyond, where signals at the fb
level are potentially observable.

The discovery contours for integrated luminosities between
2--30~$fb^{-1}$ are summarized in Figs.~\ref{fig:contour2},
\ref{fig:contour10} and \ref{fig:contour3}. We see that for low
values of $\tan\beta$, experiments at Run~II may probe $m_{1/2}$
values~\footnote{The soft SUSY breaking $SU(2)$ gaugino mass is $\sim
0.8m_{1/2}$ and, as long as $|\mu|$ is not small, $m_{\tw_1}
\sim(0.7-0.8)m_{1/2}$.} beyond 250~GeV at the 5$\sigma$ level if other
parameters are in a favourable region, while at Run~III this reach may
exceed 275~GeV (corresponding to a gluino of almost 700~GeV). The
discovery potential is sensitive to $\tan\beta$, but even for
$\tan\beta=35$ experiments at Run~II will probe beyond the reach of LEP
2, whereas with an integrated luminosity of 30~fb$^{-1}$ SUSY discovery
for $m_{1/2}$ values up to 180~GeV may be possible. There remain,
however, other ranges of parameters where the signal remains
undetectable in this channel even if charginos are just beyond current
experimental limits, so that sparticles may evade detection even if they
are relatively light.

Another interesting strategy for sparticle detection at large
$\tan\beta$ explored~\cite{lm} at the Workshop entails direct detection
of taus via their hadronic decays. This is especially interesting, as
observation of an excess of $\tau$ leptons in SUSY events over
corresponding $e$ and $\mu$ signals, would suggest Yukawa interaction
effects, and may thus serve to indicate that $\tan\beta$ is large. A
particularly challenging scenario with $2m_{\tz_1} \sim (4/3)m_{\ttau_1}
\sim m_{\tw_1}$ (with other sparticles heavy), so that $\tw_1$ and
$\tz_2$ almost exclusively decay via $\tw_1 \to \ttau_1\nu$ and$\tz_2
\to \ttau_1\tau$, respectively, was examined using TAUOLA and PYTHIA
interfaced with the SHW detector simulation. The signatures consist of
events with `tau jets' and/or soft leptons from secondary decays of
$\tau$s. It was confirmed that the usual trilepton signal is
unobservable (at the $3\sigma$ level) even at Run~IIb unless charginos are
lighter than $\sim$110~GeV, so that SUSY has to be searched for via
channels with identified $\tau$s. The misidentification of QCD jets as
taus is then an important (detector-dependent) background.  Nonetheless,
it was shown that SUSY signals in $\ell\ell\tau_h$ and
$\ell^{\pm}\ell^{\pm}\tau_h$ channels (here, $\ell =e$ or $\mu$, and
$\tau_h$ denotes a tau tagged via its hadronic decay) would be
observable (Fig.~\ref{sigeff}) at the $3\sigma$ level for integrated
luminosities of a few to $\sim$30~fb$^{-1}$, for a chargino mass up to
140~GeV.  The same-sign dilepton plus tau channel has the better signal
to background ratio, but suffers from low rates. The development of new
tau triggers was not crucial to this analysis. The observability of the
`tau jets' is helped by the fact that $\ttau_1$ is dominantly $\ttau_R$
(in many models, as well as in this analysis) since the polarization of
the daughter taus then results in the hadronic decay products being
preferentially emitted along the tau direction, and so leads to harder
`jets' which are, of course, easier to detect: $\tau_h$ signals in
(fortunately, unconventional) models where $\ttau_1 \sim \ttau_L$ would
be more difficult to identify.

Large Yukawa couplings of the top quark, and if $\tan\beta$ is large,
also of the bottom quark, make the corresponding squark lighter than
other squarks. The reach of the Tevatron for stops~\cite{regina} and
sbottoms~\cite{sbottoms,regina} is sensitive to how these decay and is
summarized in Figs.~\ref{sen:st}--\ref{re:sb2}. At Run IIa (Run~IIb), experiments should be sensitive to stops as heavy as 180--200~GeV (250~GeV) if $\tt_1 \to
b\tw_1$ or $\tt_1 \to b\ell\tnu$, and is $\sim 40$~GeV smaller if $\tt_1
\to c\tz_1$. Experiments at Run~IIa (Run~IIb) should be sensitive to
$b$-squarks heavier than 200~GeV (240~GeV) if they decay via $\tb_1 \to
b\tz_1$. The degradation of the reach is expected to be smaller than
30--40~GeV even if $\tb_1$ decays via modes which make the signal more
difficult to detect~\cite{sbottoms}.

While most of the focus of our Group was on Tevatron signals within the
mSUGRA framework, we recognize that our conclusions about the Tevatron
reach are sensitive to untested underlying assumptions about the
symmetries of physics at higher energies. It could, however, be that these
assumptions turn out to be incorrect, and the lightest neutralino decays
into photons (as in some gauge-mediated SUSY breaking
models~\cite{summ-gmsb}), or into leptons (as in some $R$-parity violating
models~\cite{rpv}), which then provide additional handles to beat down
Standard Model backgrounds, and hence, enhancing the reach of the
Tevatron. On the down side, it is also possible that the lightest
neutralino decays hadronically, and for one reason or another, leptons
from cascade decays of sparticles are either soft or even entirely
absent. In this case, as discussed in Sec.~\ref{hidden-susy}, SUSY may well remain hidden at the Tevatron even if sparticles are light.

To sum up, within the mSUGRA framework, the reach of luminosity upgrades
of the Tevatron is mainly determined by signals from $\tww$ and
$\tw_1\tz_2$ production. For favorable values of parameters,
experiments in Run~II should be able to probe $m_{1/2}$ values as high
as 250~GeV corresponding to a gluino mass beyond 600~GeV. At Run~IIb, it
may be possible to probe $m_{1/2}$ up to $\sim$280~GeV. Even in the
problematic large $\tan\beta$ region, it may be possible to access
$m_{1/2}$ as large as 140 (180)~GeV at Run~IIa (Run~IIb) via the trilepton
channel with soft cuts on the leptons. It may be possible to further
extend this reach via novel signatures involving $\tau$ leptons (and,
perhaps, also $b$-jets), but work on this is in its infancy. The most
promising signals in these channels appear to be rate-limited, and call
for accumulating the highest possible integrated luminosity. It may also
be possible to discover third generation squarks
if these are relatively light.  Thus while it is possible that
experiments at the Tevatron may indeed discover supersymmetry, it should
be remembered that there are other ranges of parameters where $\tz_2$
decays to leptons (including taus) are strongly suppressed. For these
unfavorable regions of parameter space, it may prove very difficult to
find supersymmetry at luminosity upgrades of the Tevatron.

The Fermilab Tevatron will be the energy frontier until LHC operation is
well under way. While supercolliders that directly probe the TeV scale
appear essential for decisively probing weak scale supersymmetry,
experiments at luminosity upgrades of the Tevatron (starting with
Run~II) provide an opportunity for a discovery that could result in a
paradigm shift in our thinking about the physics of elementary particles.
Considering the importance of the physics that might be accessible in
these experiments, we urge that these be continued and the Tevatron
be operated  to accumulate as much
integrated luminosity as possible.

%% file: sugra.bbl
\begin{thebibliography}{99}
\bibitem{FN2} In the extreme case where the branching fraction for
  $\tilde{b}_1 \to b\tilde{\chi}^0_2$ is 100\% when this decay is allowed, and
  if $\tilde{\chi}^0_2$ only decays invisibly, the reach can once again be
  read off from Fig.\ \ref{re:sb}.
\bibitem{FN3} We expect that this leads to a lower $\etm$ than in
the case when it decays invisibly.
\end{thebibliography}
